\documentclass[a4paper,plainchapterheads,yschapters,twoside,truedoublelespace,openright]{iitbthesis}
\def\units#1{\hbox{ $\,{\rm #1}$}}

\usepackage{epsfig}
\usepackage{times}
\usepackage[T1]{fontenc}
\usepackage[utf8]{inputenc}
\usepackage{mathptmx}
\usepackage{amssymb}
\usepackage{amsmath,epsfig}
\usepackage{graphicx,graphics}
\usepackage{url}
\usepackage{hyperref}
\usepackage[capitalise]{cleveref}
\usepackage{textcomp}
\usepackage{natbib}
\usepackage{pdflscape}
\renewcommand\bibname{References}
\usepackage{titlesec}
\usepackage{bm,bbm}
\usepackage{float,lscape}
\usepackage{subfig}

\usepackage{epstopdf}
\usepackage{tabu}
\usepackage{multirow}
\usepackage{array}
\usepackage{booktabs}
\usepackage{graphicx}
\usepackage{caption}
\usepackage{lettrine}
\usepackage{lscape}
\usepackage{rotating}
\usepackage{booktabs}
\usepackage{longtable}
\usepackage{mathptmx}
\usepackage{anyfontsize}
\usepackage{t1enc}
\usepackage{relsize}
\usepackage{placeins}
\usepackage{nomencl}
\usepackage{rotating}
\usepackage{glossaries}
\usepackage{scrlayer}
\usepackage{enumitem}
\usepackage{xcolor}
\usepackage{ulem}
\usepackage{cancel}
\usepackage{mathrsfs}

\usepackage[titletoc]{appendix}
\usepackage[nottoc]{tocbibind}

\usepackage{multirow}

\begin{document}
%=====================================================================
% Include the prelude for Title page, abstract, table of contents, etc
% You need to modify it to contain your details
\abovedisplayskip=8mm
\abovedisplayshortskip=8mm
\belowdisplayskip=8mm
\belowdisplayshortskip=8mm

% prelude.tex
%   - titlepage
%   - dedication (optional)
%   - approval sheet
%   - course certificate
%   - table of contents, list of tables and list of figures
%   - nomenclature
%   - abstract
%============================================================================

%\clearpage\pagenumbering{roman}  % This makes the page numbers Roman (i, ii, etc)

\pagenumbering{gobble}

% TITLE PAGE
%   - define \title{} \author{} \date{}
\title{\Huge{Cosmic-ray propagation and production of secondary particles in the Galaxy}}
\author{Pedro de la Torre Luque}
%\date{XXXIII ciclo, 2017-2020}
\date{Dottorato di ricerca in fisica, XXXIII ciclo, 2017-2020}

%  - Roll number, required for title page, approval sheet, and
%    certificate of course work 
%\rollnum{} 

%   - The default degree is ``Doctor of Philosophy''
%     (unless the document style msthesis is specified
%      and then the default degree is ``Master of Science'')
%     Degree can be changed using the command \iitbdegree{}
\iitbdegree{Doctor Philosophiae}

%   - The default report type is preliminary report.
%      * for a PhD thesis, specify \thesis
%\thesis
%      * for a M.Tech./M.Phil./M.Des./M.S. dissertation, specify \dissertation
%\dissertation
%      * for a DIIT/B.Tech./M.Sc.project report, specify \project
%\project
%      * for any other type, use  \reporttype{}
\reporttype{}

%   - The default department is ``Unknown Department''
%     The department can be changed using the command \department{}
\department{Dipartimento Interateneo di Fisica "M. Merlin"}
%#Universitá degli Studi di Bari Aldo Moro

%    - Set the guide's name
\setguide{Prof. Mazziotta, Mario Nicola \hspace{3.3 cm} Prof. Iaselli, Giuseppe \\ Prof. Gargano, Fabio \hspace{10cm} \  } 
%    - Set the coguide's name (if you have one)
%\setcoguide{Prof. Gulab Singh}
%    - Set external guide (if you have one)
%\setexguide{Prof External Guide}

%   - once the above are defined, use \maketitle to generate the titlepage
\maketitle 

\newpage\null\thispagestyle{empty}\newpage % % For blank pages

%--------------------------------------------------------------------%
% DEDICATION
%   Dedications, if any, must be first page after title page.
%\begin{dedication}
%\large{Dedicated to my beloved parents.}
%\end{dedication}

%\newpage\null\thispagestyle{empty}\newpage % % For blank pages

%--------------------------------------------------------------------%
% APPROVAL SHEET
%   - for final thesis, you need Approval Sheet. So, uncomment the
%     \makeapproval command.

%\input{approval.tex}

%\newpage\null\thispagestyle{empty}\newpage % % For blank pages

%     it should come after dedication, if dedication is
%     present. Otherwise it is the first page after title page.
%\makeapproval
%\input{Acceptancecertificate}
%\addcontentsline{toc}{chapter}{Acceptance Certificate}
\newpage
\textbf{\large Posterior note.}
\newline

This thesis was presented in Bari, on 29/03/2021, and the work behind it was awarded with the Cum Laude credit. 

The main chapters have been published, mostly in JCAP, and correspond to:
\begin{itemize}
    \item Chapter 2: ``Implications of current nuclear cross sections on secondary cosmic rays with the upcoming DRAGON2 code'' JCAP03 (2021) 099 
    
    \item Chapter 3: ``FLUKA cross sections for cosmic-ray interactions with the DRAGON2 code'' (2022) ArXiv:2202.03559
    
    \item Chapter 4: ``Markov chain Monte Carlo analyses of the flux ratios of B, Be and Li with the DRAGON2 code'' JCAP07 (2021) 010
    
    \item Chapter 5: ``Combined analyses of the antiproton production from cosmic-ray interactions and its possible dark matter origin'' JCAP11 (2021) 018
\end{itemize}

\newpage

\clearpage
\thispagestyle{empty}

\begin{center}
\Large  {\bf Declaration }
\end{center}
\vspace{-6in}
I declare that this written submission represents my ideas in my own words and where others ideas or words have been included, I have adequately cited and referenced the original sources. I also declare that I have adhered to all principles of academic honesty and integrity and have not misrepresented or fabricated or falsified any idea/data/fact/source in my submission. I understand that any violation of the above will be cause for disciplinary action by the Institute and can also evoke penal action from the sources which have thus not been properly cited or from whom proper permission has not been taken when needed.
\vspace{0.5in}

%\begin{flushleft}
%Date:
%\end{flushleft}
%\begin{flushright}
%Surendar M\\
%(Roll. No. 114310007)
%\end{flushright}

%
 \begin{table}[h]
 \begin{flushleft}

\vspace{-3.2in} 
 \begin{tabular}{ccccc}
 % %\hline 	\rule[5ex]{0pt}{-10ex} &&(Signature) (Date) && \\ 
 \rule[5ex]{0pt}{-10ex}&& Date: && \\ 
 \end{tabular}
\end{flushleft}

\vspace{-0.5in} 
\begin{flushright}
 \begin{tabular}{ccccc}
 % %\hline 	\rule[5ex]{0pt}{-10ex} &&(Signature) (Date) && \\ 
 
 \hline 	\rule[5ex]{0pt}{-10ex}&& Pedro de la Torre Luque && \\ \\ 
 %\rule[5ex]{0pt}{-10ex}&& Roll No. 19XXXX&& \\ \\
 \end{tabular}
\end{flushright}
\end{table}

\pagebreak
%\addcontentsline{toc}{chapter}{Declaration}

%--------------------------------------------------------------------%
% COPYRIGHT PAGE
%   - To include a copyright page use \copyrightpage
%\copyrightpage
%\newpage\null\thispagestyle{empty}\newpage % % For blank pages
%--------------------------------------------------------------------%
% ABSTRACT
\clearpage
\pagenumbering{roman}
\begin{abstract}
  \renewcommand{\thepage}{\roman{page}} \setcounter{page}{1}

Cosmic rays are nowadays a crucial tool to study the astrophysics of extreme objects in the Universe, the cosmic environmental plasma (both Galactic and extra-galactic), the physics of hadronic interactions or the properties of elementary particles at very high energies and even cosmological problems such as the dark matter puzzle. 

In this thesis, the phenomenology on the transport of Galactic cosmic rays is studied in light of the most recent experimental data in the field and new analyses are presented in order to obtain better constraints. Throughout the thesis, secondary particles produced from collisions of cosmic rays with the interstellar gas, such as secondary cosmic ray nuclei (B, Be and Li), antiprotons and gamma rays, are treated in order to adjust and test our models and probe different scenarios, such as possible signatures of dark matter decay or annihilation. A preliminary version of the upcoming DRAGON2 code has been used to perform the propagation computations.

New technological and theoretical advances have made possible precision studies that must be carefully performed. The cross sections of secondary cosmic-ray production (spallation cross sections) are currently the main concern in the community, since the lack of experimental data and the amount of interaction channels involved in the generation of each isotope make difficult to precisely determine the fluxes of these cosmic-ray species. In this dissertation, we have deeply studied the most recent cross sections parametrisations (namely GALPROP and DRAGON2 cross sections) as well as developed a new set of cross sections from the FLUKA Monte Carlo nuclear code, whose computations do not depend on existing data and can be, thus, fully extended to different species and energies.

We have applied a new way to evaluate the impact of the cross sections uncertainties on the flux of the secondary cosmic rays B, Be and Li independently of the chosen diffusion coefficient and proposed different ways to refine these cross sections and their inclusion in cosmic ray propagation analyses. A Markov-chain Monte Carlo analysis has been implemented, intended to reproduce experimental flux ratios between Li, Be and B and their ratios to the primary cosmic-ray nuclei C and O, taking into consideration the uncertainties related to cross sections when combining these ratios. This is the first time this kind of analysis has been successfully implemented with a numerical code dedicated to cosmic rays. Furthermore, the analyses are performed including the very recent AMS-02 data for Ne, Mg and Si, which are important to evaluate the fluxes of these secondary cosmic rays.

In addition to develop new analyses to fully benefit from recent cosmic ray data, the theoretical frame of their diffusion has been discussed in order to choose suitable parametrisations for the diffusion coefficient.

%An extensive comparison of the predictions obtained for B, Be and Li fluxes with the most updated spallation cross sections sets is presented in order to estimate the propagation parameters that best explain experimental data, such as the halo size or the alfven velocity, resulting in a set of diffusion models which are then evaluated by the study of other secondary emissions, namely antiprotons and gamma rays.

Then, the derived diffusion models are used to infer the antiproton spectrum and study them in light of an excess of data over predictions, peaking around $10 \units{GeV}$, that has been recently reported and associated to a possible signature of a dark matter particle with mass around $30$-$90 \units{GeV}$. This feature is found for all the different predicted diffusion models although the full uncertainties involved are calculated to be of the order of 30\% at $10 \units{GeV}$. In this calculation, the uncertainties coming from the cross sections of secondary cosmic ray production are taken into consideration for first time.

Finally, the Galactic gamma-ray emission is also investigated for these diffusion models. The gamma-ray intensity are computed with the GammaSky code implementing the gamma-ray production cross sections from the FLUKA code. We show the calculated gamma-ray sky maps and compare the different diffusion models for the average gamma-ray intensity in the central region of the Galaxy and the outer sky region. This comparison is also performed for the local gamma-ray emissivity since this magnitude is exempted from the uncertainties derived associated to the gas distribution in the sky. In addition, the gamma-ray lines expected to be produced at very low energy ($\units{MeV}$) are also shown to underline the advantage of using the FLUKA nuclear code in order to calculate cross sections.

\end{abstract}
\chapter*{Sintesi della tesi}
\label{ch:abs_it}
\addcontentsline{toc}{chapter}{\nameref{ch:abs_it}}

I raggi cosmici sono oggigiorno uno strumento cruciale per studiare l'astrofisica di oggetti estremi nell'Universo, il plasma ambientale cosmico (sia galattico che extra-galattico), la fisica delle interazioni adroniche o le proprietà delle particelle elementari ad energie molto elevate e persino problemi cosmologici come il puzzle della materia oscura.

In questa tesi si è studiata la fenomenologia del trasporto dei raggi cosmici galattici alla luce dei più recenti dati sperimentali e viene presentata una nuova analisi che ha permesso di migliorare i limiti sui parametri utilizzati nei modelli. Nella tesi vengono studiate le particelle  secondarie prodotti nelle collisioni dei raggi cosmici con il gas interstellare, tra cui i nuclei dei raggi cosmici secondari (B, Be e Li), gli antiprotoni e raggi gamma, al fine di ottimizzare i modelli di propagazione e testare diversi scenari che prevedono possibili segnali di decadimento o annichilazione di materia oscura. Per simulare la  propagazione si è utilizzata una versione preliminare del codice DRAGON2. 

I recenti sviluppi sperimentali hanno permesso misure di elevata precisione, che richiedono un'adeguata interpretazione. Le sezioni d'urto che descrivono la produzione secondaria di raggi cosmici (sezioni d'urto di spallazione) rappresentano attualmente il problema principale  per la comunità scientifica che lavora sui raggi cosmici, poiché la mancanza di dati sperimentali e la quantità di canali di interazione coinvolti nella generazione di ciascun isotopo rendono difficile determinare con precisione i flussi di queste specie di raggi cosmici. In questa tesi si sono studiate a fondo le più recenti parametrizzazioni delle sezioni d'urto basate su dati sperimentali (GALPROP e DRAGON2) e si è sviluppato un nuovo set di sezioni d'urto utilizzando il codice FLUKA. In questo caso si è ottenuto un set consistente e completo di sezioni d'urto su un ampio intervallo di energie. 

Nel lavoro di tesi si è valutato l'effetto delle incertezze sulle sezioni d'urto sul flusso dei raggi cosmici secondari B, Be e Li indipendentemente dai parametri di diffusione scelti. Si sono quindi proposti vari modi per ottimizzare le sezioni d'urto e includerle nelle simulazioni sulla propagazione dei raggi cosmici. È stata implementata un'analisi di tipo "Monte Carlo Markov chain", intesa a riprodurre sia i rapporti tra i flussi misurati di Li, Be e B che quelli con i nuclei primari di C e O, tenendo conto delle incertezze relative alle sezioni d'urto. Questo tipo di analisi è stato implementato per la prima volta con successo in un codice numerico dedicato ai raggi cosmici. Le analisi sono state inoltre eseguite includendo i recenti dati di AMS-02 per Ne, Mg e Si, che sono importanti per valutare i flussi di questi raggi cosmici secondari.

Oltre a sviluppare nuove analisi per sfruttare al massimo i recenti dati sui raggi cosmici, si è discusso il quadro teorico sul processo di diffusione nella galassia, in modo da scegliere opportune parametrizzazioni per il coefficiente di diffusione.

I modelli di diffusione derivati sono stati quindi utilizzati per dedurre lo spettro degli antiprotoni. Recentemente, alcuni autori hanno riportato un possibile eccesso del flusso di antiprotoni rispetto alle previsioni, nella regione intorno ai 10 GeV, che è stato associato a un possibile segnale di paricelle di materia oscura con massa intorno ai 30-90 GeV. Questo eccesso è previsto da diversi modelli di diffusione, sebbene le incertezze sui flussi calcolati siano di circa il 30\% a 10 GeV. In questo lavoro di tesi sono state prese in considerazione, per la prima volta, le incertezze derivanti dalle sezioni d'urto della produzione dei raggi cosmici secondari e si è dimostrato che l'eccesso intorno ai 10 GeV è compatibile con i flussi previsti senza introdurre sorgenti esotiche.

Infine, per questi modelli di diffusione, si è anche calcolata l'emissione di raggi gamma galattici. L'intensità dei raggi gamma viene calcolata con il codice GammaSky implementando le sezioni d'urto di produzione gamma ottenute da FLUKA. Nella tesi vengono calcolate le mappe celesti dei raggi gamma e sono confrontati i diversi modelli di diffusione per l'intensità media dei raggi gamma nel centro della galassia e nella regione esterna. Questo confronto viene eseguito anche per l'emissività locale dei raggi gamma, poiché questa grandezza è esente dalle incertezze associate alla distribuzione del gas nel cielo. Inoltre vengono mostrate le linee negli spettri dei raggi gamma che dovrebbero essere prodotte a energia molto bassa (MeV), che sono calcolate grazie alle sezioni d'urto ottenute da FLUKA.

%--------------------------------------------------------------------%
% CONTENTS, TABLES, FIGURES
\tableofcontents
%\listoftables
%\listoffigures

% Here is the file for Abbrivations
%\input{abbreviations}

% To automate abbriavtions using Nomencluture  package.
% Comment the  \input{abbreviations}
% Then include NOMEN...... package 
% Refer for this q31_nom_tex 

%\addcontentsline{toc}{chapter}{Abbreviations and Nomenclature}
%--------------------------------------------------------------------%
% NOMENCLATURE
%\begin{nomenclature}
%\begin{description}
%\item{\makebox[0.75in][l]{$C_1$}} Constant 1
%
%\item{\makebox[0.75in][l]{$V$}}    Voltage 
%
%\item{\makebox[0.75in][l]{\$}}     US Dollars
%\end{description}
%\end{nomenclature}

%\cleardoublepage\pagenumbering{arabic} % Make the page numbers Arabic (1, 2, etc)

\chapter*{List of Abbreviations}
\label{ch:ListOfAbr}
\addcontentsline{toc}{chapter}{\nameref{ch:ListOfAbr}}

\makeatletter
\newcommand{\tocfill}{\cleaders\hbox{}\hfill}
\makeatother
\newcommand{\abbrlabel}[1]{\makebox[3cm][l]{\textbf{#1}\ \tocfill}}
\newenvironment{abbreviations}{\begin{list}{}{\renewcommand{\makelabel}{\abbrlabel}
\setlength{\itemsep}{0pt}}}{\end{list}}
%\begin{abbreviations}[labelsep=1em,font=\bfseries]
\begin{abbreviations}
\item[CMB]	Cosmic Microwave Background
\item[CRE] Cosmic Ray Electrons
\item[DM]	Dark Matter 
\item[EBL] Extra-galactic Background Light
%\item[DSA] Diffusive Shock Acceleration
\item[GCR]	Galactic Cosmic Rays
\item[GMF] Galactic Magnetic Field
\item[ISM] Interstellar Medium
\item[ISRF] Interstellar Radiation Field
\item[LIS]	Local Intersterllar Spectrum
\item[LISM]	Local Intersterllar Medium
%\item[AMS]  Alpha Magnetic Spectrometer Experiment
%\item[ACE]	Advanced Composition Explorer
%\item[GLAST] Gamma-ray Large Area Telescope
\item[MC] Monter Carlo
\item[MCMC] Markov-Chain Monter Carlo
\item[MHD]	Magnetohydrodynamics 
\item[PDF]	Probability Distribution Function 
\item[PWN]	Pulsar Wind Nebula 
\item[SN] 	Supernovae
\item[SNR] 	Supernova Remnant
\item[UHECR] Ultra-High energy cosmic rays
\item[XSec] Cross section

%CGRO: Compton GammaRay Obser-vatory, EGRET: Energetic gamma-ray telescope, GALPROP:Galactic Propagation code, H.E.S.S.:High Energy Stereoscopic System, :  field, PLD: path-length distribution; Inverse Compton scattering
\end{abbreviations}

\chapter*{Glossary of keywords}
\label{ch:Gloss}
\addcontentsline{toc}{chapter}{Glossary of Keywords} 

\textbf{\hspace{0.82cm}Primary cosmic ray:}	Those cosmic rays whose flux is mostly the one injected by the sources. For example, oxygen, carbon, neon.

\textbf{Injection spectrum:} Energy spectrum at which a cosmic ray element exits from the acceleration source and is injected to the interstellar medium.

\textbf{Secondary cosmic ray:} Those cosmic rays which principally are formed via spallation reactions. In general, a secondary particle is created from the collision of a primary particle and a pure secondary nuclei is not injected in the sources. Typical examples are B, Be and Li.

\textbf{Spallation reaction:} Hadronic, inelastic reactions between a projectile CR and nuclei from the ISM (as target) which generate and eject at least a product nucleus. Typically spallation interactions are associated to production of new particles (also called inclusive cross sections) and inelastic cross sections to the destruction of particles.

%\textbf{Inelastic reaction:}

\textbf{Grammage:} Amount of matter by cubic centimeter traversed by cosmic rays in their propagation path. Its dimensions are [$g/cm^2$].

\textbf{Ghost nuclei:} Those unstable nuclei, generated via spallation reactions, whose lifetime is negligible compared to typical propagation times, and their contribution is directly added up to the formation of their daughter nuclei.

\textbf{Direct spallation cross sections:} Spallation cross sections of prompt production of a cosmic ray nuclei.

\textbf{Cumulative spallation cross sections:} 
Spallation cross sections which add the contribution from the decay of a ghost nucleus up to the prompt production. While a direct reaction is $^{12}C + p \longrightarrow ^{11}C + X$, the cumulative reaction for $^{11}B$ production is $^{12}C + p \longrightarrow ^{11}C + X \hspace{0.5 cm} {\xrightarrow{\makebox[1.5cm]{fast decay}}} \hspace{0.4cm} ^{11}B$.

\textbf{Main spallation channels:} Those spallation reaction channels which are dominant for the production of a particular isotope of a secondary nuclei. Spallation reactions with $^{12}C$ and $^{16}O$ as projectiles, since taking just these channels account for more than 50\% of the total flux of B, Be and Li.

\textbf{Secondary spallation channels:} Those spallation reaction channels from primary cosmic ray isotopes other than $^{12}C$ and $^{16}O$, which account usually by less than $\sim5\%$ to the total production of a particular secondary particle.

\textbf{Tertiary spallation channels:} Those spallation reaction channels with a secondary cosmic ray isotope as projectile producing another cosmic ray isotope.

\textbf{Local cosmic ray spectrum of cosmic rays:} Spectrum of cosmic rays measured in the Earth's vicinity.

\chapter*{Motivation and goals}
\label{ch:sym}
\section*{Motivation}
\label{sec:goals}

This dissertation is aimed at the study of the transport of Galactic cosmic rays and the importance of their interactions with the gas in the interstellar medium, in order to obtain a complete and consistent picture that agrees with the most recent and precise observations. This leads to refinement of former models and the development of new ones, by means of new analysis techniques, which, besides to provide a better comprehension on their phenomenology, offer the possibility of using these predictions for probing the properties of the interstellar medium (combined to other messenger emissions) and test new physics hypothesis in the fields of particle physics, astrophysics and cosmology.

\section*{Thesis outline}
	The subject matter of the thesis is presented in the following chapters, 
\begin{enumerate}[label=\checkmark]
\item	Chapter 1 provides an overview of the main concepts that have derived into the current understanding of the phenomenology on Galactic cosmic rays. It also presents the main public codes for simulations on the topic as well as a short description of the current measurements and detectors.% In addition, the motivation of this research and objectives are emphasized at the end.
\item	Chapter 2 is aimed at analysing, using a preliminary version of the upcoming DRAGON2 code, the main parametrisations of spallation cross sections nowadays used and their implications in deriving diffusion parameters and in the determination of the halo size. Furthermore, possible primary origins of the secondary cosmic rays lithium, boron or beryllium are proven and a new strategy to get rid of systematic uncertainties from the spallation cross sections is proposed. 
\item	Chapter 3 describes the development of a new set of cross sections, derived from the Monte Carlo package FLUKA, with specific treatments for the interaction and transport of particles and nuclei in matter. These cross sections are compared to data and the other parametrisations analysed in chapter 2 and are implemented in the DRAGON code to analyse their compatibility to reproduce cosmic ray data.
\item In chapter 4 detailed Markov-chain Monte Carlo analyses of the optimal diffusion parameters derived from the secondary lithium, boron and beryllium and their flux ratios to the primary carbon and oxygen are performed comparing the results of different cross sections parametrisations and having into account the cross sections uncertainties. In addition to the independent analyses of the ratios and their combined analyses, a new strategy is followed to infer the diffusion parameters in a more reliable way. Here, the diffusion break hypothesis for the hardening at a few hundreds of $\units{GeV}$ in the cosmic-ray spectra is also compared to the source injection break hypothesis.
\item	Chapter 5 shows the consequences of the investigated diffusion models in leptons, gamma-rays and antiprotons spectra. The hypothesis of possible dark matter annihilation to explain the antiproton spectrum is revisited too. Moreover, the gamma-ray sky emission and the local emissivity spectrum is deeply studied.
\item In Chapter 6, the proposed methodologies and the discussions of the results, including the important findings along the performed studies are summarized. Future extensions of these research works are proposed successively, following the conclusion on the basis of important extracts and understanding of Galactic cosmic-ray physics.

\end{enumerate}

\setlength{\parskip}{2.5mm}
\titlespacing{\chapter}{0cm}{55mm}{10mm}
\titleformat{\chapter}[display]
  {\normalfont\huge\bfseries\centering}
  {\chaptertitlename\ \thechapter}{20pt}{\Huge}
  
  \titlespacing*{\section}
  {0pt}{8mm}{8mm}
  \titlespacing*{\subsection}
  {0pt}{8mm}{8mm}
\pagebreak
%\newpage  
%\cleardoublepage\pagenumbering{arabic}

%\include{Glosary}
\pagenumbering{arabic}

\makeatletter
\def\cleardoublepage{\clearpage\if@twoside \ifodd\c@page\else
	\hbox{}
	\vspace*{\fill}
	\begin{center}
		This page was intentionally left blank.
	\end{center}
	\vspace{\fill}
	\thispagestyle{empty}
	\newpage
	\if@twocolumn\hbox{}\newpage\fi\fi\fi}
\makeatother
%=====================================================================
% Include the technical part of the report
%%\include{chap_intro}             % Chapter 1: Introduction

\newpage
\pagebreak
\cleardoublepage
\chapter{Introduction: Galactic cosmic rays}
\label{sec:1}
\section{Background}
\label{sec:background}

%\lettrine[findent=2pt]{\textbf{W}}{}rite some intro here. How to cite a figure. 

Cosmic rays (CRs) are those particles arriving to Earth from the outer-space, which constitute a sea of energetic emissions intimately related to powerful single astrophysical events. They comprise a background of radiation permeating the Milky Way and extra-galactic space, which has kept roughly constant the levels of unstable isotopes formed in the atmosphere for at least the past 100000 years. Therefore, cosmic ray physics couples the phenomenology of macrophysics at galactic and extra-galactic scales with the complex microphysics of single particle interactions in diverse environments.

They mainly consist of nuclei spanning several orders of magnitude in energy, which, due to the interactions they undergo while travelling the space, reach homogeneously Earth. CRs are composed by protons, alpha particles ($\sim9\%$), heavier nuclei ($<1\%$), leptons ($<1\%$) and gamma rays or even antimatter particles.

\begin{figure}[!b]
	\centering
	\includegraphics[width=0.6\textwidth, height=0.31\textheight]{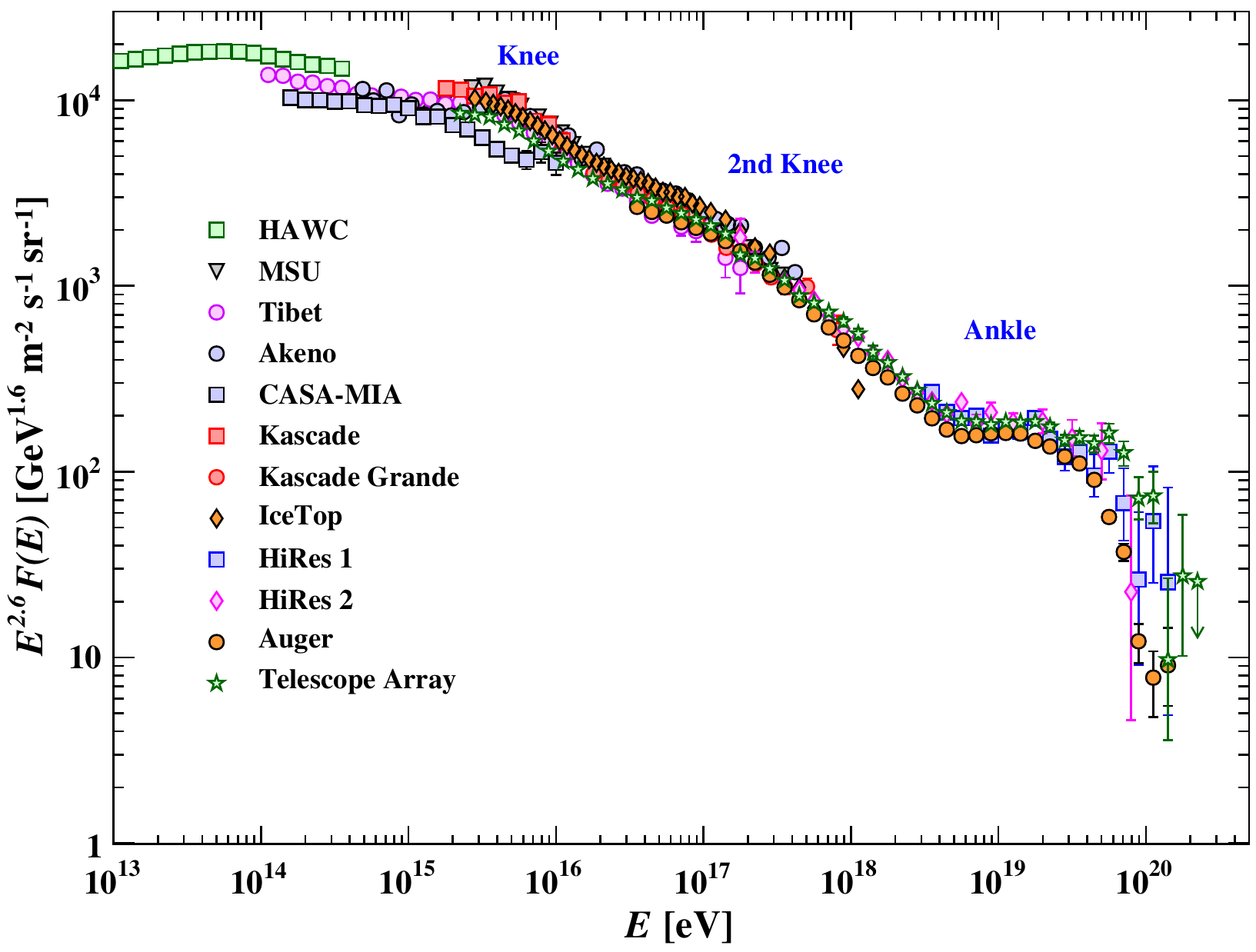}
	\hspace{1mm}
	\caption{\footnotesize All particle differential spectrum in energy, from~\cite{cosmic2002pdg}. It reflects there are two clear changes in the power-law trend, the knee and the ankle, supposed to be due to different origin (acceleration mechanism).} 
	\label{fig:Allparticles_spect}
\end{figure}

Figure~\ref{fig:Allparticles_spect} shows the differential flux (flux of particles reaching Earth per unit time, unit surface, unit energy interval, and unit solid angle) of all the nuclear species (i.e. the all-particle spectrum). This spectrum shows that the full energy range is close to be described by a single power law. There are two clear steepenings of the spectrum, labeled in the plot as ``knee'' (at energy $\sim3\times 10^{15} \units{eV}$) and ``ankle'' ($\sim2\times10^{18} \units{eV}$). CRs with energies lower than those displayed in Fig.~\ref{fig:Allparticles_spect} produced in the Sun can also reach the Earth. On the other hand, low-energy CRs of extra-solar origin are not allowed to reach the Earth due to the shielding of the solar magnetic field.

Measurements on these particles provide valuable information on their astrophysical origins and their propagation through interstellar and inter-galactic media. As we will see, the models we build for their origin and transport are based on the interpretation of several different pieces of observations within a unified frame \cite{BLASIRev}.

\subsection{Historical notes}
\label{Hist_notes}

Quoting the Slovene Pierre Auger Observatory website \cite{slovPAOwp}, "the history of cosmic ray research is a romantic story of scientific adventure. For three quarters of a century, cosmic ray researchers have climbed mountains, ridden hot-air balloons, and traveled to the far corners of the earth in the quest to understand these fast-moving particles from space".

The sudden discharge of electroscopes, now known to be due the ionization of the air induced by CRs, was observed even before the eighteenth century. One of the first reports about the mystery of CRs dates back in 1785, when Charles-Augustine de Coulomb found that the air is a weak conductor (chemically a good insulator) because the electroscopes spontaneously discharge by the action of the air. This remained an open question, attributing the effect to a bad insulation of the device, until Michael Faraday (1835) confirmed the same observation as Coulomb using better insulation technologies. 

In 1879, Crookes observes that the speed of discharge of an electroscope decreases when pressure is reduced, concluding that the agent responsible for the discharge is the ionized air. However, Wilson and Geitel, in 1900, used a container with metallic insulation and concluded (as many researches with similar experiments) that the air is ionized by an external agent. At this point, after the discovery of radioactivity in 1896, the cause of the air ionization was presumed to be radiation. 
In 1901, Nikola Tesla patented (US patent $685957/8$) an "Apparatus for the Utilization of Radiant Energy", assuming that radiations are originated from sources like the Sun (even though his purpose was not the study of CRs), but, later, Mache compared the variations of the radioactivity with diurnal variations finding no significant reduction.

Then, McLennan and Burton (1903) observed a $40\%$ reduction of the air ionization when shielding the container with a 120 cm wide barrier of water. In the same year, Rutherford and McLennan noticed that spontaneous signals appeared in their highly shielded detectors, which means that this radiation must be very energetic. Kurz and Cline, in 1909, pointed to the radiation from the Earth's crust as the responsible.

Some time before (1907$-$1908), Eve performed measurements over the Atlantic Ocean, observing the same radioactivity over the centre of the ocean as he had observed in England and Montreal. 
Nevertheless, in 1908, Elster and Geitel observed a drop of $28\%$ when the apparatus was taken from the surface down to the bottom of a salt mine. They concluded that, in agreement with the literature, the Earth is the source of the penetrating radiation and that certain waters, soils and salt deposits are comparatively free from radioactive substances, and can thus act as efficient screens. 

At the moment, contradictory results remarked that the electroscope needed more sensitivity and Theodor Wolf improved its accuracy. With this apparatus, he went to the Eiffel Tower, in 1910, expecting an exponential reduction of the radiation level. What he found is that the radiation intensity decreased very slightly, supporting the idea of the Earth as the origin of CRs.

Domenico Pacini performed careful studies of the radiation levels in the Bracciano lake and at the Livorno coast. He reported a small reduction of the radiation levels in the middle of the lake while an important difference appeared when being 3 m underwater, results that were published in the journal “Nuovo Cimento” $VI/3$, in 1912, with the conclusion that these results can not be explained by a radiation coming from Earth.

On the other hand, the expeditions on board balloons started around the 1909, with Bergwitz and Gockel, but the results were inconclusive until the masterpiece study performed by Victor Hess in 1911$-$1912 clarified the situation. He was awarded with the Nobel Prize for the discovery of CRs.
Posterior explorations on balloons, as those of Kolhörster, confirmed Hess's conclusions.

Meanwhile, the debate on the nature of CR particles was open. Millikan, in his studies around 1925, considered the hypothesis of neutral radiation from outer-space given its penetrating power and coined the term "Cosmic Rays", starting the debate about the nature of the radiation. Some scientists considered CRs as the residuals of the Big-Bang \cite{hillas1998cosmic}, born few years earlier. At the end, measurements carried out in 1927$-$28 showed geomagnetic effects on CRs intensity and let to conclude in favour of charged particles (with Compton as the leader of this idea) against the gamma-ray hypothesis \cite{compton1936recent, compton1933geographic} and new detectors confirmed the corpuscular nature of radiation: using a newly invent cloud chamber, Dimitry Skobelzyn observed the first ghost tracks left by cosmic rays in 1929.

This gave rise to the particle physics era from cosmic rays, leading to the discovery of elementary particles like the positron (Carl Anderson, 1932) or the muon (Neddermeyer and Anderson, 1937) along with dozens of new particles discovered by the 70s (G.Hooft dedicated a chapter called ``The zoo of elementary particles before 1970'' on his book ``elementary particles'').

In 1938, Pierre Auger positioned particle detectors high in the Alps, noticing that two detectors located many meters apart both signaled the arrival of particles at exactly the same time. Auger had discovered the "extensive air showers," showers of subatomic particles formed by the collision of primary high-energy particles with air molecules. On the basis of his measurements, Auger concluded that he had observed showers with energies of $10^{15} \units{eV}$ — ten million times higher than any known before.

For details on these historical remarks one can find precious information in many books and reviews as: \cite{carlson2011nationalism}, \cite{de2015introduction}, \cite{walter2012early}, \cite{de1991discovery}, \cite{de2011domenico}, \cite{hoffmann1921experimental}, \cite{carlson2013discovery}, \cite{dorman2004cosmic}, \cite{schein1941nature} and the wonderful references therein.

\subsection{CR basic features}
\label{CRstudy}

\subsection*{Cosmic-ray acceleration}
\label{intro_accel}
Cosmic ray physics has undergone an important evolution in the last century and we have realised they play a crucial role in, e.g. the formation of clouds in the atmosphere (therefore, in weather; \cite{palle2004possible}; \cite{brumfiel2011cloud}; \cite{veretenenko2018galactic}), the DNA auto-reparation mechanisms under mutations \cite{blakely2000biological, atri2014cosmic, todd1994cosmic} and the modelling of structures in galaxies \cite{jubelgas2008cosmic, booth2013simulations} among many other topics  \cite{medvedev2007extragalactic, griebetameier2005cosmic}. In addition, we are starting to benefit from them in archaeology, vulcanology and spatial missions for the study of ground topology, \cite{morishima2017discovery, wohl2007scientist}, security storage and material identification \cite{morris2012particle, morris2012obtaining, kudryavtsev2012monitoring}, etc. 

Nowadays, we have gathered a bunch of experimental observations with increasing precision allowing us to, at least qualitatively, explain most of the general features involving cosmic rays.

\begin{figure}[!b]
	\centering
	a)
	\includegraphics[width=0.45\textwidth, height=0.32\textheight]{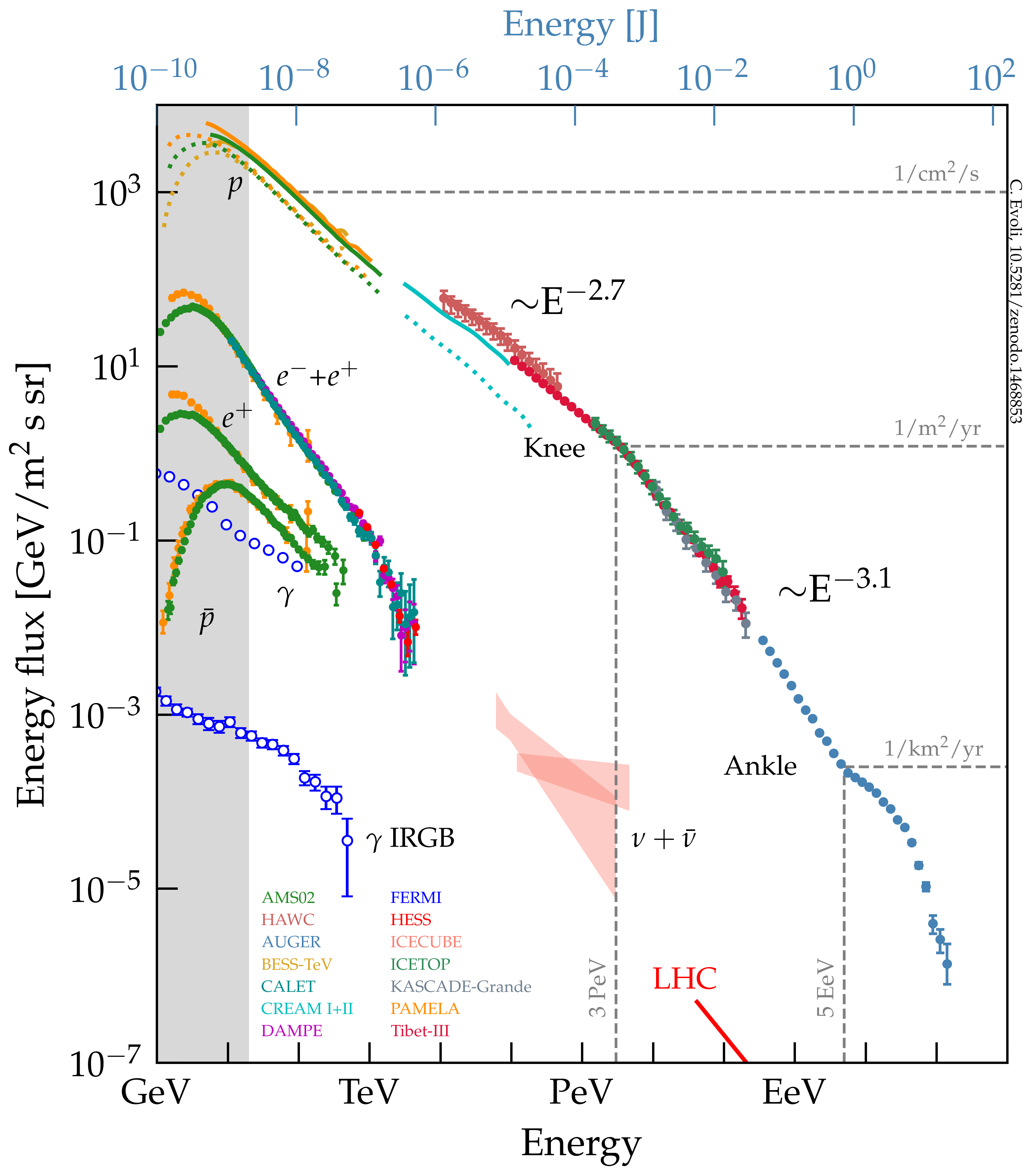}
	\hspace{4mm}
	b)
	\includegraphics[width=0.45\textwidth, height=0.31\textheight]{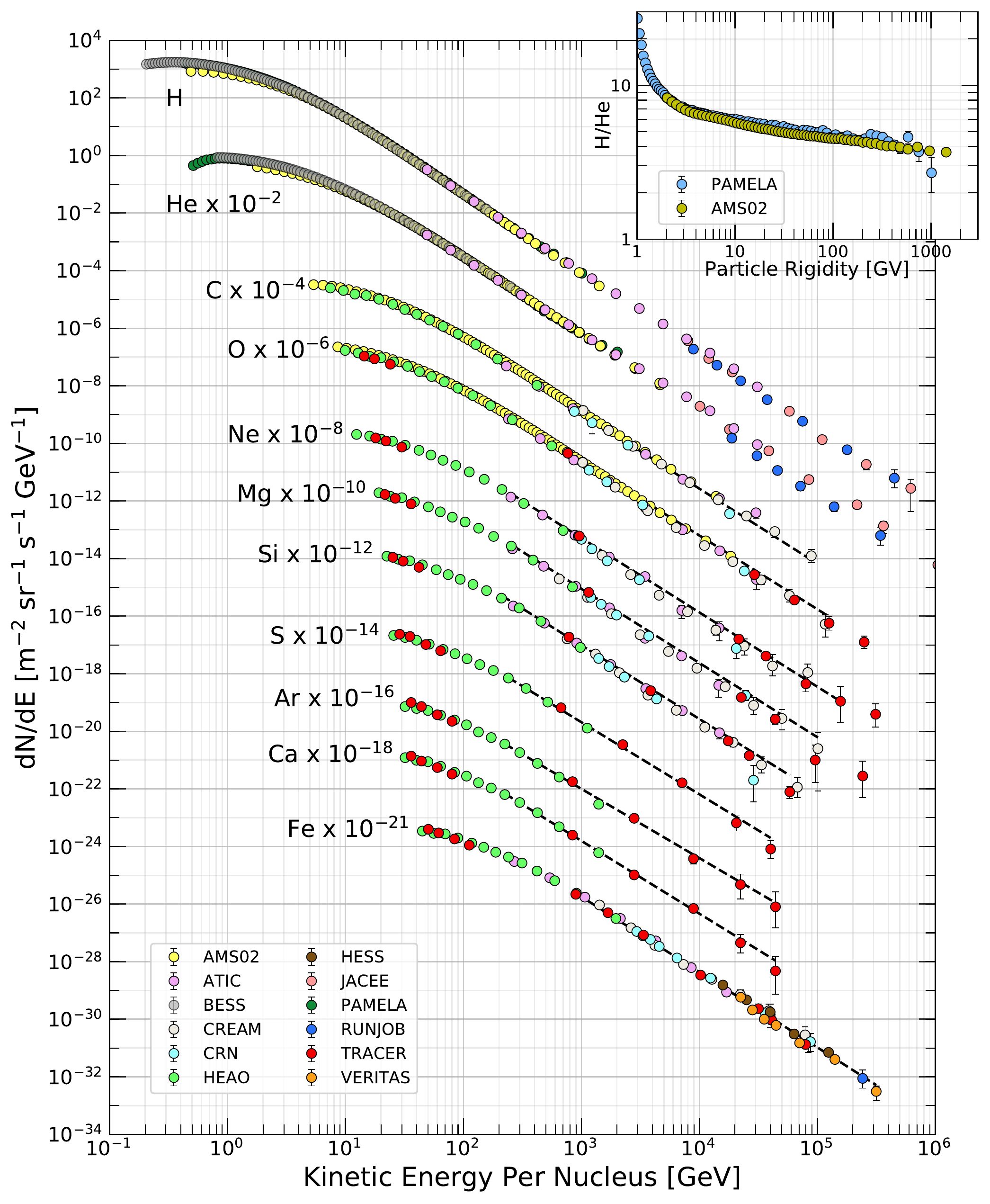}
	\caption{\footnotesize a) spectra of different CR components, from ref.~\cite{evoli_carmelo_2018_2360277}, including also leptons and antiprotons. b) comparison among the fluxes of different CR nuclei, taken from~\cite{cosmic2002pdg}. %The upper right window shows that, from very precise and recent data, these trends are not completely equal for He and protons.
	} 
	\label{fig:CR_components}
\end{figure}

The origin of cosmic rays is one of the main puzzles physicist must solve. Both plots in Figure~\ref{fig:CR_components} show that all different species share the power-law behaviour ($J(E) \propto E^{-p}$, with p $\sim2.7-3.3$) as already mentioned. This may imply they share a common origin, i.e. the mechanism for them to acquire such energies must be essentially the same. 

In 1934 the visionary Fritz Zwicky pointed out that the collapse of heavy enough stars at the end of their lives would produce explosions of cosmic rays, leaving behind neutron stars. An argument in favor is that the luminosity released by supernovae (of the order of $10^{51}$ $ergs/s$, assuming a rate of explosions of $1/30$  $years^{-1}$) is about ten times larger than the energy per unit second we measure for CRs, in agreement with the efficiency one would expect for CR acceleration. The detection of synchrotron emission from cosmic ray electrons, later in the 1950s, \cite{ginzburg1965cosmic, ginzburg2013origin} was shifting the paradigm from supernovae (SN) to supernova remnants (SNRs). 

Currently, the idea of SNRs as the sources (accelerating sites) of cosmic rays is widely accepted and supported by gamma ray observations \cite{BLASIRev} associated to CR interactions from SNRs close to molecular clouds \cite{ackermann2013detection, tavani2010direct} and by the gamma ray emission detected from the Tycho SNR \cite{giordano2011fermi, acciari2011discovery, morlino2012strong, berezhko2012nature}, although no direct and decisive proof has been found yet.

\begin{figure}[!b]
	\centering
	\includegraphics[width=0.37\textwidth, height=0.24\textheight]{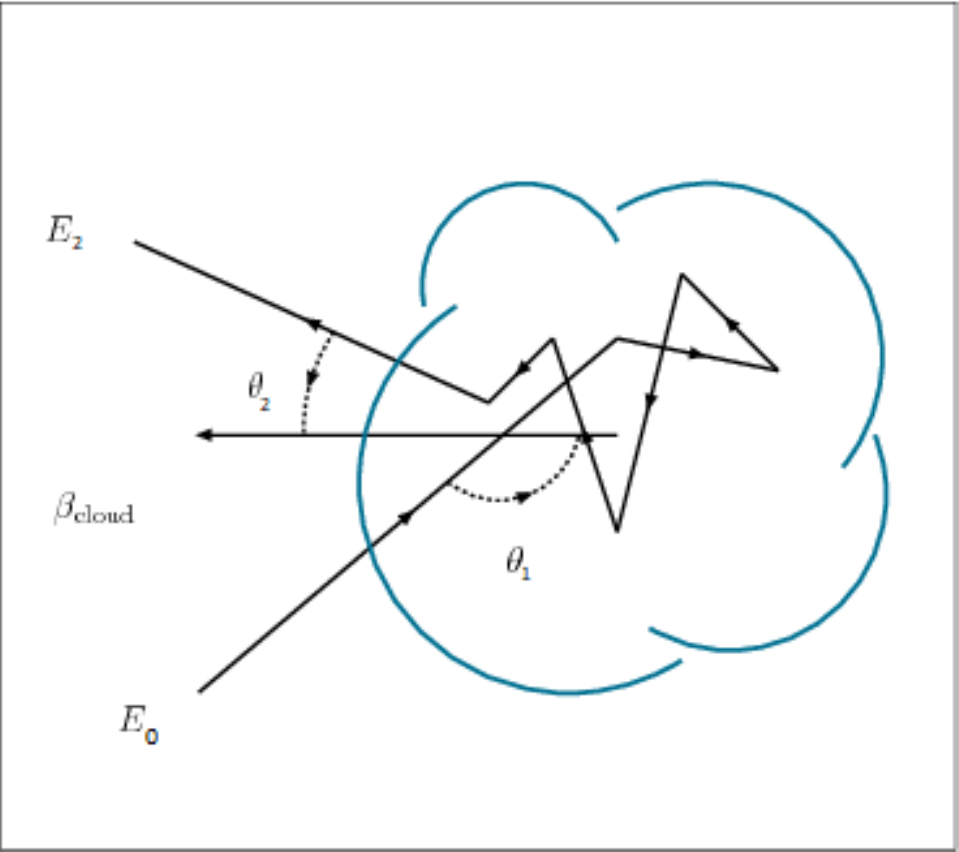}
	\caption{\footnotesize Sketch of the "magnetised clouds" able to successively accelerate charged particles according to the Fermi second order mechanism. Initial and final particle energies and entrance and exit angles are depicted too.} 
	\label{fig:2ndFermi}
\end{figure}

In fact, in the 1950s, Enrico Fermi pointed to the idea of an iterative acceleration process that would successfully give rise to the power-law spectrum via the interaction with moving "magnetic clouds" \cite{fermi1949origin, fermi1954galactic}. These magnetic clouds are randomly moving clouds of gas with embedded magnetic fields. CRs can exchange energy and momentum as they scatter with clouds (see Fig.~\ref{fig:2ndFermi}). In the clouds' reference frame, the energy of the particle does not change, since it scatters via Lorentz force with the magnetic field inside the cloud. The energy of the particle in this frame can be found with a Lorentz transformation as $E' = \gamma (E_0 + \beta p \mu_1)$ where $E_0$ is the initial energy in the laboratory frame, $\mu_1$ is the cosine of the angle $\theta_1$ between the particle and the cloud directions of motion at the entrance (see Fig.~\ref{fig:2ndFermi}) and $\beta$ is the cloud's speed in units of speed of light. 

After the scattering, the energy remains the same in the cloud's frame and the final energy in the laboratory reference frame is given by $E'' = \gamma E'(1 + \beta \mu'_2)$ where $\mu'_2$ is the cosine of the exit angle $\theta_2$. The fraction of energy change is:
\begin{equation}
    \frac{E''-E_0}{E_0} = \frac{1 - \beta \mu_1 + \beta \mu'_2 - \beta^2 \mu_1 \mu'_2}{1-\beta^2}  - 1.
\end{equation}
For non-relativistic clouds (the cloud velocity is tens of km/s), $\beta << 1$ and assuming that the exit angle $\theta_2$ is random (i.e. $\left< \mu'_2 \right>= 0$) the previous result reduces to: 
\begin{equation}
\left<\frac{E''-E_0}{E_0}\right> \approx \frac{1-\beta \left< \mu_1 \right> }{1-\beta^2} -1.    
\end{equation}
 Since head-on collisions are more probable than tail-in ones, $-1 <$ $\left< \mu_1 \right>$ $< 0$ and this results into an energy gain. To average on the scattering angle, we should consider the probability distribution of angles, which is proportional to the relative speed, i.e. $P(\mu) \propto (1-\beta \mu)$. After calculating $\left< \mu_1 \right>$, the previous equation yields:
\begin{equation}
    \zeta \equiv \left< \frac{E_2-E_0}{E_0}\right>_{\mu_1}  \sim \frac{4}{3}\beta ^2 .
    \label{Fermi2nd}
\end{equation} 

We stress here that the energy transfer is due to the moving clouds and not to their magnetic fields. If these encounters occur repeatedly, the average particle energy after the $n$-th collision will be given by $E_{n} = E_{n-1} + \zeta E_{n-1}$, where $E_{n-1}$ is the average energy after the $(n-1)$-th collision. If the initial particle energy is $E_0$, the energy after $n$ collisions will be:
\begin{equation}
E_n = (1+\zeta)^n E_0.   
\end{equation}
The previous equation allows us to calculate the number of encounters needed for the CR particle to reach an energy $E$:
\begin{equation}
n = \frac{log(\frac{E}{E_0})}{log(1 + \zeta)}.
\label{ncycles}
\end{equation}

Indicating with $P_{esc}$ the probability for the particle to abandon the acceleration region after each scattering, the integral energy spectrum, i.e. the fraction of particles with energy $E>E_n$ will be given by:
\begin{equation}
f(E>E_n) = \sum_{m=n}^{\infty}(1-P_{esc})^m = \frac{(1-P_{esc})^n}{P_{esc}}
    \label{fEn}.
\end{equation}

Using eq.~\ref{ncycles}, we get that 
\begin{equation}
(1-P_{esc})^n = \left( \frac{E_n}{E_0} \right) ^{-\gamma}
\end{equation}
with $\gamma = - \frac{log(1-P_{esc})}{log(1 + \zeta)}\approx \frac{P_{esc}}{\zeta}$ in the limit of $P_{esc}, \zeta << 1$. 

This mechanism explains a power law in the energy spectrum, but the fact that the speed of the clouds is non-relativistic ($\beta << 1$) and the small cloud dimensions ($\sim$1 pc) lead to an inefficient mechanism that would need tens of Gigayears to cause an appreciable acceleration (except in specific cases, see for example \cite{winchen2018energy}). For some time this mechanism was supposed to occur during all the CR life, but this hypothesis was discarded when more precise measurements were performed \cite{cowsik1986sporadic, giler1987continuous, ferrando1987propagation}.

\begin{figure}[!bht]
	\centering
	\includegraphics[width=0.42\textwidth, height=0.31\textheight]{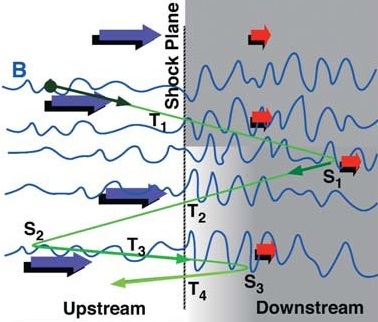}
	\caption{\footnotesize Sketch of the shock front movement in the plasma under the shock wave reference's frame, taken from \cite{mobius}. The green lines represent the movement of a test particle from upstream to downstream region and vice versa.} 
	\label{fig:1stFermi}
\end{figure}

Nevertheless, this idea evolved in another iterative process that seems to be effective in SNRs. 
It is based on the idea of a shock wave moving in a hydromagnetic environment such as the one generated by a SNR. A shock wave (analogously to the ones created in fluids) is a discontinuity in a given property of the ambient that is propagating out into a smooth medium. A sketch of this phenomenon is shown in Fig.~\ref{fig:1stFermi}. It depicts two zones separated by the shock plane which are originated by the discontinuity in the density of the medium where the shock propagates. The important point of this picture is that the downstream is left as a turbulent medium, which means, for a hydromagnetic medium, the creation of turbulent magnetic waves and magnetic instabilities that will interact with the charged particles making their path chaotic (random or diffusive, indeed). For an extensive discussion see, e.g. \cite{baring1997diffusive}.

An external observer would see it as a shock moving with a speed $v_{shock}$ towards the calmer upstream region. The shock causes the downstream region to have a density $\rho_d$ and the upstream one with density $\rho_{ups}$ and this implies a mismatch in the velocity of particles between the two zones, $v_d$ and $v_{ups}$. 
The continuity equation,  $\frac{\partial\rho}{\partial t} + \Vec{\nabla}(\rho \vec{v}) = 0$, holds across the shock front and, when integrating from one side of the shock to the other side we get:
\begin{equation}
    v_d  \rho_d = v_{ups}  \rho_{ups}
\end{equation} 

Hence the compression ratio will be:
\begin{equation}
    R = \frac{\rho_d}{\rho_{ups}} = \frac{v_d}{v_{ups}} = \frac{\tilde{\gamma} + 1}{\tilde{\gamma} - 1 + 2/M_{ups}^2}
\end{equation} 
where $\tilde{\gamma}$ is the adiabatic index ($\tilde{\gamma} = 5/3$ in case of monoatomic gases) and $M_{ups}$ is the Mach number of the upstream region defined as $M_{ups} = v_{ups}/v_{sound}$. Typical velocities for the ejecta are around $30000$ km/s, while the sound speed in these media is $v_{sound}\sim 10(\frac{T}{10^4K})^{1/2}$ km/s \cite{BLASIRev}. Therefore, in the so-called strong shock limit, $M \gg 1$ and, considering monoatomic gas in the shock surroundings, the compression ratio becomes $R=4$.

In the shock's frame, the particles will be going back and forth from the downstream to the upstream regions due to the magnetic interactions with the plasma instabilities, acting as a magnetic mirror. This is shown by the green line in Fig.~\ref{fig:1stFermi}. In this frame, $v_{ups} = -v_{shock}$ and $v_d = v_{ups}/R$. The energy of a particle flowing towards the downstream region, in this reference frame, will be $E' = \gamma E(1+\beta\mu)$, where $\mu$ is the cosine of the incidence angle with respect to the shock front direction (the so-called pitch angle) and $\beta = \frac{v_{ups} - v_d}{c}$, where $c$ is the speed of light.

When a particle bounces back to the upstream region, its energy is conserved in the downstream reference frame since the magnetic interactions do not change particles energy. For an external observer it will return to the upstream region with an energy $E'' = E'\gamma(1-\beta\mu')$, since $E'$ is conserved in the interaction. This gives a fractional energy change:
\begin{equation}
    \zeta = \frac{\Delta E}{E} = \gamma^2 (1+\mu\beta)(1-\mu' \beta) - 1
    \label{Fermi1_efficiency_def}
\end{equation}  

We must average eq.~\ref{Fermi1_efficiency_def} under pitch angles $\mu$ and $\mu'$, taking into account that the probability of entering (or exiting) with a given direction; this is just the probability of crossing a wall ($\frac{dP}{d\Omega} \propto \mu$):
\begin{equation}
    \left<\zeta\right> = \left<\frac{\Delta E}{E}\right>_{\mu, \mu'} = \frac{4\beta}{3} \hspace{3mm}\text{, for } \beta \ll 1
    \label{Fermi1_efficiency}
\end{equation}  

Now, to determine the probability of undergoing a cycle let us again work in the frame of the shock. First, one needs to calculate the flux of CR particles returning to the shock upstream from downstream as $J_{-} = \int_{down\rightarrow ups} d\Omega \frac{n c}{4\pi} \mu = \int_0^{2\pi} d\phi \int_{0}^{1} d\mu \frac{nc}{4\pi} \mu = \frac{nc}{4}$, with n as the particle number density in the vicinity of the shock and considering the CR speed to be near the speed of light to be able to cross back the shock. Then, as no particle escapes from the shock upstream the shock conservation of CR flux requires that the flux of CR particles entering the shock from downstream ($J_{+}$) can be calculated as the sum of CR particle flux that escapes into the far downstream (not returning flux), $J_{\infty}$ and $J_{-}$. Therefore, the escape probability can be written as $P_{esc} = \frac{J_-}{J_+} = \frac{J_-}{J_- + J_{\infty}} = 1 - P_{cycle}$. Finally, as the CR particle distribution in the downstream frame is assumed to be isotropic, the escaping flux is simply given by $J_{\infty} \sim n v_d $. %file:///C:/Users/pedro/AppData/Local/Temp/Handout11_FermiAccl.pdf
With this, the probability of completing a cycle is:
\begin{equation}
    P_{cycle} = 1 - P_{esc} = 4 \frac{v_d}{c}
    \label{Fermi1st_prob}
\end{equation}  
So, in each cycle a particle will gain a fraction of energy $\zeta$ (eq.~\ref{Fermi1_efficiency}) transferred from the shock's energy with a probability of repeating the cycle $P_{cycle}$ (eq.~\ref{Fermi1st_prob}). After n cycles, a particle with initial energy $E_0$ will get an energy $E_n = (1+\zeta)^n E_0$.

Thus, obtaining, from eq.~\ref{ncycles}, that: 
\begin{equation}
(P_{cycle})^n = (1-P_{esc})^{\frac{log(\frac{E_n}{E_0})}{log(1 + \zeta)}} = \left( \frac{E_n}{E_0} \right)^{-\gamma}
\end{equation}
with $\gamma = - \frac{log(1-P_{esc})}{log(1 + \zeta)}\approx \frac{P_{esc}}{\zeta}$ in the limit of $P_{esc}, \zeta << 1$.

This means that the differential flux in energy ($ = \frac{df(E>E_n)}{dE}$) of accelerated particles under this process has the form $Q(E) \propto E^{-\gamma - 1} = E^{\alpha}$, which reproduces the power-law behaviour the CR spectrum exhibit. The exponent is $\alpha = \frac{P_{esc}}{\zeta} + 1 = \frac{4\frac{v_d}{c}}{\frac{4}{3}\frac{v_{ups} - v_d}{c}} + 1 = \frac{3}{R - 1} + 1 = 2$, in the strong shock limit ($R=4$). This value for $\alpha$ may be larger not considering the strong shock limit ($R\leq4$), as happens in the case of old SNRs and cosmic rays that re-accelerate later during their propagation \cite{wandel1988supernova}.

However this mechanism finds problems to explain the full spectrum when computing the maximum achievable energy in a time consistent with the expansion phase of SNRs (the Sedov blast-wave phase; see \cite{chevalier1977interaction}, \cite{jones19981051} or \cite{reynolds2008supernova} among others). 

This mechanism is able to roughly reproduce the shape of the particles spectra but, how to explain the breaks at the knee and ankle energies? are those the result of different kind of populations of SNRs? To answer this question one needs to see how effective this process is in accelerating particles.

The acceleration rate can be easily calculated by: 
\begin{equation}
\frac{dE}{dt} = \frac{\Delta E}{\tau_{cycle}}
\label{1stEcycle}
\end{equation}
where $\tau_{cycle}$ is the time needed to complete a cycle of acceleration. This time can be estimated considering that a particle going from the downstream region to the to upstream one would need a time $t_d$ to meet back the shock front, travelling a distance $l_d$ by performing a diffusive motion whose diffusion coefficient is $D$ from the point it exited from the downstream region. They are related by $l_d \sim \sqrt{D t_d}$. In that time, as the shock is also moving, it would have travelled the same distance, $l_d = v_{shock} t_d$. In this way, the diffusion time can be taken as an order-of-magnitude estimate of $\tau_{cycle}$ and takes the form $\tau_{cycle} \sim t_d \sim D/v_{shock}^2$. This stresses the fact that the more energetic the particle is, the longer it takes to complete a cycle. 

Then, taking the diffusion length, $l_d$, to be of the order of the particle's gyroradius, $r_g$ (i.e. Larmor radius), the diffusion coefficient must be $D \sim \frac{r_g c}{3}$, approximating the particle velocity to c. The gyroradius is $r_g = p/ZeB \approx E/ZeBc$, with $p$ as the momentum component perpendicular to the magnetic field $B$ and $Ze$ the particle charge. Plugging up these terms in eq.~\ref{1stEcycle} and using eq.~\ref{Fermi1_efficiency} to calculate $\Delta E$ we get
\begin{equation}
\frac{dE}{dt} \sim \frac{\zeta E v_{shock}^2}{\frac{r_g c}{3}} = \frac{3 \zeta E v_{shock}^2}{\frac{E}{ZeB}} = 4 \beta v_{shock}^2 ZeB
\end{equation}

Then, assuming that the maximum allowed time for a cycle is the SNR expanding phase (Sedov-Taylor phase), the maximum energy can be computed as $E_{max} = \frac{dE}{dt} t_{ST}$. Supposing shock velocities of $3\times10^9$ cm/s, $t_{ST}$ of the order of thousand years and $\beta \sim 10^{-2}$, the maximum energy becomes $E_{max} \sim  Z \left(\frac{B}{3 \mu G}\right) 6\times10^{13} \units{eV}$, written in this way since the typical intensity of the magnetic field of $3\mu G$.%Formula from BLASI_REVIEW

With this estimate, the maximum energy achievable for iron nuclei ($Z=26$) will be $E_{max} \sim 1.6 \times 10^{15} \units{eV}$. This is indeed a very optimistic estimation \cite{lagage1983maximum}, that can be even an order of magnitude below a more sophisticated estimation \cite{bell2018cosmic, bell2013cosmic, bednarz1996acceleration}. 

%It is interesting to point that the diffusion coefficient, $D$, is directly proportional to the energy of the particle (the bigger the energy the particle has the easier for it to avoid interacting with irregularities), which means that the more  energetic particles being accelerated the longer the cycle is. Assuming the diffusion coefficient behaves as $D = D_0 E^{\Delta}$ , we get

%\begin{equation}
%E_{max} = \left(\frac{v_{shock}^2 t_{age}}{D_0}\right)^{1/\Delta}
%\label{MaxEn}
%\end{equation} 

The fact that this model (so-called Fermi 1st order mechanism) is able to explain the CR spectra below the knee suggests that the breaks are due to changes in the acceleration mechanisms, i.e. different sources. 

Several different mechanisms able to accelerate particles in the Galaxy have been proposed (e.g. \cite{hayakawa1964part} or \cite{ramaty1973cosmic}). For example, particles accelerated at the Sun via magnetic reconnections are expected or fast varying magnetic fields \cite{priest2007magnetic, sakai1988particle, olinto1999galactic}. This happens because the fast variation in the magnetic field in, for instance, a sunspot produces intense electric fields. The energy gained in this process can be easily calculated as $\Delta E = \mid eU \mid = e \pi R^2 \frac{dB}{dt}$, where we suppose such spots are circular with radius R, and the magnetic field B perpendicular to the spot. Assuming typical radii of $10^7m$ and a magnetic field strength B=2000 G for a sunspot of one day lifetime \cite{grupen2005astroparticle}, the energy gain is $\Delta E = 0.73\times 10^9 \units{eV}$. Hence our Sun is another source of charged particles, but it cannot account for the highest CR energies. %At most, particles of $10^{11} \units{eV}$ have been observed from the Sun \cite{Sun_1011}.

Pulsars are another interesting site of CR acceleration \cite{beskin2000particle, gunn1969acceleration}. They are extreme objects with high rotation speed and huge magnetic fields. They show a "spin-down" that is caused by dipolar emission. Therefore, a fraction of this rotational energy loss is expected to contribute to particle acceleration. The estimations make them able to accelerate particles up to knee energies and beyond, but their contribution for heavy nuclei is not considerable. In turn, they play a key role in the lepton flux as it seems to be recently demonstrated \cite{manconi2020contribution, manconi2019multi, di2019evidences, di2019detection}.

However, the modern theory of \textit{diffusive shock acceleration} (DSA) explains, in the frame of magneto-hydrodynamics \cite{blandford1978particle, bell1978acceleration, krymskii1977regular, axford1982structure}, that acceleration efficiency in SNRs can change depending on the magnetic field orientation \cite{bell2011cosmic, ellison2004diffusive} and that the magnetic field experiences amplification \cite{schure2012diffusive, zirakashvili2008modeling, amato2009kinetic}. In conclusion, the maximum energy a particle can reach by magnetic field amplification within the DSA theory is $< 1 \units{EeV}$.
Moreover, there are interesting evidences in favour of young massive stellar clusters (OB associations and, in general, SN occurring in superbubbles) as likely sources of CRs with$\units{PeV}$ energies \cite{bykov2014nonthermal, ackermann2011cocoon}. See \cite{hillas2005can} for a review on the topic.

It seems that there are no galactic sources able to account for the very high energy CRs at the ankle (see \cite{aloisio2012transition} for a deep discussion of the transition from galactic to extragalactic sources of CRs). Which other sources are able to accelerate particles to such energies can be easily understood following the Hillas' argument \cite{ptitsyna2010physical, hillas1984origin}. It is based on the idea that for a particle to be iteratively accelerated in a source, the size of this source must be, at least, the same as the particle's gyroradius. This imposes the condition $E_{max} \leq ZBR_{source}$.

\begin{figure}[!ht]
	\centering
	\includegraphics[width=0.56\textwidth, height=0.42\textheight]{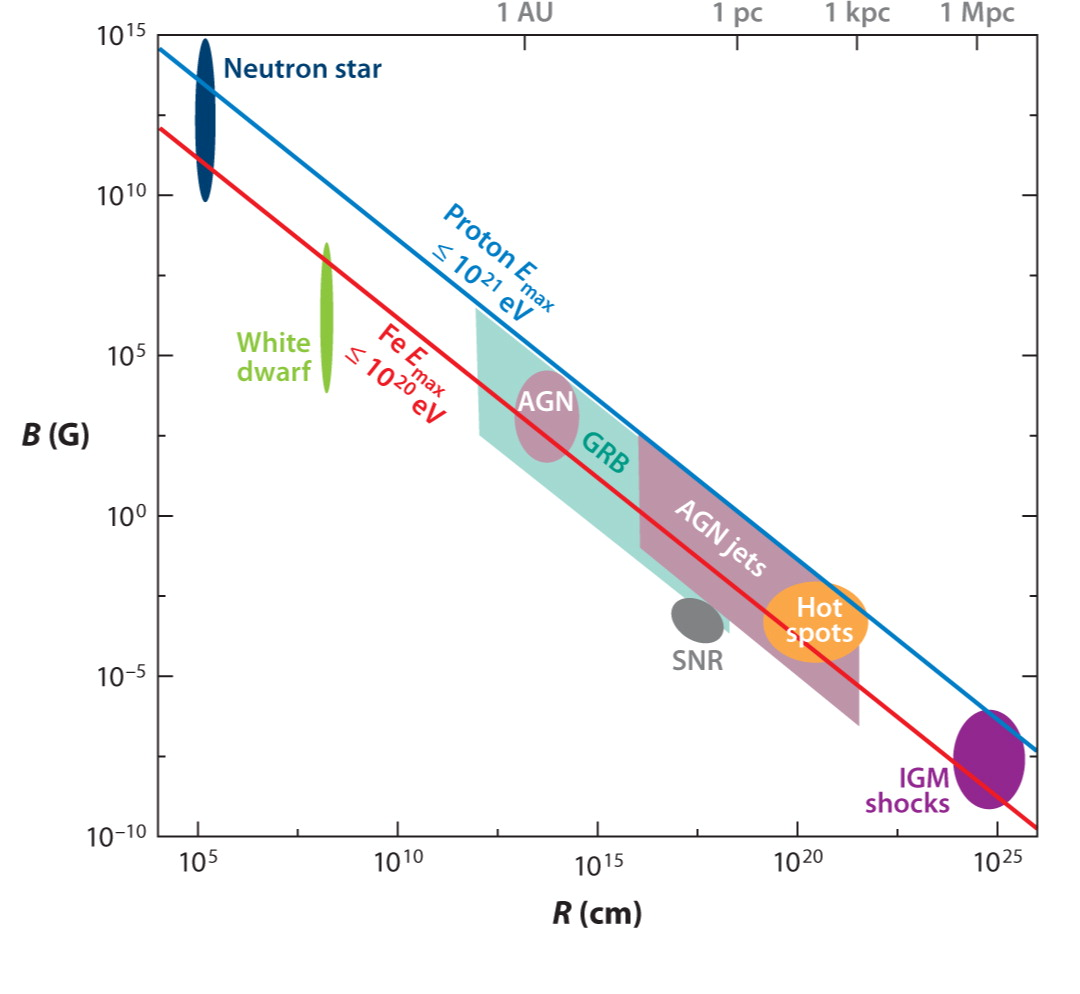}
	\caption{\footnotesize Hillas plot, taken from \cite{doi:10.1146/annurev-astro-081710-102620}. It shows the delimiting lines for protons and iron acceleration until $10^{20} \units{eV}$ to reflect that all possible sources in the region between the lines are able to accelerate particles until such high energy.} 
	\label{fig:hillasplot}
\end{figure}

From this condition Hillas drew a plot (similar to Fig.~\ref{fig:hillasplot}) for the relation $\log B = \log E_{max} - \log Z - \log R_{source}$, showing that the logarithm of the maximum energy is linearly correlated with the logarithm of the magnetic field. In the Hillas plot it is possible to draw the lines corresponding to energies of $10^{20} \units{eV}$ to see which sources gather the Hillas condition (see \cite{aloisio2017acceleration} for a deep discussion).

As it can be seen in the Hillas plot, only extra-galactic sources can account for these ultra-high energy cosmic rays (UHE-CRs), separating the all-CR-particles spectrum into Galactic and extra-galactic CRs, as shown in Fig.~\ref{fig:CR_components} a. In fact, it has been experimentally found direct evidence of associations of AGNs (Active Galactic Nuclei) and GRBs (Gamma-Ray Bursts) with CRs \cite{ansoldi2018blazar, icecube2018neutrino}. The presence of the knee (and another feature called second knee, less evident, but present at $4-7 \times10^{17} \units{eV}$) can be explained in several ways (see \cite{aloisio2012transition}) as the transition from light CRs ($\sim$ protons) to heavy CRs ($\sim$ iron).

\subsection*{Cosmic-ray interactions}
\label{intro_interact}

The CR showers reaching the ground, discovered by Pierre Auger, are mainly generated by UHE-CRs. This can be explained as the result of the inelastic interactions of very energetic CR particles with target particles in air (the molecules in the atmosphere, mainly $N_2$), resulting into hadronic showers. The more energetic is the particle, the deeper the hadronic shower will penetrate in the atmosphere, eventually reaching the ground. Therefore, most of the detected CRs at ground are not the original ones, i.e. they are secondary cosmic rays. 

It has, indeed, been observed that the detected cosmic ray flux peaks at about 15 km in altitude and then drops sharply. This kind of variation was discovered by Pfotzer in 1936 and suggests that the detection method used was mainly detecting secondary particles rather than the primary particles reaching the Earth from space (see \cite{HyperP} and its illustration of this effect). Thanks to these interactions we could discover antimatter and other strange (in the sense of strange hadrons) particles that, otherwise, would have either decayed or interacted before reaching us. Nowadays, these secondary particles represent an intense background in dark matter or neutrino experiments and have been widely studied \cite{schonert2009vetoing, cecchini2012atmospheric, gaisser2012spectrum}.

From this point of view, one may wonder which other interactions CRs experience during their journey before reaching the Earth. As an example, UHE-CRs propagating in the interstellar space can undergo photonuclear reactions with the background photons of the CMB \cite{greisen, zatse}, giving rise to the GZK (Greisen-Zatsepin-Kuz'min) cut-off in the energy spectrum at $\sim 10^{20} \units{eV}$ for protons and at higher energies for heavier CR primaries, or interactions with the extra-galactic background light \cite{puget1975photonuclear, aloisio2007dip}, producing characteristic features in their spectrum.

As a matter of fact, the majority of CRs are Galactic and they travel across the interstellar gas and clouds of the Milky Way, undergoing inelastic interactions (called spallation reactions) that may lead to the creation of secondary CRs.

\begin{figure}[!b]
	\centering
	\includegraphics[width=0.6\textwidth, height=0.44\textheight]{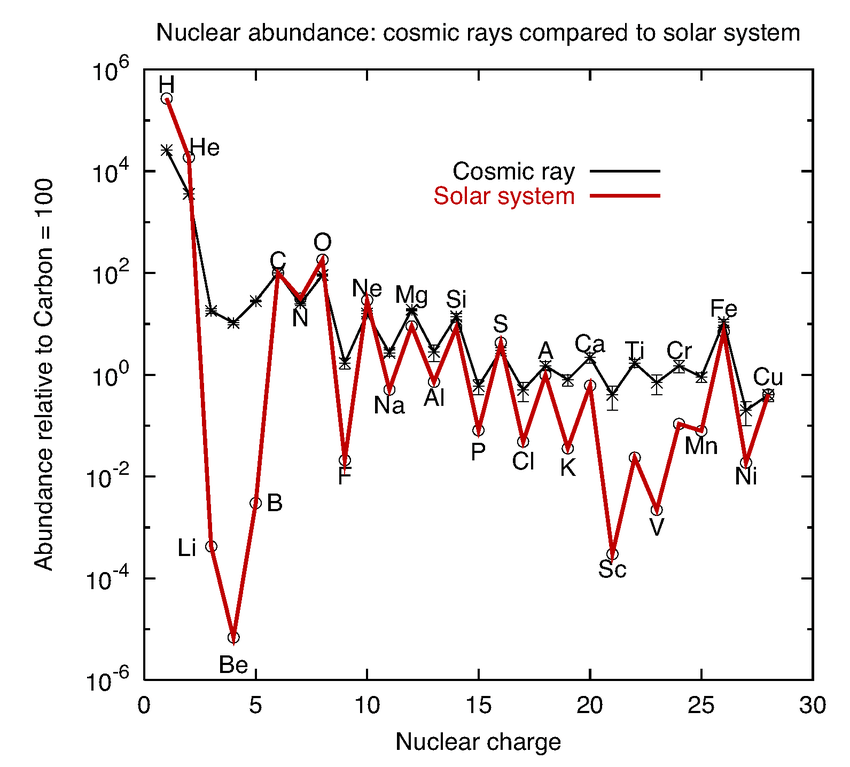}
	\caption{\footnotesize Composition of CRs compared to the solar system composition, taken from \cite{fatimathesis}. The difference in the Li, Be, B and sub-iron species is significant, which can be explained by spallation processes CRs undergo.} 
	\label{fig:composition}
\end{figure}
% Downloaded from https://www.researchgate.net/figure/Nuclear-abundance-cosmic-rays-compared-to-the-solar-system-Source_fig1_310831062

In Fig.~\ref{fig:composition} a comparison between the mean composition of the Solar System (similar to the solar composition) with the mean composition of CRs arriving Earth, at $10\units{GeV}$, is displayed. It shows a good agreement in the abundance of those species which are created in the nucleo-synthesis phase of the Big Bang (protons and helium), those created in the massive stars (carbon, oxygen and heavier, until iron) and those created during SN explosions (heavier than iron), but there are are large discrepancies with Li, B, Be (LiBeB group) and sub-Iron species (Sc, Ti, V). It is easy to realise that, if cosmic-ray spallation reactions with molecules in the Earth's atmosphere are responsible for the formation of unstable nuclei like $^7$Be or $^{14}$C, not natural at Earth, this anomaly may be explained by the same argument, so that Li, Be, B and sub-iron species are secondary CRs.

It can be found that, if we consider primary and secondary cosmic rays to be coupled, the path length travelled by cosmic rays until reaching the Earth is directly connected to the amount of secondary particles formed.

Let's write a simple equation relating a primary (P) species (say carbon) steadily fragmented by spallation in a secondary (S) species (say boron). The path length is usually expressed in terms of grammage, measured in units of $g/cm^2$ and defined as:
\begin{equation}
    X = \int dl \rho (l) = \int dt \rho (l) v(t)
\end{equation}
where $\rho$ is the mass density of the interstellar medium (ISM) and $v$ is the CR speed. The density $\rho$ depends on the position in the Galaxy, as the interstellar medium is not uniform, while the CR velocity can change in time due to energy losses or other effects as reacceleration.

Supposing that the only allowed spallation reaction is $P \rightarrow S$ and that there is no new generation of primary CRs (no source term for the primary P) we can write the following equations:
\begin{subequations}
  \begin{equation}
    \label{Gramm1}
        \frac{dN_p(X)}{dX} = -\frac{N_p(X)}{X_p}
  \end{equation}
  \begin{equation}
    \label{Gramm2}
        \frac{dN_s(X)}{dX} = -\frac{N_s(X)}{X_s} + \frac{P_{ps}}{X_p}N_p(X)
  \end{equation}
\end{subequations}

Here, $X_p$ and $X_s$ are the interaction path lengths of primaries and secondaries, i.e. the mean path length needed for each particle to have an interaction with the gas. This interaction length is $X_i = \frac{\Bar{m}}{\sigma_{inel}}$, where $\Bar{m}$ is the mean mass per particle in the gas (in the case of the interstellar medium it is roughly the proton mass) and $\sigma_{inel}$ is the total inelastic cross section of the interaction of the particle with the interstellar gas. The last term in the r.h.s. of equation~\ref{Gramm2} is a source term for the secondary CRs and contains a factor $P_{ps}$ which is the probability of generating a particle S from the spallation reaction of a particle P with the interstellar gas. This probability usually takes the form $P_{ps} = \frac{\sigma_{p \rightarrow S}}{\sigma_{inel}}$, where $\sigma_{P \rightarrow S}$ is the inclusive cross section of the interaction.

The solution of eq.~\ref{Gramm1} is just: 
  \begin{equation}
    \label{solution1}
        N_p(X) = N_p(0)\exp{(-X/X_p)}
  \end{equation}
where $N_p(0)$ is the initial number of primaries.

The solution of eq.~\ref{Gramm2} can be obtained by multiplying both sides for the factor $\exp{(X/X_s)}$ and is given by:
  \begin{equation}
        N_s(X) = N_p(X)\frac{P_{ps} X_s}{X_s - X_p}[\exp{(X/X_s - X/X_p)} -1]
        \label{solution2}
  \end{equation}
demonstrating that the amount of secondary CRs depends on the grammage traversed by the primary CRs. 

Apart from nuclei, these interactions can yield other particles, mainly unstable hadrons that decay into neutrinos and gamma rays, whose emission is directly correlated with the CR flux. This has opened the window for the multimessenger era, which combines information from different species of particles to keep track of the processes that generate them \cite{kadler2015tanami, telescope2018multimessenger, nakamura2016multimessenger}. Spallation reactions also create leptons (electrons and positrons mainly) and antimatter (antiprotons, antideuterons, helium-3, etc).

In addition, cosmic rays are deflected by interstellar magnetic fields, which give them high homogeneity and isotropy in arrival directions to Earth. The Larmor radius of a $10 \units{GeV}$ proton inside the Milky Way ($B \sim  3 \mu G$) is of the order of 1 $\mu pc$, which means that after a small distance from the source, the particle loses all the information of its original direction. For UHE protons (with energies $10^{20} \units{eV}$) it is possible to find some correlation with their source directions, as the gyroradius is of the order hundreds of kpc. This means that a dipolar anisotropy should be observed when looking at CRs of very high energies \cite{aab2017observation, abbasi2004study}. 

% 3.336e+6 cm * (p/GeV/c)*(1/Z)*(G/B) --> 10e+13 cm; 1km = 3,24078e-14 pc --> 1cm = 3.24078e-19 pc;         see in https://w3.iihe.ac.be/~aguilar/PHYS-467/PA3.html --> Superposition principle 

% B in Gauss == 10e-8 Volt*s*cm^{-2}

On the contrary, galactic cosmic rays should not exhibit dipolar anisotropies. A well-known (from the 1930s) an apparent anisotropy is the Compton-Getting effect \cite{compton1936recent, gleeson1968compton}. This is simply due to the Earth's relative motion around the Sun (analogous to the CMB dipolar anisotropy).

\begin{figure}[t]
	\centering
	\includegraphics[width=0.4\textwidth, height=0.29\textheight]{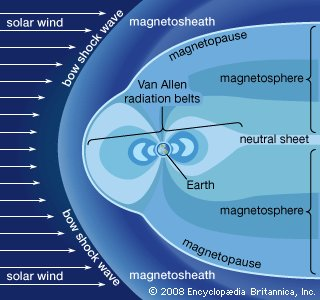}
	\hspace{1mm}
	\includegraphics[width=0.5\textwidth, height=0.24\textheight]{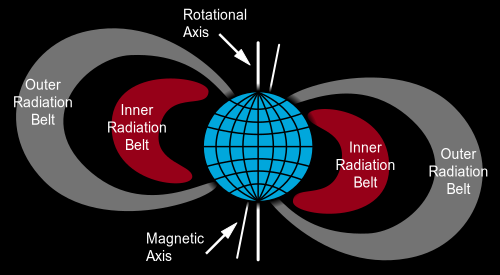}
	\caption{\footnotesize Left: detailed sketch of the Van Allen belts. Credits: Encyclopaedia Britannica~(\url{www.britannica.com/science/Van-Allen-radiation-belt\#/media/1/622563/60532}). Right: a view of the Earth magnetic field (similar to a dipole at small distances), depicting the magnetic and geographic axes (taken from Wikipedia).} 
	\label{fig:Earth_B}
\end{figure}

Moreover, as mentioned above, due to the fact the bulk of CRs are positively charged, an asymmetry must appear in the flux coming from the east and west, the East-West effect. The flux of low-energy CRs from east is lower than that of CRs from west for an observer located at the equator. Since the Earth's magnetic field is approximately dipolar (very good approximation near the surface; see Figure~\ref{fig:Earth_B}, right), the field at the equator is orthogonal to the equatorial plane, allowing positive particles to reach the Earth from the west (allowed clockwise circular trajectories) and negative ones from the east (anticlockwise). At the equator's surface, the geomagnetic field presents a roughly constant value of $0.31 G$ \cite{lipari2000east}.

Another consequence of the dipolar structure of the geomagnetic field is that equatorial zones are slightly shielded from low-energy particles, since they follow the magnetic lines towards the poles. Hence, the net effect of the geomagnetic field are latitude and longitude variations of the CR intensity. In addition, the field intensity varies non uniformly in another points around the Earth, which induces further anomalies (e.g. the South Atlantic anomaly; \cite{pavon2016south}; \cite{trivedi2005geomagnetic}). Nevertheless, the geomagnetic field looks different at larger scales, as the solar wind modifies its shape, populating the Van Allen belts as illustrated in the left panel of Figure~\ref{fig:Earth_B}. 

The solar magnetic field creates an environment known as heliosphere, a bubble permeated with the solar winds that extends far beyond the Solar System Planets. %($\sim 500 $ AU along equatorial regions and $\sim250$ along polar regions). 
The heliosphere bends low-energy particle trajectories, reducing the intensity of cosmic radiation inside it. In fact, the inner heliosphere is not only a region where cosmic rays and solar wind interact, but magneto-hydrodynamics (MHD) and gas dynamics models show that the heliospheric magnetic field is strongly amplified near the edge of the heliopause because of flow deceleration \cite{florinski2003galactic}, constituting a region strongly influenced by the turbulent motions of the diffuse matter within the galaxy \cite{fermi1954galactic}. Therefore, a heliopause as the region separating the solar and interstellar plasmas is predicted and considered as an outer modulation boundary, with a heliosheat region between the heliopause and the bow (termination) shock (Fig.~\ref{fig:Earth_B}). Furthermore, the  presence of a cycle in the Sun's activity (that may be tracked by the number of spots or Wolf number) of 11 years, from which the solar field changes from (approximately) a dipole at minimum activity to higher orders during maximum activity, affects these regions. At solar maximum, the field reverses sign and direction to lead to the 22-year polarity cycles \cite{potgieter2013long}. 

Galactic cosmic rays experience convection and diffusion when they penetrate the heliosphere, since they find a moving solar wind in a highly turbulent plasma, reaching minimum levels of intensity at solar maxima. In addition, they experience important drifts and energy losses, due to energy exchange with the ambient plasma, during their diffusion inside. This is called the solar modulation, whose effects stop being important at energies $> 30 \units{GeV}$. Excellent reviews of these ideas are \cite{potgieter2013solar} and \cite{jokipii2000galactic}.  

The basic transport equation for galactic CRs in the heliosphere was derived by Parker \cite{parker1965passage} and it has the following form:
  \begin{equation}
        \frac{\partial f}{\partial t} = - (\vec{V} + \left<\vec{v}_d\right>) \cdot \vec{\nabla} f + \vec{\nabla} \cdot (D_s \vec{\nabla} f) + \frac{\alpha}{3} (\vec{\nabla} \cdot \vec{V}) \frac{\partial f}{\partial ln P}
        \label{Parker}
  \end{equation}
where $f(\vec{r}, P, t)$ is the particle distribution function  dependent on the position and rigidity of the particle and on time. Here $\vec{V}$ corresponds to the solar wind velocity, $\left<\vec{v}_d\right>$ is the average drift velocity of particles, $D_s$ the symmetric diffusion coefficient and $\alpha = \frac{E_k + 2m}{E_k + m}$ where $E_k$ is the particle kinetic energy \cite{Jokipii:1971sx}. 

While the first term in the r.h.s. involves CR convection and drift, the second term describes the diffusion. The last term describes the adiabatic energy losses, and takes the form $-\frac{\alpha P}{3} \vec{\nabla} \cdot \vec{V}$ \cite{ruffolo1994effect}. Other energy losses are the ionization and inelastic interactions, but they are subdominant in the heliosphere. In the next section all these terms will be quantitatively described.

\section{Transport of galactic cosmic rays}
\label{sec:diff}

While the basic features of cosmic rays can be easily explained by the simple arguments given in the last section, there are a few pieces missing to complete this puzzle.

Plugging experimental values \cite{PDB_2018} for the constants in equation~\ref{solution2}, in the case of a primary C, N or O ($X_{CNO} \sim 6.7$ $g/cm^2$) and secondary Li, Be or B ($X_{LiBeB} \sim 10$ $g/cm^2$ and $P_{CNO \rightarrow LiBeB} = 0.35$), we are left with a direct relation between the primary-to-secondary CR ratio and the mean amount of traversed matter. Taking this relation from any current experiment (see section~\ref{sec:Exps}, or, e.g., from Figure~\ref{fig:composition}), this ratio, at $10 \units{GeV}$, is  $\frac{n_{LiBeB}}{n_{CNO}}\sim 0.2$, which implies a mean amount of grammage travelled by the primaries $X \sim 4.3$ $g/cm^2$ \cite{freier1948evidence}. 

%A historical event was the arrival of satellite measurements of isotopic Li, Be, B in the 1970’s (45). Since then the subject has expanded enormously with models of increasing degrees of sophistication. The simple observation that the observed composition of CR is different from that of solar, in that rare solar-system nuclei like Boron are abundant in CR, proves the importance of propagation in the interstellar medium. The canonical ‘few g cm−2’ of traversed material is one of the widest-known facts of cosmic-ray physics.

Then, to compare this with any reference value, first it is necessary to think of the picture of a CR particle moving along the galactic disc, with approximate thickness of $h \sim 200 pc$ and eventually interacting with the surrounding gas. The disc has a well known average density $n_H \approx 1 cm^{-3}$. If the primary CR is travelling in the disc along a ballistic trajectory, the grammage traversed would be: $X_{ballistic} \sim m_p n_H h \sim 10^{-3}$ $g/cm^2$. This is a value three order of magnitude smaller than the one derived from experimental data. Even considering the CRs moving completely radially in the disc the value one gets is always smaller than the observed one. For the ballistic trajectory, the propagation time will roughly be $t_{prop} \sim h/c \sim 1$ $ky$, while from the grammage calculated with the measurements one gets $t_{prop} \sim \frac{X_{traverse}}{X_{ballistic}\frac{h}{c}} \sim 5$ $My$.

The conclusion from these numbers is that the propagation region must be much thicker to justify the grammage inferred from secondary over primary ratios. In this way, CRs must wander around the  galactic disc crossing it several times, remaining long time above and below the disc, which also explains the detection of synchrotron radiation coming from high galactic latitudes, apparently radiated by CR leptons \cite{haslam1981galactic}. Another proof for this assumption comes from the amount of the isotope $^{10}$Be detected \cite{hayakawa1958origin}. This is a radioactive nucleus with a lifetime $\sim 1.4\units{My}$~\cite{chmeleff2010determination}. One would expect to find an amount of this isotope similar to that of other Be isotopes or at least in proportion to its production (inclusive) cross section from primary CRs. Nevertheless, the amount of $^{10}$Be hardly represents the 10\% of the total Be flux. In case the diffusion time corresponds to a ballistic motion, this particle would not be disintegrated at all, being present in larger proportions (for an extensive discussion, see~\cite{simpson1988cosmic}).

\begin{figure}[!b]
	\centering
	\includegraphics[width=0.37\textwidth, height=0.20\textheight]{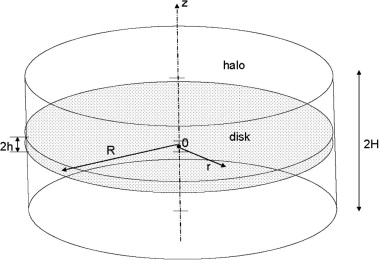}
	\hspace{1mm}
	\caption{\footnotesize Representation of the 2D model (often called 2-zones model) of the Galaxy used for CR computations since it catches the main features of the Milky Way, from  \cite{ptuskin2012propagation}. It is characterised for having a disc region homogeneously permeated with gas (thickness of 2h) and a magnetised halo region over and below the disc (with half-thickness of H).} 
	\label{fig:2Dmodel}
\end{figure}

We can therefore introduce a simplified model of the Galaxy, called ``two-zone model''. It reflects the most essential features of the real system we are considering, but with a simplified geometry, and was presented for first time by \cite{ginzburg1976origin}. In this model (sketched in Figure~\ref{fig:2Dmodel}), the Galaxy is assumed to have the shape of a cylinder with radius $ R \sim20$ kpc and a height 2H (H $\sim$ few$\units{kpc}$). The sources of CRs are distributed, usually following the distribution of SNRs in the galaxy as in~\cite{ferriere2001interstellar} or in~\cite{case1996revisiting}, based in the distribution of pulsar and progenitor star surveys, as in~\cite{evoli2007diffuse} or based on gamma ray data as in~\cite{strong1996gradient}) within a thin disc in the galactic equatorial plane with a typical thickness $h \sim 200\units{pc}$. The Sun is located at a distance of $(8.3 - 8.5)\units{kpc}$ from the galactic center. 

The idea of incorporating a halo arose in the 1970s, explaining the, already mentioned, radio observations of electrons' synchroton emissions (see \cite{ginzburg1980origin} and references therein).
At the moment, its height may be constrained by different kinds of observations, as radio emission \cite{bringmann2012radio}, X and gamma ray emission \cite{biswas2018constraining}, CR nuclei and CR leptons \cite{moskalenko2000diffuse} or even from antiprotons \cite{jin2015cosmic}.
Nevertheless, its exact value remains undetermined, ranging from $3$ to $10 \units{kpc}$.

It was soon realized that, in order to reconcile experimental data with this framework, some leakage or probability to escape from the halo (see pioneering ideas of \cite{cowsik19663} and \cite{shapiro1970heavy}) should be included in the model. This approach was later called ``Leaky Box'' and was able to explain most of the experimental results.

Within the Leaky Box model, the equation that describes stable (particles with infinite lifetime) and unstable (it has finite lifetime:$\tau^{dec} \neq 0$) CR particles is:
\begin{equation}
    \frac{dN_i}{dt} = Q_i (E) - \frac{N_i}{\gamma_i \tau^{dec}_i} - n \sigma_i \beta_i c N_i - \frac{N_i}{\tau^{esc}}.
    \label{leaky_prim}
\end{equation}
Here the index $i$ refers to the CR primary species, $\beta_i$ and $\gamma_i$ are the velocity (in units of $c$) and the Lorentz factor of the $i$-th primary, $\tau_i$ is its lifetime, $\sigma_i$ is its inelastic scattering cross section and $Q_i(E)$ the source term. Finally $n$ is the mean density traversed by CRs and $1/\tau^{esc}$ is the rate of escape of CRs from the Galaxy. As we can see, the inelastic cross sections are inserted to account for destruction of CR species from interactions with the ISM gas.

For the secondary unstable CR of the $j$-th species, the equation takes the following form:
\begin{equation}
    \frac{dN_j}{dt} = \sum_i n \sigma_{i \rightarrow j} \beta_i c N_i + \sum_i \frac{N_i}{\gamma_i \tau^{dec}_{i \rightarrow j}}  - \frac{N_j}{\tau^{esc}} - \frac{N_j}{\gamma_j \tau^{dec}_j}
    \label{leaky_unstable}
\end{equation}
where $\sigma_{i \rightarrow j}$ is the cross section of production of particles of the $j$-th species from CR primaries of the $i$-th species and $\tau_{i \rightarrow j}$ is the lifetime of CR primaries of the $i$-th species decaying into secondaries of the $j$-th species. 

To find a solution, equilibrium or steady-state is assumed (i.e. $\frac{dN}{dt} = 0$) for all species. Let's write eq.~\ref{leaky_unstable} in terms of characteristic grammage, since it has been already estimated. The characteristic escape grammage is defined as $\Lambda^{esc} = n \beta c \tau^{esc}$ (here we assume $\beta=1$ for all particles) and the characteristic decay grammage of primaries as $\Lambda^{dec}_{ij} = m_H n \beta c \gamma_i B_{ij} \tau^{dec}_i$, where the term $B_{ij}$ is the branching ratio, that tells the probability of particle $i$ decaying into particle $j$. The characteristic decay grammage for the secondary CRs of the species $j$ is $\Lambda^{dec}_j = m_H n \beta c \gamma_j \tau^{dec}_j$. Then, the characteristic spallation grammage is $\Lambda^{spall}_{ij} = m_H / \sigma_{ij}$. Finally, the density of the $j$-th CR species can be written in the form:
\begin{equation}
    N_j = (\Lambda^{esc} + \Lambda^{dec}_j) \sum_i N_i \left ( \frac{1}{\Lambda^{spall}_{ij}} + \frac{1}{\Lambda^{dec}_{ij}} \right)
    \label{leaky_Be10}
\end{equation}

One of the most interesting consequences of the Leaky Box model is that the escape time (or grammage) must be a function of energy to reproduce the observational data. This can be easily seen considering equation~\ref{leaky_prim} for a stable particle (no decay term), whose steady state solution is: 
\begin{equation}
    N_i = \left( \frac{1}{\Lambda^{spall}_i} + \frac{1}{\Lambda^{esc}} \right)^{-1}  \frac{q_i}{\beta_i c}
    \label{leaky_estable}
\end{equation}
with $q_i = Q_i m_H n$. 

Assuming now that inelastic scattering cross sections follow $\sigma_{inel} \approx 45 \cdot A^{0.7}$ mb, we can calculate the spallation term for protons to be $\Lambda^{spall} \sim 1.7\cdot 10^3$ $g/cm^{2}$ and realise the $\frac{1}{\Lambda^{spall}_i}$ term can be neglected at high energies (at low energy resonances may appear and this relation does not hold) since $\frac{1}{\Lambda^{esc}} \sim \frac{1}{4.3}$ $g/cm^{2}$ for particles at $10 \units{GeV}$, as calculated above. Therefore, using a power law source term, $q_i \propto E^{-2}$, equation~\ref{leaky_estable} can be approximated as $N_i \sim K E^{-2} \Lambda^{esc} \sim K E^{-2.7}$, for galactic CRs. It follows that the energy dependence of the characteristic grammage of escape must be a power law of the form $\Lambda^{esc} \propto E^{- \delta}$.

However, this model does not give an explanation to the energy dependence of this escape time. Quoting \cite{codino2008misleading}: "these models (Leaky Box models) exploit the notion of equilibrium between creation and destruction processes of cosmic ions in an undifferentiated arbitrary volume representing the Galaxy, ignoring the galactic magnetic field, the size of the Galaxy, the position of the solar cavity, the spatial distribution of the sources, the space variation of the interstellar matter and other pertinent observations". In addition to this, its predictions started to disagree when more precise experimental measurements were performed, as happened with the predicted secondary-over-primary ratios or the spectrum of radioactive species \cite{codino2008misleading}, becoming just an approximation only good for certain studies \cite{ptuskin2009leaky}.

Thus, they were updated to be more consistent with data and an interesting upgrade was the incorporation of a constant extra grammage associated to the source. These are called nested Leaky Box models \cite{cowsik1975nested}. They assume that the characteristic escape grammage has the form $\Lambda^{esc} = \Lambda_{source} + \Lambda_0 \beta E^{-\delta}$. Here, the $\beta$ term is embedded in the term of grammage (see equation~\ref{leaky_estable}).

However, the knowledge on the plasma environment of the interstellar medium leads physicists to explain this non-ballistic motion of cosmic rays as a diffusion generated from the combined effect of regular magnetic fields plus turbulent magnetic fields, created locally due to the magneto-hydrodynamics waves. The observed small degree of linear polarization of the synchrotron radiation from galactic CR electrons indicates that the galactic magnetic field contains a large turbulent component. The galactic magnetic field intensity can be expressed as $B = B_0 + \delta B$ where $B_0$ is the steady component and $\delta B$ is the turbulent one. 

Turbulence is observed in everyday fluids, like smoke from cigarettes or fast flowing rivers, being a common and not well understood phenomenon in fluid dynamics. In the presence of turbulence (generated by energy-momentum exchange or interactions of some external agent with the ambient plasma) in magnetized plasmas, MHD waves arise. These are magnetic transverse waves propagating along the magnetic field lines (Alfvén waves) or longitudinal waves propagating across it (compressional or magnetosonic). While the Alfvén waves are relatively undamped (neutrals in the medium or Landau damping are the main damping mechanisms), magnetosonic waves are highly damped by the medium. The Alfvén velocity is defined as $V_A = \sqrt{\frac{B}{4\pi \rho}} \sim 2.18\cdot 10^5 \frac{B}{1 \mu G}\left(
\frac{1 cm^{-3}}{n}\right)^{1/2}$ $cm/s$. In the same way the onset of turbulence can be predicted by the Reynolds number, there is a magnetic Reynolds number in the MHD context as well.

At this point the picture is slightly more complete: CR diffusion is isotropic along the full halo, until the particles escape freely through its boundaries into intergalactic space, where the CR density is assumed to be negligible. The most likely scattering mechanism of CRs is pitch-angle scattering by the turbulent magnetic field fluctuations, as Coulomb scatterings with the ISM particles are far too slow due to the low ISM gas density.

Charged CR particles are considered to follow the guiding center principle in the ambient magnetic field, i.e. they will be spiraling around the lines of the regular magnetic field. They undergo an acceleration perpendicular to the magnetic field, as $d\vec{p}/dt = q \vec{v} \times \vec{B}/c$, as no large scale electric field can exist since plasmas are electrically neutral. Considering a regular magnetic field of the form $\vec{B} = B \hat{e}_z$, the x and y components of velocity have the form $v_x = v_{\bot} cos(\Omega t + \phi)$ and $v_y = v_{\bot} sin(\Omega t + \phi)$, while the z component will remain constant. Here $\Omega = \frac{q B_0}{m c \gamma} $ is the Larmor gyrofrequency. In the framework of the quasi-linear theory (QLT) a small perturbation may appear such that $\delta B \ll B_0$ and $\delta B$ $\bot$ $B_0$, changing the equation of motion into $d\vec{p}/dt = q \vec{v} \times (\vec{B}+\delta \vec{B})/c$. The momentum parallel to the magnetic field is just $p_{\parallel}=|\vec{p}| cos (\hat{p} \times \vec{B}) = |\vec{p}| \mu$, where $\mu$ represents the cosine of the angle between particle velocity and magnetic field orientation, the so-called pitch angle. We can then write the following equation:  
\begin{equation}
    \frac{d p_{\parallel} }{dt} = \frac{|\vec{p}| d \mu }{dt} = q \frac{|\vec{v} \times \delta \vec{B}|}{c} = \frac{q}{c} \frac{|\vec{p}|}{m\gamma} \sqrt{1- \mu^2} \left[ cos(\Omega t) \delta B_y - sin(\Omega t) \delta Bx \right]
\label{Loretz_parallel}    
\end{equation}
since the Lorentz force which does not change the magnitude of the momentum, and were we have set $\phi=0$ for simplicity and used $v_{\bot} = |\vec{v}|\sqrt{1 - \mu^2}$.  The perturbation terms can be better expressed as $ \delta B_y = |\vec{\delta B}| cos (kz + \psi)$ and $\delta B_x = |\vec{\delta B}| sin (kz + \psi)$. Hence, recalling the identity $cosA cosB + sinA sinB = cos(A-B)$, this becomes:
\begin{equation}
    \frac{d \mu }{dt} = \frac{q|\vec{\delta B}|}{m\gamma c} \sqrt{1- \mu^2} cos(\Omega t - kz -\psi)
    \label{mut}
\end{equation}

In addition, the z coordinate can be rewritten in terms of the particle's velocity in the z direction as $z = v \mu t$ (to simplify the notation we set $|\vec{v}|=v$) to have:
\begin{equation}
    \frac{d \mu }{dt} = \frac{q|\vec{\delta B}|}{m\gamma c} \sqrt{1- \mu^2} cos[\left(\Omega - v\mu k) t - \psi)\right]
    \label{mut2}
\end{equation}

Integrating eq. \ref{mut2} we get:
\begin{equation}
    \Delta \mu = \frac{q|\vec{\delta B}|}{m\gamma c} \sqrt{1- \mu^2} 
     \int dt' cos[\left(\Omega - v\mu k -\psi) t' \right] 
     \label{deltamu}
\end{equation}
At this point, averaging equation~\ref{deltamu} over wave phase makes the average pitch angle change in time vanish ($< \Delta \mu >_{\psi} = 0$). On the other hand, the variance is given by: 
\begin{equation}
    \Delta \mu \Delta \mu = \left(\frac{q|\vec{\delta B}|}{m\gamma c}\right)^2 (1- \mu^2) \int dt' cos[\left(\Omega - v\mu k) t' - \psi)\right] \int dt'' cos[\left(\Omega - v\mu k) t'' - \psi)\right]
    \label{varmu}
\end{equation}
which, using the property $cos A cos B = \frac{1}{2} [cos(A+B) + cos(A-B)]$ to simplify it, becomes:
\begin{equation}
    \Delta \mu \Delta \mu = \left(\frac{q|\vec{\delta B}|}{2m\gamma c}\right)^2 (1- \mu^2) \int dt'\int dt'' \left\{ cos[\left(\Omega - v\mu k) (t'+t'') - 2\psi  \right] +  cos[\left(\Omega - v\mu k) (t' - t'')\right] \right\}
    \label{varmu2}
\end{equation}
Here we are assuming that the variation of $\mu$ after a collision is small ($\left<\Delta \mu\right> \sim 0$), so that this term was taken out from the integral in time. 

The first cosine term in equation~\ref{varmu2} averages over phases to 0, while the second cosine term can be replaced by a delta function since $\delta(a) = \frac{1}{2 \pi} \int_{-\infty}^{\infty} dt e^{iat}$ and $cos(at) = \frac{e^{iat} + e^{-iat}}{2}$, obtaining:
\begin{equation}
\left<\Delta \mu \Delta \mu\right>_{\psi} = \left(\frac{q|\vec{\delta B}|}{2m\gamma c}\right)^2 (1- \mu^2) 2\pi \delta(\Omega - \mu v k) \Delta t 
    \label{muvarave}
\end{equation}

Finally, defining the resonance wavenumber as $k_{res} \equiv \Omega / \mu v = 1 / \mu r_g$, where $r_g$ is the Larmor radius, and since $\delta (\Omega - v \mu k) = \frac{1}{v\mu} \delta (k - k_{res})$, we get: 
\begin{equation}
\left<\frac{\Delta \mu \Delta \mu}{\Delta t}\right>_{\psi} = \frac{\pi q^2 v (1-\mu^2)}{c^2 p^2 \mu} |\vec{\delta B}|^2 \delta(k - k_{res}) = \pi (1-\mu^2) \frac{|\vec{\delta B}|^2}{B_0^2} k_{res} \Omega \delta (k-k_{res})
    \label{mutime}
\end{equation}
This quantity is often expressed as the pitch angle diffusion coefficient, $D_{\mu \mu} = \frac{1}{2}\left<\frac{\Delta \mu \Delta \mu}{\Delta t}\right>_{\psi}$, but it is customary to use the rate of scattering in pitch angle:
\begin{equation}
\nu = \frac{1}{2}\left<\frac{\Delta \theta \Delta \theta}{\Delta t}\right>_{\psi} = \pi\frac{|\vec{\delta B}|^2}{B_0^2} k_{res} \Omega \delta (k-k_{res})
\label{ratescat}
\end{equation}
This equation tells us that the only waves which really matter are those with the resonant wavenumber, i.e. when the particle's gyrofrequency and wave interact resonantly. Now we can introduce the power spectrum of $k$ modes, defined as $P(k)dk = \frac{\delta B^2(k)/4\pi}{B_0^2/8\pi}$ to have the correct normalization, whose meaning is the wave power density in the range $dk$. 

Introducing the power spectrum in eq.~\ref{ratescat} we get: 
\begin{equation}
2\nu = \pi \Omega k_{res}\int dk \frac{P(k)}{2} \delta(k-k_{res}) = \frac{\pi}{2} \Omega k_{res} P(k_{res})
\label{ratepower}
\end{equation}

The time required for a particle to effectively scatter is $\tau \sim \frac{1}{\nu} \sim (k_{res} \Omega P(k_{res}))^{-1}$ which, in agreement with experimental data, should be a power law in energy, as stated above when discussing $\tau^{esc}$ in the Leaky Box approximation. The associated spatial diffusion coefficient (parallel to the magnetic field lines) to this interaction rate is $D \simeq \frac{1}{3} v \lambda_{mfp} = \frac{1}{3} v (v \tau) \propto  E^{\delta}$. To this extent, the power spectrum of magnetic turbulence must be a power law in wavenumber, i.e. $P(k) \propto k^{-\delta'}$, which was predicted in the 1940s with the important advances introduced by Kolmogorov and expanded to magnetised fluids later \cite{frisch1991global}. 

These are simple, heuristic, arguments to show how these pitch-angle interactions can turn out to lead to a diffusion process effective for GCRs. More refined and complex calculations bring to the same formulas. However this process is far from being completely understood since the non-linear regime is needed to understand turbulent phenomena.

Single CR-wave interactions can be seen, from the particle's reference frame, as a wave moving towards it with a shifted wavenumber $v\mu k$, so that when the particle gyrofrequency matches the shifted wave frequency, the CR sees a constant magnetic field for long time (compared with the wave frequency), hence interacting more effectively than if there is a varying field. These random encounters will result in a diffusive walk, which is dominated by resonant interactions.

Besides, due to high electrical conductivity of the plasma, acceleration is not possible from any large-scale electrical field, but only locally, leaving the possibility of stochastic acceleration from these encounters. This translates into a diffusion in momentum space, due to electromagnetic waves, whose coefficient is $D_{pp}$. While this coefficient can be theoretically inferred, for instance from the Fokker-Planck equation or kinetic theory in general, it is usually derived from its relation with the spatial diffusion coefficient. Heuristically, this relation can be found from the fact that while for the spatial diffusion coefficient $D \simeq \frac{1}{3} v^2 \nu^{-1}$, for the momentum diffusion coefficient $D_{pp} \simeq \frac{1}{3} \Delta p^2 / \tau \sim  \nu p^2 \frac{V_A^2}{v^2 \tau}$, so that $D_{pp} D \approx \frac{1}{9}p^2v_A^2$.

A correct derivation of this relation needs to take into account forwards and backwards scattering wave rates ($\nu_+$ and $\nu_-$, respectively). Following the treatment of \cite{osborne1987cosmic} and  \cite{seo1994stochastic}, the diffusion coefficient in momentum space takes the form:
\begin{equation}
    D_{pp} = \frac{4}{3}\frac{1}{\delta (4 - \delta^2)(4 - \delta)} \frac{v_A^2 p^2}{D}
    \label{kpkx}
\end{equation}
where $\delta$ comes from the form of the energy density spectrum they use: $P(k) \propto k^{-4 + \delta}$ 

In conclusion, the propagation of CRs in the Galaxy can be treated as a diffusive process due to the collisionless interactions with plasma waves. A complete description of their transport and interactions requires the combination of the source distribution function, interstellar radiation fields and gas density distributions, knowledge on the turbulent and static magnetic field as well as on spallation and inelastic cross sections, and boundary conditions for all CR species. This description can be enclosed in a series of coupled transport equations for every of the CRs involved and must describe diffusion, convection by the hypothetical galactic wind, energy losses (mainly coulomb interactions) and reacceleration process, in addition to the nuclear collisions with interstellar gas atoms and the decay of radioactive isotopes to account for the full CR nuclei network. \cite{gabici2019origin} and \cite{Grenier:2015egx} are wonderful reviews of the basic phenomenology described above.

\subsection{The diffusion equation for galactic cosmic rays}
\label{diff_eq}

From the picture we have described above, the full steady state $\left(\frac{\partial N}{\partial t} = 0\right)$ transport equation, for any given species $i$, is typically expressed as
\begin{equation}
\begin{split}
\begin{gathered}
\vec{\nabla}\cdot(\vec{J}_i - \vec{v}_{\omega}N_i) + \frac{\partial}{\partial p} \left[p^2D_{pp}\frac{\partial}{\partial p} \left( \frac{N_i}{p^2}\right) \right] %\\
= Q_i + \frac{\partial}{\partial p}  \left[\dot{p} N_i - \frac{p}{3}\left( \vec{\nabla}\cdot \vec{v}_{\omega} N_i \right)\right] \\  - \frac{N_i}{\tau^f_i} + \sum_j \Gamma^s_{j\rightarrow i}(N_j) - \frac{N_i}{\tau^r_i} + \sum_j \frac{N_j}{\tau^r_{j\rightarrow i}}
\label{eq:caprate}
\end{gathered}
\end{split}
\end{equation}
where the flux term $\vec{J_i}$, contains the information on the spatial diffusion by means of the Fick's Law: $\vec{J}_i = -D_{ke} \vec{\nabla} N_i$, where e and k represents spatial components of the diffusion tensor (see equation~\ref{eq:Dtensor}). Here $N_i$ is the density per unit momentum (sometimes expressed as $f(\vec{x}, p)$ too). The second term in the l.h.s. of the equation accounts for diffusion in momentum space (important at low energies), $\vec{v}_{\omega}$ is the advection speed (convection, which is important for low energies, comparable with the wind kinetic energy). The first term in the r.h.s., $Q_i \equiv Q(E, \vec{x})$, represents the distribution and energy spectra of CR sources as function of position, energy and time; the second term describes the momentum (energy) losses; finally, the last four terms are due to fragmentation and decays. The subscript $i$ indicates the CR species, while $j$ indicates any other CR species which can be related with $i$ by these processes. The term $\Gamma^s$ can be written, for any nuclei $j$ and $i$ as $\Gamma^s_{j\rightarrow i} = \beta_j c n_H \sigma_{j\rightarrow i}N_j$, where $n_H$ is the ISM density of H nuclei, $\beta_j c$ is the velocity of the particle j and $\sigma_{j\rightarrow i}$ is the inclusive cross section for the formation of nuclei $i$ from $j$.

Since the full transport equation has no exact solution, approximate solutions can only be found under certain conditions. For example, for high energy protons, for which energy losses, convection and reacceleration are negligible effects and inelastic interactions hardly affect their density function, the diffusion equation is
\begin{equation}
    \frac{\partial N}{\partial t} - D\nabla^2 N = Q(E, \vec{x})
    \label{approx1}
\end{equation}
where the diffusion coefficient was supposed to be independent of position. The usual way to solve it is by means of the Green functions vanishing at the boundaries, i.e. at $r=R$ and at $z=\pm H$, and defined by the following equation (in 1 dimension):
\begin{equation}
        \begin{split}
           \begin{gathered}
    \frac{\partial G(x, x_0, t)}{\partial t} - D\nabla^2 G(x, x_0, t) = \delta(x-x_0) \\
    N(x, t) = \int^{\infty}_{-\infty} G(x, x_0, t) Q(E, x) dx
        \end{gathered}
        \end{split}
    \label{Green}
\end{equation}
% $G(x, 0) = \delta(x)$)
with $x_0$ as the source position. The Green function can be easily calculated by taking its Fourier transform into the differential equation~\ref{Green} and transforming it back again, to get
\begin{equation}
    G(x, x_0, t) = \frac{1}{\sqrt{4\pi Dt}} \exp\left[ -\frac{(x-x_0)^2}{4Dt} \right]
    \label{Greensol}
\end{equation}

Notice that steady state solutions are expected for GCRs, since after long time the Galaxy is stable and reaches equilibrium and there is no evidence of intensity change with time. The non-steady case can happen when there are sudden particle emissions, like in solar flares. In the steady case, equation~\ref{eq:caprate} reduces (approximately) to the Leaky Box solution discussed around equation~\ref{leaky_estable}.

Obviously, the full network can be solved via numeric integration techniques (see section~\ref{sec:Codes}). Equation \ref{eq:caprate} may be approximated to one dimension (direction perpendicular to the galactic plane), supposing complete symmetry in the radial dimension, to obtain exact solutions by means of the Weighted Slab method \cite{ptuskin1999modified, jones1991best, jones2001modified}. It is based on the idea of separating equation~\ref{eq:caprate} in two equations, one for the particle distribution function (density per unit momentum) depending on particle's position, and another for the nuclear interactions occurring in a portion (slab) of the Galaxy, so that the distribution function at each point of the Galaxy can be calculated by ''weighting'' the solutions of the slab equations. The two points that make this technique to be just an approximation are that it neglects low energy losses and that propagation is supposed to depend on the energy per nucleon instead of rigidity (electromagnetic interactions operate on rigidity). However, in ref.~\cite{ptuskin1996using} the authors reported that it could be also made exact for Galactic propagation models in which energy gains and losses are proportional to the same mass density that determines nuclear fragmentation and time-dependent processes.

To illustrate it, let's solve a steady-state equation with the inclusion of energy losses and inelastic interactions:
\begin{equation}
-\frac{\partial}{\partial z} D \frac{\partial N}{\partial z} + \frac{\mu_S v \sigma}{m_H} \delta(z)N + \frac{1}{p^2} \frac{\partial}{\partial p}\left[p^2\left(\frac{dp}{dt}\right)_{ion}N\right] = Q(E)\delta(z)
\label{WSeq}
\end{equation}
The parameter $\mu_S$ is the surface mass density of the galactic disc \cite{ferriere1998global}, $\mu_S \sim 2.4\cdot 10^{-3} \units{g/cm^2}$. The solution for the density function for $z \neq 0$ is 
\begin{equation}
N = N_0 \frac{H - |z|}{H}.
\label{fsol}
\end{equation}

The ``slab equation'' would be:
\begin{equation}
\frac{2D}{H}N_0 + \frac{\mu_S v \sigma}{m_H}N_0 +  \frac{1}{p^2} \frac{\partial}{\partial p}\left[p^2 \mu_S \frac{b_0}{m} N_0\right] = Q_0(E)
\label{slabsol}
\end{equation}
after an integration of eq.~\ref{WSeq} around $z \pm 0^{\pm}$ where the ionization losses where considered to be $\left(\frac{dp}{dt}\right)_{ion} = \frac{\mu_S b_0(p)}{m_H}\delta(z)$. Solving equation~\ref{slabsol} in terms of CR density as function of kinetic energy ($N_0 = I(E_k) A/p^2$, with $A$ as the atomic number) we get a simplified equation of the form:
\begin{equation}
\frac{I(E_k)}{X_{esc}} + \frac{d}{dE_k}\left[\left(\frac{dE_k}{dx}\right)_{ion}I(E_K)\right] + \frac{\sigma}{m_H}I(E_K)  = Q_0(E_k)
\label{WSeq2}
\end{equation}
whose solution can be easily found (the energy loss term is proportional to $I(E_k)$ since it has power law form) for specific ionization losses processes. The term $X_{esc} = \frac{\mu_S v H}{2D}$ is the mean grammage traversed at $z=0$. The fact that the energy per nucleon is conserved in spallation reactions (see \cite{Tan_1983}, to understand the extent of this approximation) makes more convenient to express the intensity as a function of energy per nucleon.

\begin{figure*}
\begin{center}
\includegraphics[width=0.9\textwidth, height=0.235\textheight]{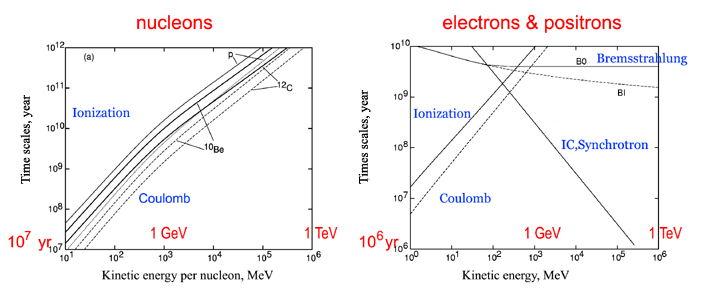}    
\end{center}
\caption{\footnotesize Time scales for energy losses for nucleons, for which the main loss are due to ionization and Coulomb scattering, and leptons, which suffer also from interactions with the magnetic and radiation fields. Ionization losses are represented by solid lines and Coulomb losses by dashed lines. The left panel shows nucleon timescales (protons, $^{10}$Be and $^{12}$C) while the right panel illustrates the timescale losses for electrons, including also IC, synchrotron (as solid lines) and bremsstrahlung (in neutral, denoted as B0, and ionized, denoted as BI, hydrogen) energy losses. Taken from ref.~\cite{Strong:1998pw}.}
\label{fig:Elosses}
\end{figure*}

Concerning energy losses, charged particles may undergo bremsstrahlung interactions with interstellar gas in addition to ionization and Coulomb losses ($dE/dx \propto \beta^{-2}log(E^2) $) and adiabatic expansion losses (discussed above). They are not important at high energies (>$1\units{GeV}$), although bremsstrahlung losses can be important for leptons in the MeV region (as we show in section~\ref{sec:emissivity} of the chapter~\ref{sec:5}). Moreover, magnetic and radiation fields play crucial roles in leptons (due to the low electron mass) synchrotron and inverse Compton interactions.
Figure~\ref{fig:Elosses} shows the time scales of energy loss ($E \cdot (\frac{dE}{dt})^{-1}$) of the discussed processes, from \cite{Strong:1998pw}, to have an idea of their relative importance at different energies.

To solve these previous equations, the diffusion coefficient has been considered to be isotropic. Nonetheless, anisotropic diffusion is physically motivated by the presence of the large scale Galactic magnetic field, which introduces a preferred direction. A compact way to describe the diffusion coefficient as a tensor that%, under the assumption of azimuthal symmetry of the system, 
can be expressed as \cite{ptuskin1993diffusion}:
\begin{equation}
D_{ij} \equiv D_{\bot}\delta_{ij} + (D_{\parallel} - D_{\bot}) b_ib_j + D_A(x)\epsilon_{ijk}b_k
\label{eq:Dtensor}
\end{equation}
where $b_i$ is the magnetic field versor along the $i$ direction, i.e. $b_i = \frac{B_i}{|\vec{B}|}$, the symmetric components $D_{\parallel}$ and $D_{\bot}$ are the diffusion coefficients along and across the regular field, and the term $D_A$ is the antisymmetric diffusion coefficient, which takes into account CR drifts (only important at energies $E > PeV$ for GCRs and when carefully treating the solar modulation in the heliosphere).

Although the term $D_{\bot}$ is not totally well understood so far, its relation with $D_{\parallel}$ is, from quasi-linear theory, $\frac{D_{\bot}}{D_{\parallel}} \sim P(k)^2 << 1$. Monte Carlo MHD simulations showed that $D_{\bot} \sim 0.1 D_{\parallel}$ \cite{candia2004diffusion, de2007numerical}, which is expected to be due to the fact that the total power of waves perpendicular to magnetic field lines is about a $10\%$ of the total.

When considering the Galactic magnetic field to be purely azimuthal, the parallel diffusion vanishes (by the symmetric configuration of the galaxy in the z direction). Neglecting the drifts leaves the diffusion tensor as an homogeneous diffusion coefficient, $D \rightarrow D_{\bot}$, which means that the main confinement mechanism in the Galaxy is the perpendicular diffusion (for more details see \cite{Cerri:2017joy}). 

The diffusion coefficient is usually assumed to obey a power law in energy, which most general form is:
\begin{equation}
D = D_0 \beta^{\eta}\left(\frac{R}{R_0} \right)^{\delta} F(\vec{r}, z)
\label{eq:depdiff}
\end{equation}
where the spatial function term, $F(\vec{r}, z)$, is not well known and seems to be $\neq 1$. If this spatial function is not constant, the diffusion term, $\vec{\nabla} J_i$, of equation~\ref{eq:caprate} can be also written as $\vec{\nabla} (D \vec{\nabla} N) = D [N] + \vec{u}_D \cdot \vec{\nabla} N$, with the diffusion velocity defined as $\vec{u}_D \equiv \vec{\nabla} D$. The $\eta$ parameter describes complicated physical effects that may play a major role at low energy as the dissipation of Alfvén waves. In addition, phenomenological parametrisations can present breaks in the diffusion spectral index, $\delta$. At low energies, a break can be justified by wave damping, making the particles to diffuse more easily \cite{Ptuskin_2006}. At high energies, this is a feature observed experimentally (see section~\ref{sec:Exps}) and has different possible explanations, as, for example, the transition from self-generated turbulence to preexisting turbulence \cite{Blasi:2012yr, Aloisio:2013zia, Aloisio:2013tda, Evoli:2018nmb}.

Secondary CRs allow us to tune the diffusion coefficient parameters ($D_0$, $\eta$ and $\delta$) as well as reacceleration processes (by means of the effective Alfvén waves speed, $V_A$) and possible convective winds ($v_w$). Inclusive cross sections play a key role in modeling the secondary spectra, while inelastic cross sections, since the spallation term is not dominant at medium-high energies, play a secondary role for all CR species. Primary nuclei are not significantly affected by these parameters, so they are used to determine the source term. On the contrary, long-lived radioactive secondaries can be studied in order to check the halo size, H. Also lepton data provide constraints for halo size, magnetic field intensities and the interstellar radiation fields when combined with gamma-ray data. K-capture processes, undergone by heavy nuclei, specially in the sub-iron species, bring valuable information about the gas density and Galaxy structure. Nonetheless, combined analysis become more necessary to find precise answers, and are now possible thanks to the current experimental precision.

\section{CR detection and current experiments}
\label{sec:Exps}

In order to build the theory presented above, direct and indirect measurements of all CR species spanning several order of magnitudes in energy have been exploited during the last two centuries. While for GCRs, below the knee, direct detection is possible by means of balloon-borne detectors, satellites or detectors allocated in the International Space Station (ISS), UHE-CRs are indirectly detected at Earth via the study of the generated particle showers. While the precision for GCR detection is smaller than 10\%, larger uncertainties are present in the measurements of UHE-CRs. 

Facilities devoted to above-knee-CRs can take up areas of square kilometers, and basically study the initial CR particle by the features of the shower developed after its interaction with atmospheric nuclei. They can be based on the direct detection of the secondary particles generated (Telescope Array~\cite{Fukushima:2003ig}; HAWC~ \cite{1352672}; ARGO~\cite{Camarri:2017uuv}; KASKADE-Grande~\cite{inproceedings}), on the detection of Cherenkov light emitted by the showers (IACT telescopes, like HESS~ \cite{de2019hess} or MAGIC~\cite{Cortina:2009jj}) or on the combination of secondary particle detection with atmospheric fluorescence, like in the Pierre Auger Observatory (PAO)~\cite{Mockler:2019ujr} or the High Resolution Fly's Eye (HiRes)~\cite{hires}. A complete review of the recent results and implications on UHE-CRs is~\cite{10.1093/ptep/ptx054}.

In turn, while for UHE-CRs enormous areas are needed given the small particle fluxes, GCR detectors are very reduced in size, since they must operate in the top of the atmosphere (balloons) or in space. These detectors are generally equipped with a tracker system embedded in a magnetic field along with a calorimeter and a trigger system. 

Balloon experiments were carried out since the discovery of CRs. Scientific balloon payloads have been flown for periods of 1–2 days since large polyethylene balloons were first introduced in the 1950s. Current balloon experiments can carry payloads up to 3600 kg and fly at altitudes up to 42 km~\cite{Seo:2012pw}. These experiments are able to detect CRs up to the knee region, due to their small size compared to ground based detectors. Current balloon experiments are more focused on the detection of antimatter (like the CAPRICE~\cite{Weber:1997zwa} and the HEAT~\cite{DuVernois:2001bb}  missions), using very constraining trigger conditions. The most representative examples of this kind of detectors are the Balloon Experiment with Superconducting Spectrometer (BESS) ~\cite{abe2008measurement}, the cosmic ray energetic and mass (CREAM)~\cite{ahn2008measurements} and high-altitude balloon and Advanced Thin Ionization Calorimeter (ATIC)~ \cite{Panov:2011ak}.

\begin{figure}[!th]
	\centering
	a)
	\includegraphics[width=0.42\textwidth, height=0.235\textheight]{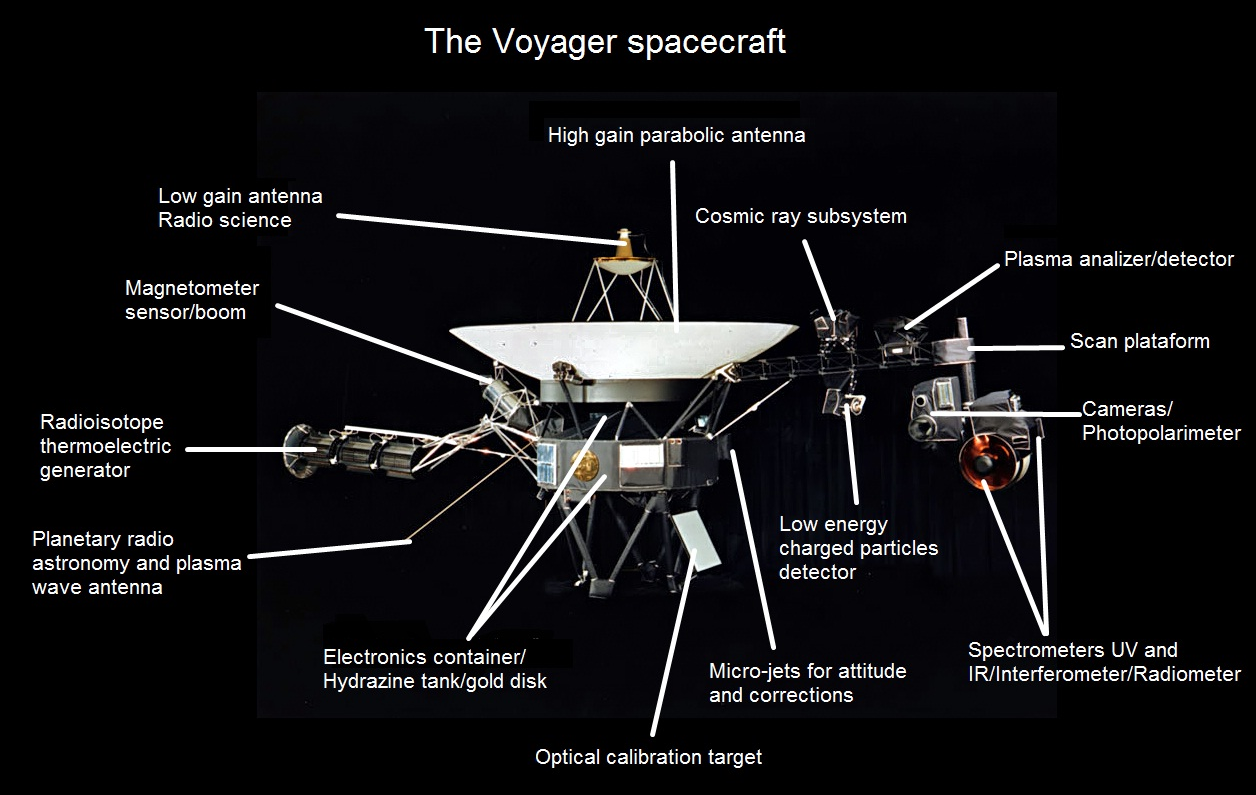}
	\hspace{8mm}
	b)
	\includegraphics[width=0.42\textwidth, height=0.235\textheight]{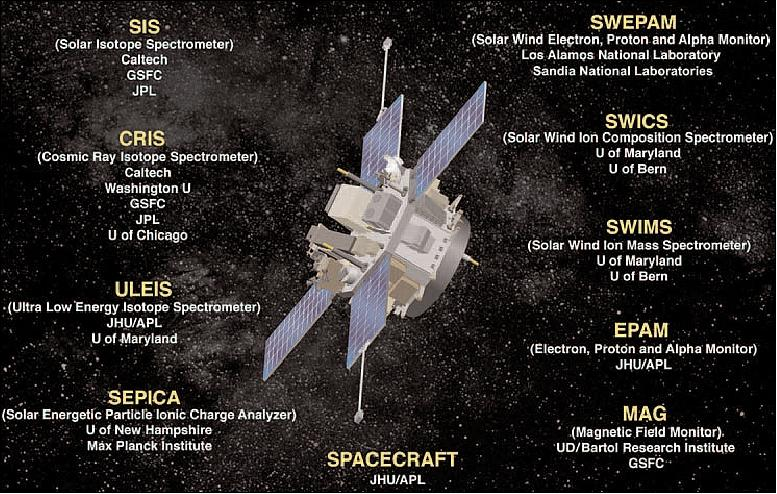}
	
	\vspace{5mm}
	
	c)
	\includegraphics[width=0.44\textwidth, height=0.24\textheight]{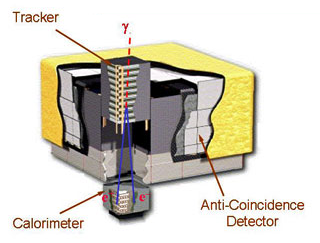}
	\hspace{9.5mm}
	d)
	\includegraphics[width=0.41\textwidth, height=0.20\textheight]{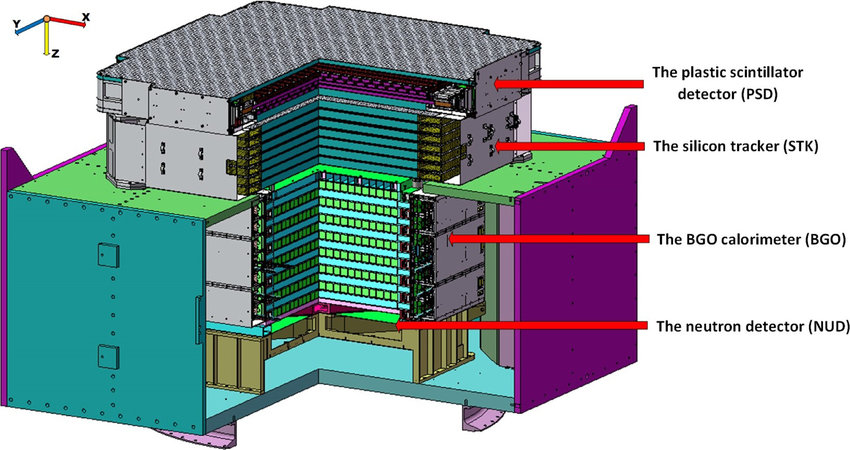}
	\caption{\footnotesize Sketches of various important detectors for the study of GCRs: a) diagram of the Voyager-1 components (Credits: Wikipedia); b) instruments onboard the ACE satellite to study the particles' environmental medium and CRs (credits: \url{https://directory.eoportal.org/web/eoportal/satellite-missions/a/ace}); c) representation of the Fermi-LAT detector, intended to study diffuse gamma rays and leptons. Plot taken from~\cite{Space_crafts}; d) the DAMPE instrument, which offers the highest accuracy in lepton detection. Taken from~\cite{CHANG20176}.}
	\label{fig:ACE_Voy_FERMI_DAMPE}
\end{figure}

On the other hand, space missions are launched in rockets, either to put the satellite in its orbit, or to place the detector near the ISS to be attached there. One of the first missions that are bringing valuable information is the Voyager program~\cite{cummings2018galactic, krimigis1977low, heacock1980voyager}. It consists of two probes (Voyager 1 and Voyager 2) launched in 1977 that are now exploring the outer boundary of the heliosphere. This means that they are out of the influence of the solar magnetosphere, which enables to take data exempt of solar modulation effects. Although these probes were not intended at precise CR studies (more on the investigation of the plasma properties and outer-planets view), the possibility of taking data out of the solar modulation makes them essential nowadays, as they can take low-energy data with a precision around 10\%. A sketch of the Voyager 1 spacecraft can be seen in Fig.~\ref{fig:ACE_Voy_FERMI_DAMPE} a). 

% Fermi plot and also PAMELA from https://www.researchgate.net/figure/Schematic-view-of-PAMELA-left-Fermi-LAT-center-and-AMS-02-right-detectors_fig1_303715112

In Ref.~\cite{neronov2017cosmic}, the authors review the consequences of the comparison of Voyager 1 data with measurements made by other missions at higher energies, in the energy region between 1 and 10 \units{GV}. This comparison clearly points to the fact that, at low energies, ISM CR data are not compatible with the local measurements (inside solar magnetosphere) unless there is (at least) one break in the CR spectrum at around $8-10 \units{GV}$. The same is observed in several studies of gamma-ray emissions from SNR (for instance, \cite{malkov2011mechanism}). This must be due to the fact that, when the SNR shock expands out approaching the ISM gas, it mixes with neutrals that produce a damping on low frequency Alfvén waves ($w \sim k v_A$). As resonant interactions depends on the wave frequencies ($k p_{\parallel}/m_H = w_{cyclotron}$), those particles with high parallel (with respect to the magnetic field lines) momentum easily escape \cite{kulsrud1969effect, zweibel1982confinement}. As a consequence, the source term in the diffusion models incorporates a break around $8\units{GV}$ in their energy dependence.

To highlight another satellite mission taking precise data at the low energy part of the spectrum (but in the local ISM) we must mention the ACE (Advanced Composition Explorer) probe, whose main purposes are the study of the origin and isotopic composition of CRs, formation of the solar corona and acceleration of particles at the solar wind~\cite{stone1998advanced, stone1998cosmic}. All the technical information of this space craft can be taken from \url{http://www.srl.caltech.edu/ACE/}. Fig.~\ref{fig:ACE_Voy_FERMI_DAMPE} b) shows a sketch of its main detectors. The Cosmic Ray Isotope Spectrometer (CRIS) is intended to make measurements on the isotopic composition of CR with excellent isotopic resolution up to Z=30 (Zinc) and beyond. The measurement of the $^{10}$Be flux is a key piece for the determination of the halo size, since the mean lifetime of the element and this apparatus provides the most precise measurement of the isotope's flux, although only at low energy. 

%\newpage

\begin{figure}[!th]
	\centering
	a)
	\includegraphics[width=0.37\textwidth, height=0.27\textheight]{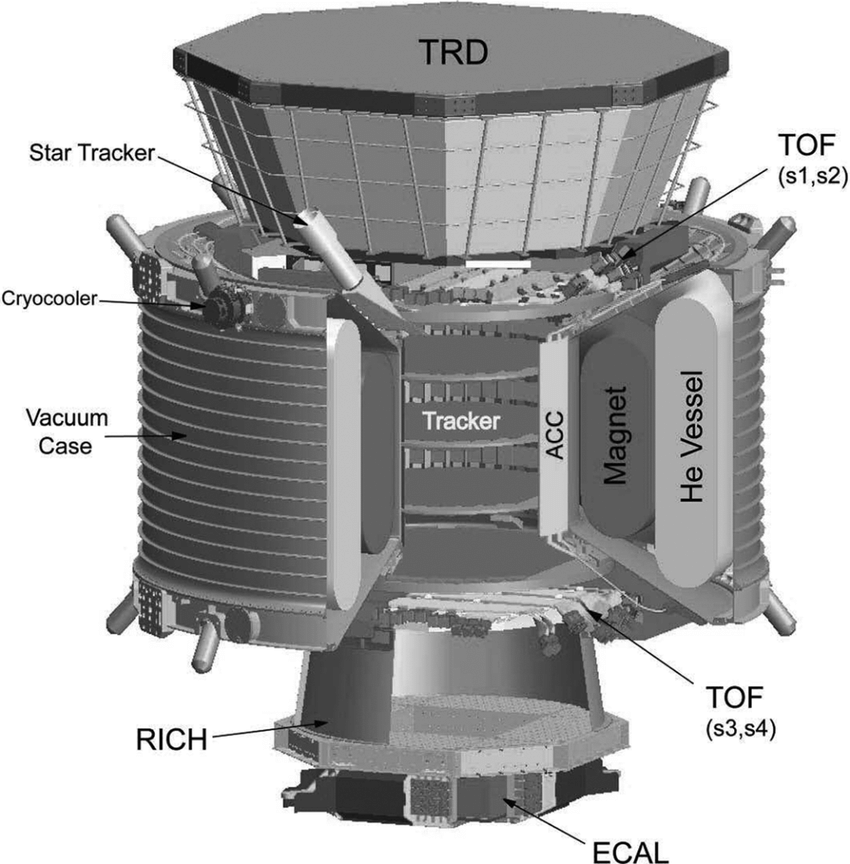}
	\hspace{7mm}
	b)
	\includegraphics[width=0.49\textwidth, height=0.27\textheight]{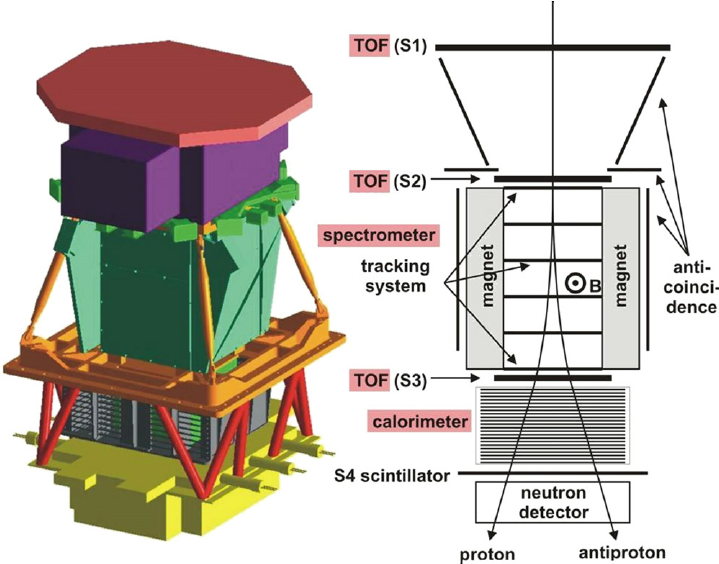}
	\caption{ \footnotesize The recent most important detectors for GCRs: a) AMS detector (\url{https://ams02.space/}) which has unveiled non-expected features of the physics of GCRs thanks to its unprecedented accuracy; b) scheme of the PAMELA instrument (\url{https://pamela.roma2.infn.it/}), highlighting its main parts. It started a new era of precise measurements for the study of GCRs.} 
	\label{fig:CR_experiments}
\end{figure}

\begin{figure}[!th]
	\centering
	a)
	\includegraphics[width=0.415\textwidth, height=0.22\textheight]{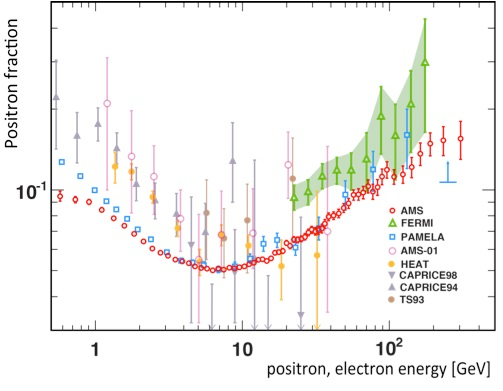}
	\hspace{7mm}
	b)
	\includegraphics[width=0.415\textwidth, height=0.22\textheight]{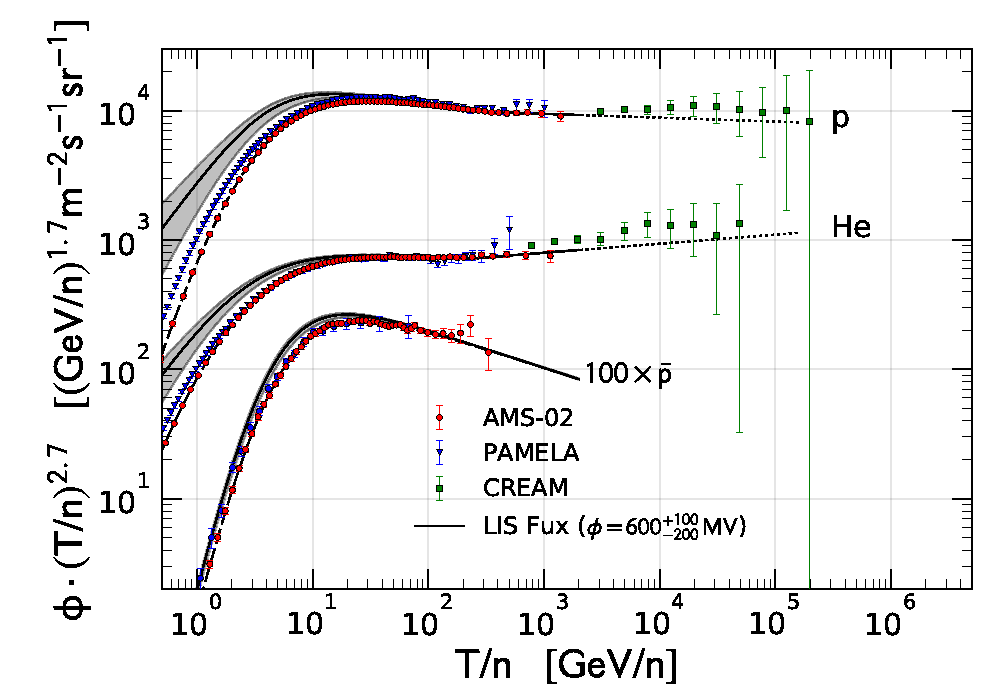}
		\vspace{5mm}
	c)
	\includegraphics[width=0.415\textwidth, height=0.22\textheight]{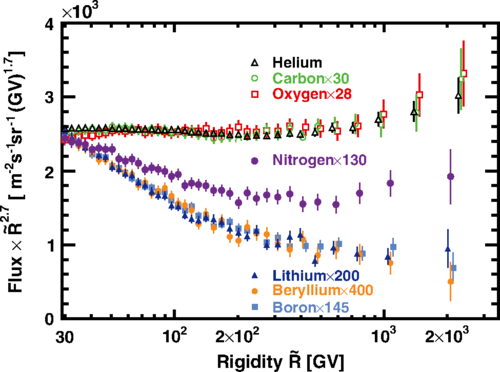}
	\hspace{9.5mm}
	d)
	\includegraphics[width=0.415\textwidth, height=0.22\textheight]{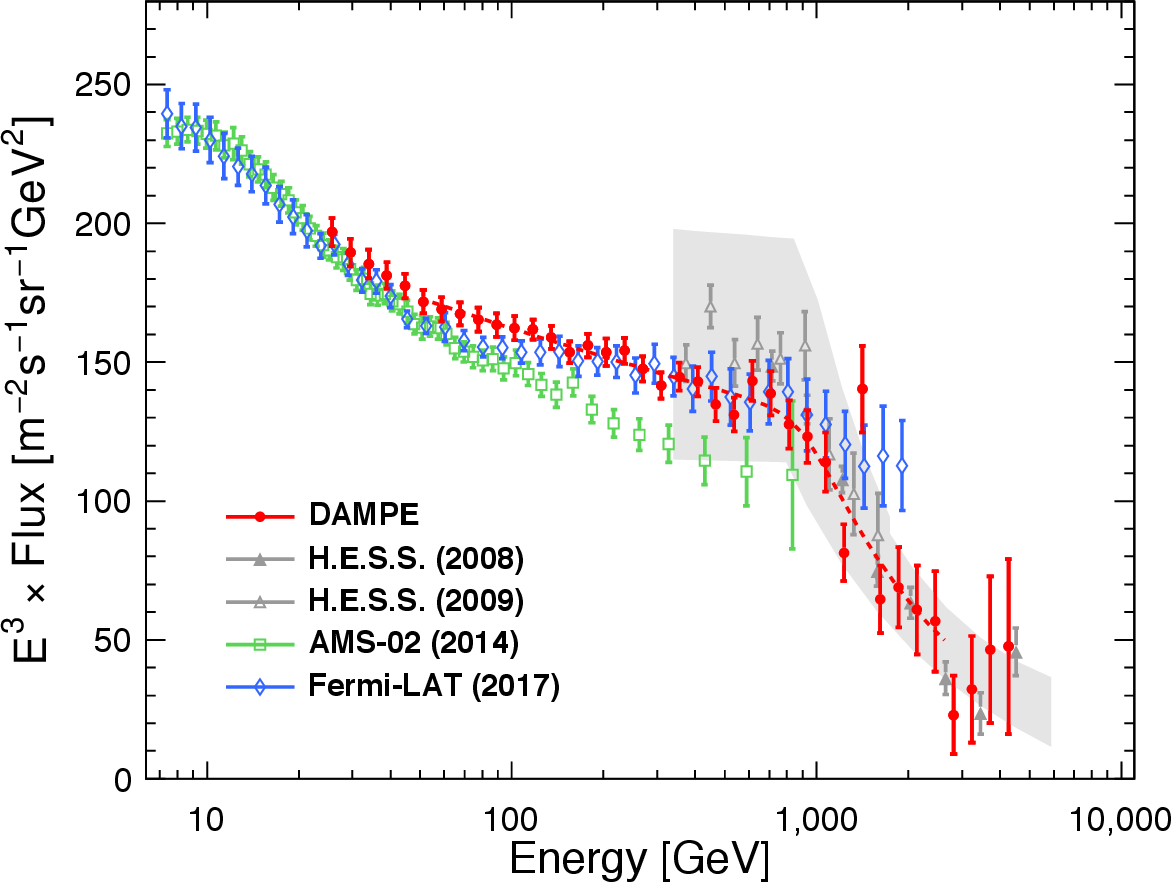}
	\caption{\footnotesize Compilation of recent measurements for leptons, antiprotons and CR nuclei to highlight their main features. a): Ratio of positrons over (positrons + electrons) presented in \cite{Aguilar:2019ksn}, that shows a rising trend not expected considering the positrons as pure secondary CRS. b): Protons, helium and antiprotons spectra showing the different trends they present. It seems that the hardening (at a rigidity around $300 \units{GV}$) is more pronounced for helium (as well as for heavier nuclei)  than for nuclei \cite{Donato:2017ywo}. c): Comparison of the AMS measurements for the trends of pure primary cosmic ray fluxes, pure secondary CR fluxes and nitrogen (which is a nearly even admixture of primary and secondary CRs), taken from \cite{Aguilar:2018keu}. The break can be seen at the same rigidity for all species, being doubly pronounced in the secondary nuclei (the change in the spectral index at this rigidity is double that for primary CRs). d): The CRE spectrum \cite{Ambrosi:2017wek}, which seems to evidence a sharp cut-off at around $1 \units{TeV}$.} 
	\label{fig:positrons_PAM}
\end{figure}

On the other side, the precision of CR spectra measurements did not allow precise tests of the diffusion theory until the PAMELA experiment~\cite{Adriani:2002sd, Adriani:2014pza} came out. A sketch of the instrument is shown in Fig.~\ref{fig:CR_experiments} b). In particular, its determination of the boron-over-carbon spectrum enabled to rule out old theories and to show up new features. On top of this, first accurate results on positrons~\cite{Adriani:2008zr} and antiprotons~\cite{Adriani:2008zq} were possible at higher energies than ever, unveiling the unexpected behaviour they present at high energies. Dark matter explanations for both excesses were searched, but the necessary cross sections and masses of dark matter candidates are way too large to reproduce the positron data (Fig.~\ref{fig:positrons_PAM}). As explained above, pulsars are possible candidates to solve the positron puzzle as well as other primary sources~\cite{DeRujula:2019wbk}. On the other hand, the antiproton spectrum is not totally well reproduced yet, and many works are performed to study whether dark matter can explain it \cite{Smorra:2019qfx, Cuoco:2019kuu} but several other studies (for example,~\cite{boudaud2020ams}) show that it is possible to explain their spectrum inside the current total uncertainties. However, the main problem for antiprotons is the accurate determination of their production cross sections~\cite{Korsmeier:2018gcy}.

Nonetheless, the most striking feature discovered by the PAMELA mission was the apparent change in the high energy behaviour of all CR species at energies $\sim300\units{GV}$~\cite{adriani2011pamela}, previously observed for protons and helium by the ATIC-2~\cite{Panov:2011ak} and CREAM~\cite{yoon2011cosmic} missions. In addition to this hardening, the propagated proton spectrum seems to be softer than the helium spectrum at all energies and the heavier primary nuclei (mainly carbon and oxygen) behave very similarly to helium.

%Very nice AMS pictures from from Cosmic rays continue to confound Cern courier S.Ting, September 2016

Recently, the Alpha Magnetic Spectrometer (AMS-02) detector~\cite{duranti2019ams}, on board the ISS, reported accurate CR flux measurements up to 3 TV, confirming the spectral features pointed out by PAMELA~\cite{Weng:2018muy}, with much more precision (around 1\% uncertainties for protons and around 3\% for heavier species). A scheme of the detector is shown in the Figure~\ref{fig:CR_experiments} a). AMS-02 was installed on 19 May 2011 and started to report initial results in March 2013. It incorporates a permanent Neodymium-Iron-Boron magnet, which leads to unprecedented charge discrimination, being able to determine the charge of CRs up to iron (Z=26). Another novelty is the transition radiation detector (TRD), which identifies electrons and positrons with a very high rejection factor against protons.  %Infor from https://www.nasa.gov/mission_pages/station/research/news/ams_how_it_works.html

As discussed in the previous section, the change in the spectral behaviour of nuclei at high energies is not well understood yet. The most credited theories (see~\cite{Vladimirov:2011rn}) are a (already mentioned) diffusion break related to the self-generated turbulence of CRs in the ISM plasma, the possibility of a multi-component flux, where the break is due to different populations of CR sources injecting particles with different spectral indexes, and the hypothesis of local sources with different spectra than the average distant sources. One of the consequences of the diffusion scenario is that it forecasts a rigidity break doubly pronounced for secondary CRs that that for primaries (twice the change in spectral index with respect to that of primary CRs at the break position), while the other scenarios imply same spectral index change. Thanks to the very precise AMS-02 data this feature was checked, finding exactly the behaviour expected from the diffusion scenario~\cite{Jia:2018aze}. However, the other scenarios can not be totally discarded yet, since a combination of these scenarios would also explain the differences found in the position of the break for different nuclei (see Fig.~\ref{fig:positrons_PAM}).

Similar breaks have been also observed in the all-lepton spectrum (electrons+positron, also called CRE) with very good precision by the Fermi-LAT telescope~\cite{Ackermann:2014ula}. Fermi (see Figure~\ref{fig:ACE_Voy_FERMI_DAMPE} c)) is a space observatory launched on 11 June 2008 which is mainly equipped with two instruments: the Gamma-ray Burst Monitor (GBM), intended at the detection of sudden flares of gamma rays, and the Large Area Telescope (LAT), used to perform all-sky surveys studying astrophysical and cosmological gamma-ray related phenomena in the range from $20\units{MeV}$ to $300\units{GeV}$. 

Another important step forward in the knowledge of the very-high-energy spectrum of leptons and the (TeV) gamma-ray sky was the High Energy Stereoscopic System (H.E.S.S.; \url{https://www.mpi-hd.mpg.de/hfm/HESS/}) \cite{Hofmann:2001pj}, located in Namibia. H.E.S.S. constitutes a system of ground-based telescopes, in operation from about the last 17 years, that makes use of Cherenkov techniques that study atmospheric particle showers to obtain information about very-high-energy CR leptons and gamma rays.

Finally, it is worthwhile to mention another very recent instrument aimed at being an extension of the Fermi and AMS-02 missions and complementing measurements from CALET: the Dark Matter Explorer (DAMPE) mission (Figure~\ref{fig:ACE_Voy_FERMI_DAMPE}, d), shows an sketch of the full detector), which is a satellite launched at the end of 2015. Its main scientific goal is the detection of electrons and photons with unprecedented energy resolution in order to identify possible signatures of dark matter. As it can be seen in Figure~\ref{fig:positrons_PAM}, d), there is some discrepancy in the CRE spectrum between satellite-based experiments and AMS-02 (in general instruments settled in the ISS, as CALET) that is not well understood at the moment, but apparently is just due to systematic uncertainties (associated to flux calibrations) not well taken into account.

\section{Current simulation codes}
\label{sec:Codes}

The new generation of instruments studying GCR has revealed that more sophisticated approaches are necessary to face CR propagation. Since we are far from having full MHD simulations combined with interactions of CRs with Alfvén waves, the description of their transport as a diffusive movement with the inclusion of convection, reacceleration, decay and spallation can catch the overall picture. In general, the exact solution for the propagation of galactic cosmic rays (equation~\ref{eq:caprate}) can only be found if some simplifications are assumed. Leaky Box and Weighted Slab models approximate the problem to obtain, despite the inter-dependencies of the many nuclei involved or complicated spatial effects, an effective grammage traversed by CRs. Several codes for solving the equation as accurately as possible are currently public.

Another technique that has been employed to solve the galactic propagation equation is the Monte Carlo method~\cite{Batista:2019nbw, 1997AdSpR..19..817W, farahat2008cosmic, webber1997monte}. They are based on a system of stochastic differential equations~\cite{strauss2017hitch} with the evolution of particles taken as a Markov process. While this approach is frequently used for UHE-CRs, its adoption in GCRs is difficult because of the amount of particles which need to be followed one by one to have low statistical errors and considering the inter-connections of the whole nuclei network (very computationally-demanding).

Equation~\ref{eq:caprate} can be simplified to be solved in a semi-analytical way as it is done in the USINE package~\cite{Maurin:2018rmm}. This code solves the partial derivatives involving spatial coordinates  analytically~\cite{lerche1988energy, Webber:1992dks}, while partial derivatives involving energy coordinates are solved via numerical integration. Its main advantage is to be faster than the rest of the current propagation codes, although the analytical solutions it uses often require simplifying the problem and simpler geometries. The two-zone diffusion model already explained in 1D or 2D can be used. The code is written in C++ and all the technical information about it can be found in \url{https://dmaurin.gitlab.io/USINE/}. 

On the other hand, it is more common to solve the galactic transport equation relying on finite difference numerical schemes (various numerical differentiation techniques can be used, offering very small error, like the Runge-Kutta method used in~\cite{stephens1998cosmic} and more advanced ones). This technique needs the solver to be numerically stable and is slower than the semi-analytical recipe, although it takes no simplification of the problem. 

The GALPROP code~\cite{Strong:1998pw} (\url{https://galprop.stanford.edu/}) was the first code to be publicly available back in 1990s, becoming the most popular code even nowadays. Written in C++ (with some FORTRAN-77 routines incorporated), it makes use of the Crank-Nicholson approach~\cite{crank1947practical} to solve numerically the transport equation and provides detailed description of the geometry of the Galaxy. The time-integration steps are an important aspect for the convergence of the simulation, since for very small time-steps the solution can take even years. In GALPROP the time-steps are controlled by the configuration file. 

Regarding the CR source distribution, a model developed in~\cite{Strong:2004de} is used, which explains the distribution of gamma rays in the Milky Way and is in agreement with pulsar distribution. The interstellar gas distribution, consisting of molecular, atomic and ionized gas, are imported as tables (fits files), as well as the energy densities of the CMB, infrarred and radio radiation fields. 

\begin{figure}[!ht]
	\centering
	\includegraphics[width=0.9\textwidth, height=0.3\textheight]{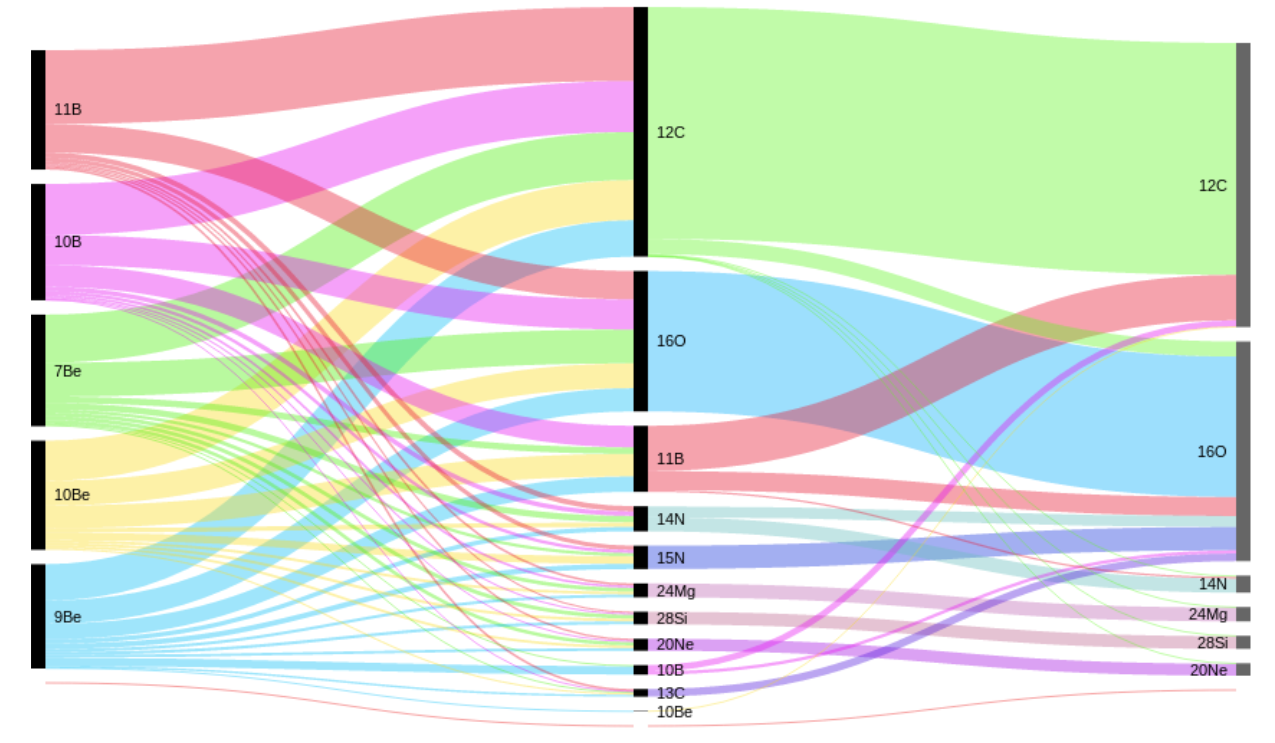}
	\hspace{1mm}
	\caption{\footnotesize Example of a 2-step network for the production of Be and B isotopes, taken from \cite{Tomassetti:2017hbe}. It is useful to visualize the complexity the reaction network presents when all primary and secondary CRs are taken into account. It also shows that important contribution to the flux of secondary cosmic rays come from 2-step (and many-steps, in general) reactions.} 
	\label{fig:XS_net}
\end{figure}

A relevant contribution of the GALPROP team is the detailed parametrisations performed for the nuclear spallation cross-sections~\cite{Moskalenko:2001qm, Strong:2001fu} that are often updated. Since the secondary CRs are the best known probes to keep track of the diffusion parameters, nuclear spallation is a key component of the models (see chapter~\ref{sec:XSecs}). The production of each isotope depends on many reaction channels and, while some of them can be simplified as a direct reaction, a big amount of the isotopes fluxes come from indirect channels from intermediate CRs or from tertiary reactions (e.g. Be, B + ISM $\rightarrow$ B). This is why we refer to the full reaction chain as spallation network (see Figure~\ref{fig:XS_net}, which shows a representation of the complexity of the network). The inelastic cross sections, in turn, are simpler to be calculated. In the GALPROP code, inelastic cross sections are parametrised from the CROSEC code, developed by Barashenkov-Polanski~\cite{unknown}: The hadron-nucleus cross-sections obtained from an interpolation between evaluated experimental data at target mass numbers $A \geq 4$ and energies from $14 \units{MeV}$ up to $1 \units{TeV}$. The nucleus-nucleus cross sections are calculated with the help of an approximated formula with fitted coefficients at energies above several \units{MeV/nucleon}.

\begin{figure}[!th]
	\centering
	\includegraphics[width=0.7\textwidth, height=0.3\textheight]{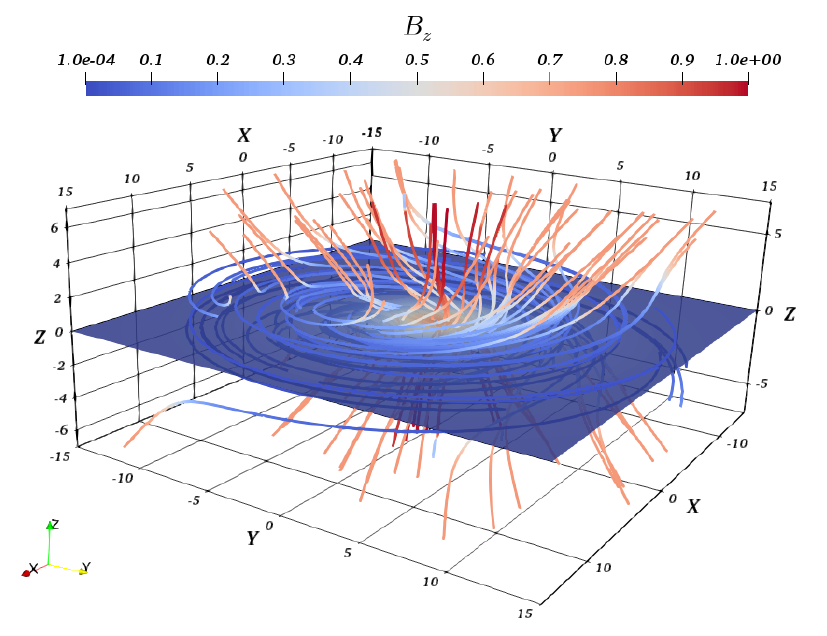}
	\hspace{1mm}
	\caption{\footnotesize Example of the complicated geometry of the magnetic field intensity that can be implemented in DRAGON, taken from \cite{Cerri:2017joy}.}
	\label{fig:B_field3D}
\end{figure}

More recently, the DRAGON (Diffusion of cosmic RAys in Galaxy modelizatiON) team (\url{www.dragonproject.org}) released a propagation code~\cite{Evoli:2008dv}, initially similar to the GALPROP one, but devised to be able to use anisotropic propagation~\cite{Cerri:2017joy}), more realistic 3D spatial structures and switch options to set the cross sections, gas and source distributions and magnetic field details to be used and make the code easily customizable. The code is written in C++ with some FORTRAN routines that can be used for nuclear interactions. The code has been used to perform novel studies like: leptons spectrum in a spatial arm structure for the source distribution~\cite{Gaggero:2013rya}, antiprotons origin related to Dark matter ~\cite{Evoli:2015vaa, Evoli:2011id}, to explain the CR gradient in terms of anisotropic diffusion \cite{Cerri:2017joy}; see Figure~\ref{fig:B_field3D}, where the complicated magnetic field used in DRAGON for this work is displayed) or to reproduce neutrino and gamma ray diffusive emission at very high energies~\cite{Gaggero:2015xza}. The last release of DRAGON, the, still evolving, DRAGON2~\cite{Evoli:2017vim, Evoli:2016xgn} is aimed at covering the most complex and relevant processes involving GCRs and secondary emissions produced during their propagation comprising a wide energy range.

It considers the reaction network, as in the GALPROP code, starting by the heaviest nuclei (the list of nuclei to be propagated can be chosen in DRAGON), solving the propagation and generation of secondary CRs and then moving to the next nuclei (with $A-1$) to complete the network. It takes care of the decays by means of the isotope list. These decays are, in majority, $\beta^{\pm}$ decays, although, for heavy nuclei around the sub-iron group, the electron capture reaction is the main decay process. The code is prepared to switch the order of the reaction network in the presence of $\beta^{-}$ decay, since the resulting nucleus has atomic number $A+1$. Nevertheless, only nuclei with a considerable lifetime are propagated (the minimum lifetime for propagation is also a parameter that can be modified).

Finally, the most recent public code using the finite difference numerical techniques for the solver is the PICARD code~\cite{Kissmann:2014sia, Kissmann:2017ehy} (\url{https://astro-staff.uibk.ac.at/~kissmrbu/Picard.html}), released around 2013. PICARD is able to get high resolutions with reasonable computer resources, using fully 3D implementations~\cite{Kissmann:2015kaa} and anisotropic diffusion~\cite{Kissmann:2017kop}. PICARD, as well as DRAGON, standard time-step sizes (dt) are implemented in the code to be successively reduced until a smallest dt is reached, which should reflect the characteristic time scale of the problem.

\newpage

\chapter{Implications of the nuclear cross sections used in CR propagation codes on the light of the secondary CRs Li, Be and B}%\chapter{Implications of current nuclear cross sections on the light of the secondary cosmic rays Li, Be and B}
\label{sec:XSecs}

Given the current precision of CR data new instruments are achieving, diffusion models can be highly constrained and tested with high precision nowadays. Nevertheless, in order to build an accurate model we must estimate (or get rid of) the uncertainties associated to the many inter-related pieces we need to deal with. As a matter of fact, to find a set of diffusion parameters that reproduce the observed secondary-over-primary ratios we need to take into account the gas density (while in a 2D model this reduces to a mean gas density in the disk, for a 3D model its spatial distribution is also important), the amount of helium in the gas (widely accepted to be around 0.1-0.11), and, essentially, the spallation cross sections. The latter is, obviously, the most important connection between the fluxes of primary and secondary CRs (for which high precision data are now available from AMS-02). Nevertheless, the amount of channels present in the nuclear network and the difficulty in performing cross sections  measurements makes this observable the most uncertain. Uncertainties related to inelastic cross sections are also present, although there are much more existing measurements (and way more precise) given these experiments are simpler to be carried out than the measurements of the spallation cross sections. Since the parameters we use to build our diffusion models are tuned based on the information we get via secondary CRs, with additional information coming from, e.g., gamma rays and synchrotron radiation, there is a clear need of having more precise and abundant (spanning various decades in energy) cross sections measurements.

The measurement of inelastic cross section of a nucleus in a medium consists of determining its mean free path (in units of grammage) $\Lambda_{mfp}(E) = m_{nuc}/ \sigma_{ine}(E)$, by means of the surviving nuclei after traversing a thickness of material, i.e. $\Lambda_{mfp}(E) = X_t/ln(N_0/N(X_t))$, where $X_t$ is the traversed thickness of the target in units of grammage. Here, $N(X_t)$ and $N_0$ are the number of nuclei at the position $X_t$ and the initial number of nuclei, respectively. %Likewise, the survival probability can be simply expressed as $P_{surv}(X_t) = \frac{N(X_t)}{N_0} = exp\left(- n_{t}\sigma_{inel} \right)$, where $n_t$ is the number of target nuclei per unit of area.  
To have a complete set of mean free path measurements with energy this must be repeated for different energies of the incoming nuclei. This procedure must be performed for different traversed thicknesses to reduce statistical uncertainties, and different target materials may require substantially different thicknesses.  An example of these systems can be found in \cite{Webber:1990ka}. 
In turn, determining the inclusive cross sections requires the measurements of multiplicity (n) of different daughter nuclei by counting the number of produced nuclei. Moreover, while for heavy projectiles the energy per unit nucleon is conserved in spallation reactions, the lightest particles deviate from this rule \cite{Tan_1983}, which makes these measurements more laborious. The inclusive cross sections is easily calculated using:
\begin{equation}
\sigma_{inc}(E, E') = \sigma_{ine}(E)  \left<n(E, E')\right>
\label{eq:Inc_Ine}
\end{equation}
where both, inclusive cross sections and average multiplicity, are functions of the kinetic energies of the projectile particle (E) and of the daughter particle (E').

% Inelastic XSecs AMS-02 https://indico.cern.ch/event/820869/contributions/3573608/attachments/1944589/3225879/XSCRC2019XS_QiYan.pdf

On top of this, the complexity of preparing the targets must be considered. For example, measurements of spallation in hydrogen targets are usually performed using solid polyethilene ($CH_2$), thus requiring the subtraction of the carbon cross sections, e.g.: $\sigma_H = \frac{\sigma_{CH_2} - \sigma_C}{2}$.

In addition, little side applications are found for performing these measurements. This explains why, in general, the current knowledge on spallation cross sections is poor and based on a few and scarce data points that hardly reach $\units{GeV}$ energies, which is the most interesting region for CR studies. Indeed, inclusive cross sections (cross sections) are crucial since the best way to constrain propagation parameters (namely, the diffusion coefficient) is by means of the secondary-over-primary CR ratios \cite{reinert2018precision, derome2019fitting}. Slight deviations of certain isotopes might lead to a significant error in the determination of the propagation model or in the resolution of quantities that may mask subtle effects such as CR origins (dark matter annihilation, \cite{genolini2017refined} and other primary origins), acceleration sources (e.g. the ratio $^{22}$Ne/$^{20}$Ne points that a 20\% of the GCRs sources are Wolf-Rayet stars, \cite{Castellina:2011gn}) or other unexpected features \cite{donato2002beta}.

The usual approach is, thus, the use of parametrisations of the inclusive (often called production) and inelastic (often also called destruction) cross sections through empirical or semi-empirical formulae and fitting the exiting data from collision experiments (beam of protons impinging on target) available at the time and for each channel to catch their energy dependence.

\begin{figure}[!ht]
	\centering
	\includegraphics[width=0.5\textwidth, height=0.42\textheight]{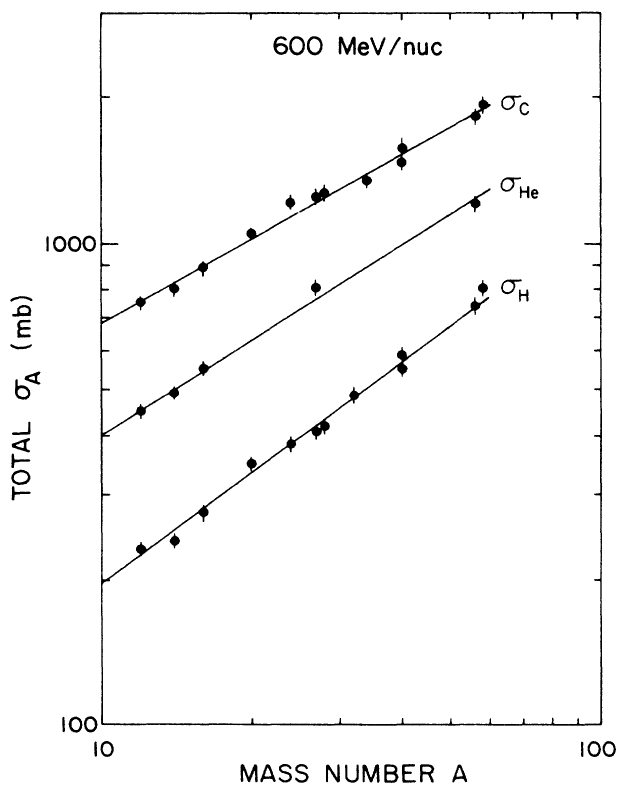}
	\hspace{1mm}
	\caption{\footnotesize proton-target spallation cross sections at $600 \units{MeV}/nucleon$ as function of the mass number of the daughter particle for H, He, and C targets, as measured and fitted in \cite{Webber:1990ka}}
	\label{fig:Webber_A}
\end{figure}

Most of the time, however, there is no available data, so analytic parametrisations are implemented (empirically, semi-empirically or combining both) to relate the cross sections from different channels. Figure~\ref{fig:Webber_A} shows a comparison of the cross sections for different nuclear targets. The use of Monte Carlo event generators, very common for accelerator particle physics, is another option. They simulate nucleus-nucleus collisions and keep track of the fragmentation probabilities and daughter particles formed in the interaction. Nevertheless, they still need to be calibrated with data and, nowadays, these theories need to be improved to be able to reproduce experimental measurements with the needed precision.

Additionally, the nucleus-helium cross sections are essentially absent, so the use of an extension of the proton cross sections is usually taken, by means of a parametrisation of the ratio $\frac{\sigma_{H}}{\sigma_{He}}$ \cite{ferrando1988measurement}.

One of the first works using empirical formulae for fitting cross sections is ref. \cite{letaw1983proton}. The authors have parametrised the proton-nucleus inelastic cross sections reaching a precision $< 2\%$ for $E_k > 2 \units{GeV}/nucleon$ and progressively degrading at lower energies until reaching a $\sim 10\%$ precision level at around $100$  $\units{MeV}/nucleon$. 

Next updates came when new data were available, in \cite{bauhoff1986tables}, only for low energy data, and \cite{wellisch1996total} for an extended energy range and nuclei with $Z > 4$. 

Later on, a team from NASA developed a general parametrisation for nucleus-nucleus total inelastic cross sections collected in ref. \cite{tripathi1999accurate}, taking into account some new theoretical considerations as Pauli blocking. It improved the quality and availability of tables and was used in CR studies.

The simplest parametrisation that can be employed in the estimation of cross sections assumes that the cross sections are proportional to the geometrical area of the colliding nuclei, $\sigma \propto R^2$, where $R^2 \sim \sum_i^N r_i^2$ ,with N and $r_i$ as the number of nucleons and their radius, respectively. While this is a good approximation for the target nucleus, the projectile effective area can be more appropriately estimated as its de Broglie length. The empirical relation between the radius of a nucleus and its number of nucleons is $R \sim A^{1/3}$, so that the inelastic cross sections may be written as: $\sigma_{ine} = \pi r_0^2 (A^{1/3}_{proj} +A^{1/3}_{target} - b_0)^2$. Here, $r_0$ acts as an effective nucleon collisional area and $b_0$ is a screening factor that reduces the interaction probability. This simple formula can be used to fit the data as it was done in \cite{bradt1950heavy}, where they generalized it even more taking $r_0$ and $b_0$ as functions of energy. See \cite{Sihver:1993pc} for detailed and more updated formulations and tables.

The authors of refs.\cite{silberberg1973partial} and \cite{silberberg1973partialb} introduced more complete and sophisticated formulations able to describe the most important channels needed in CR studies. Then, in the 1990s, W.R. Webber carried out several campaigns in order to improve and extend (in number of channels, precision and energy) the existing measurements at that time \cite{webber1990formula, Webber:1998ex3, Webber:1998ex2, Webber:1998ex}, and served to develop new empirical formulations. These extended data were also employed in ref. \cite{silberberg1998updated} with renovated formulations and remain in use even nowadays. The last update of semi-empirical formulae with the Webber cross sections was made public in ref. \cite{webber2003updated} and has been the reference until very recently (until the GALPROP cross sections came into scene).

Another point to recall is that the propagation codes use the ``cumulative'' cross sections of the nuclei instead of the direct ones. This means that, instead of propagating all the possible particles, these codes propagate only those nuclei with a significantly long lifetime, directly incorporating their decay products into the long-lived nuclei. These very short-lived nuclei are often referred to as ghost nuclei (see section~\ref{sec:3} for a discussion on their contributions to the main cross sections channels).

This chapter is aimed at studying and analysing different cross section data-sets publicly available and compare their effects in the production of the light secondary CRs boron, beryllium and lithium, commonly assumed to be pure secondary species, in order to evaluate the associated uncertainties. 

In section~\ref{sec:model} the set-up of the simulations is explained in detail and widely discussed. The results of the simulations are illustrated in section~\ref{sec:secondaries}, where the fluxes of these secondary species computed with the different cross sections are shown, and some relevant conclusions are drawn. The effects of variations on the cross sections is then tested in detail by studying secondary-over-secondary ratios in section~\ref{sec:uncert}. These ratios are proposed as a tool able to evaluate the overall cross sections efficacy (i.e. compatibility with CR data) since they are extremely sensitive to cross sections changes. In this section, the cross sections uncertainties are propagated to these ratios to evaluate their impact on the secondary fluxes; a fit on these ratios is also performed by simultaneously adjusting their cross sections. In addition, they will be used to understand the effects of cross sections uncertainties in the predicted fluxes of B, Be and Li and to look for possible imprints of any extra primary source producing them.
Furthermore, in section~\ref{sec:size}, the effect of halo size in the Be isotopes ratios with the different cross sections models is explored to get a more robust best value for the halo size. Finally, the conclusions are drawn in section~\ref{sec:conc}.

This chapter involves a work~\cite{Luque:2021joz}, currently under revision by scientific journal.

\section{Propagation setup and cross section data-sets}
\label{sec:model}

Simulations of CR propagation are performed using the 2-Dimensional model of the Galaxy offered by the DRAGON code, already discussed. The galaxy is modeled as a thin disc with thickness around $200 \units{pc}$ with gas and CR sources distributed along it and the Sun located at $8.3 \units{kpc}$ from its center.

In these simulations, we assume a geometry with azimuthal symmetry, so that CR particles propagate in a (R, z) plane where the radius of the Galaxy is set to be $20 \units{kpc}$ and the halo size will be adjusted by studying the unstable isotope $^{10}$Be (and will depend on the choice of cross sections model used; see section~\ref{sec:size}). The source spatial distribution is based on pulsar distributions and takes the form \cite{Pulsar_source}:
\begin{equation}
S(R, z) = \left(\frac{R}{R_{\odot}}\right)^a exp \left(-b\frac{R - R_{\odot}}{R_{\odot}} - \frac{|z|}{z_0}\right)
\label{eq:spatialsource} 
\end{equation}
where $a = 1.9$, $b=5$, $z_0 = 0.2 \units{kpc}$ and $R_{\odot} = 8.3 \units{kpc}$.

A diffusion-reacceleration model is used in the propagation of cosmic rays, without implicit convection ($\vec{v}_{\omega} = 0$, in equation~\ref{eq:caprate}). Nevertheless, for the study undergone in this chapter, the choice of the diffusion coefficient model and other parameters as advection speed or Alfvén speed have a secondary role, since the procedure followed here allows the conclusions drawn on the cross sections to be model-independent. 

In this study, we neglect the spatial diffusion coefficient parallel to the regular magnetic field lines because of the azimuthal symmetry assumed in the Galaxy structure \cite{Evoli:2008dv} and  assumed the most general form of the spatial diffusion coefficient perpendicular to the magnetic field lines, without including any spatial dependence (i.e. homogeneous spatial coefficient):A general form of the spatial diffusion coefficient without including any spatial dependence (it does not depends on galactic coordinates) is assumed:
\begin{equation}
 D(R) = D_0 \beta^{\eta}\left(\frac{R}{R_0} \right)^{\delta} \quad \quad
\label{eq:indepdiff}
\end{equation}
It is the same as in equation~\ref{diff_eq}, with $D_0$ chosen at the reference rigidity $R_0 = 4\units{GV}$ and assuming the diffusion to be isotropic and only perpendicular to the galactic magnetic field (as expected for Alfvenic waves under azimuthal geometry). 

These phenomenological formulas for the diffusion coefficient have been widely used, usually setting $\eta = 0$ or $\eta = 1$ \cite{Jaupart, fujii2003cosmic}, but $\eta$ is here treated as a free parameter, since also negative values of this parameter are unexplored and seem to be expected because of wave damping (see ref.  \cite{Ptuskin_2006}).

An exhaustive discussion on the different models of the diffusion coefficient will be left for section~\ref{sec:4}. This diffusion is adopted since these kind of models has successfully reproduced the shape of ratios of secondaries over primaries fluxes (see, for example, \cite{Jaupart}, \cite{Mazziot}, \cite{jones2001modified}). Indeed the inclusion of stochastic reacceleration seems to be necessary to naturally explain the shapes of these ratios. 

The expression used to parametrise the injection spectra of primary nuclei is a doubly broken power law, i.e. including a change for the slope at high energy to reproduce the hardening observed by recent experiments (again, this choice, instead of choosing a high energy break in the diffusion, plays no role for the oncoming discussion and will be widely treated in section~\ref{sec:4}). This expression takes the form: \begin{equation}
Q^{source} = \begin{cases}
     k_1 \times\left(\frac{R}{R_0}\right)^{\gamma_1} & \text{for $R < R_{1}$} \\
     k_2 \times\left( \frac{R}{R_0}\right)^{\gamma_2} & \text{for $R_{1} < R < R_{2}$} \\
     k_3 \times\left( \frac{R}{R_0}\right)^{\gamma_3} & \text{for $R > R_{2}$}
  \end{cases} 
\label{eq:powerlaw} 
\end{equation}
where $Q^{source}$ is the differential energy flux in units of $(m^2$ $s$ $sr\units{GV})^{-1}$, $R_{1,2}$ are the rigidity breaks, $\gamma_{1,2,3}$ are the logarithmic slopes below and above each break and the parameter $k$ sets the normalization of the flux. The low-energy break was set to $R_1=8 \units{GV}$ for all nuclei, while the high-energy break was set to $R_2=335 \units{GV}$ for protons and $R_2=200 \units{GV}$ for the heavier nuclei. In the simulation, we injected $^1$H, $^{4}$He, $^{12}$C, $^{14}$N, $^{16}$O, $^{20}$Ne, $^{40}$Mg and $^{28}$Si as primary nuclei with the spatial distribution of sources following the model developed in \cite{Lorimer_2006}.

In fact, we would like to stress that, while the accuracy showed in the last experimental results of the AMS-02 collaboration is of about 1-5\% (for the main nuclei involved in the creation of light secondary CRs), the uncertainties on the cross sections reach levels of 20-50\% in some channels (see \cite{Genoliniranking}), what makes clear the necessity of estimating the uncertainties we are dealing with and to handle them appropriately before reaching improper conclusions. 

The exact determination of these uncertainties is not easy for most of the channels, since sometimes data from different experiments are difficult to reconcile. This fact, along with the lack of data, leads to the conclusion that the uncertainties in the fluxes of secondary CRs are actually larger than the uncertainties in the cross sections data: properly propagating these errors needs to include all the nuclear chain.
Also the ghost-nuclei can have a sizeable effect in the estimation of the production of secondary nuclei, even more when handling precise CR measurements. At the end, the extensive compilations and parametrisations cross sections are based on extrapolations of the (often, poorly) known main production channels from the primaries C and O. These are the main progenitors of B, Be and Li, but their impact in the boron spectrum is greater than in the Li and Be fluxes (i.e. the contribution of other progenitors in the Be and Li spectra are considerably more important), what leads to the conclusion that the boron spectra is less uncertain and can be better described by our cross sections parametrisations \cite{Strong:2007nh}.

Currently, several cross sections parametrisations are publicly available.  The DRAGON code has been recently modified in order to implement the most widespread cross sections parametrisations and the propagation equations are numerically solved with a new version of the DRAGON code \footnote{A public version is available in \url{https://github.com/cosmicrays/DRAGON2-Beta\_version}}. This new version of the code, apart from incorporating these spallation cross sections parametrisations, updates the library used for the lepton and antiprotons cross sections and the inelastic ones and updates the list of isotopes and decays in order to ensure a suitable ordering of the nuclear chain and the proper decays. 

The first reliable phenomenological parametrisations comes from the measurements of W. R. Webber during the 80s, that ultimately lead to the semi-empirical WNEW code \cite{webber1990formula} with a last update in 2003 \cite{webber2003updated}. Then, there were important efforts to expand the known experimental measurements to other channels, turning out in the semi-empirical parametrisations by Silberberg and Tsao \cite{tsao1993scaling, silberberg1973partial, silberberg1985improved} that converged in the YIELDX code \cite{tsao1998partial, silberberg1998updated}. 

These codes have been extensively used and taken as reference to build the GALPROP cross sections parametrisations. The GALPROP team developed a set of routines, available in their webpage \footnote{\footnotesize{ \url{https://galprop.standford.edu}}}, that make use of a combination of semi-empirical formulae \cite{moskalenko2005propagation, Moskalenko:2001qm, strong2007cosmic} and has become the most ubiquitous. 

The nuclear reaction network they implemented is
mainly based on the Nuclear Data Sheets and the cross sections are built using the T16 Los Alamos database \cite{Mashnik:1997ht, Mashnik:1998sm, Mashnik:2002uj} and the CEM2k and LAQGSM codes \cite{1998nucl.th..12071M} by fitting particular production channels ($^2$H, $^3$H, $^3$He, Li, Be, B, Al, Cl, Sc, Ti, V, Mn), including some phenomenological approximations adapted from the YIELDX and WNEW codes (code $WNEWTR.FOR$, version of 1993 and updated version of 2003 and/or Silberberg and Tsao code, $YIELDX\_011000.FOR,$ version of 2000) eventually renormalized to the available data. See \cite{Moskalenko:2001qm}, \cite{2003ICRCMoska}, \cite{2003ApJMosk} and \cite{GalpProc} for a more complete list of references.

The GALPROP routines incorporated in DRAGON come from the package v54 (see \url{https://github.com/cosmicrays/DRAGON/blob/master/README.md} for details). However, the current version of the GALPROP code is now in its version v56 and the authors recommend the use of the GALPROP-22 cross sections package (although there are no differences with respect to the cross sections used in the version v54), which is also used by (and publicly available in) the USINE code.

A couple of years ago, a new set of cross sections derived from different parametrisations of data for every individual channel was presented as the default algorithm for cross sections for the incoming DRAGON2 code (see \cite{Evoli:2017vim} and \cite{Evoli:2019wwu} for a complete explanation). They have been successfully used in other studies as \cite{CarmeloBeB} and are fully available in the github repository of the new DRAGON version. 
To parametrise the inclusive cross sections the recipe introduced in \cite{reinert2018precision} is used, which has the form:
\begin{equation}
\sigma_{j + H \rightarrow i} = \sigma_0 \frac{\Lambda^2 (T - E_{th})^2}{(T^2 - M^2)^2 + \Lambda^2M^2} + \sigma_1 \left(1-\frac{E_{th}}{T}\right)^\zeta \left(1+\frac{\Delta}{(E_{th}/T)^2}\right)
\label{eq:CarmeloXS_parametrization} 
\end{equation}

Here, $E_{th}$ is the energy threshold above which the cross sections show a resonance peak, which is controlled by the parameters $\sigma_0$, M and $\Lambda$ (normalization, peak position and width of the peak, respectively). When the resonance peak is not visible for a given channel, the term $\sigma_0$ is set to 0, as well as when the addition of this resonance term does not improve the fit significantly. The energy thresholds are taken from the online database of the  National Nuclear Data Centre (NNDC) and based
on the experimental results reported in \cite{Wang_2017}.

%These cross sections are available in \urlhttps{https://zenodo.org/record/3529761#.XjBN8SMo82y} in the form of table.
 
These three cross sections models will be considered in this chapter. Rigorously speaking, both GALPROP and DRAGON2 cross sections use, as starting point, the Webber parametrisations and upgrade it with more data and corrections, expecting a better agreement of these models with the current cross sections measurements. All of them account for the cross sections on He target using the parametrisation in \cite{ferrando1988measurement}. 

Furthermore, the DRAGON code incorporates the inelastic cross sections from the CROSEC code (often also called CRN6 code, \cite{Barashenkov:1994cp}) down to He and using the parametrisations taken from \cite{Kafexhiu:2014cua} for protons. They do not differ too much from other inelastic cross sections parametrisations as those of Tripathi, Letaw or Wellisch as it is studied in section 3.1 of ref. \cite{genolini2017refined}, where the authors find an uncertainty around 3\% for the C, N and O fluxes. They state that the uncertainties for heavier elements (e.g. Ne, Si, Mg, Fe) seem to be larger, what translates into more uncertainty for the Li and Be predicted fluxes with respect to the B flux. Even so, the uncertainties in the inclusive cross sections introduce the largest errors in the secondary CR fluxes, and it is clear the need to estimate these uncertainties in a general way, beyond directly comparing the performance of different parametrisations.

In the present simulations, we are propagating particles up to $Z=14$ (silicon nuclei), which implies that we can fully describe the generation of secondary nitrogen and boron (which is considered to be fully secondary, as well as Be and Li). Nevertheless, the missing iron (mainly $^{56}Fe$ but also its isotopes) and, in very low proportion $^{32}S$, makes us underestimate the total amount of Li and Be in a 3.22\% and 3.7\% in average respectively (see tables $IV$ and $V$ in \cite{Genoliniranking}). %Only few individual channels involving $^{56}Fe$ reactions add more than 1\% to the total flux of Be and Li (at an energy of 10 GeV, \cite{Genolini}).
In order to save computational time for the various simulations needed, we compensate the missing source terms (primary Fe and S) adding just a multiplicative factor to the Li and Be spectra. %(in fact this small percent produce almost unnoticeable changes and does not modify the shape of the spectra). 
% You already said this
In this way, we can fully describe the light secondaries Li, Be and B and study the effect of cross section changes in their spectra. The gas profile used is the same as in GALPROP, with a modification in the center of the Galaxy to adopt the $H_2$ and HI density profiles from \cite{ferriere2007spatial}.

\begin{figure}[!ht]
\centering
\includegraphics[width=0.56\textwidth,height=0.24\textheight,clip]{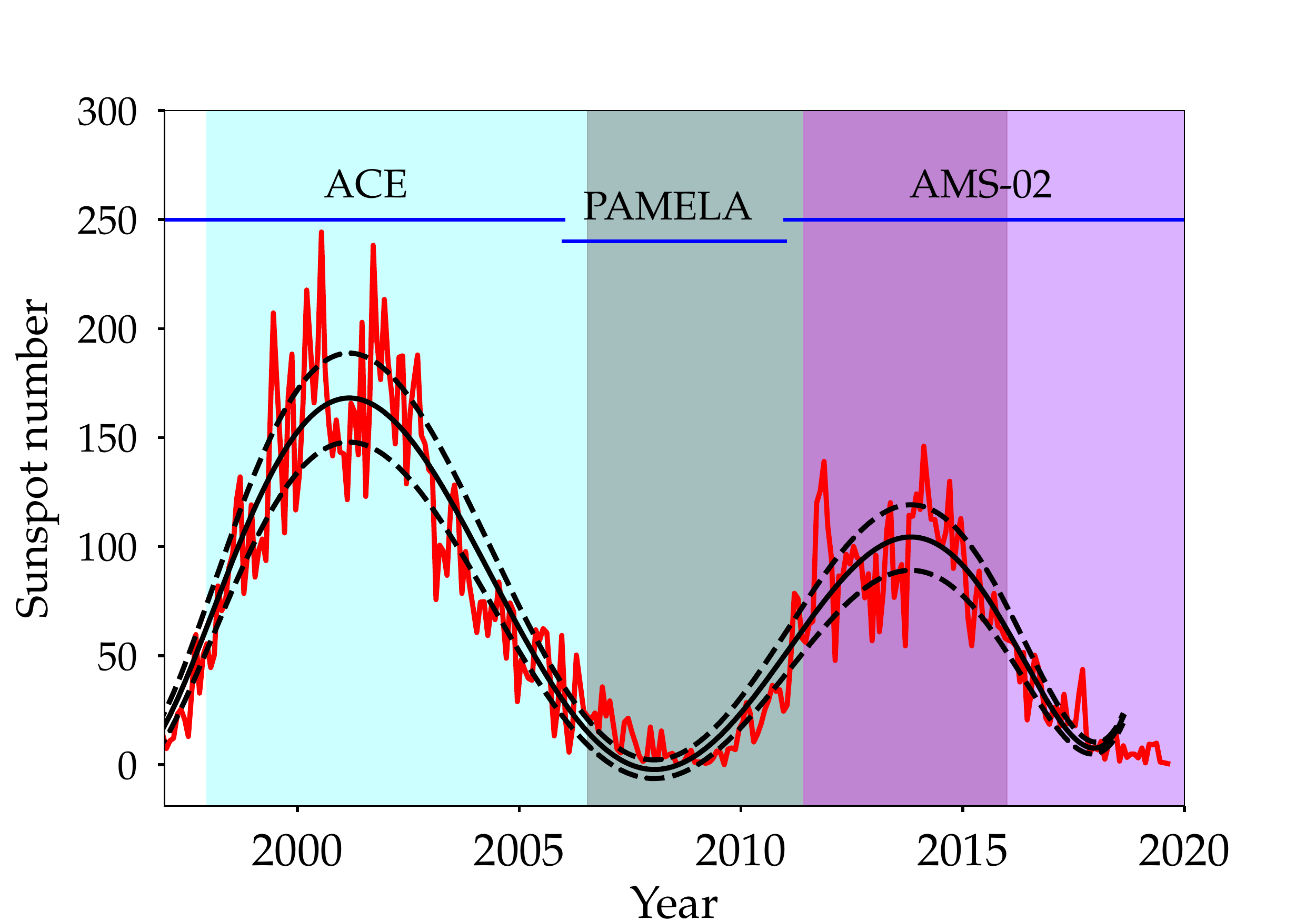}
\caption{\footnotesize Solar activity evolution with time in terms of sunspot number. The solid and dashed lines show the mean sunspot numbers and the 90\% intervals according to the monitored data \cite{ghelfi2017neutron}. The shaded regions represents the periods of data taking for PAMELA (2006-2016), AMS-02 (2011-currently) and ACE (1997-currently) experiments. Sunspot data taken from \url{http://sidc.be/silso/datafiles}, \cite{sidc}.}%(shifted leftwards by one year considering the possible delay of modulation effect compared with the solar activity) by ACE (for 10Be=9Be), PAMELA (for protons), and AMS-02 (for all species) detectors.}
\label{fig:Modulation}
\end{figure}

Finally, in our set-up the solar modulation is modeled by using the Parker equations and the force-field approximation for the Sun's magnetic field \cite{forcefield}. Modulation depends on the solar activity and, therefore, on the epoch in which the experimental data were taken (see Figure~\ref{fig:Modulation}). 

The variable used in the force-field approximation to characterize the solar modulation is the Fisk potential $\phi$, and the flux of CRs reaching a detector at Earth is related to that in the Local Interstellar medium (LIS) by the following equation:
\begin{equation}
\Phi_{obs} (E_{obs})=  \left( \frac{2mE_{obs} + E_{obs}^2}{2mE_{LIS} + E_{LIS}^2}\right) \Phi_{LIS}(E_{LIS})
\label{eq:potmod} 
\end{equation}
with $E_{LIS} = E_{obs} + e \phi |Z|/A$. In these expressions, $E$ is the kinetic energy per nucleon of the particle.

The solar modulation totally reshapes the spectrum at low energies and it can obscure low-energy features of the spectra if it is not well accounted for.

The usual way to determine the Fisk potential is by means of fitting the unmodulated data (CR data outside the Solar magnetosphere). Voyager-1 left the solar magnetosphere in 2010, supplying precious data for the low-energy fluxes outside the Solar System. For the Voyager-2 data, which left the solar magnetosphere at the end of 2018, we need to wait a couple of more years yet. 

There are a few data points from Voyager-1 for light nuclei, with a precision at a level around 10\%. However, as the systematic uncertainties are not well accounted for, these data lead to important uncertainties in the determination of the Fisk potential (and, in general, the modeling of the solar modulation). In fact, the difficulty to reconcile the modulated spectra at low energies with the unmodulated one has been often manifested and eventually corrected by adding breaks in the source term. In addition, the fact that the cross sections resonances are expected to be in the energy region close to the Voyager-1 data (around $100 \units{MeV}$) leads to some extra uncertainties difficult to prevent.

In this study, the solar potential value was chosen according to the data from the neutron monitor experiments that better reproduce the low-energy part of the spectra measured with the Voyager-1 data points, maintaining the fit to the AMS-02 data. These constraints made us choose the NEWK (9-NM64 Newark, in Swarthmore, USA) neutron monitor experiment\footnote{\footnotesize{ \url{http://www.nmdb.eu/station/newk/}}} as the one with a solar potential value, for the period of AMS-02 data we are using (2011-2016), that better matches the two conditions. This value is $\phi = 0.66 \pm 0.05\units{GV}$, having a best fit value for the simulations of $\phi = 0.61\units{GV}$. This value has been also used in other works like \cite{Mazziot}. The calculation of these potentials come from \cite{ghelfi2017neutron} and \cite{ghelfi2016non} and they are publicly available in \url{https://lpsc.in2p3.fr/cosmic-rays-db/}. Nevertheless, one must mention that different neutron monitors show large discrepancies between each other, meaning that there are systematics not well accounted for. %this instrument gives the minimum solar potential value from the list of experiments in the cited webpage (these experiments actually average to a value $ \phi = 0.70 \pm 0.1$ GV).

There are other dedicated codes to simulate the solar modulation numerically integrating equation~\ref{Parker} as HELMOD \cite{Boschini:2017gic}, HELIOPROP \cite{2014Gaggeroheliosphere} or SOLARPROP \cite{2016Solarprop}. Nevertheless, the lack of knowledge about diffusion at so low energies makes the force-field approximation to suffice for this study, more based on the region at higher energies.

Besides the uncertainties of Voyager-1 and neutron monitor experiments data, the difficulty of having a full theory of CR-wave scattering at low energy (see~\ref{sec:4}) and the approximations made to simulate the solar modulation effect, we must consider that the spallation cross sections parametrisations in the region around $100 \units{MeV}$ can be significantly uncertain, since this is the energy region governed by resonances. 

Different spallation cross sections parametrisations can lead to discrepancies of around 10\% at $100 \units{MeV/n}$ in the B/C spectrum when using the diffusion coefficient that fits the AMS-02 data (Fig.~\ref{fig:BCComp}) and about 20\% when using the same diffusion coefficient for the different parametrisations. %These uncertainties are more important for the secondary CRs, since the importance of the diffusion coefficient is greater.

\section{Cross sections impact on the secondary cosmic rays}
\label{sec:secondaries}

\begin{figure*}[!hbt]
\includegraphics[width=0.495\textwidth,height=0.243\textheight,clip] {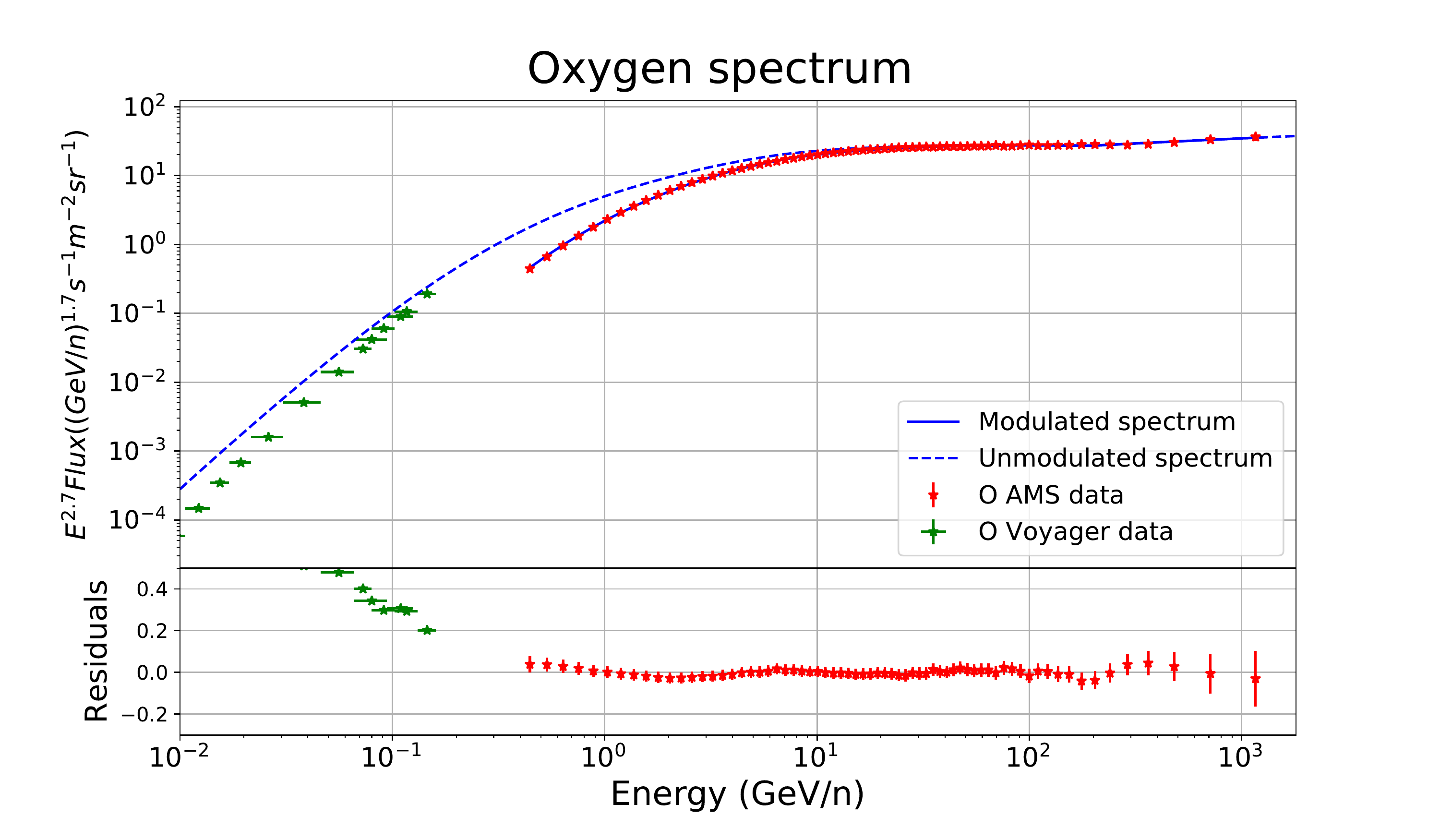} %\hspace{0.5cm}
\includegraphics[width=0.495\textwidth,height=0.243\textheight,clip] {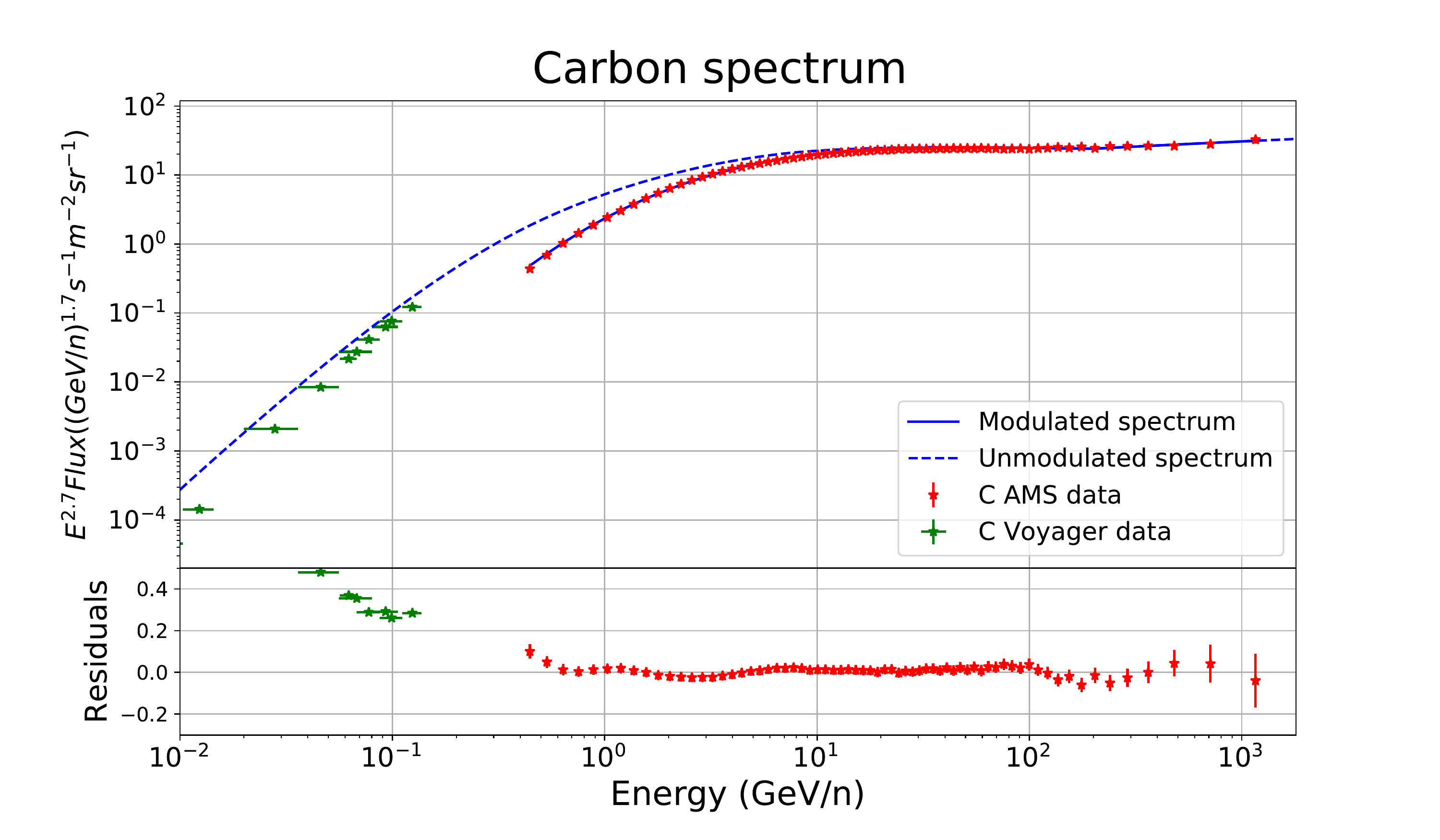} 
%\vspace{0.5cm}

%\includegraphics[width=0.45\textwidth,height=0.23\textheight,clip] {./Figure/ProtonComparison.png} \hspace{0.5cm}
%\includegraphics[width=0.45\textwidth,height=0.23\textheight,clip] {./Figure/HeliumComparison.png}
%\vspace{0.5cm}
\includegraphics[width=0.495\textwidth,height=0.243\textheight,clip] {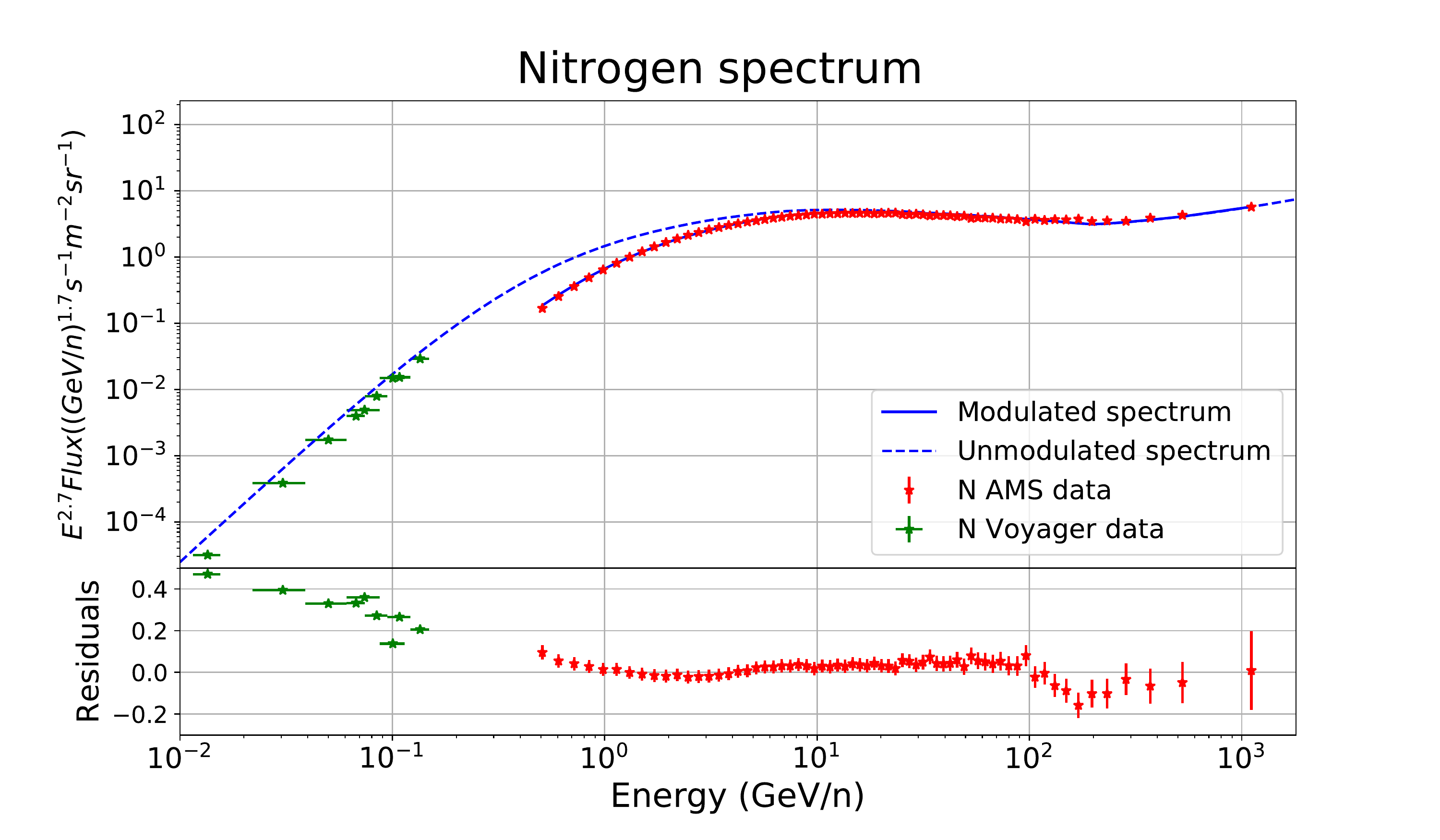} %\hspace{0.5cm}
\includegraphics[width=0.495\textwidth,height=0.243\textheight,clip] {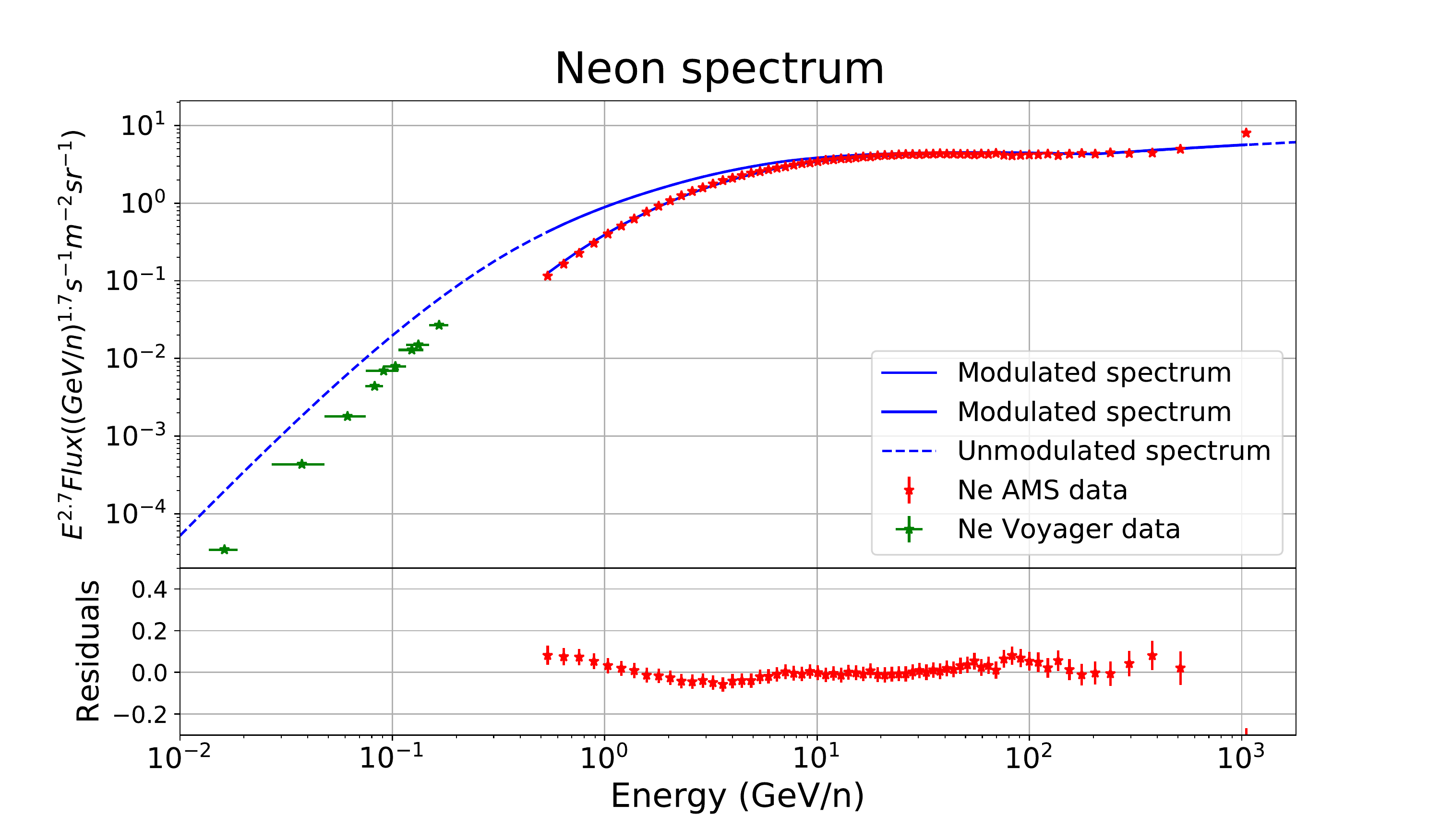}
%\vspace{0.5cm}

\includegraphics[width=0.495\textwidth,height=0.243\textheight,clip] {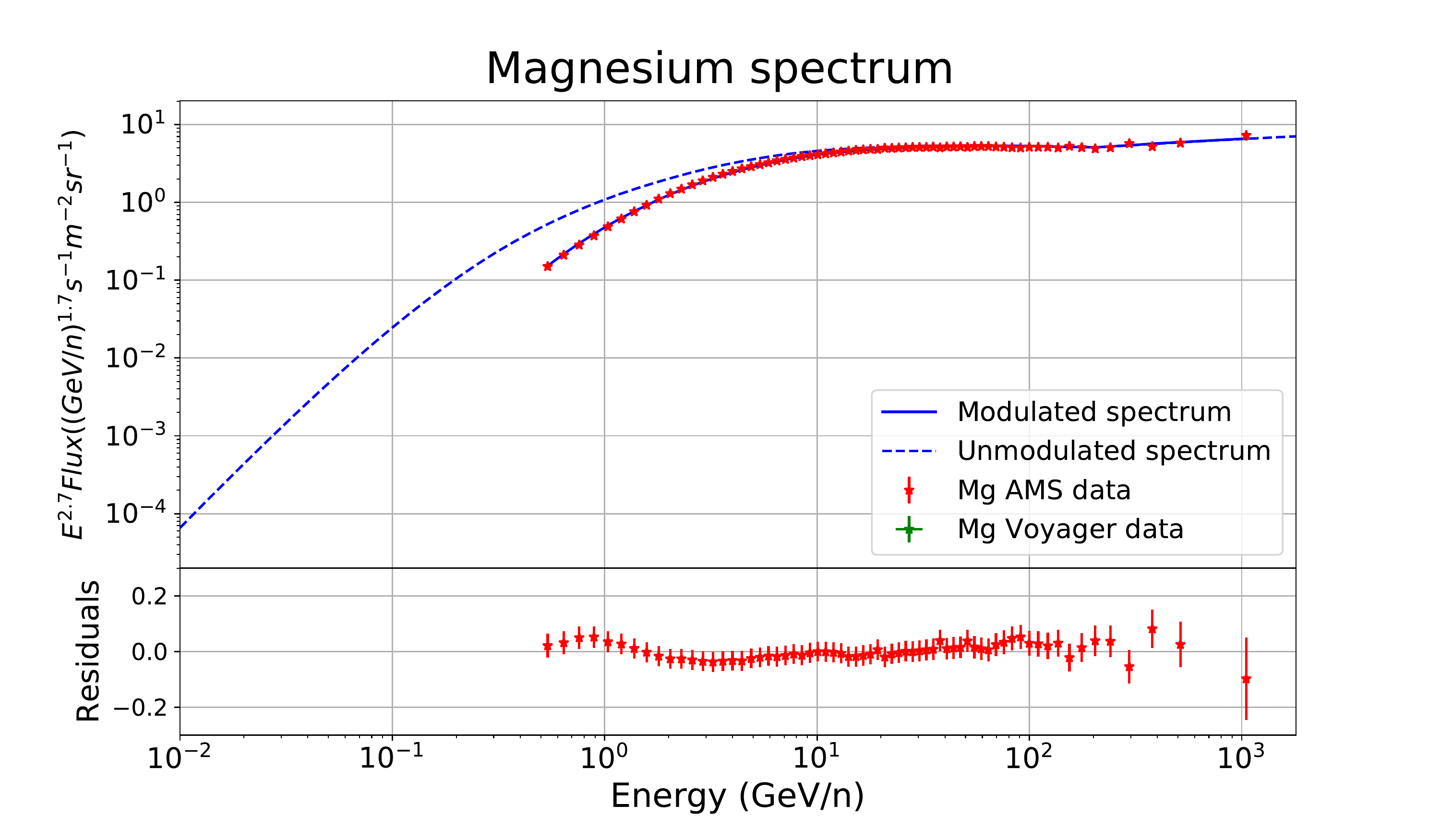} %\hspace{0.5cm}
\includegraphics[width=0.495\textwidth,height=0.243\textheight,clip] {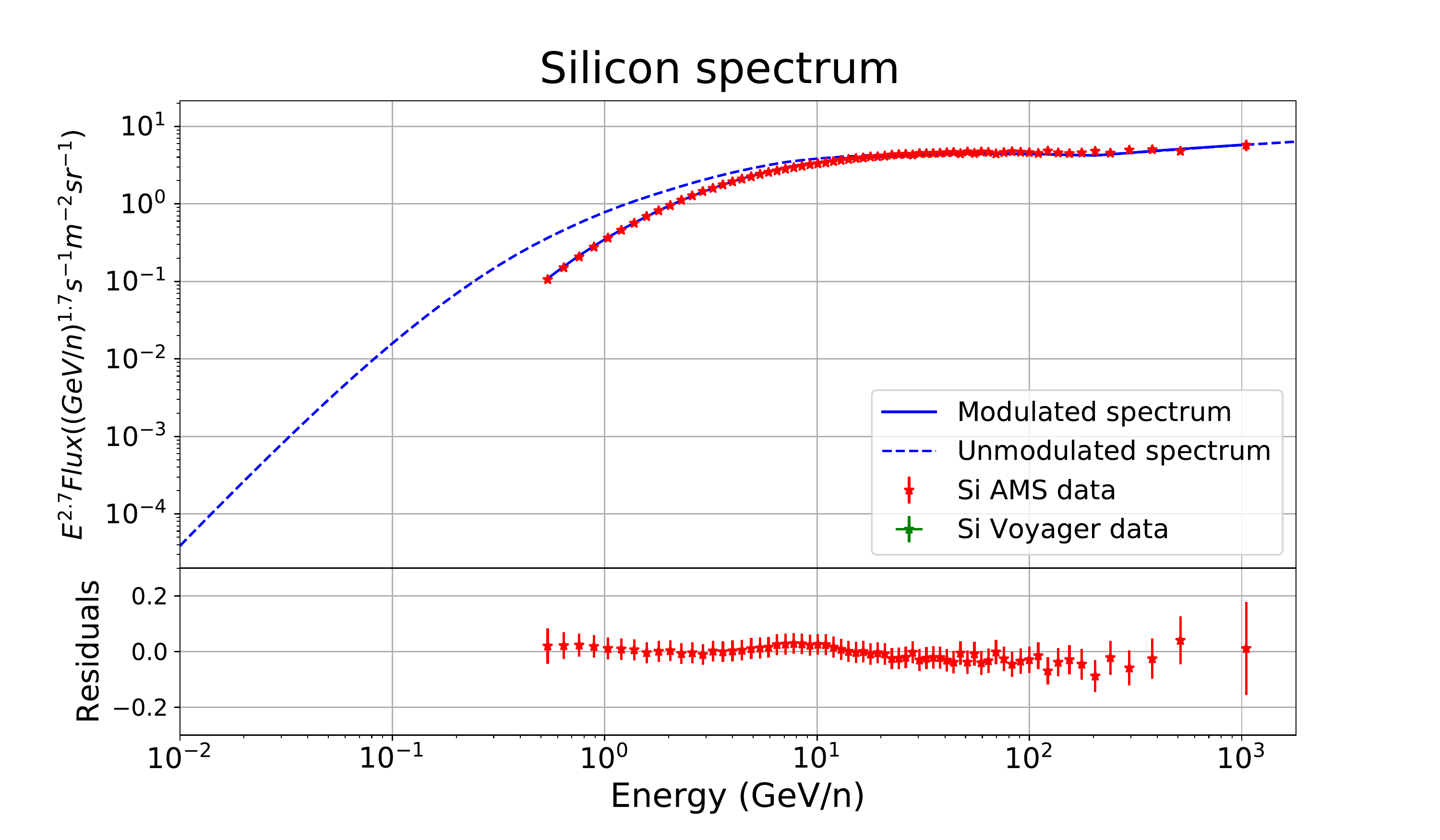}

\caption{\footnotesize Fits of the primary CRs obtained in the simulations, using the DRAGON2 cross sections. All the cross sections models reproduce the primary fluxes inside the experimental uncertainties, as shown. Data from the Voyager-1 mission and AMS-02 are plotted, including the very recent Ne, Mg and Si data from AMS-02. Voyager-1 is outside the heliosphere, which means no solar modulation effect, while the Fisk potential value used for to fit AMS-02 data is $\phi = 0.61$.}
\label{fig:primFit}
\end{figure*}

With the set-up above described we proceeded to study the secondary CRs generated and compared their fluxes to experimental data. The fits of the primary spectra are similarly well reproduced under all cross sections models, as expected, and are shown in Figure~\ref{fig:primFit} for the DRAGON2 cross sections.
%One should notice the tension between the unmodulated and modulated spectra, which seems to be growing with decreasing energy. 
%As pointed out in the previous section, there are differences between the modulated and unmodulated spectra, which increase with decreasing energy.

%In addition, we see how the modulated low energy points (coming from experiments as ACE \cite{ACE}) are in good agreement with the simulations reproducing the AMS-02 data, which reflects the fact that the Force Field approximation suffices in reproducing the modulated experimental data within uncertainties.

\begin{figure*}[!ht]
\begin{center}
\begin{tabular}{c c}
\textbf{\underline{Webber}} &  \textbf{\underline{Galprop}}  \\
\includegraphics[width=0.41\textwidth,height=0.19\textheight] {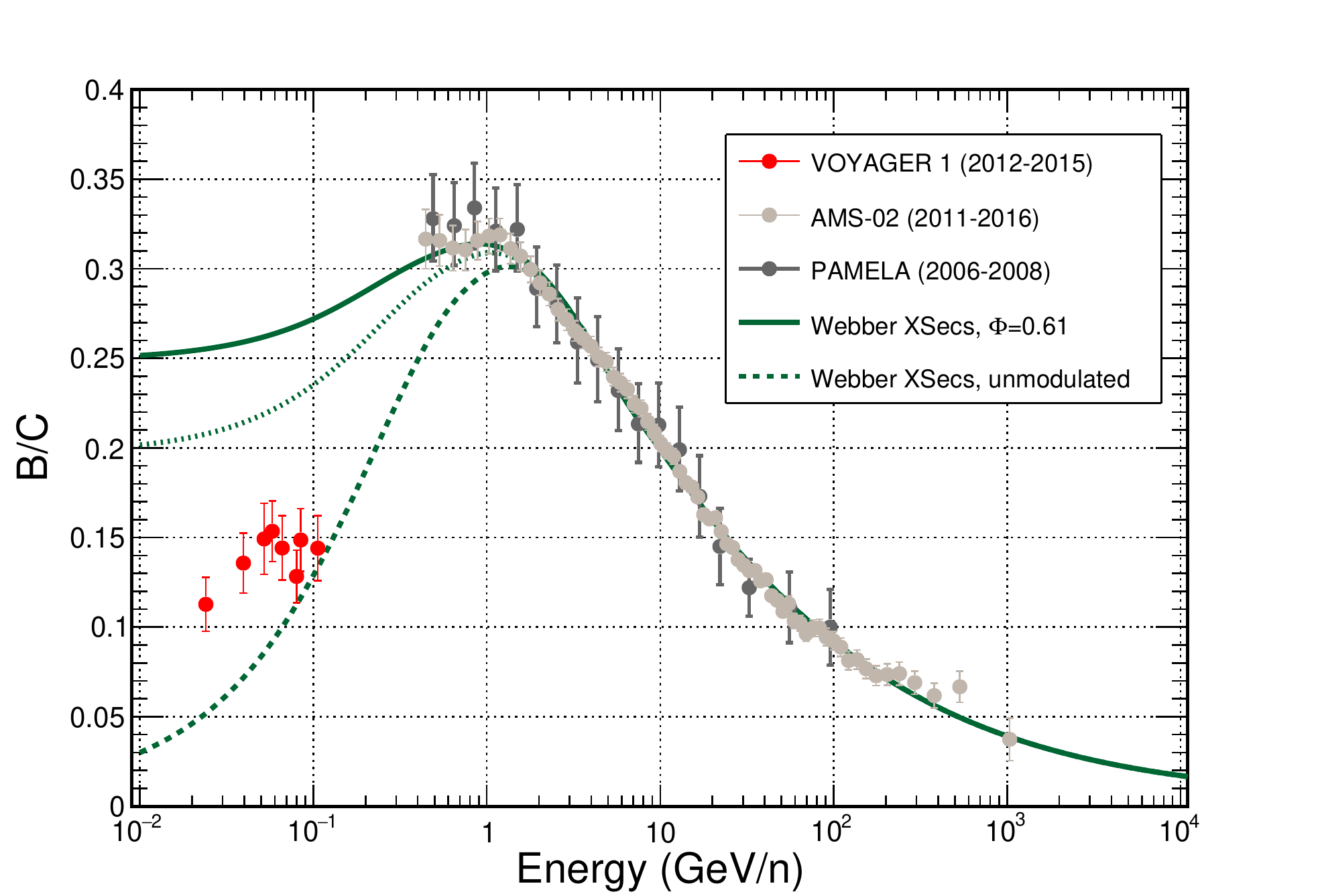} &
\includegraphics[width=0.41\textwidth,height=0.19\textheight] {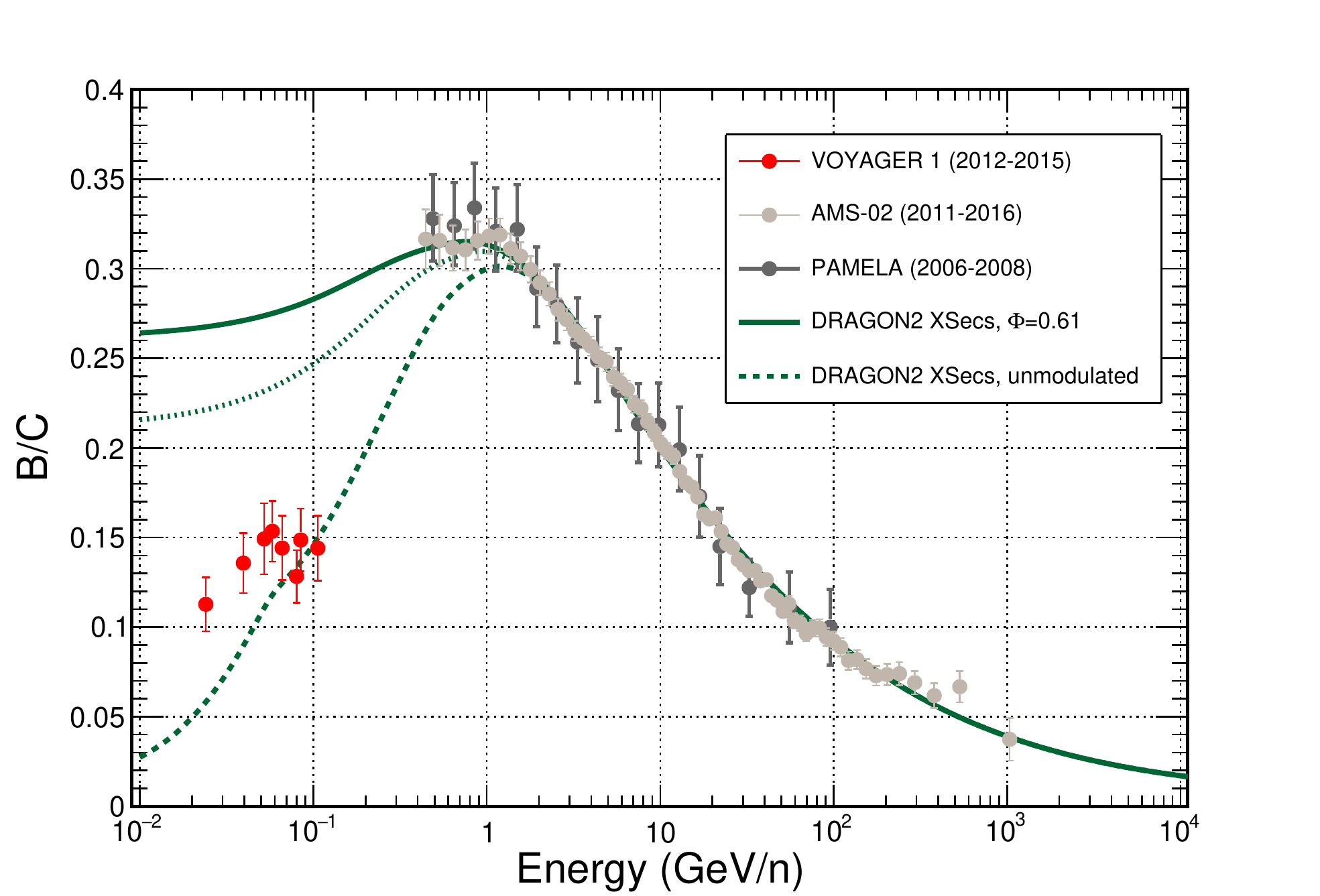}%&
    \end{tabular}
\textbf{\underline{DRAGON2}} \\
%\hspace{-0.7cm}
\includegraphics[width=0.41\textwidth,height=0.19\textheight] {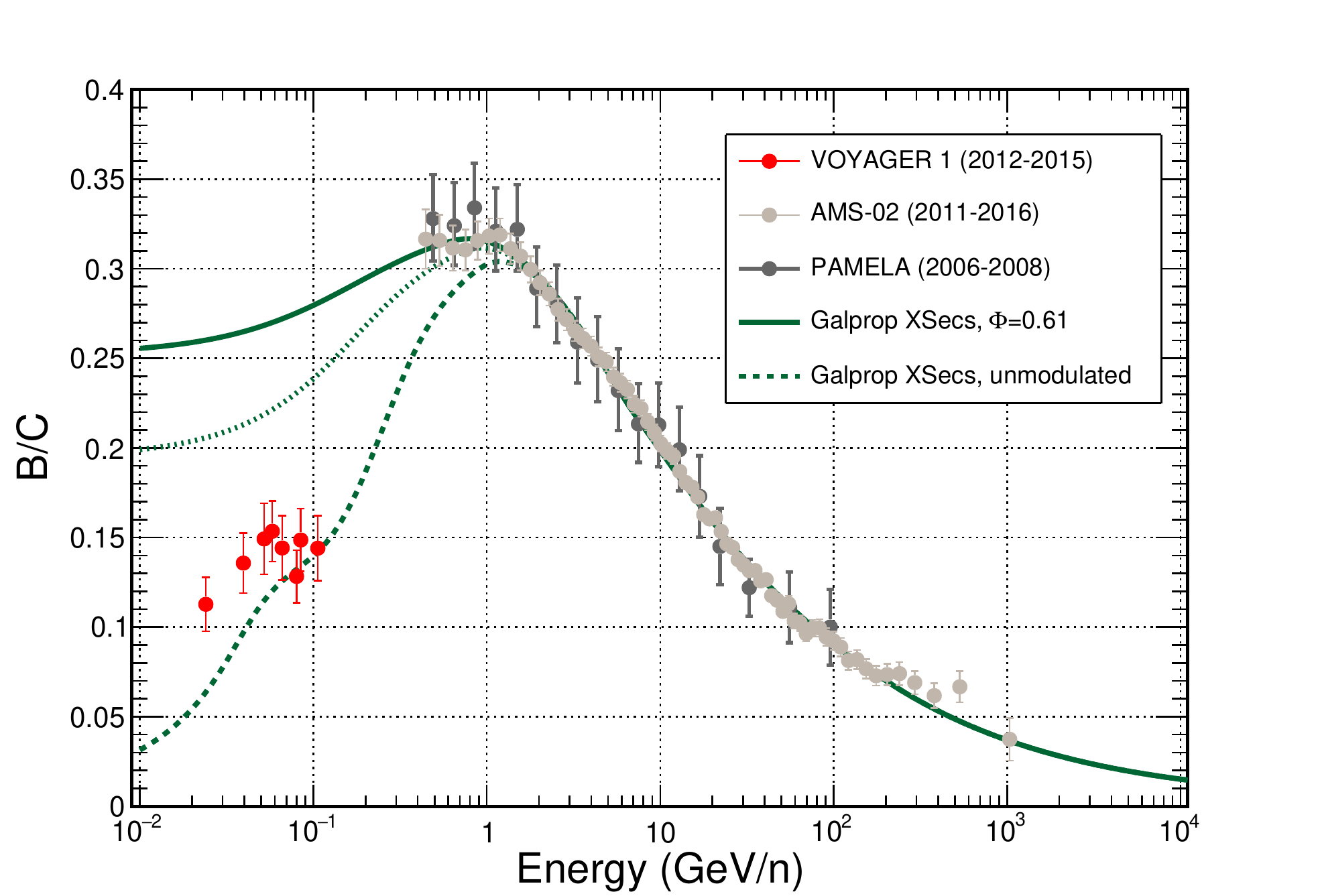}
\end{center}
\caption{\footnotesize Fits of the boron-over-carbon ratios for the three cross sections models considered. A line with a solar modulation $\phi = 0.4 \units{GV}$ is added for completeness.}
\label{fig:BCComp}
\end{figure*}

For each cross sections model we start selecting the parameters for the spatial diffusion coefficient that reproduce the boron-over-carbon data (see Figure~\ref{fig:BCComp}) of AMS-02 and using the optimal halo size found from the procedure explained in section~\ref{sec:size}. Here we do not perform a dedicated study of the diffusion parameters that yield the best fit of the boron-over-carbon spectrum, but a faster fit of the spectrum that satisfies the condition  $\chi^2/n_{points} < 1.05$, below which the fit is considered to be very good. 
The fits of all the primary and secondary species are performed mainly by the comparison to the most recent AMS-02 data \cite{Aguilar:2018wmi, Aguilar:2018njt, Aguilar:2018keu, Aguilar:2015ooa, AMS_Ne, Aguilar:2017hno}. See \cite{jia2018measurement} and \cite{Jia:2018aze} for the description on the measurements of secondary CRs in AMS-02. Furthermore, we also use Voyager-1 data for the unmodulated CR spectra \cite{stone2013voyager, Cummings_2016}. In addition, PAMELA, ACE and other's data are also used for completeness. All experimental data from CR experiments were taken from \url{//https://lpsc.in2p3.fr/crdb/} \cite{Maurin_db1, Maurin_db2} and \url{https://tools.ssdc.asi.it/CosmicRays/} \cite{ssdc}

%This consists of reproducing the boron-over-carbon ratio, below $3.5\%$ of mean error and $\chi^2 < 40$, for each model and then comparing the spectra of Li, B and Be.  
More dedicated studies on how the different parameters in the diffusion coefficient can change for different cross sections models have already been performed in other works (see, for example, \cite{maurin2010systematic}) and they, in general, agree on the variability of the diffusion parameters found here (see table~\ref{tab:diff_params}), so we will not enter in this discussion in this chapter.

%That's also why we are not comparing how the different parameters in the diffusion coefficient can change for different cross sections models, since these studies have already been performed in other works (see, for example, \cite{maurin2010systematic}) and they agree on the variability of the diffusion parameters found here, which are of the order of the $10-15\%$ for each parameter. 

%In definitive, the scope is to study how the spectra of different secondary CRs change with the different cross section models, assuming the same diffusion mechanism. For each cross section model we take as reference the diffusion coefficient parameters which reproduce the AMS-02 B/C data (see Fig.~\ref{fig:BCComp}).

\begin{table}[!hptb]
\centering
\resizebox*{0.52\columnwidth}{0.14\textheight}{
\begin{tabular}{|lccc|}
  \multicolumn{4}{c}{\hspace{0.3cm}\large} \\ \hline & \textbf{ Webber}  & \hspace{0.2 cm}\textbf{GALPROP} & \hspace{0.2 cm}\textbf{DRAGON2}\\ %\toprule
  \hline
{$D_0$} ($10^{28} cm^{2} s^{-1}$) & 2.3 & 6.65 & 7.1\\
{$v_A$} ($km/s$) & 29.9 & 25.5 & 27.7\\
{$\eta$}    & -0.25 & -0.55 & -0.6\\      
{$\delta$}    & 0.42 & 0.44 & 0.42\\   
{H} ($\units{kpc}$)    & 2.07 & 6.93 & 6.76\\  
\hline
\end{tabular}
}
\vspace{0.1cm}
\caption{\footnotesize Diffusion parameters used in the cosmic ray propagation with the different cross section parametrisations. These values have been obtained from a fit of the B/C data from AMS-02 and of the $^{10}$Be/$^9$Be data from various experiments (see Figure~\ref{fig:sizes}).}
\label{tab:diff_params}
\end{table}

%This chapter focuses on the light secondary nuclei production (Li, Be and B). 
Very few works have recently studied the secondary Li and Be (see \cite{primsec, Weinrich:2020cmw}, for example) after the AMS-02 data release, while nearly all studies just use boron and its ratios to a primary CR (usually C) to develop their models. This is due to the fact that, before AMS-02, the data samples on Li and Be were poor and to the fact that the uncertainty on the predicted fluxes from the cross sections parametrisations in the boron channels is expected to be the smallest~\cite{de2006observations}. In fact, typical uncertainties on the predicted Li and Be fluxes are around 20-30\%, and 15-25\% respectively, as usually quoted \cite{de2006observations}, while the uncertainties on the predicted B flux are around 10\%~\cite{Genoliniranking}, although these values could be larger in some cases (this can happen, for instance, with a poor estimation of the diffusion coefficient). However these uncertainties are very difficult to evaluate, given the current cross sections measurements and missing channels (see appendix~\ref{sec:appendixA} to take a look at the experimental data in the main spallation channels and discussed parametrisations, and in, e.g., \cite{Genoliniranking} for other rarer channels). 

\begin{figure*}[!tb]
\begin{center}
\includegraphics[width=0.345\textwidth,height=0.20\textheight,clip] {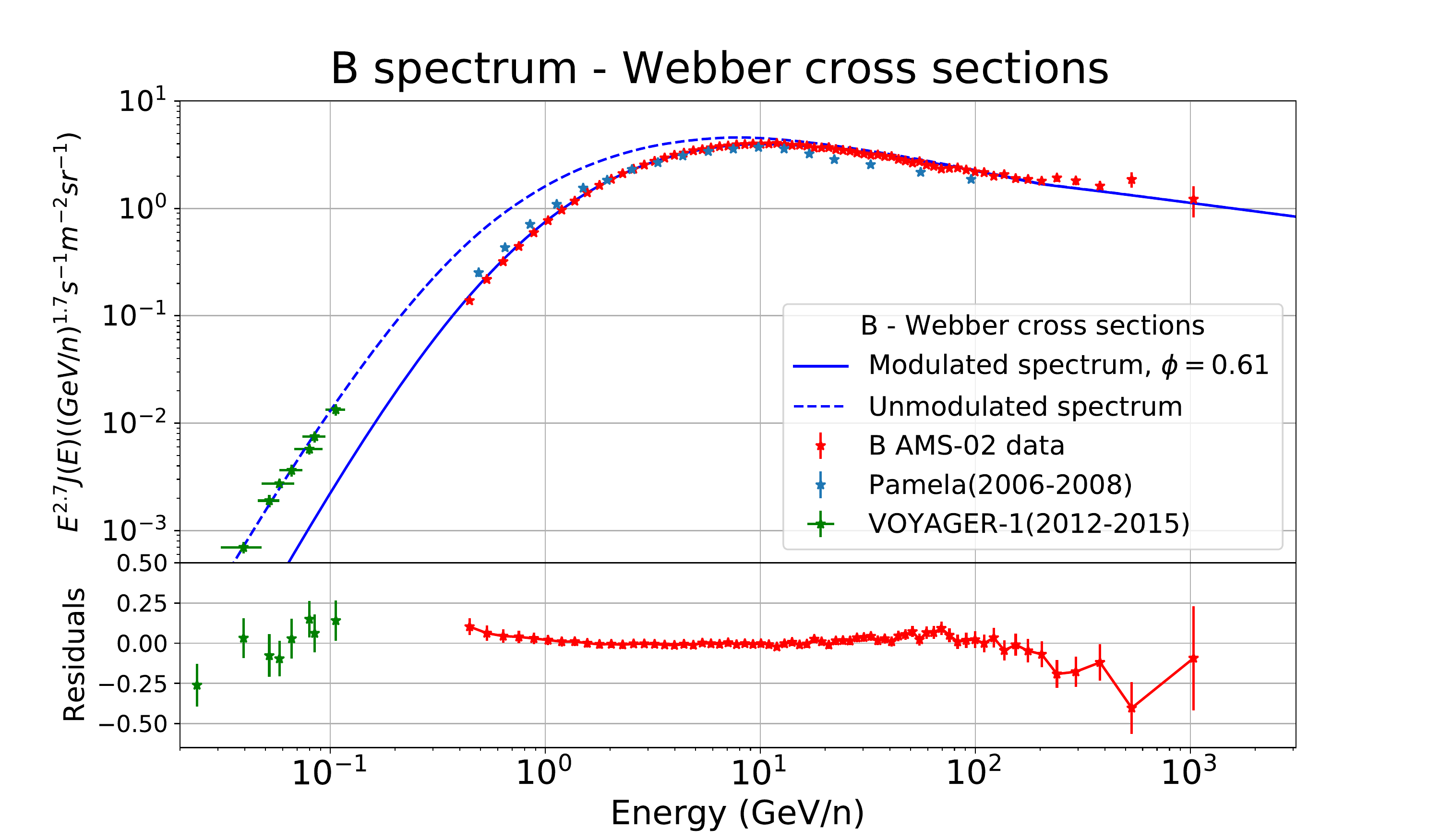} \hspace{-0.6cm}
\includegraphics[width=0.345\textwidth,height=0.2\textheight,clip] {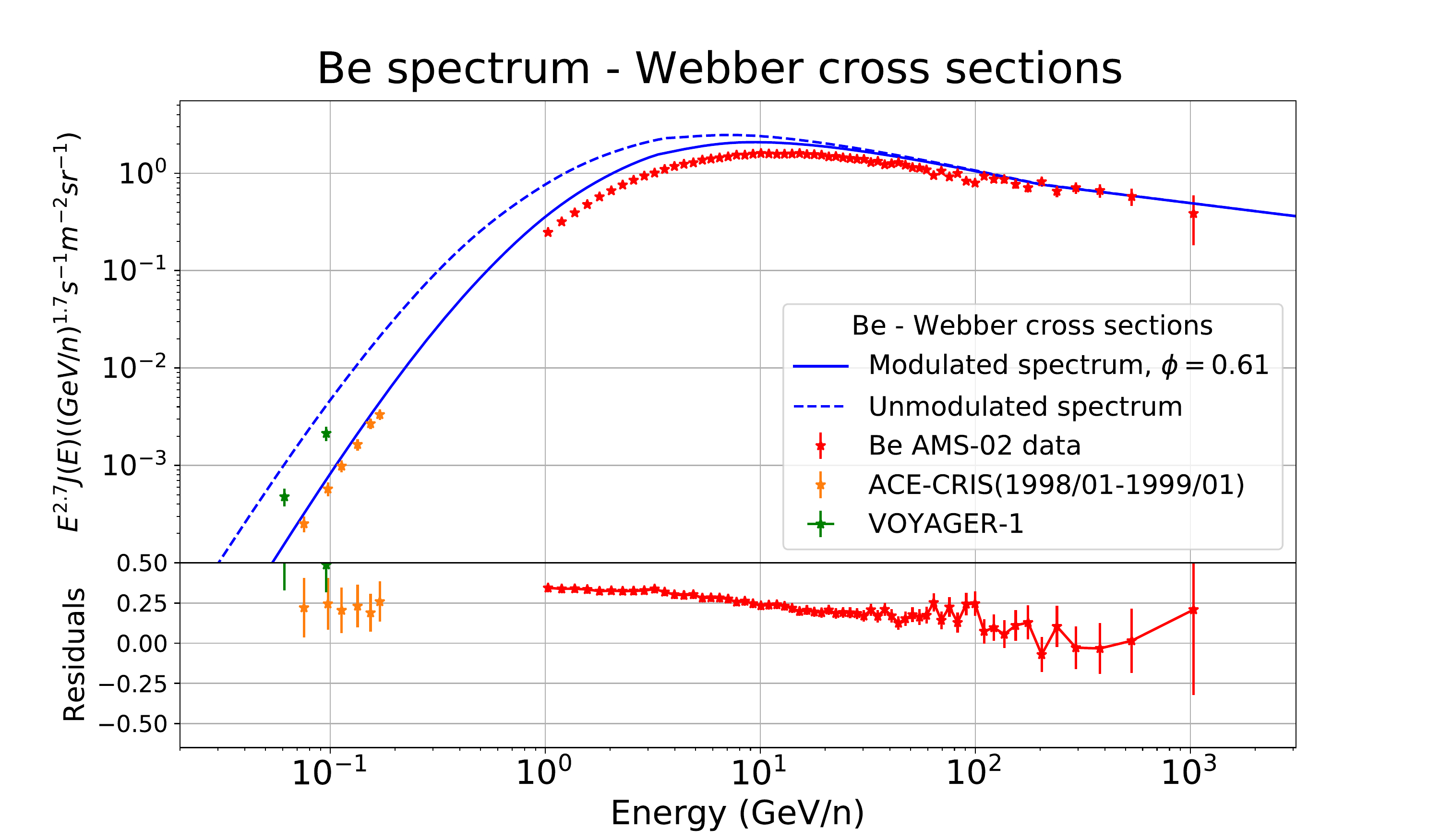}
\hspace{-0.6cm}
\includegraphics[width=0.345\textwidth,height=0.2\textheight,clip] {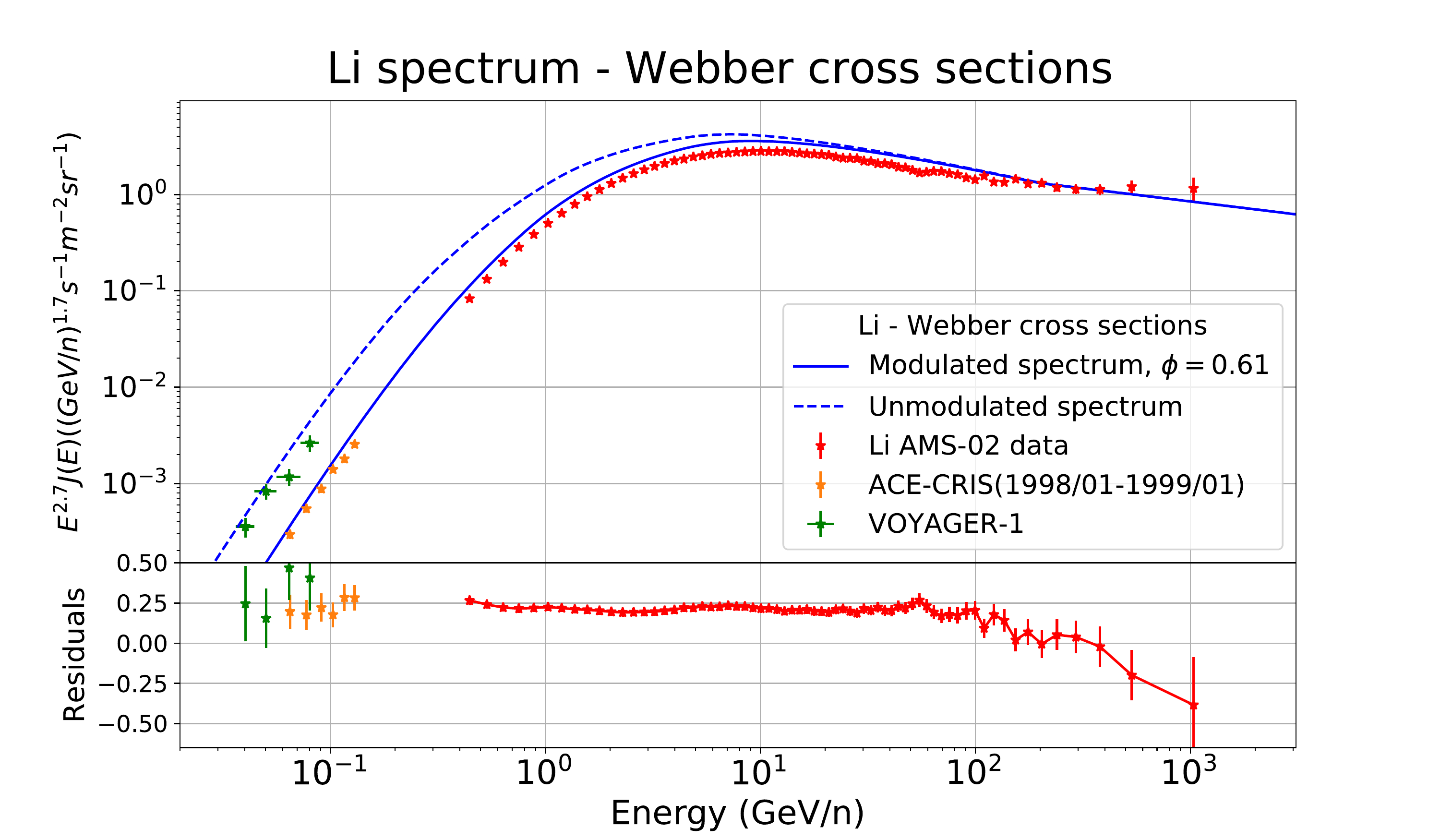}
\vspace{0.2cm}
\includegraphics[width=0.345\textwidth,height=0.20\textheight,clip]{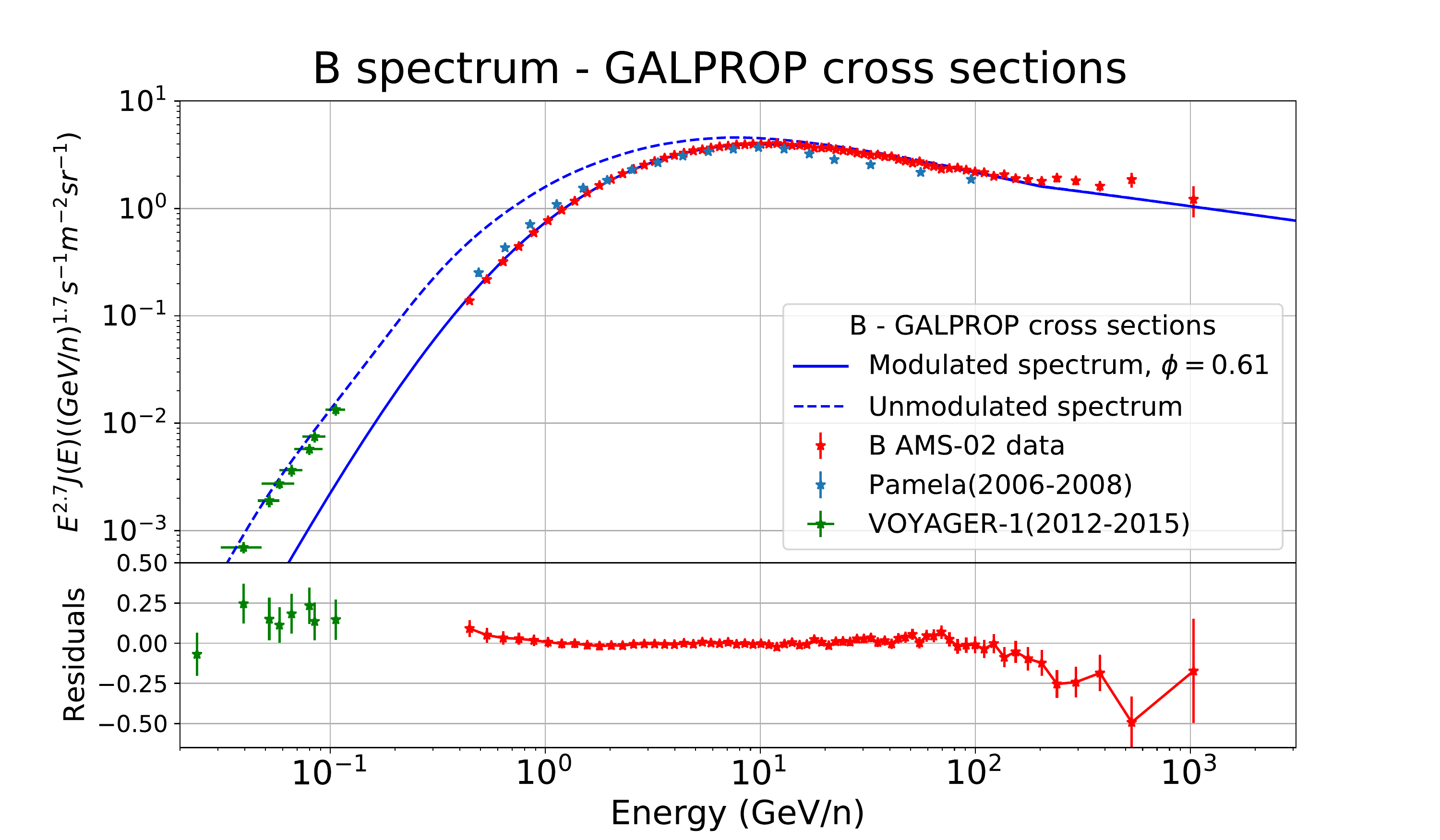} \hspace{-0.6cm}
\includegraphics[width=0.345\textwidth,height=0.20\textheight,clip]{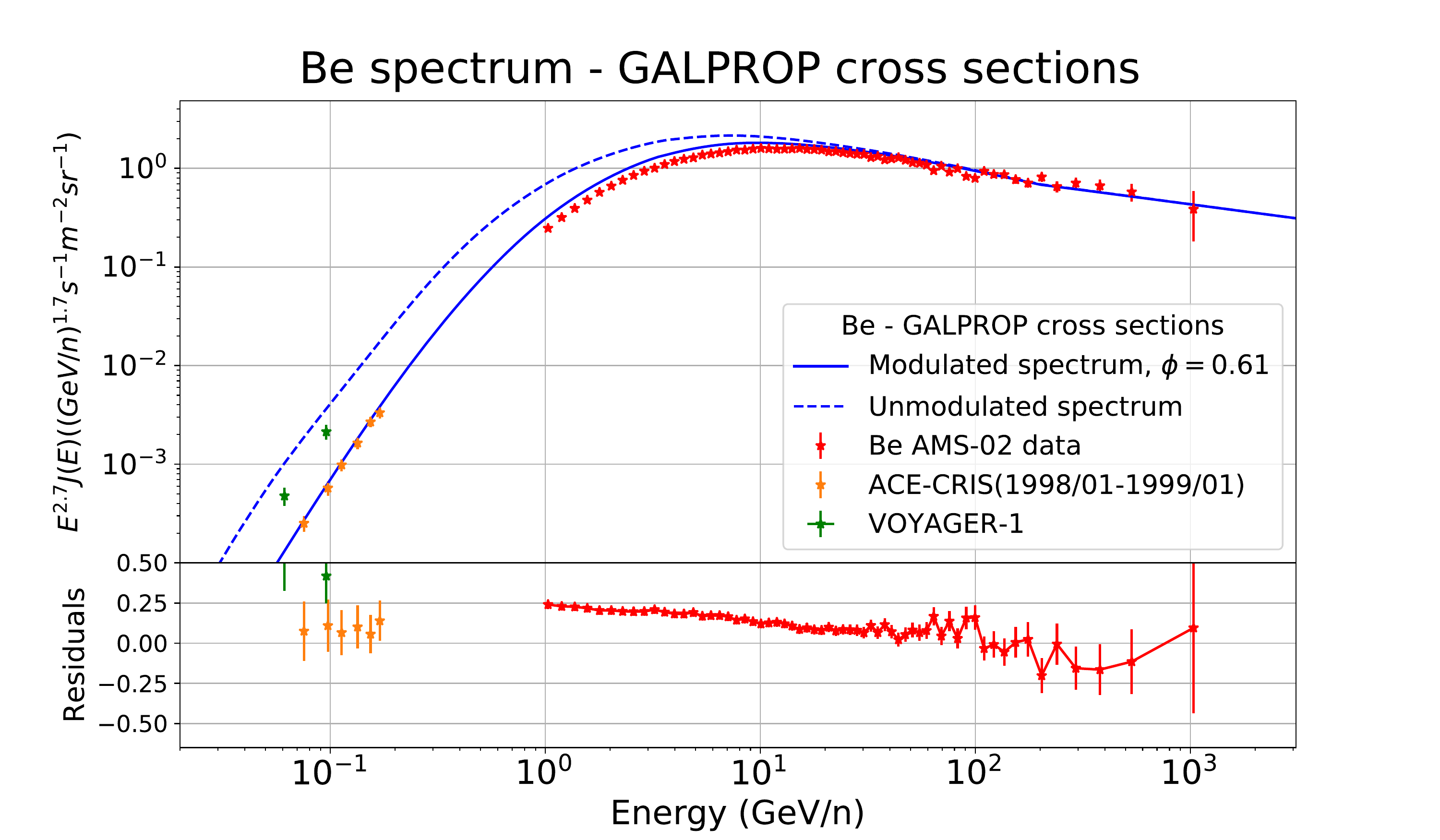} 
\hspace{-0.6cm}
\includegraphics[width=0.345\textwidth,height=0.20\textheight,clip]{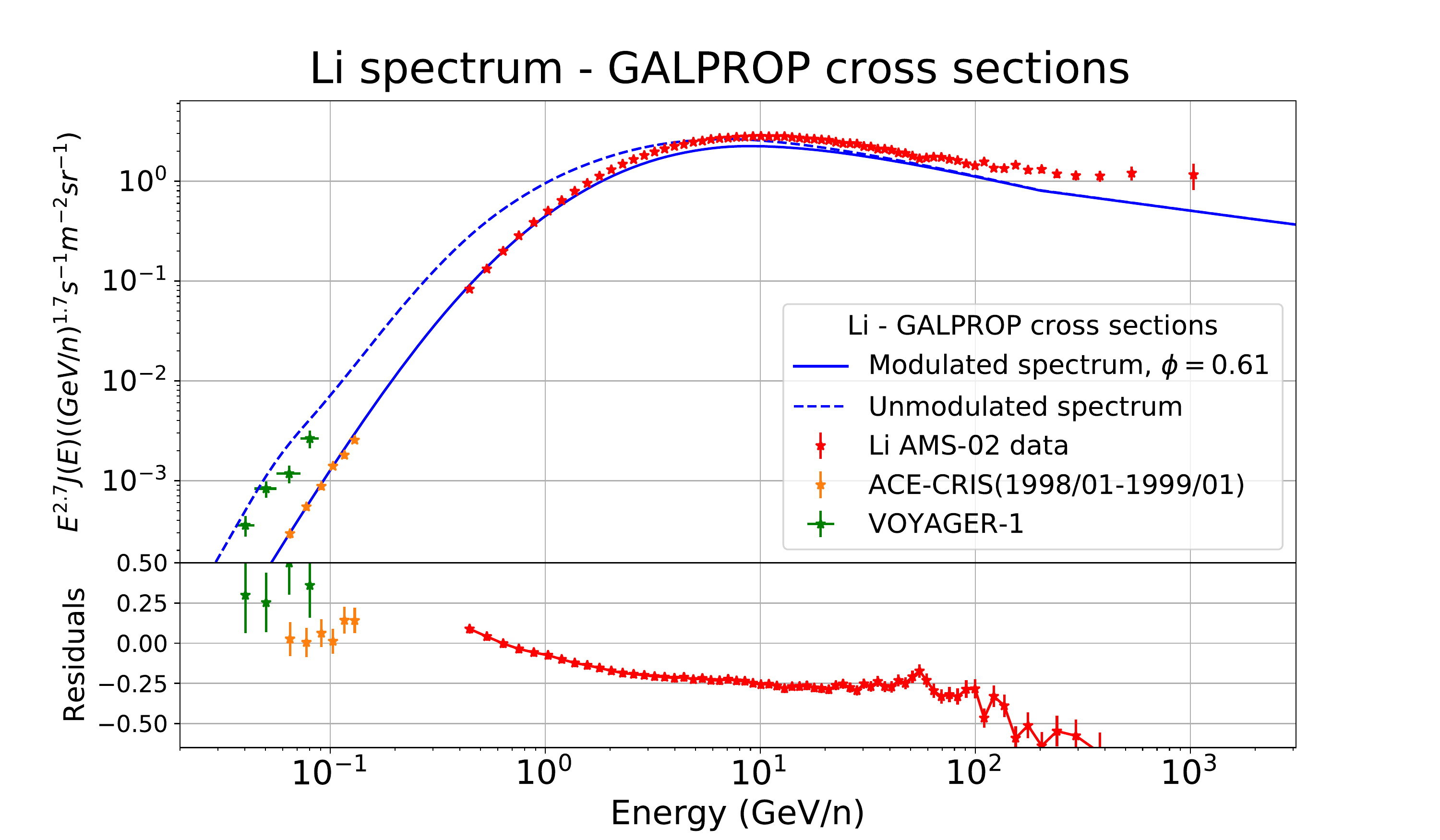}
\vspace{0.1cm}
\includegraphics[width=0.345\textwidth,height=0.20\textheight,clip]{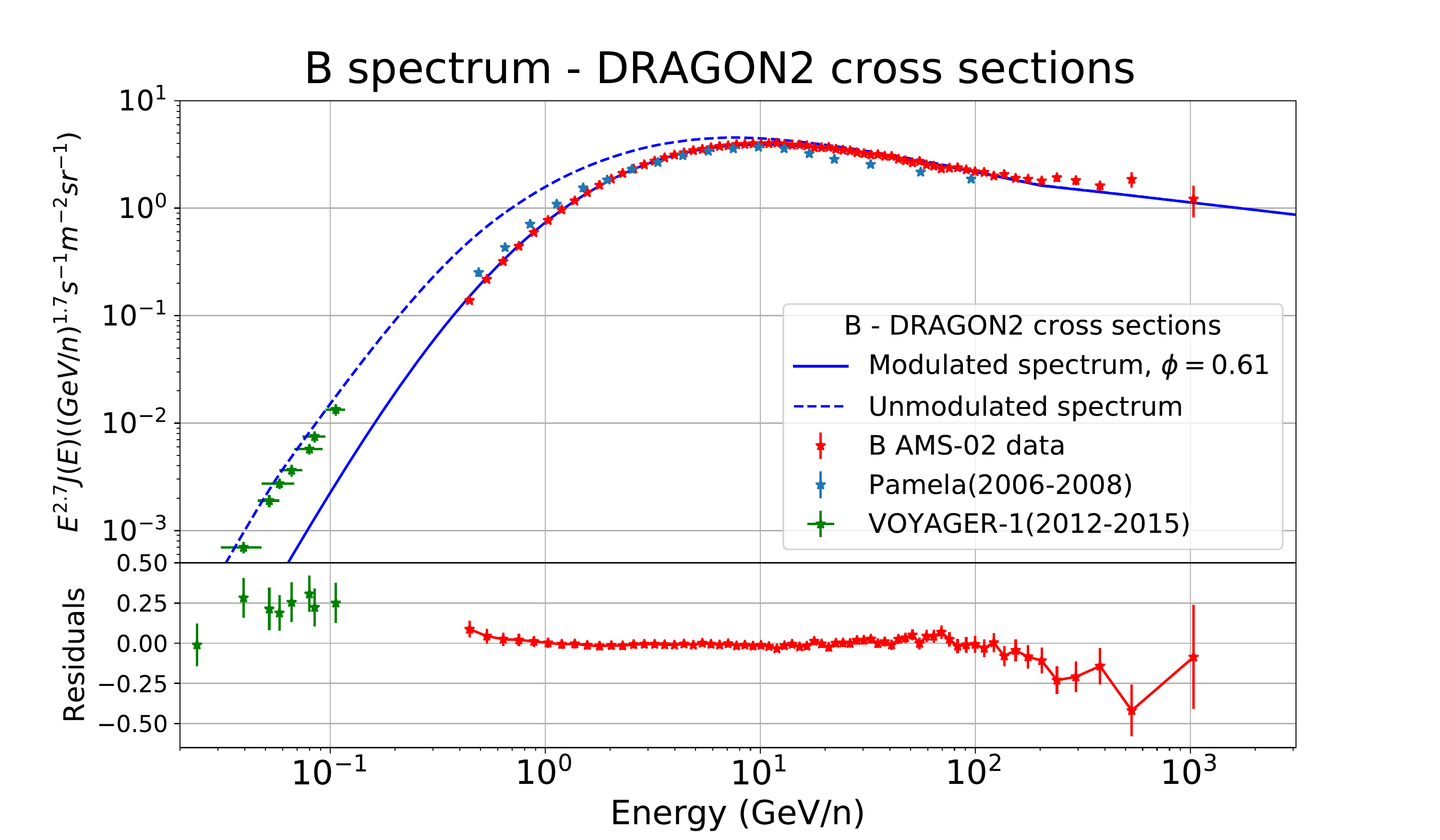} \hspace{-0.6cm}
\includegraphics[width=0.345\textwidth,height=0.20\textheight,clip]{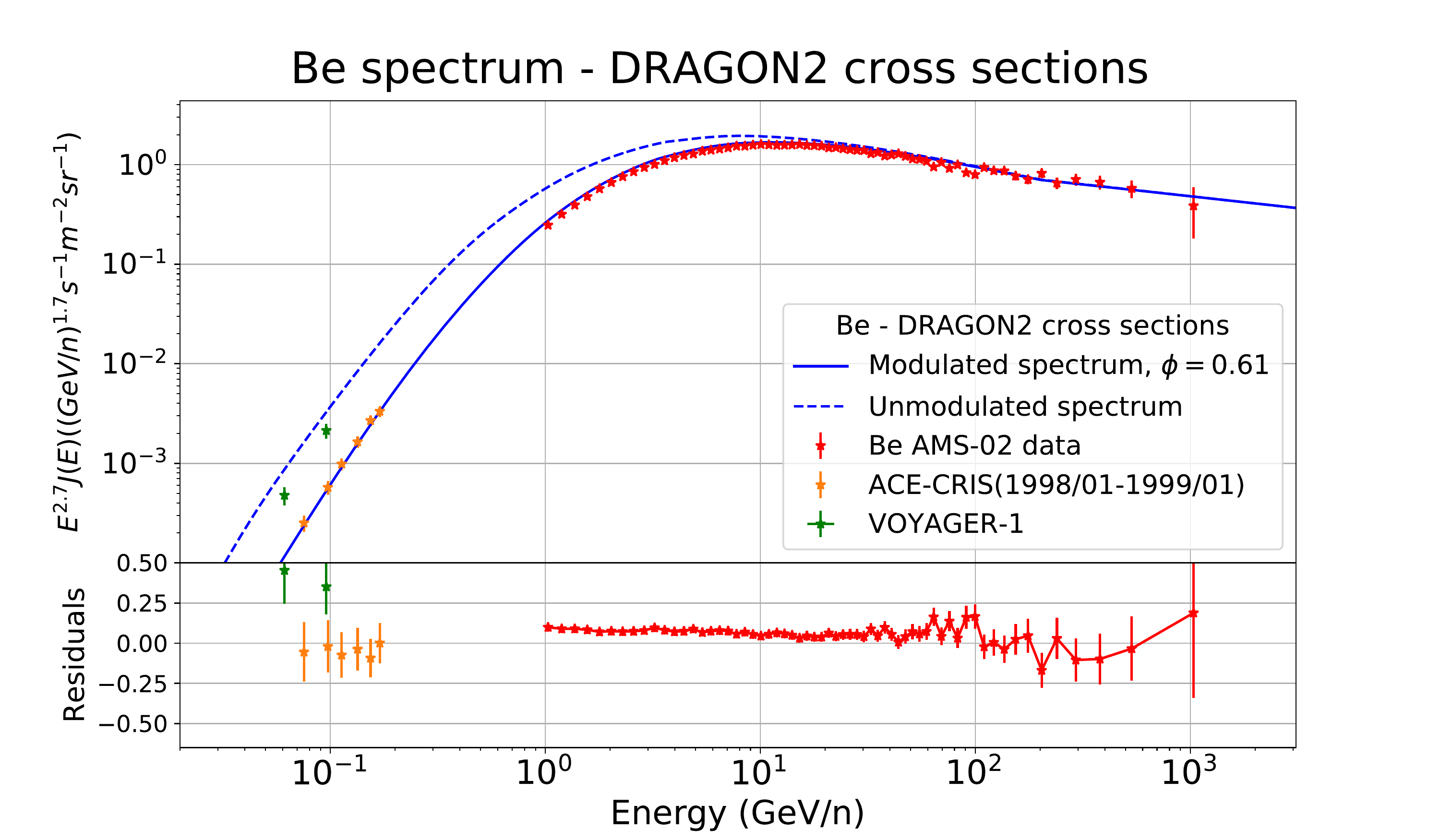} 
\hspace{-0.6cm}
\includegraphics[width=0.35\textwidth,height=0.20\textheight,clip]{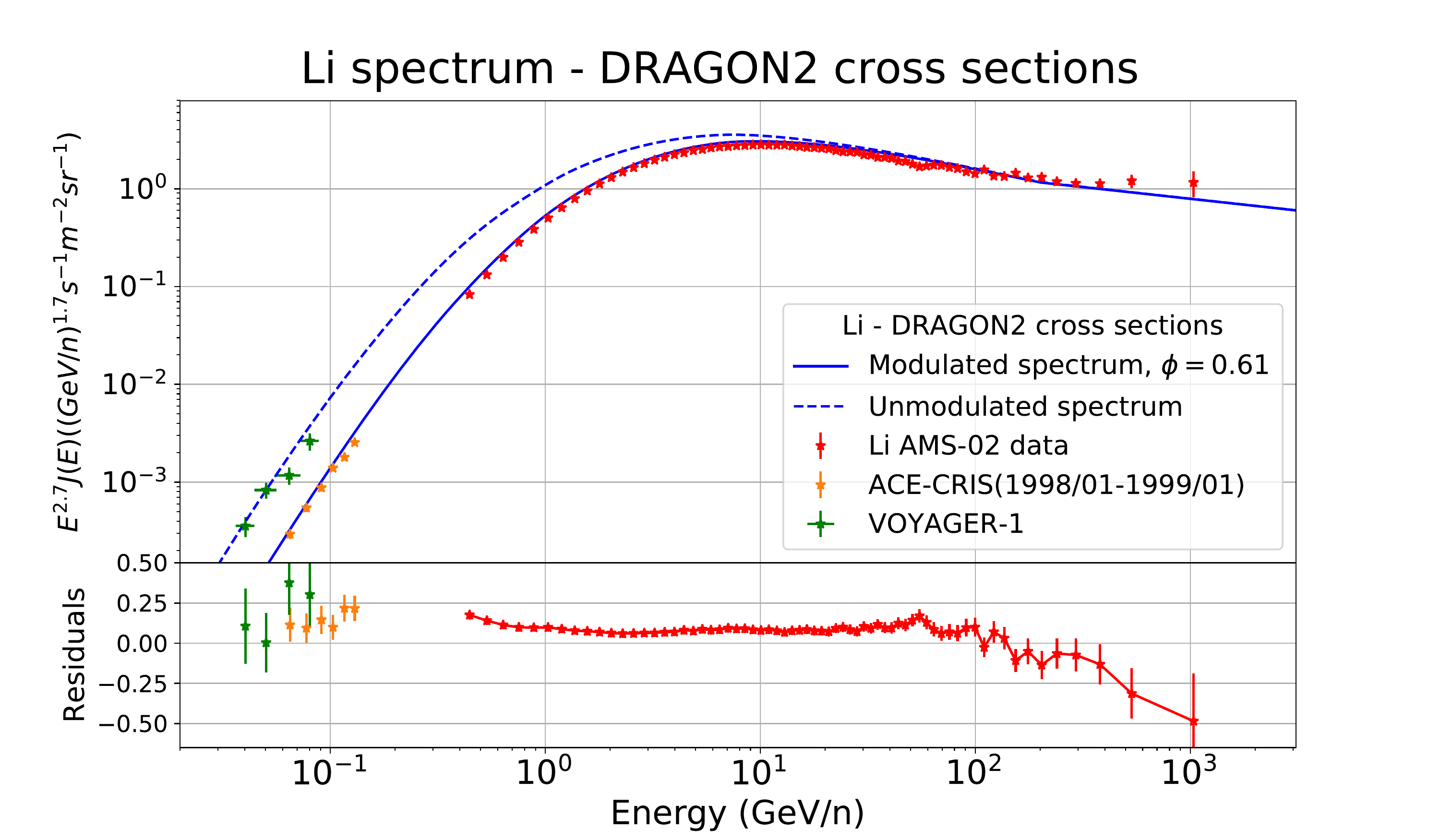}    
\end{center}
\caption{\footnotesize Spectra of the light secondary nuclei obtained with the diffusion parameters fitting the B/C spectrum using the Webber cross sections (upper row), the GALPROP parametrisations (middle row) and the DRAGON2 model (bottom row). The residuals (defined as (model-data)/model throughout all the thesis) are also shown to have an idea about the discrepancies for the different cross sections models. %Experimental data were taken from \url{//https://lpsc.in2p3.fr/crdb/}.
}
\label{fig:secfit_Webber}
\end{figure*} 

Figure~\ref{fig:secfit_Webber} shows the spectra of B, Be and Li for the three cross section models adopted in the present work. The boron spectrum is well fitted, since we use the diffusion parameters which reproduce the B/C ratio. These diffusion parameters are shown in table~\ref{tab:diff_params}. It must be highlighted here the importance of adding the new released data of AMS-02 \cite{AMS_Ne} for the Mg, Si and Ne fluxes, which appear to be significantly different from the earlier data (at a level up to $\sim 50\%$). This is mainly important for Be and Li, since it may imply a change larger than $10 \%$ in their predicted fluxes.

From the figure we see some discrepancies at low energies, probably due to our simplified description of solar modulation. In addition, in the very high-energy region, above $200 \units{GeV}/n$, we see that the measured fluxes of Li, Be and B are slightly higher than the predicted fluxes for all the three cross section parametrisations. This feature suggests the necessity of introducing a break in the energy dependence of the diffusion coefficient (e.g., \cite{genolini2017indications}) rather than in the source CR injection spectra.

In the case of the Webber cross sections, we see that the Li and Be fluxes follow a similar trend, being overestimated with respect to the AMS-02 data with residuals generally below 25\%. In turn, for the GALPROP parametrisations, we see an opposite behaviour for Li, with the simulation underestimating the experimental data of about 22\%, while in the case of Be the fit is closer to the experimental data, with discrepancies less than 15\% on average, after $10 \units{GeV}$. Finally, the DRAGON2 default cross sections seem to succeed reproducing all the secondary CRs at the same time within 10\% discrepancies in the full energy range. 

%\newpage

\subsection{Secondary-over-secondary ratios}
\label{sec:simplesec}

A comparison among the predicted Li, Be and B fluxes for the three models shows the effects of the limited knowledge on spallation cross sections due to missing data and poorly known reaction channels, in particular for Li. Further uncertainties on the fluxes due to the total inelastic cross section with the ISM have also been studied in~\cite{derome2019fitting} and are roughly 3\%. Nevertheless, these fluxes can compensate the cross sections defects by modifying the parameters in the diffusion coefficient, when studying them independently \cite{Weinrich:2020cmw}.

Unlike secondary-over-primary flux ratios, ratios among secondary CRs have demonstrated to be roughly unaffected by the parametrisation of the diffusion coefficient at high energies (see section~\ref{sec:theory_sec}) while they show a mild dependence on the spectra of primary CRs. This, indeed, makes a comparison between the predicted secondary fluxes easier.

However, there is another quantity which can significantly affect these ratios among secondaries: the size of the galactic halo. As mentioned above, different unstable CR species have different decay lengths and, depending on the path length they travel until reaching the Earth, different fractions of unstable nuclei can decay, thus influencing the ratios. In particular, the Be/B ratio is highly sensitive to the halo size, due to the presence of the radioactive isotope $^{10}$Be (see \cite{CarmeloBeB},  where the authors discuss the halo size effects on the Be/B ratio at low energy), while the ratios involving other species of secondaries seem to be quite insensitive above a few $\units{GeV}$. The Li/B spectrum exhibits a soft dependence on the halo size, since $^{10}$Be isotopes can decay into $^{10}$B. We point out here that the Be flux at low energies is also affected by the gas density distribution of the Galaxy, %The gas profile in the z direction is crucial, as it determines the source term for secondary CRs. 
since the $^{10}$Be decay length at low energy is of the order of a few hundreds of parsecs. Changing the gas distribution can, therefore, change the amount of predicted $^{10}$Be and, thus, the predicted secondary ratios involving Be, as it is shown in appendix~\ref{sec:appendixB}. 

Generally, it can be observed that the secondary spectra are strongly related to the cross sections, and above a few tens of $\units{GeV}$ they have very little dependence on any other parameter. In fact, they are so sensitive to cross sections that they could, in principle, be used to constrain their parametrisations. Within the current accuracy of spallation measurements, errors even larger than $50\%$ in these simulated ratios, with respect to experimental data, may be expected.  

\begin{figure*}[!hbt]
\begin{center}
\includegraphics[width=0.34\textwidth,height=0.175\textheight,clip] {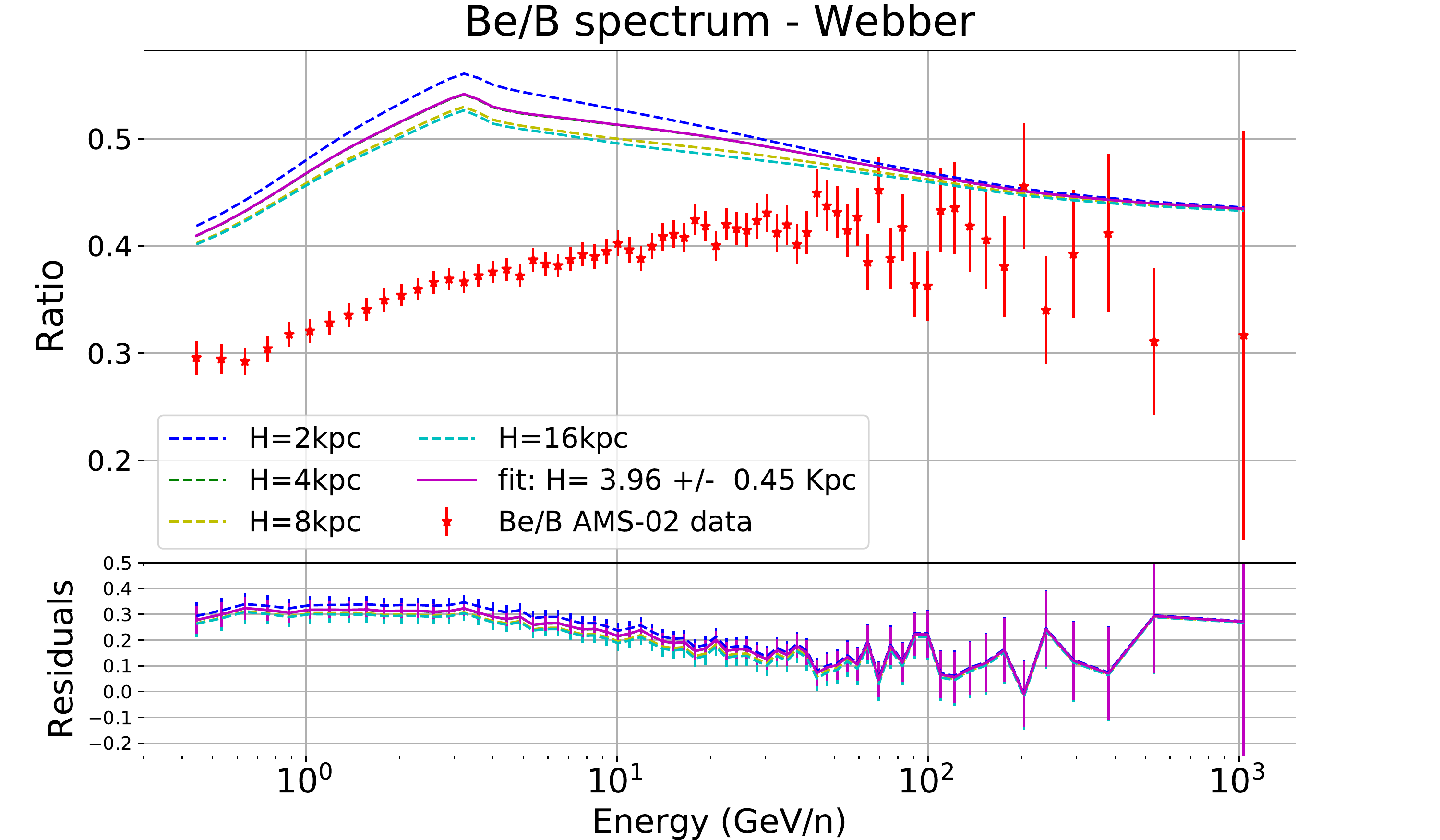} \hspace{-0.4cm}
\includegraphics[width=0.34\textwidth,height=0.175\textheight,clip] {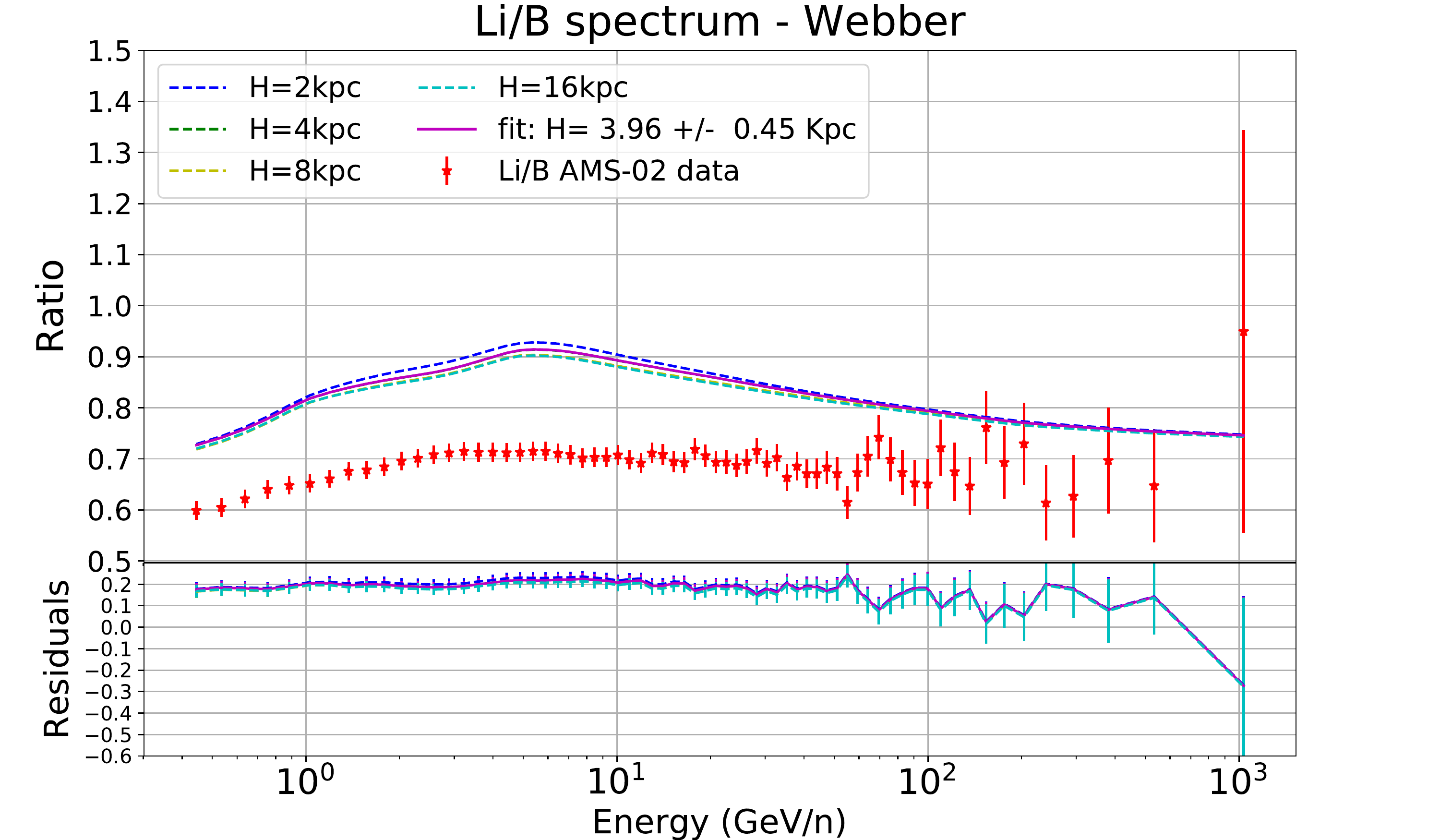} 
\hspace{-0.4cm}
\includegraphics[width=0.34\textwidth,height=0.175\textheight,clip] {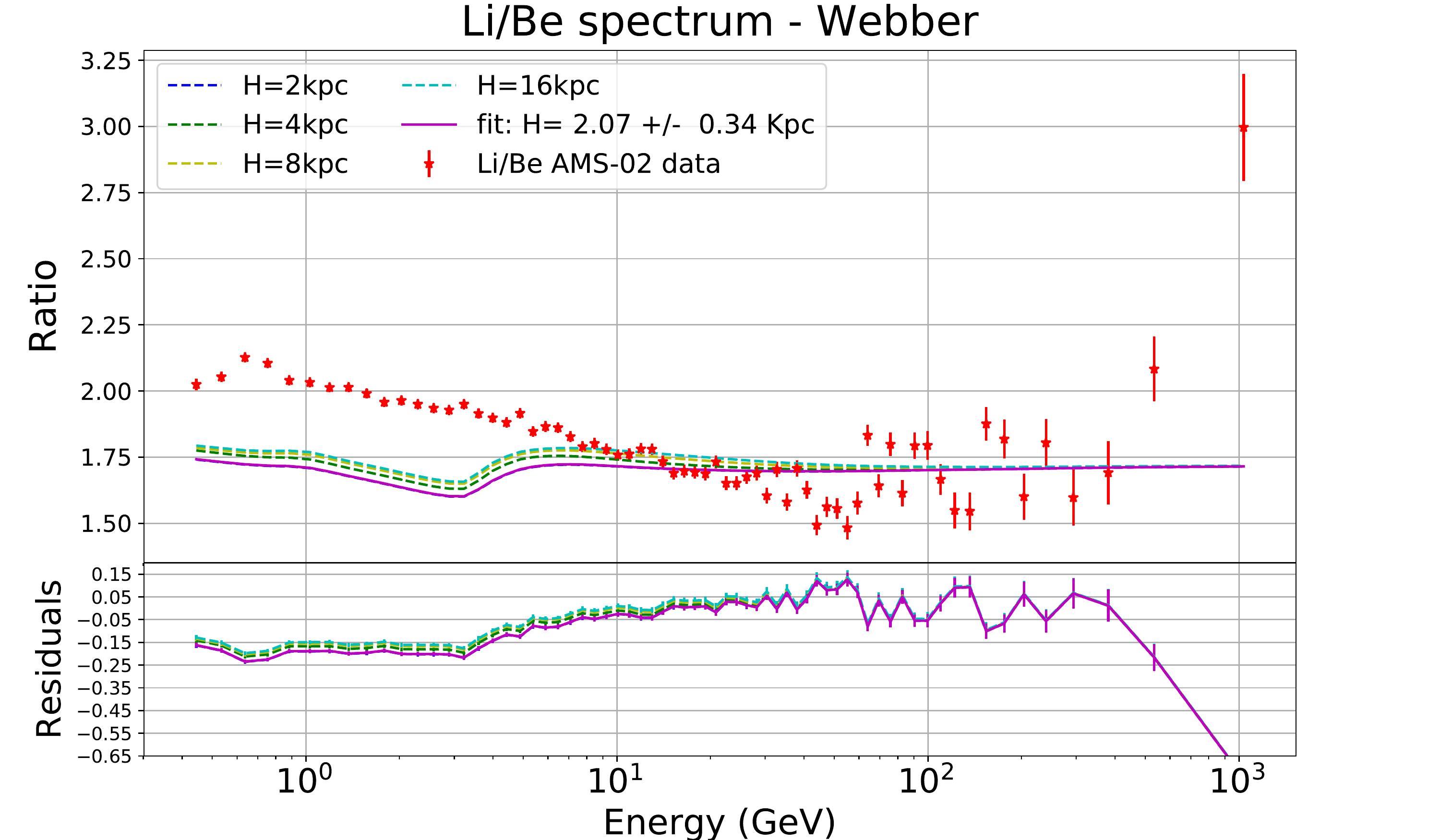}

\includegraphics[width=0.34\textwidth,height=0.175\textheight,clip] {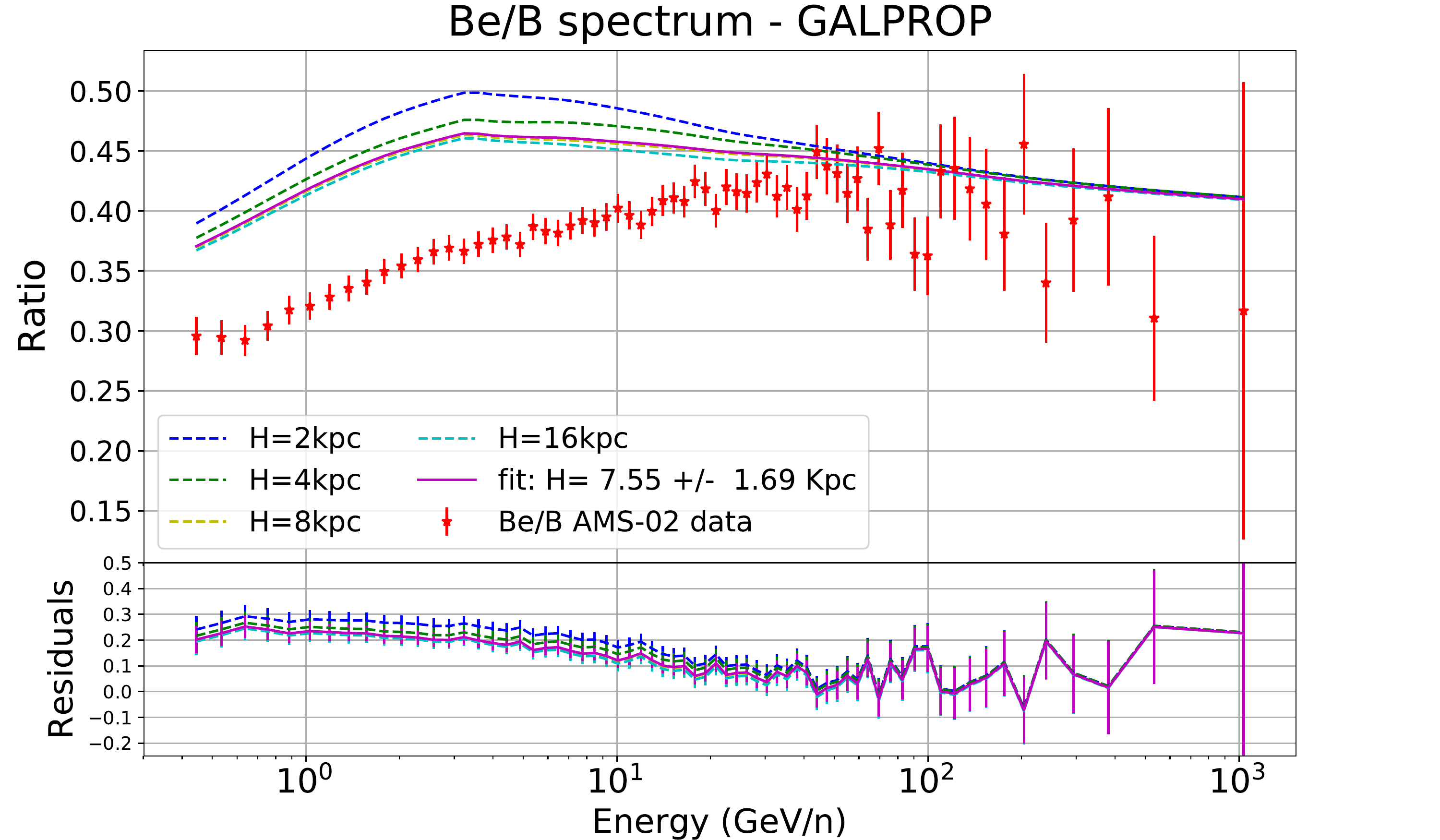} \hspace{-0.4cm}
\includegraphics[width=0.34\textwidth,height=0.175\textheight,clip] {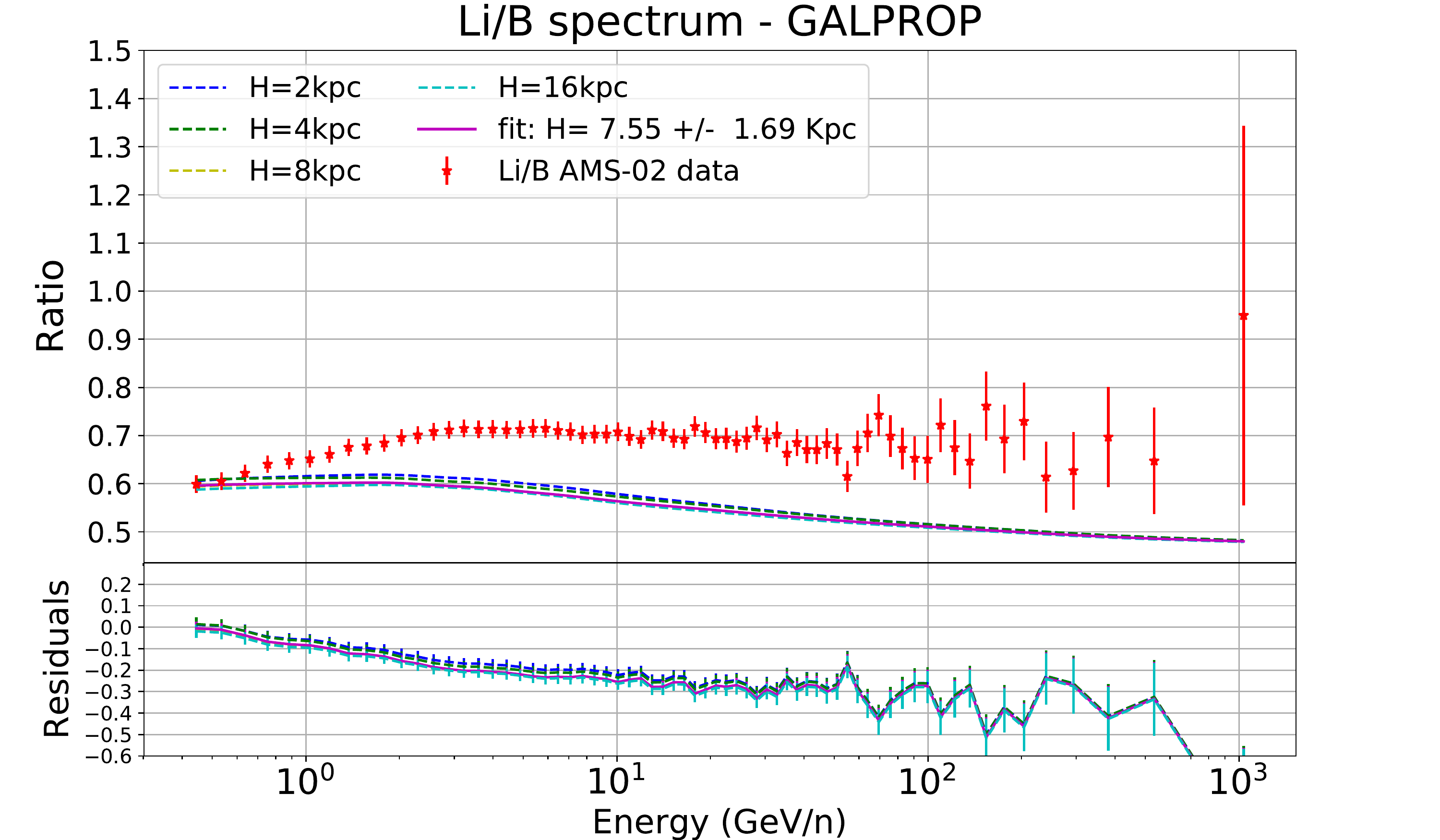}
\hspace{-0.4cm}
\includegraphics[width=0.34\textwidth,height=0.175\textheight,clip] {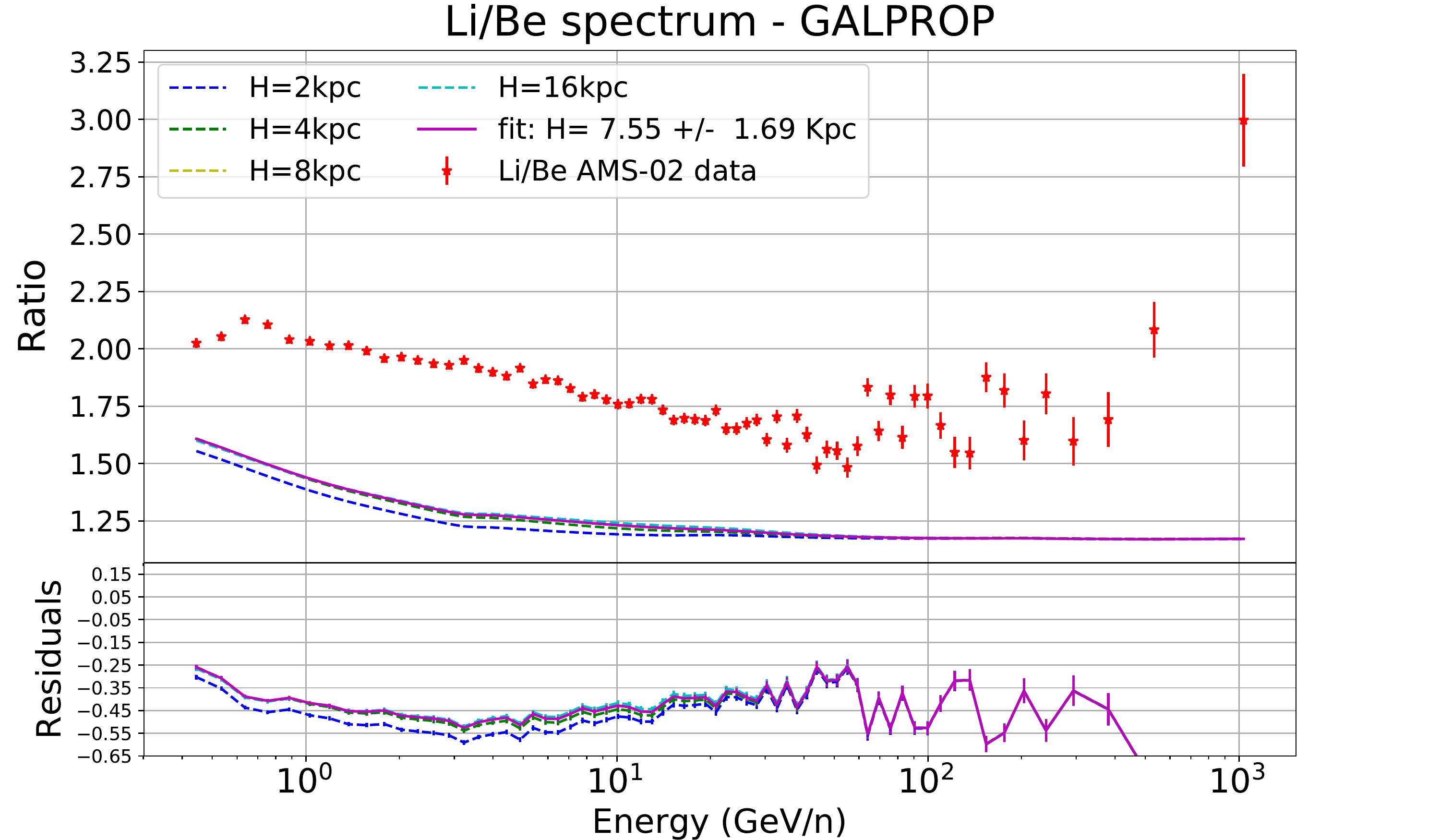}

%\vspace{0.1cm}
\includegraphics[width=0.34\textwidth,height=0.175\textheight,clip] {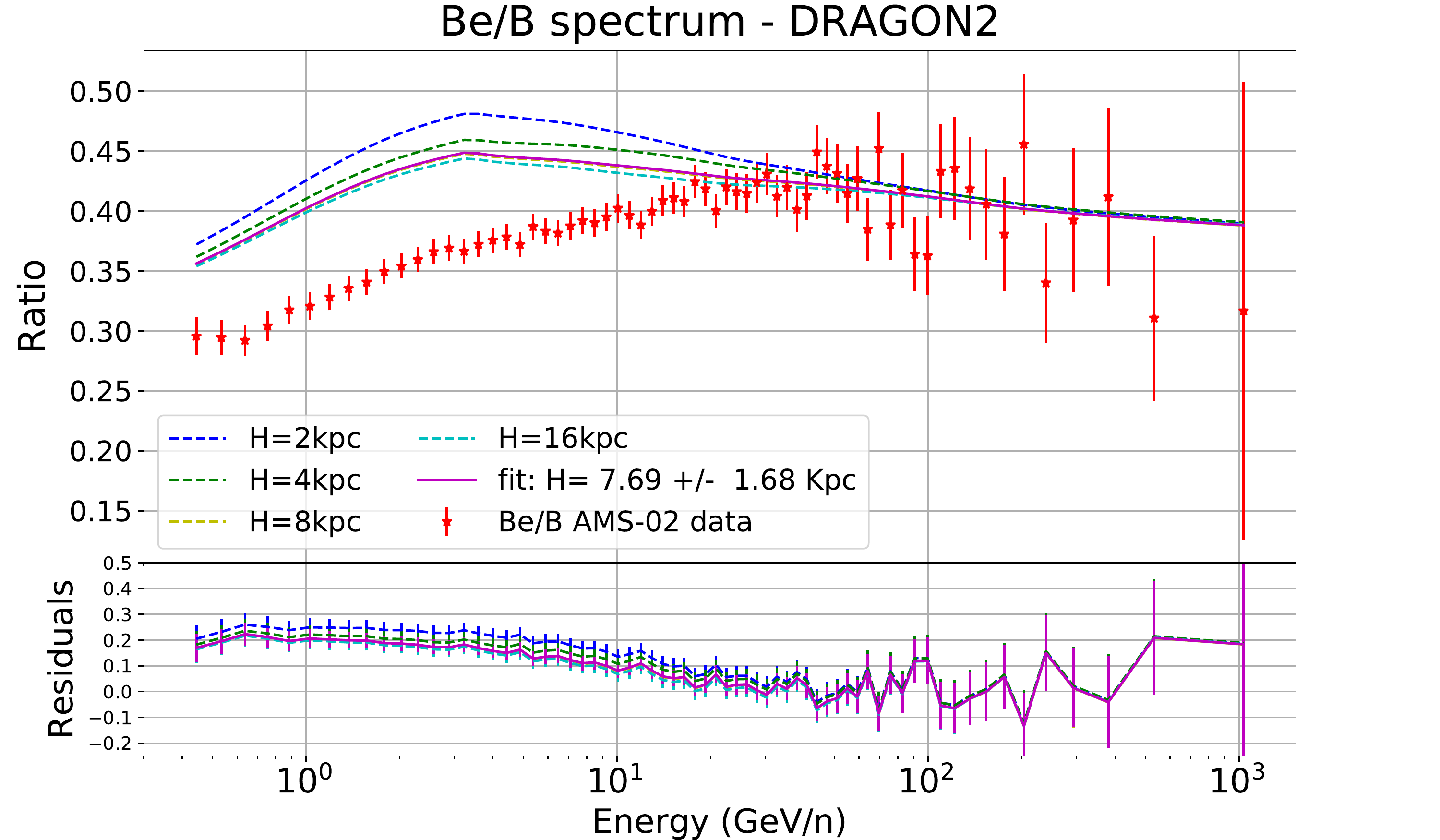} \hspace{-0.4cm}
\includegraphics[width=0.34\textwidth,height=0.175\textheight,clip] {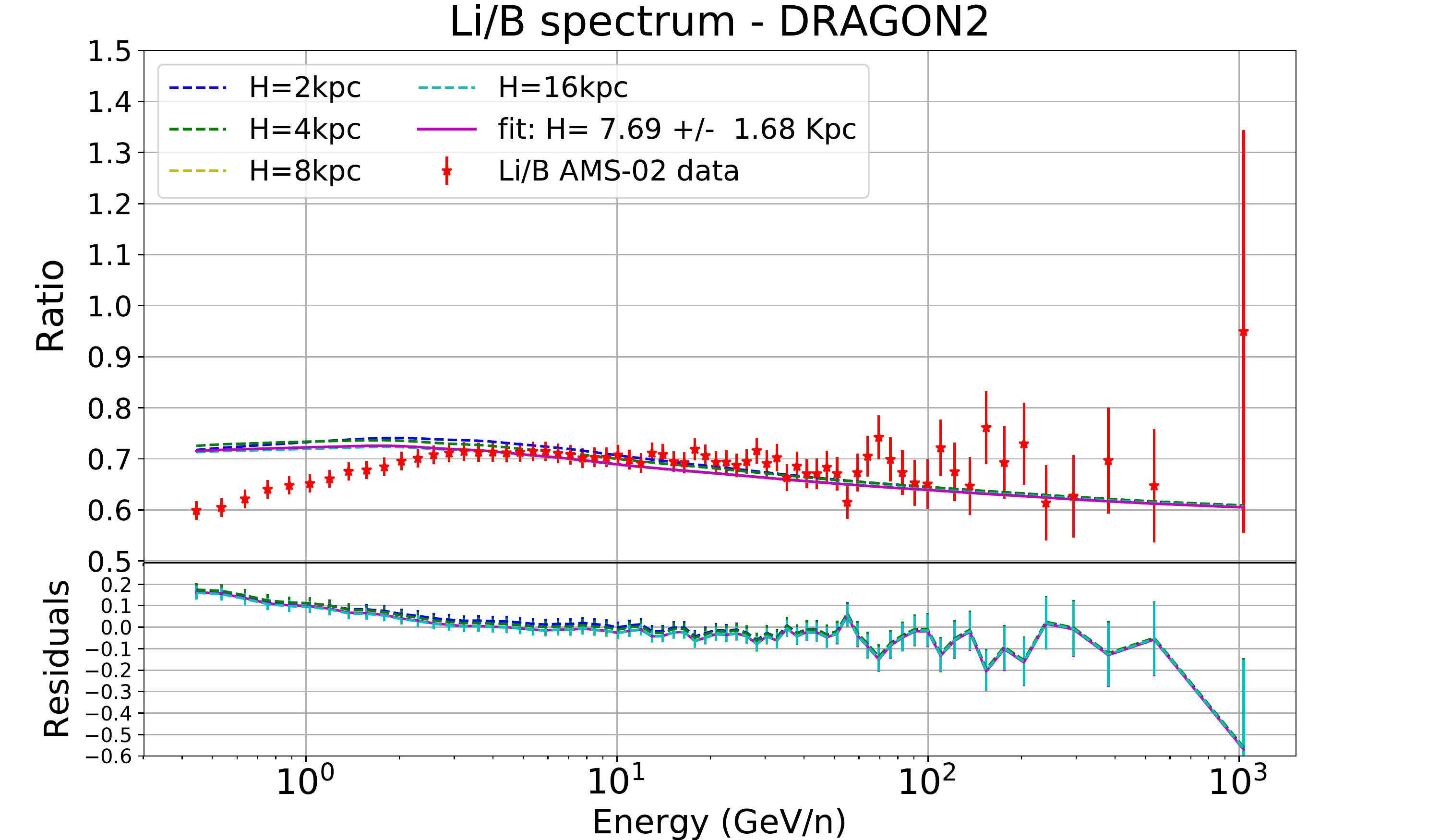}
\hspace{-0.4cm}
\includegraphics[width=0.34\textwidth,height=0.175\textheight,clip] {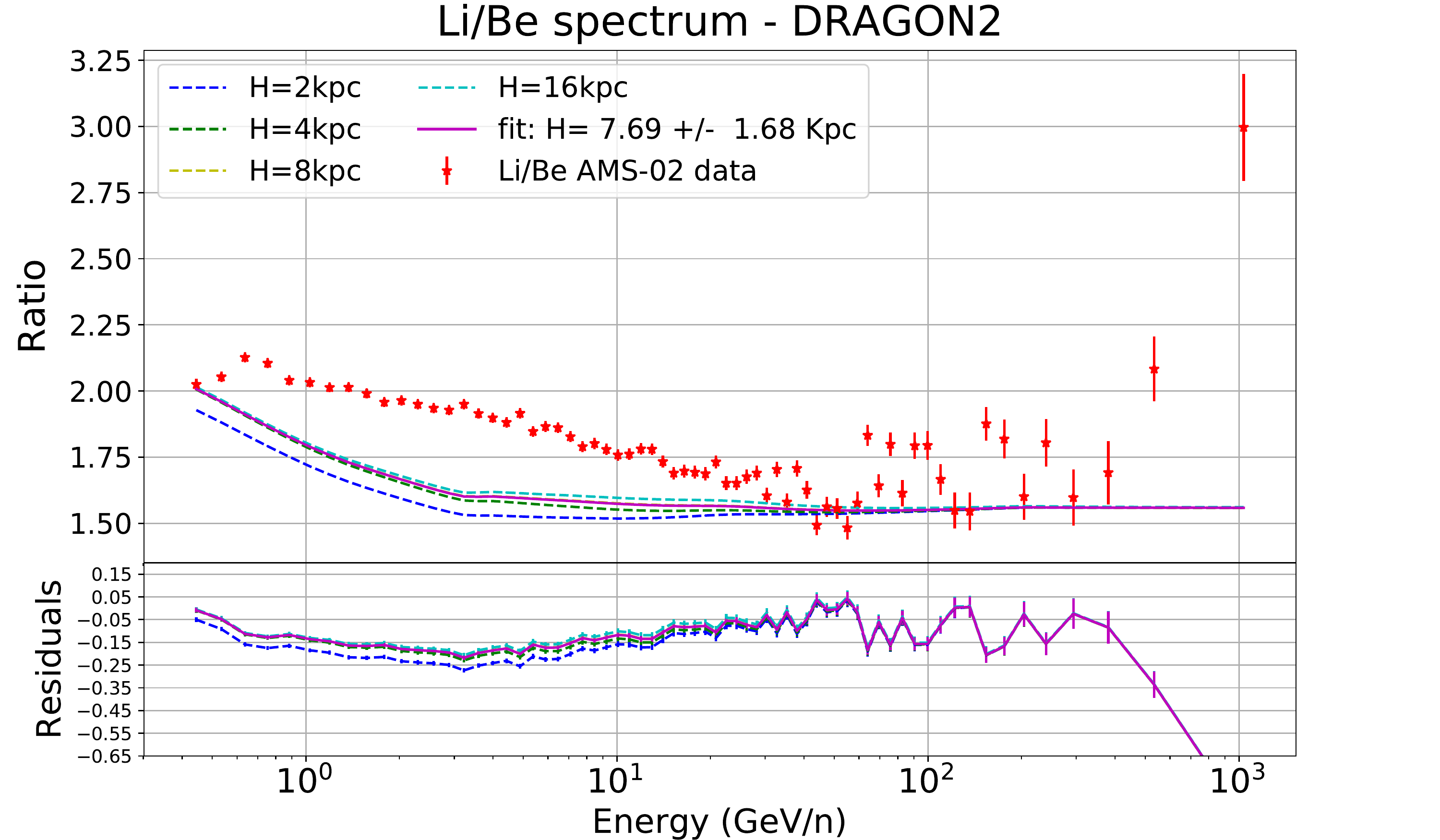}
\end{center}
\caption{\footnotesize Secondary-to-secondary ratios of the light secondary CRs for the Webber, GALPROP and DRAGON2 cross sections models. The residuals are also shown to better illustrate the discrepancies between simulations and data. These plots include the simulated spectra for various halo sizes, since the presence of $^{10}$Be and its beta decay to $^{10}$B modifies the shape of the spectra at low energies. The simulated spectrum for the halo size that best fits the spectra of ratios of Be isotopes, as explained in the section~\ref{sec:size}, is also included.}
\label{fig:secsec_Webber}
\end{figure*}

Figure~\ref{fig:secsec_Webber} shows the Be/B, Li/B and Li/Be ratios predicted using the three cross section models (see table~\ref{tab:diff_params}) assuming different sizes for the galactic halo, from $2$ to $16 \units{kpc}$, compared with the experimental data in the energy range from $500 \units{MeV/n}$ up to $1 \units{TeV/n}$. In the figure we also show the model corresponding to the best fit of the halo size, which will be discussed in section~\ref{sec:size}. In the case of the Li/B and Li/Be ratios, variations of the halo size yield variations of the ratios of no larger than 5\%. On the other hand, in the case of the Be/B ratio, variations of the halo size yield variations in the spectra up to 10\% in the low energy region.

From this figure, we can see that the largest residuals are usually found at low energies, since this part is dependent on many propagation parameters. In particular, the shape of the ratios at low energy is not well reproduced for the Webber cross sections, with residuals up to 40\% in the low-energy region. The shapes of the GALPROP and DRAGON2 parametrisations are very similar, finding that the largest discrepancies for the former are those in the ratios involving Li. However, with the cross sections of the DRAGON2 code, the residuals are the smallest, with discrepancies less than 10\% above $5 \units{GeV}$. It is worth emphasizing the importance of the use of the new AMS-02 data for Ne, Mg and Si, which appreciably changes the shapes and residuals with respect to old data. 

%In general, the maximum level of residuals we see above 10 GeV is below 40$\%$, which is inside the expected uncertainties due to the cross sections parametrisations. But, comparing different models (namely, Webber and Galprop), it may be seen that the lithium spectra are different by around a $40\%$ and those of Be less than $20\%$ (as the B flux is the result of the fit, it remains constant, so we can easily compare different cross sections models from the comparison of the B ratios). 
%Again here, the parametrisations of the DRAGON2 default model for cross sections success in reproducing these ratios with very small discrepancy (smaller than 10\%).

\begin{figure*}[!hbt]
\begin{center}
\includegraphics[width=0.52\textwidth,height=0.27\textheight,clip] {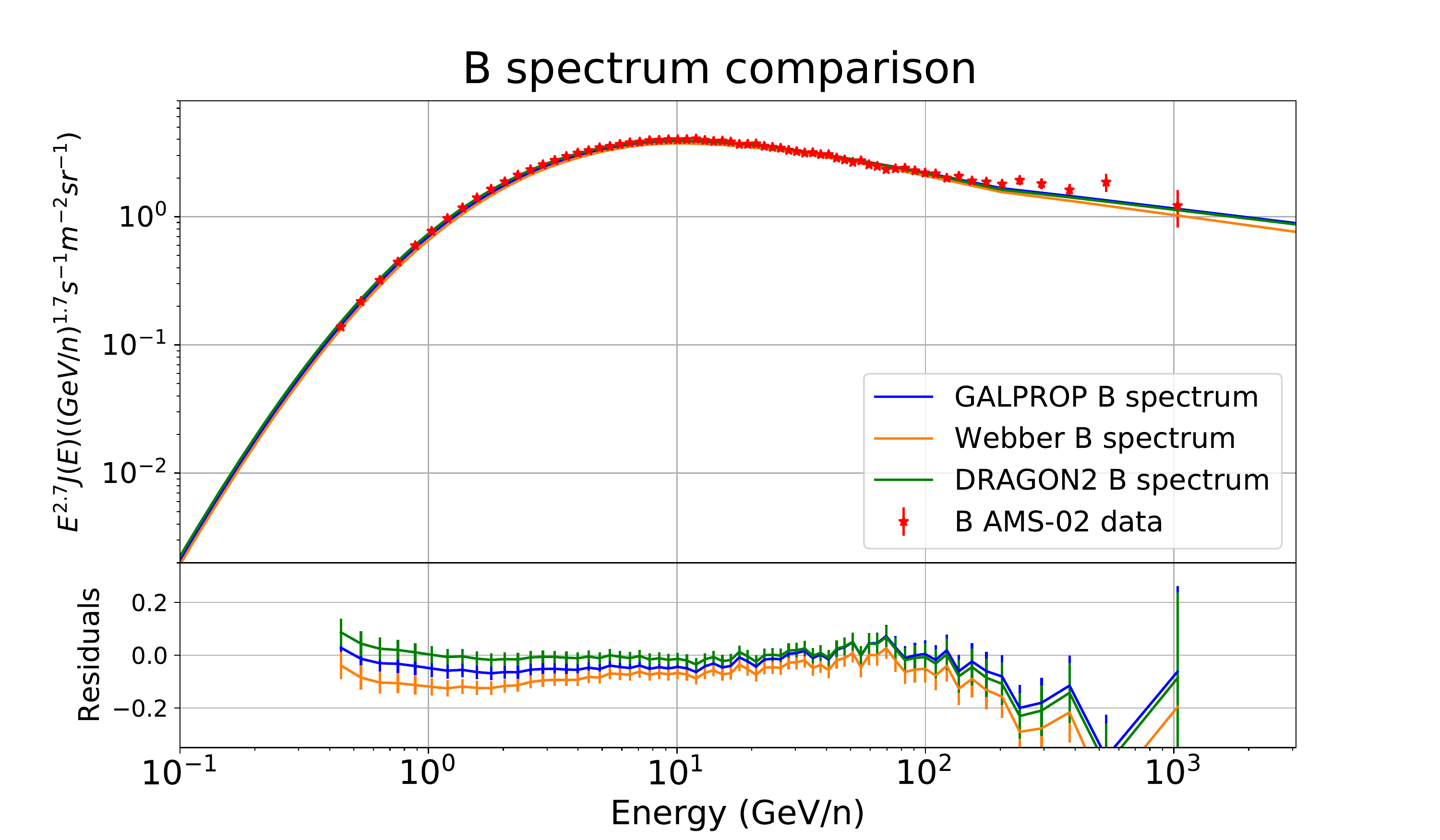} \end{center}
\caption{\footnotesize Comparison of the B spectra simulated with fixed diffusion coefficient for the different cross sections parametrisations used in this work. The values of the diffusion parameters are those obtained in the fit of the boron-to-carbon ratio using the DRAGON2 model.}    
\label{fig:BComps}
\end{figure*}

Since figures~\ref{fig:secsec_Webber} and~\ref{fig:secfit_Webber} show the fluxes predicted when fixing the B/C flux ratio to the AMS-02 data, it is worth also comparing the predictions of the different parametrisations with fixed diffusion parameters. This is shown in Fig.~\ref{fig:BComps}, where the simulations have been performed for the three parametrisations using the diffusion parameters found for the DRAGON2 model. We can see here that differences less than $10\%$ are found between the GALPROP and DRAGON2 parametrisations, while these differences are slightly larger when comparing to the Webber cross sections.
In general, these differences become less important above $10 \units{GeV}$.

\subsection{Theory on secondary-over-secondary ratios}
\label{sec:theory_sec}
%\vspace{0.4cm}

In this subsection we will give an overview on the basic theory about the secondary-over-secondary flux ratios. Here we demonstrate that these ratios are mainly dependent on the cross sections of secondary CR production. 
Roughly speaking, if one considers that the flux of a primary CR species at Earth is $J_{\alpha}(E) \propto q_{\alpha}\times\tau_{diff}(E)$ and $\tau_{diff}(E) = H^2/D(E)$, it will take the form $J_{\alpha}(E) \sim C_{\alpha}E^{-\gamma_{\alpha} - \delta}$ at high energies (more precisely, around a few$\units{GeVs}$, when $1/\sigma_{\alpha}^{ine}(E) \gg n_g hc \tau_{diff}(E)/H$). The secondary CR fluxes take the form $J_{i}(E) \propto \sum^{\alpha} J_{\alpha}(E)cn_g\sigma_{\alpha
\rightarrow
i}(E)\times\tau_{diff}(E)$ whose ratio leaves the expression:
\begin{equation}
\label{eq:sectosec}
\begin{split}
%\centering
\frac{J_k}{J_j}\Bigl(E\Bigr) \propto \frac{\sum^{\alpha \rightarrow k}J_{\alpha}(E)\sigma_{\alpha \rightarrow k}(E)}{\sum^{\alpha \rightarrow j}J_{\alpha}(E)\sigma_{\alpha \rightarrow j}(E)} \hspace{0.8cm} {\xrightarrow{\makebox[1cm]{high energies}}} \hspace{0.6cm} \sim \frac{\sum^{\alpha \rightarrow k}C_{\alpha}E^{-\gamma_{\alpha}}\sigma_{\alpha \rightarrow k}(E)}{\sum^{\alpha \rightarrow j}C_{\alpha}E^{-\gamma_{\alpha}}\sigma_{\alpha \rightarrow j}(E)}
\end{split}
\end{equation}
%\vspace{0.6cm}
Here we point out that the factor $E^{-\delta}$ is common to all the terms in the summations at the numerator and at the denominator, thus cancelling out. From equation~\ref{eq:sectosec} one can see that these ratios have a direct dependence on the local spectrum of primary nuclei ($\alpha$) and the on the overall spallation cross sections. Nonetheless, the local primary spectrum may have some dependence on the diffusion parameters at low energies and reacceleration can also slightly contribute. The total repercussion of these parameters in the low energy region can be estimated to be at a level smaller than 10\%. Then, due to the presence of the radioactive $^{10}$Be, the most important contributor to the low-energy uncertainties is the halo size, along with the gas profile, being able to introduce variations of more than 10\% and a few percent, respectively. At the end, these ratios demonstrate to be mainly dependent on the source term of primary CRs and on their spallation cross sections at high energies. As the AMS-02 data for the local spectrum of primary CRs are highly precise, the largest uncertainty at high energy lies on the spallation cross sections and this fact makes them so suitable for adjusting the overall cross sections (not individual channels) with high precision (AMS-02 precision, indeed).

\begin{figure*}[!tbp]
\begin{center}
\includegraphics[width=0.57\textwidth,height=0.25\textheight,clip] {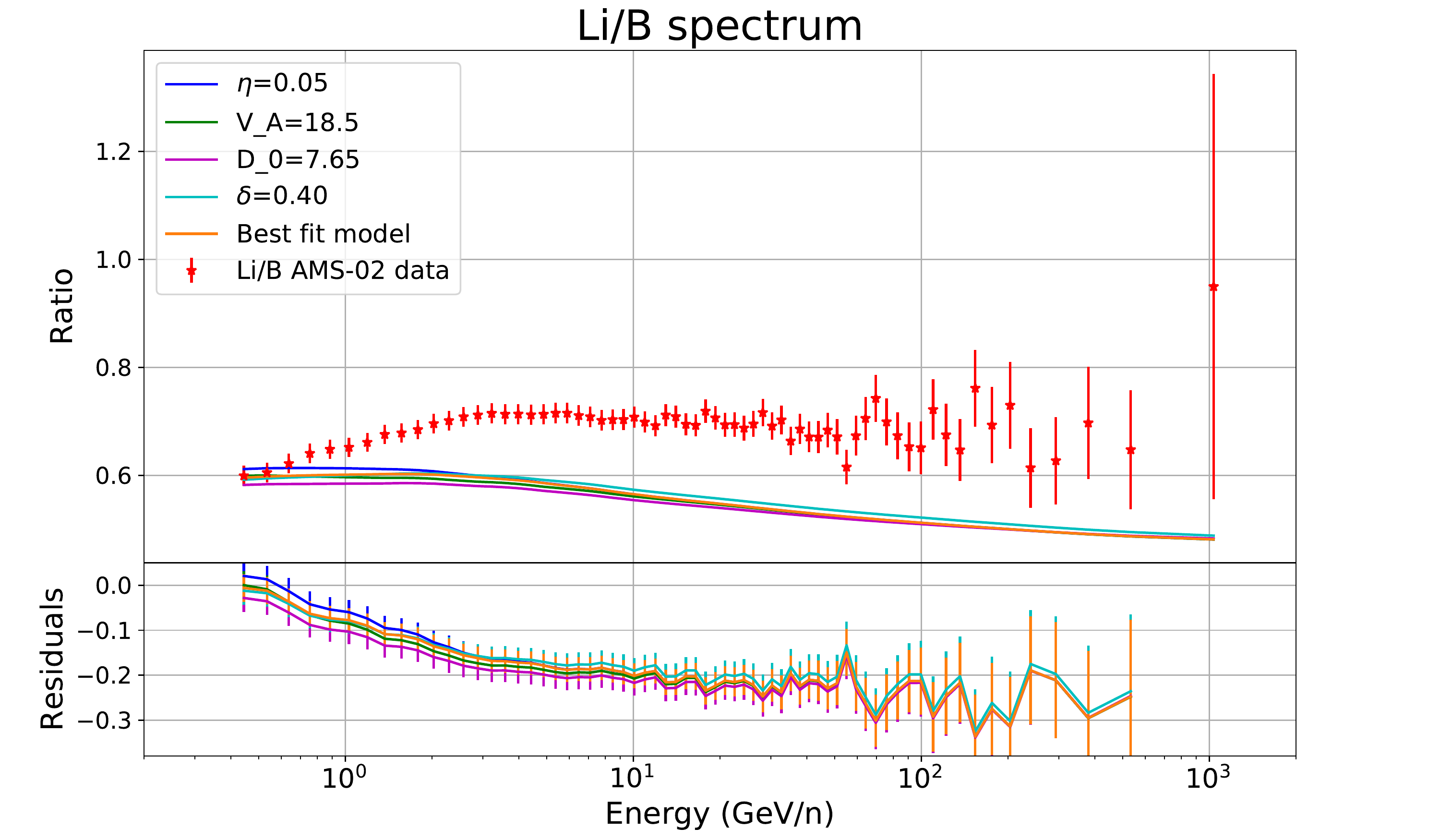}     
\end{center}
\caption{\footnotesize Simulated ratios with same primary source term (except for the simulation with change in the diffusion spectral index $\delta$, in which the source term was changed to hold same $\alpha + \delta$ value) and changing diffusion parameters. The best fit model has been shown in table~\ref{tab:diff_params} and the simulations change by $\Delta\eta \sim 0.5$, $\Delta V_A \sim 7 \units{km/s}$, $\Delta D_0 \sim 1 \times 10^{28} \units{cm^2/s}$, $\Delta \delta\sim0.04$, respectively, much higher than the usual uncertainties quoted in the best fits of these parameters.}
\label{fig:Rat_mod}
\end{figure*}

Figure~\ref{fig:Rat_mod} shows the effect of these changes on the Li/B ratio (since the importance of the halo size and gas profile is negligible).
%\vspace{0.25cm}

As we see, at $10 \units{GeV}$ the maximum difference between the predictions is around 4\%, while at $30 \units{GeV}$ it goes down to 3\% for variations of the diffusion parameters larger than the usual uncertainties related to their determination (as can be seen from the tables in appendix~\ref{sec:appendixC}). This largest variation between predictions corresponds to the $\sim15\%$ change in the $D_0$ parameter, which translates into $\sim 18\%$ variation in the fluxes of secondary species, while changes in the other parameters do not involve variations larger than 2\%.
Nonetheless, the differences at high energy are not due to the change on the diffusion parameters by itself, but are mostly due to the subtle differences in the primary fluxes used in each simulation and to the change in the amount of secondary C (roughly $1/5$ of its flux is secondary) and N ($\sim 73\%$ of it is secondary at $10 \units{GeV}$), which leads to sizeable changes in the secondary CRs. This may also mean that the $\sim 2-3\%$ uncertainties in the AMS-02 primary fluxes imply a maximum of $\sim 4\%$ uncertainty for adjusting the cross sections normalizations under this procedure. 

\section{Cross sections uncertainties assessment}
\label{sec:uncert}
Given the predictions from the diverse parametrisations, each cross section model may lead to a different interpretation of the CR data. As an example, from Fig.~\ref{fig:secfit_Webber} we see that the Webber cross section parametrisation yields a $\sim20\%$ excess in the Be and Li fluxes in comparison to the B flux. One could correct this discrepancy either by adding a primary component of boron (injecting boron from the source) and reducing the total grammage traversed by CRs (in order to fit the B/C ratio) or by rescaling the cross sections of B production (which would have the same effect on the grammage necessary to fit the B/C ratio), such that the three fluxes will reproduce AMS-02 data at the same time. On the other hand, the GALPROP parametrisation yields a $\sim 25\%$ deficit in the Li flux. As a consequence of this result, one could think of the necessity of adding an additional component of Li generated at the source (see \cite{Boschini:2019gow}). Nevertheless, given the current uncertainties in the cross sections, one could reproduce the predicted ratios from the GALPROP cross sections also for a certain rescaling on the production cross sections.

In conclusion, one has to take into account that, as explained above, more than 20-40\% of discrepancy can come from our limited knowledge in the cross sections, specially on the Li and Be channels. %In fact, this hypothesis leaves other problems, as why there is a significant amount of primary boron while no primary Li and Be.

%Something similar could happen when trying to explain the Li discrepancy with the GALPROP cross sections model (again, assuming very small uncertainties in its XS). This was first spotted by \cite{Boschini:2019gow}, suggesting the necessity of some amount of primary Li coming from novae. However, one has to be a little conservative, taking into account we don't even know the exact level of uncertainties our parametrisations have: for lithium spallation, in most of the channels experimental data is almost non-constraining. In fact, using other parametrisations do not lead to underestimation (rather, the opposite is the case with Webber XS) and the figures in appendix~\ref{sec:appendixA} show that GALPROP parametrisations are below the others for most lithium channels.

\begin{figure*}[!hptb]
\begin{center}
\includegraphics[width=0.42\textwidth,height=0.20\textheight,clip] {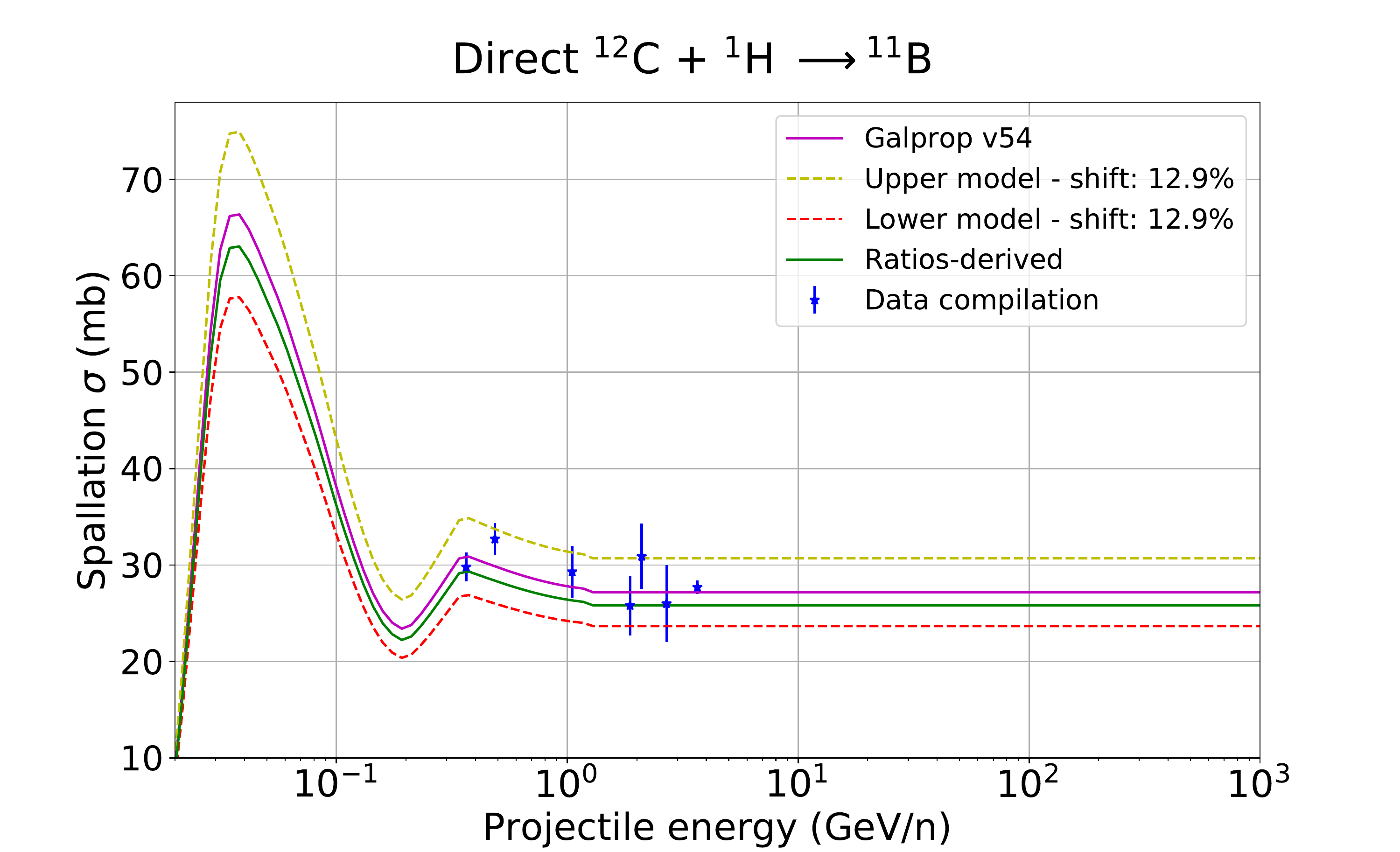} \hspace{0.5cm}
\includegraphics[width=0.42\textwidth,height=0.20\textheight,clip] {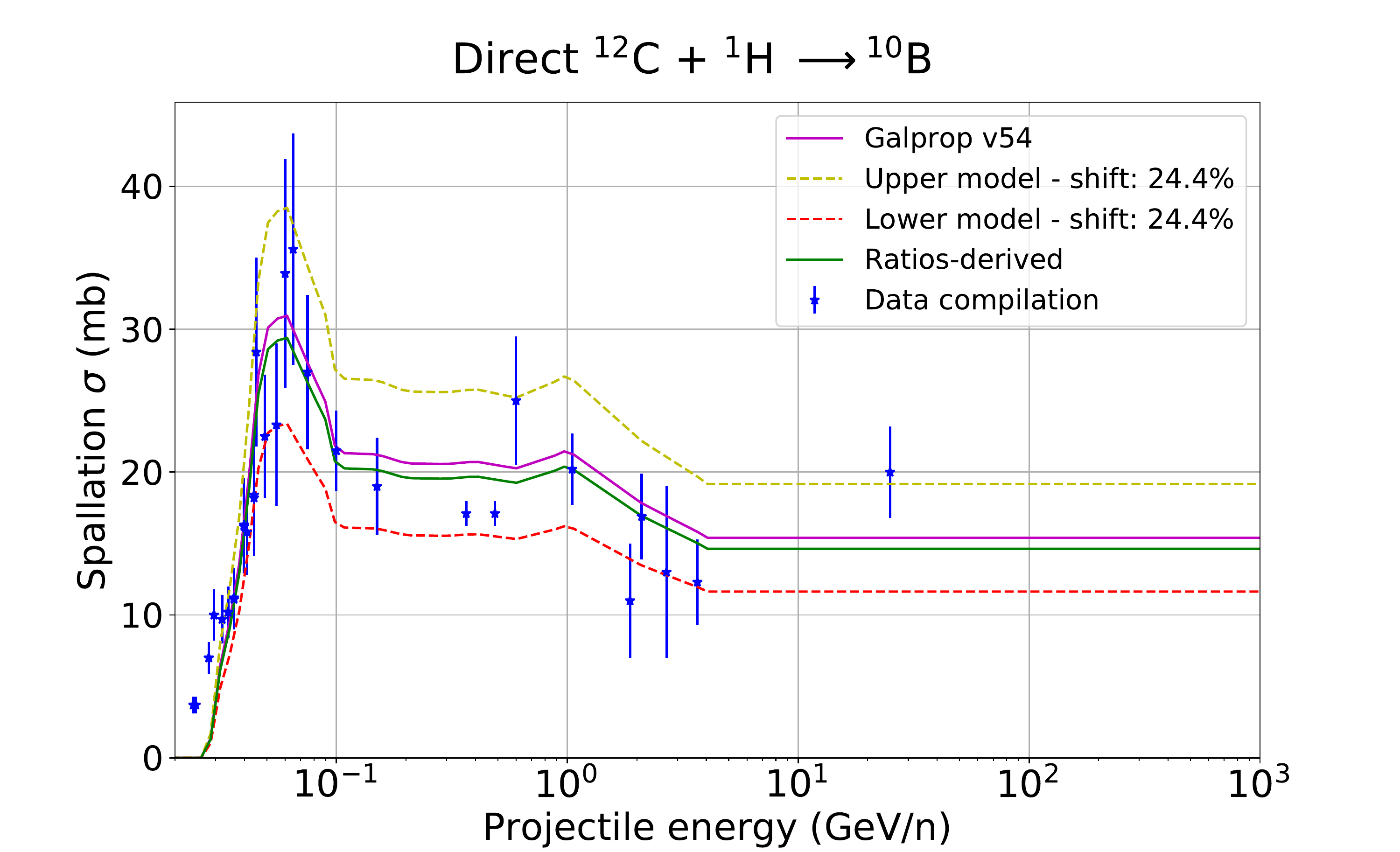} 

%\vspace{0.5cm}
\includegraphics[width=0.325\textwidth,height=0.18\textheight,clip] {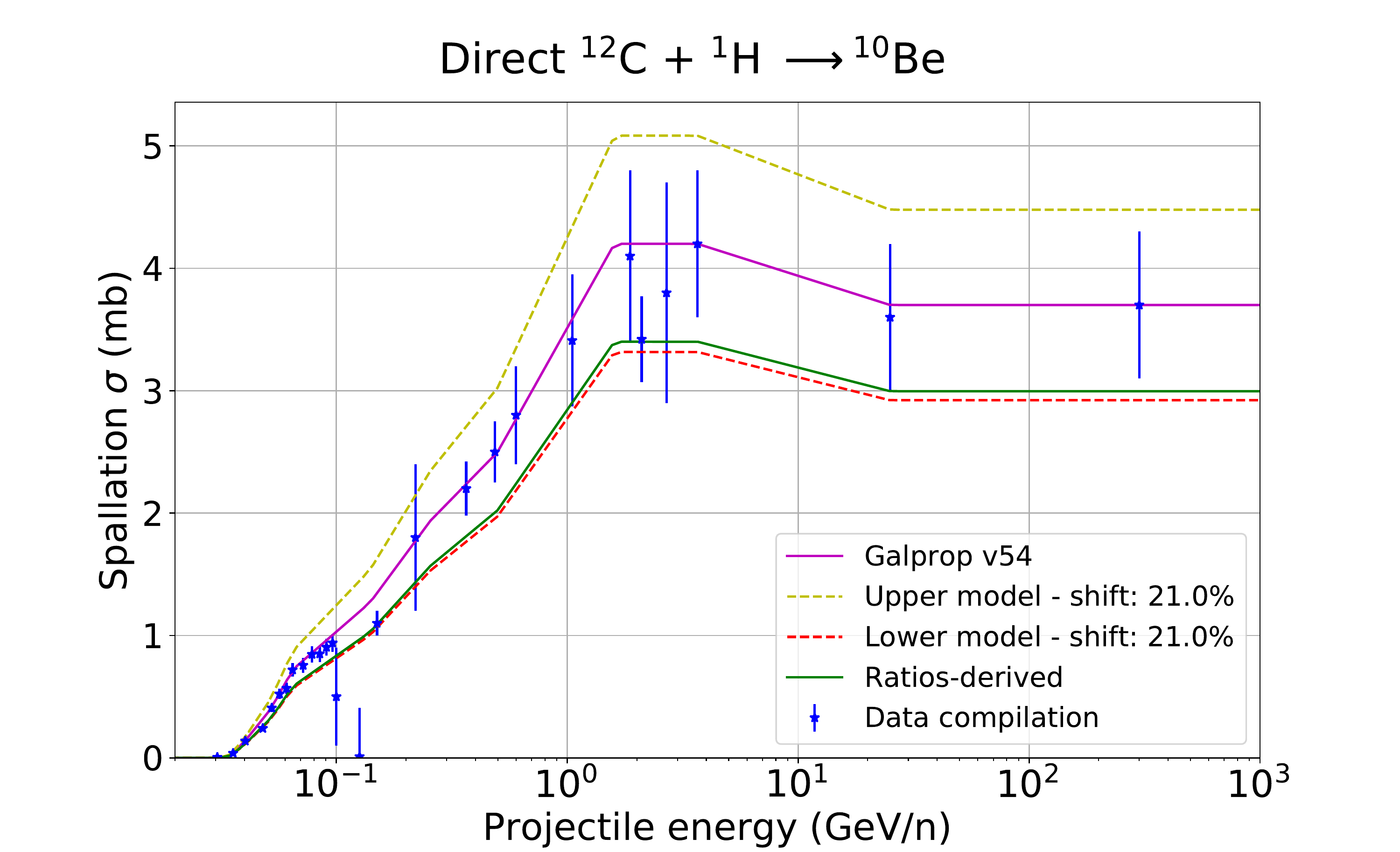} %\hspace{0.5cm}
\includegraphics[width=0.325\textwidth,height=0.18\textheight,clip] {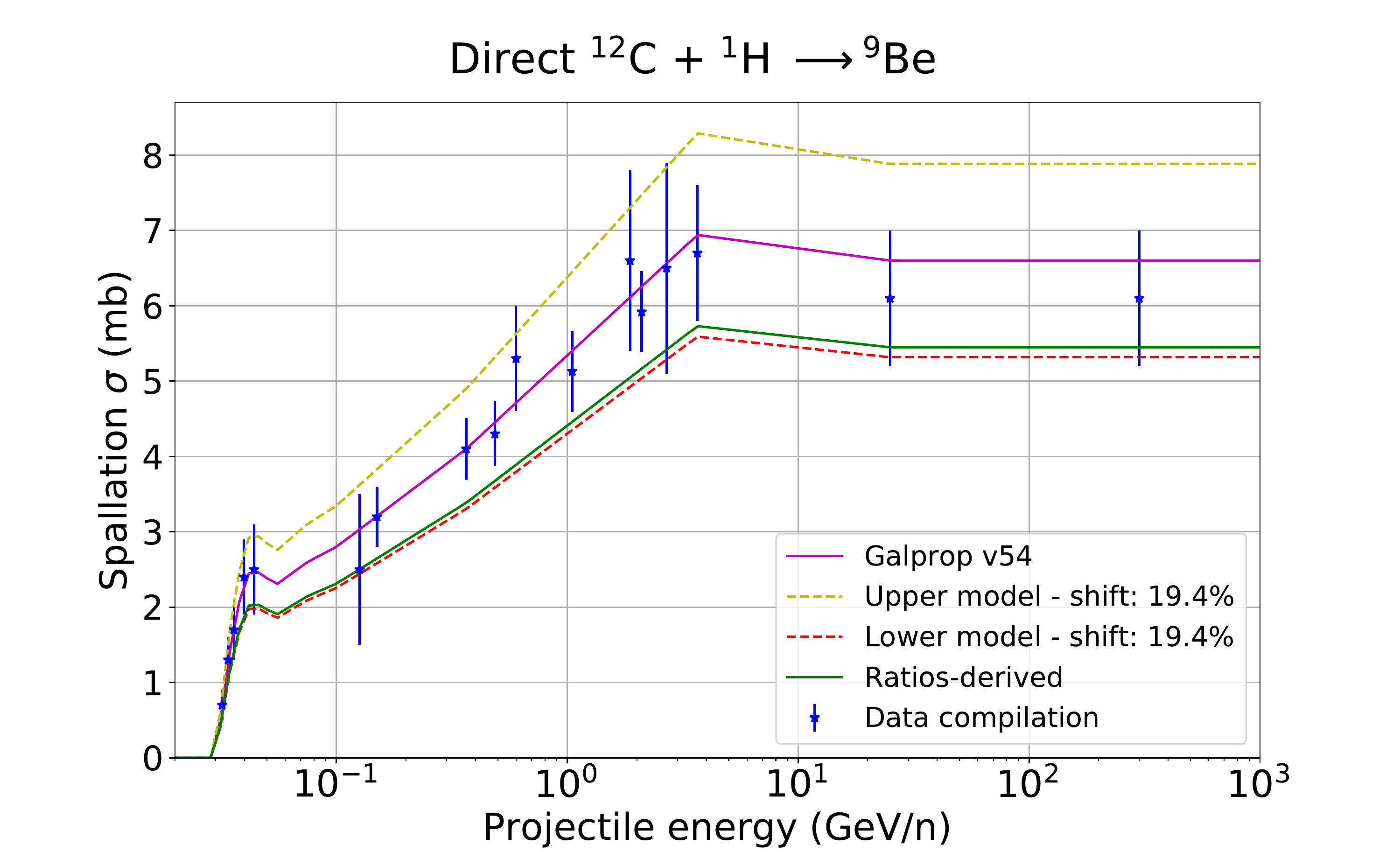}
\includegraphics[width=0.325\textwidth,height=0.18\textheight,clip] {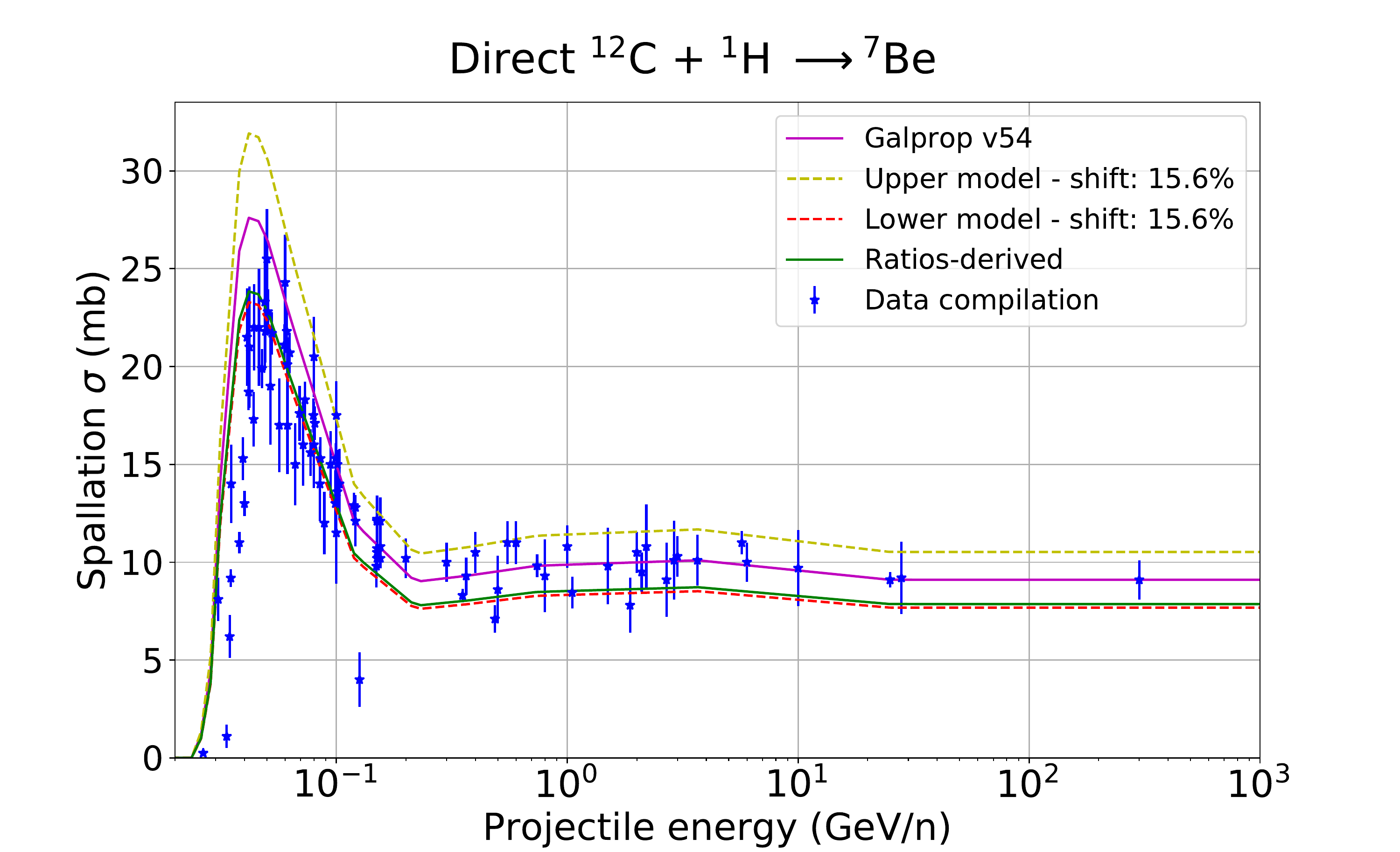}
%\vspace{0.5cm}

\includegraphics[width=0.42\textwidth,height=0.20\textheight,clip] {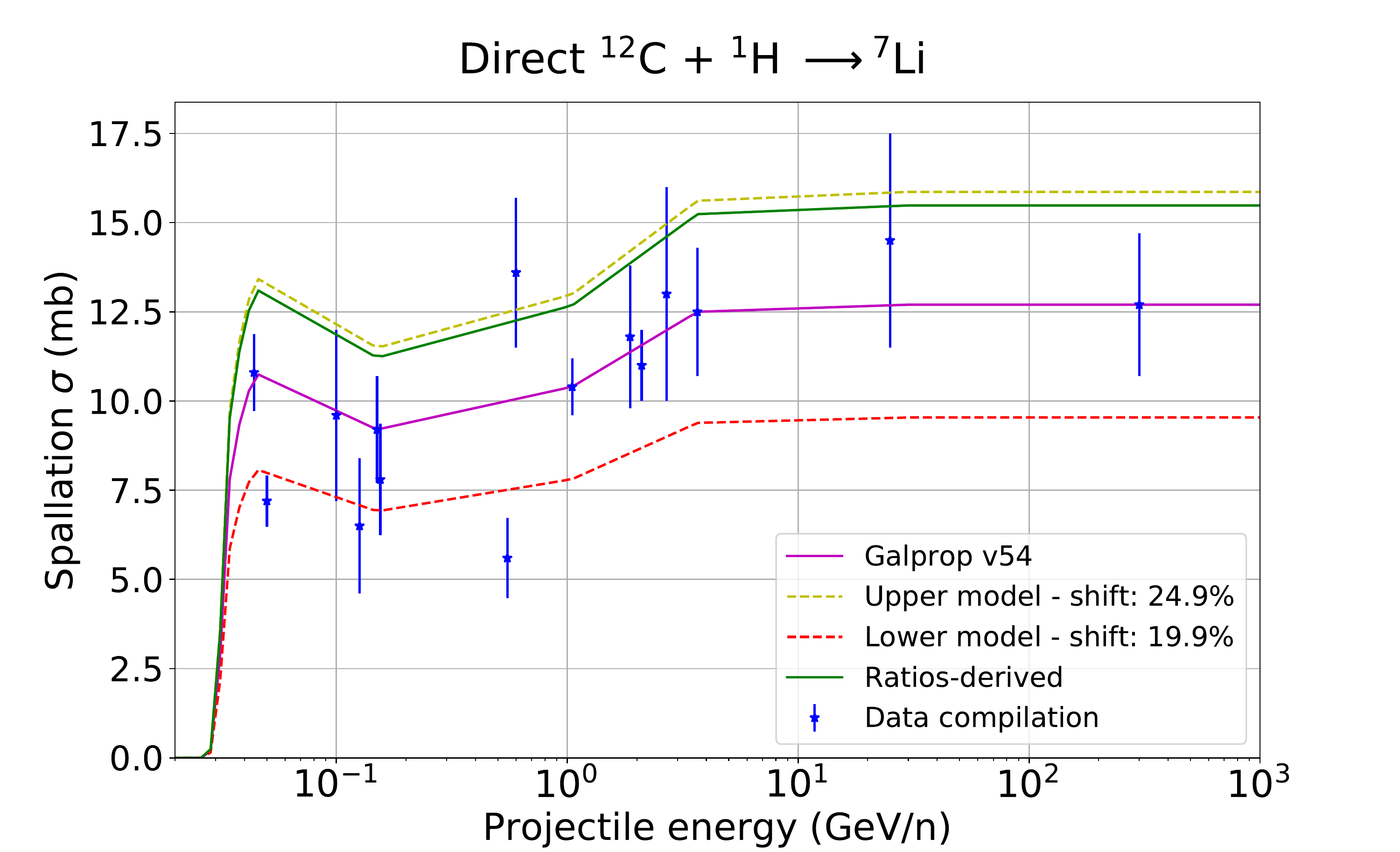} \hspace{0.5cm}
\includegraphics[width=0.42\textwidth,height=0.20\textheight,clip] {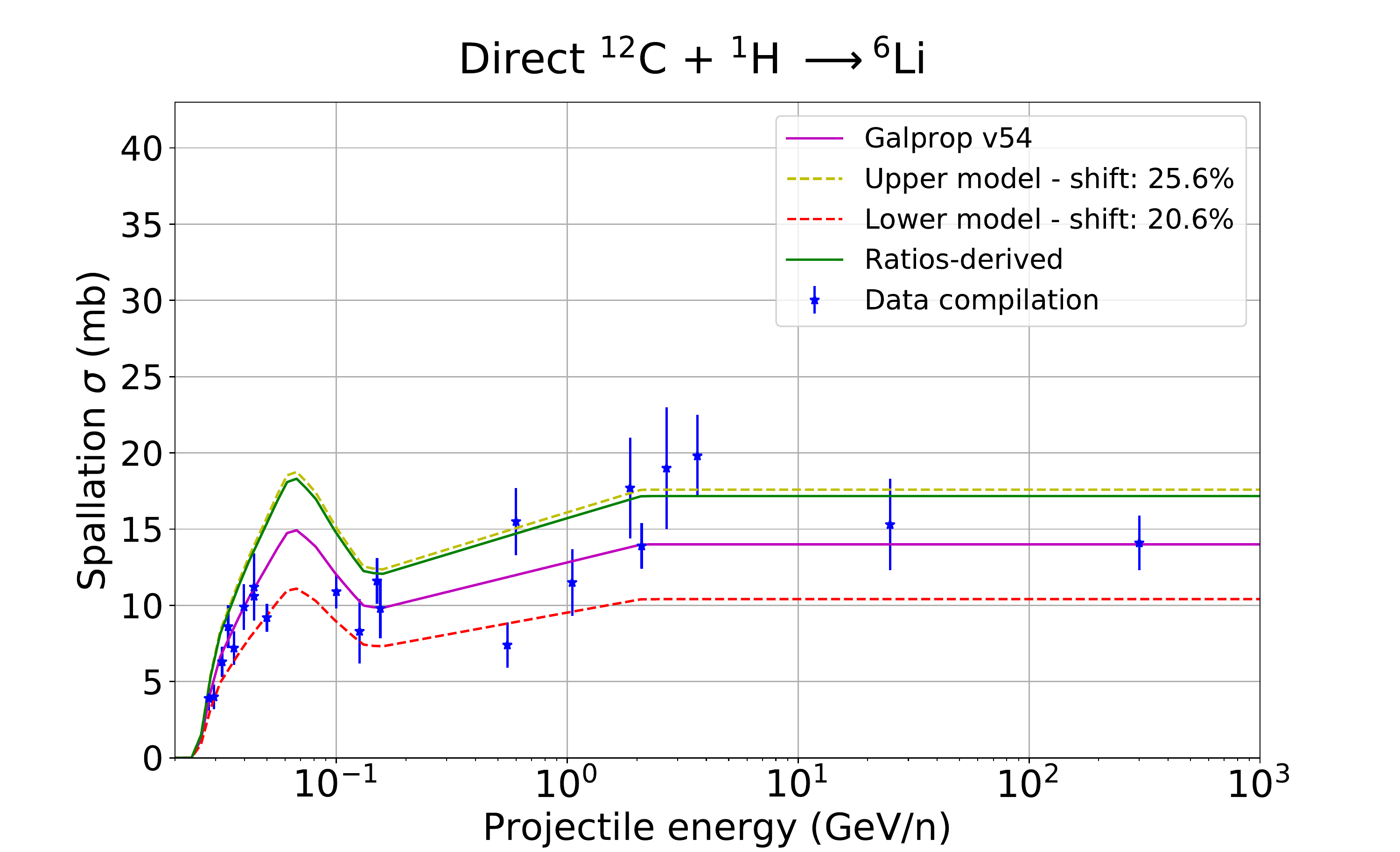}
\end{center}
\caption{\footnotesize Secondary nuclei from $^{12}$C: bracketing models of spallation cross sections. They give an idea on the limiting cases  in which our parametrisations can be renormalized remaining credible with respect to experimental data. The percentage of renormalization is indicated in the legends. The cross sections derived by matching the secondary-over-secondary ratios are also shown and compared with the GALPROP spallation cross sections. }% 
\label{fig:LUmodelsC}
\end{figure*}

\begin{figure*}[!hptb]
\begin{center}
\includegraphics[width=0.42\textwidth,height=0.20\textheight,clip] {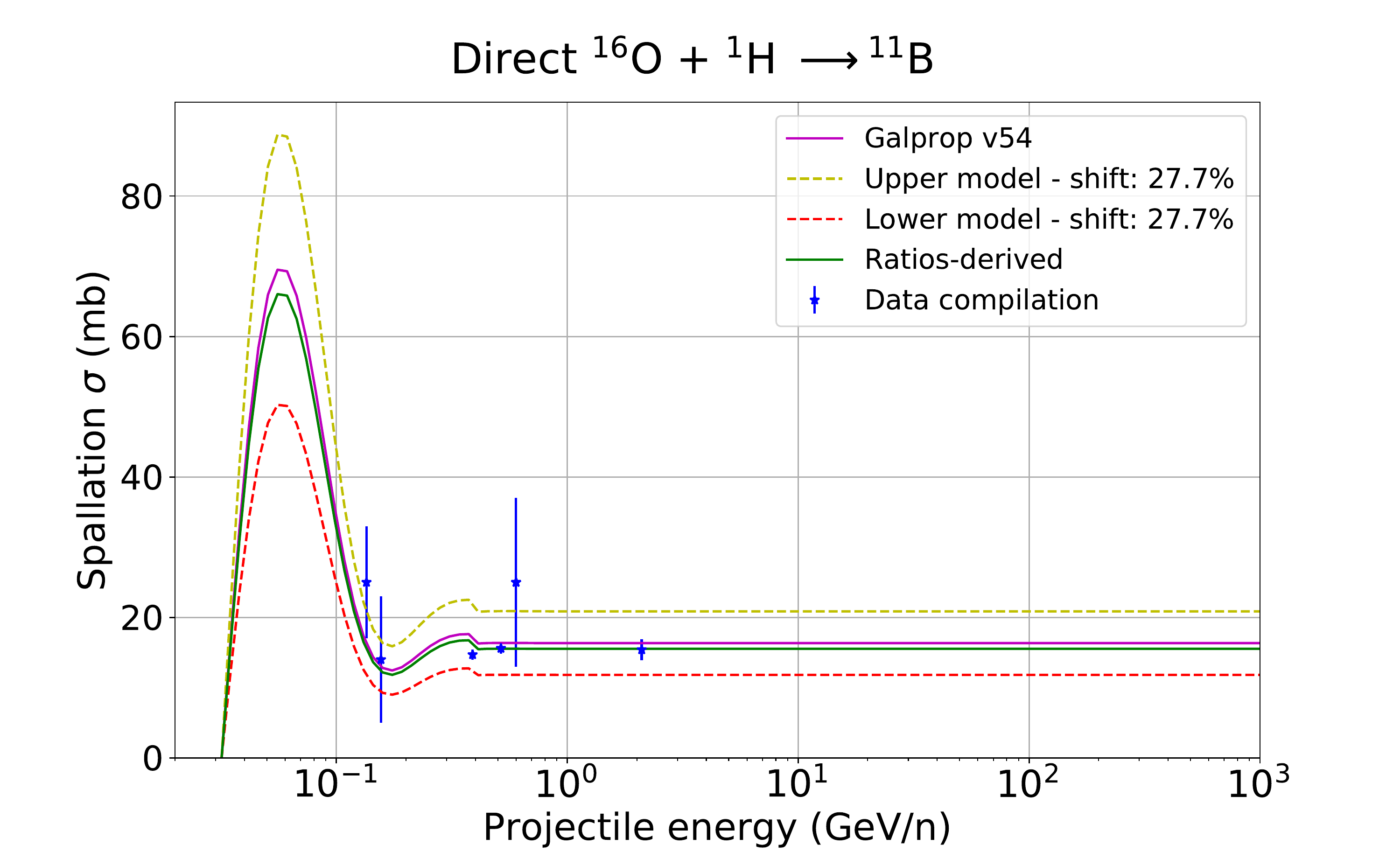} \hspace{0.5cm}
\includegraphics[width=0.42\textwidth,height=0.20\textheight,clip] {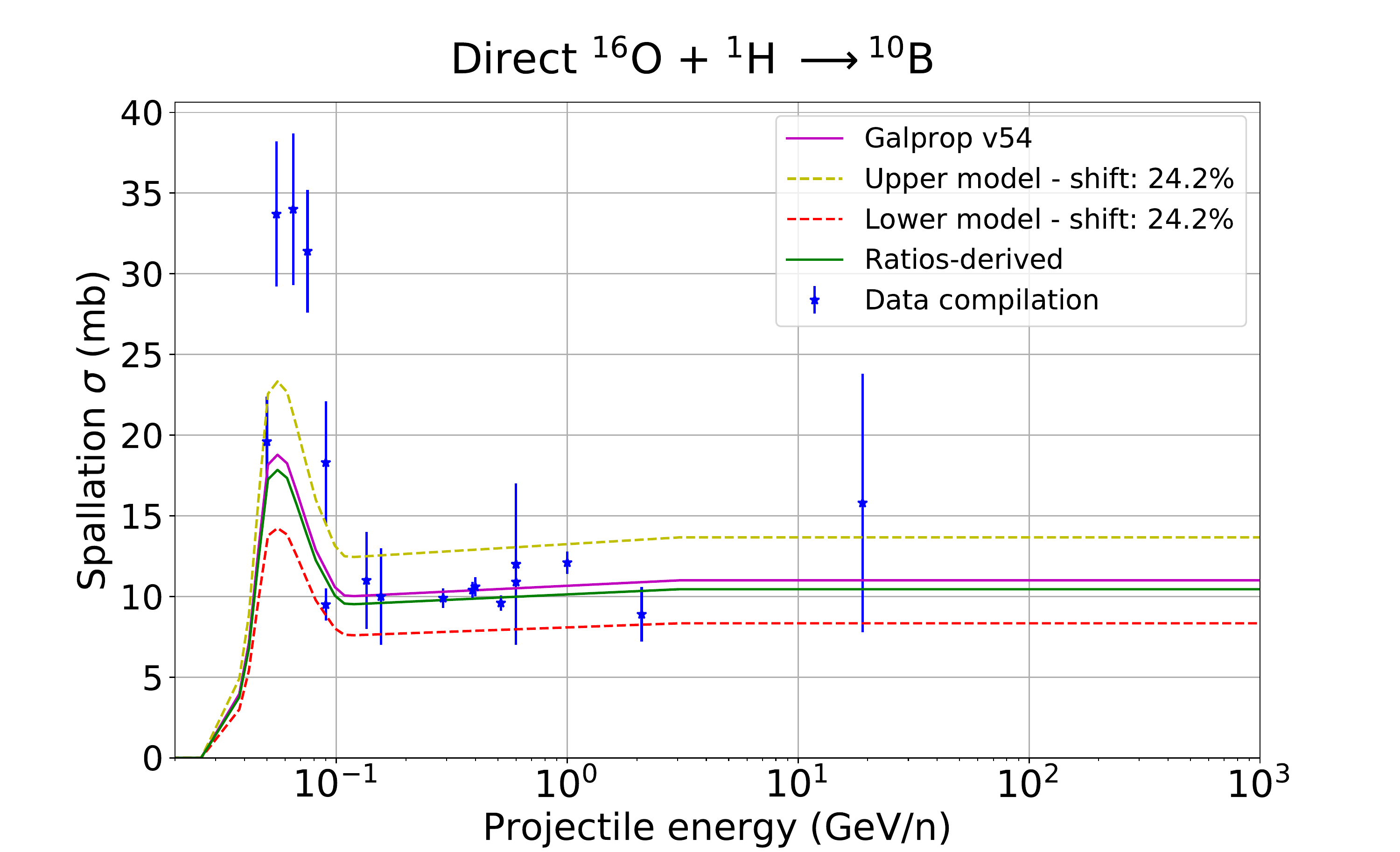}

%\vspace{0.5cm}
\includegraphics[width=0.325\textwidth,height=0.18\textheight,clip] {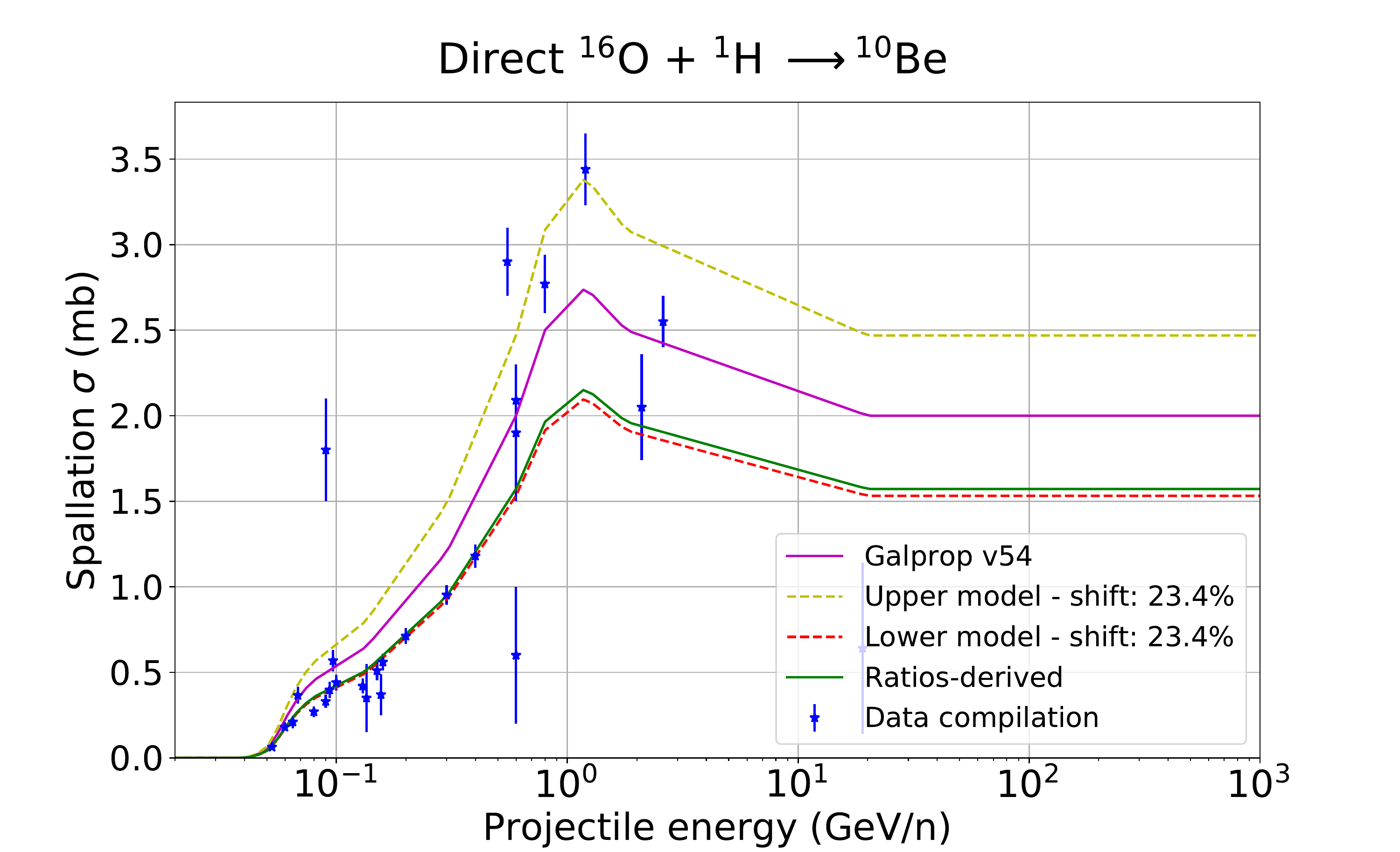} 
\includegraphics[width=0.325\textwidth,height=0.18\textheight,clip] {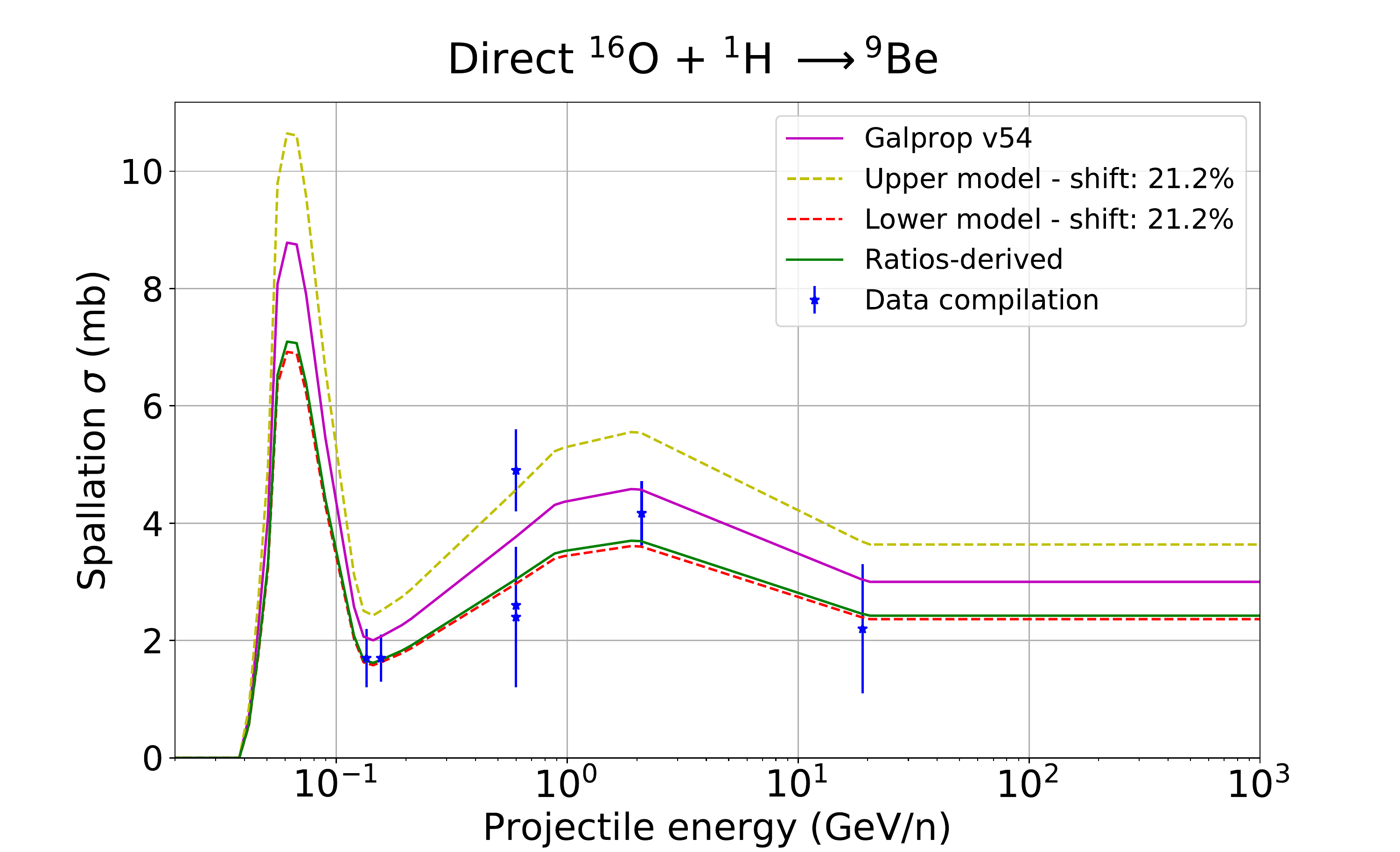}
\includegraphics[width=0.325\textwidth,height=0.18\textheight,clip] {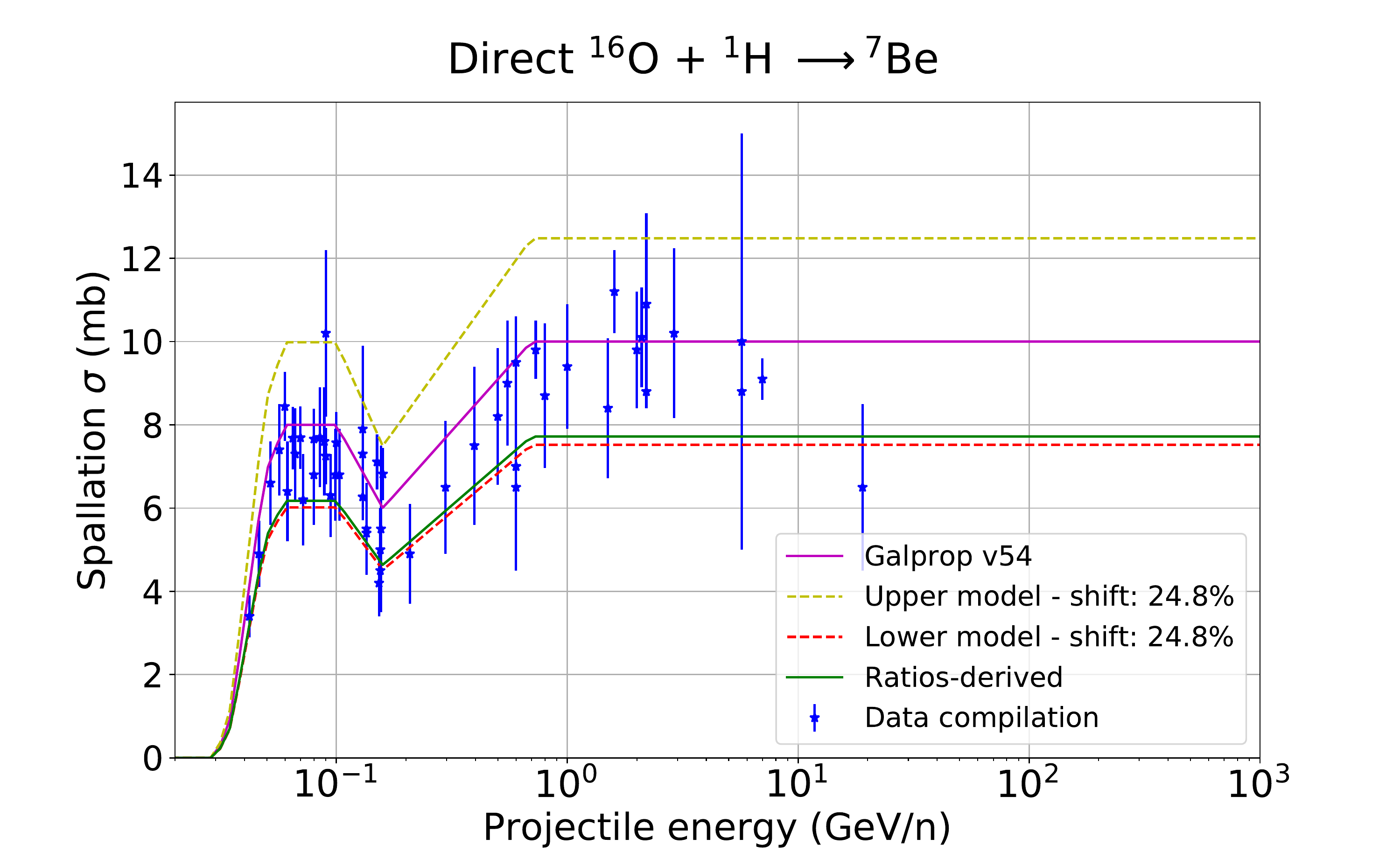}
%\vspace{0.5cm}

\includegraphics[width=0.42\textwidth,height=0.20\textheight,clip] {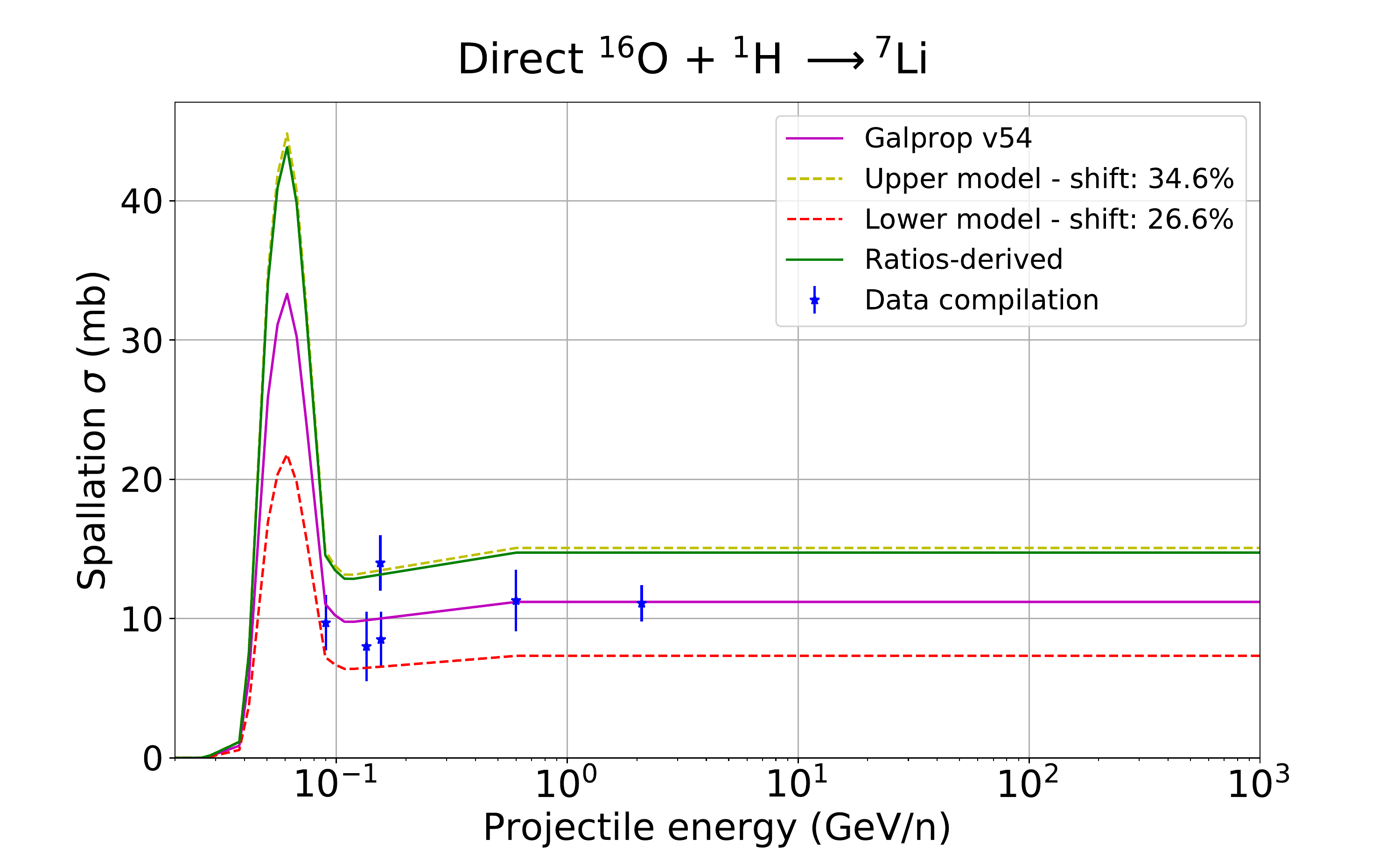} \hspace{0.5cm}
\includegraphics[width=0.42\textwidth,height=0.20\textheight,clip] {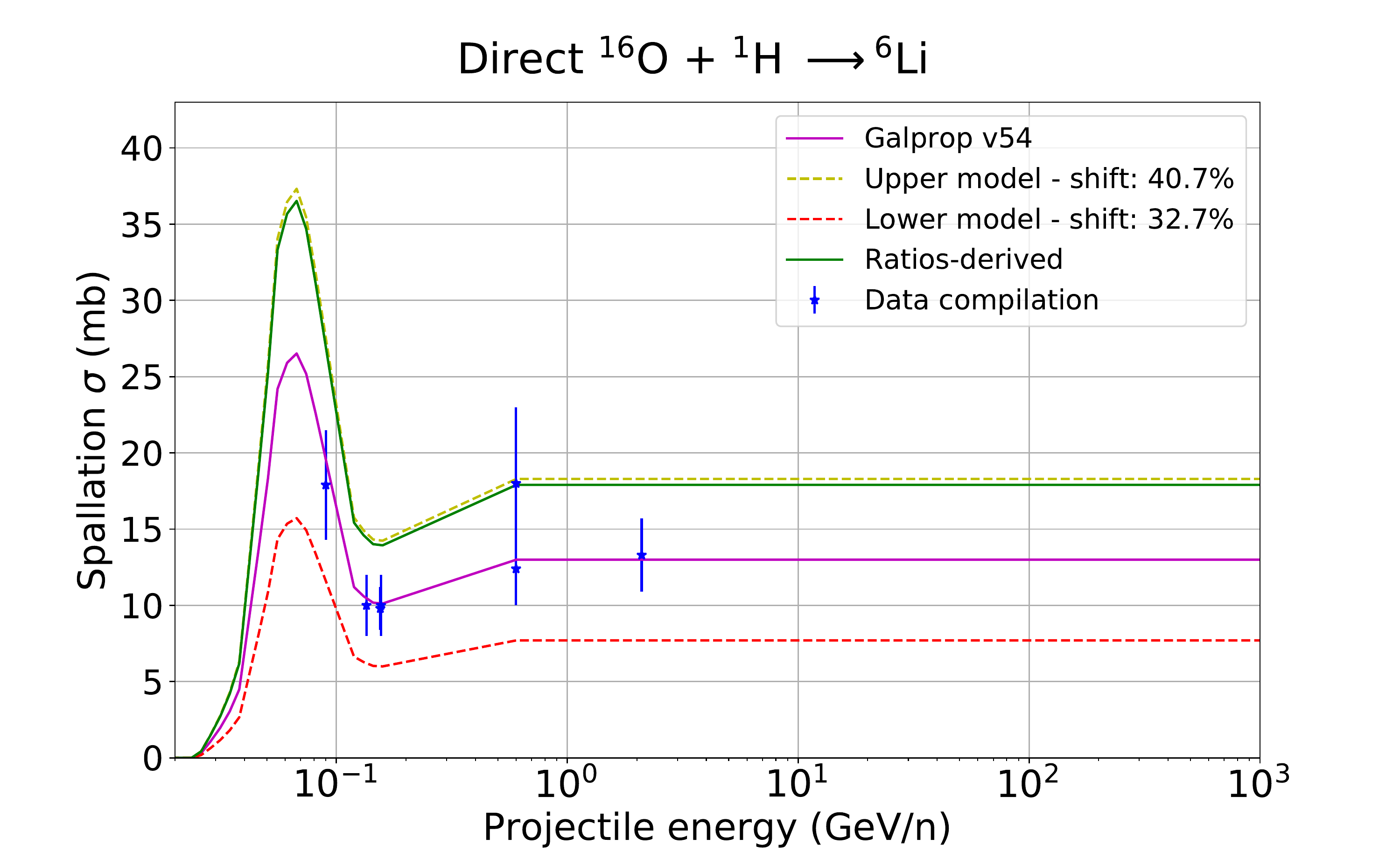}
%\vspace{0.2cm}
\end{center}
\caption{\footnotesize Same as in Figure~\ref{fig:LUmodelsC} but for secondary nuclei produced from $^{16}$O collisions with the ISM.}
\label{fig:LUmodelsO}
\end{figure*}

For the GALPROP cross sections a deeper research has been done: first, it can be considered that the uncertainties of these parametrisations are directly related with the data uncertainties. Then, we have defined two limiting models from the GALPROP parametrisation, in which the spallation cross sections for each interaction channel with $^{12}$C and $^{16}$O as projectile on H target have been shifted up or down using a scaling factor corresponding to the average uncertainties on the cross section experimental data at $\pm 1 \sigma$ level. This is motivated by the fact that the energy dependence of the cross sections is supposed to be well known, while their normalization is not so precisely determined \cite{Evoli:2019wwu}. These bracketing models are tested to be compatible with the cross sections data measurements and to reasonably contain all of them. The result of this procedure can be seen in the figures~\ref{fig:LUmodelsC} and~\ref{fig:LUmodelsO}. 

A couple of channels were differently rescaled to better contain the cross section data. In the case of the Li production channels from $^{16}$O and $^{12}$C, extra shift of 12\% and 7\%, respectively, were applied to the upper model, since a few data points exhibit larger excursions with respect to the nominal model. In the channels of $^{16}$O going to $^{9}$Be and $^{10}$Be the shift was taken to be half of the experimental uncertainties.

The bracketing models are shown in figures~\ref{fig:LUmodelsC} and~\ref{fig:LUmodelsO}, where the shifts, typically corresponding to a $\pm 20\%$ variation from the cross sections normalization, are shown in the legends for each channel. We see that the Be channels are those with smaller uncertainties and more data points, while the Li channels are those with the higher uncertainties and less data points. We also see that the channels coming from the spallation of $^{16}$O exhibit significantly larger experimental errors and less data points than those from the spallation of $^{12}$C. The GALPROP cross sections are added to see relative changes. The fit on the secondary-over-secondary ratios is shown in Figure~\ref{fig:secsec_Uncert}. In addition, a set of cross sections, scaled from the original GALPROP ones, that simultaneously fit the secondary-over-secondary flux ratios was derived and it is shown to be compared with experimental cross sections data. This corresponds to a combined correction (rescaling) of the B, Be and Li production cross sections that allow us to simultaneously reproduce their high energy spectra within the experimental uncertainties, as explained below.

%The upper and lower models shown in figures~\ref{fig:LUmodelsO} and~\ref{fig:LUmodelsC} show that the Be channels are the ones with less uncertainty and more data points, while Li channels are the least complete ones. In addition, generally speaking, the channels coming from the spallation of $^{16}O$ show larger experimental errors for every isotope and less number of experimental data points than the $^{12}C$ channels, which indicates that more data and more precise are necessary in the O channels.

The next step here consists of the evaluation the spectra of secondary CRs for these bracketing models. Changing just these main channels yields variations (according to ref.~\cite{Genoliniranking}) of $\sim 70.7\%$ of the total B flux, of $\sim 53.5\%$ of the total Be flux and of $\sim 52.6\%$ of the total Li flux (at $10 \units{GeV}$). This means that, handling only these main channels, we are able to modify the fluxes of secondary CRs by, at most, those percentages, shifting them up and down.

We point out here that we are not changing the cross sections from the spallation reactions of other nuclei than $^{12}$C and $^{16}$O (the main channels). Most of the channels with minor importance have no or very few data, which constitutes the largest source of uncertainties on the total production cross sections of the isotopes we are studying. Although the individual contribution to the secondary CR fluxes from each of these channels is small, the sum of all their contributions is relevant ($\sim 25\%$ for B and $\sim 50\%$ for Li and Be) and the use of the very recent AMS-02 data for Mg, Si and Ne is also very important. 

%\vspace{0.7cm}

\newpage
\subsection{Uncertainty effects and cross sections evaluation from the secondary-over-secondary ratios}
\label{sec:model_secratios}

In this section we derive the secondary-over-secondary flux ratios using the two bracketing models for the cross sections, in order to show the uncertainty band arising from the cross sections uncertainties and to verify that the AMS-02 data of the ratios lie inside this band. %This means that all the three light secondary CRs can be successfully explained at the same time inside the given uncertainties. 

\begin{figure*}[!bh]
\begin{center}
\includegraphics[width=0.333\textwidth,height=0.17\textheight,clip] {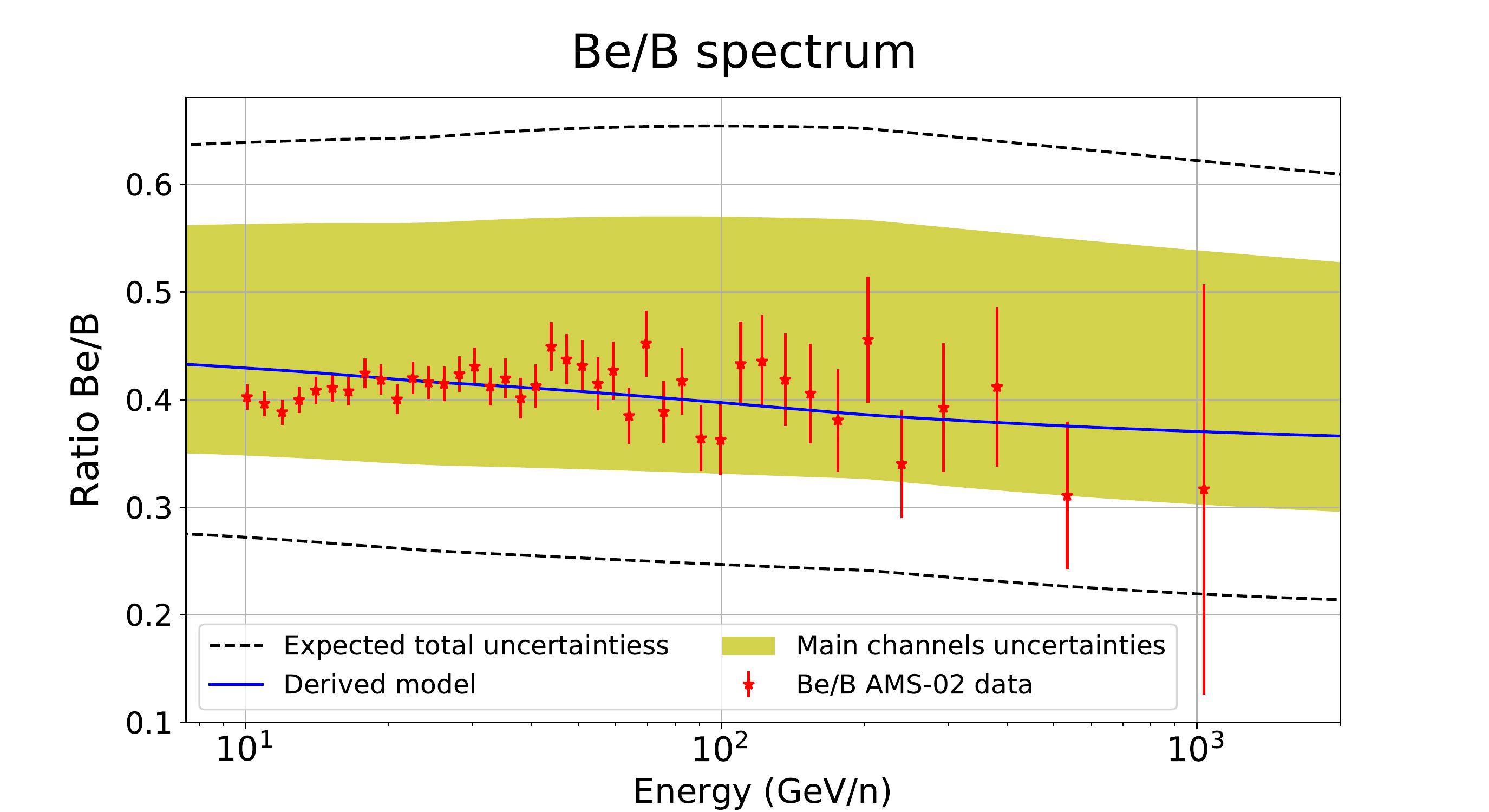} \hspace{-0.2cm}
\includegraphics[width=0.333\textwidth,height=0.17\textheight,clip] {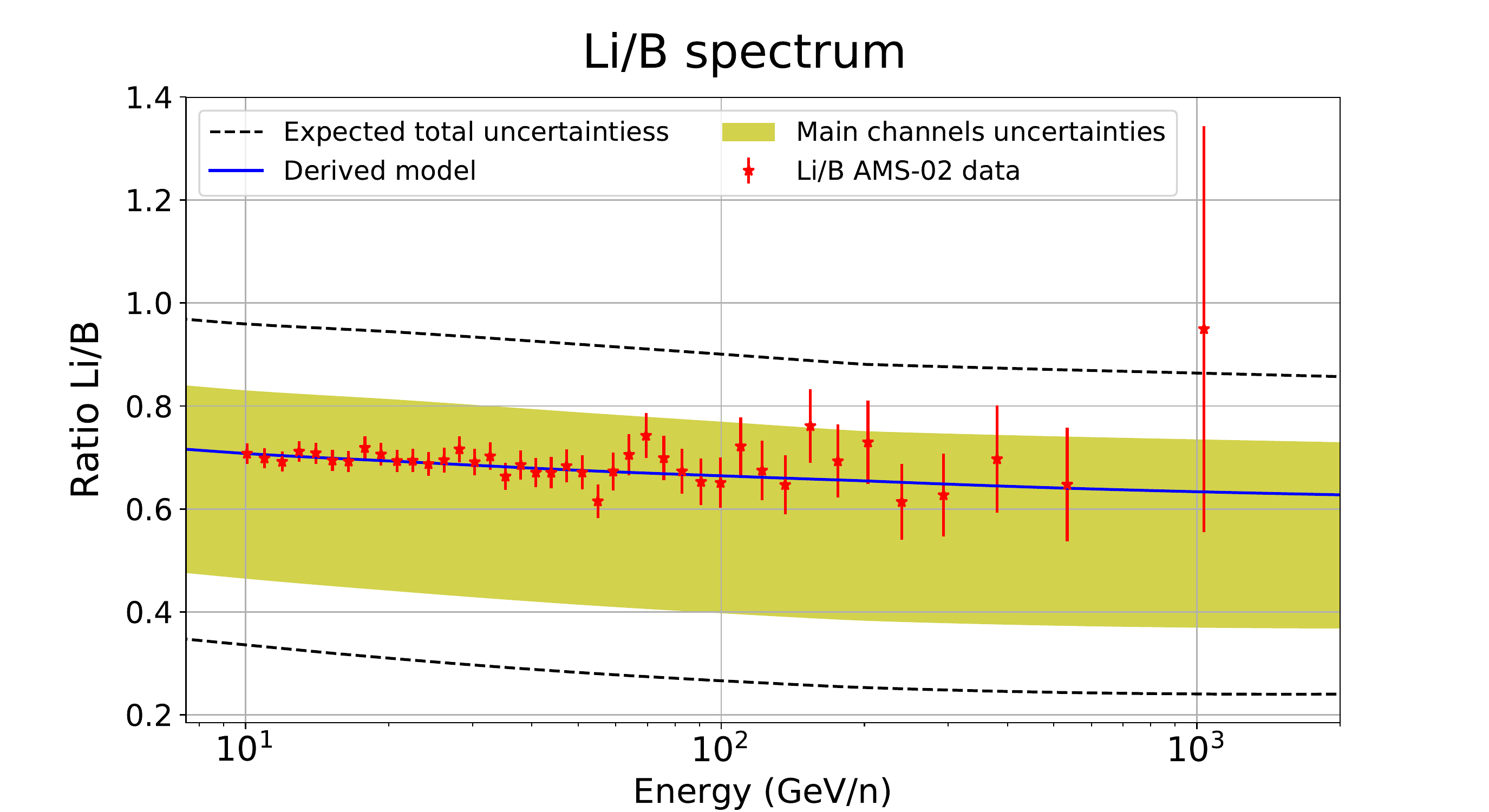} 
\hspace{-0.2cm}
\includegraphics[width=0.333\textwidth,height=0.17\textheight,clip] {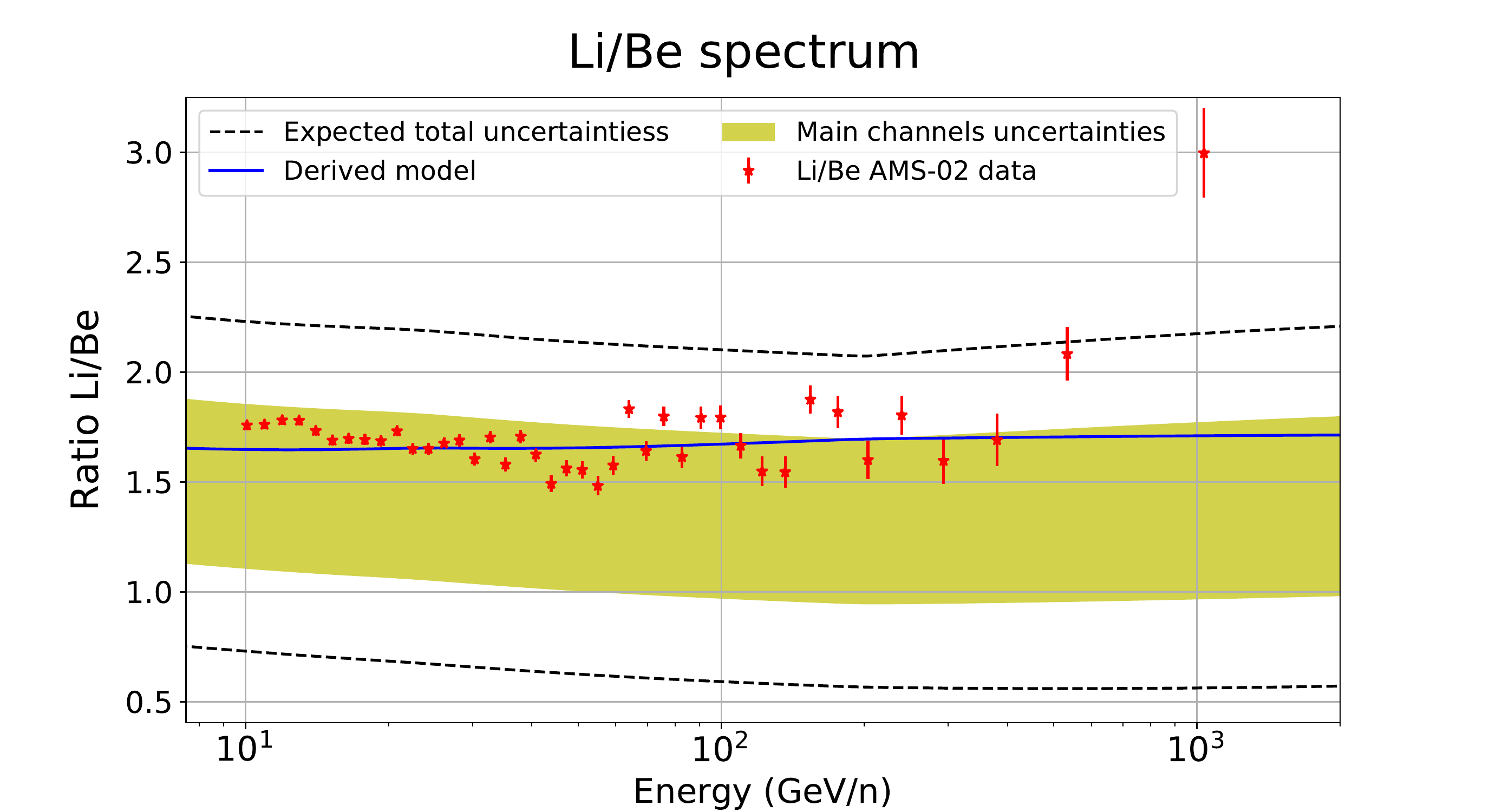}
\includegraphics[width=0.339\textwidth,height=0.185\textheight,clip] {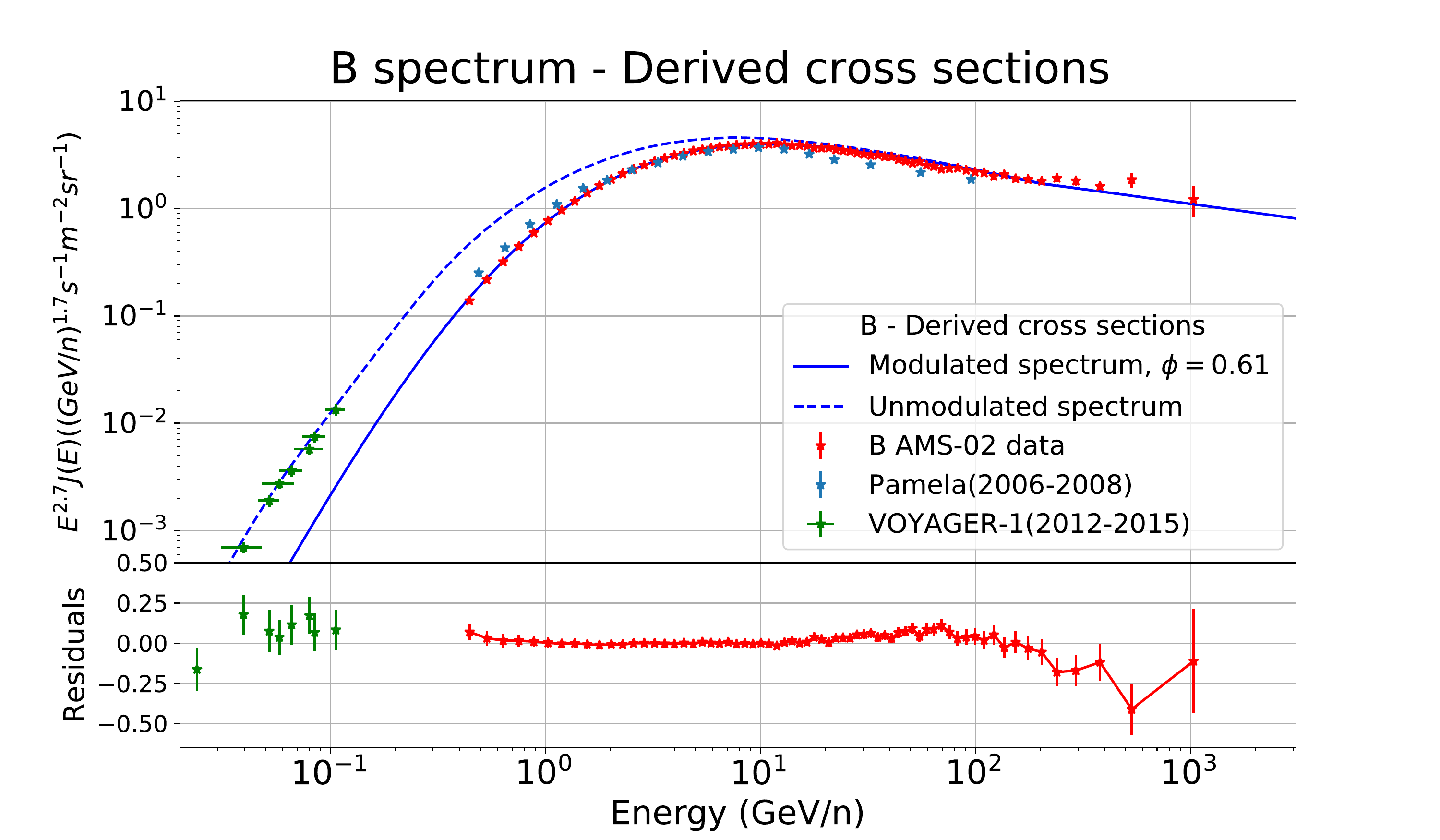} \hspace{-0.35cm}
\includegraphics[width=0.339\textwidth,height=0.185\textheight,clip] {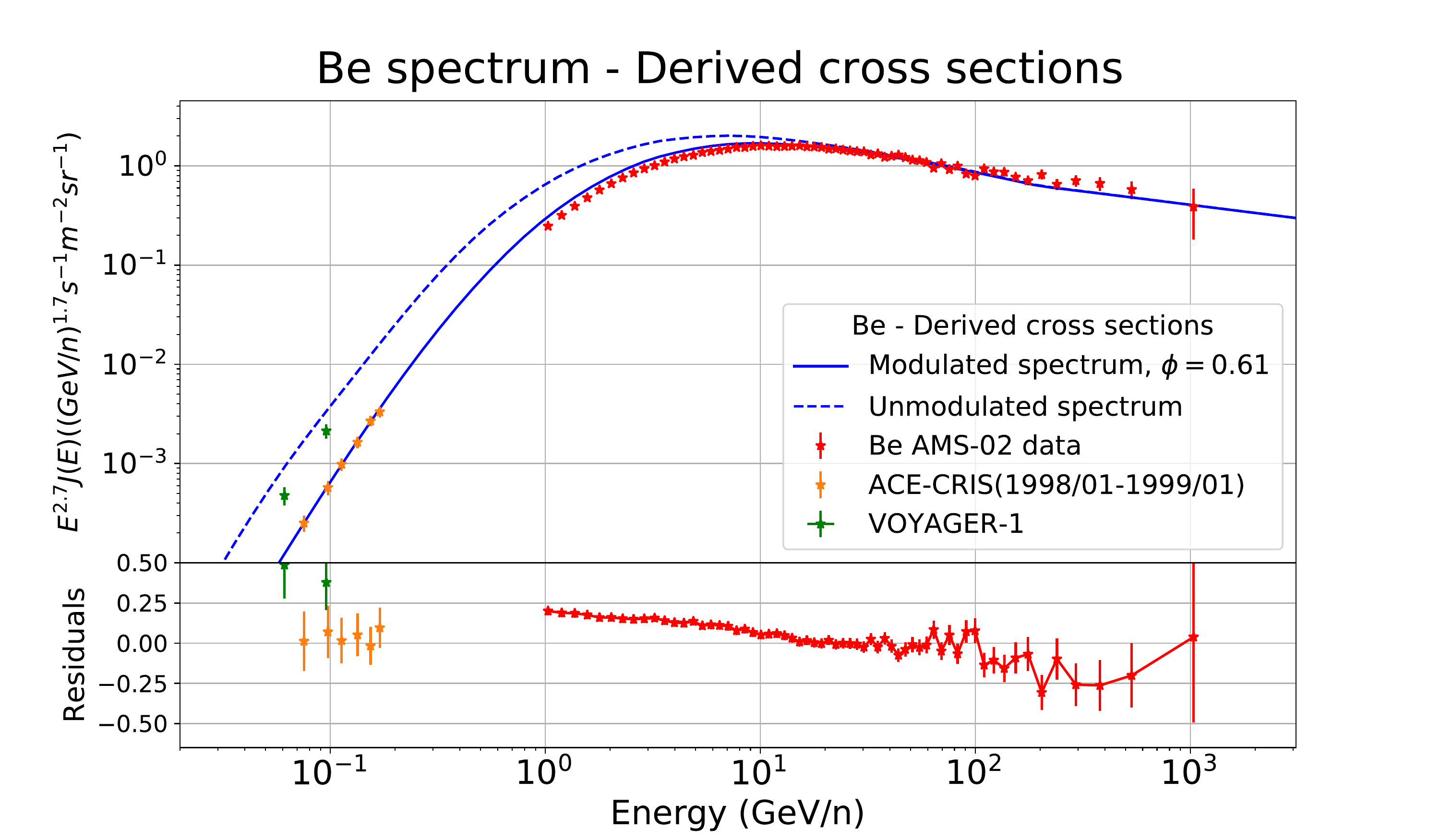}
\hspace{-0.35cm}
\includegraphics[width=0.339\textwidth,height=0.185\textheight,clip] {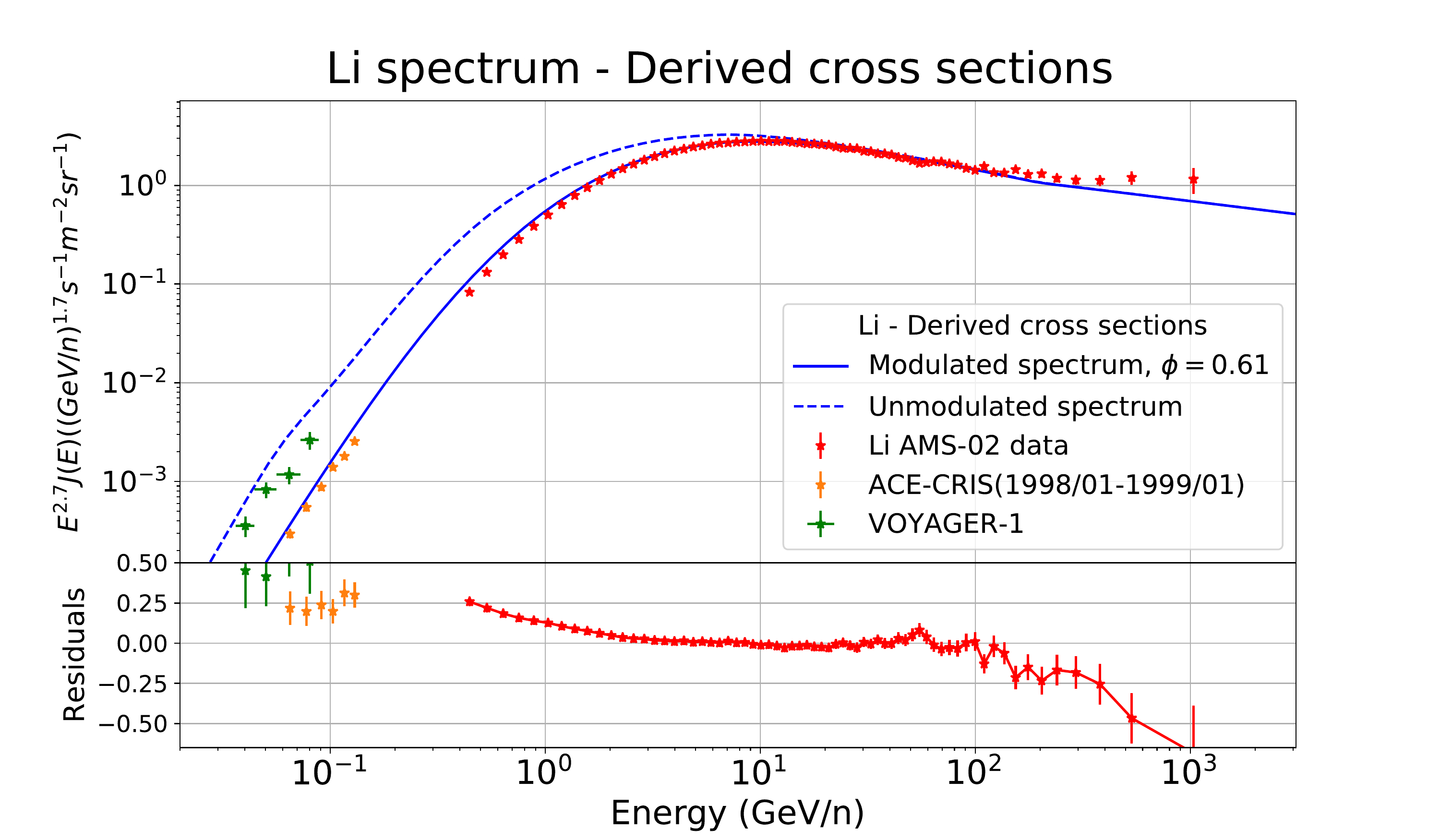}
\end{center}
\caption{\footnotesize Secondary-over-secondary spectra for Li, Be and B showing the effect of using the bracketing cross sections models. The yellow band gives an idea of the effect of cross sections variations on the secondary flux ratios. It shows that cross uncertainties can explain the fluxes of all secondary species at the same time, without need of invoking primary sources for these CRs. The lower plots are the results of the derived model of cross sections for the B, Be and Li spectra, with diffusion coefficients fitted to the B/C AMS-02 data.}
\label{fig:secsec_Uncert}
\end{figure*}

This is shown in Figure~\ref{fig:secsec_Uncert}, where the yellow bands represent the values of the ratios within the two limiting models. The upper bound of each band corresponds to the situations in which the numerator is taken from the upper bracketing model and the denominator from the lower bracketing model (thus maximizing the ratio) and vice versa for the lower bound of the band. Showing the band on the secondary-over-secondary flux ratios has the advantage that it hardly depends on the diffusion coefficient used, while projecting these bands on the flux of a secondary CR would have a strong dependence on the diffusion coefficient chosen.

%In fact, the variation we see here do not cover the full band, since only the main channels were varied. For the Li/Be ratio this band can be even 4 times larger ($\sim 1/(0.535\times0.526)$) and in the ratios involving B of the order of 2.8 times larger ($\sim 1/(0.53\times0.707)$).  This is wrong!!!
The full band can be as large as twice for the Li/Be ratio (since around half of the flux of these species comes from other channels) and around 75\% larger for the other ratios. This is represented by the black dashed lines. In any case, just varying these channels suffices to confirm that there is no clear necessity of adding a primary component of B, Be or Li (although it still can exist and, probably, it does at very low proportions).

Furthermore, as a comment, we point out that, while this evaluation of the errors on the ratios allows us to determine whether there is clear need of a primary source of any of these CR species (in case cross sections uncertainties are not enough to explain the data), they would remain unchanged if there is an extra (primary) component at similar proportions for all of them (what can be actually expected). %coming from the so-called cocoons (see~\cite{Evoli:2019wwu}), since this extra component will be formed in agreement to the cross sections ratios (and so, to the sec/sec ratios).

On the other hand, it should be stressed that the use of these secondary-over-secondary ratios can help improving the spallation cross sections and, to demonstrate it, a set of cross sections were derived from a fit to the high energy part of these ratios. The simultaneous adjustment of the studied ratios does not have an exact solution (there can be degenerate solutions), so that finding the correct relation between the cross sections and the ratios is not an easy task.

In the analysis of the GALPROP model, it is obvious that an increase of Li cross sections is needed, but just a change of this cross sections does not account for the Be/B and Li/B discrepancies. This means that making variations just in the Be and Li cross sections (or whichever pair of secondaries) independently will never reproduce the three ratios inside the bracketing models at the same time, implying that a simultaneous adjustment of all the three fluxes is needed. These combined adjustments, thus, would modify our predictions on the secondary-over-primary ratios for all these secondaries, improving the current determination of the diffusion coefficient parameters too.

The strategy followed here consists of a progressive shift of the cross sections, for each of the channels, from the original parametrisation until we find a configuration that matches the three ratios at the same time. %Renormalizations with smaller "overall distance" (understanding "overall distance" as $\sqrt{\Delta Li_{ren}^2 + \Delta Be_{ren}^2 + \Delta B_{ren}^2 }$ and $\Delta X$ as the amount of renormalization) are reasonably preferred since this balances the ratios evenly for the three species. 
Indeed, in this case we see that the fit converges when Li and Be cross sections are very close to touch the depicted limits (upper model for Li and lower for Be), which means that there is very narrow margin to find other degenerate solutions. This means that there are other possible solutions allowed by the three ratios, but they lie very near or outside the limiting cross sections models (and are therefore forbidden). This makes the solution found more robust and subject to less uncertainties due to other degenerate solutions. Furthermore, the uncertainties on this derived model are mainly due to the AMS-02 data uncertainties in the CR primary and secondary fluxes and, as stated in subsection~\ref{sec:theory_sec}, must be, at most, roughly of 3-4\%. At the end, the renormalization to the main channels of the GALPROP cross sections was a reduction of 5\% for the boron flux, %(constant for all the channels) 
a reduction of $18\%$ for the beryllium flux ($\sim16\%$ in the $^{12}$C channels and $\sim 20$ in the $^{16}$O channels in average) and an increase around $28\%$ for the lithium flux (22\% for $^{12}$C channels and $\sim34\%$ $^{16}$O channels).

As expected, the scaling needed in the cross sections of the B channels with respect to the original GALPROP parametrisation is very small, while the Li and Be main channels show larger shifts, in agreement with the uncertainties one would expect. These differences are also related to the fact that the impact of a change in these main channels for B is more than twice the impact of the minor channels (those with an impact in the B flux below 2\%). In turn, the impact of the main channels for Li and Be fluxes is approximately the same as the impact of the rest of the minor channels, which means larger level of unknowns.

However, as already argued, we are not searching for the "real" cross sections (to do this, we would also need to have data on the isotopic fluxes), since we just tuned the main channels in order to compensate the deficit (or surplus) there is in the cross sections of those channels with little effects in the secondary species (mainly N, Ne, Mg, Si and Fe). This combination of flux ratios among secondary CRs may be considered as a strategy to take care of the deficiencies on the extrapolations typically used for the rest of the minor channels by slightly varying the main channels. 

These arguments suggest that these ratios can be used as a tool to improve the present cross sections parametrisations by combining the cross sections current measurements with the very precise AMS-02 data. These statements must be true only in the case we consider the shape of our cross sections and, therefore, the shape of the simulated ratios are correct and we only need to apply some normalization. 

Moreover, from the lower plots of Figure~\ref{fig:secsec_Uncert} we see that the fluxes of these three light secondary CRs are in agreement with AMS-02 data at the same time in the intermediate-energy region. The common discrepancy at high energies is due to the choice of diffusion coefficient, as already mentioned, which can be also the reason for the discrepancy of Li below $2 \units{GeV}$. The low energy part of the Be spectrum is extremely influenced by the halo size, which may explain its deviation from experimental data, as stated in~\cite{Aguilar:2018njt}.

In conclusion, we have demonstrated that we can tune the normalization of the spallation cross sections of secondary species to reproduce the secondary-over-secondary flux ratios and overcoming the lack of knowledge we have in the cross sections parametrisations, specially for those channels with little effects on the formation of secondary CRs. This is crucial in order to determine the diffusion parameters with better accuracy. Having a correct balance of the secondary CRs we could, in principle, use also Li and Be in addition to B to determine the diffusion coefficient parameters and this correct balance would even ensure better cross section estimations for boron formation, as we will try to do in chapter~\ref{sec:4}.

%\newpage

\section{Implications of the cross sections on the halo size determination}
\label{sec:size}

While no differences have been observed when changing the radius R of the galaxy model in the fluxes of CRs \cite{evoli2008cosmic}, the halo height H plays an important role. It does not usually receive too much attention in the study of stable secondary species since it is degenerated with the normalization of the diffusion coefficient for the secondary-to-primary ratios ($N_{B/C}(E) \propto H/D(E)$, at medium-high energies), but it plays an important role in the study of leptons (since radiative energy loss rates are of the order of their diffusion time), antiprotons and, as mentioned above, unstable isotopes, like $^{10}$Be, $^{14}$C, $^{22}$Na, etc.

To overcome the problem of the degeneracy of H with $D_0$, many researchers highlighted the need of using additional observables depending on the confinement time of CRs in the galaxy (for instance, see \cite{Weinrich:2020ftb}). This time is $\sim H^2/D(E)$ and it is of the order of tens of My (depending on the particle's energy).

The usual way to constrain the halo size is by means of the study of the ratios of $^{10}$Be to the total Be flux or to the $^9$Be flux \cite{UlysesBe, moskaBe}. The spectrum of the isotope $^{10}$Be depends on an interplay between the diffusion time of the primary CRs (diffusion time, $\tau \propto E^{-\delta}$) and the decay time of this isotope (which is around $1.4 \units{My}$, as determined in \cite{chmeleff2010determination}). Other methods have been used to perform this calculation, as from radio observations of leptons' synchrotron emission \cite{bringmann2012radio}, X-ray and gamma-ray studies \cite{biswas2018constraining}, CR leptons and other heavy nuclei \cite{moskalenko2000diffuse} and even antiprotons \cite{jin2015cosmic}.

\begin{figure*}[!ht]
\begin{center}
\includegraphics[width=0.46\textwidth,height=0.245\textheight,clip] {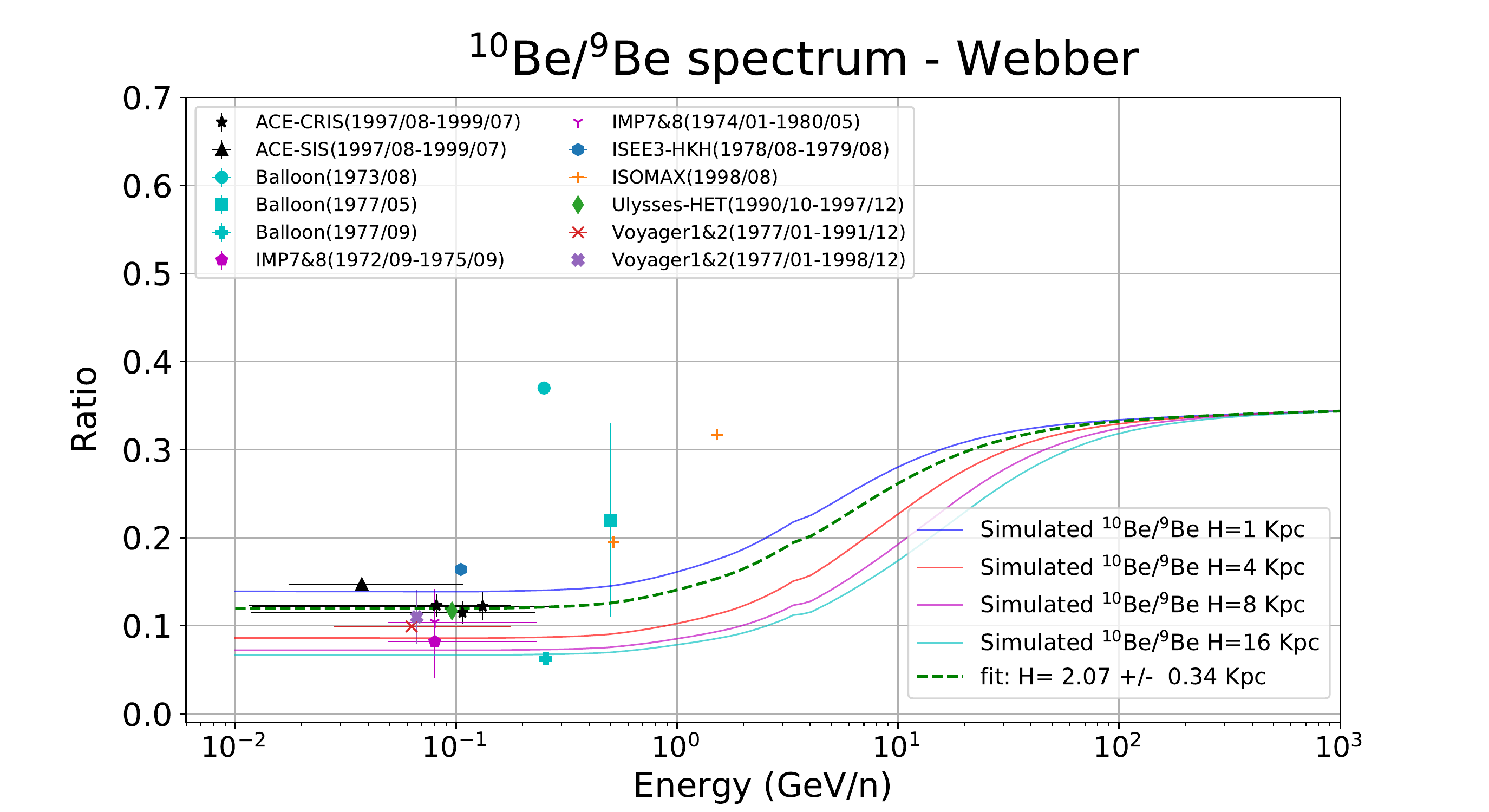} \hspace{0.1cm}
\includegraphics[width=0.46\textwidth,height=0.245\textheight,clip] {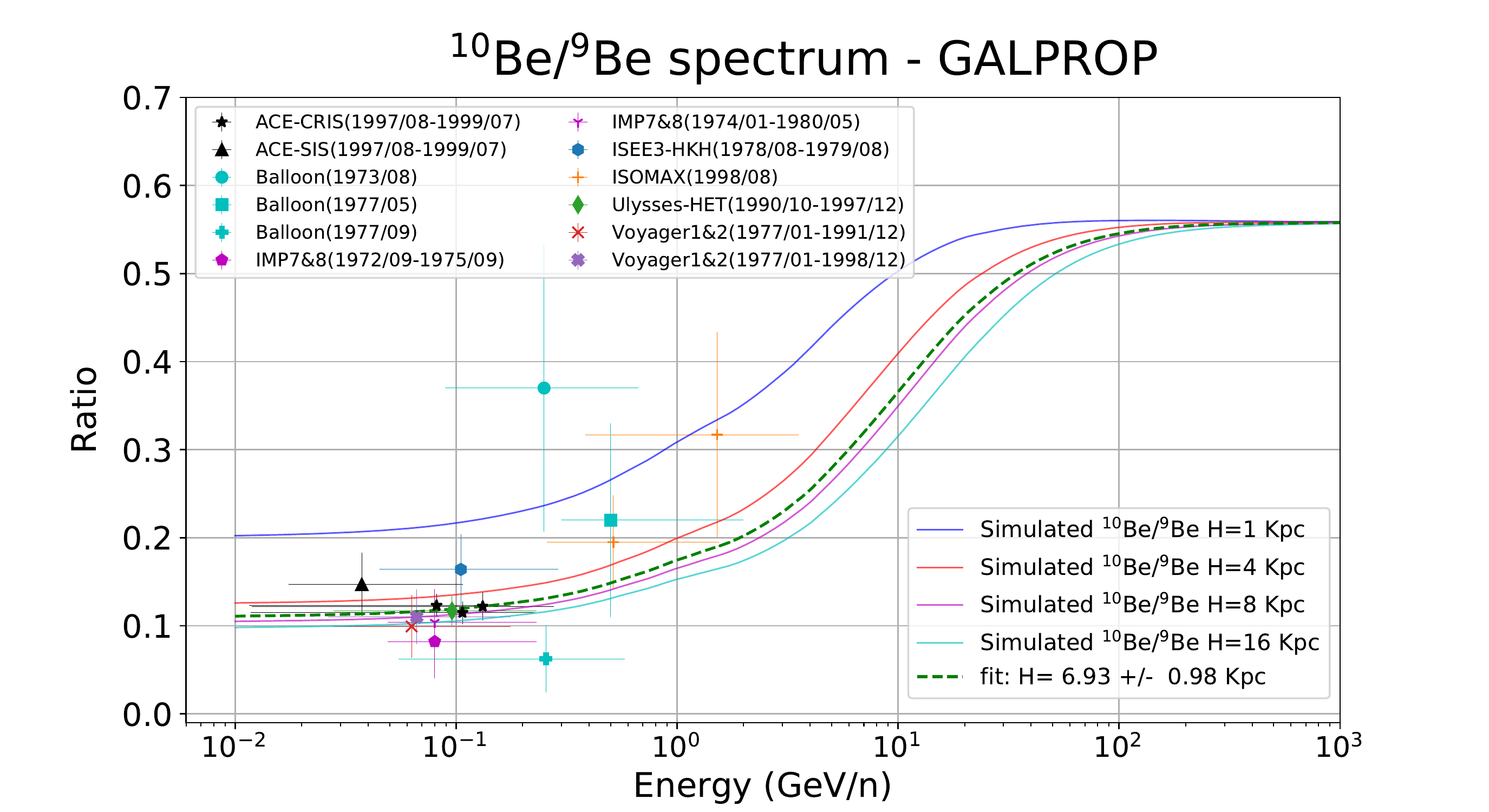} 
\hspace{0.1cm}
\includegraphics[width=0.46\textwidth,height=0.245\textheight,clip] {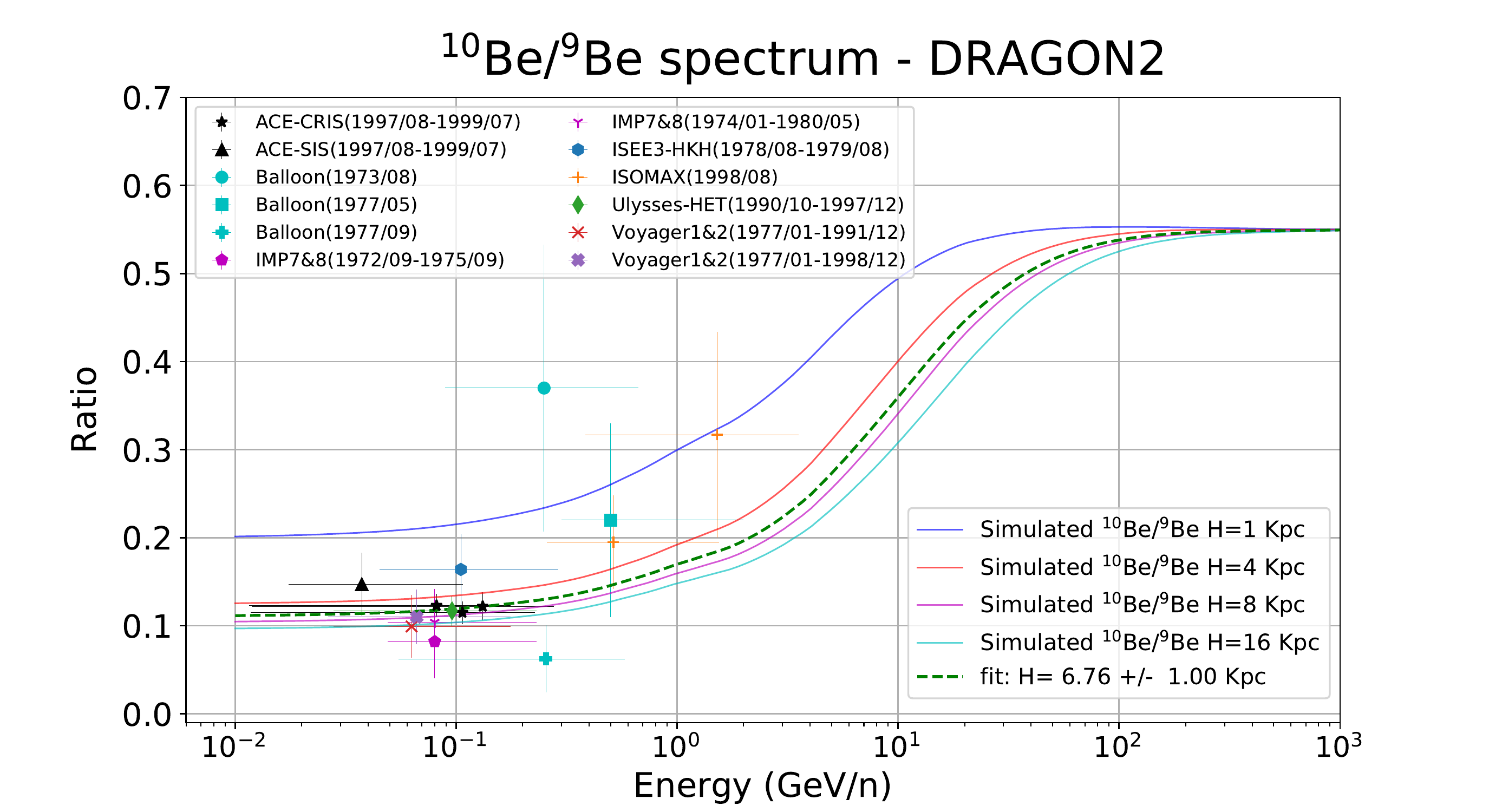}
\hspace{0.1cm}
\includegraphics[width=0.46\textwidth,height=0.245\textheight,clip] {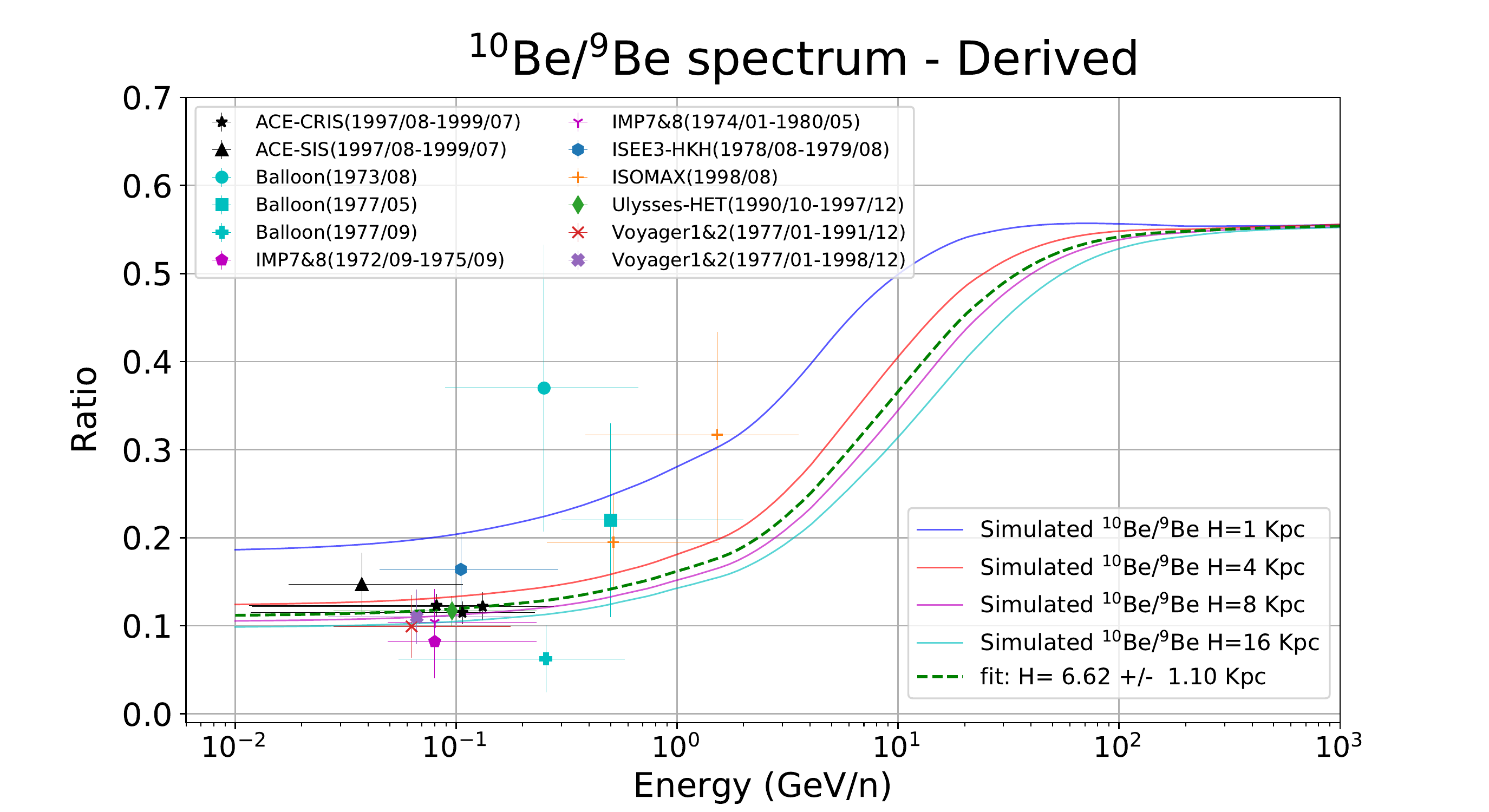}
\vspace{0.1cm}    
\end{center}
\caption{\footnotesize $^{10}$Be/$^{9}$Be spectra compared to all experimental data available and for every cross sections parametrisation studied here. For each model various simulations with different halo sizes are shown, including the simulation that yields the best fit value.}
\label{fig:sizes}
\end{figure*}

One of the obvious consequences of using secondary isotopes to determine any feature of the propagation is that we will be highly influenced by the cross section model used. To take into account these uncertainties in the determination of the halo size value we proceed to compare different predictions coming from different cross sections parametrisations (DRAGON2, Webber, GALPROP and the derived cross sections, in this case), as shown in Fig.~\ref{fig:sizes}. In order to perform this comparison, different halo sizes (from $1 \units{kpc}$ to $16 \units{kpc}$) are tested for all the cross sections models and are interpolated using a 2D interpolation with the tool \textit{RegularGridInterpolator} \footnote{\url{ https://docs.scipy.org/doc/scipy/reference/generated/ scipy.interpolate.RegularGridInterpolator.html}}. The error introduced by the interpolation is smaller than $1\%$ for every energy bin. Then, the fit is performed with the \textit{$curve\_fit$} package from the \textit{$scipy.optimize$} library.
As mentioned above, in each simulation (i.e. for every halo size value) the diffusion coefficient is chosen to fit the B/C flux ratio measured by the AMS-02 experiment.

Here we stress the need of more experimental data points above $1 \units{GeV}$, since they are very few and with large error bars. Projects intended to perform isotopic measurements, like the future Helix mission \cite{HELIX} are proposed and will give a key help in these determinations. So far, measurements of the flux ratios between Be isotopes with relatively low error bars are only available below $300 \units{MeV}$ and the uncertainties in the predictions at these energies can be large because of the lack of knowledge due to the solar modulation and even the form diffusion coefficient. In addition, as mentioned above, the effect of the gas profile along the z direction plays an important role at low energies, since the isotopes can be formed far away from the disc with respect to their decay length.

\begin{figure}[!hptb]
\begin{center}
\includegraphics[width=0.43\textwidth,height=0.22\textheight,clip] {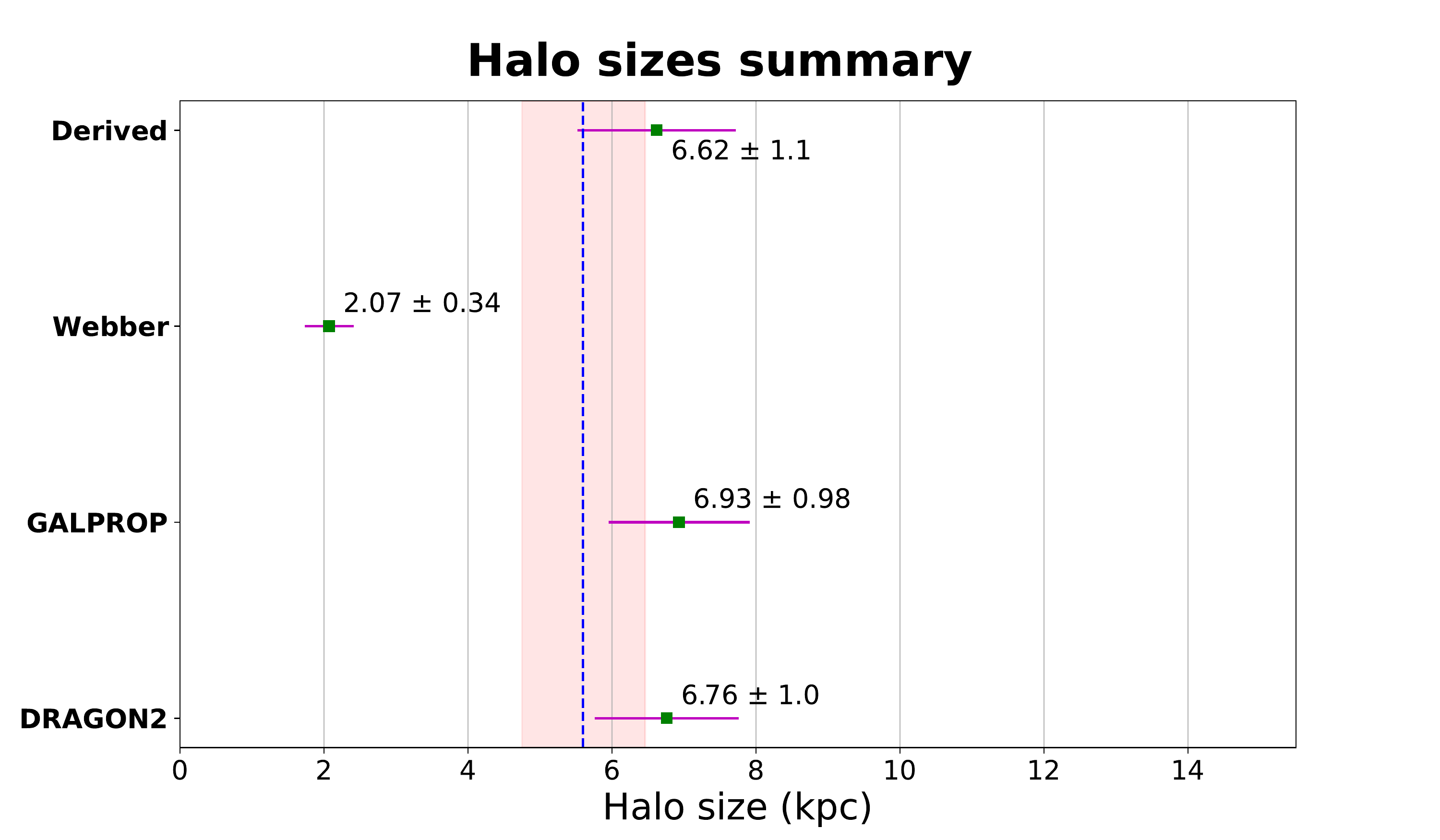} 
\end{center}
\caption{\footnotesize Summary of the results obtained from the fits of the halo size to the $^{10}$Be/$^{9}$Be experimental data. Also the values obtained %doing the fit for the Be/B spectrum for the DRAGON2 cross sections and
with the derived cross sections are shown for a broader comparison. Error bars reflect only statistical uncertainties as explained above. The dashed blue line represents the mean value of the halo best fit values, and the red band the uncertainties on the mean ($1 \sigma$), calculated as the mean of the halo best fit values $\pm 1 \sigma$.}
\label{fig:size_sum}
\end{figure}

The value of the halo size that best fits the experimental data for each model is accompanied by the uncertainties coming from its determination. These values are summarized in Figure~\ref{fig:size_sum} together with a line indicating the mean halo size value. It must be highlighted that the error bars associated to the determination of H by each model come just from the fit of the data and their errors. A change in the halo size, for low halo size values, causes very large differences in the values of the ratios (see the differences from $1$ to $2 \units{kpc}$ in these plots), while for high halo sizes the values of the ratios become progressively closer (ratios are very similar for halo sizes larger than $\sim 8 \units{kpc}$). This explains the error bars in Figure~\ref{fig:size_sum}, where the halo size determined with the Webber parametrisations shows a very small error bar. Then, we take the mean of these best fit values to have an approximated value of the halo size and of its uncertainty related to different cross sections.

We see that the derived cross sections cast a value of the halo size very similar to the DRAGON2 and GALPROP predictions, of around $6.7 \units{kpc}$ (very similar to the value found in ref.~\cite{CarmeloBeB}). The small value found for the Webber parametrisations can be explained having a look to the cross sections around hundreds $\units{MeV}$. The most important channels generally show a good agreement with cross sections data, but in the case of the $^{12}C \rightarrow$ $^{10}$Be channel, the discrepancy is considerable. Therefore, from this underestimation of the cross section, the halo size prediction is expected to be also underestimated (less $^{10}$Be production implies a lower halo size to avoid $^{10}$Be decaying more).

%The small value found for the Webber parametrisations can be explained looking at the cross sections around hundreds $\units{MeV}$. The most important channels generally show a good agreement with cross sections data, but in the case of the $^{12}C \rightarrow$ $^{10}$Be channel the discrepancy is considerable. From this underestimation of the cross section we expect the halo size prediction to be also underestimated (we are biased with less $^{10}$Be production, which implies a lower halo size to avoid $^{10}$Be decaying more).

In any case, the value obtained using the Webber parametrisations seems to be out of the standard picture of a halo size between $3$ and $10 \units{kpc}$ height found in measurements of diffuse gamma-ray background \cite{zaharijas2012fermi} and with other radio observations \cite{di2013cosmic,beuermann1985radio,orlando2013galactic}. The mean of these best fit values (including the determination of the halo size using the Webber parametrisation) is exactly $5.6^{+0.86}_{-0.85} \units{kpc}$, and the value obtained using the most updated cross sections is around $6.8 \pm 1\units{kpc}$.
%Mean values: 6.36, 8.39, 2.25, 6.62 -- > 5.9
%uncertainties: 1.1, 1.05, 5.25, 0.33
%Mean of the fit values +- their uncertainties
%13.64, 2.58, 7.72, 7.41 --> 7.56    5.9+1.66
%1.87, 5.52, 5.31, 3.14 --> 3.96      5.9-1.94

\newpage
\section{Final Conclusions}
\label{sec:conc}
We are in a very exciting period of high quality CR data thanks to the AMS-02 experiment, which allows making precise analyses on the nature of CRs and their propagation throughout the Galaxy. This implies that we can accurately test our propagation models. 

However, while the astrophysical information seems to be accurate enough to face new physical challenges, we are missing information from a key ingredient: spallation cross sections. Since the study of secondary-over-primary CRs is the best tool to test the diffusion coefficient parametrisations we need to have high precision on the secondary CRs fluxes and in their production (inclusive) cross sections.

The amount of channels that matters in the spallation network and the difficulty to perform the measurements lead to parametrisations that extend the energy range of the actual measurements and expand to all important nuclei (even those without experimental data to compare with). While we have some insight in the shape of cross sections distribution with energy, the normalization is still somewhat uncertain. 

In this chapter, we have deeply discussed different cross sections parametrisations and we have investigated their direct impact in the fluxes of Li, Be and B with an upcoming version of the DRAGON2 code. We have highlighted the immediate relation of the secondary-over-secondary ratios with the spallation cross sections used. In this way, it has been shown that predictions from different cross sections models can differ in the amount of Be and Li fluxes by a level of 20\% to more than 40\%, emphasizing how uncertain we are in the determination of secondary CRs spectra, although the uncertainties in the B flux seem to be remarkably smaller, since its production is essentially due to the contribution of C and O channels (the ones with more experimental data). To visualize the differences from the cross sections models in the B spectrum, a comparison of their prediction for fixed diffusion parameters is shown in Figure~\ref{fig:BComps}, spotting a variation in the predictions smaller than 10\%. In turn, we have demonstrated the good performance achieved in the simulations when using the default DRAGON2 cross sections model, being able to reproduce every observable within very small error.

It has been discussed that there is no need of invoking primary sources of the studied secondary CRs as the cross sections uncertainties can largely account for the discrepancies found with respect to data. To better show this, two models of cross sections have been considered as limiting cases. These bracketing models represent minimum and maximum credible renormalizations of the GALPROP model of cross sections, maintaining the same energy dependence. At the end, the propagation of their uncertainties into the fluxes of secondary CR species means that the minimum uncertainty in the flux of a secondary element is constrained by the cross sections channel with minimum error bars. As we have seen, the channels with more data points and less uncertainty (by far) are those of Be, involving mean relative errors at the level of 16\%.

We have investigated how these secondary-over-secondary ratios would behave under the combination of the bracketing models to test whether, between these two limiting cases, there is a space of cross sections that can reproduce all secondary CRs above $10 \units{GeV}$ at the same time.

To go further, a set of cross sections was derived by matching these ratios at high energy to demonstrate that they are an excellent tool to constrain and even adjust the cross sections parametrisations. This derived cross sections model balances the deficiencies on the description of those channels which are very poorly known and gives some insight on how much our cross sections network is biased for every secondary species. This showed that, while the boron channels should not be renormalized by more than 5\% to reproduce the ratios, the beryllium and lithium ones needed an overall renormalization of around 18\% and 28\% (in average), respectively. This follows the expectations of the uncertainties for the determination of fluxes. 

In fact, this means that we have already been able to combine the information arising from secondary CRs to get rid of the systematic uncertainties related to the spallation cross sections. This strategy must be taken as a more precise way to adjust the overall effect of spallation cross sections since it mostly relies on the very precise AMS-02 data.

We argue that this combined tuning of the normalization on a cross sections parametrisation, such that they reproduce these spectra, serves to improve the determination of the diffusion coefficient parameters as well, and it is important in order to be consistent with all the observables at the same time.

Finally, these ratios were also shown for different halo sizes in every cross sections set, since the $^{10}$Be isotope has a lifetime close to the propagation time of CRs in the Galaxy. Together with this, the ratio $^{10}$Be/$^{9}$Be was studied for each of the models. This allowed the discussion on the wide repercussion of the cross sections on the determination of the halo size too. The implications of the diffusion time are relevant for the study of leptons currently. The determination of the halo size from the most updated parametrisations gives a mean value around $6.8\pm1 \units{kpc}$, which is in agreement with most of the values obtained in other works.

In conclusion, this chapter comprises a fist step towards a better determination of the phenomenology behind the CR propagation and the reduction of systematic uncertainties coming from the cross sections used. Once this main concern (actually the most important concern for achieving accurate propagation models) is understood, we can proceed to properly study different propagation models and their diffusion coefficient parameters.

\chapter{New set of Monte Carlo based cross sections: the FLUKA inelastic and inclusive cross sections for cosmic-ray interactions.}
\label{sec:3}

Due to the underlined necessity of having precise cross sections parametrisations and the lack of data in most of the channels, there have been several attempts to develop the cross sections network purely from physical principles. 
These computations are difficult to perform because of the non-linear nature of these quantum-interaction processes and usually rely on Monte Carlo simulations and event generators. These simulation codes have experienced a positive boost in the last years, usually driven by radiological and medical applications, which need to accurately describe the transport of ions in different materials and their interactions. Examples of these codes are PHITS \cite{Sato:2015iqw, Sato:2017pja, Niita:2006zz}, SHIELD-HIT \cite{Sobolevsky:2002zla, Sobolevsky:2015khp} and the GEANT4 code \cite{Pia:2003cj, Moralles:2009zz}, widely used in particle physics to simulate detector responses, as well.

Despite this progress, the accuracy on the computation of interaction cross sections has demonstrated to be below that of the dedicated semi-analytical codes previously discussed, at least in the range of energies interesting for Galactic cosmic rays. In general, past nuclear codes have not been able to reproduce any of the channels with accuracy better than 5\% at best \cite{Genoliniranking}. 

This chapter aims at computing the full cross sections network (including inelastic and inclusive cross sections), needed to be implemented in CR propagation codes with one of the most recently updated Monte Carlo nuclear codes, FLUKA \cite{Ferrari:2005zk}, and to demonstrate its compatibility with data. In addition, a careful study of the effect of ghost nuclei, helium and hydrogen target cross sections is performed. Finally, the cross sections are implemented in the DRAGON code in order to demonstrate that they can be used to reproduce the new and accurate CR data at a similar level than the semi-analytical parametrisations.

The main feature expected in the FLUKA cross sections is their steep rise with energy, that is reliably observed in particle physics experiments \cite{PhysRevD_rise, Block_2011}, but which is not incorporated in any of the current relevant parametrisations. In addition, another relevant difference concerns the helium target cross sections, which are usually calculated by an empirical relation from the hydrogen target cross sections. The possibility of transporting all isotopes also enables the cumulative cross sections to be more complete, which is a relevant difference with respect to old calculations with nuclear codes.

The work performed in this chapter is inspired by \cite{Mazziot}, where the secondary hadrons, leptons, gamma rays and neutrino yields produced by the inelastic interactions between several species of long-lived CR projectiles and different target gas nuclei are calculated, although this chapter is focused on GCR studies and to directly test whether the FLUKA cross sections can be competitive with the current parametrisations.

A related study has been carried out in \cite{FlukaSun}, where the CR interactions with the Sun's atmosphere were simulated to predict the yields of gamma rays, electrons, positrons, neutrons and neutrinos to be detected at Earth. This study is one of the main works performed by the thesis' author and it is briefly reviewed in appendix \ref{sec:appendixFluka} as one of the astrophysical applications of the FLUKA code. Other examples of astro-particle studies making use of FLUKA are \cite{Battistoni:2005pd, Tusnski:2019rpd, Heinbockel:2011zz}.

This chapter is organized as follows: we first give a small overview on the basics of nucleus-nucleus reactions in the energy range of interest for GCR, in section~\ref{sec:nucinter}; then, in section~\ref{sec:Flukaintro} the FLUKA code is presented together with a brief overview of the theories implemented to simulate these reactions. An explanation of the computations of cross sections (both inelastic and inclusive) is done in~\ref{sec:FXSecs}, where the results are fully shown in comparison with data and, more importantly, with the parametrisations described in chapter~\ref{sec:XSecs}, putting special emphasis in the resonance positions and high energy behaviour. Hydrogen and helium targets cross sections are examined and contrasted with semi-empirical formulas too. In addition, a discussion on the channels that are more affected by ghost nuclei is carried out. %From https://www.linguee.com/spanish-english/translation/la+discusi%C3%B3n+se+llevar%C3%A1+a+cabo.html  --> Held, made or carried out are the options
Finally, once these cross sections are implemented in the DRAGON code, in section~\ref{sec:FDRAGON} we examine the compatibility of the simulations against the most recent AMS-02 data. Here it is demonstrated that every observable can be reproduced with this cross sections data-set in a similar way the other cross sections parametrisations do, inside a reasonable range of diffusion parameters. The secondary-over-secondary ratios are also studied finding them to be plausible as well. In section~\ref{sec:Fconc}, the main conclusions are highlighted.

\section{Basics on hadron-nucleus and nucleus-nucleus interactions}
\label{sec:nucinter}

The complexity of interactions involving nuclei is due to their composite nature: a nucleus is a group of nucleons, which are bound with some energy that depends on their quantum properties (spin, isospin, angular momentum, charge, etc...) and continuously exchange momentum. These internal interactions between nucleons make the nuclear interaction very complex to be described, while the description of single nucleon interactions is simpler. Nevertheless, when the energy of a beam of primary hadrons exceeds a few tens of$\units{MeV}$, inelastic interactions start to play a major role, and secondaries have enough energies to trigger further interactions, giving rise to hadronic showers. The most abundant secondary particle produced is the pion (the threshold kinetic energy for the projectile in order to produce a pion in a proton-proton interactions is around $290 \units{MeV}$). The decay of pions, and mesons in general, causes that, as the energy of interaction increases, an increasing fraction of the energy is transferred from the hadronic shower to the electromagnetic sector.

\begin{figure*}[!hbt]
\begin{center}
\includegraphics[width=0.36\textwidth,height=0.27\textheight,clip] {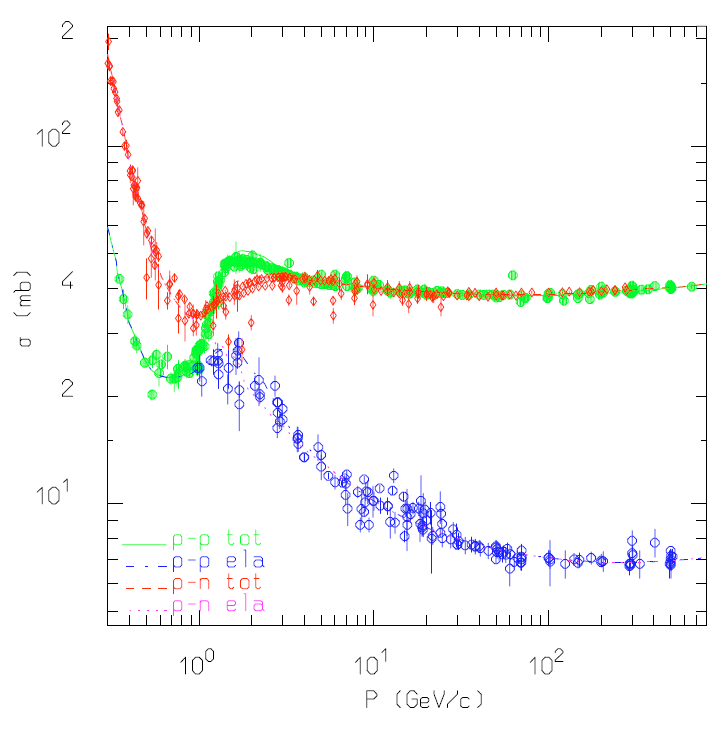} \hspace{0.8cm}
\includegraphics[width=0.36\textwidth,height=0.27\textheight,clip] {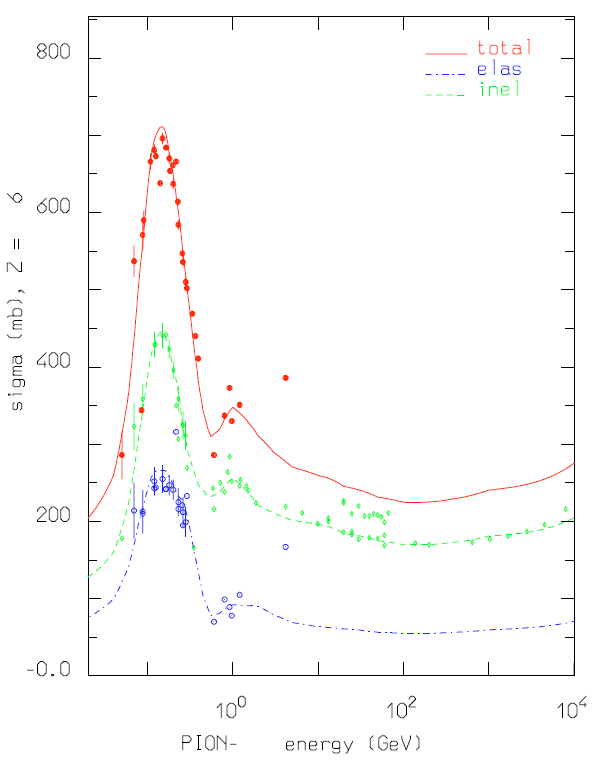}
\end{center}
\caption{\footnotesize Examples of interaction cross sections. In the left panel, the total and elastic cross sections for p-p and p-n collisions simulated with FLUKA are compared to experimental data. They describe nucleon-nucleon collisions. In the right panel, the total, inelastic and elastic cross sections of an example of hadron-nucleus interaction, the $^{12}$C + $\pi^-$ reaction, are shown. Both plots were taken from \cite{interactions_review}.}
\label{fig:nuc_inter}
\end{figure*}

Below the pion production threshold, the dominant processes are the elastic interactions between neutrons and protons in the nuclei. The left panel of figure~\ref{fig:nuc_inter} shows the p-p and p-n elastic and total interaction cross sections implemented in FLUKA compared with experimental data. Both p-n and p-p cross sections exhibit a fast increase with decreasing energy below $1 \units{GeV}$. We also see a factor 3 of difference in this energy region (where elastic interactions are dominant) between the n-p and p-p cross sections, as expected by symmetry and isospin considerations (see section 4.2 of \cite{interactions_review}). The right panel of Figure \ref{fig:nuc_inter}) shows the $\pi^- - $C cross section. 
When the energy is larger than the pion creation threshold, pion-nucleon elastic collisions and charge exchange scatterings (i.e. quasi-elastic reactions of the type: $p + \pi ^- \longrightarrow n + \pi ^0$ and $n + \pi^+ \longrightarrow p + \pi^0$) are dominant up to a pion laboratory energy of about $250 \units{MeV}$, where inelastic processes start to take over. 
(Inelastic) production of particles starts when additional pions can be created ($N_1 + N_2 \longrightarrow N'_1 + N'_2 + \pi$, whose threshold energy in the lab frame is $290 \units{MeV}$, and $\pi + N \longrightarrow \pi' + \pi'' + N'$, whose threshold energy is around $150 \units{MeV}$). The isobar model (reactions creating particles inelastically proceed through an intermediate state containing at least one resonance) is supposed to describe these interactions well, motivated by the dominance of the $\Lambda$  and the $N^*$ resonances at higher energies, in the $\pi$-N channel. 

However, as the incident hadron energy exceeds $3-4 \units{GeV}$, the description of non-elastic interactions via quasi two-body reactions with formation and decay of resonances starts to become very difficult and hadron-nucleon interactions at high energy are more easily described by models adopting basic concepts from QCD (which are effective only for large transverse momentum, though). In this regime, hadrons are essentially described by their distributions of valence and sea quarks and interacting strings (quarks held together by the gluon-gluon interaction into the form of a string), following the Fermi statistics and the Pauli exclusion principle.

At this regime, one of the most successful models is the Dual Parton Model (DPM), developed in 1979 (see ref. \cite{Capella:1983ie}), In this framework, a hadron is a low-lying excitation of a string with quarks, diquarks and antiquarks sitting at its ends. Interactions are described following the reggeon-pomeron calculus in the framework of perturbative Regge field theory \cite{collins_1977}. A simplified view on the interactions under this theory is that each hadron is split into two coloured partons, which combine into two colourless chains, hadronizing into two jets of particles. 

Once suitable models for describing hadron-nucleon interactions are available, the high energy regime can be properly handled describing  multiple primary collisions according to the Glauber approach (see, e.g. \cite{Kaidalov:1990ha}). Models which include these features are often referred to as Glauber cascade approaches (see refs. \cite{Ferrari_1996} and \cite{Ferrari2_1996}). V.N. Gribov first showed the way to incorporate the Glauber model \cite{Glauber:1970wr} for interactions of hadrons with nuclei into a general framework of relativistic quantum field theory. In the Glauber model, the hadron-nucleus interactions are described by successive re-scatterings of an initial hadron on the nucleons composing the nucleus. The Gribov modification implies that simultaneous interactions of the nucleus and hadron can be treated as waves, leading to the same results as the Glauber theory for elastic interactions (demonstrating elastic interactions can be successfully treated by elastic re-scatterings) plus all possible diffractive excitations of the initial hadron. At energies around $100 \units{GeV}$ these terms lead to corrections to the Glauber approximation of the level of $10-20\%$ for the total hadron-nucleon cross section \cite{Kaidalov:1973hh, Kaidalov:1973bb}.

In the nucleus rest frame, a high-energy projectile approaches with a small effective size, corresponding to its De Broglie wavelength. Thus, for energies high enough, the projectile will interact with individual nucleons, leaving the other ones unaltered, justifying the already mentioned straight-ahead approximation, where the kinetic energy per nucleon is conserved during the collision. Eventually, the projectile undergoes multiple scatterings in the nucleus, with a number of interactions roughly proportional to the number of nucleons along the line of sight ($\sim A^{1/3}$). The interacting nucleons generally leave a sea of partons which hadronize into jets going away from the nucleus, leaving it in a high-energy excited state. 

The models that treat these complex interactions consider the nucleus as a sphere (droplet model) with a density modeled by a suitable parametrisation (namely, the Woods-Saxon density) or as a harmonic oscillator shell in the case of light isotopes ($A < 16$), which are divided into several radial zones of isodensity \cite{density, Hofstadter2092, charge_density}. The impinging projectile and the interaction daughters are transported inside the region of the nucleus according to their nuclear mean field and to the Coulomb potential.
In this treatment the Fermi motion of the particles in the nucleus must be taken into account in the interaction potential. Many possible shapes of this potential have been proposed, either on the basis of self-consistency (Hartree-Fock like) or of computational usefulness to reproduce experimental quantities (harmonic oscillator,
Nilsson anharmonic oscillator, Woods-Saxon, all with spin-orbit coupling). At the end, the calculation consists of solving the Schrodinger equation with a complex potential that is different for different hadrons, $V(\Vec{r}) = - [U(\Vec{r}) + iW(\Vec{r})]$, which is called the optical potential. It can be derived from the convolution of the interactions between the projectile and all the nucleons in the nucleus, or can be fitted to the scattering data with preliminary assumptions about its radial dependence. The real part of the optical potential is also sometimes used to describe bound states.

After the primary collisions, a stage of interactions of part of the particles produced inside the nucleus follows, implying successive  re-interactions in the nucleus. This is the basic assumption of the models called IntraNuclear Cascades (INCs) \cite{CUGNON1982191, Armstrong1980}, which can be applied for both elastic and inelastic interactions.

Once these interactions cease, there is a residual nucleus left, which is typically unstable because of nucleon imbalance (unstable distributions of neutrons and protons) and may fragment or evaporate into lighter nuclei (pre-equilibrium phase) that usually need to pass another equilibrium stage (eventually requiring several steps) to finally de-excite via gamma-ray emission (implying relaxation to a lower energetic level). Already in 1966, Griffin described the spectra following nucleon-induced reactions in terms of a pre-equilibrium model \cite{GriffinPhysRevLett.17.478}, i.e. a transition between the first step of the reaction and the final thermalization. The two leading approaches are the quantum-mechanical multi-step treatment, which has a very good theoretical background \cite{BONETTI19941}, although it is complex, and the exciton model \cite{preq_dec} which relies on statistical assumptions. The pre-equilibrium process in the exciton model is described as a chain of steps, each corresponding to a certain number of excitons (understanding an exciton as either a particle above the Fermi surface or a hole below the Fermi surface). 

At the end of the exciton chain, the residual nucleus is supposed to be left in an equilibrium state, in which the excitation energy is shared among a large number of nucleons. Such compound nucleus is supposed to be characterized by its basic properties (mass, charge and excitation energy), with no further memory of the steps which led to its formation. The excitation energy can be higher than the separation energy, thus nucleons and light fragments can be still emitted (see, e.g. \cite{BARASHENKOV1974462}). The emission process can be well described as an evaporation from a hot nucleus, relying on the formula of Weisskopf \cite{Weisskopf}, which links the probabilities per unit time of capturing a particle and forming a composite nucleus.

The evaporation process continues till it is energetically allowed, and generally leaves the residual nucleus in an excited state. The residual excitation energy is then dissipated by emission of gamma rays. Actually, the gamma emission occurs even during evaporation, as a competing process, but with a small branching ratio. Gamma de-excitation proceeds through a cascade of consecutive photon emissions, until the ground state is finally reached.

\section{The FLUKA code}
\label{sec:Flukaintro}

FLUKA (FLUktuierende KAskade) is a Monte Carlo package written in Fortran 77, extensively used at beam facilities (specially at CERN) for interactions, radioprotection and facility design calculations, optimized for an energy range extending from below the$\units{keV}$ (in electromagnetic interactions) or$\units{MeV}$ (for hadronic interactions) up to thousands of$\units{TeV}$, both for particles and ions. it was mostly aimed at pure shielding calculation, and evolved to a multipurpose one since 1989. %A part of it was included in GEANT3 around 1993.

FLUKA has been jointly developed by the European Laboratory for Particle Physics (CERN), and the Italian National Institute for Nuclear Physics (INFN) in the framework of an international collaboration until the end of 2019, when the collaboration agreement was terminated. An interesting review on the evolution of the FLUKA code can be found in: \url{http://www.fluka.org/fluka.php?id=man_onl&sub=103}

FLUKA is a general purpose tool that can be used to transport particles in arbitrarily complex geometries, including magnetic fields, through the FLUKA combinatorial geometry. This makes it very useful in accelerator shielding \cite{Lari:2019jlo}, calorimetry \cite{Onorati:2020zne, Dissertori:2020yoq} or detector design \cite{BrizMonago:2018pef, Appleby:2018bkg}, but it is mainly intended to be used in collider physics \cite{Mokhov:2018hjs}, hadron therapy (e.g. \cite{Arico:2019pcz}) and dosimetry \cite{Bewer:2020qqc, Amin:2018ukd}. It is also designed to be used in astroparticle physics \cite{Battistoni:2008hga, Andersenarticle}, raising high expectations.

Transport of charged particles is performed through a treatment of multiple Coulomb scattering and of ionisation fluctuations which allows the code to handle accurately some challenging problems such as electron back-scattering and energy deposition in thin layers even in the few$\units{keV}$ energy range. Particle transport also includes time-dependence. Energy losses of charged particles are described according to the Bethe equation, with optional delta-ray production and transport, accounting for spin effects and ionization fluctuations. Shell and other low-energy corrections are derived from Ziegler \cite{Ziegler} and the density effect is included according to Sternheimer \cite{STERNHEIMER1984261}. For all charged particles a special transport algorithm, based on Moliere's theory of multiple Coulomb scattering improved by Bethe \cite{Moliere-Bethe, striganov}, accounts for correlations between lateral and longitudinal displacements, between projected deflection angles and between projected step length and total deflection.

The possible energy range covered for hadron-hadron and hadron-nucleus interactions in FLUKA extends from a threshold (depending of the hadron, but $\sim 10 \units{MeV}$) up to $10^4 \units{TeV}$, while electromagnetic and muon interactions can be treated from $1 \units{keV}$ ($100 \units{eV}$ for photons) up to $10^4 \units{TeV}$. Nucleus-nucleus interactions are also supported from $10 \units{MeV/n}$ up to $10^4 \units{TeV/n}$. Neutron transport and interactions below $20 \units{MeV}$ down to thermal energies are treated in the framework of a multi-group approach (see, e.g., \cite{Hebert2010}), with cross sections data sets developed by FLUKA starting from standard databases (mostly ENDF/B-VII, JENDL and JEFF). 

Hadron elastic scattering is described by means of parametrised nucleon-nucleon cross sections, tabulated nucleon-nucleus cross sections and tabulated phase shift data for pion-proton and kaon-proton scattering. Hadron-nucleon inelastic collisions are described in terms of resonance production and decay model up to $3-5 \units{GeV}$, including hundreds of resonances and possible decays, and the Dual-Parton model for higher energies.

The extension from hadron-nucleon to hadron-nucleus interactions is achieved by an extended PEANUT (PreEquilibrium Approach to NUclear Thermalization) model \cite{FassFLUKAP, Ferrari:1993xr}, which has proven to be a precise and reliable tool for intermediate energy reactions. At momenta below $3-5 \units{GeV/c}$ the PEANUT package includes a very detailed Generalised Intra-Nuclear Cascade (GINC) and a preequilibrium stage, while at high energies the Gribov-Glauber multiple collision mechanism is included in a less refined GINC. Both modules are followed by equilibrium processes: evaporation, fission, Fermi break-up, gamma de-excitation.

For heavy ion interactions above $5 \units{GeV/n}$, as well as for hadron-hadron and hadron-nucleus interactions above $20 \units{TeV}$, FLUKA interfaces to DPMJET-3, the nuclear framework around the PHOJET \cite{Engel:1994vs, Engel:1995yda} event generator for hadron-hadron, photon-hadron and photon-photon collisions. See Ref.\cite{changes_fluka-models}, for more information.

\begin{figure*}[!hbt]
\begin{center}
\includegraphics[width=0.86\textwidth,height=0.25\textheight,clip] {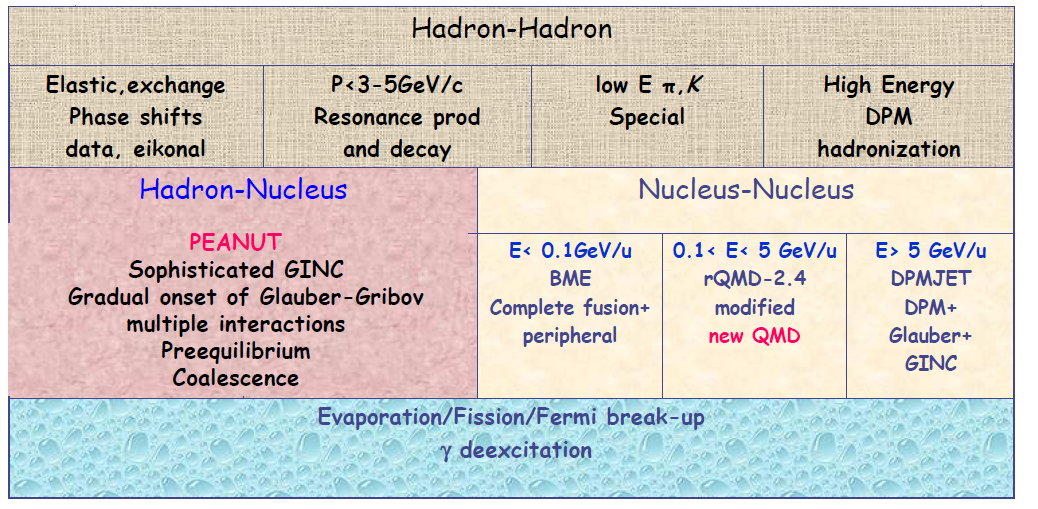} 
\end{center}
\caption{\footnotesize Scheme of the models used in the different energy regimes for the different hadronic interactions. Taken from the FLUKA presentation at the Hadronic Shower Simulation Workshop held in 2006 (see refs. \cite{Waters:2007zz} and \cite{Battistoni:2007zzb}).}
\label{fig:Fluka_models}
\end{figure*} 

All nuclear interaction models, including those generated by ions, share parts of the common PEANUT framework. In particular, all nuclear fragments, irrespective of the originating reaction, are de-excited through the same evaporation/fragmentation and gamma production chain. The final steps of the reaction include evaporation in competition with fission and gamma de-excitation.

The exciton formalism employed in PEANUT follows that of M. Blann and collaborators \cite{Blann71, Blann72, Blann75, Blann83, Blann83b} called Geometry Dependent Hybrid Model (GHD).

In the early treatments of the nucleus-nucleus interactions, FLUKA used a upgraded version (eq.~\ref{eq:tripathi_inel}) of the cross section parametrisation developed by Tripathi \cite{tripathi1999accurate}, which has the form:
\begin{equation}
\sigma_R = \pi r_0^2 (A_p^{1/3} + A_t^{1/3} + \delta_E)^2 (1-R_CB/E_{cm})X_m
\label{eq:tripathi_inel} 
\end{equation}
where $r_0 = 1.1$ fm, $\delta_E$ is a complex expression that depends on energy and includes the effects of Pauli blocking and transparency, among others, $E_{cm}$ is the center-of-mass energy of the target and projectile system, $R_C$ is the Coulomb multiplier, $B$ is the energy dependent Coulomb interaction barrier (equations 3 and 4 of ref. \cite{tripathi1999accurate}) and $X_m$ is a low energy multiplier that accounts for the strength of the optical interaction model. The Coulomb interaction term (second to last term in the r.h.s.) is important at low energies ($\sim$ a few tens of$\units{MeV}$), while the Coulomb multiplier is needed in order to have the same formalism for the absorption cross sections in light and heavy colliding systems (see ref. \cite{tripathi1999accurate}). 

\begin{figure*}[!hbt]
\begin{center}
\includegraphics[width=0.84\textwidth,height=0.19\textheight,clip] {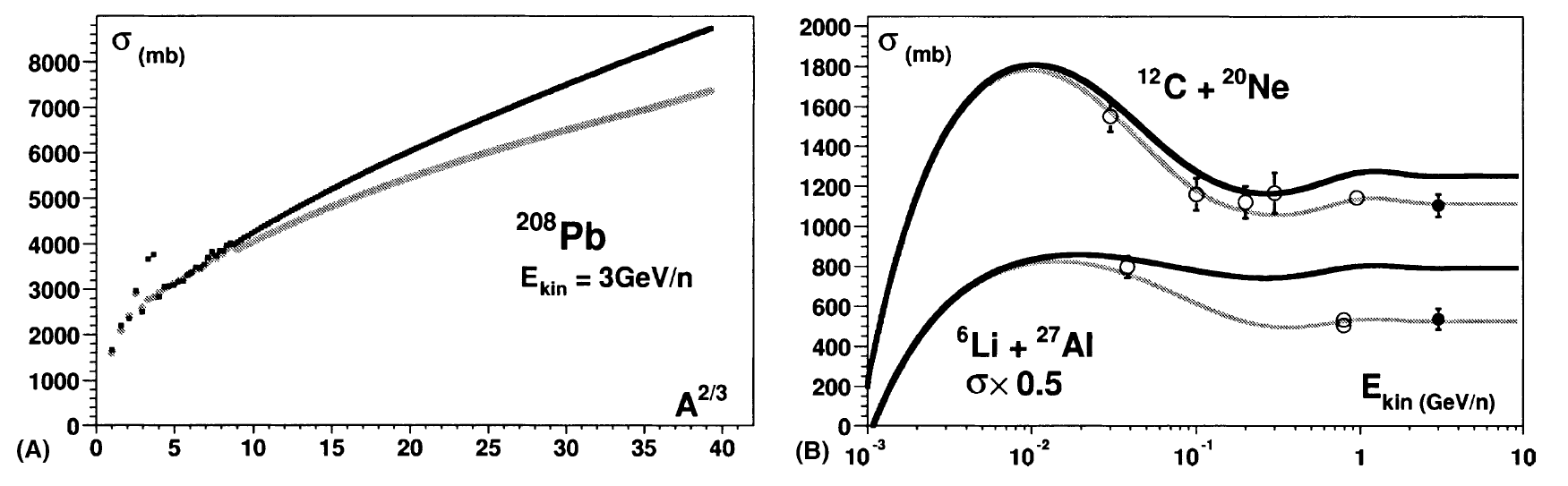} 
\end{center}
\caption{\footnotesize Total reaction cross sections calculated for: a) $^{208}$Pb at $3 \units{GeV/n}$ kinetic energy as a function of the target nucleus area $\propto A^{2/3}$. The Tripathi parametrisation (thin black curve) and the DPMJET model (grey curve) exhibit a systematic discrepancy for heavy systems ($A_P$ and $A_T$ both larger than 26); b) predictions of the original (thick black curves) and modified (thin grey curves) parametrisations as a function of kinetic energy compared to available data (open circles). The curves for $^6$Li $+$ $^{27}$Al (lower two curves) are scaled by a 0.5 factor. The solid black circles represent the DPMJET prediction at $3 \units{GeV/n}$ kinetic energy  with an assumed 5\% error. Graphs were taken from \cite{Andersenarticle}, while references for the experimental data can be found in \cite{tripathi1999accurate}.}
\label{fig:TripvsFluk}
\end{figure*}

This parametrisation was found to provide correct fits for nuclei at low energies, in general. Nevertheless, discrepancies with respect to experimental data as well as to the DPMJET predictions were found in the case of projectile kinetic energies approaching $1 \units{GeV/n}$, depending on the combination of projectile and target nuclei. The origin of the discrepancies might partially be attributed to the available experimental data which historically were biased towards light systems with a large fraction of measurements at very low energies \cite{Andersenarticle}. In addition, early measurements of nucleus-nucleus absorption (or production) cross sections suffer from systematic uncertainties since they are very model dependent. In order to address the discrepancies found, an empirically modified version of this parametrisation was implemented in FLUKA. The modifications were derived using both experimental data and the DPMJET-3 cross sections predictions at $3 \units{GeV/n}$, and then applied to the computation of $\delta_E$ only. A comparison between the quality of the original parametrisation with experimental data and the DPMJET-3 predictions at $3 \units{GeV/n}$ kinetic energy is shown in Figure~\ref{fig:TripvsFluk}, for Pb interactions and the reactions $^{12}$C $+$ $^{20}$Ne and $^6$Li $+$ $^{27}$Al.

In the latest versions of FLUKA, this procedure was replaced to use more sophisticated treatments, and nucleus-nucleus interactions are treated through interfaces to external event generators: at energies above $5 \units{GeV/n}$, DPMJET-3 (see refs. \cite{DPMJET},  \cite{DMPJET-gamma} and references therein), with a special initialization procedure, then, between $0.1$ and $5 \units{GeV/n}$, a modified Rqmd-2.4 (see refs. \cite{Schonfeld:1991ee} and \cite{Rose:1999np}) and finally, the BME (Boltzmann Master Equation;  \cite{BME}) below $0.1 \units{GeV/n}$, as shown in Figure~\ref{fig:Fluka_models}. An example of the use of FLUKA hadronic models applied to CRs can be found in \cite{BATTISTONI200888}.
%This is explained in these slides: https://indico.cern.ch/event/656336/contributions/2729624/attachments/1552506/2439700/HeavyIons_Radioactivity.pdf

Full information on the different models used by the code and its related publications and references can be found in the FLUKA webpage \footnote{\url{http://www.fluka.org/fluka.php?id=publications&mm2=3}}.

\section{Monte Carlo-based cross section calculation}
\label{sec:FXSecs}

Once we have discussed the main ingredients in FLUKA to treat nuclear interactions and their validation, we target at developing the computation of cross sections for all species needed to describe the nuclei involved in the CR network. This means, the treatment of all stable isotope until iron, and the unstable nuclei with their decay products if they do not have lifetime enough to be propagated in the DRAGON code (lifetimes smaller than $10^3$ years).

The first step in the computation consist of simulating two beams of interacting ,with one reproducing the ISM gas target and the other the projectile CR.\footnote{This is done by means of the $\it{PPSOURCE}$ option in the beam card in FLUKA} This configuration reproduces a source of secondary particles from the point of interaction of the two colliding beams, thus describing the CR-gas interactions. The calculations are performed with hydrogen and helium targets separately. Then, the total cross sections needed for the fragmentation and production terms of eq. \ref{eq:caprate} are calculated by weighting the hydrogen and helium cross sections according to their density proportions in the ISM. 

At this point, the decays of unstable particles are switched off, such that they behave just as stable ones. From this simulation, two quantities are extracted: the inelastic cross sections of the nuclei crossing the target and the multiplicity of all the species of particles produced in the collision. This information is needed to calculate spallation cross sections, following \cite{Mazziot}, as: 
\begin{equation}
\frac{d\sigma_{spall}(E_s|E)}{dE_s} = \sigma_{ine}(E) \frac{dn(E_s|E)}{dE_s}
\label{eq:spall-inel} 
\end{equation}
where $E$ and $E_s$ are the kinetic energy of the projectile and of the secondary particle produced, respectively, and $\frac{dn(E_s|E)}{dE_s}$ is the differential multiplicity spectrum of the secondary particle to be studied.

In the following step, the unstable secondary nuclei are taken separately, and their decay is simulated to keep track of the daughter nuclei and of their energies. In this way, after simulating 50,000 events to adequately sample the energy spectra of all decay products, we include the daughter particles in the evaluation of the multiplicities. In this way, the cumulative cross sections can be calculated in addition to the direct cross sections. Thanks to this approach, the effect of ghost nuclei in each of channel can be studied.

This setup has been applied to all isotopes, from protons to iron, for energies from $1 \units{MeV/n}$ to $35 \units{TeV/n}$, using 176 bins equally spaced in a logarithmic scale. The results can be summarized in a set of spallation and inelastic cross section tables, with the spallation ones calculated for both the direct and cumulative channels. 

\begin{figure*}[!hptb]
\begin{center}
\includegraphics[width=\textwidth,height=0.33\textheight,clip] {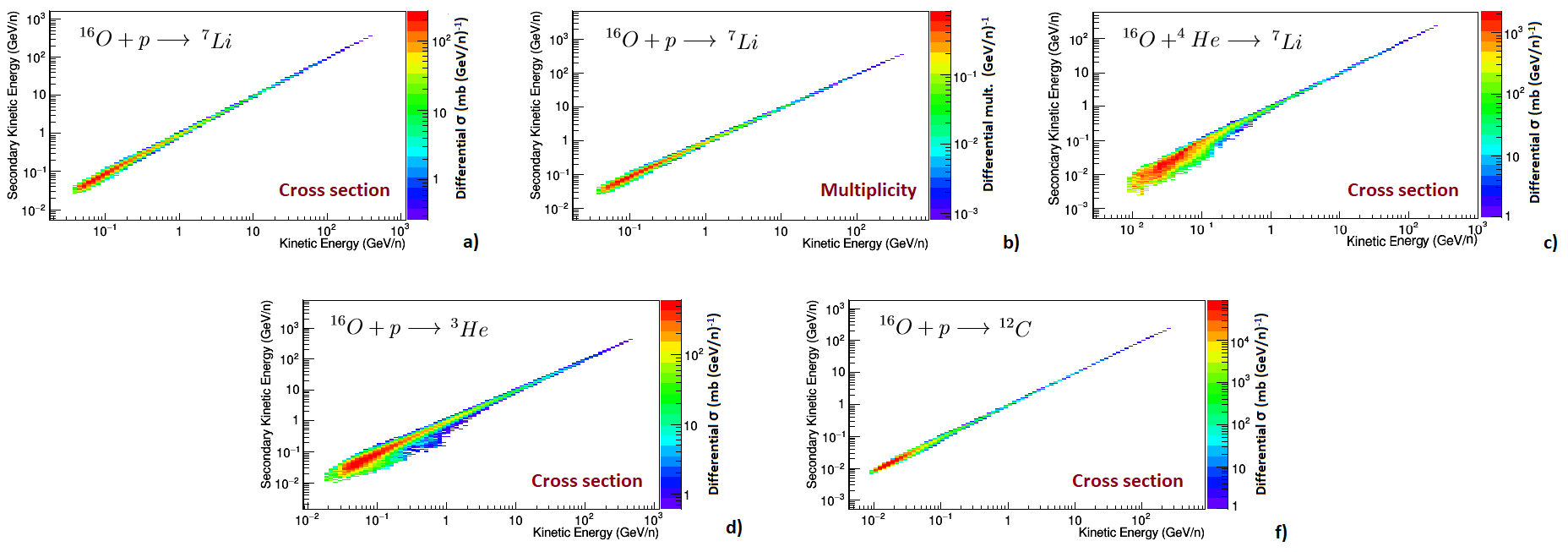}
\end{center}
\caption{\footnotesize Cross sections and multiplicity for some nuclear reactions in FLUKA as a function of the primary nucleus kinetic energy and of the secondary kinetic energy: a) cross section of the $^{16}O+p \rightarrow ^{7}$Li reaction; b) multiplicity of $^{7}$Li nuclei in the same reaction; c) cross section of the $^{16}$O$+He \rightarrow ^{7}$Li reaction; d) cross section of the $^{16}$O$+p \rightarrow ^{3}He$ reaction; e) cross section of the $^{16}$O$+p \rightarrow ^{12}$C.}
\label{fig:head_on}
\end{figure*}

As an example, some of the meaningful results for the reactions with $^{16}$O as projectile are shown in Figure~\ref{fig:head_on}. First of all, we see that the multiplicity and the spallation cross sections for all the reactions roughly follow the same behaviour, since the inelastic cross sections are roughly constant at high energies (mainly for heavy species), as it is shown in Figure 1 of \cite{Mazziot} and in Figure~\ref{fig:ineplots}. This means that the spallation cross sections at middle-high energies will follow the same behaviour as multiplicity. In addition, we see that the relation between the energy per nucleon of the projectile and that of the secondary particle is roughly linear with slope close to 1, as predicted by the head-on approximation, usually applied in CR propagation codes. Nevertheless, deviations from this linear relation are found at low energies (mainly under $100 \units{MeV/n}$), and this effect becomes relevant for the production of very light nuclei (as seen in Figure~\ref{fig:head_on}, d). To avoid these discrepancies at very low energies, the DRAGON code implements the proton-proton collisions and secondary protons depending on the energy of primary and secondary nuclei, as explained in \cite{Evoli:2017vim}. Furthermore, as it can be observed by comparing Figure~\ref{fig:head_on}, c with Figure~\ref{fig:head_on}, a, these discrepancies are larger in the interactions with helium target. This small caveat introduces some uncertainties in the low energy part of the spectrum of CRs, but it is negligible at middle-high energies for nuclei with $Z>2$ and it can be an issue to be taken into account when studying $^3He$, since it is a pure secondary CR particle (see Figure~\ref{fig:head_on}, d). 

%\newpage
\subsection{Inelastic and spallation cross sections}
\label{sec:FIne_Spa}

With the main features of the nucleus-nucleus interactions (thus of CR particles) already pointed out, the next step is the validation of both, inelastic and spallation cross sections.

\begin{figure*}[!htb]
\begin{center}
\footnotesize{a)}\includegraphics[width=0.47\textwidth,height=0.19\textheight,clip] {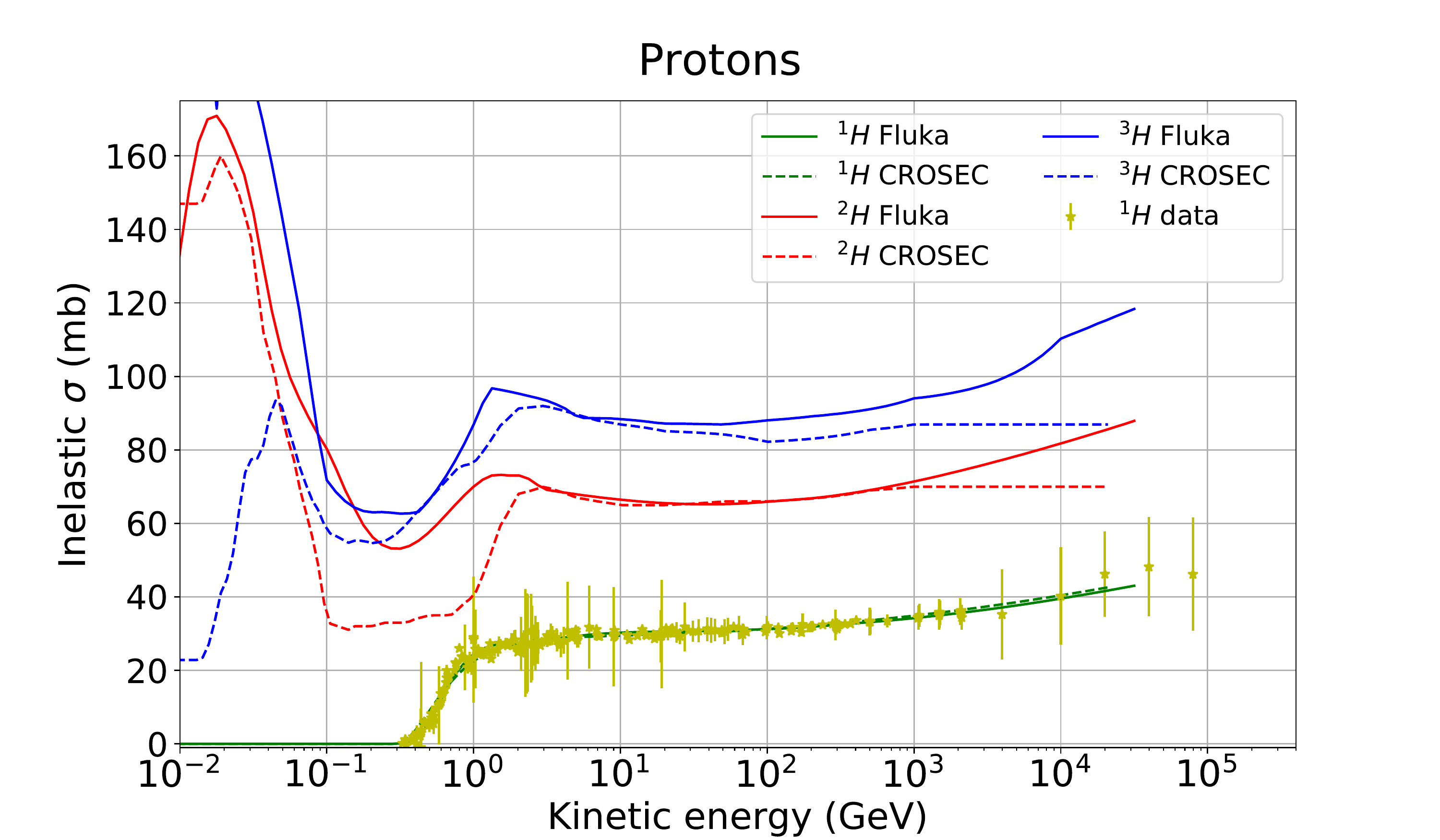} %\hspace{0.4cm}
\footnotesize{b)}\includegraphics[width=0.47\textwidth,height=0.19\textheight,clip] {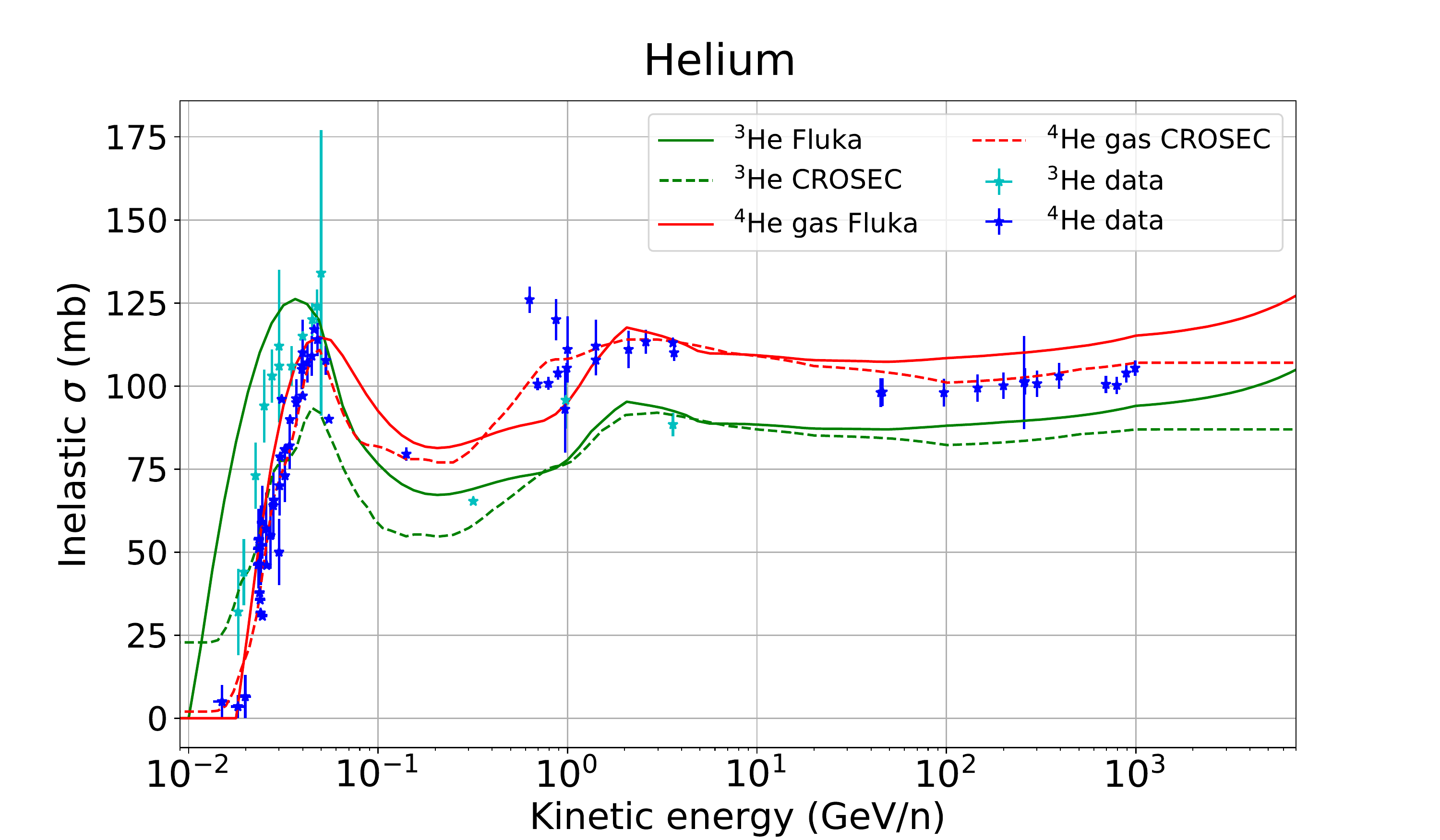}
%\vspace{0.5cm}
\footnotesize{c)}\includegraphics[width=0.47\textwidth,height=0.19\textheight,clip] {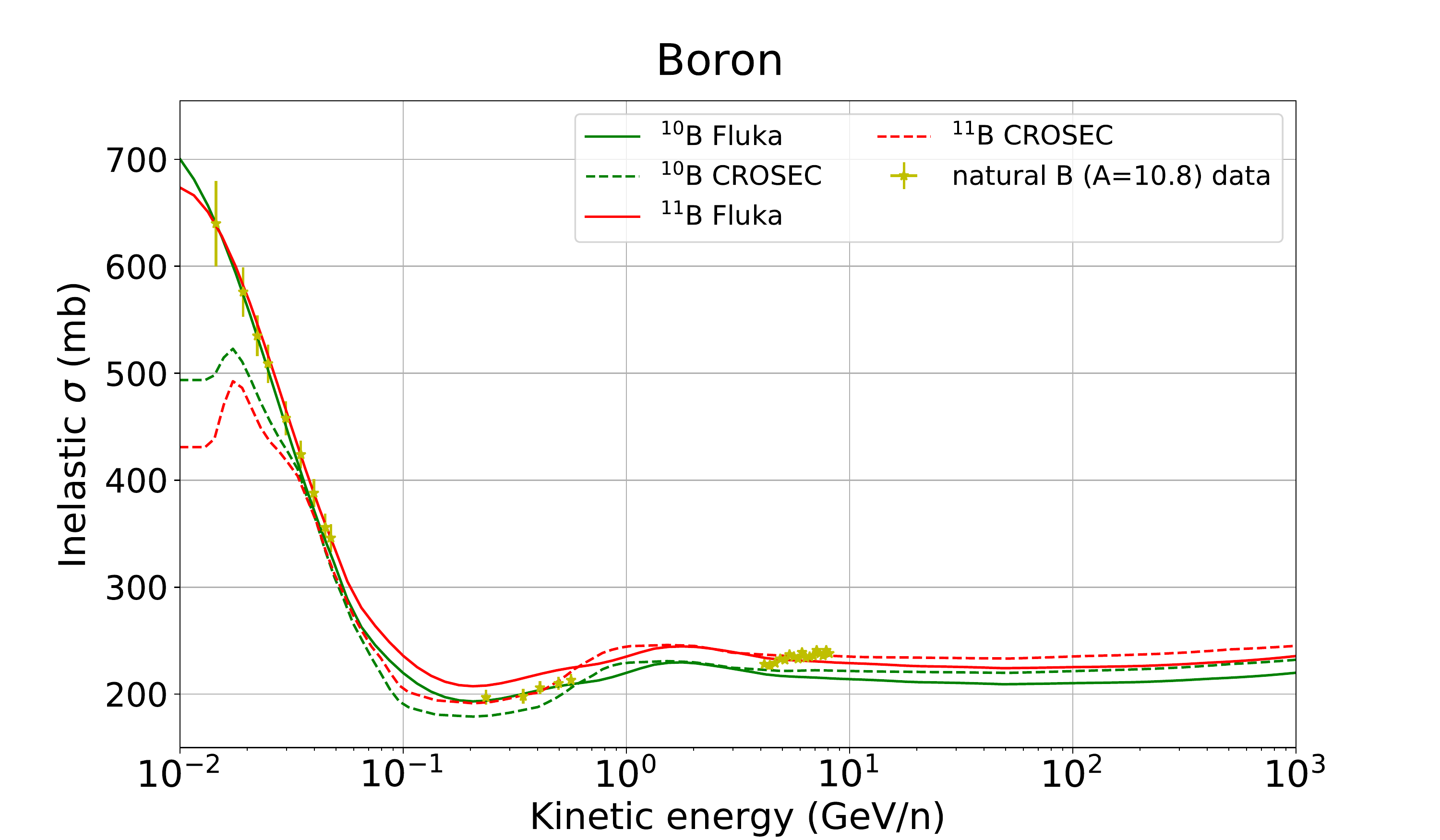} %\hspace{0.4cm}
\footnotesize{d)}\includegraphics[width=0.47\textwidth,height=0.19\textheight,clip] {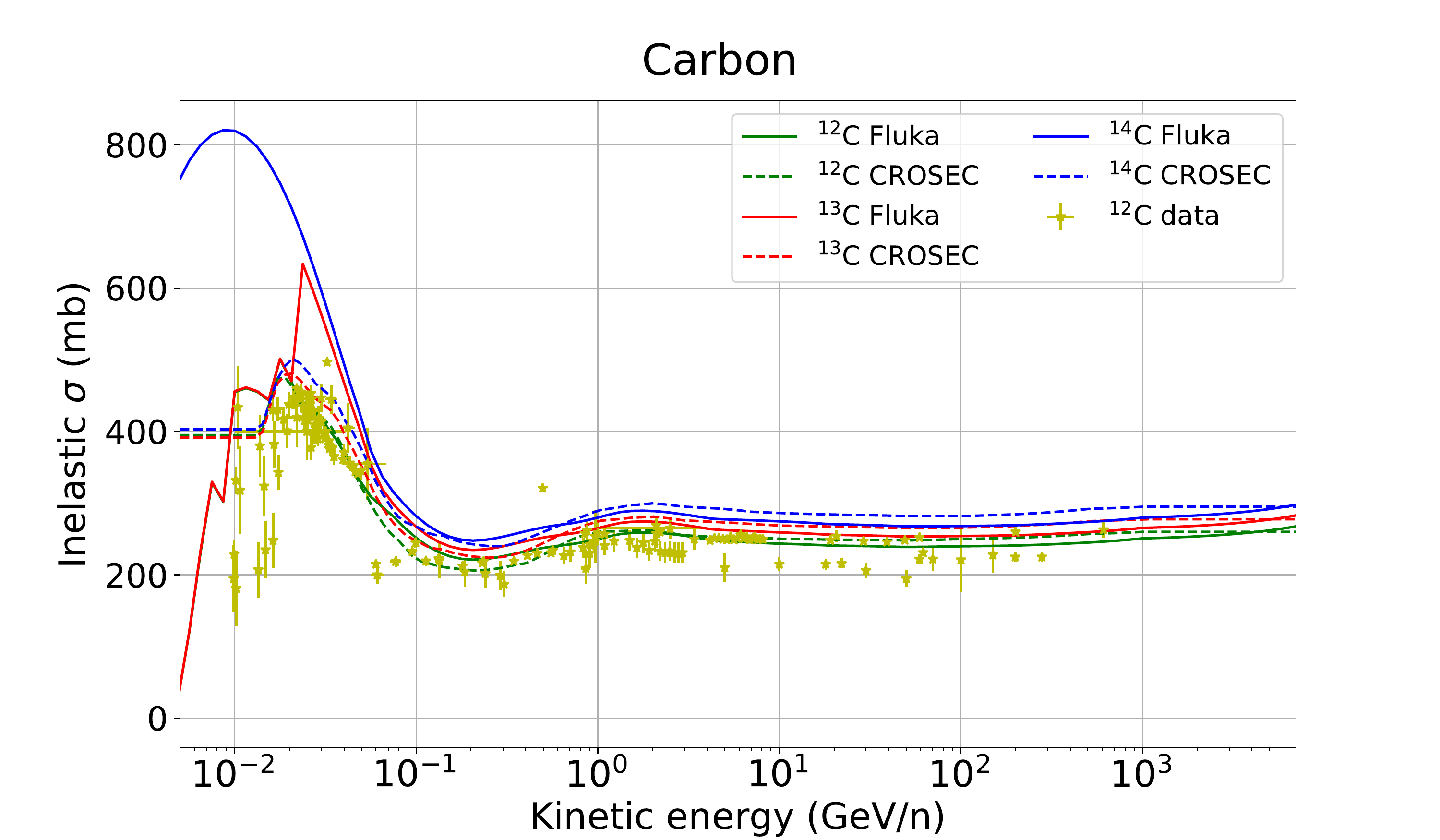}
%\vspace{0.5cm}
\footnotesize{e)}\includegraphics[width=0.47\textwidth,height=0.19\textheight,clip] {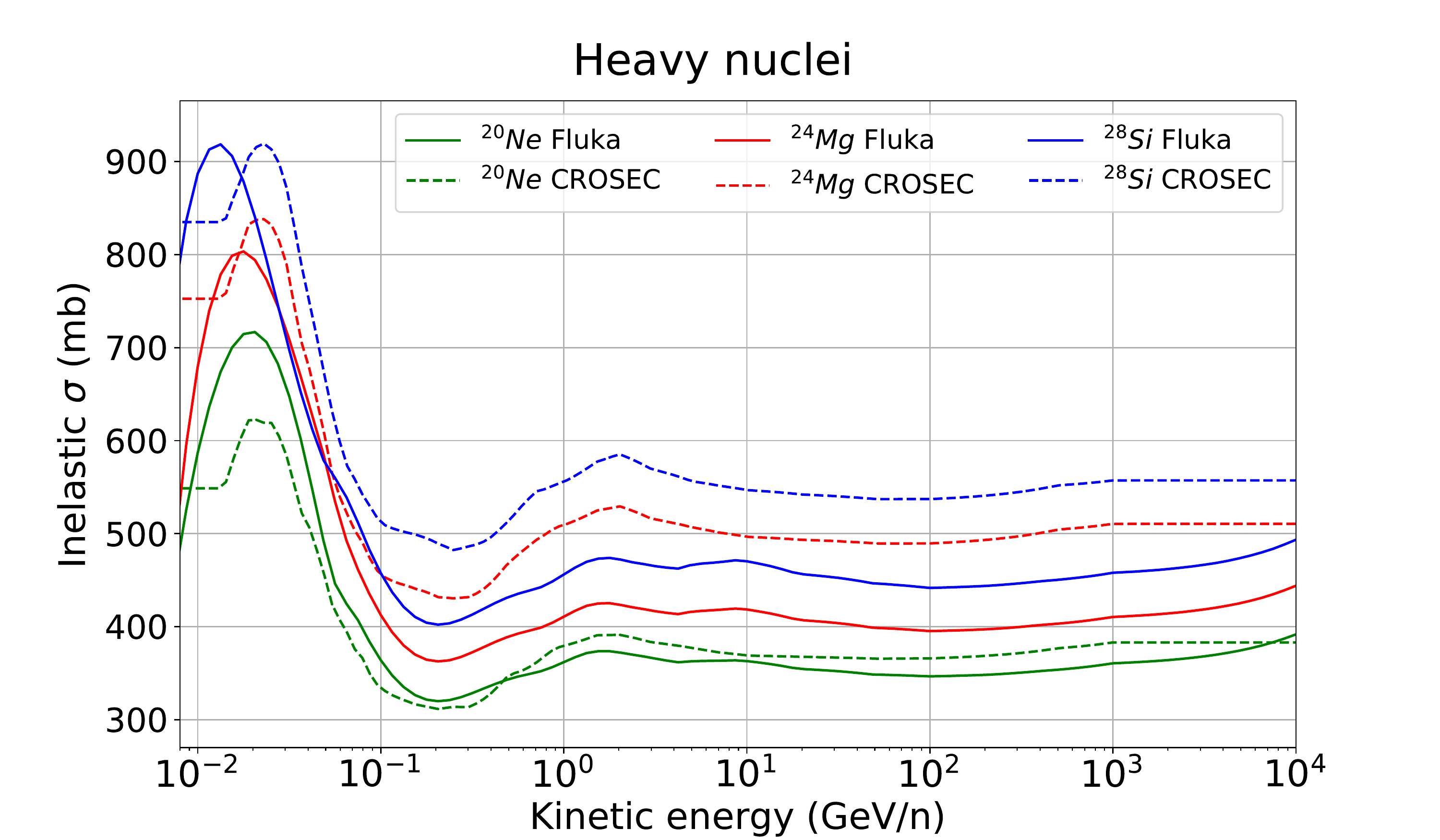} %\hspace{0.4cm}
\footnotesize{f)}\includegraphics[width=0.47\textwidth,height=0.19\textheight,clip] {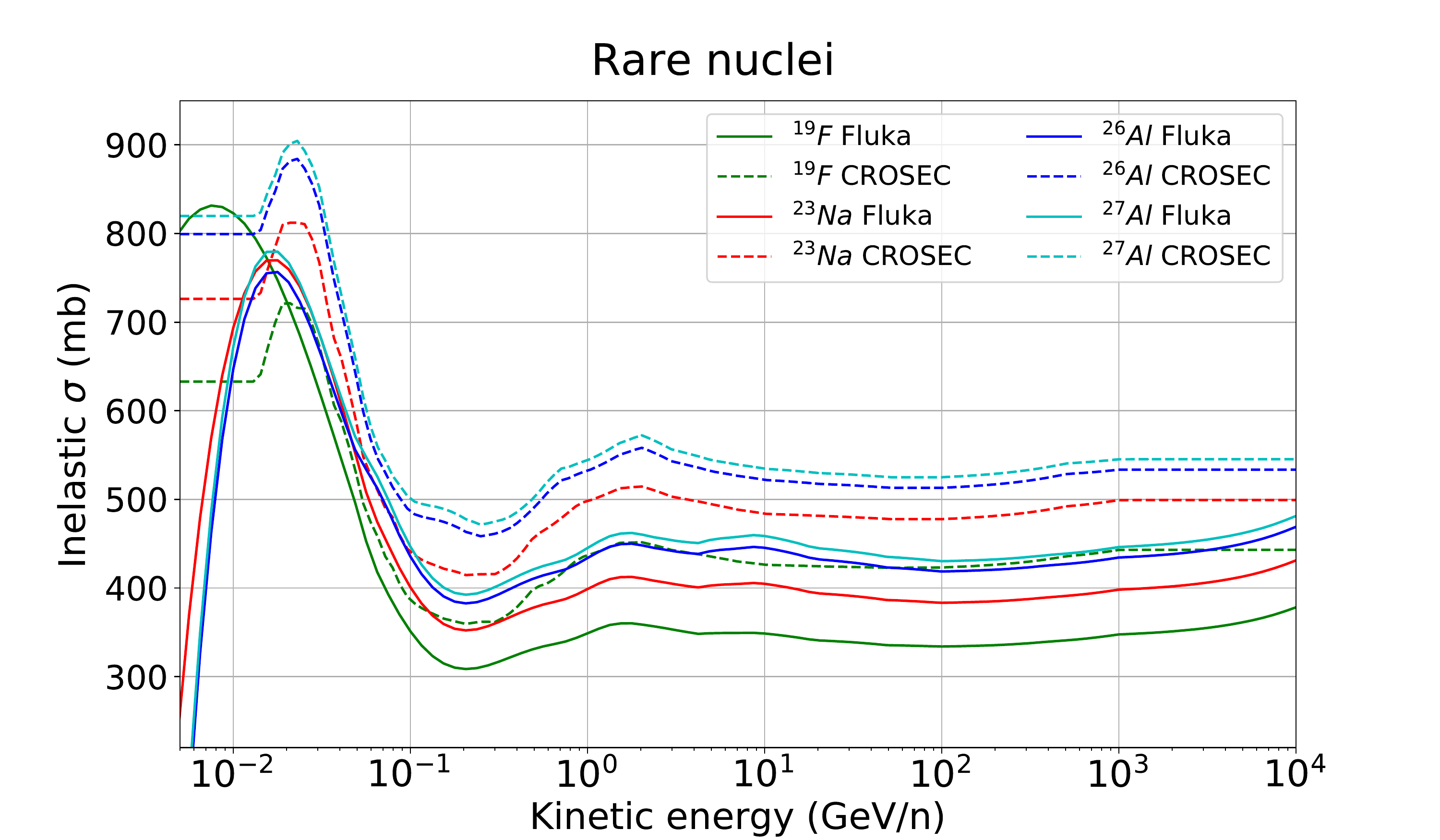}
\end{center}
\caption{\footnotesize Comparison of the CROSEC and FLUKA inelastic cross sections on hydrogen target for diverse species of projectiles relevant for CR physics: a) protons; b) helium; c) boron; d) carbon; e) other heavy nuclei relevant on CR physics (Ne, Mg and Si); f) less common nuclei, or rare nuclei, for CR studies ($^{19}F$, $^{23}Na$, $^{26}Al$, $^{27}Al$). Experimental data are also shown when they are available. Proton-induced reactions data are taken from \cite{Bobchenko:1979hp} and are assembled in tables in \url{http://www.oecd-nea.org/dbdata/bara.html}. Data for heavier nuclei come from a compilation using the EXFOR database \cite{OTUKA2014272, Zerkin_2018}. Full references can be found in \cite{Genoliniranking}.}
\label{fig:ineplots}
\end{figure*}

Some of the most important inelastic cross sections, for hydrogen as target, computed with FLUKA are shown in Figure~\ref{fig:ineplots}. It shows the proton, helium (as examples of light primary nuclei), carbon (as an example of heavy primary nuclei), boron (as an example of secondary nuclei) and other heavy elements (as a sample of other CR nuclei with very few measurements available) inelastic cross sections calculated with FLUKA and with the CROSEC code as a function of energy compared to experimental data. For protons, FLUKA adopts the parametrisation found in \cite{Kafexhiu:2014cua}, which is also the parametrisation used in DRAGON2. The CROSEC code (also called CRN6) was initially built for inelastic cross sections of heavy ions, then expanded to lighter nuclei, and it is used as the default option for inelastic collisions in DRAGON (except for proton-proton inelastic collisions). Notice that in the case of boron, the experimental data correspond to the isotopic composition (with $A\sim 10.8$). Comparisons for the primary CRs $^{20}$Ne, $^{24}$Mg and $^{28}$Si are shown to emphasize the differences found between both computations in channels where the amount of experimental data is very scarce. In fact, as in the case of the spallation cross sections, parametrisations for inelastic reaction channels with few or no data (as in Figure~\ref{fig:ineplots}, f) are just extrapolations of the better measured channels, which may significantly bias the final results. Channels above silicon turn out to be not very important in CR studies: they have an impact on the production of light secondary CRs $< 4\%$ for Be and Li, and $<1\%$ for B.

This comparison shows that both approaches exhibit a very similar behaviour and, in general, give good agreement with experimental data. However, it seems that the resonances, located at low energies, slightly favour the FLUKA computations. This is due to the fact that typically, there have been more measurements performed above $1 \units{GeV/n}$ (an issue already stated in the appendix E of \cite{Genoliniranking}). Another  main difference is that the FLUKA inelastic cross sections take into account the rise in the total and elastic nucleus-nucleus cross sections experimentally observed \cite{RiseofXS, Block_2011}, whereas most of the parametrisations of inelastic and total cross sections consider an energy-independent behaviour above a few$\units{GeV/n}$. 

In the case of heavier species (Figure~\ref{fig:ineplots}, e, f), discrepancies around 10-20 \% are found between the two computations at atomic numbers above that of Ne ($Z \geq 10$). For all these heavy nuclei, the energy dependence of their cross sections is very similar. The lack of data in these channels makes difficult to decide which computation is more reliable, but, in general, it seems that the inelastic cross sections calculated with FLUKA are at the level of precision of the most updated semi-analytical parametrisations used by the CR community, as the CROSEC ones. 

\begin{figure*}[!hbt]
\begin{center}
%\vspace{0.5cm}
\textbf{\underline{Direct spallation cross sections of reactions with hydrogen}} \\ \vspace{0.5 cm}
\includegraphics[width=0.325\textwidth,height=0.17\textheight,clip] {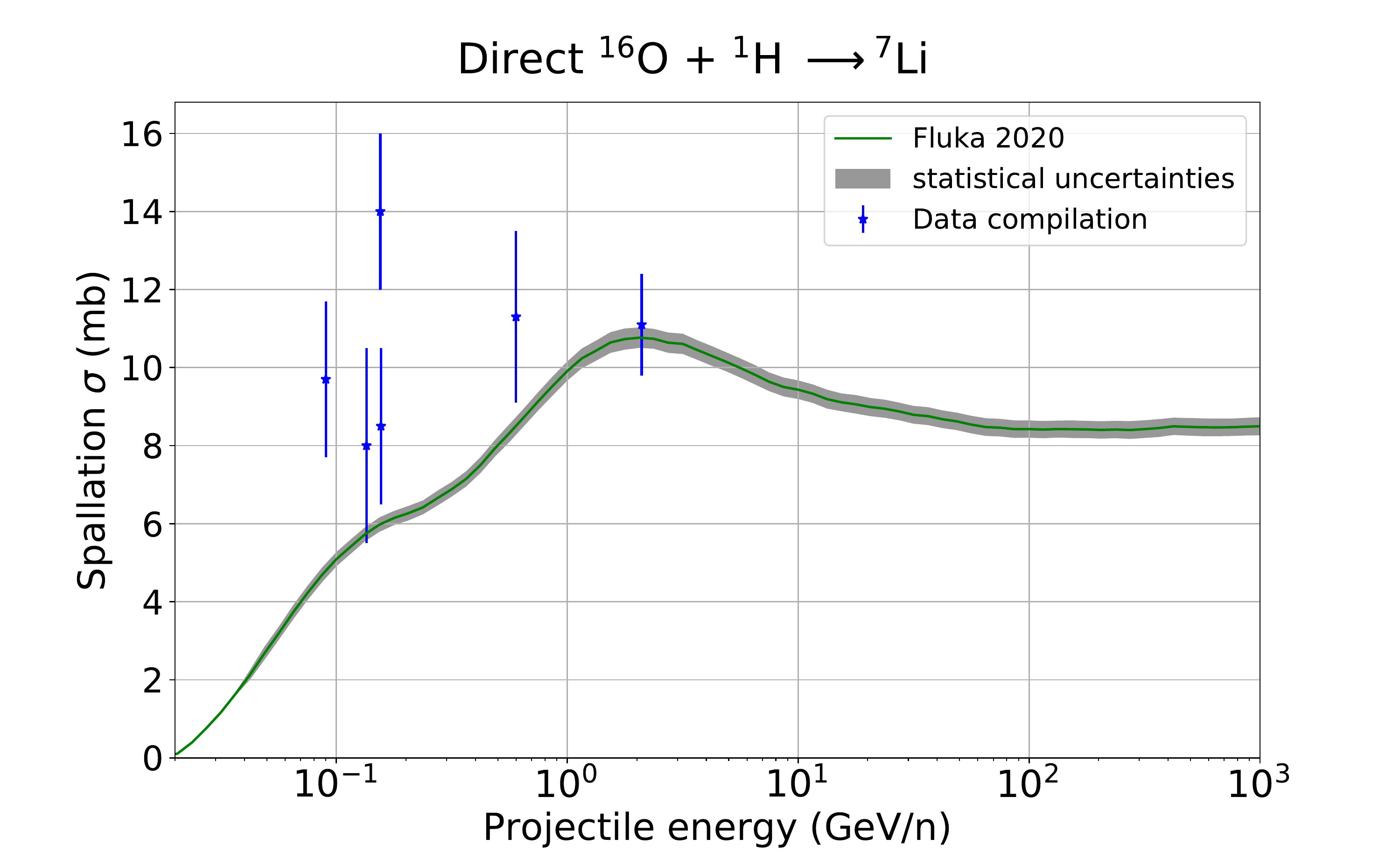}
\includegraphics[width=0.325\textwidth,height=0.17\textheight,clip] {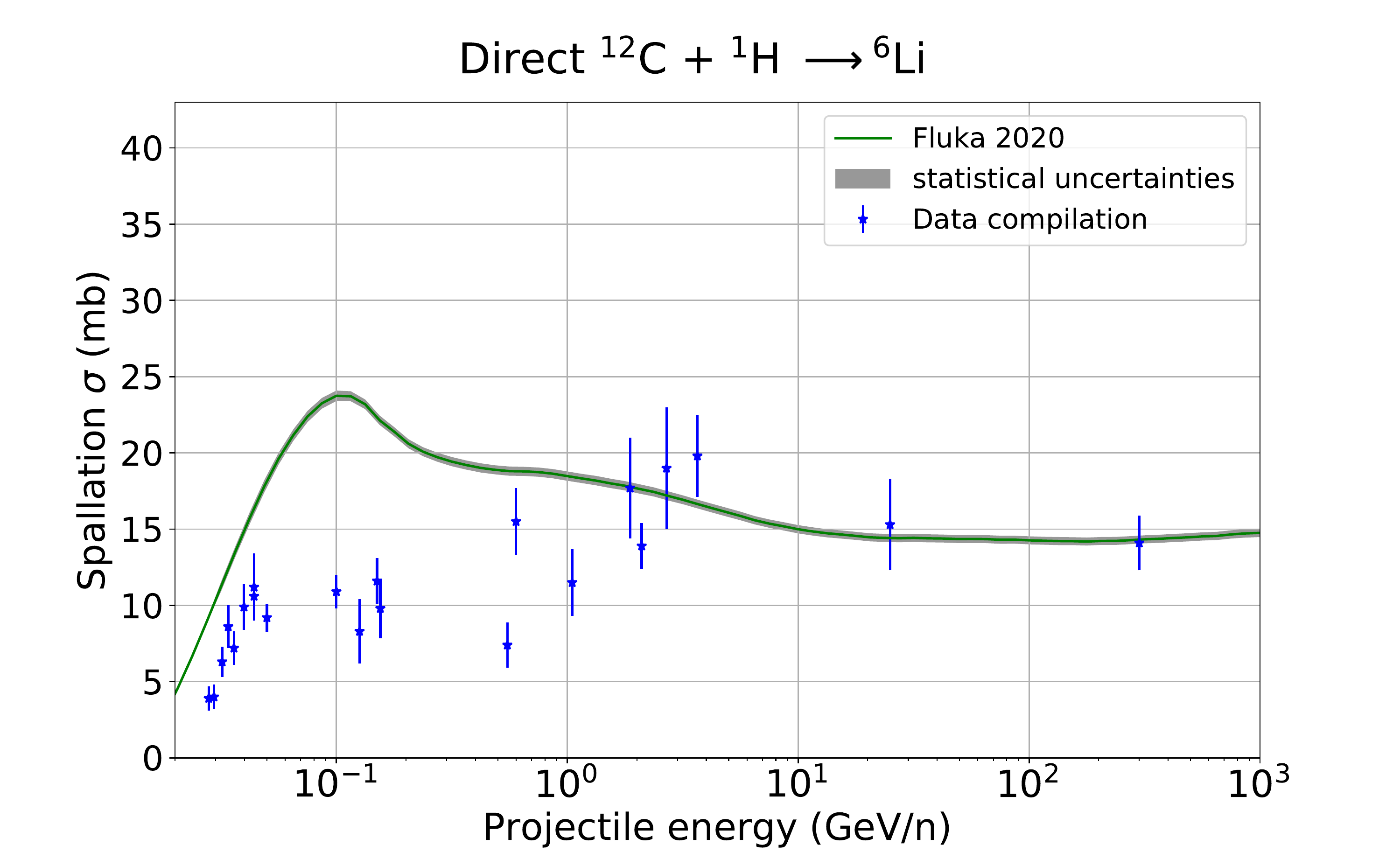}
\includegraphics[width=0.325\textwidth,height=0.17\textheight,clip] {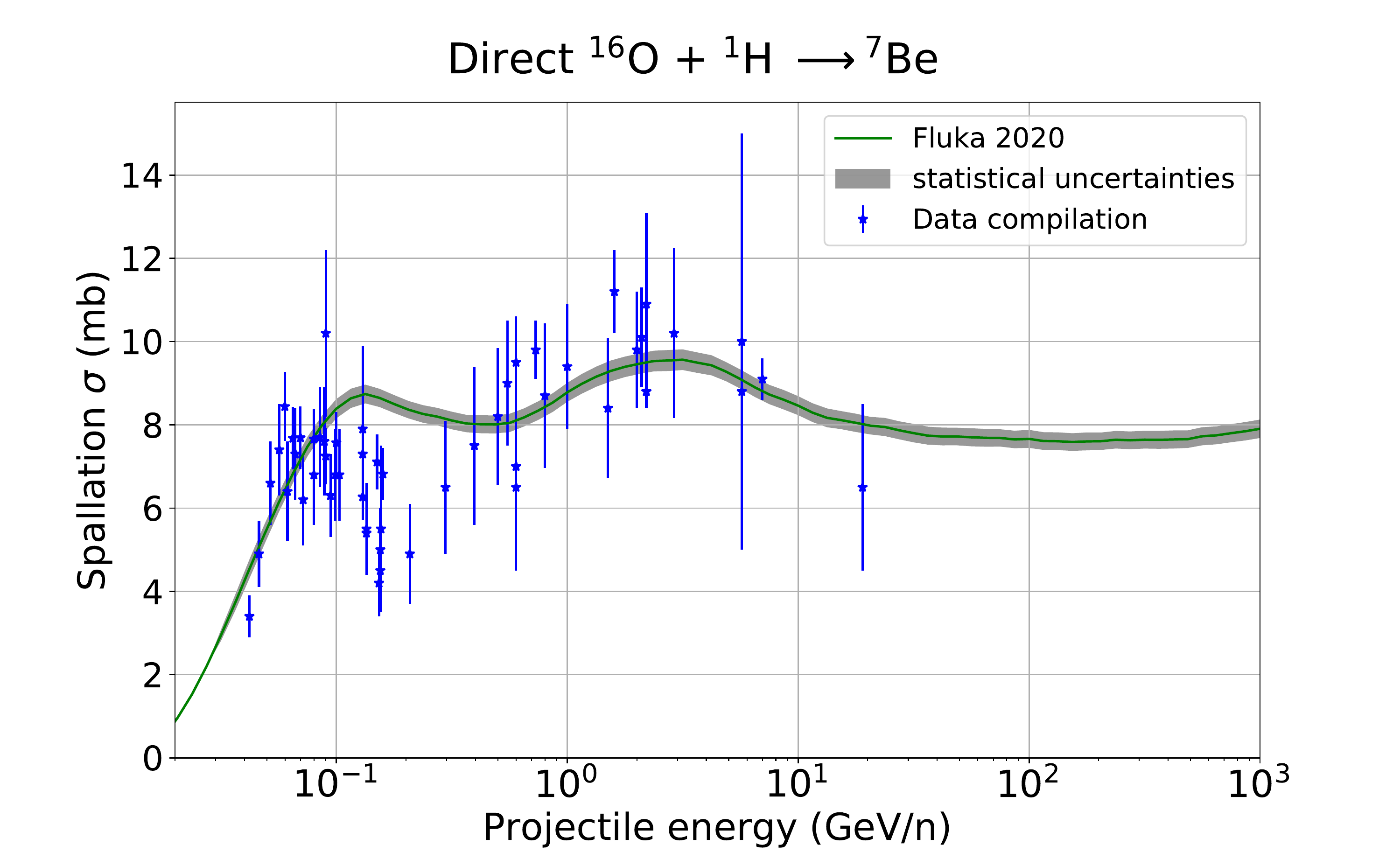} 

\includegraphics[width=0.325\textwidth,height=0.17\textheight,clip] {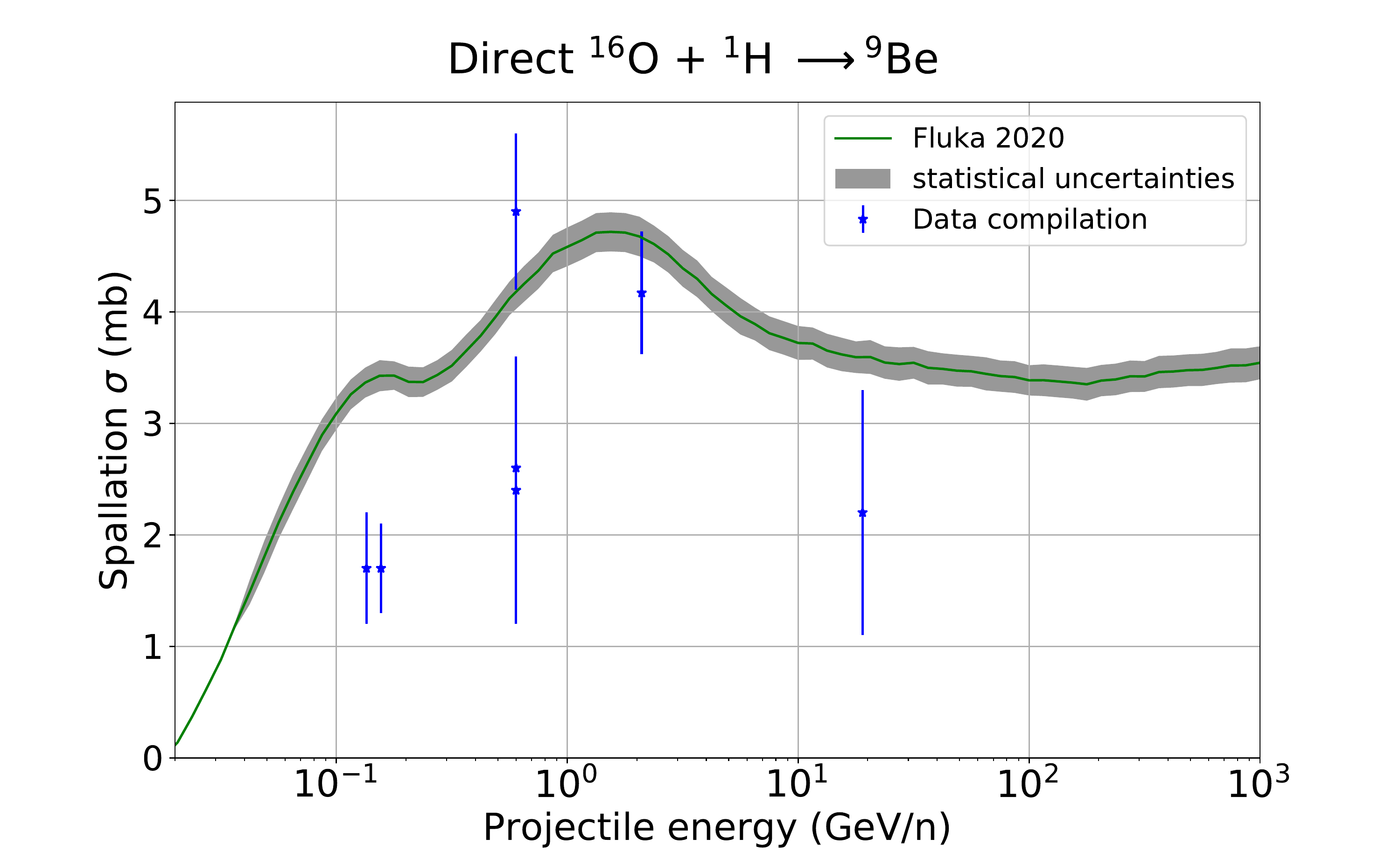} 
\includegraphics[width=0.325\textwidth,height=0.17\textheight,clip] {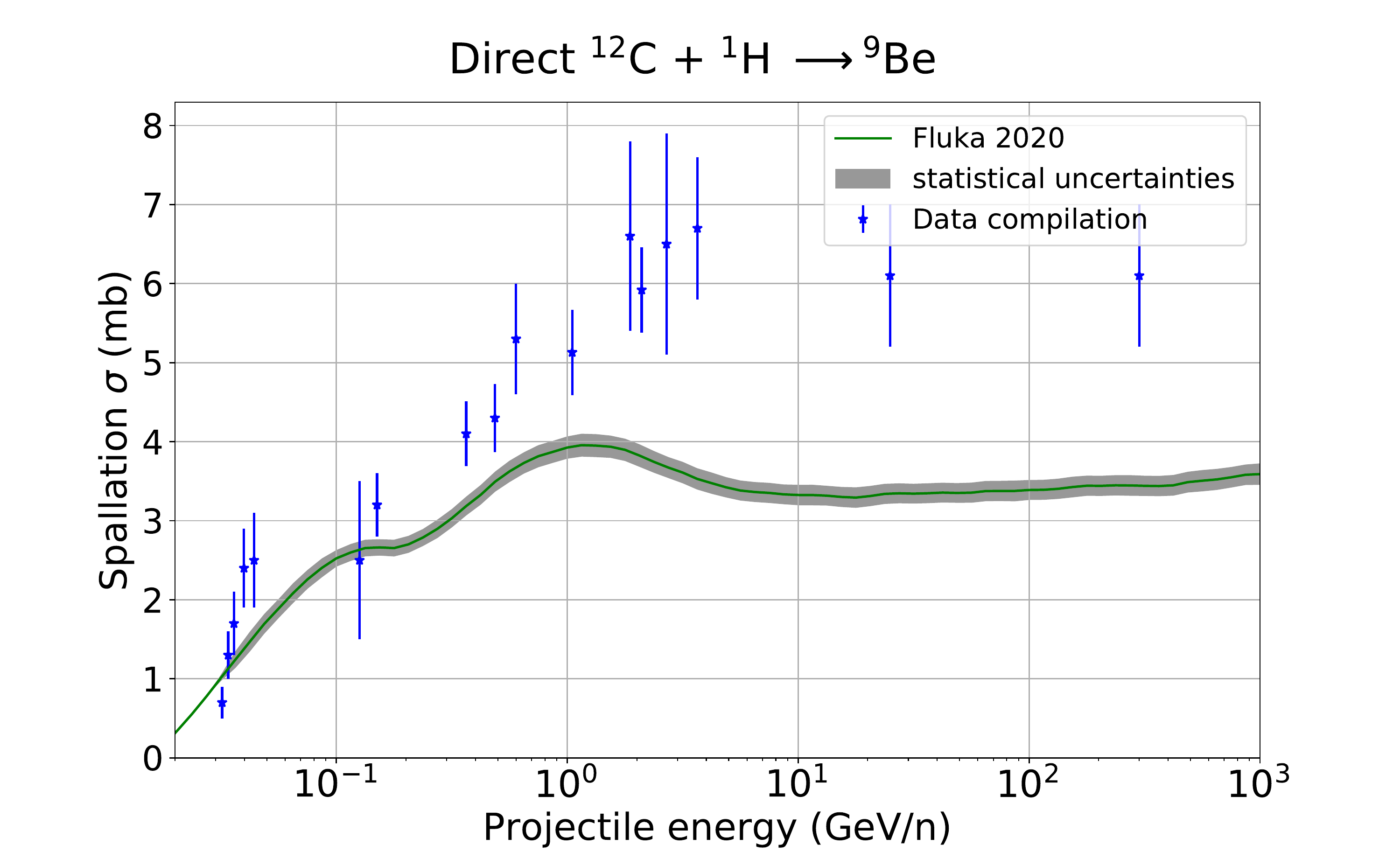}
\includegraphics[width=0.325\textwidth,height=0.17\textheight,clip] {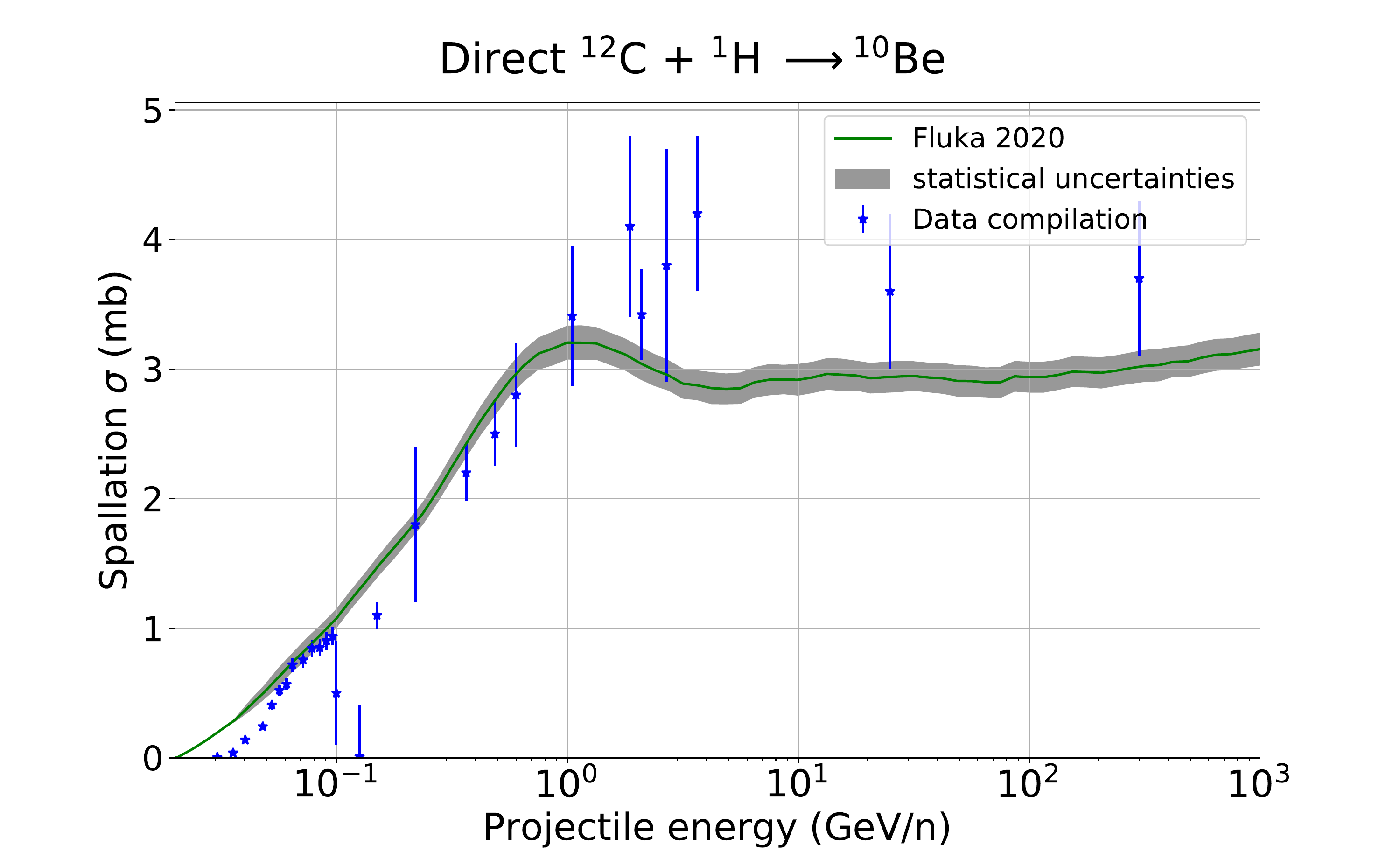}

\includegraphics[width=0.325\textwidth,height=0.17\textheight,clip] {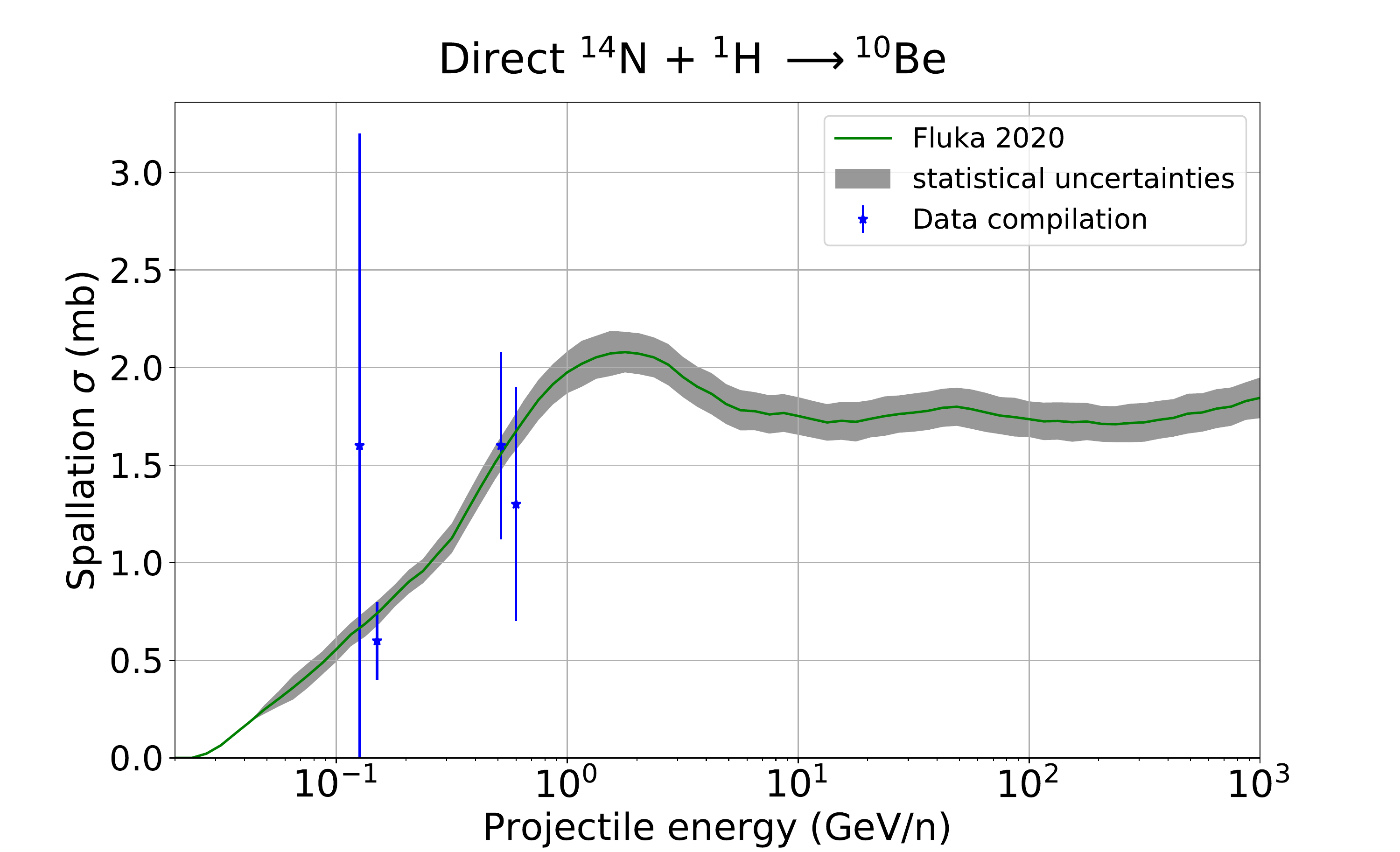}
\includegraphics[width=0.325\textwidth,height=0.17\textheight,clip] {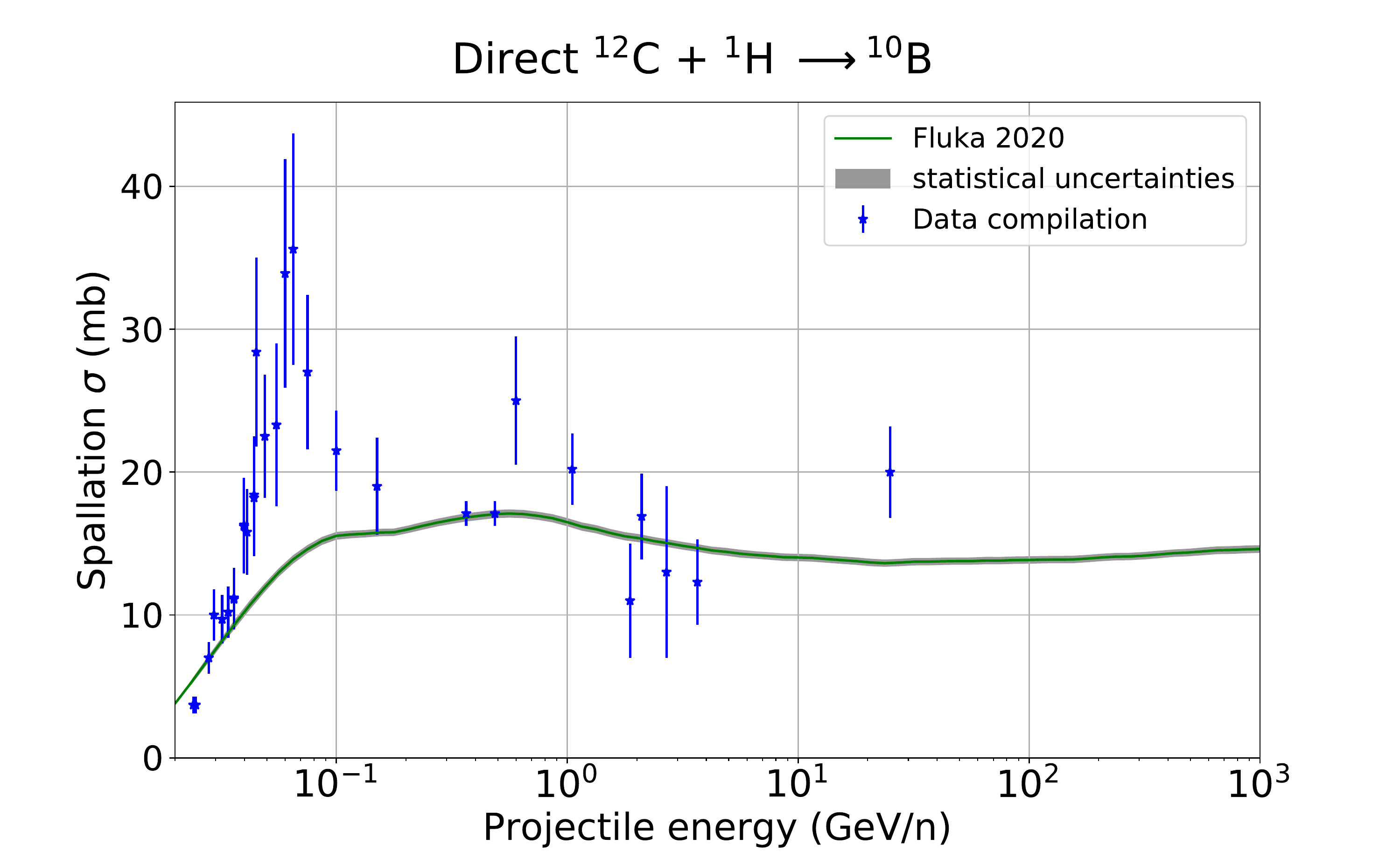}
\includegraphics[width=0.325\textwidth,height=0.17\textheight,clip] {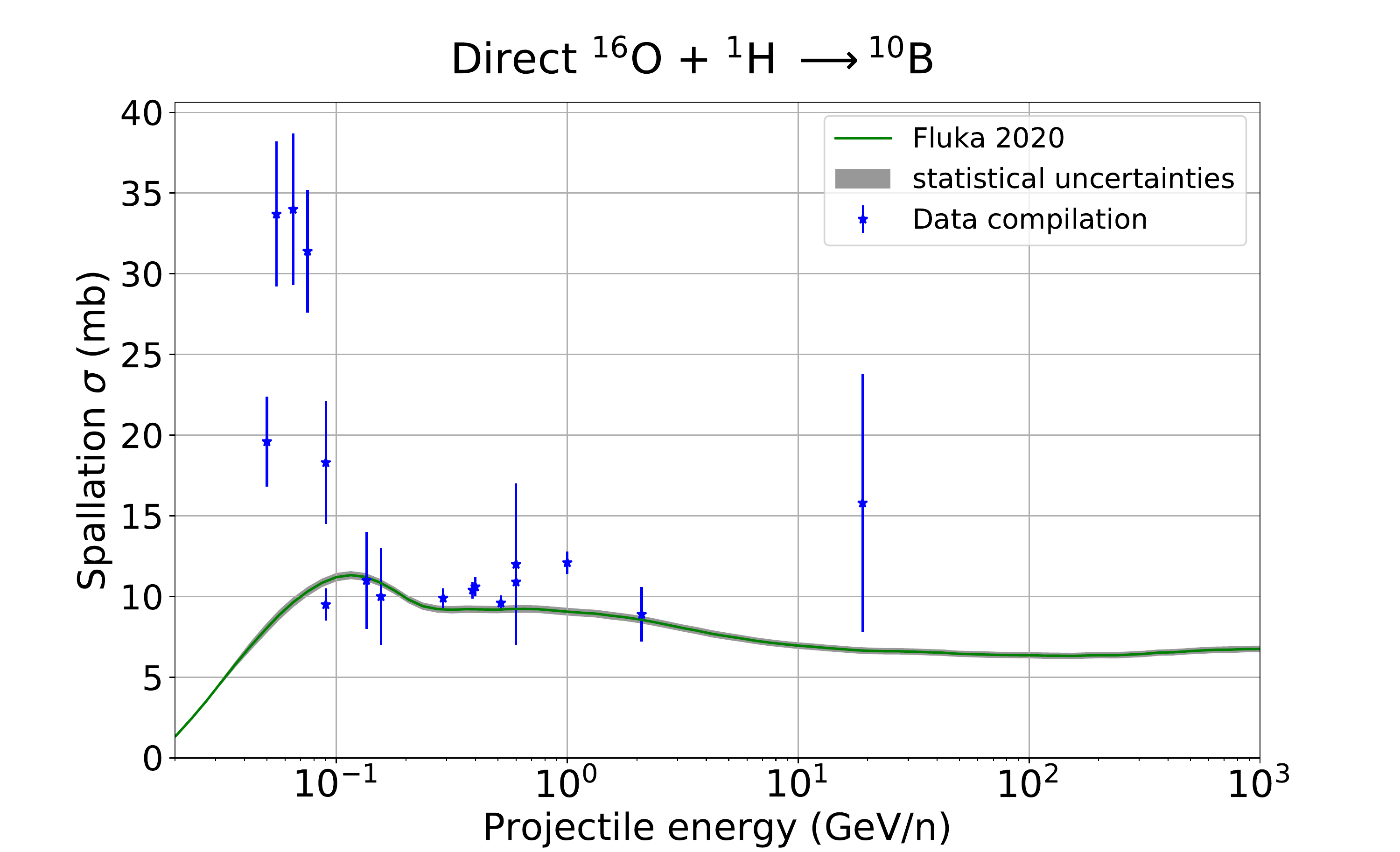}

\includegraphics[width=0.325\textwidth,height=0.17\textheight,clip] {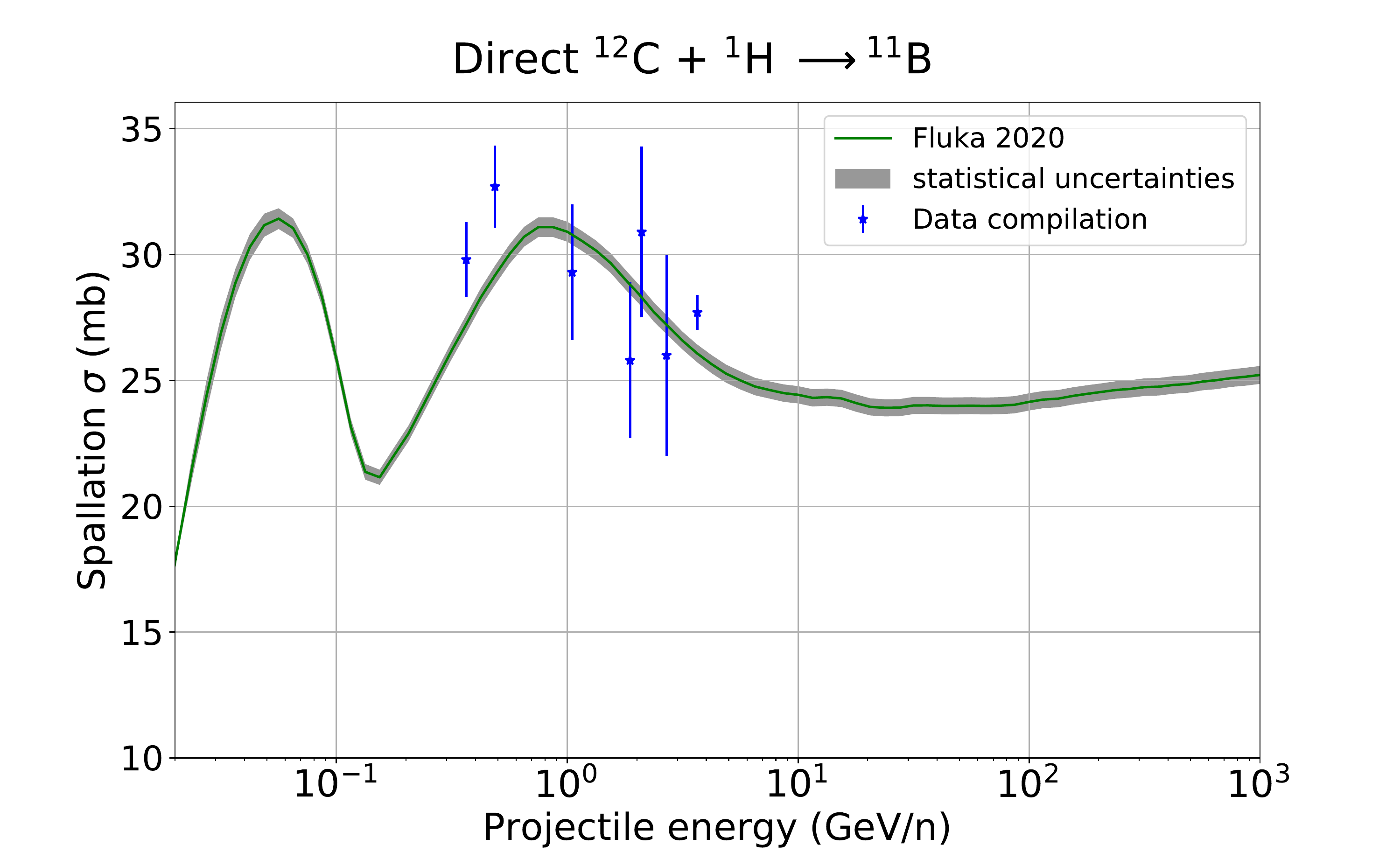}
\includegraphics[width=0.325\textwidth,height=0.17\textheight,clip] {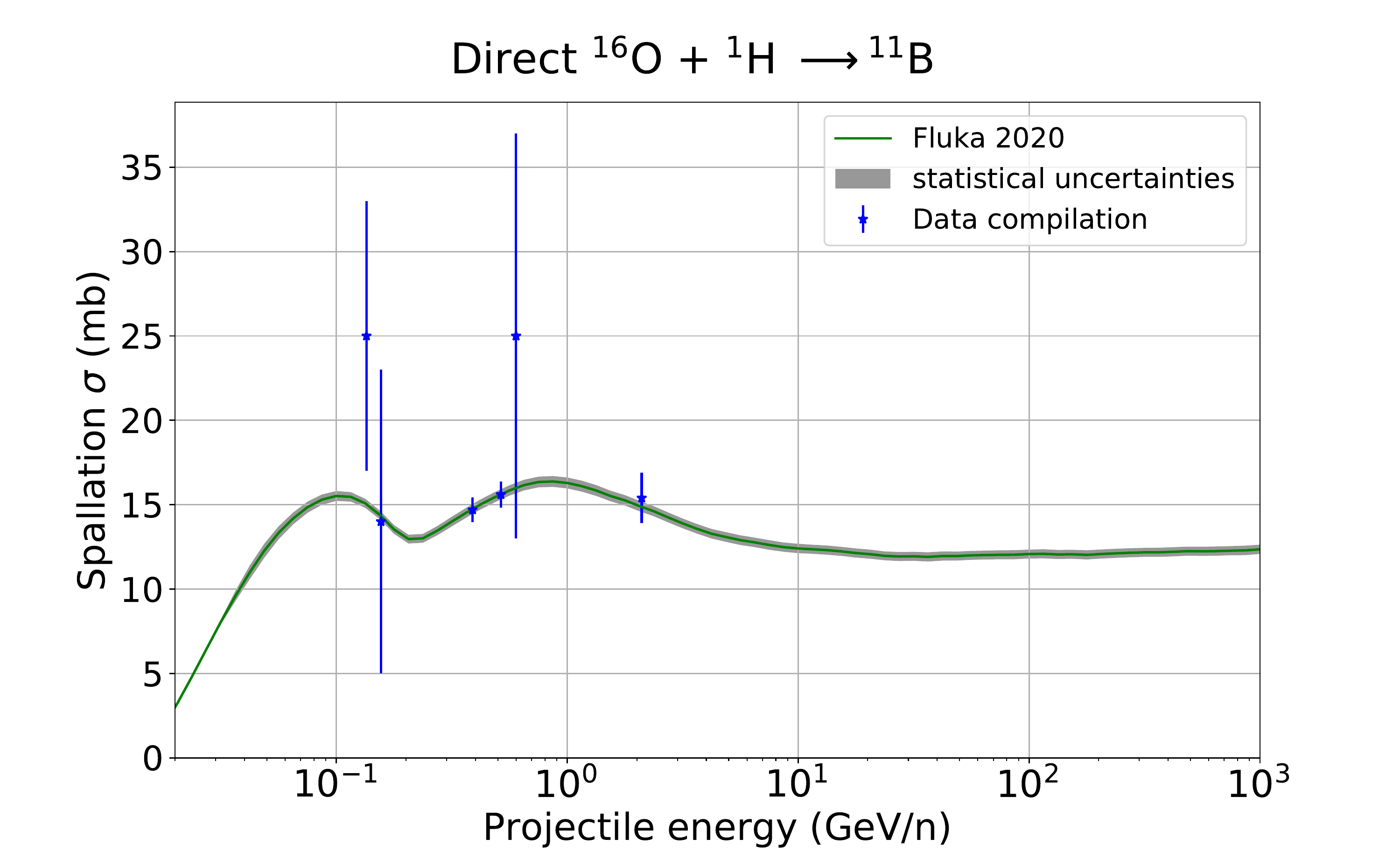}
\includegraphics[width=0.325\textwidth,height=0.17\textheight,clip] {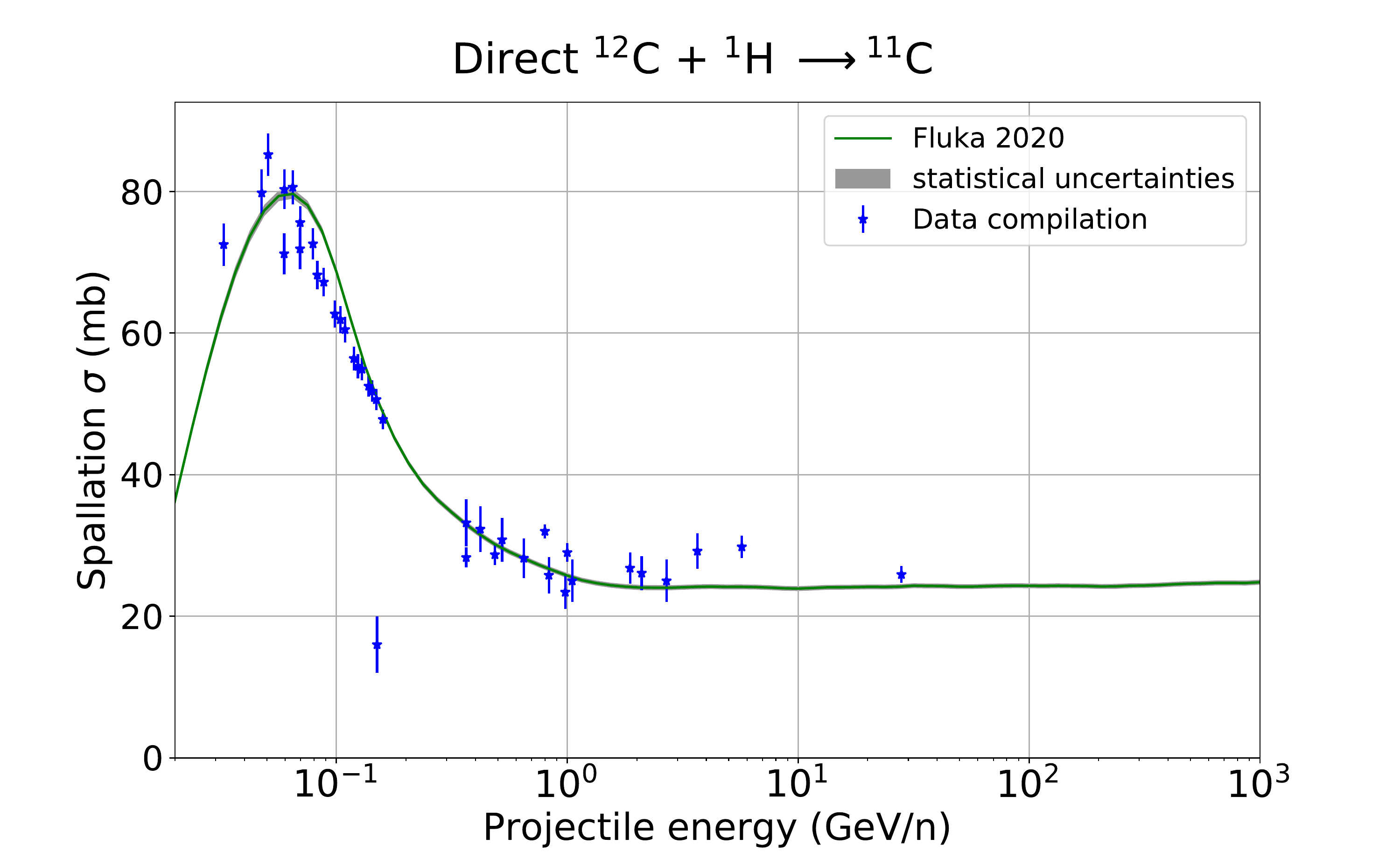}
%\vspace{0.5cm}

\end{center}
\caption{\footnotesize Some of the most important inclusive cross sections of nuclear reactions on hydrogen target calculated with the FLUKA code and compared with experimental data (see appendix \ref{sec:appendixA} for the references).}
\label{fig:direct_hyd}
\end{figure*}

On the other hand, the case of spallation cross sections is more delicate, as previously discussed, due to their importance in the production of secondary CRs and to the uncertainties related to them. Furthermore, here we must take into account the ghost nuclei too. Figure~\ref{fig:direct_hyd} shows a comparison of experimental spallation cross section data to the cross sections calculated with FLUKA for the direct reaction of hydrogen in some of the more important channels. The cross sections calculated from FLUKA have associated a small statistical uncertainty, since it is a Monte Carlo based computation. In this figure, a sample of some of the most representative reactions and channels with more experimental data points are displayed. In general, they seem to be consistent with the experimental measurements, while there are some discrepancies that is worth pointing out (that are found for all the channels of reaction, not only the ones displayed): 

\begin{itemize}
    \item There is no systematic bias observed in the different channels. In other words, the discrepancies do not exhibit any common trend with respect to data in all the channels.
    \item The most significant discrepancies are found in the channels of Be production, specially $^9$Be, as we can see from the $^{12}$C$ \longrightarrow$ $^{9}$Be reaction. On the other hand, the isotope $^{10}$Be shows agreement with data in general. Nevertheless, the large error bars in the data for the $^{10}$Be and $^{9}$Be production makes difficult to make precise statements, which is a problem when using these isotopes to predict the halo size.
    \item Resonances in the production of $^{10}$B are not accounted for by the FLUKA calculation, although this only matters for very low energy (in the range of energy of the Voyager-1 data). Nevertheless, other resonances are, in general, well accounted for, as it can be seen in the $^{12}$C $\longrightarrow$ $^{11}$C reaction.
    \item The scarcity of data at very high energies makes difficult to verify from these spectra whether the expected logarithmic rise of cross sections is really needed or the energy-independent shape of the cross sections (usually adopted in parametrisations) is adequate.
\end{itemize}

%\newpage

\begin{figure*}[!hb]
\begin{center}

%\vspace{0.5cm}
\includegraphics[width=0.40\textwidth,height=0.17\textheight,clip] {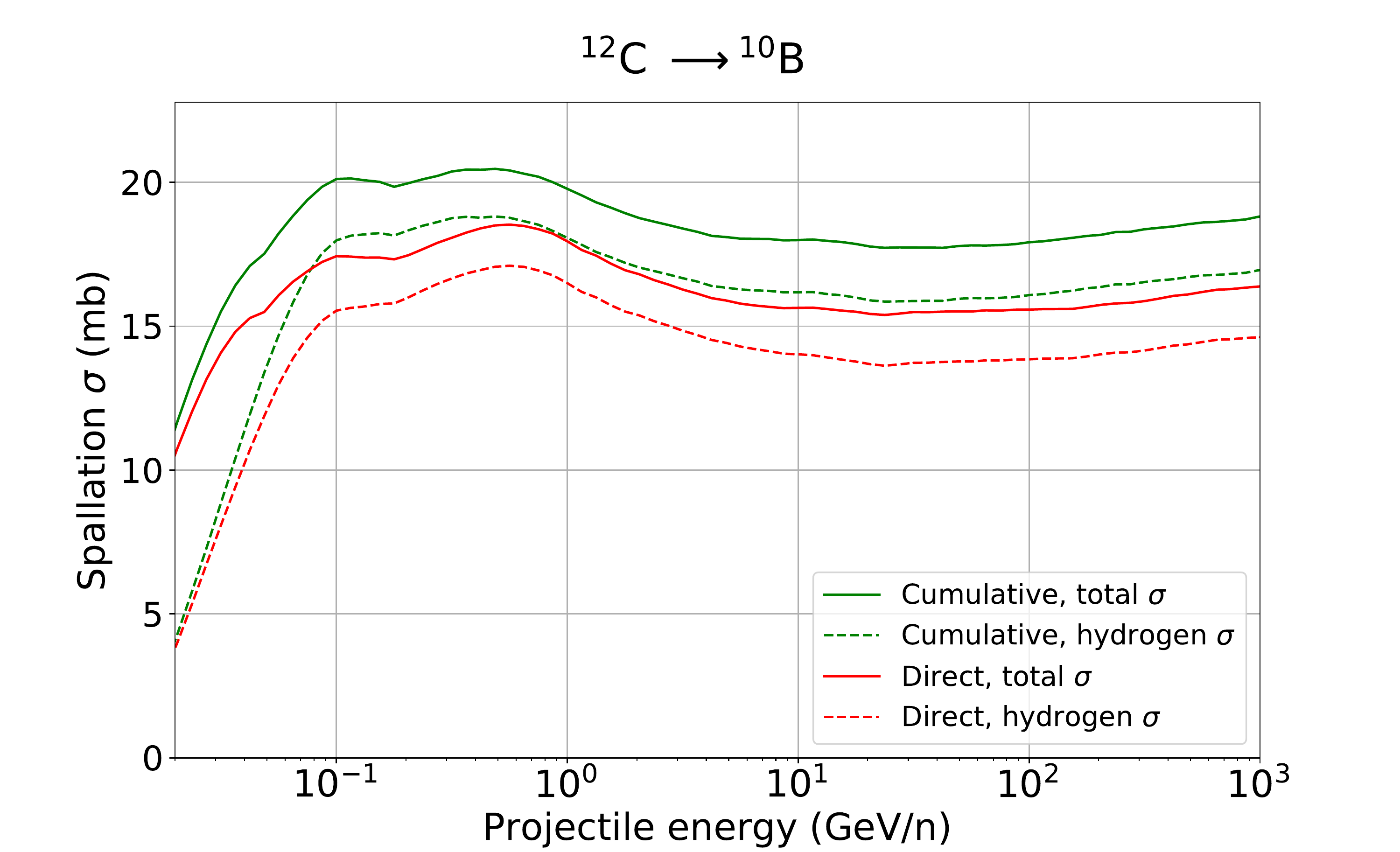}
\includegraphics[width=0.40\textwidth,height=0.17\textheight,clip] {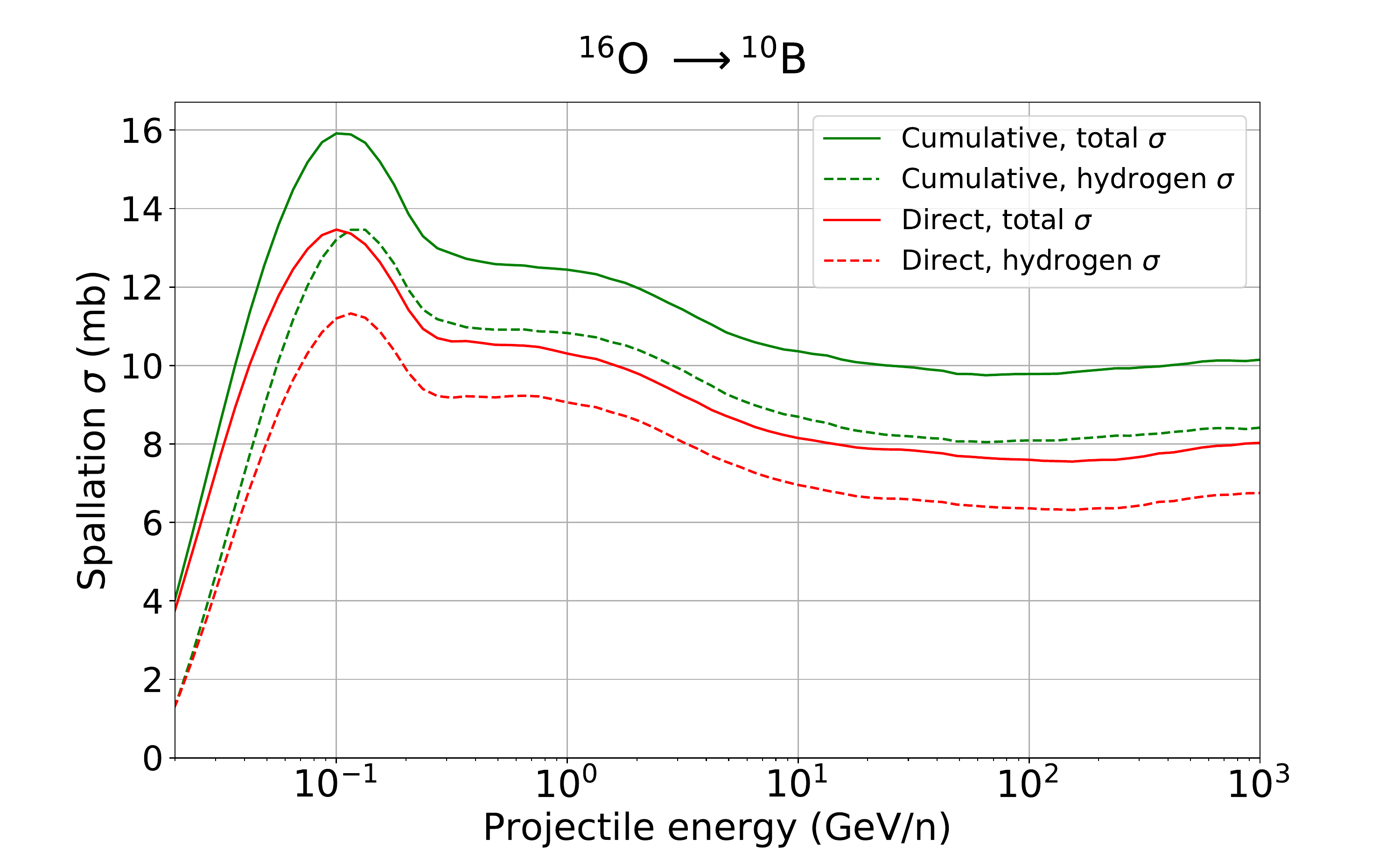}

\includegraphics[width=0.325\textwidth,height=0.165\textheight,clip] {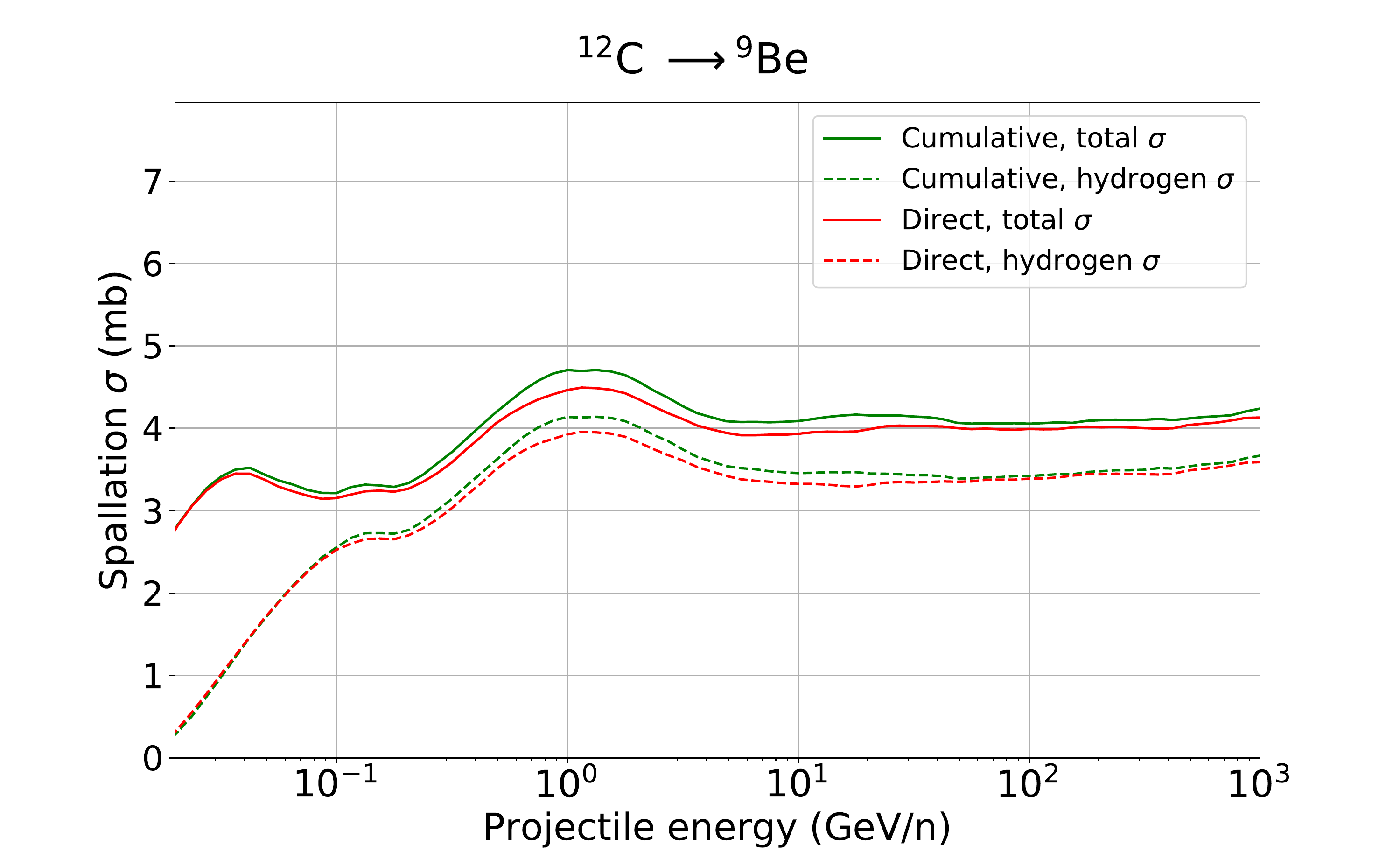} 
\includegraphics[width=0.325\textwidth,height=0.165\textheight,clip] {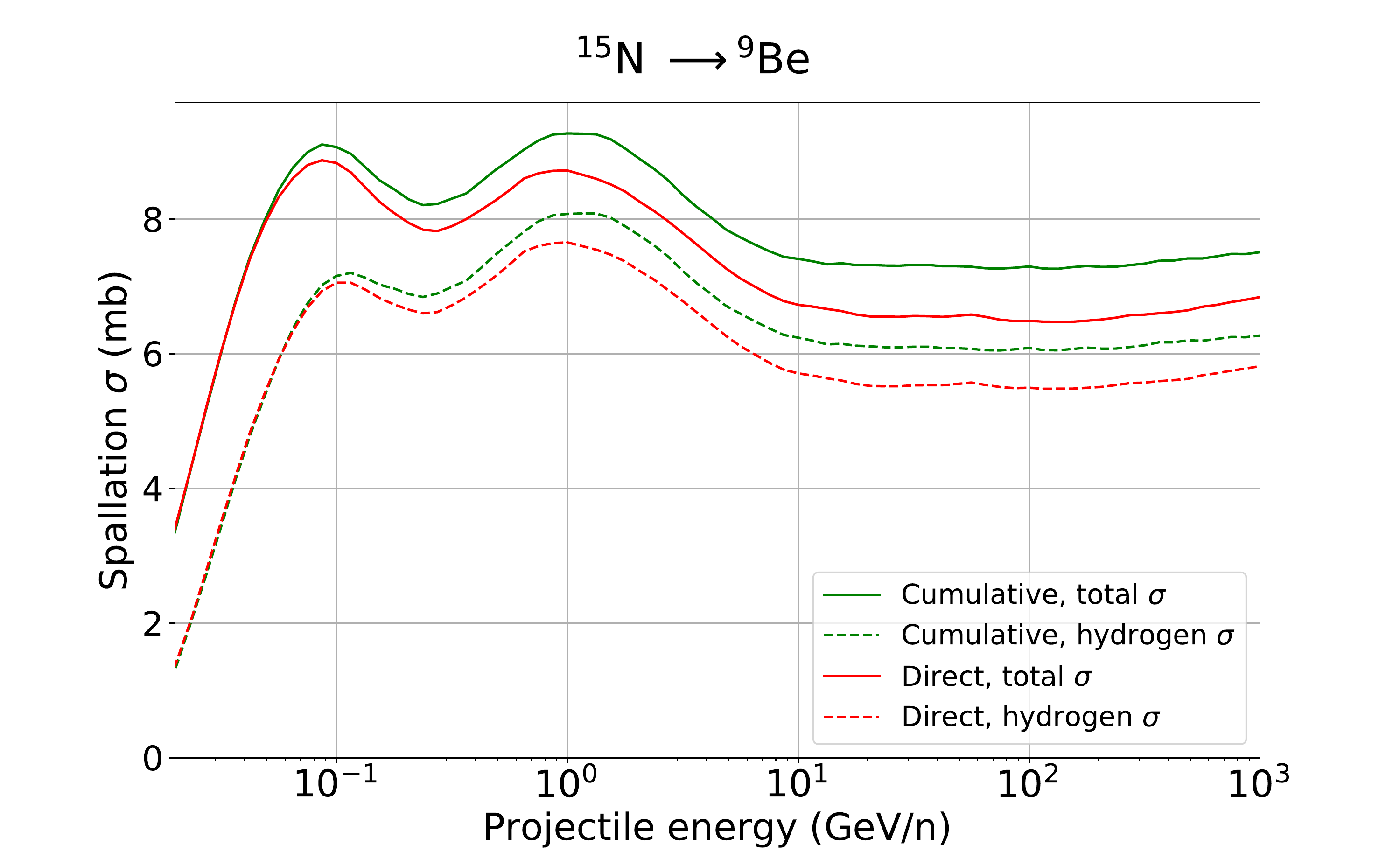}
\includegraphics[width=0.325\textwidth,height=0.165\textheight,clip] {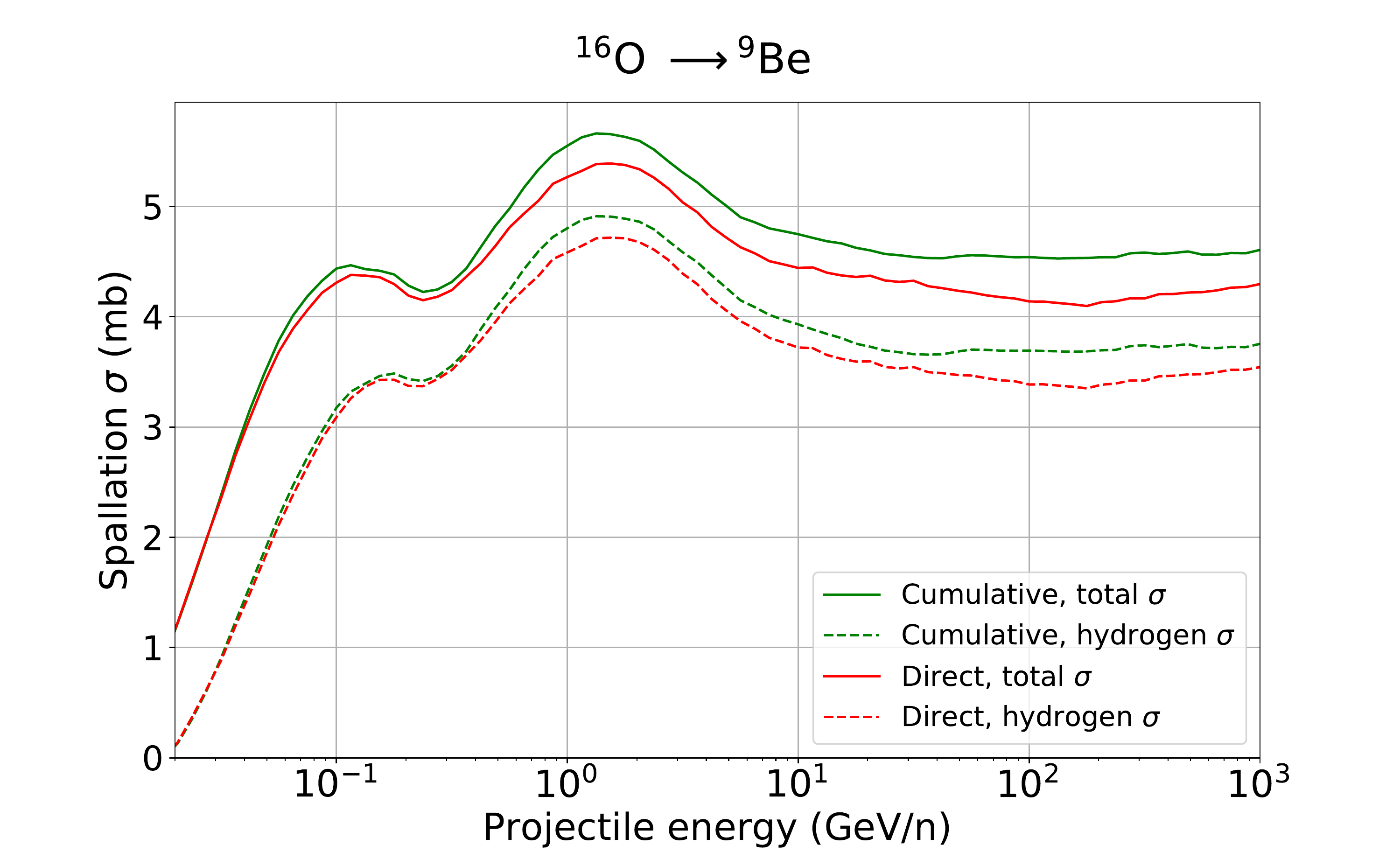}

\includegraphics[width=0.40\textwidth,height=0.17\textheight,clip] {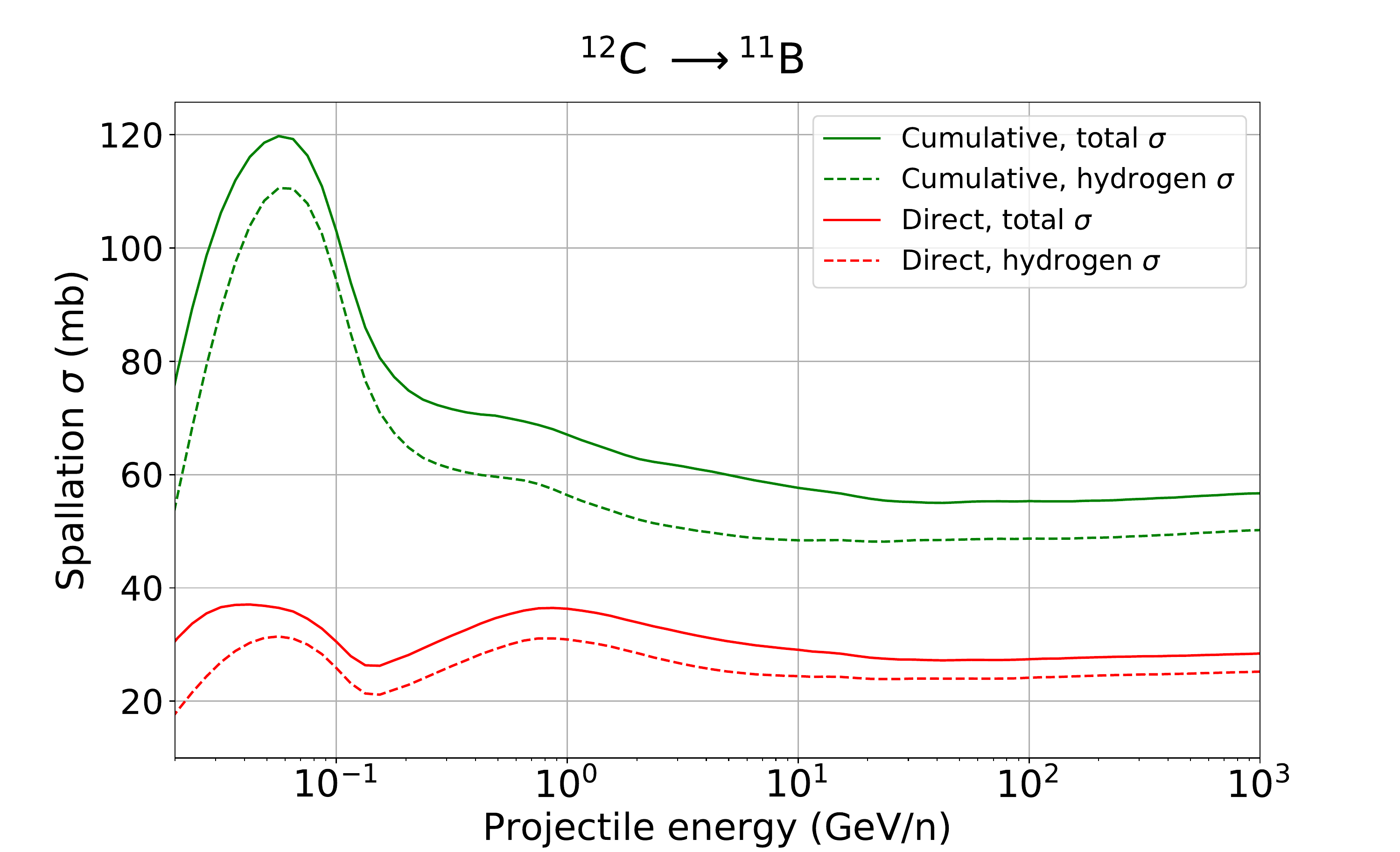}
\includegraphics[width=0.40\textwidth,height=0.17\textheight,clip] {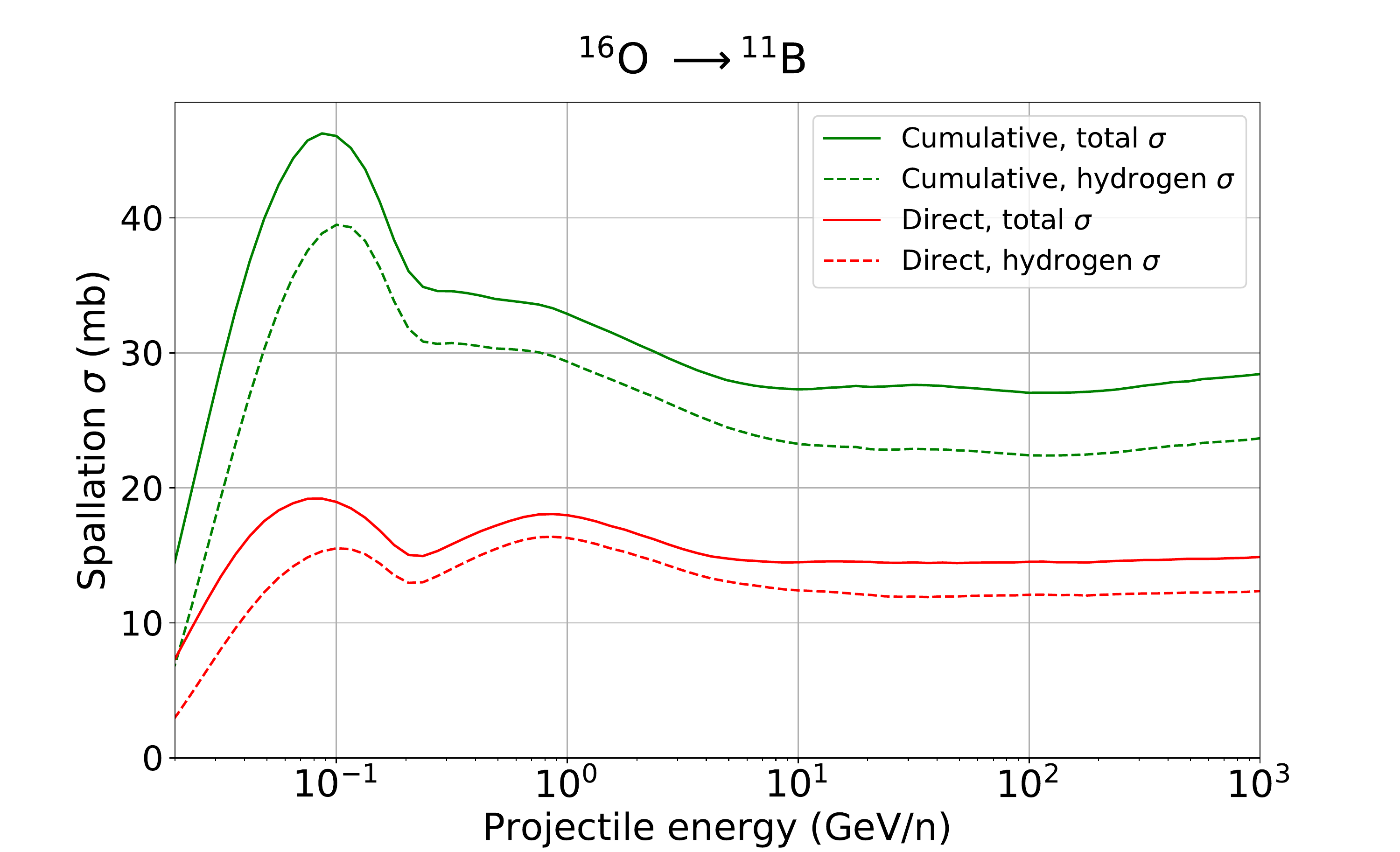}
\end{center}
\caption{\footnotesize Cross sections for the reaction channels where the generation of fast-decaying nuclei (ghost nuclei) is important for the production of secondary CRs. A comparison between the direct production and the production after taking into account the ghost nuclei (cumulative cross sections) is shown for the reactions with hydrogen as target and ISM composition as target (total).}
\label{fig:ghost_XS}
\end{figure*}

The impact of ghost nuclei has been also studied and it is shown in Figure~\ref{fig:ghost_XS}. Only the channels with any sizeable discrepancy between the cumulative and direct cross sections are displayed. In this figure, both the cross sections of the direct reaction with hydrogen and with the distribution of gas found in the in the ISM are shown. The latter are called ``total cross sections'' and are evaluated with the formula $\sigma_{Total} = \sigma_{H} + \frac{n_{He}}{n_H}\sigma_{He}$, where $\frac{n_{He}}{n_H} = 0.11$ is the fraction of helium in the gas (see ref. \cite{Evoli:2017vim}).

The most important effect of fast-decaying (ghost) nuclei is observed in the $^{11}$B and $^{10}$B isotopes. This is well known to be mainly due to the decays of $^{11}$C into $^{11}$B and $^{10}$C into $^{10}$B: \begin{center}$^{12}$C$ + gas\hspace{0.1cm} {\xrightarrow{\makebox[2cm]{spallation}}} \hspace{0.1cm}  ^{11}$C$, ^{10}$C$ \hspace{0.1cm}  {\xrightarrow{\makebox[1cm]{decay}}} \hspace{0.1cm} ^{11}$B$, ^{10}$B
\end{center}
since the production of these isotopes of C is large due to the amount of primary $^{12}$C. In fact, the decays of $^{12}$Be and $^{11}$Be have almost no impact in the production of $^{11}$B. We also see that there is another isotope for which the contribution of ghost nuclei is important, the $^{9}$Be isotope, for which the difference between the cumulative and direct cross sections can be $\sim 1\%$ at high energies in the case of $^{15}$N and $^{16}$O projectiles. There are two nuclei that may decay into it, $^{9}$Li ($\sim 50\%$ via $\beta^-$) and $^{11}$Li. We also can see how the amount of this isotope coming from ghost nuclei is increased for heavier nuclei, which may be explained since, as primary nuclei get heavier, the multiplicity of formation of secondary light nuclei like $^{9}$Li via spallation increases (as an example, the inclusive cross sections for $^{9}$Li from $^{15}$N is $\sim 2.5$ mb at high energies, which may well account for the amount of $^{9}$Be formed via decays). %Finally, there are other two isotopes, $^{7}Li$ and $^{7}Be$, in which there have been found small fluctuations between their direct and cumulative cross sections in most of their formation channels. This is, most likely, due to the decay of $^{7}Be$ into $^{7}Li$, only possible if the $^{7}Be$ isotope captures an electron, what can explain such small fluctuations.

\begin{figure*}[!hbt]
\begin{center}
\includegraphics[width=0.325\textwidth,height=0.17\textheight,clip] {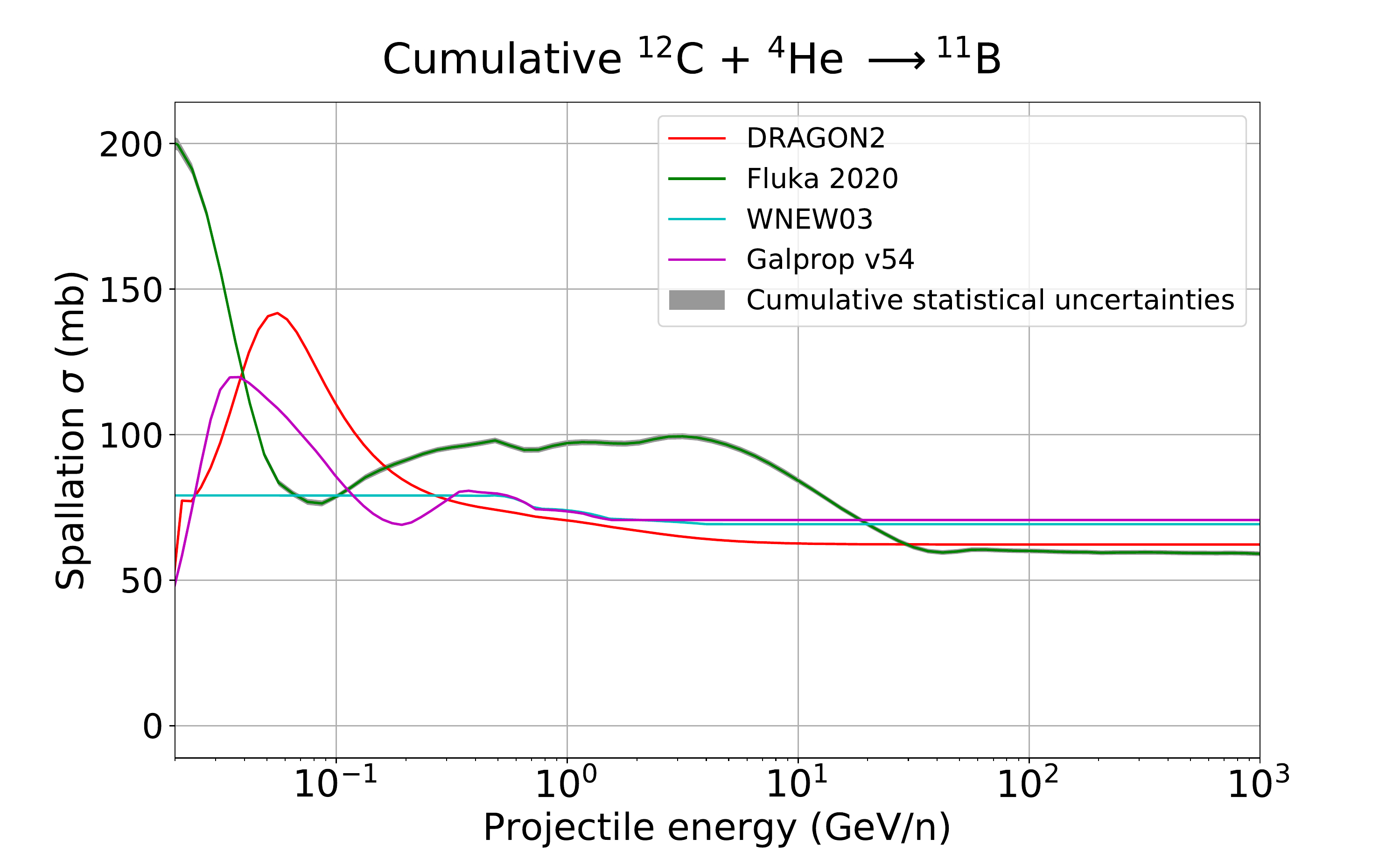} 
\includegraphics[width=0.325\textwidth,height=0.17\textheight,clip] {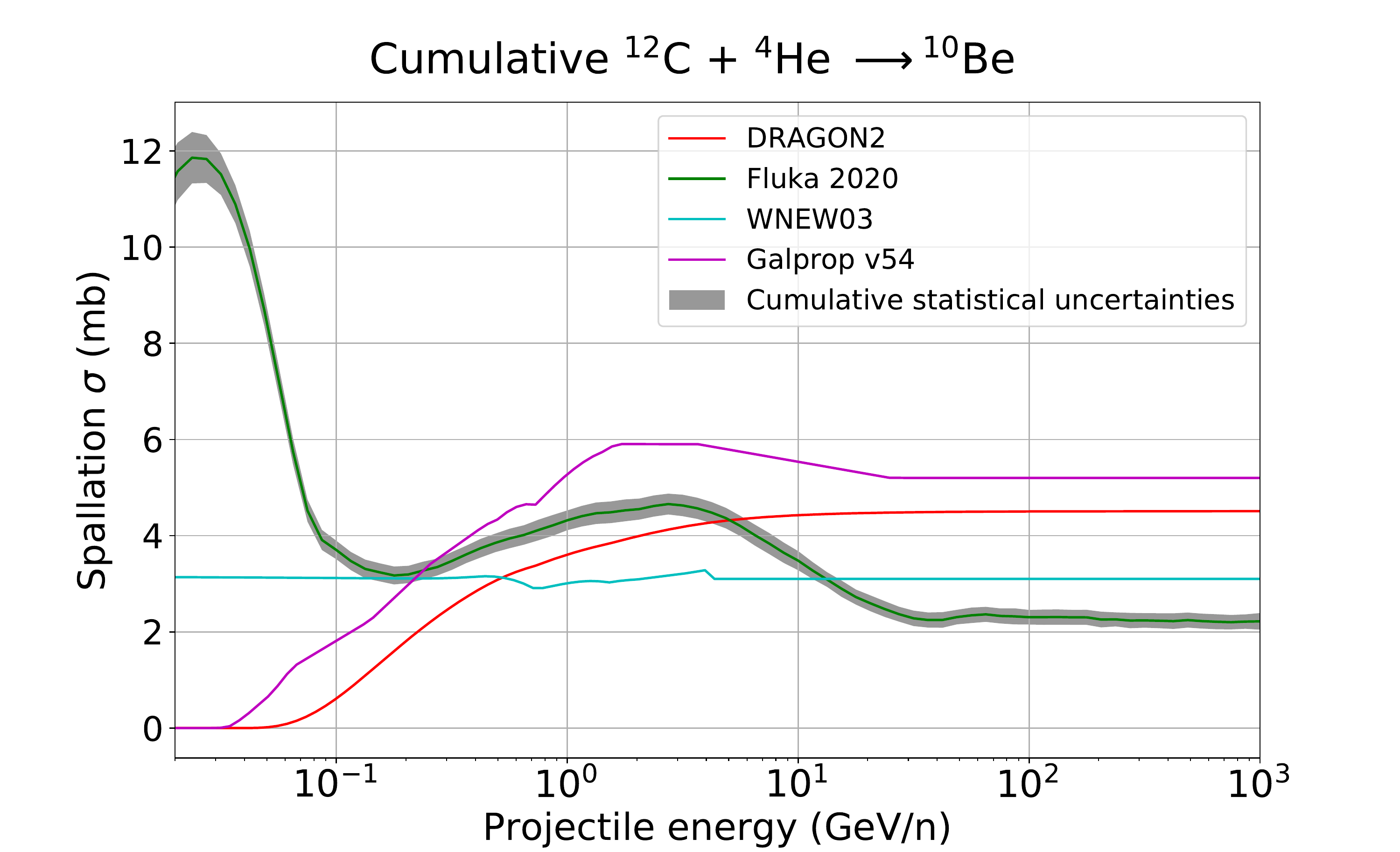}
\includegraphics[width=0.325\textwidth,height=0.17\textheight,clip] {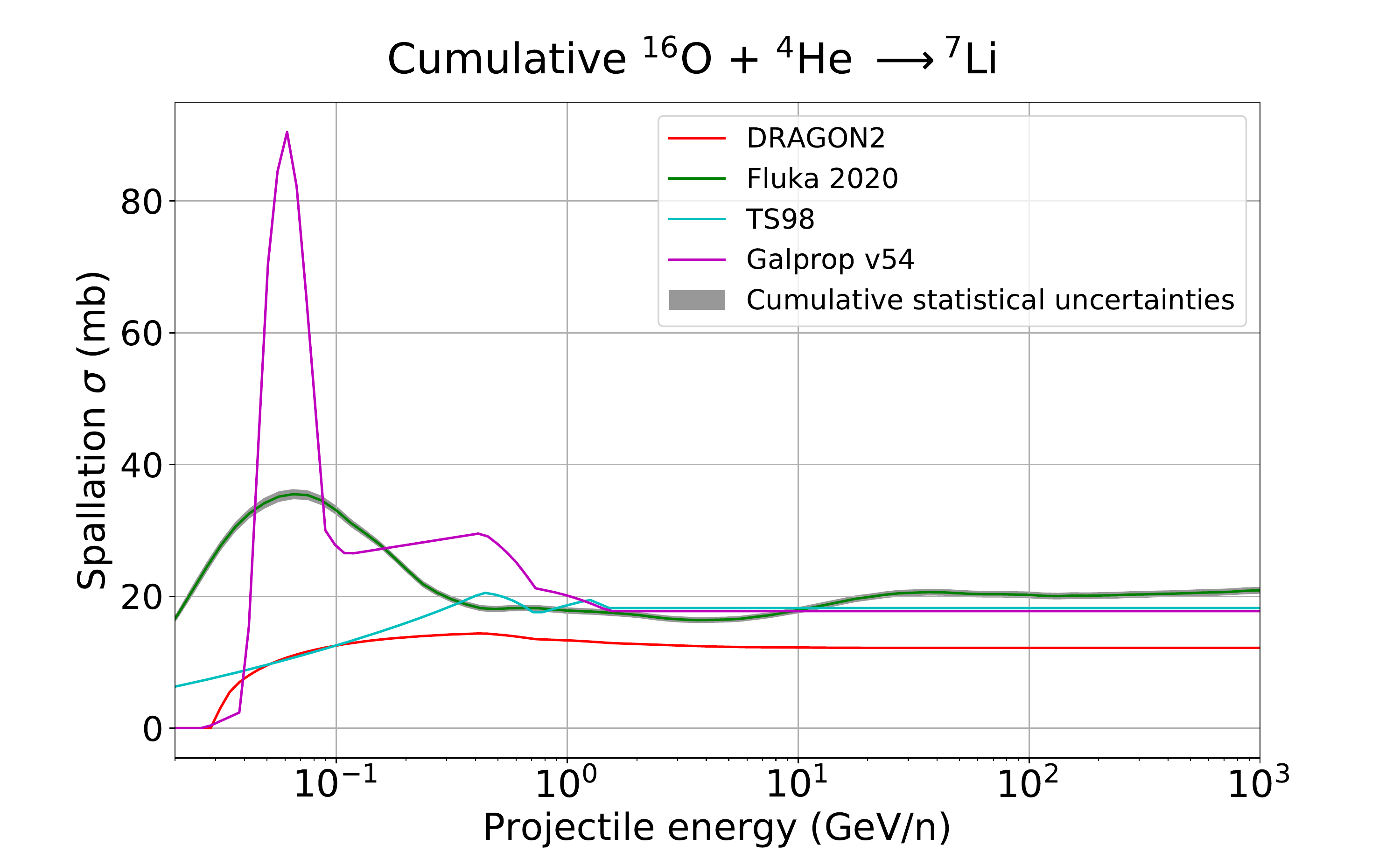}

\includegraphics[width=0.325\textwidth,height=0.17\textheight,clip] {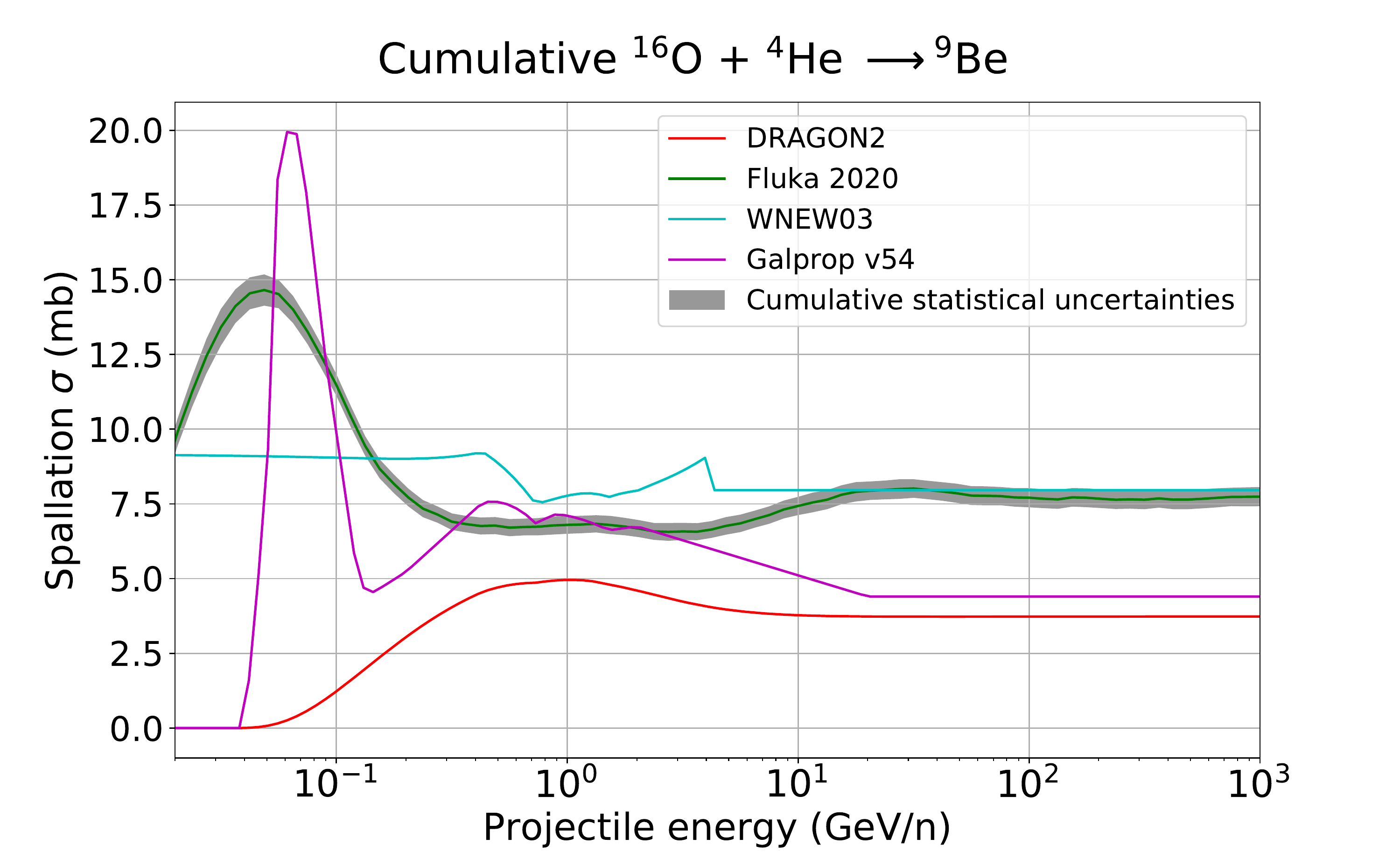}
\includegraphics[width=0.325\textwidth,height=0.17\textheight,clip] {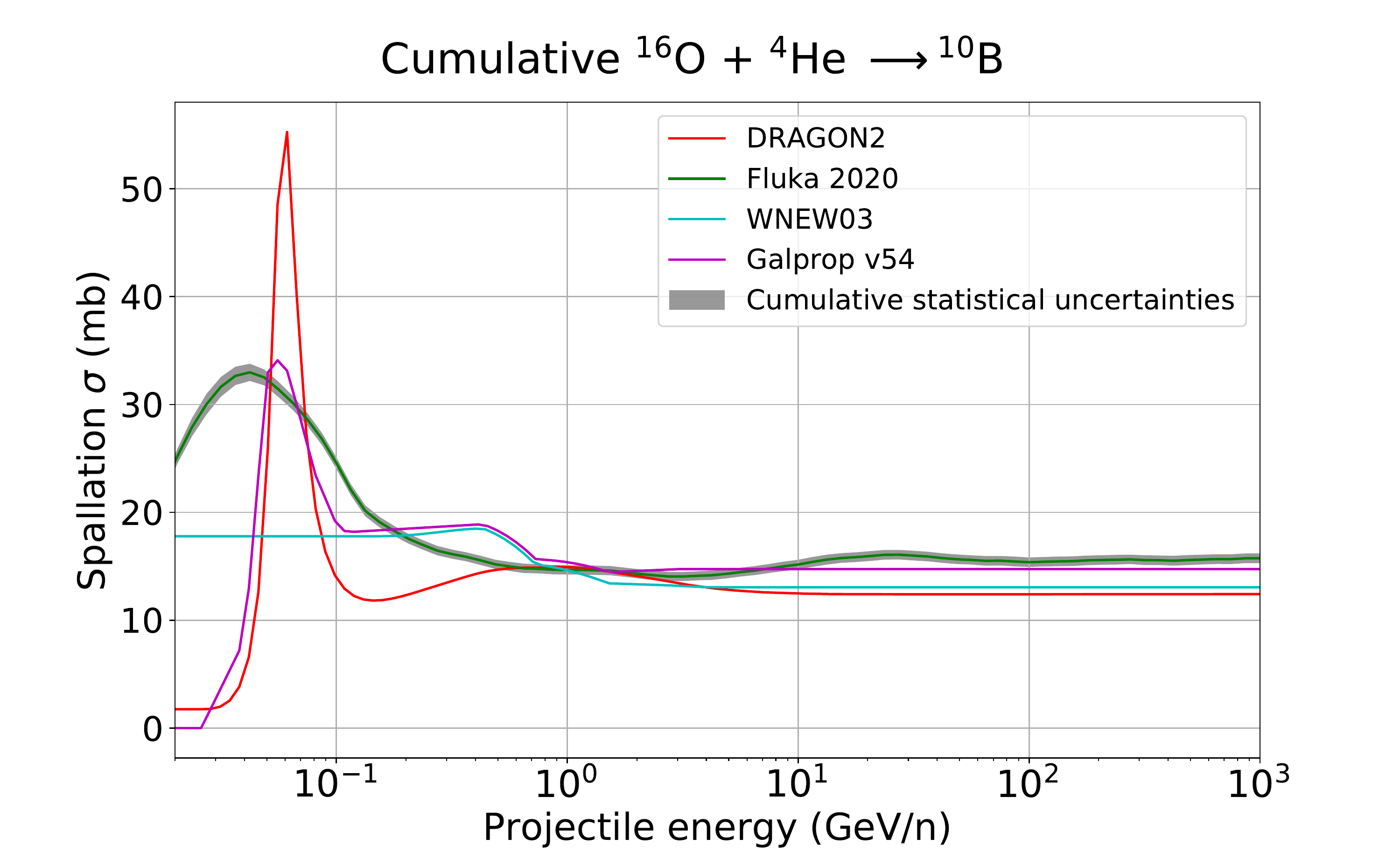}
\includegraphics[width=0.325\textwidth,height=0.17\textheight,clip] {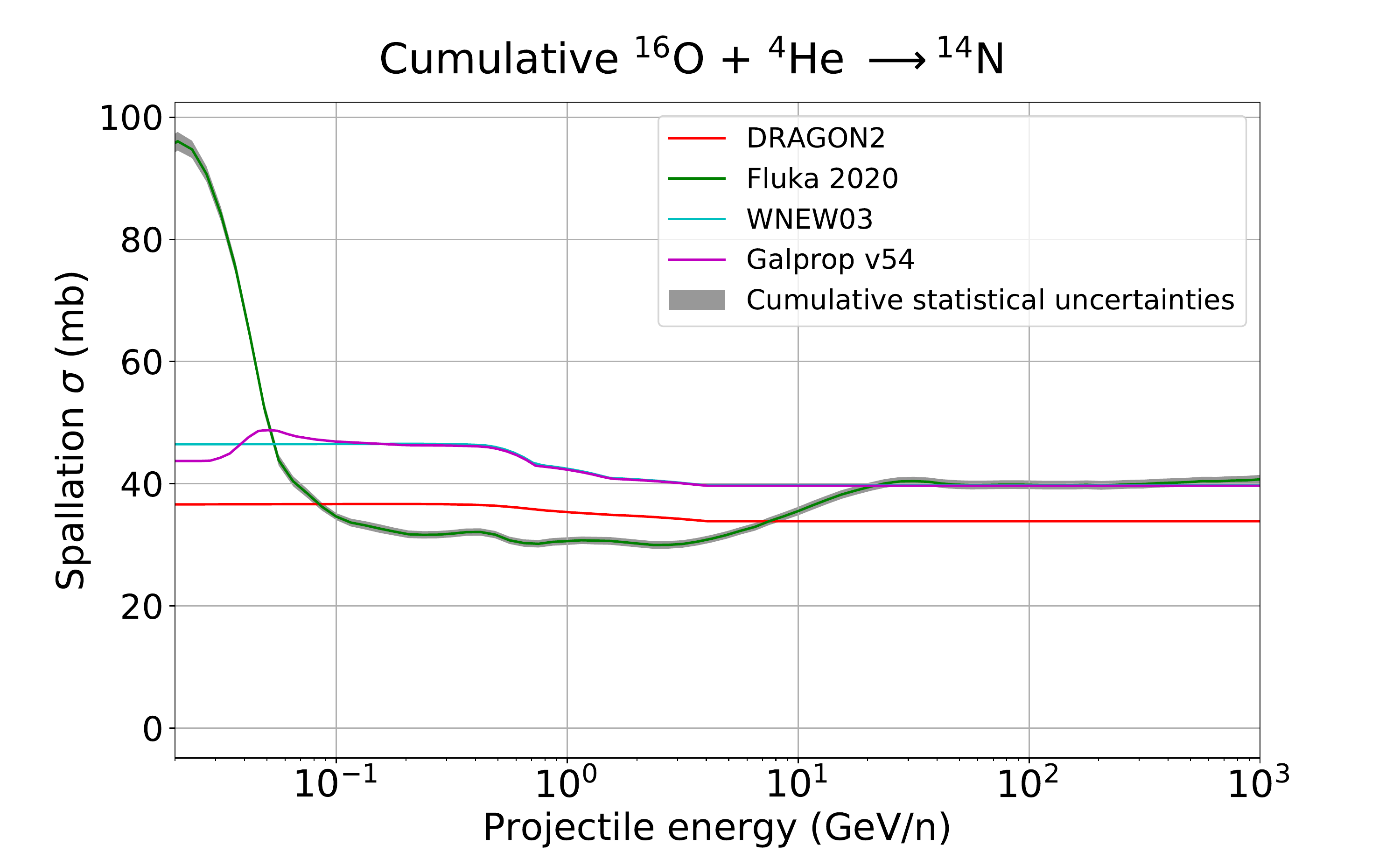}

\includegraphics[width=0.325\textwidth,height=0.17\textheight,clip] {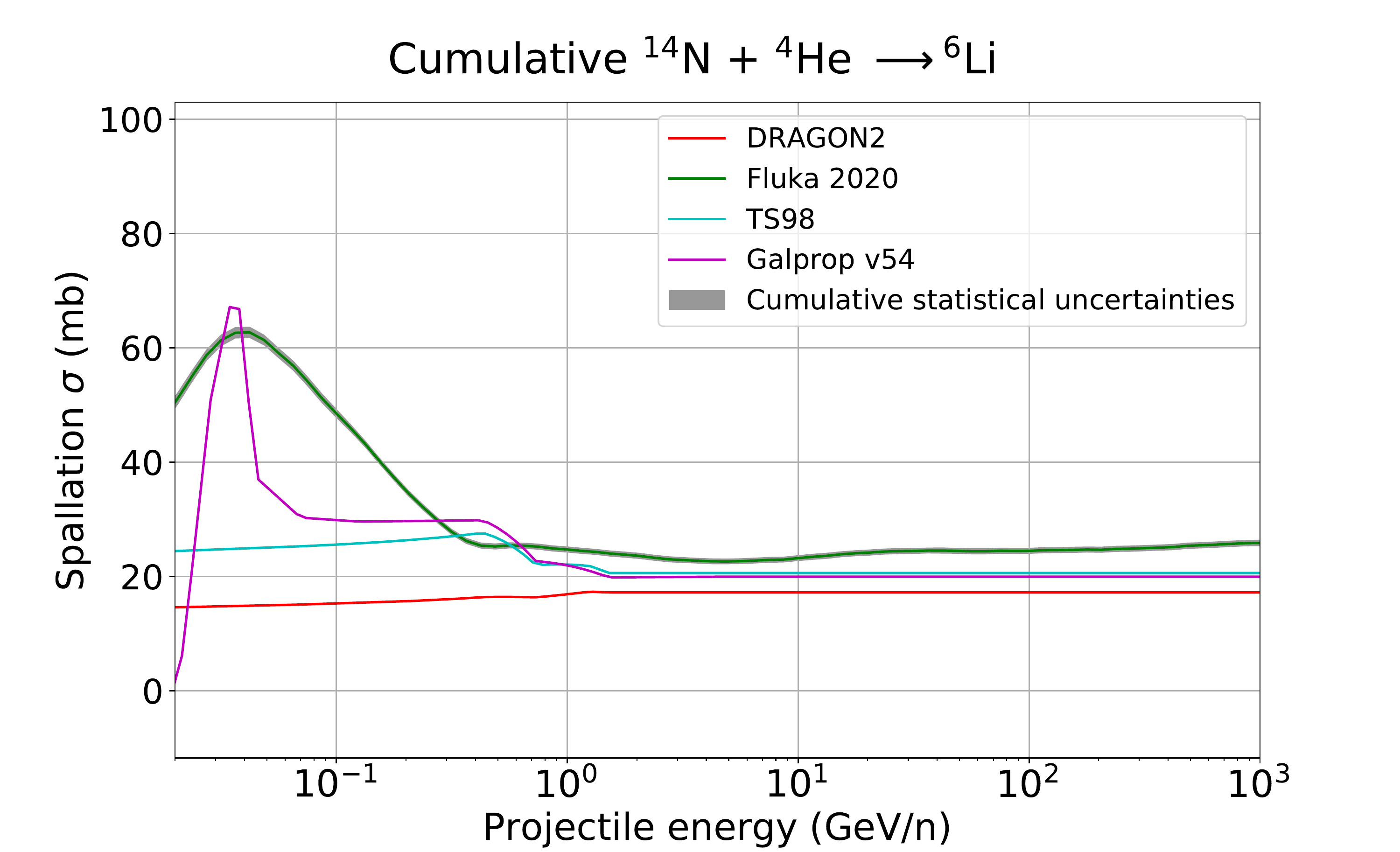}
\includegraphics[width=0.325\textwidth,height=0.17\textheight,clip] {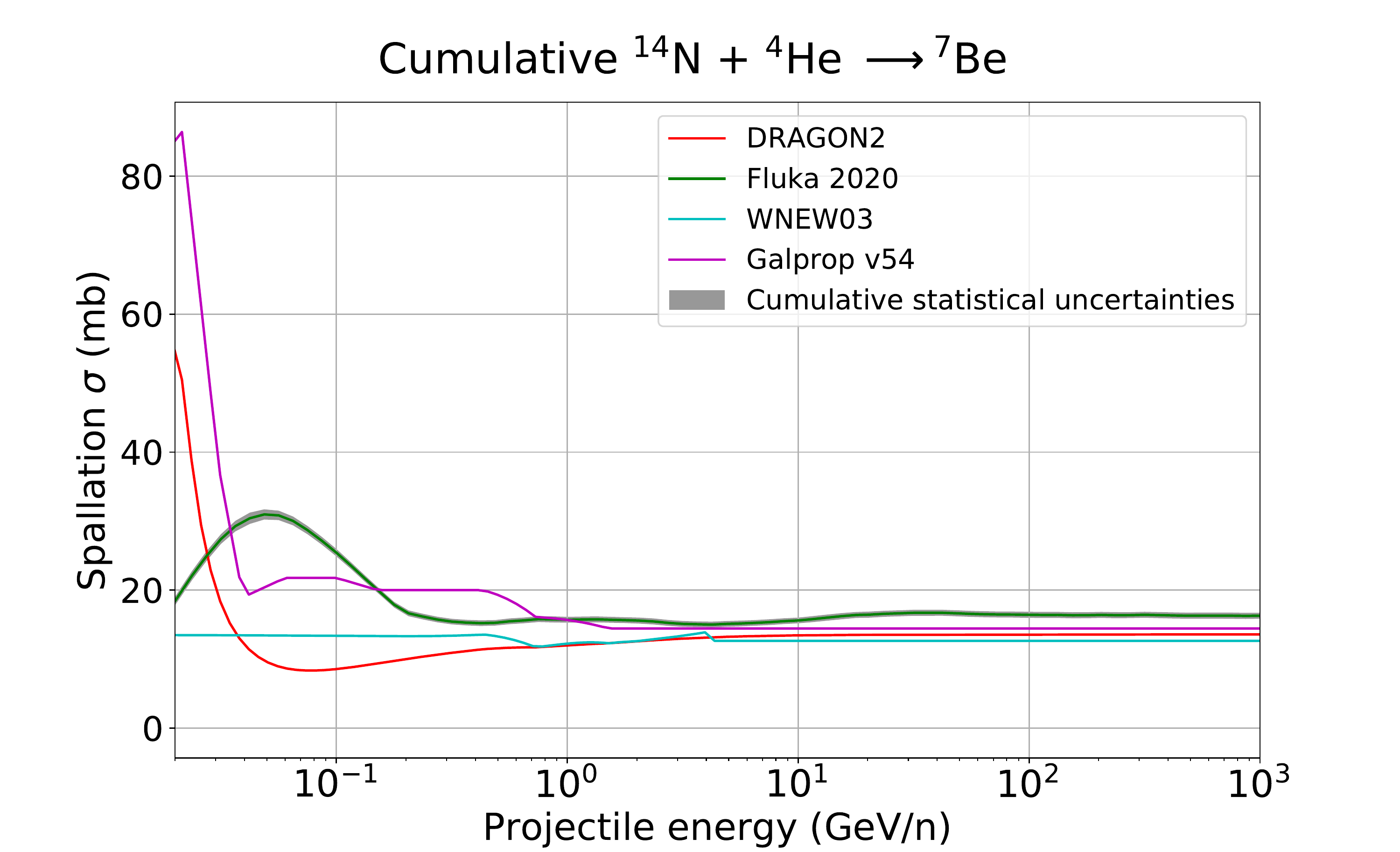}
\includegraphics[width=0.325\textwidth,height=0.17\textheight,clip] {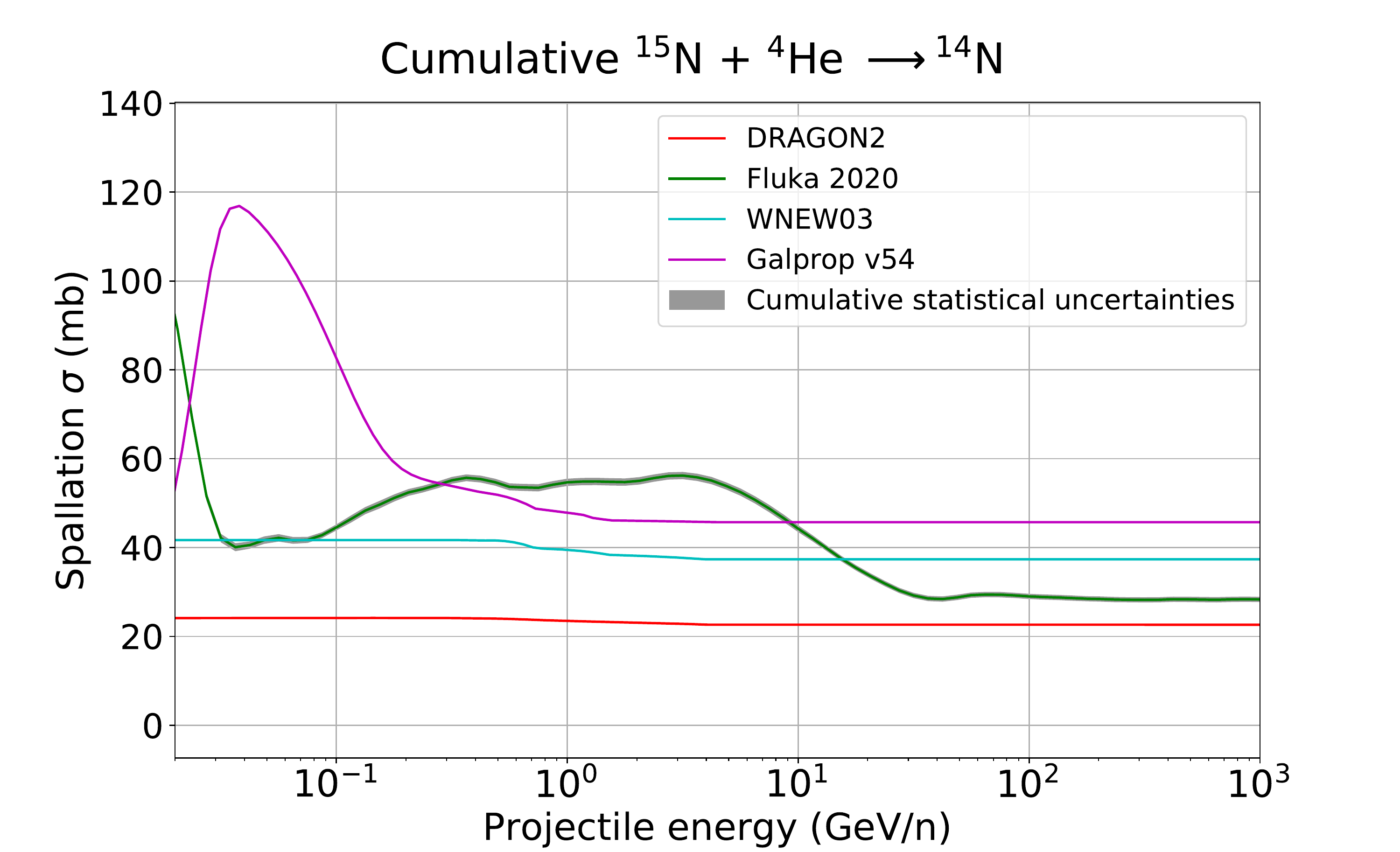} 
\end{center}
\caption{\footnotesize Comparison of the spallation cross sections for He target reactions computed with FLUKA and the parametrisations discussed in chapter~\ref{sec:XSecs}. We show the relevant channels for the production of the most important secondary CR nuclei.}
\label{fig:He_XS}
\end{figure*}

At this point, once we have the cross sections calculated for the cumulative distributions of hydrogen target and helium target reactions, we proceed to compare the helium target and total (ISM gas composition as target) cross sections with the parametrisations discussed in the chapter~\ref{sec:XSecs}. In these parametrisations, the helium + target cross sections are calculated using a simple extrapolation based on the ratio $\frac{\sigma_H}{\sigma_{He}}$ (see ref. \cite{ferrando1988measurement}) due to the lack of data for these reactions. This may result in a source of bias, which propagates the errors in the cross sections of hydrogen target channels into the helium target ones. Some of the cross sections for these reaction channels are shown in Fig.~\ref{fig:He_XS}. 

As in the case of the reactions with hydrogen target, the channels producing Be isotopes are those showing larger differences between the FLUKA predictions and the predictions of the other parametrisations. It is interesting to notice that the cross sections of production of N are very similar in all cases. There is also a relatively good agreement between the different predictions in the production of the $^{10}$B and $^{11}$B isotopes. It is found that the resonances are usually found in a similar region, except for the Webber parametrisations. It is also worth pointing to the fact that for the reactions $^{12}C \longrightarrow ^{10}$Be and $^{16}$O$ \longrightarrow ^{14}$N the FLUKA computations predict a very pronounced resonance at low energies, while the rest of parametrisations predict an almost constant cross section.

\begin{figure*}[!bph]
\begin{center}
\textbf{\underline{Cumulative spallation cross sections of reactions with insterstellar gas}} \\ \vspace{0.4 cm}
\includegraphics[width=0.325\textwidth,height=0.17\textheight,clip] {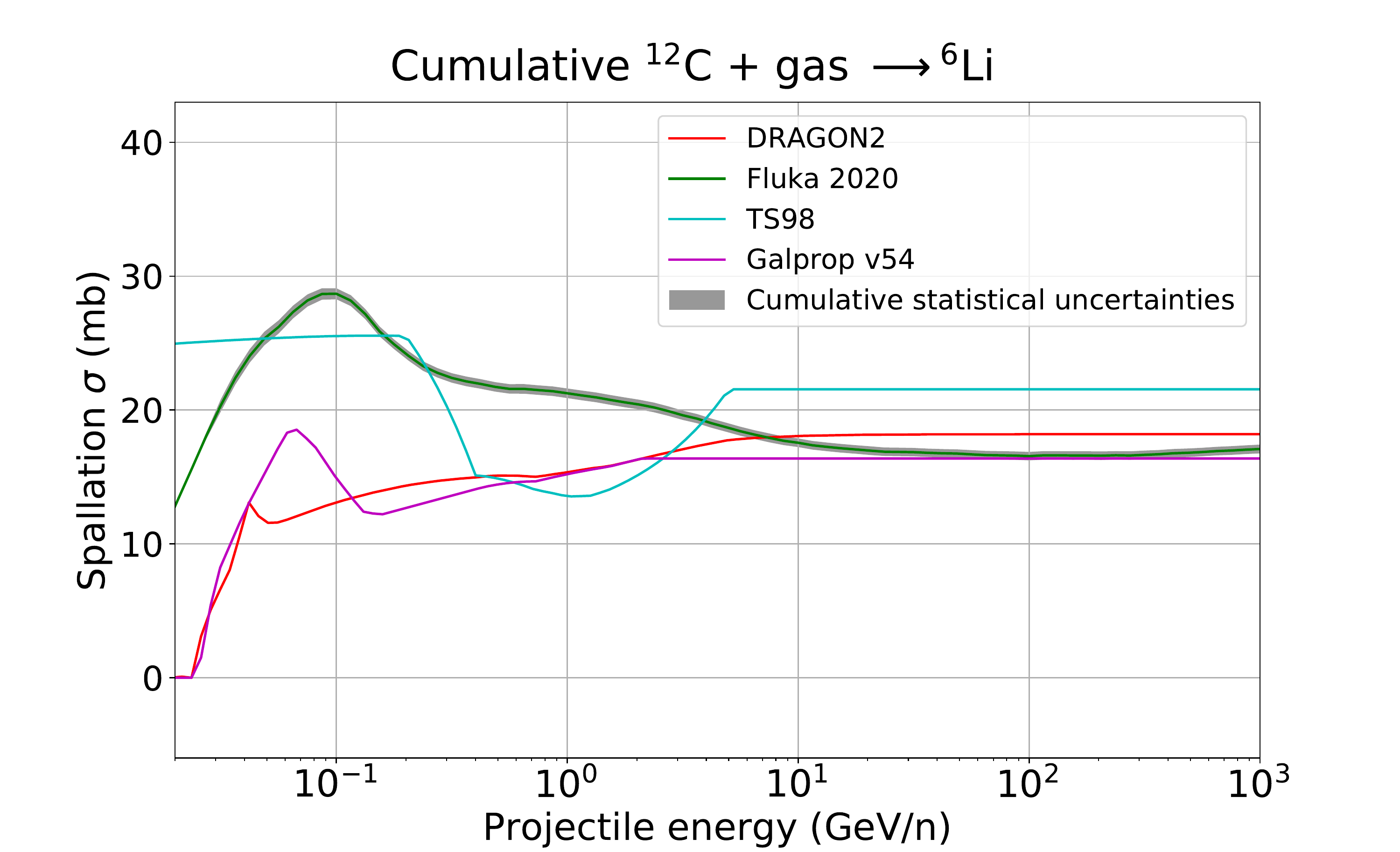}
\includegraphics[width=0.325\textwidth,height=0.17\textheight,clip] {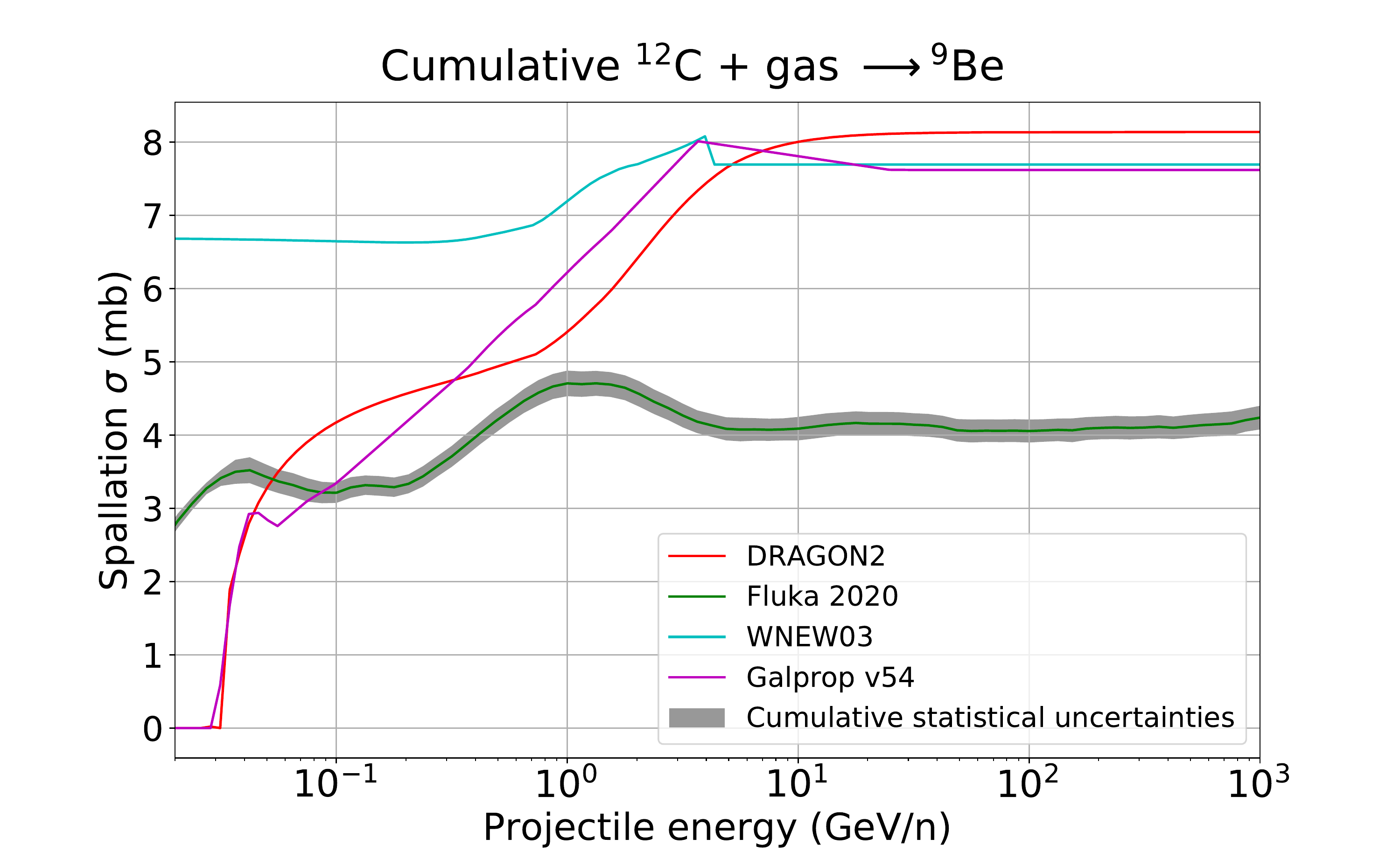}
\includegraphics[width=0.325\textwidth,height=0.17\textheight,clip] {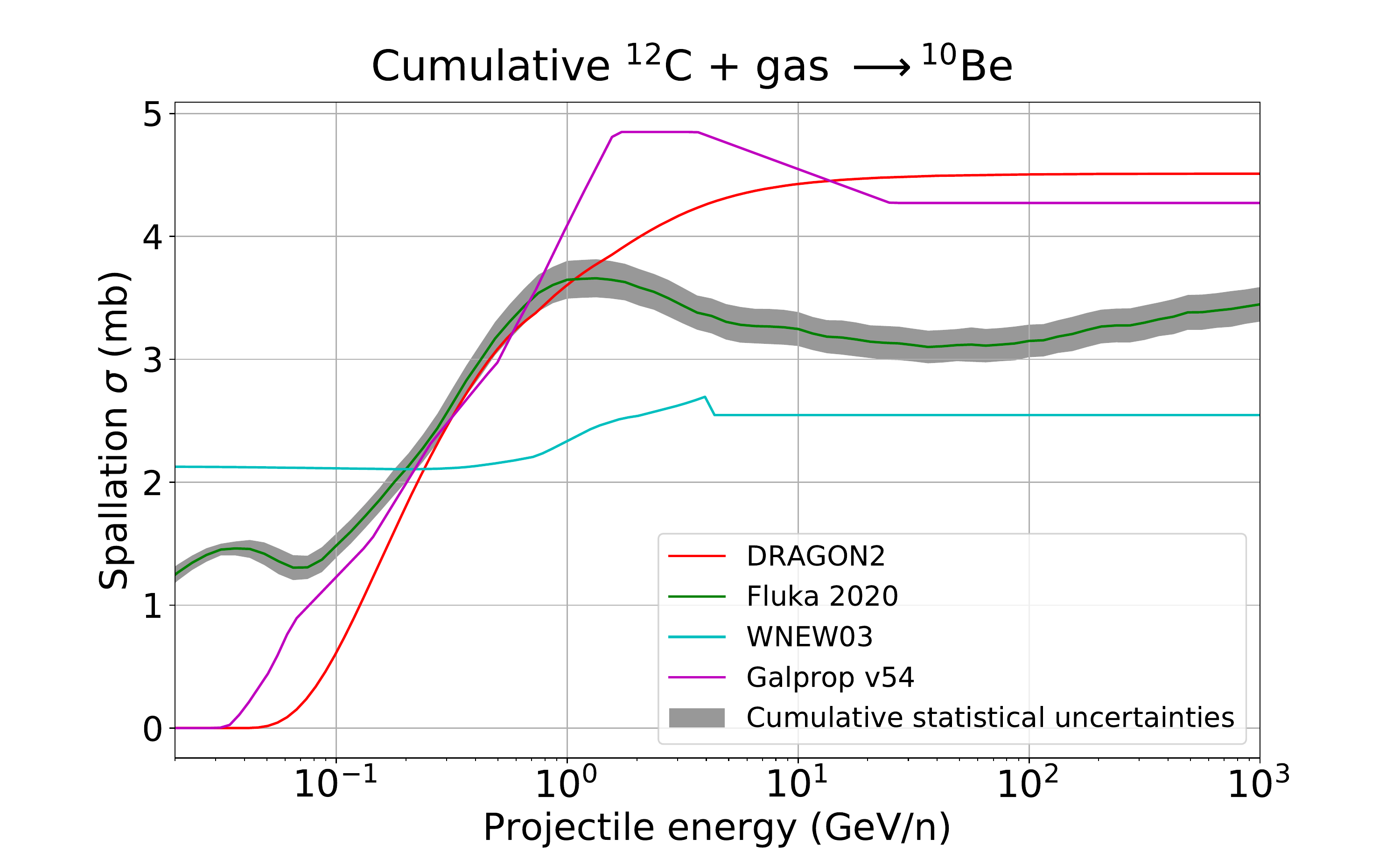}

\includegraphics[width=0.325\textwidth,height=0.17\textheight,clip] {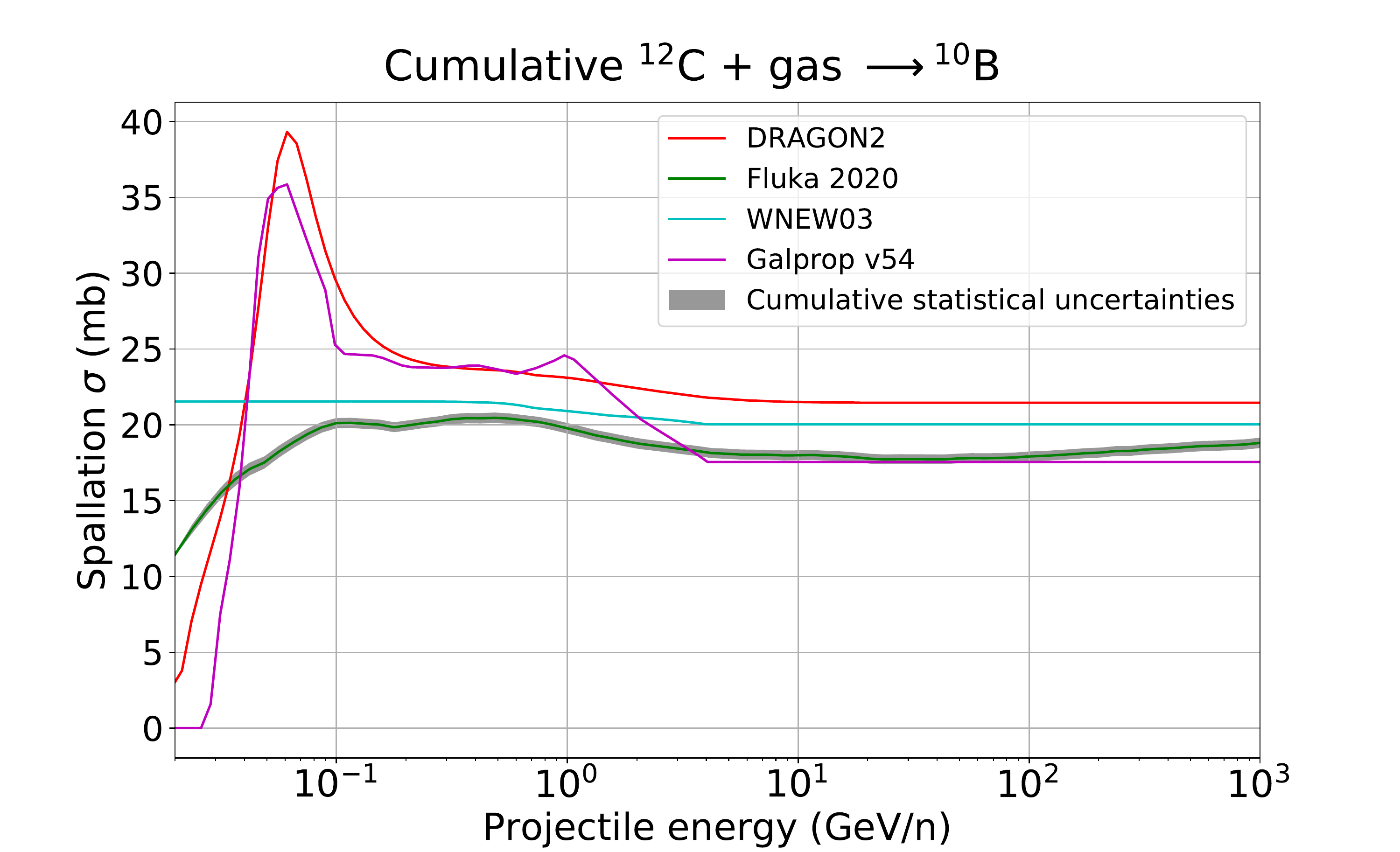}
\includegraphics[width=0.325\textwidth,height=0.17\textheight,clip] {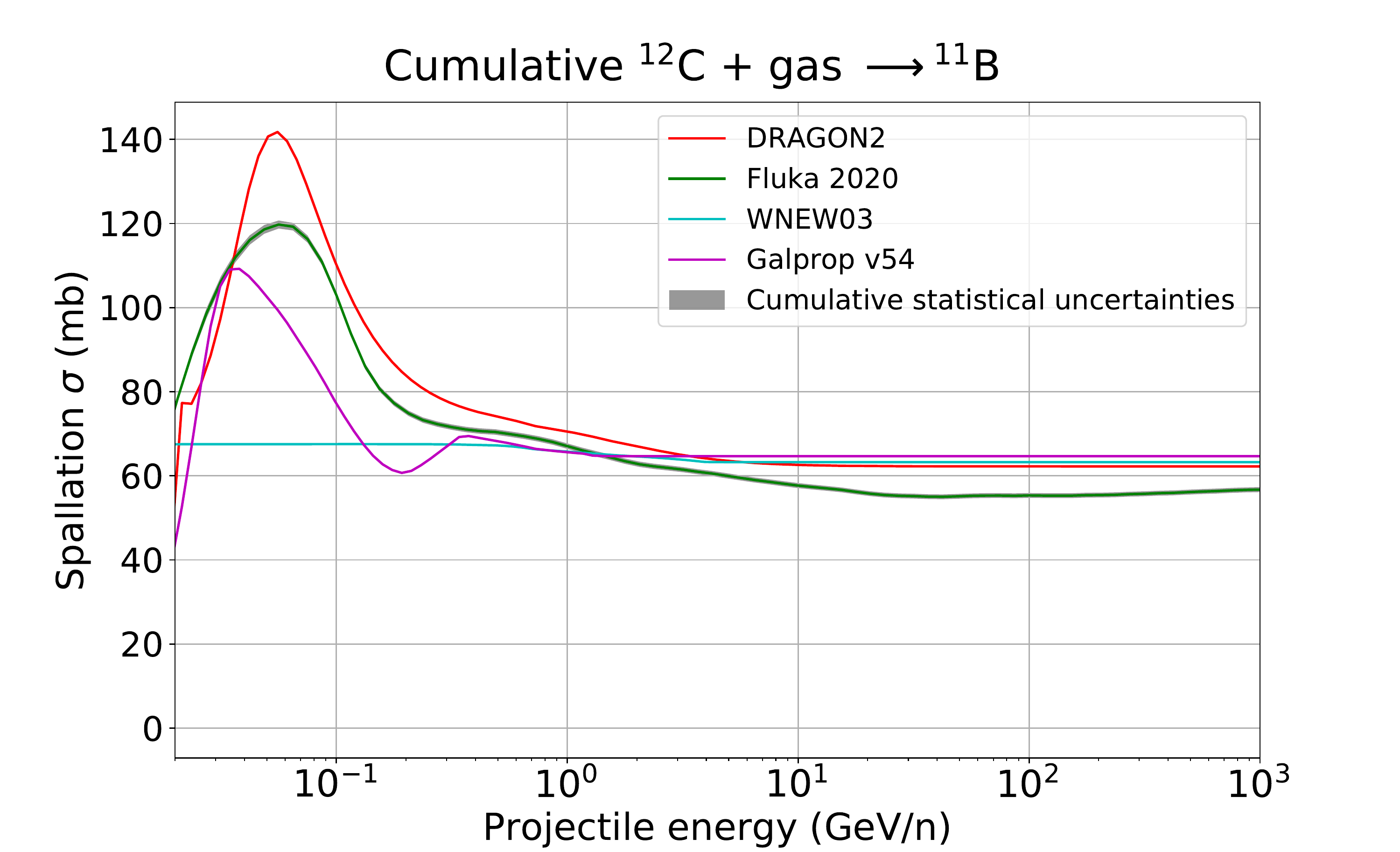}
\includegraphics[width=0.325\textwidth,height=0.17\textheight,clip] {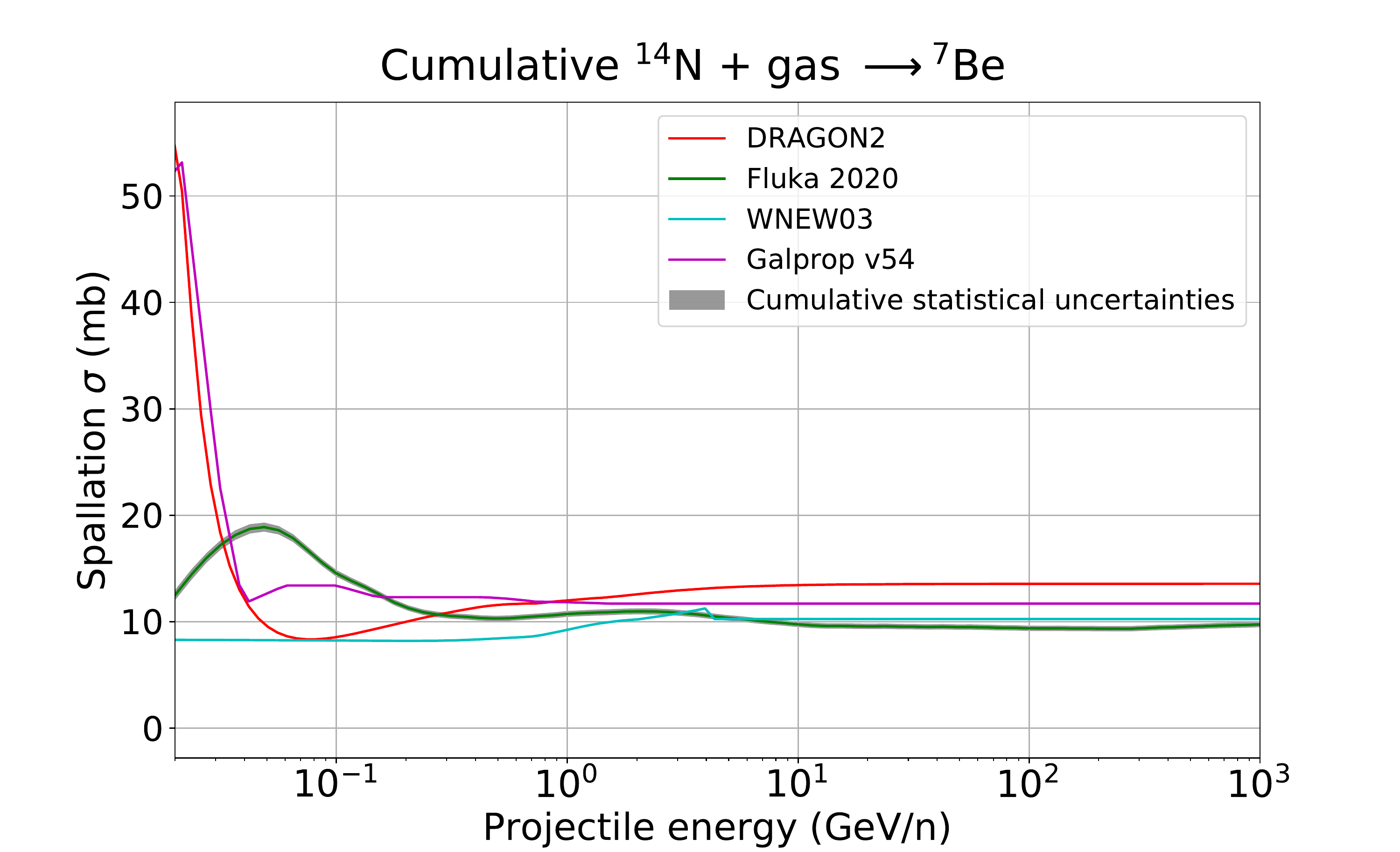}

\includegraphics[width=0.325\textwidth,height=0.17\textheight,clip] {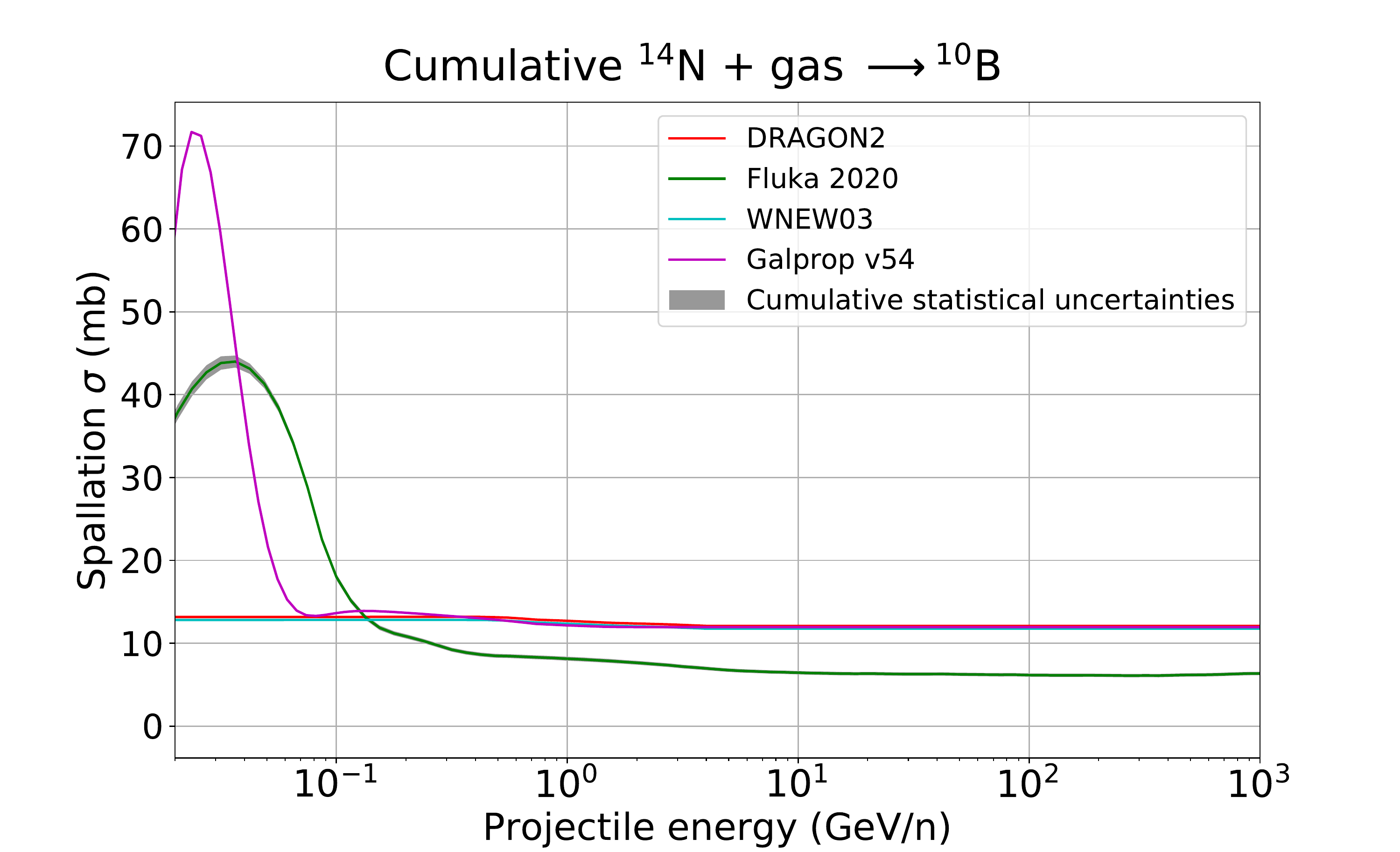}
\includegraphics[width=0.325\textwidth,height=0.17\textheight,clip] {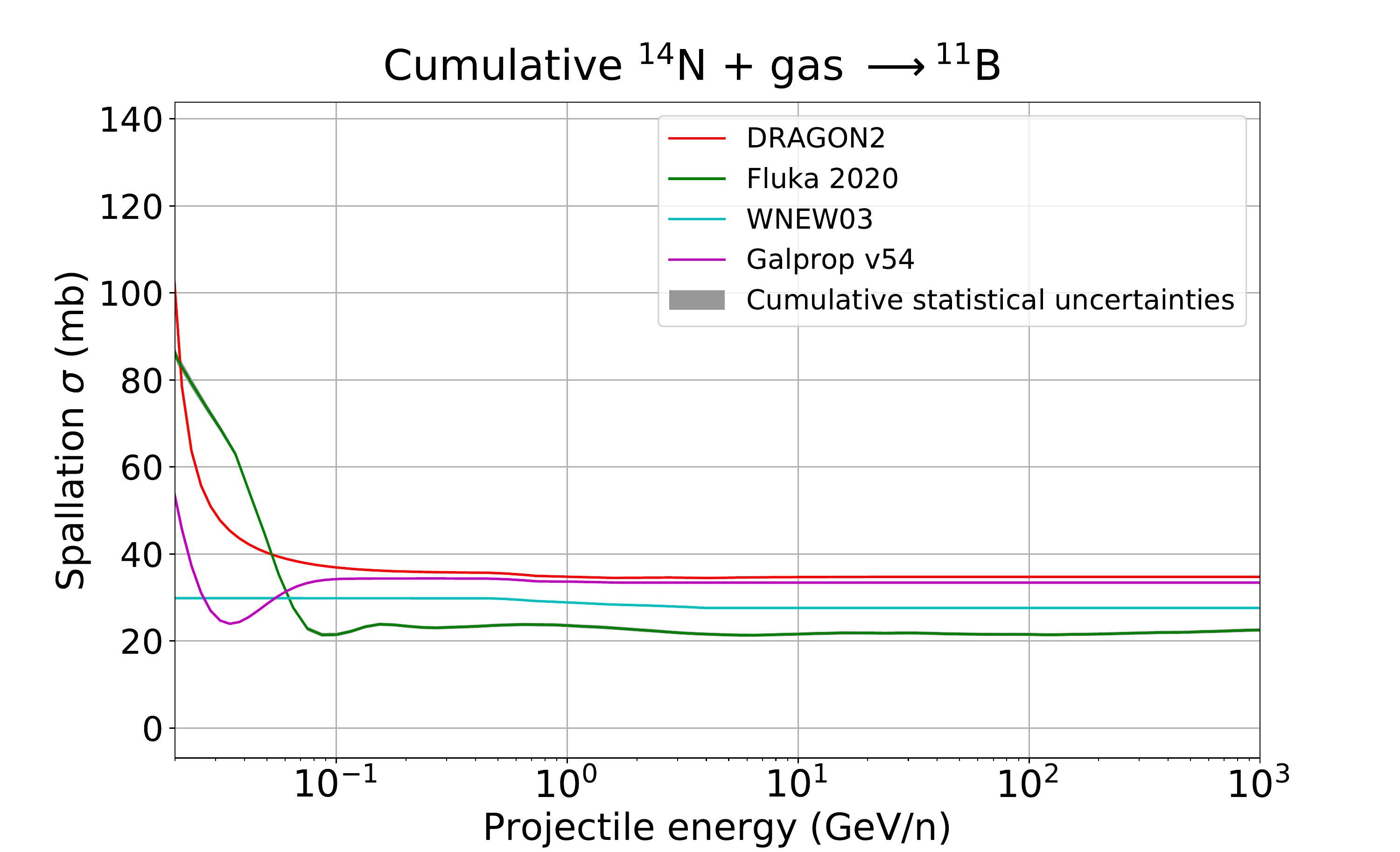}
\includegraphics[width=0.325\textwidth,height=0.17\textheight,clip] {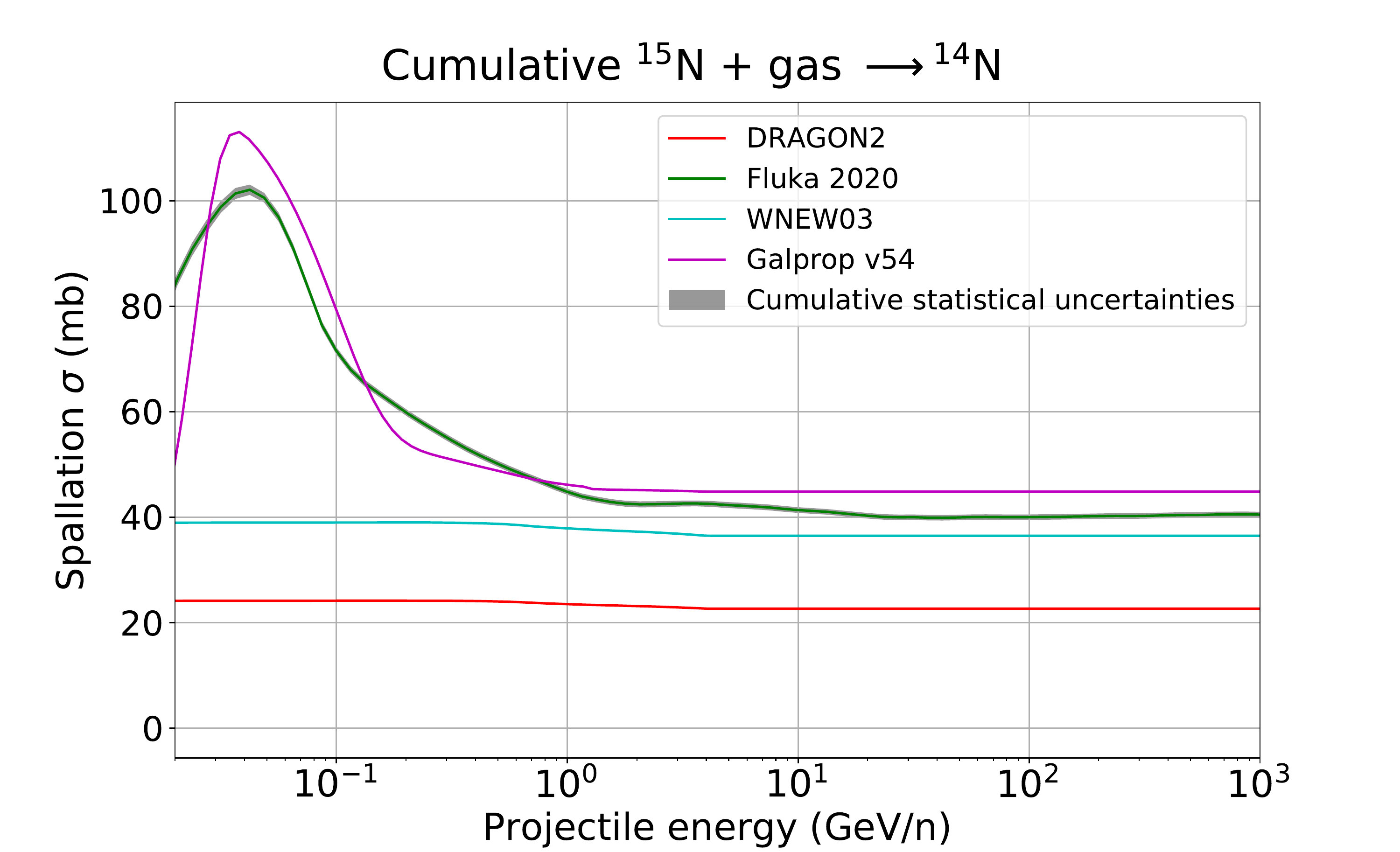}

\includegraphics[width=0.325\textwidth,height=0.17\textheight,clip] {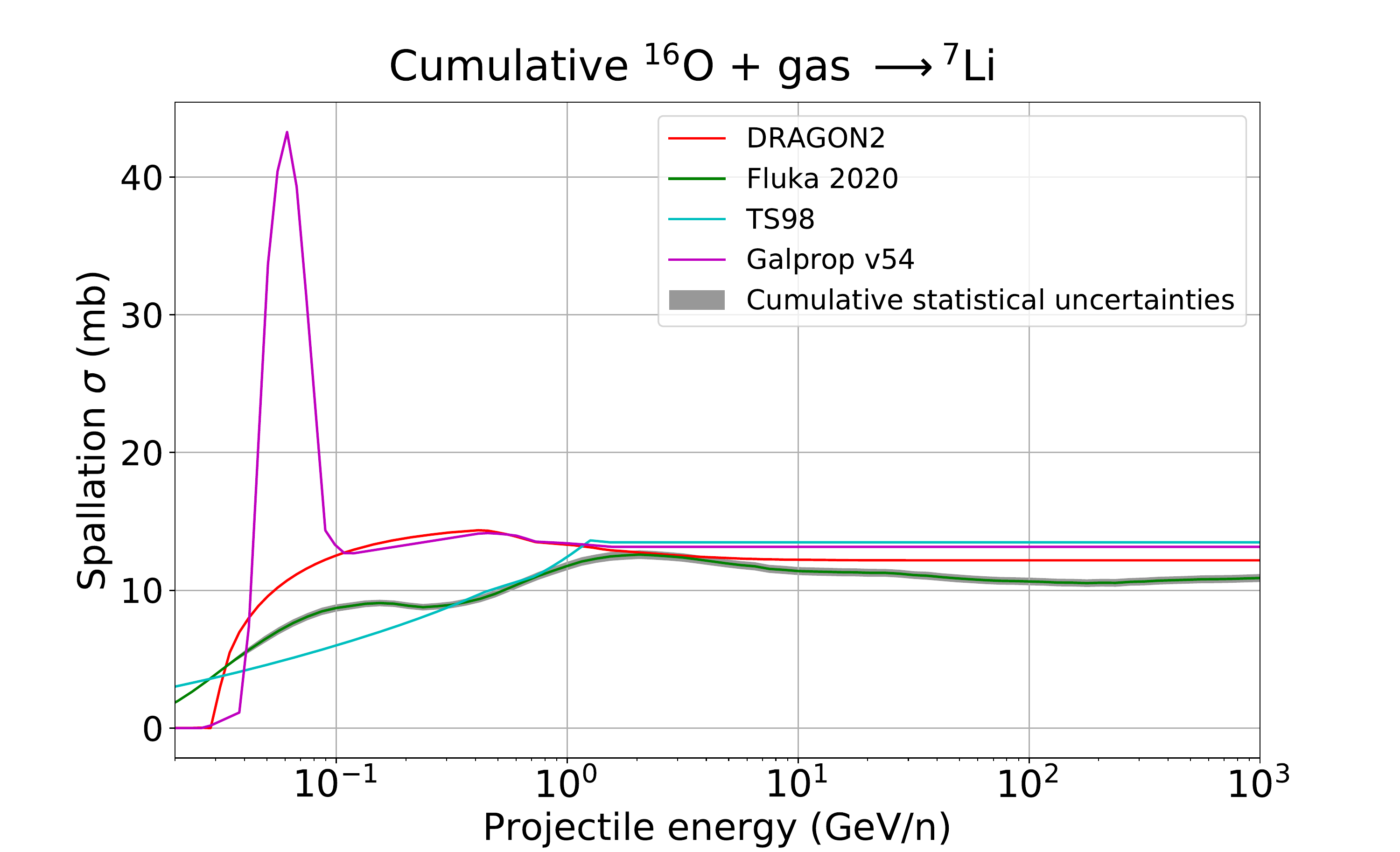}
\includegraphics[width=0.325\textwidth,height=0.17\textheight,clip] {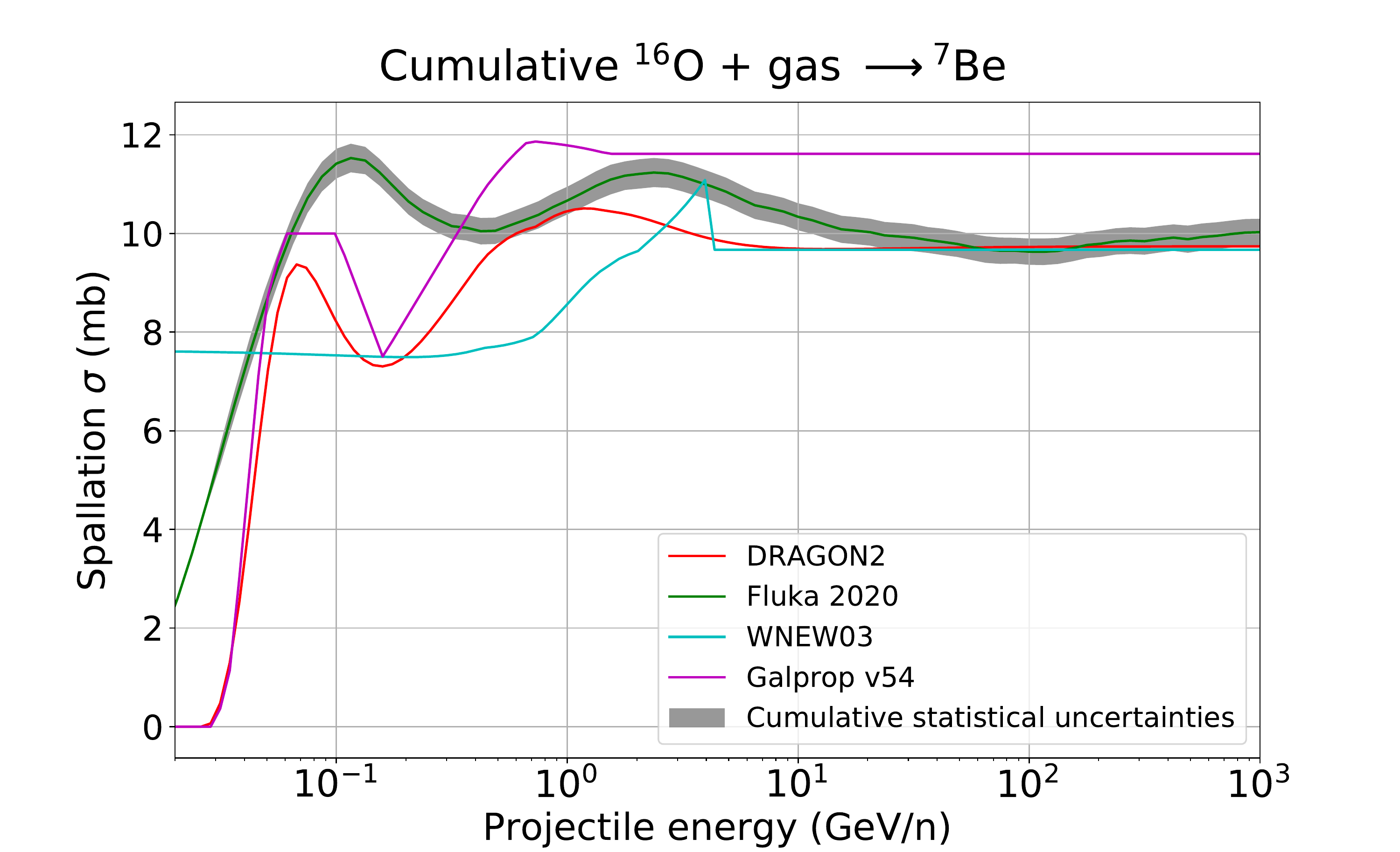}
\includegraphics[width=0.325\textwidth,height=0.17\textheight,clip] {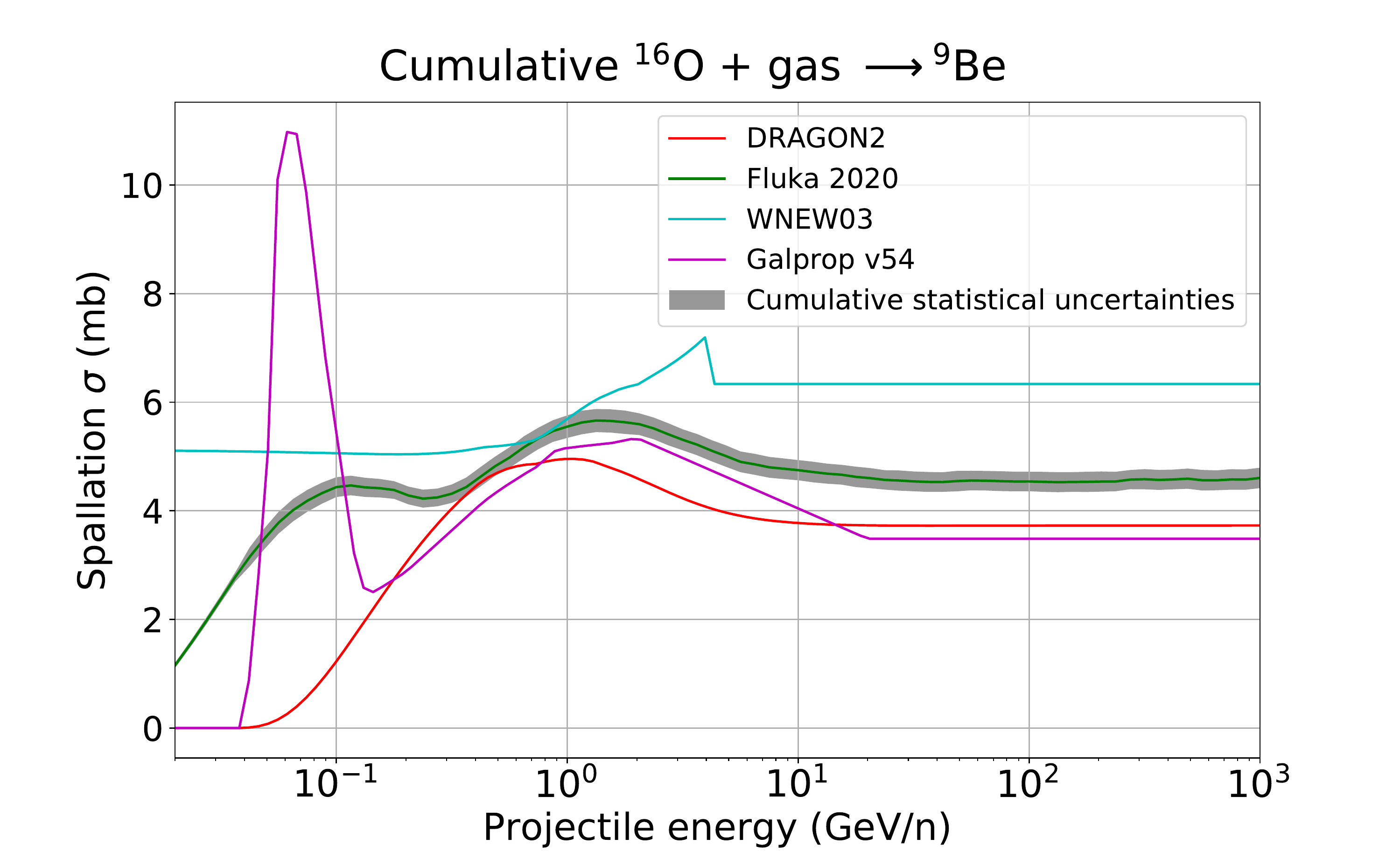}

\includegraphics[width=0.325\textwidth,height=0.17\textheight,clip] {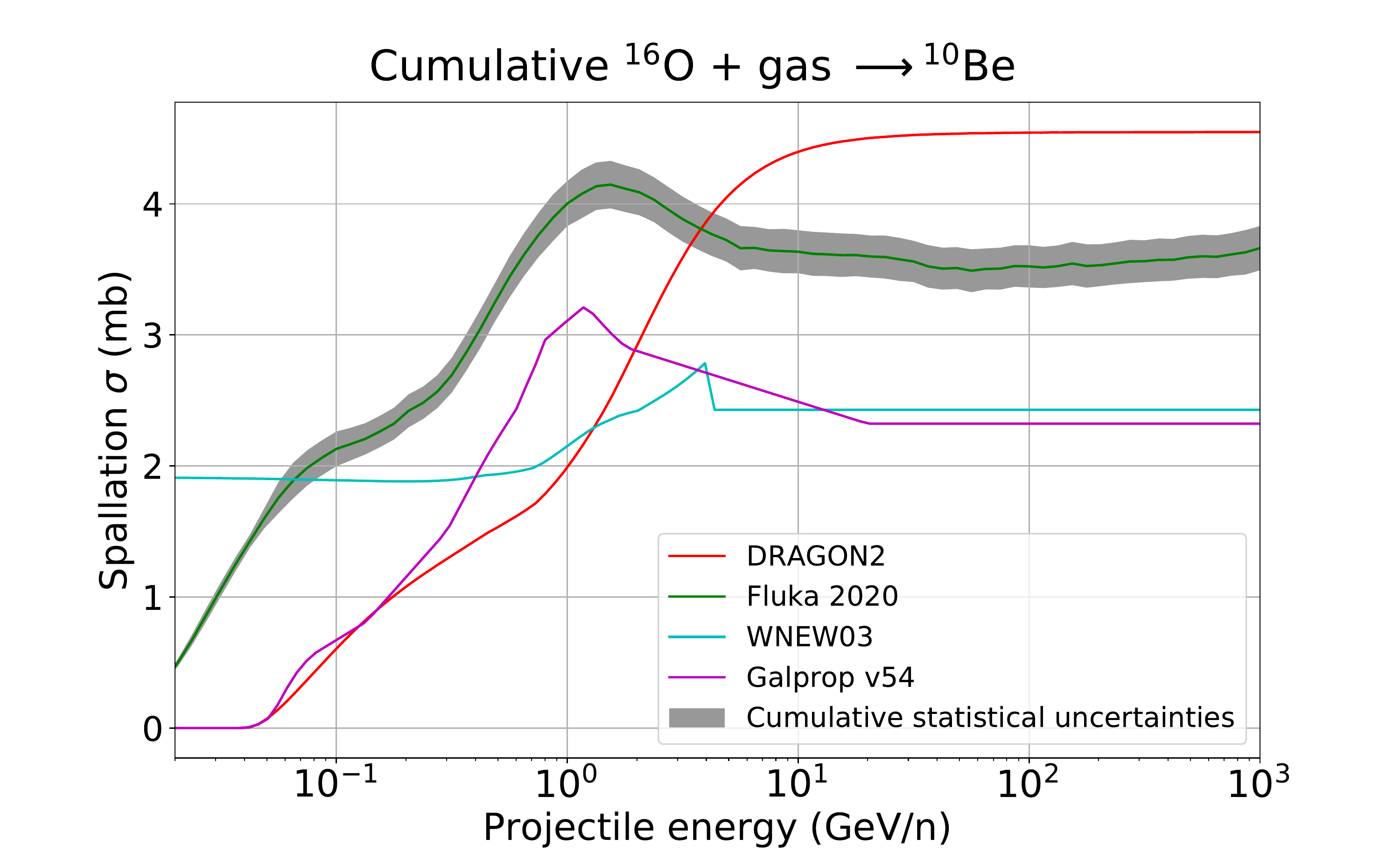}
\includegraphics[width=0.325\textwidth,height=0.17\textheight,clip] {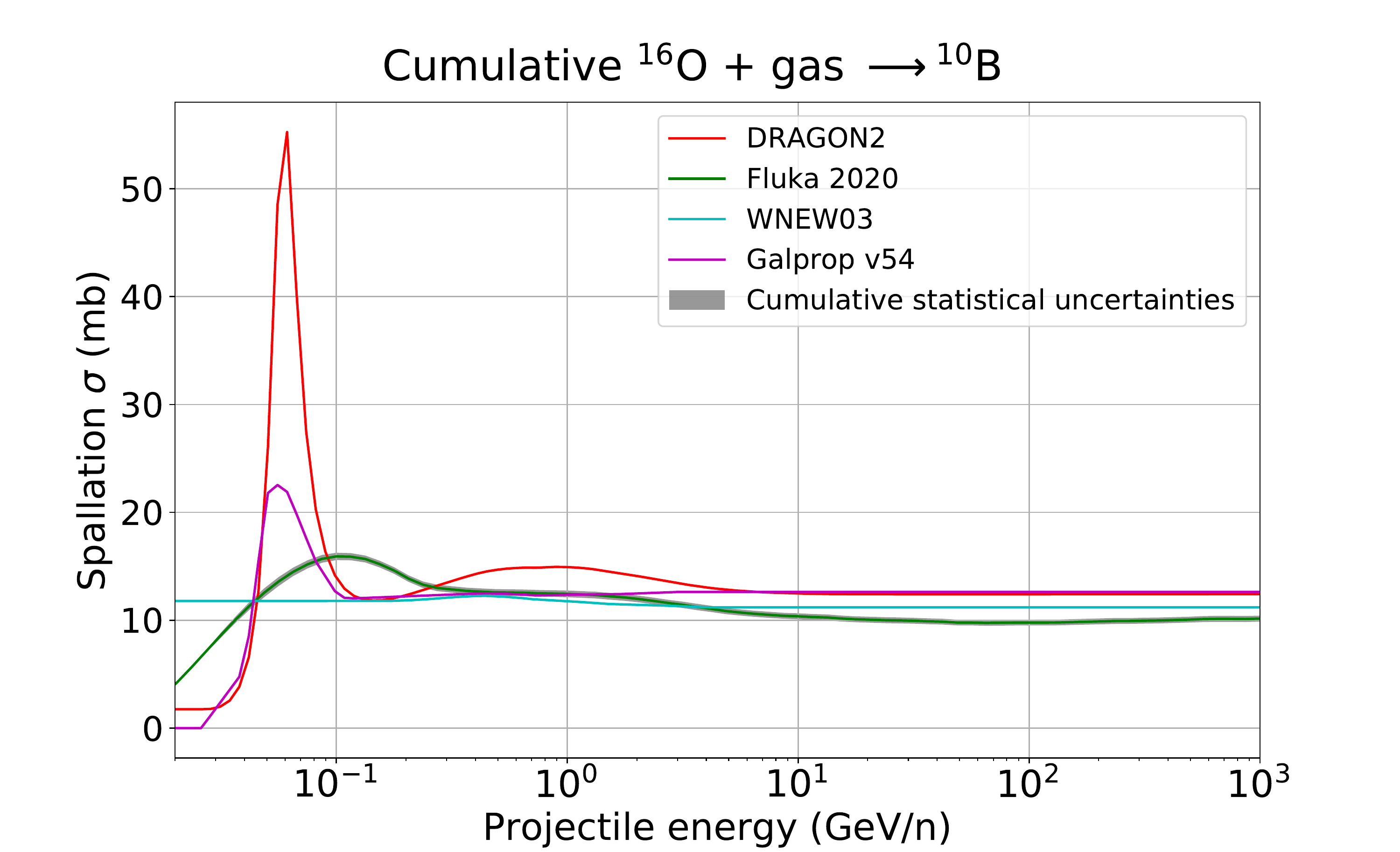}
\includegraphics[width=0.325\textwidth,height=0.17\textheight,clip] {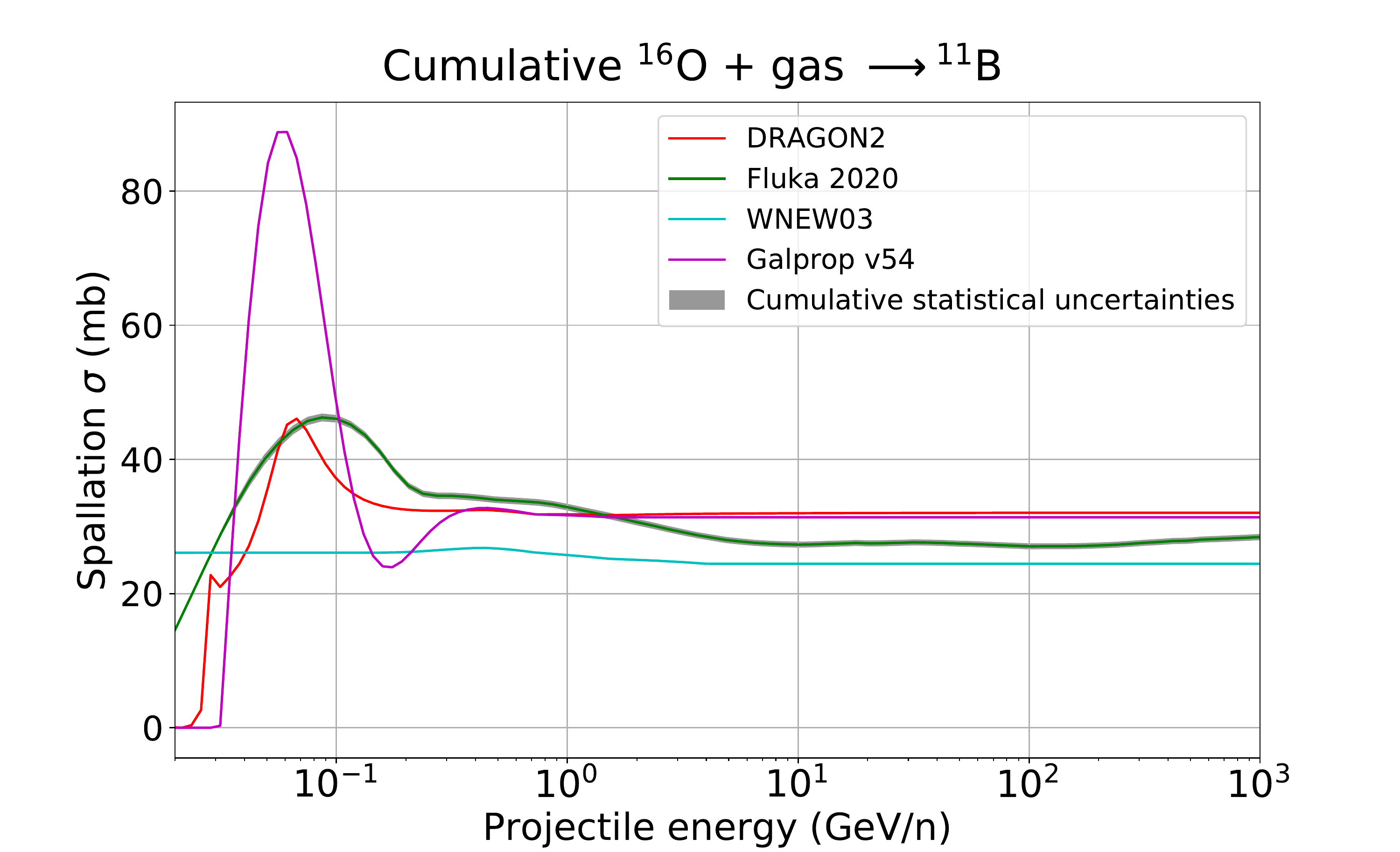}
\end{center}
\caption{\footnotesize Comparison of the spallation cross sections with ISM gas computed with FLUKA and the most widespread parametrisations.}
\label{fig:Tot_XS}
\end{figure*}

In general, these spallation cross sections parametrisations are very close to the FLUKA predictions in the channels of boron production, while the Be and Li channels usually show sizeable differences. 
Finally, the FLUKA total spallation cross sections are shown in Figure~\ref{fig:Tot_XS} in comparison with the other parametrisations. At the end, these are the cross sections that are imported into the propagation code, and are therefore the most important for CR studies. 

There is always a good agreement between all the predictions, being compatible of $\sim30\%$. The largest differences are found in the channels of $^{10}$Be production, where different computations lead to discrepancies even larger than $50\%$. The most significant discrepancies between the FLUKA cross sections and the other extrapolations are found in the production of $^9$Be (up to $\sim 50\%$ discrepancies in the $^{12}$C channel), for most of the primary nuclei as projectiles, although we do not observe any systematic discrepancy for different projectiles (i.e. no systematic excess or deficit is seen in the production of this isotope for the FLUKA predictions with respect to the other parametrisations). Nevertheless, it seems that the inclusive cross sections of the isotopes of boron do show a systematic under-prediction with respect to the other parametrisations, at the level of $10-15\%$ usually, and not only for these main channels. In fact, We stress that, while for the parametrisations based on the existing data, the most reliable secondary nucleus is boron, in the case of the FLUKA computations this is not true.

As a matter of fact, the total cumulative cross sections are not usually shown to compare different parametrisations and computations, even if they are those used in the propagation codes. Indeed, the four cross section models compared give similar predictions and follow different strategies: while the Webber and DRAGON2 parametrisations followed different phenomenological formulations (and even different experimental data sets to adjust them), the GALPROP cross sections make use of a semi-analytical approach, using Monte Carlo codes in combination with their phenomenological parametrisations and giving special attention to the cross sections measurements more reliable in case there are sizeable discrepancies between experiments in a certain energy region and FLUKA completely relies on these nuclear codes. 

In conclusion, we have shown that the FLUKA code is able to make inelastic and spallation cross sections calculations which are, in general, compatible with experimental data and also with the most used and updated parametrisations. Given the scarcity of cross sections measurements (especially for spallation reactions) these nuclear codes, usually dedicated to collider experiments, are also used for CR studies. In the next section, the capability of the FLUKA cross sections data sets to be consistent with CR data and to provide predictions on CR propagation is broadly studied.

\section{Compatibility with cosmic-ray data}
\label{sec:FDRAGON}

The implementation of these new sets of cross sections into the DRAGON code has been performed in order to test the reproducibility of the current CR data for all species. The same diffusion set-up presented in chapter~\ref{sec:XSecs} is used for the simulations performed with the FLUKA cross sections. 

\begin{figure*}[!hpbt]
\begin{center}
\includegraphics[width=0.48\textwidth,height=0.23\textheight,clip] {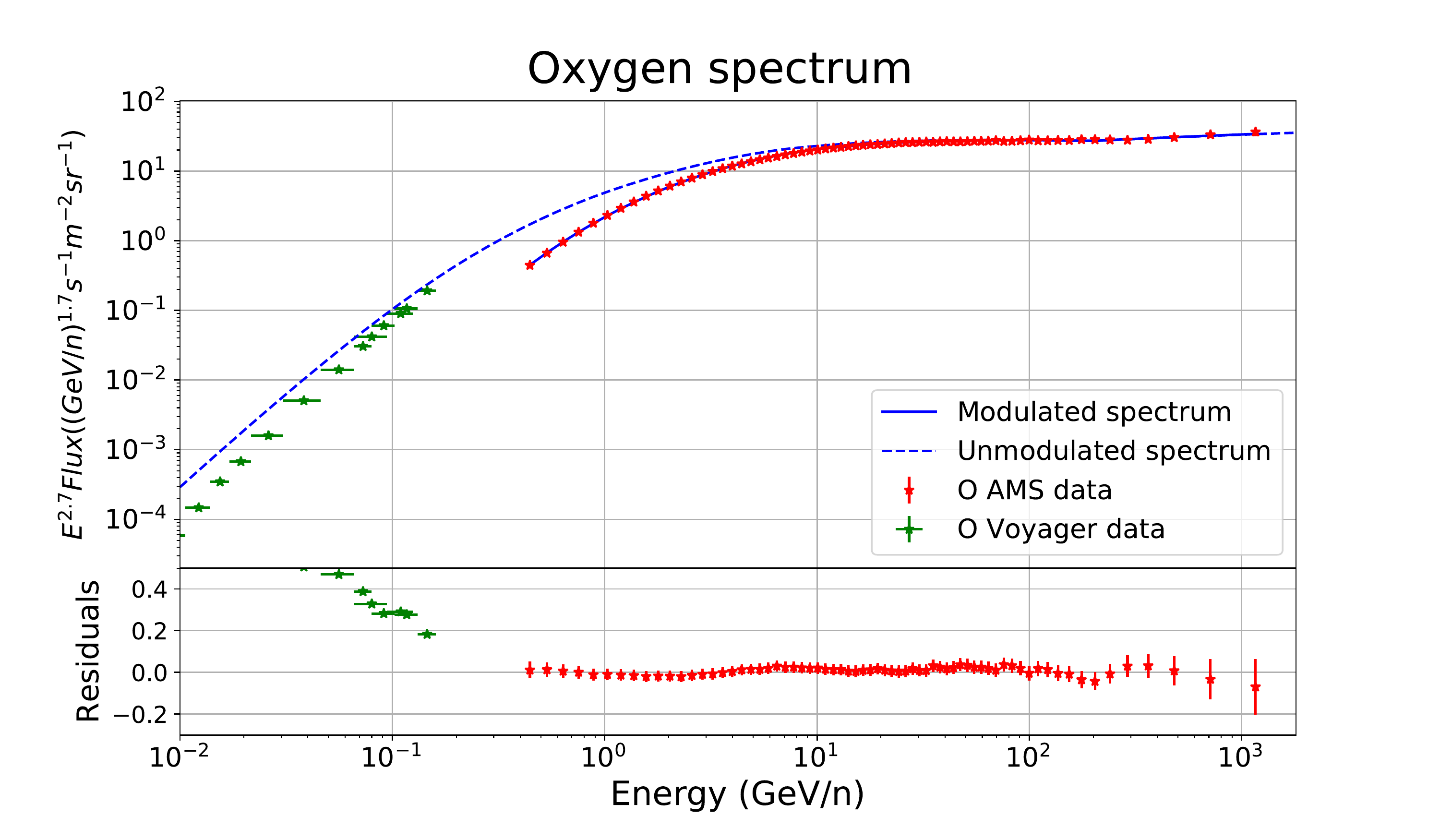} %\hspace{0.5cm}
\includegraphics[width=0.48\textwidth,height=0.23\textheight,clip] {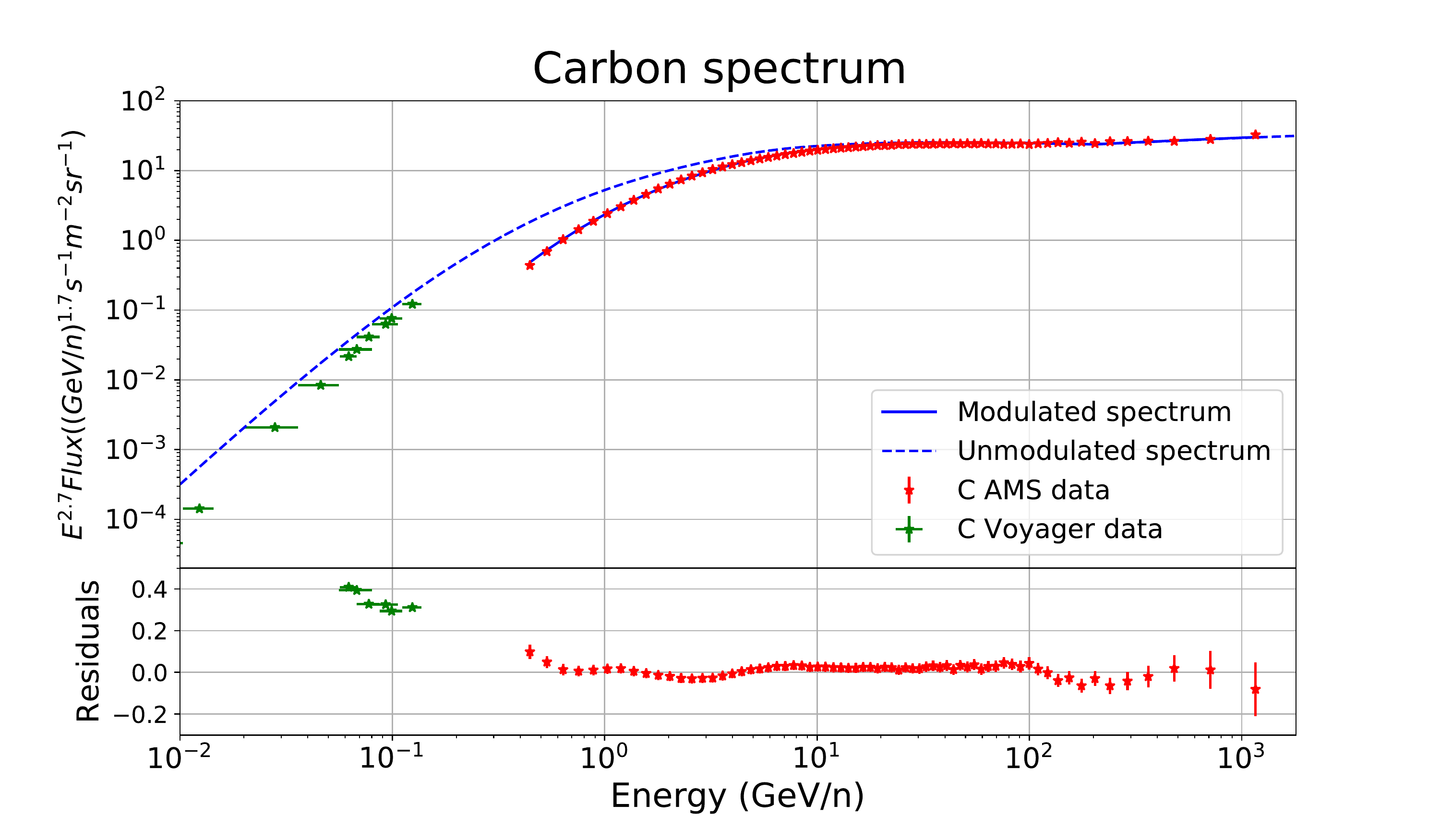} 
%\vspace{-0.2cm}

\includegraphics[width=0.48\textwidth,height=0.23\textheight,clip] {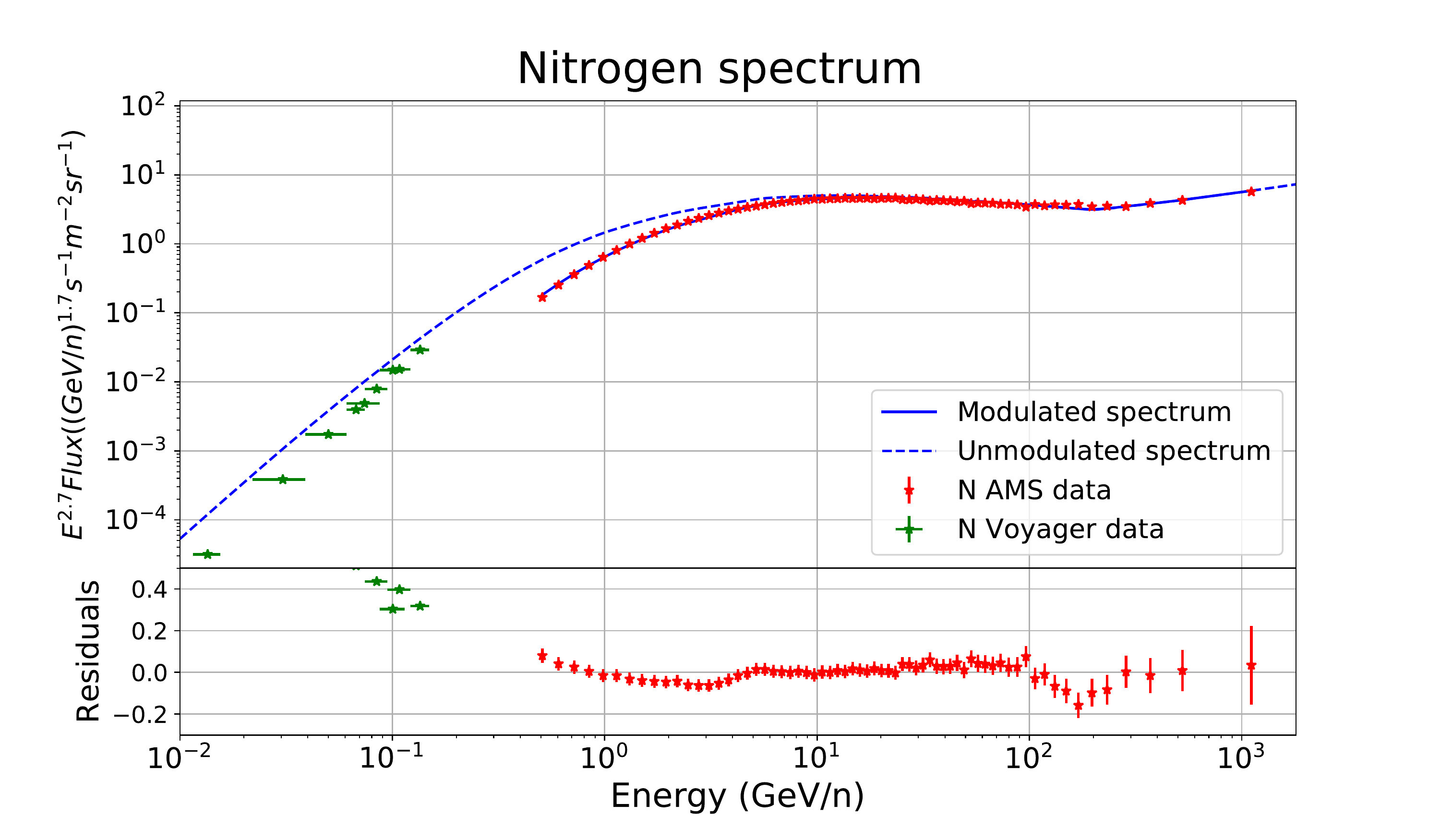} %\hspace{-0.2cm}
\includegraphics[width=0.48\textwidth,height=0.23\textheight,clip] {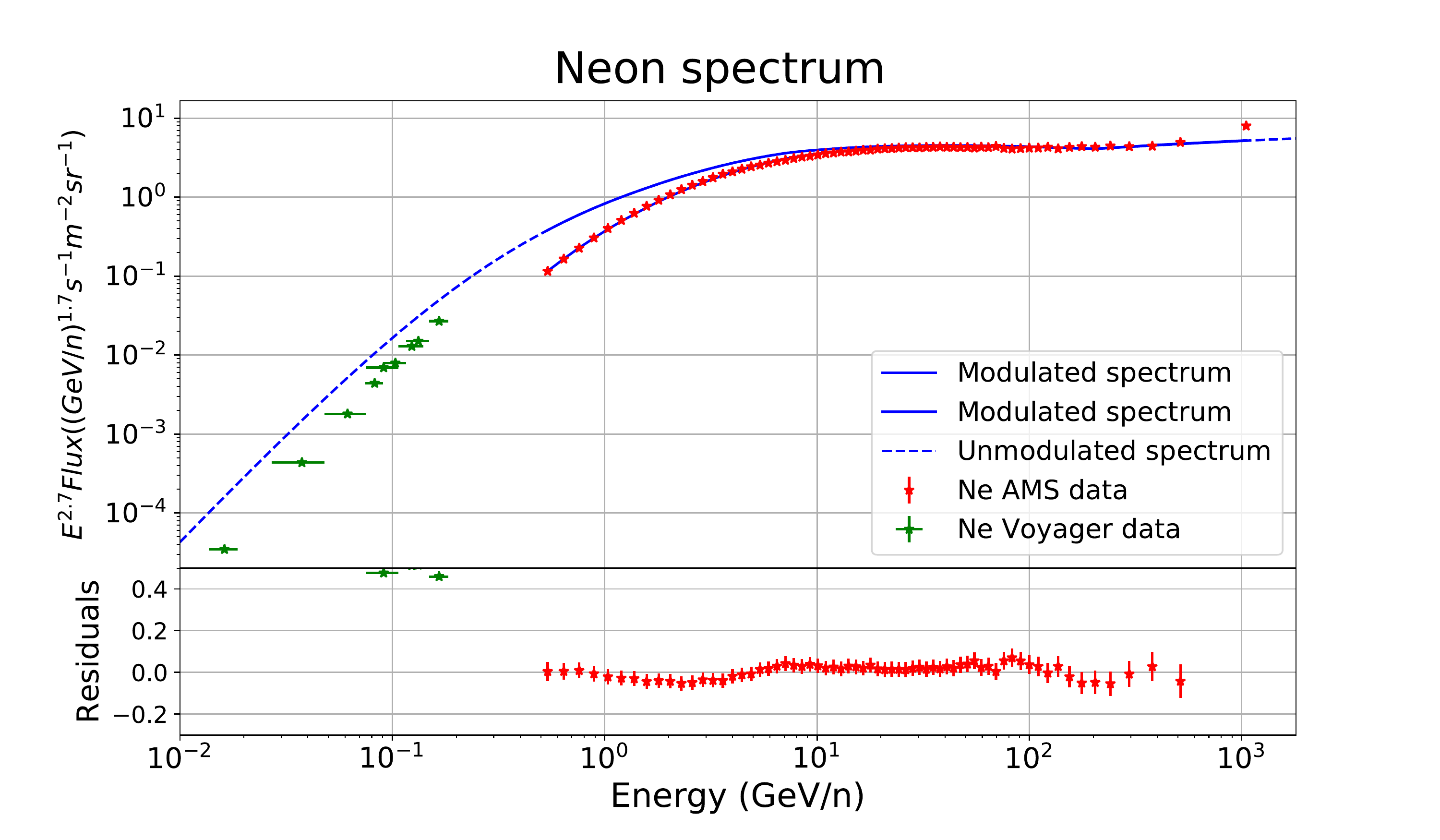}
%\vspace{0.5cm}

\includegraphics[width=0.48\textwidth,height=0.23\textheight,clip] {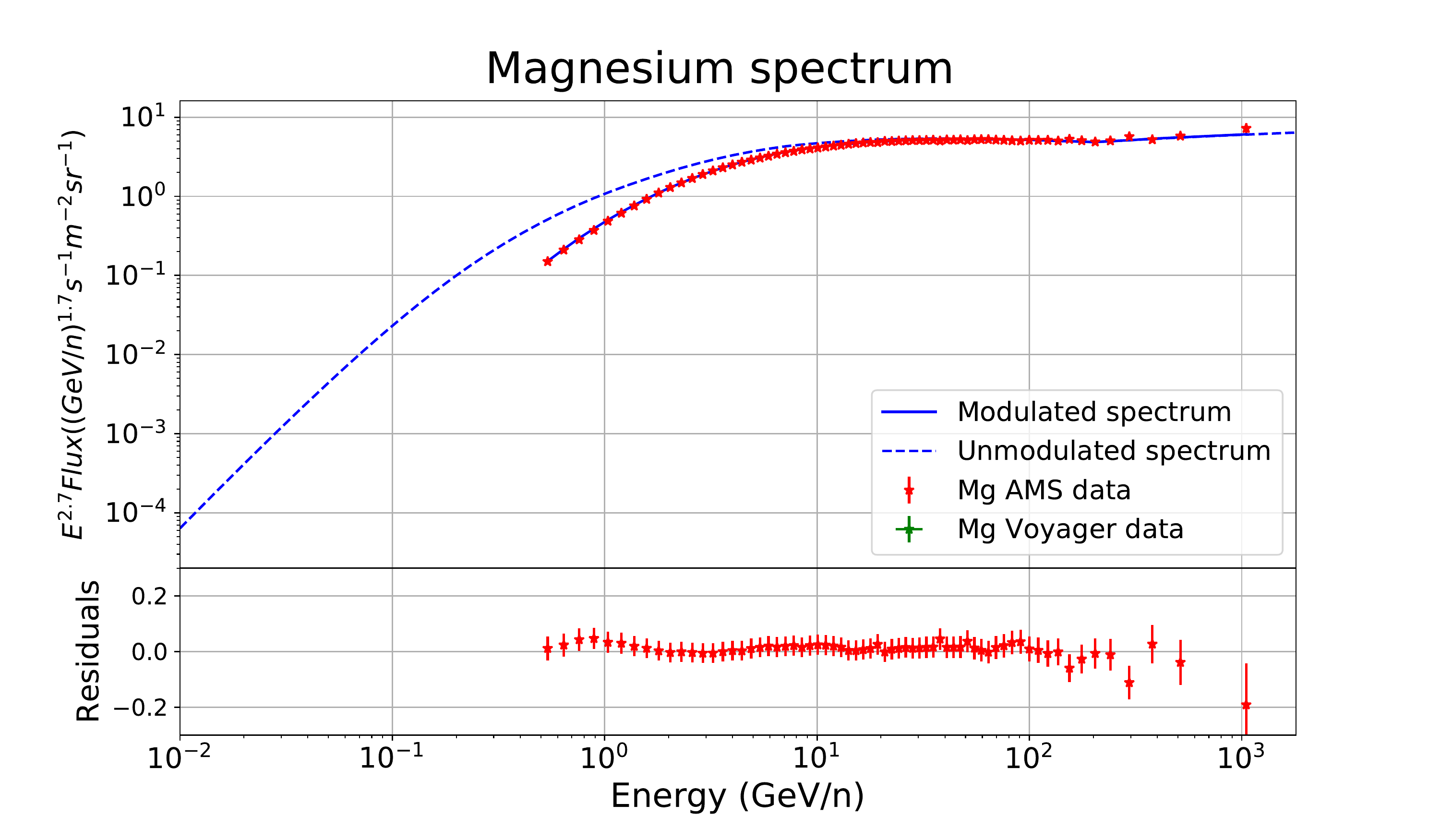} %\hspace{-0.2cm}
\includegraphics[width=0.48\textwidth,height=0.23\textheight,clip] {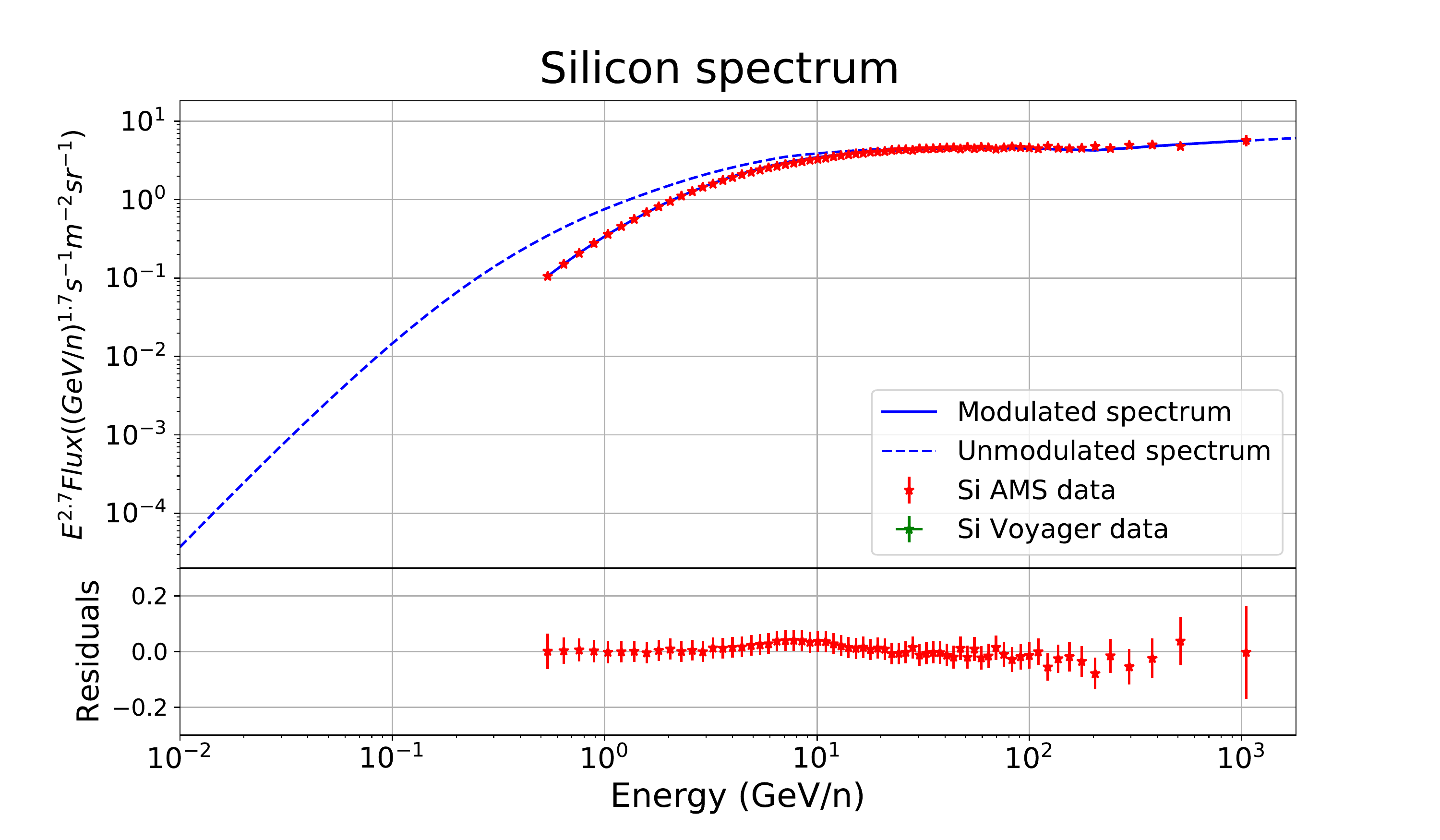}

\includegraphics[width=0.48\textwidth,height=0.23\textheight,clip] {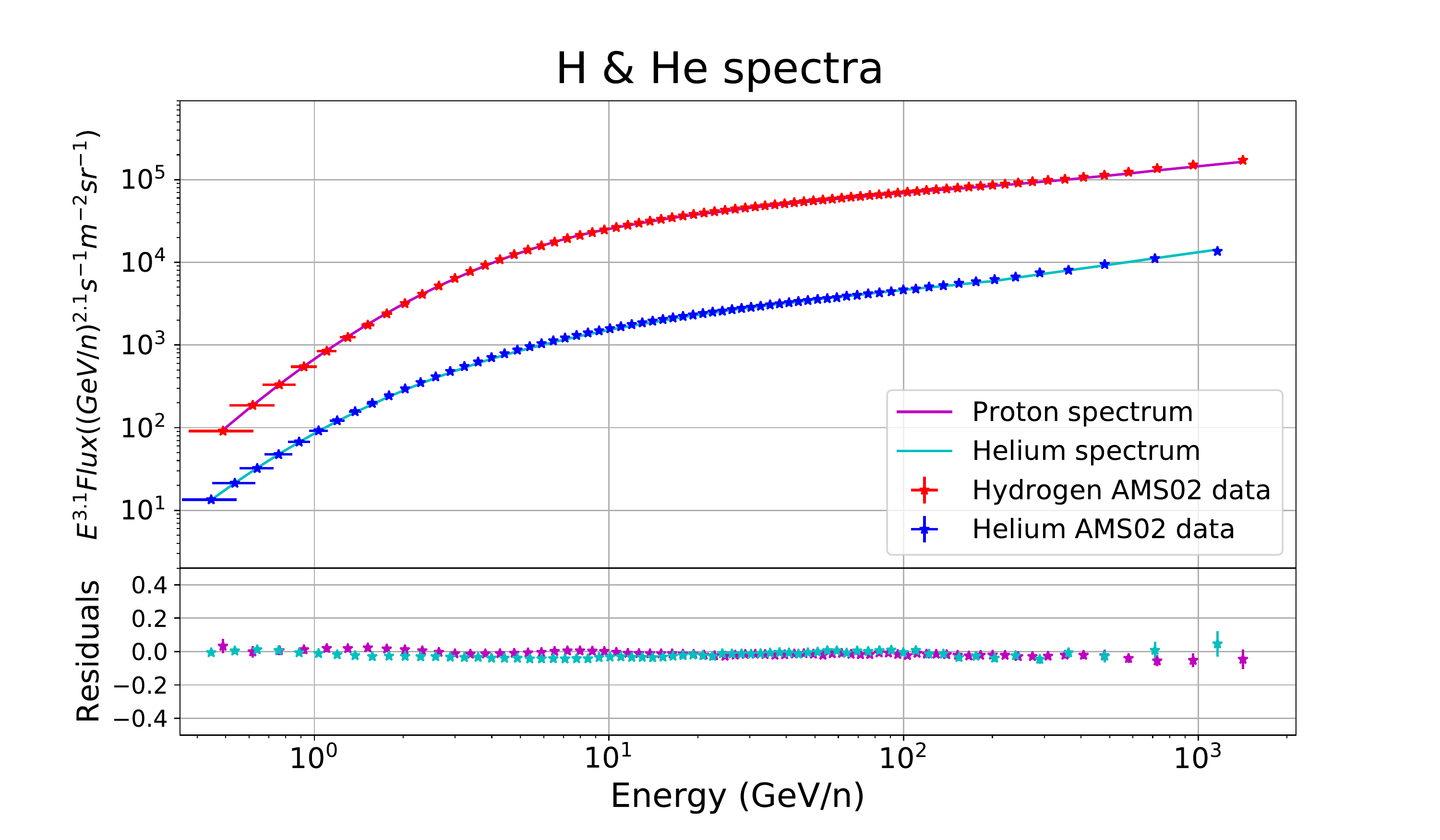}

\end{center}
\caption{\footnotesize Spectra of the injected nuclei, simulated using the FLUKA (inelastic and inclusive) cross sections, compared with the AMS-02 data.}
\label{fig:primFluk}
\end{figure*}

First of all, as we see from Figure~\ref{fig:primFluk}, the spectra of the main primary CRs measured by AMS-02 can be well reproduced. 

\begin{figure*}[!hbt]
\begin{center}
\includegraphics[width=0.335\textwidth,height=0.188\textheight,clip] {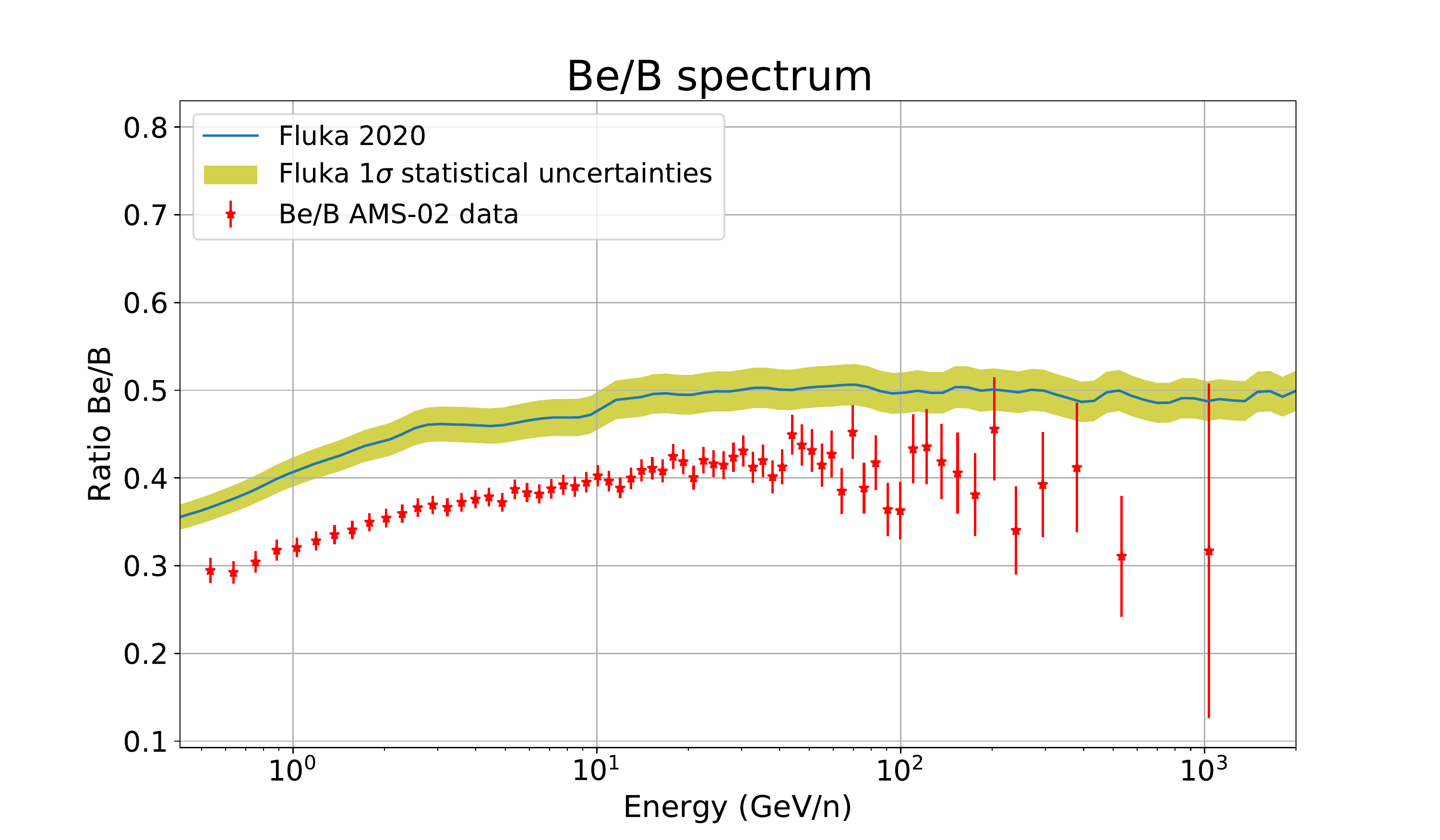} \hspace{-0.25cm}
\includegraphics[width=0.335\textwidth,height=0.188\textheight,clip] {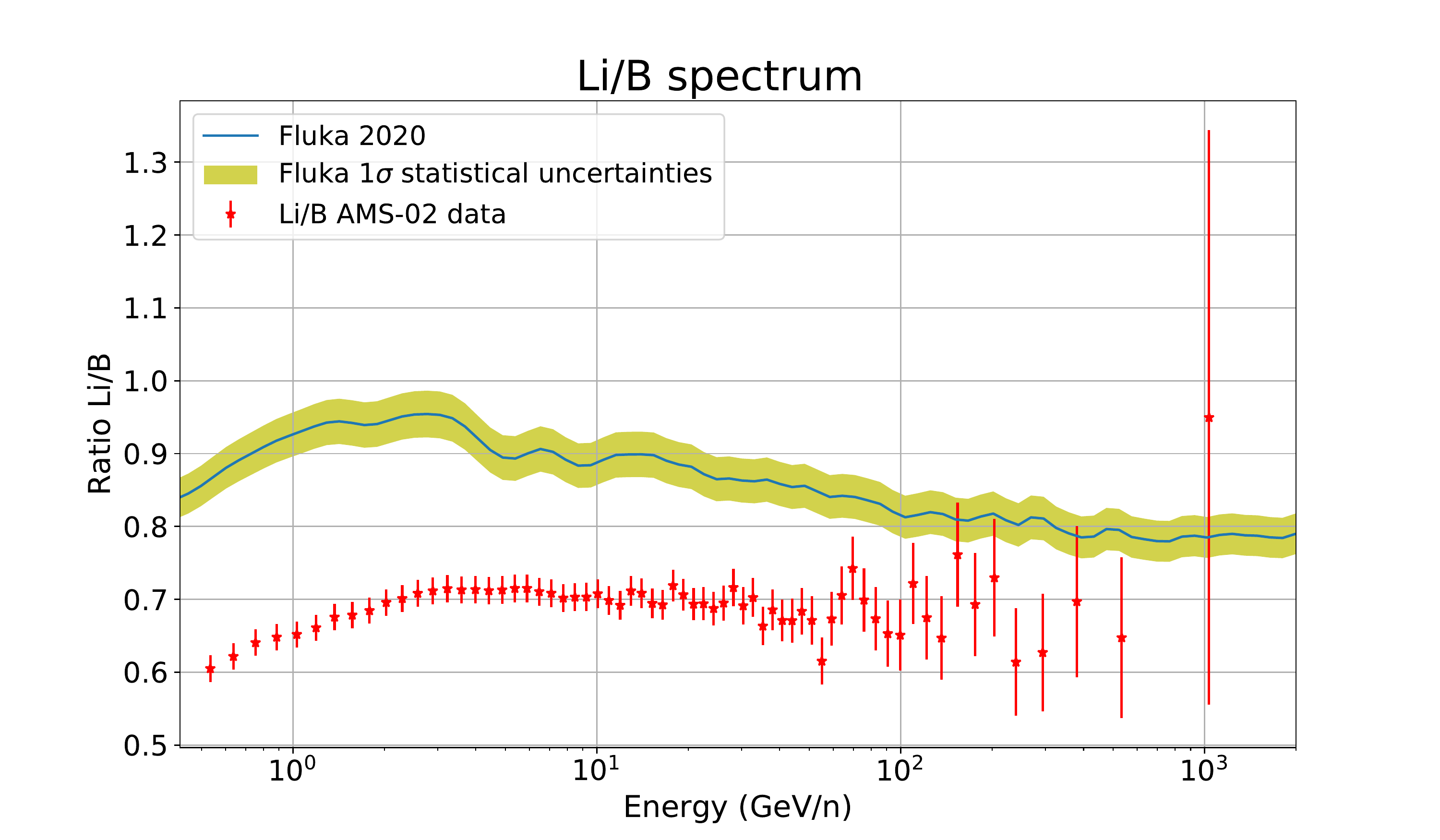} 
\hspace{-0.25cm}
\includegraphics[width=0.335\textwidth,height=0.188\textheight,clip] {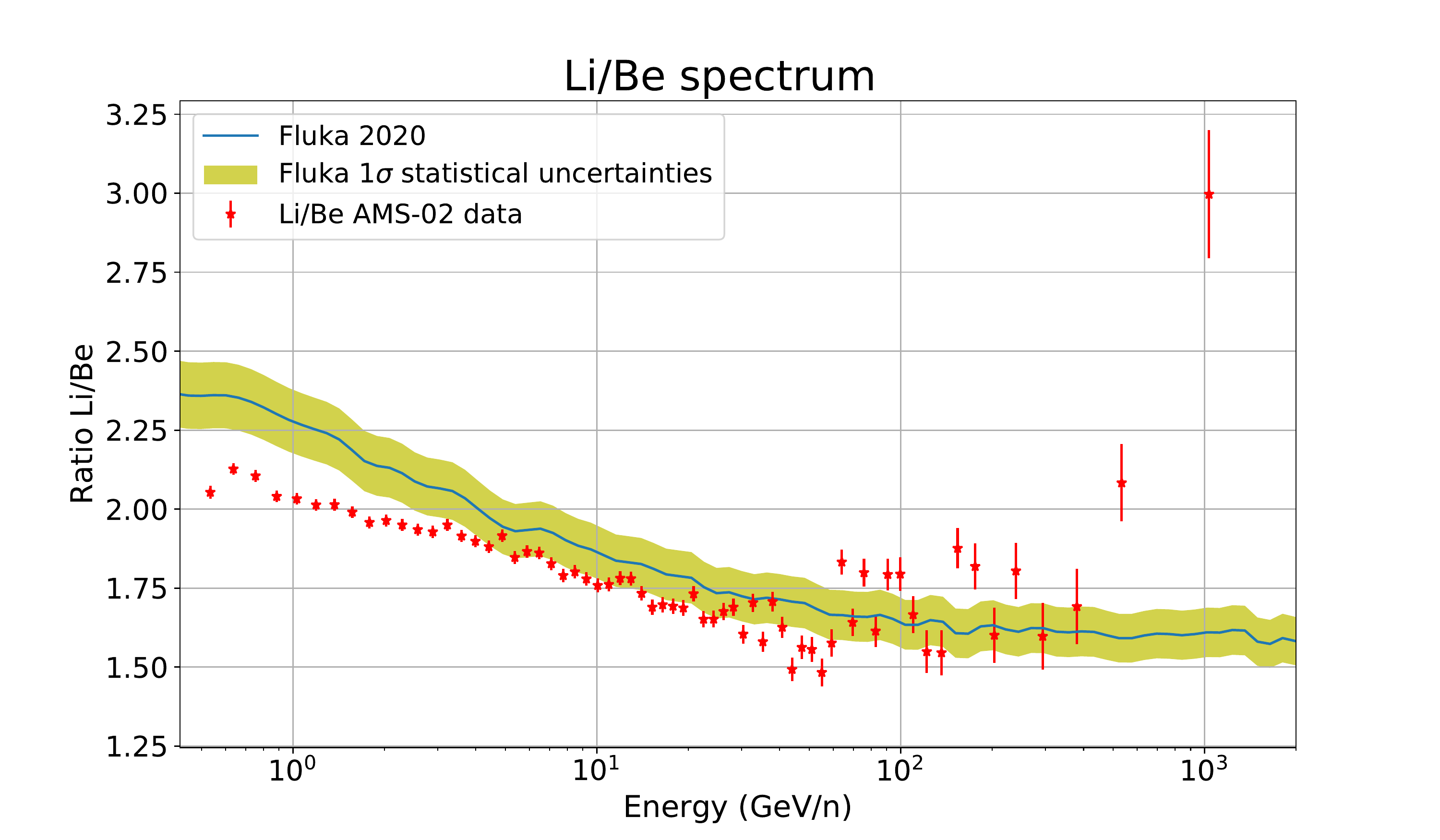}

\includegraphics[width=0.335\textwidth,height=0.189\textheight,clip] {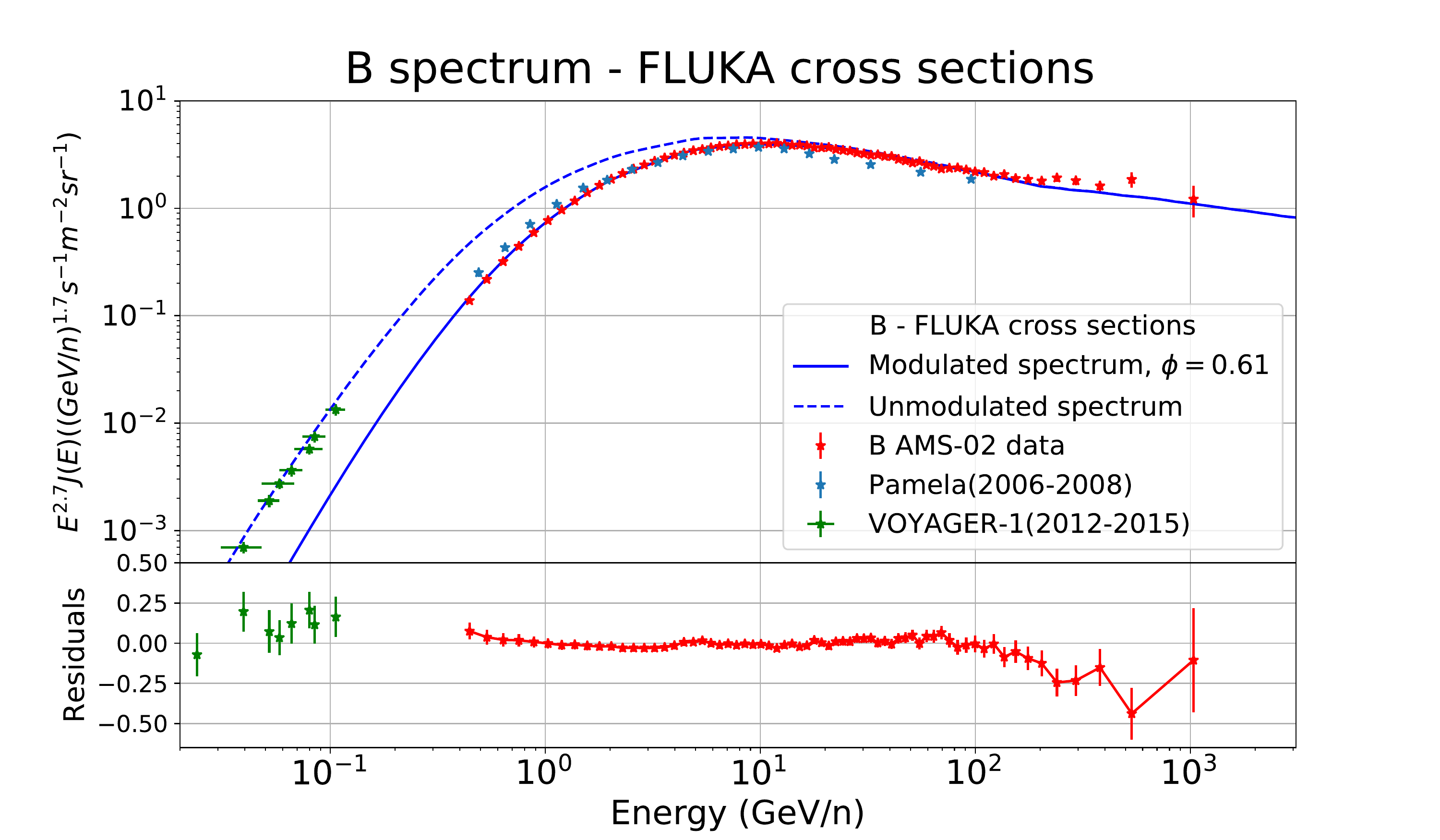} \hspace{-0.25cm}
\includegraphics[width=0.335\textwidth,height=0.189\textheight,clip] {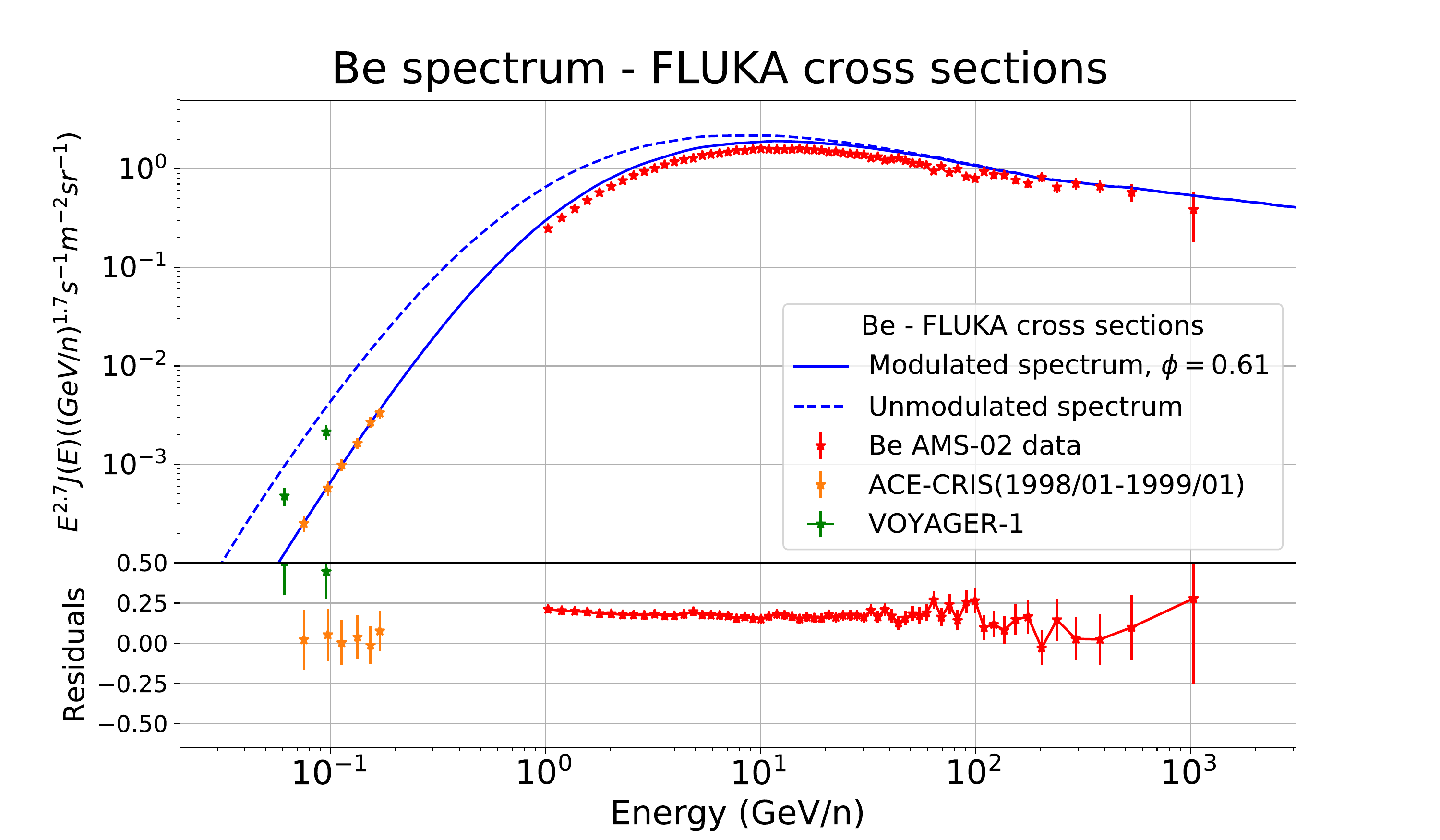}
\hspace{-0.25cm}
\includegraphics[width=0.335\textwidth,height=0.189\textheight,clip] {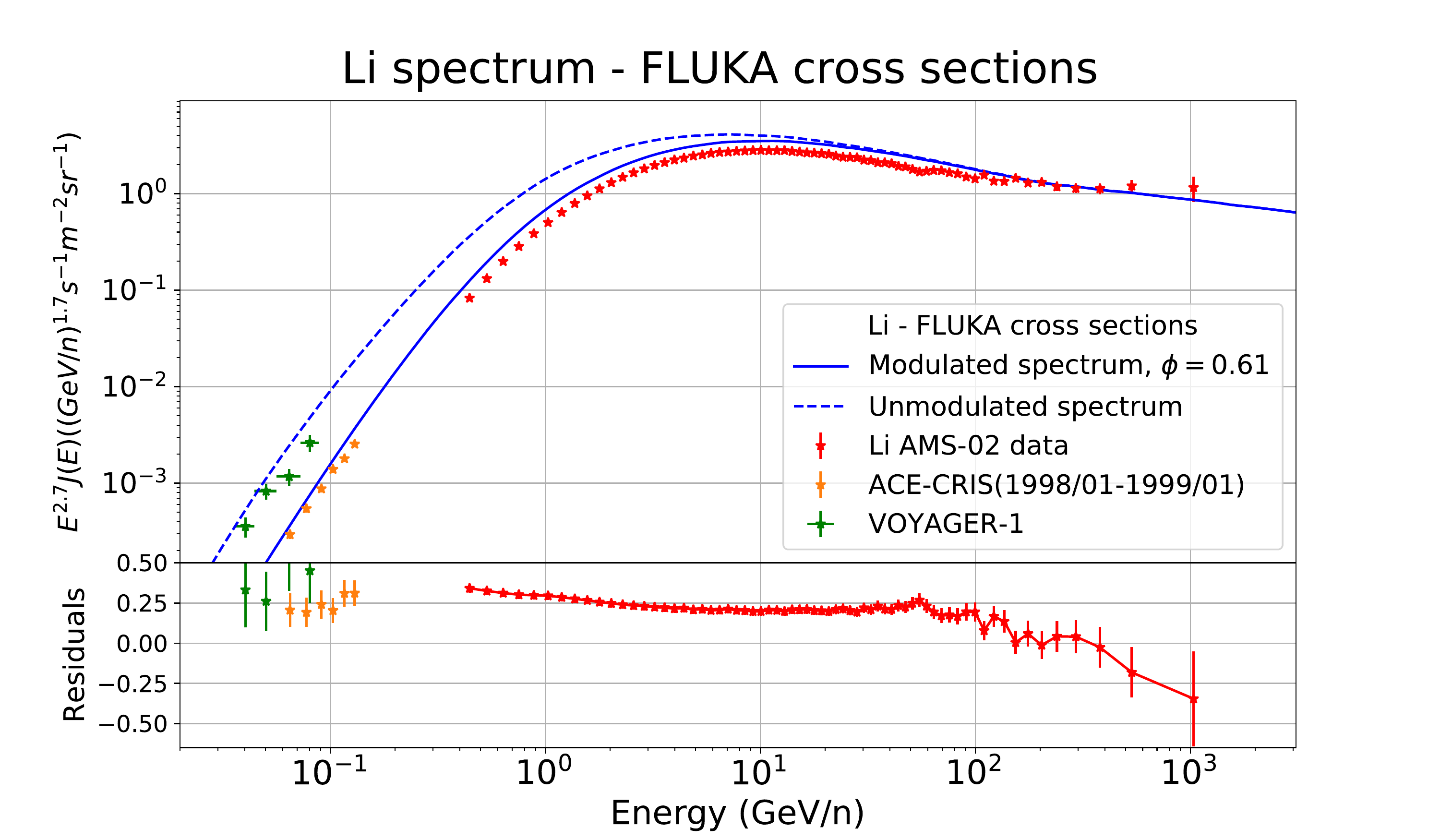}
\end{center}
\caption{\footnotesize Top row: secondary-over-secondary flux ratios for Li, Be and B obtained with the FLUKA computations, together with the band of statistical uncertainties due to the spallation cross sections calculation (since this is a MC computation). Bottom row: B, Be and Li fluxes obtained from the simulation, compared with experimental data.}
\label{fig:secsec_Fluka}
\end{figure*}

As in the previous chapter, the primary injection spectra and the diffusion parameters are all adjusted for the model to reproduce the B/C flux ratio (see Fig.~\ref{fig:Fsec_prim}) and the halo size obtained in section~\ref{sec:halo_Fluka}. The spectra of the secondary CRs Li, Be and B obtained with these diffusion parameters, together with their respective secondary-over-secondary flux ratios, are shown in Figure~\ref{fig:secsec_Fluka}.

As we can see, the predicted fluxes of Li and Be exhibit residuals of 20\% with respect to the AMS-02 data, as the flux ratios Be/B and Li/B, while the residuals found in the Li and Be spectra are smaller than 5\% (see Fig.~\ref{fig:Fhalos}). These discrepancies are reasonable and compatible with the discrepancies that arise in the simulations when using other cross sections parametrisations.

\begin{figure*}[!bhpt]
\begin{center}
\includegraphics[width=0.336\textwidth,height=0.19\textheight,clip] {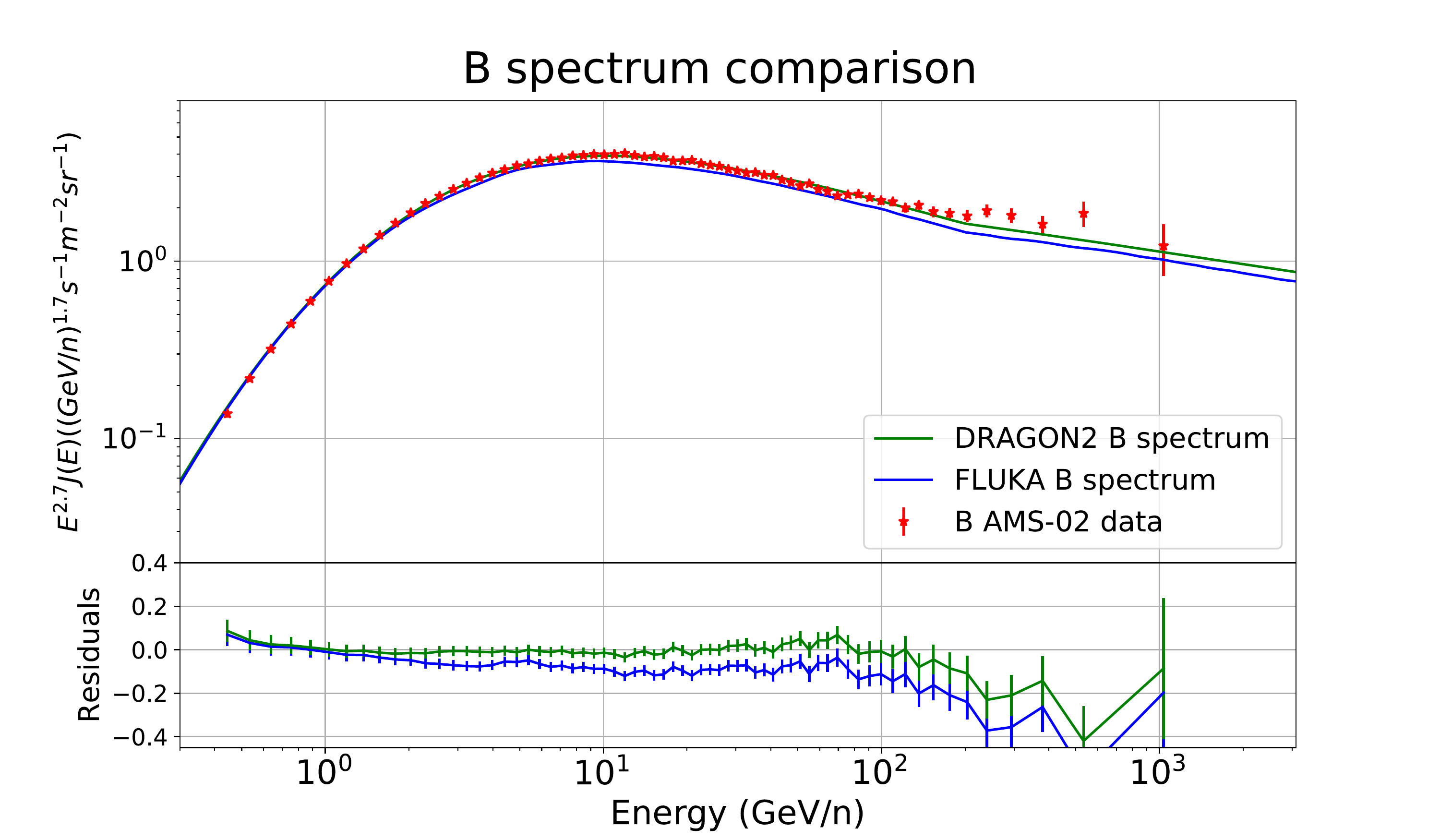} \hspace{-0.27cm}
\includegraphics[width=0.336\textwidth,height=0.19\textheight,clip] {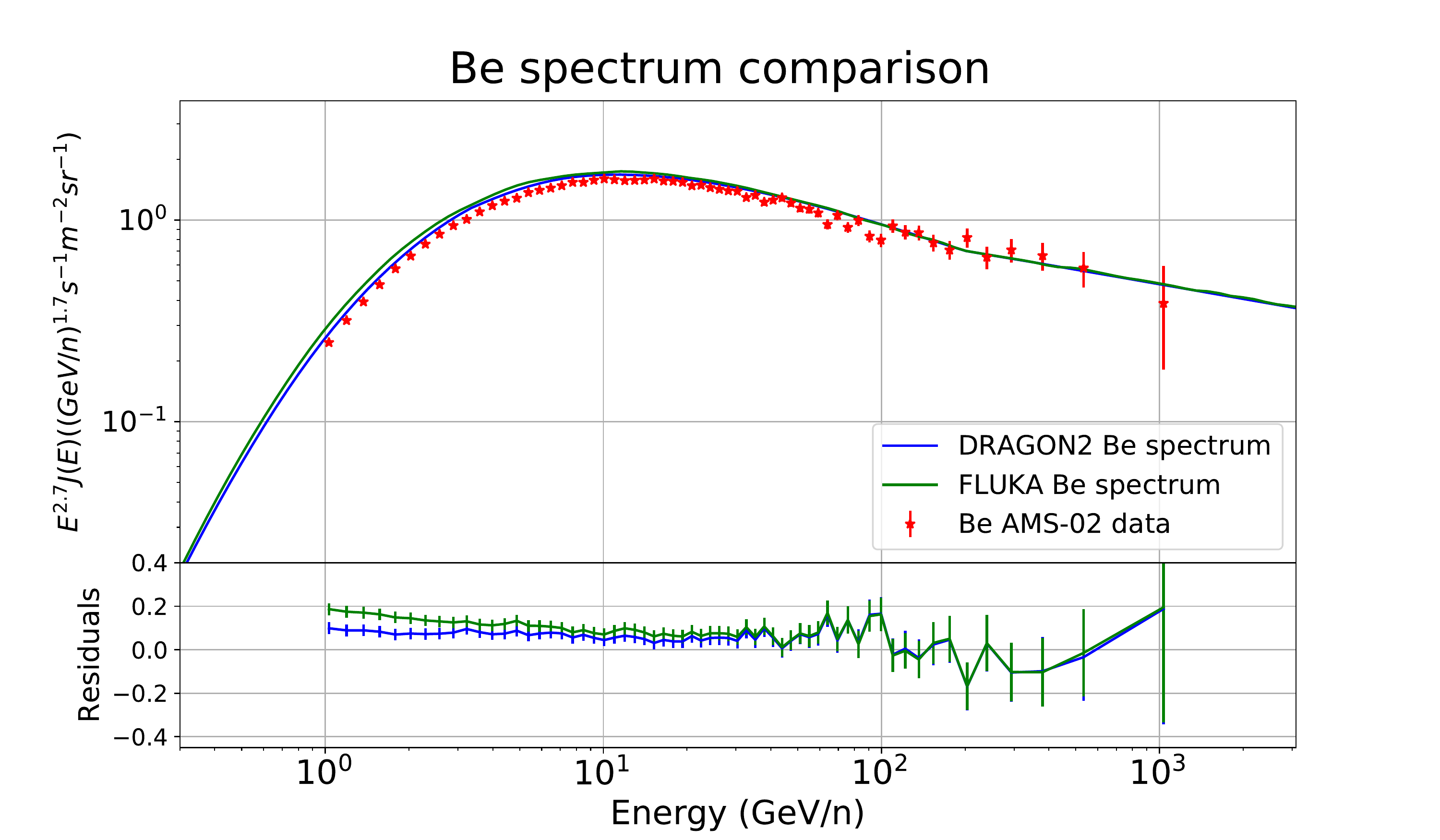} 
\hspace{-0.27cm}
\includegraphics[width=0.336\textwidth,height=0.19\textheight,clip] {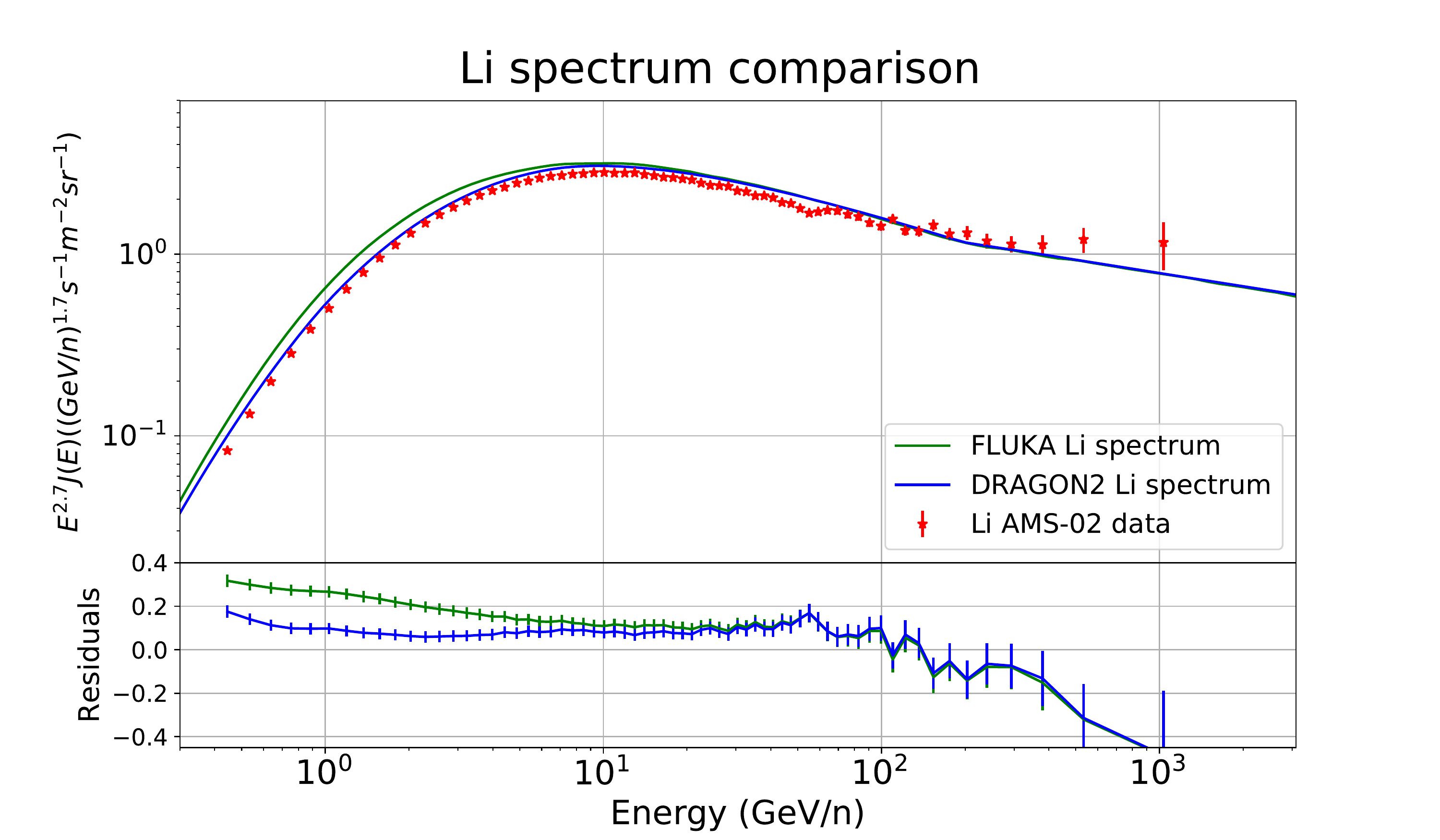}
\end{center}
\caption{\footnotesize Comparison between the FLUKA and DRAGON2 predicted fluxes of B, Li and Be using the diffusion coefficient obtained in the B/C fit with the DRAGON2 cross sections (table~\ref{tab:diff_params}). The AMS-02 data and the residuals of the predicted fluxes with respect to the data are also shown.}
\label{fig:FvsDR}
\end{figure*}

\begin{figure*}[!thb]
\begin{center}
\includegraphics[width=0.465\textwidth,height=0.21\textheight,clip] {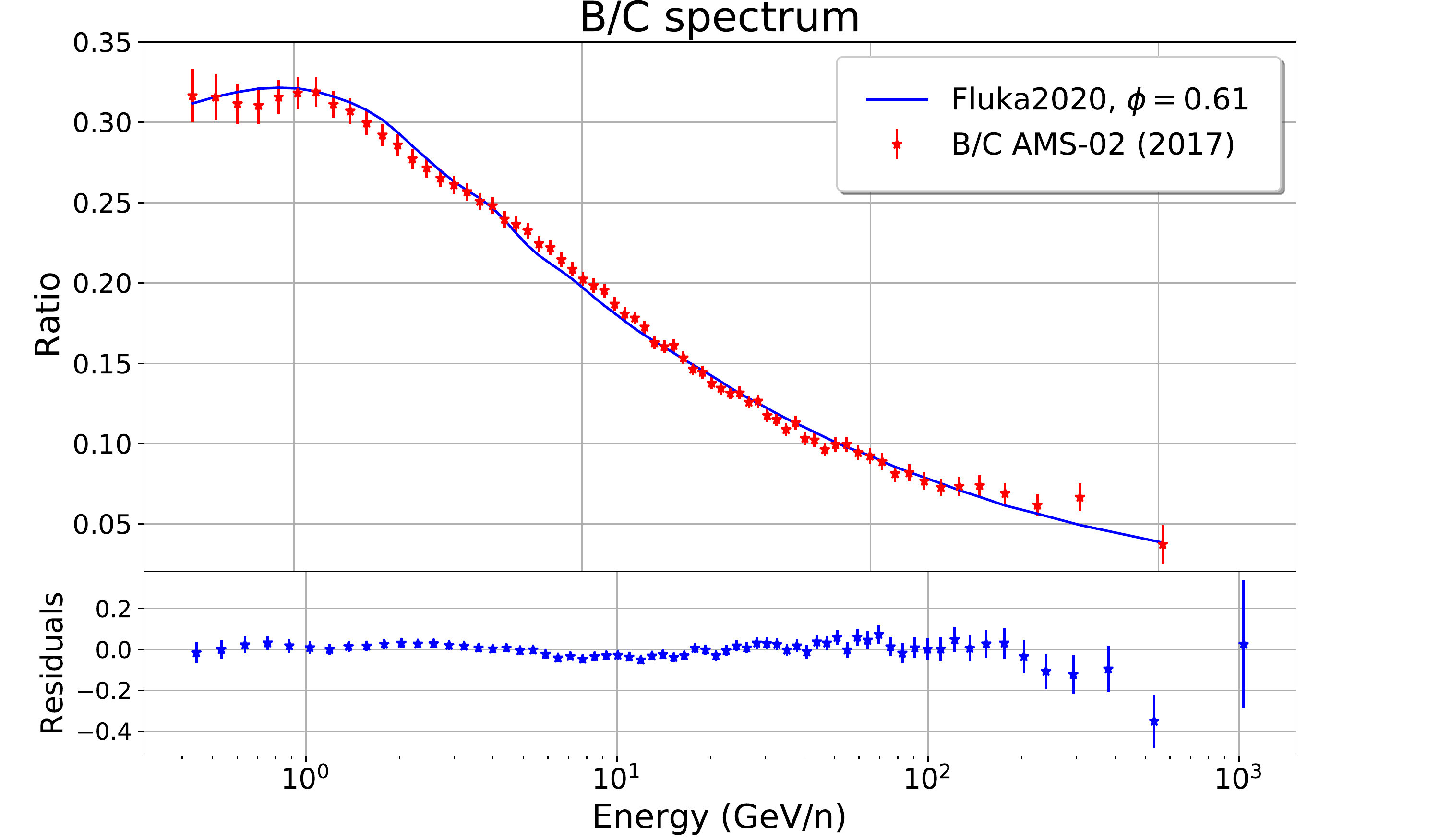} 
\includegraphics[width=0.465\textwidth,height=0.21\textheight,clip] {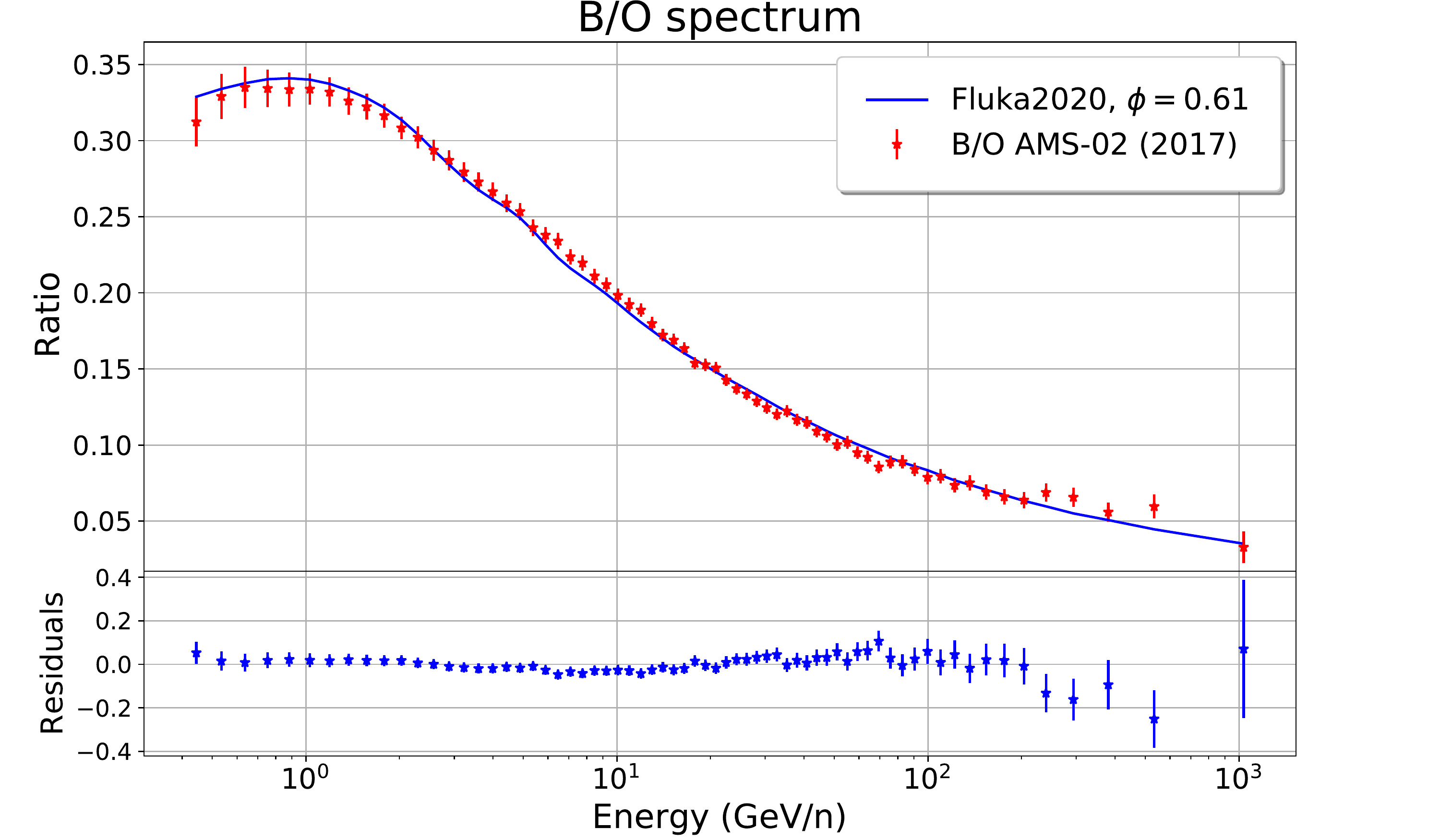} 

\vspace{0.35cm}

\includegraphics[width=0.46\textwidth,height=0.21\textheight,clip] {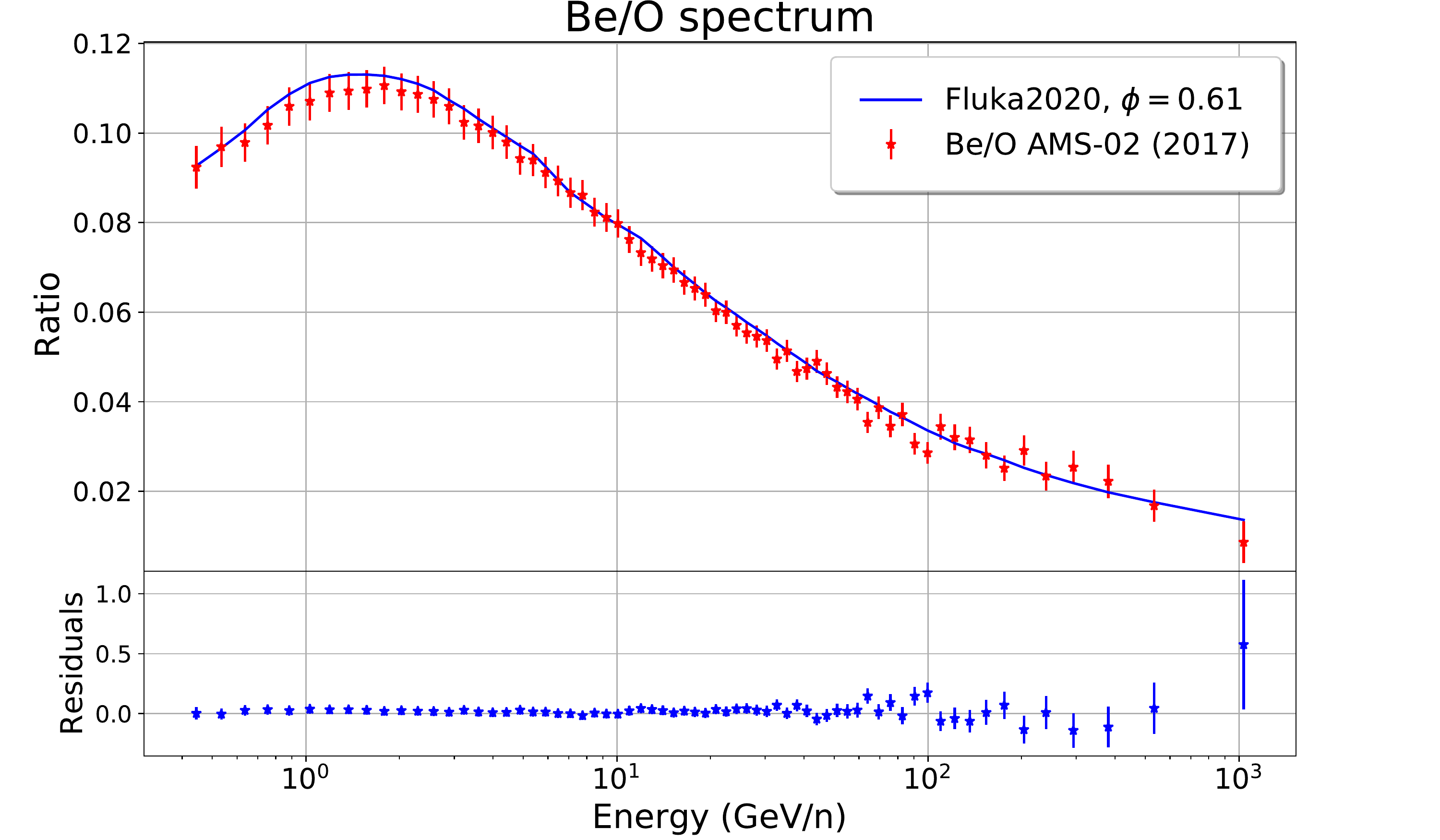}
\includegraphics[width=0.46\textwidth,height=0.21\textheight,clip] {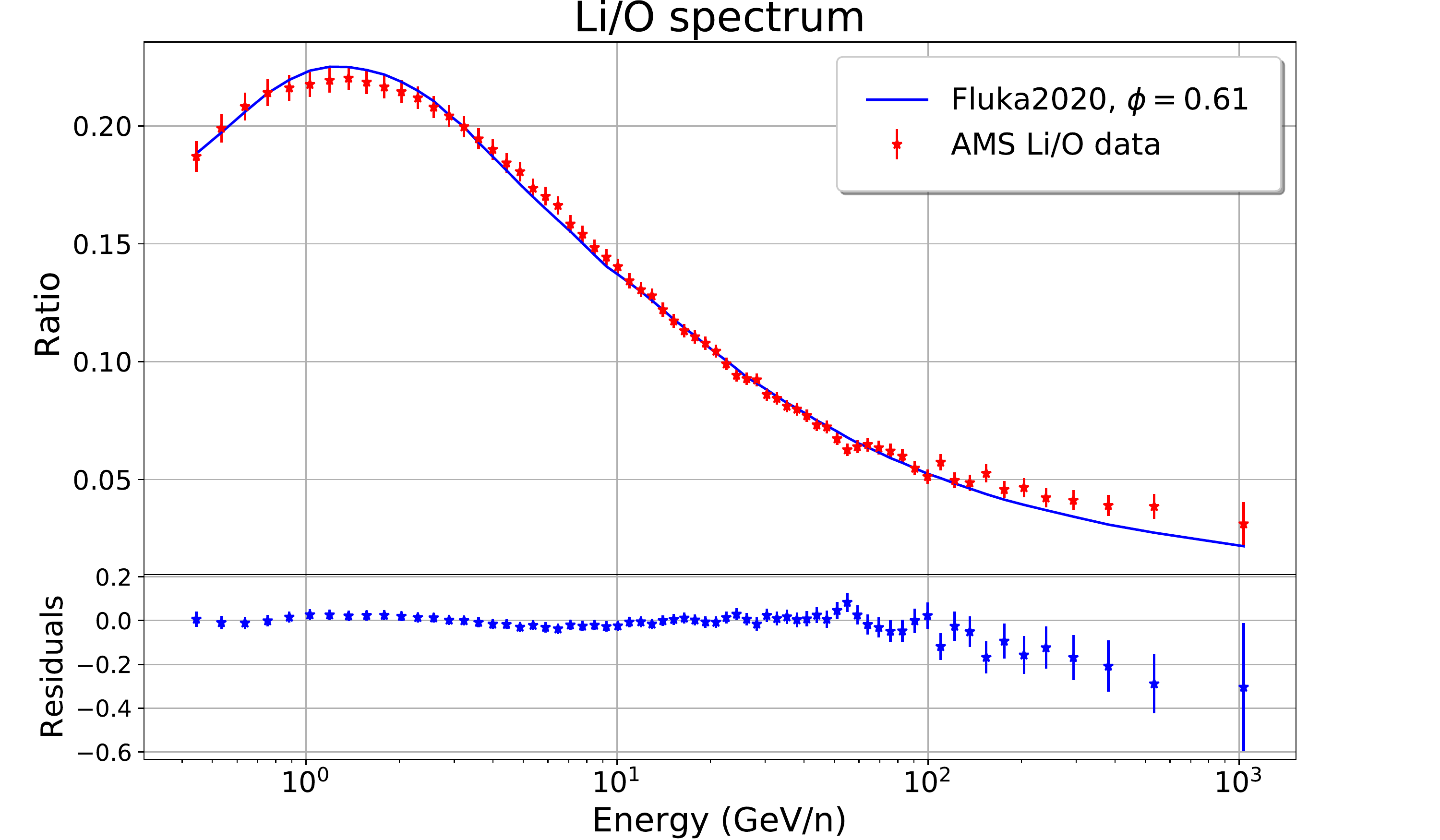}

\end{center}
\caption{\footnotesize Secondary-over-primary B/C, B/O, Be/O and Li/O spectra obtained with the FLUKA tables implemented in DRAGON compared with AMS-02 data. The same diffusion coefficient is used for the B/C and B/O spectra, while a different one is used for the Li/O and Be/O spectra.}
\label{fig:Fsec_prim}
\end{figure*}

In addition, a direct comparison between the FLUKA model against the DRAGON2 cross sections is shown for the predicted fluxes of B, Li and Be in Figure~\ref{fig:FvsDR}, using the diffusion parameters found with the DRAGON2 cross sections (see table~\ref{tab:diff_params}). Differences around 10\% are found for the B spectrum, while the Li and Be spectra show relatively good agreement above a few GeV/n. Unlike the cross sections parametrisations, which are less uncertain for the spallation channels of boron production, the FLUKA computations fully relies on theoretical models. Therefore, the discrepancy found in the boron spectrum suggests that the spallation cross sections for boron production in FLUKA are slightly underestimated, as already pointed out (see the discussion regarding Fig.~\ref{fig:Tot_XS}), that can also explain the predicted secondary-over-secondary ratios in Figure~\ref{fig:secsec_Fluka}.

Nevertheless, the most interesting observables which must be reproduced by any CR simulation are the secondary-to-primary ratios, which, as mentioned in the former chapters, depend almost exclusively on the diffusion coefficient and on the inclusive (production) cross sections ($\frac{J_{sec}}{J_{prim}}(E) \propto \sigma (E)/D(E)$). The B/C, B/O, Li/O and Be/O ratios are displayed in Figure~\ref{fig:Fsec_prim}. The same diffusion coefficient is used to fit the AMS-02 B/O and B/C ratios, while different ones are needed to reproduce the Be/O and Li/O ratios (the detailed best fit analysis of these ratios will be performed in chapter~\ref{sec:4}). The parameters of the diffusion coefficients used in each case are summarized in table \ref{tab:Fdiff_params}.

\begin{table}[!thb]
\centering
\resizebox*{0.45\columnwidth}{0.13\textheight}{
\begin{tabular}{|lccc|}
  \multicolumn{4}{c}{\hspace{0.3cm}\large} \\ \hline & \textbf{ Boron}  & \hspace{0.2 cm}\textbf{Beryllium} & \hspace{0.2 cm}\textbf{Lithium}\\ %\toprule
  \hline
{$D_0$} ($10^{28} cm^{2} s^{-1}$) & 8.35 & 10.1 & 11.7\\
{$v_A$} ($km/s$) & 19. & 12.5 & 21.\\
{$\eta$}    & -1. & -2. & -2.1\\      
{$\delta$}    & 0.43 & 0.43 & 0.41\\   
\hline
\end{tabular}
}
\vspace{0.1cm}
\caption{\footnotesize Diffusion parameters used to reproduce the AMS-02 data secondary-over-primary ratios shown in Figure~\ref{fig:Fsec_prim}. }
\label{tab:Fdiff_params}
\end{table}

\newpage
\subsection{Study of variations on the halo size with the FLUKA cross sections}
\label{sec:halo_Fluka}

\begin{figure*}[!bht]
\begin{center}
\includegraphics[width=0.33\textwidth,height=0.19\textheight,clip] {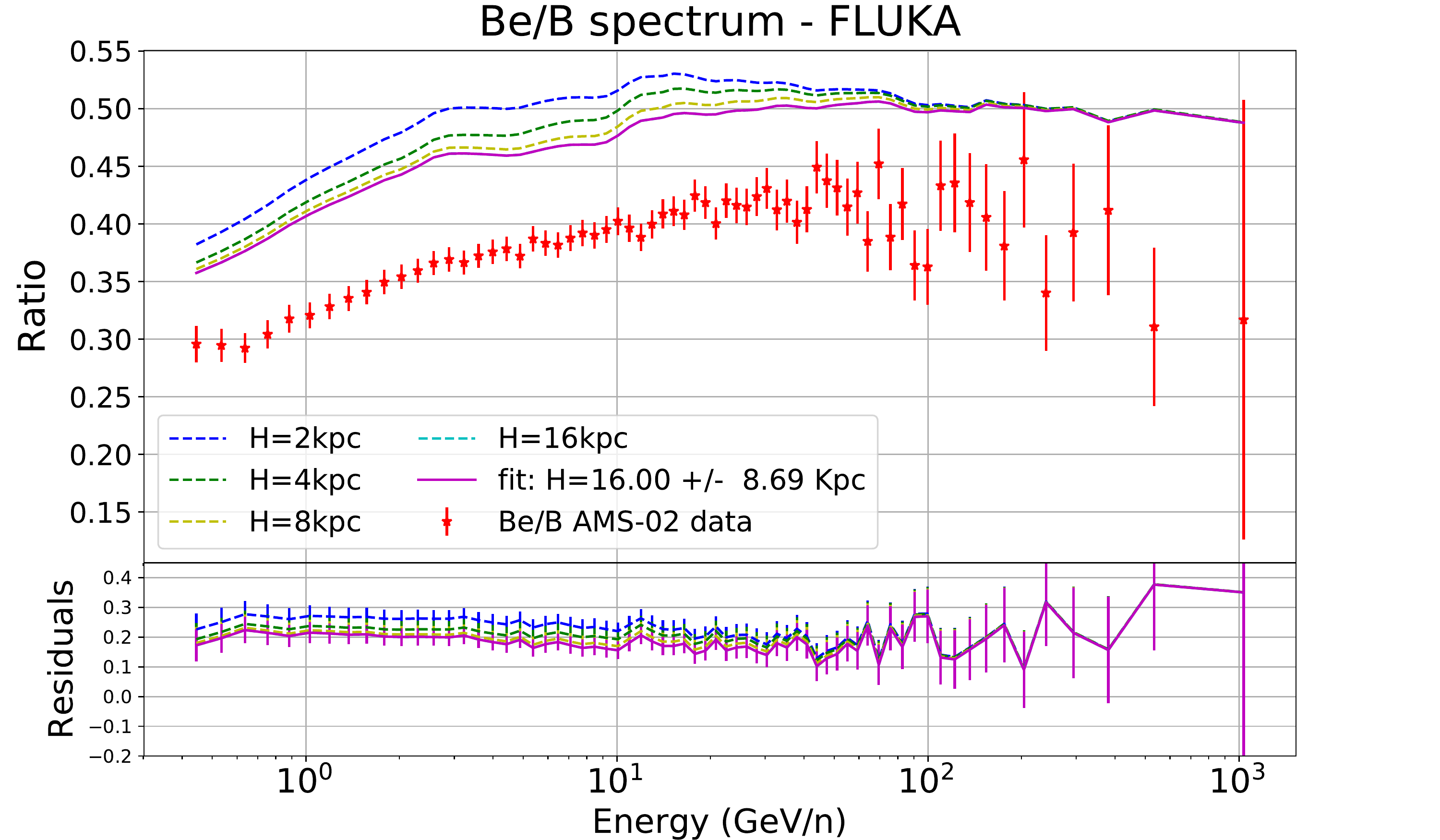} \hspace{-0.2cm}
\includegraphics[width=0.33\textwidth,height=0.19\textheight,clip] {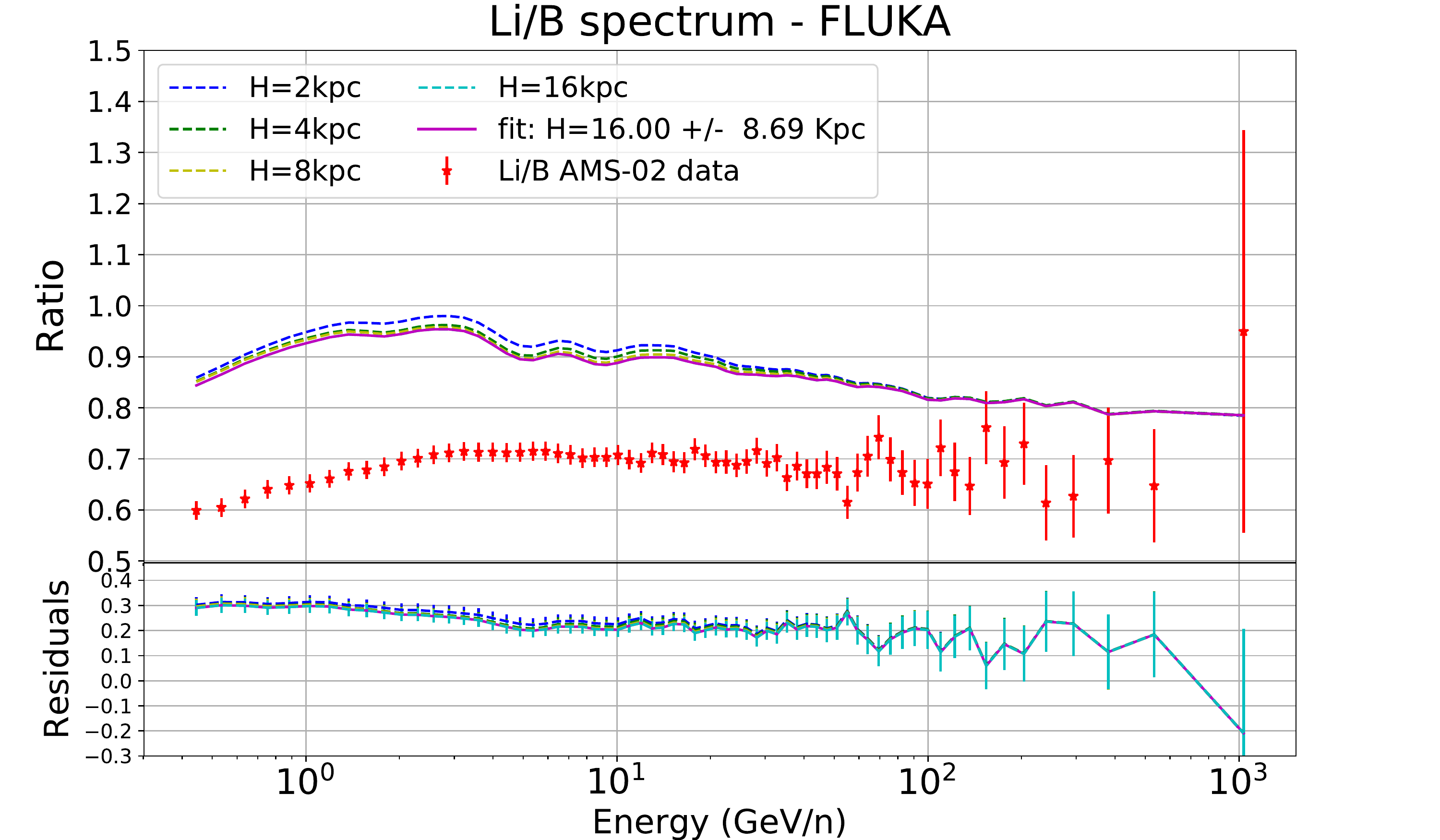} 
\hspace{-0.2cm}
\includegraphics[width=0.33\textwidth,height=0.19\textheight,clip] {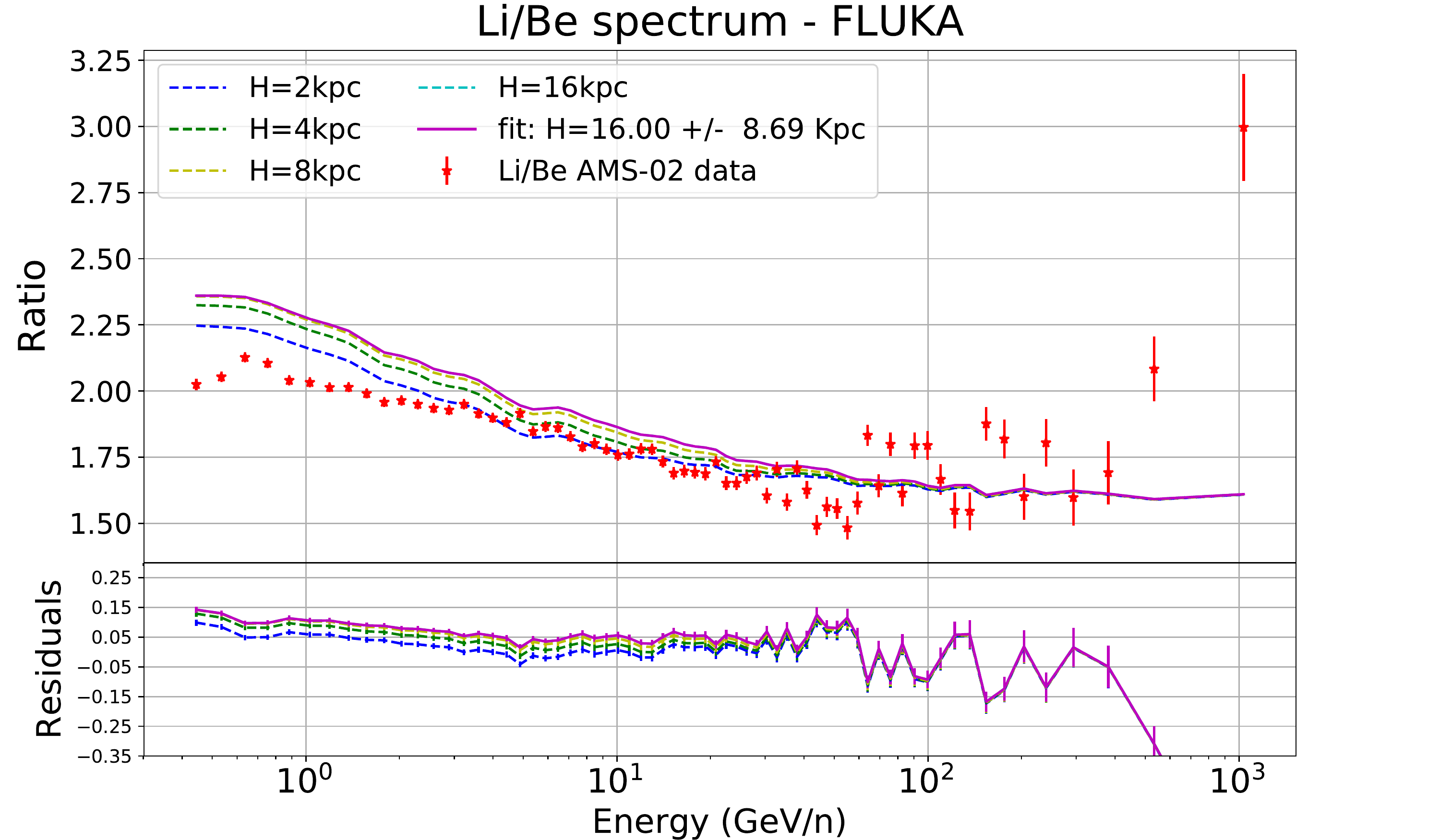}

\includegraphics[width=0.44\textwidth,height=0.20\textheight,clip] {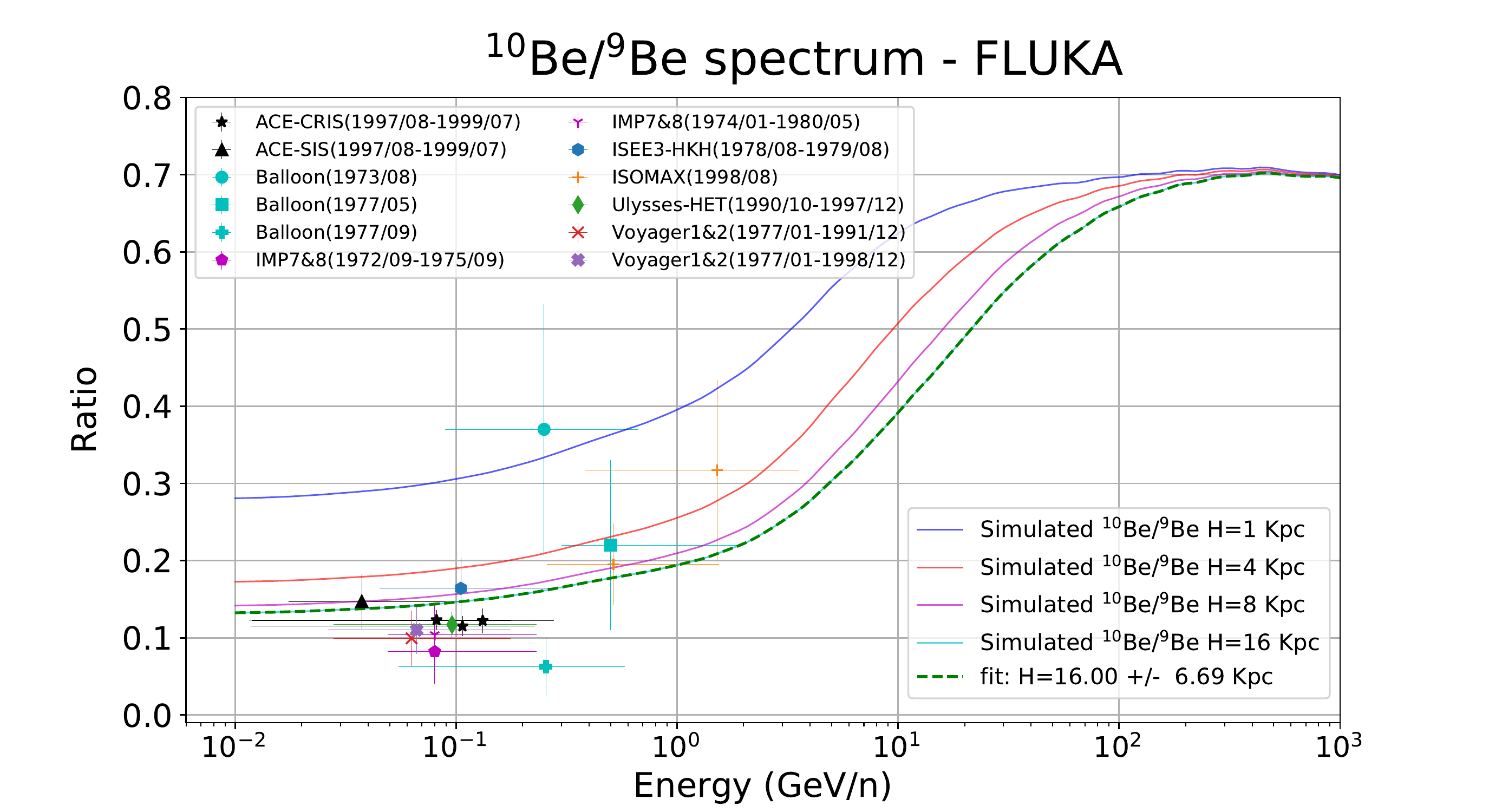}
\includegraphics[width=0.44\textwidth,height=0.20\textheight,clip] {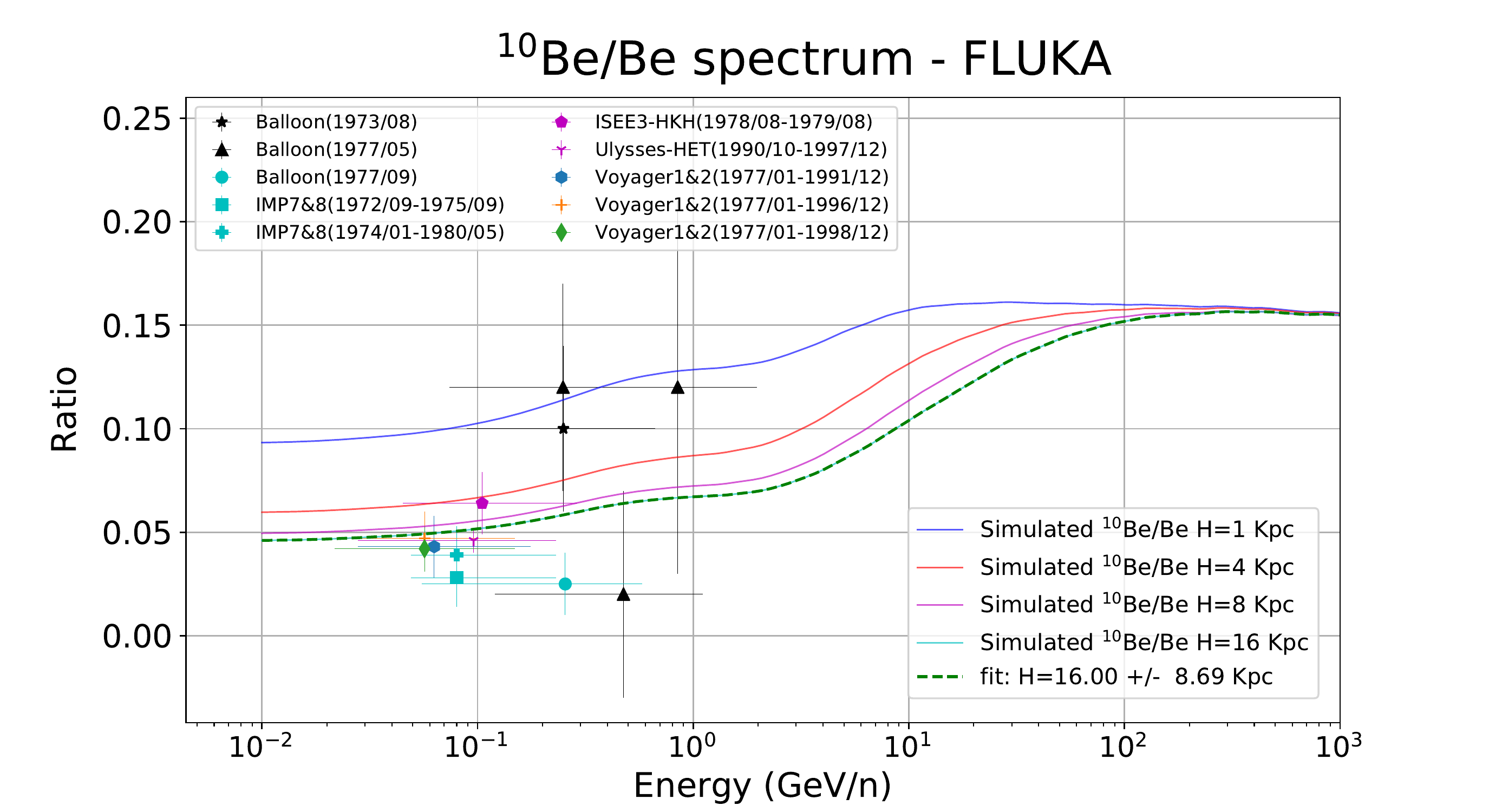}

\end{center}
\caption{\footnotesize Top panels: secondary-over-secondary flux ratios of the light secondary CRs (Li, Be, B). Bottom panels: $^{10}$Be/$^{9}$Be and $^{10}$Be/Be flux ratios. The predictions are compared to experimental data for simulations performed with various halo sizes.}
\label{fig:Fhalos}
\end{figure*}

The next step for testing whether the FLUKA cross sections allow us to reproduce CR experimental data is the study of the effect of the halo size variations in the relevant observables sensitive to this parameter, as shown in Figure~\ref{fig:Fhalos}. Here various simulations were performed for different values of the halo size with a diffusion parameter that fits the B/C AMS-02 flux ratio in every case. As in Figure~\ref{fig:sizes}, a fit on the value of the halo size that best matches the experimental data is carried out for the ratios $^{10}$Be/$^9$Be and $^{10}$Be/($^7$Be + $^9$Be + $^{10}$Be) (lower panels). The predicted secondary-over-secondary ratios are shown in the upper panels, including the halo size fit value of the $^{10}$Be/$^9$Be data. 

In this case, the fit favours large halo size values, significantly different from the value determined with the other cross sections parametrisations (see Fig.~\ref{fig:size_sum}). This is related with the already discussed discrepancies in the channels of beryllium production (Figure~\ref{fig:direct_hyd}). In fact, the need of a larger halo size is due to the slight underestimation of the $^{10}$Be cross sections and/or to an overestimation of the $^{9}$Be cross sections (this happens, for example, in the $^{16}O \longrightarrow ^9$Be channel, shown in Figure~\ref{fig:Tot_XS}).

\vspace{1cm}
\newpage
\subsection{Effect of variations in the He spallation and total inelastic cross sections}
\label{sec:Fluka_effect}

\begin{figure*}[!hbt]
\begin{center}
\includegraphics[width=0.335\textwidth,height=0.19\textheight,clip] {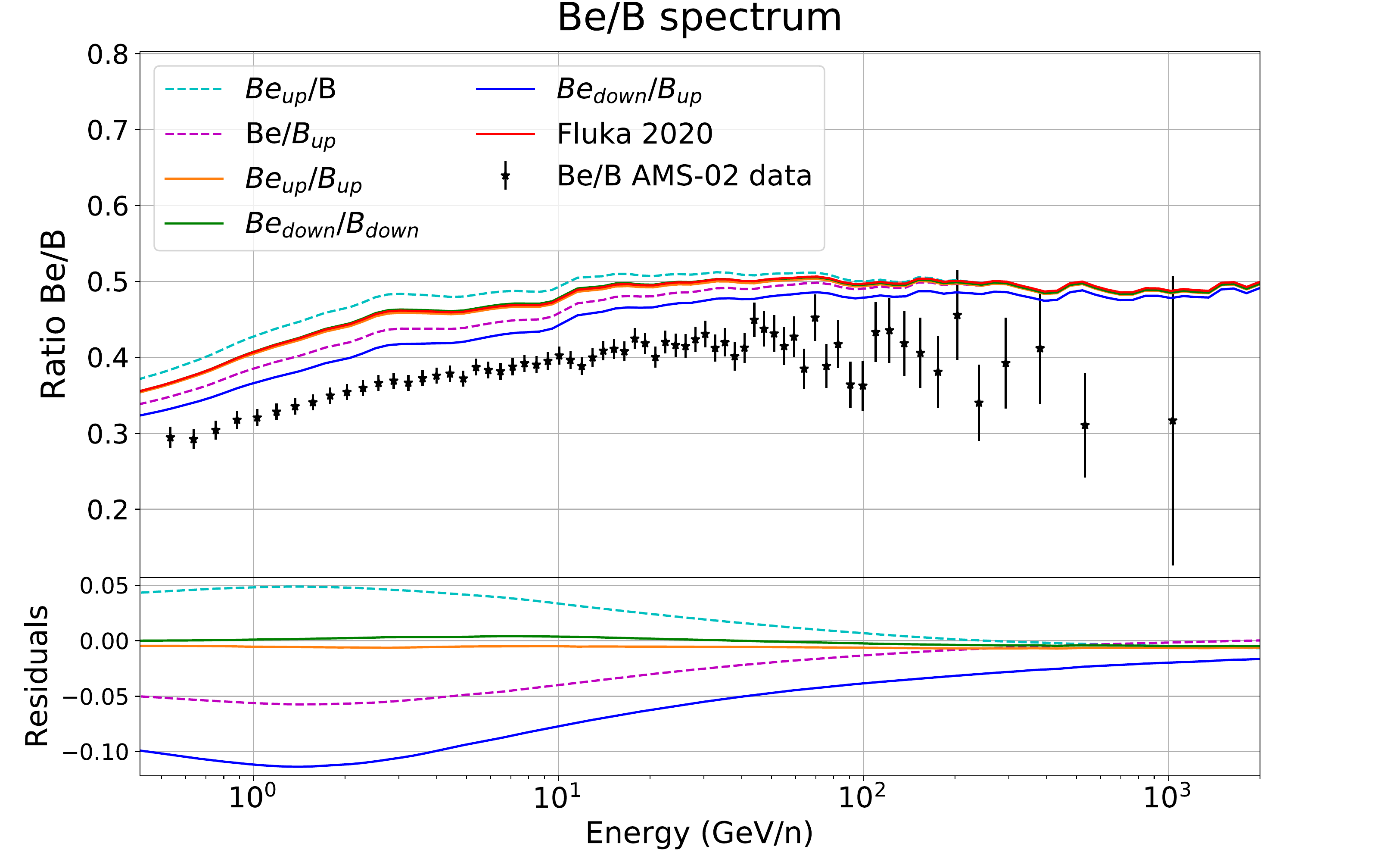} \hspace{-0.25cm}
\includegraphics[width=0.335\textwidth,height=0.19\textheight,clip] {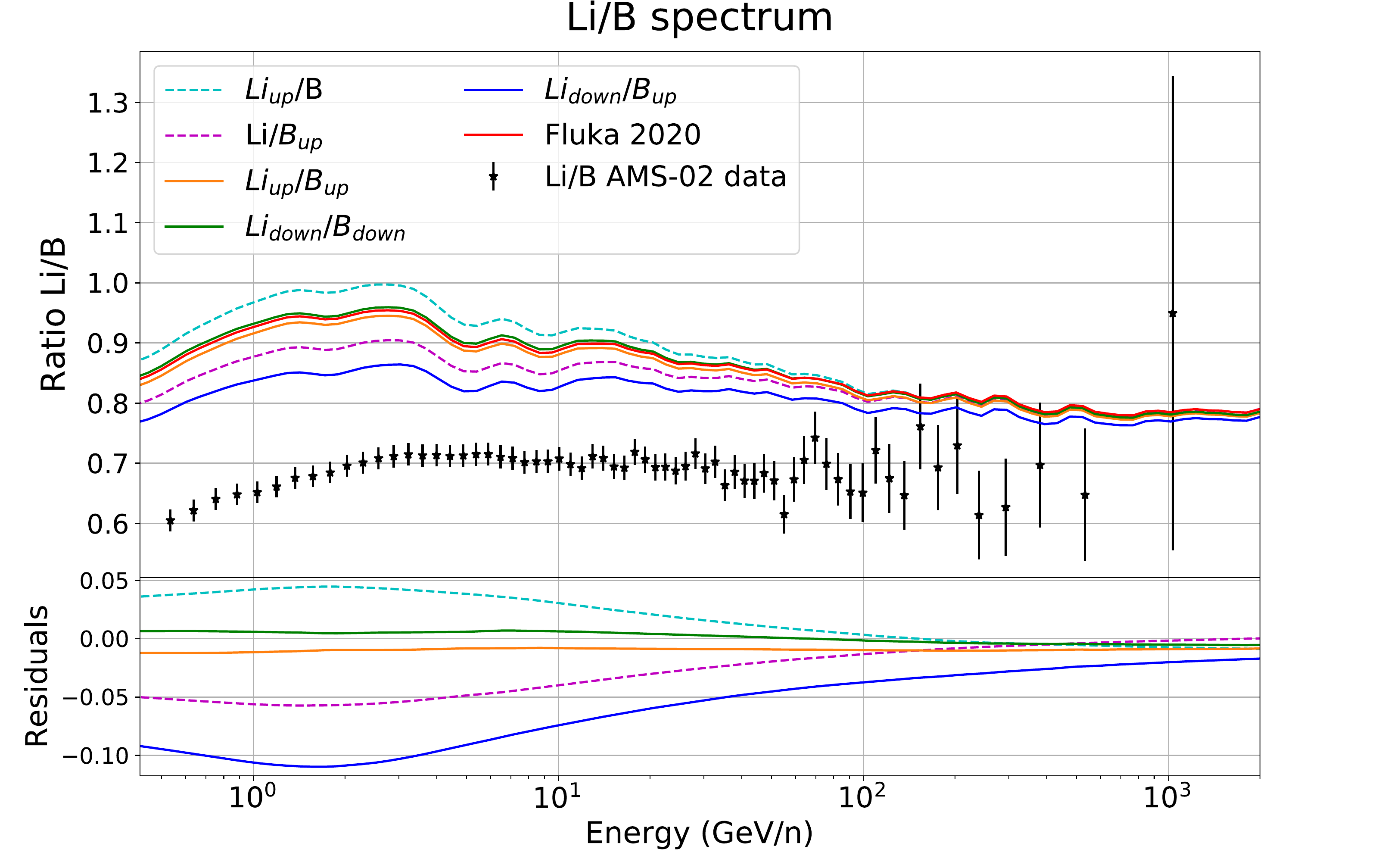} 
\hspace{-0.25cm}
\includegraphics[width=0.335\textwidth,height=0.19\textheight,clip] {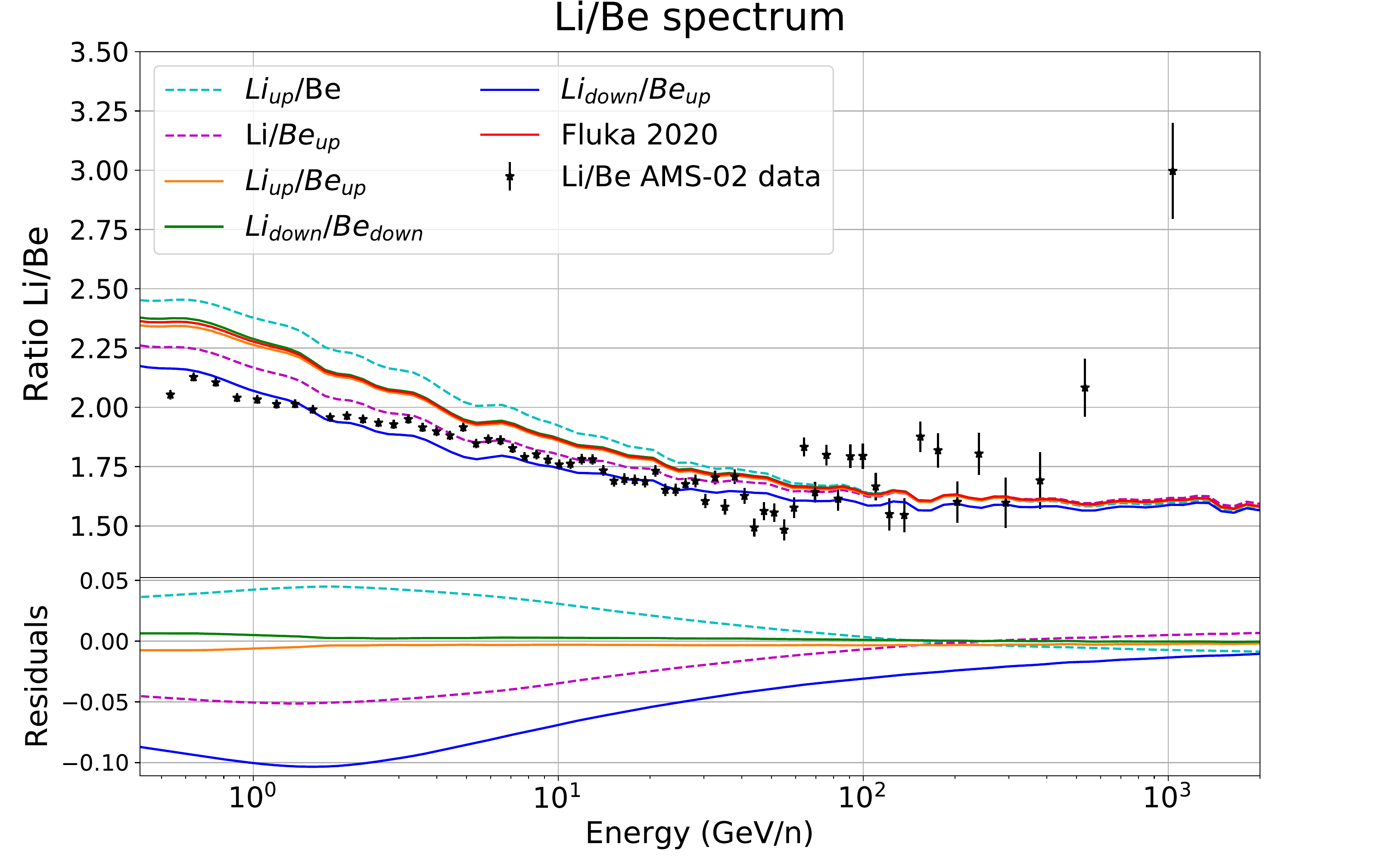}
\end{center}
\caption{\footnotesize Plots showing the effect of 10\% variations on the inelastic (fragmentation) cross sections in the light secondary CRs by means of the secondary-over-secondary flux ratios. The inelastic cross sections have been increased and decreased by 10\% for each of these species, and the results of the simulations are shown for different combinations. Residuals in this case are drawn with respect to the case in which no variation has been applied, i.e. the original prediction. The AMS-02 data are also added.}
\label{fig:Sec_IneF}
\end{figure*}

\begin{figure*}[!hbt]
\begin{center}
\includegraphics[width=0.335\textwidth,height=0.19\textheight,clip] {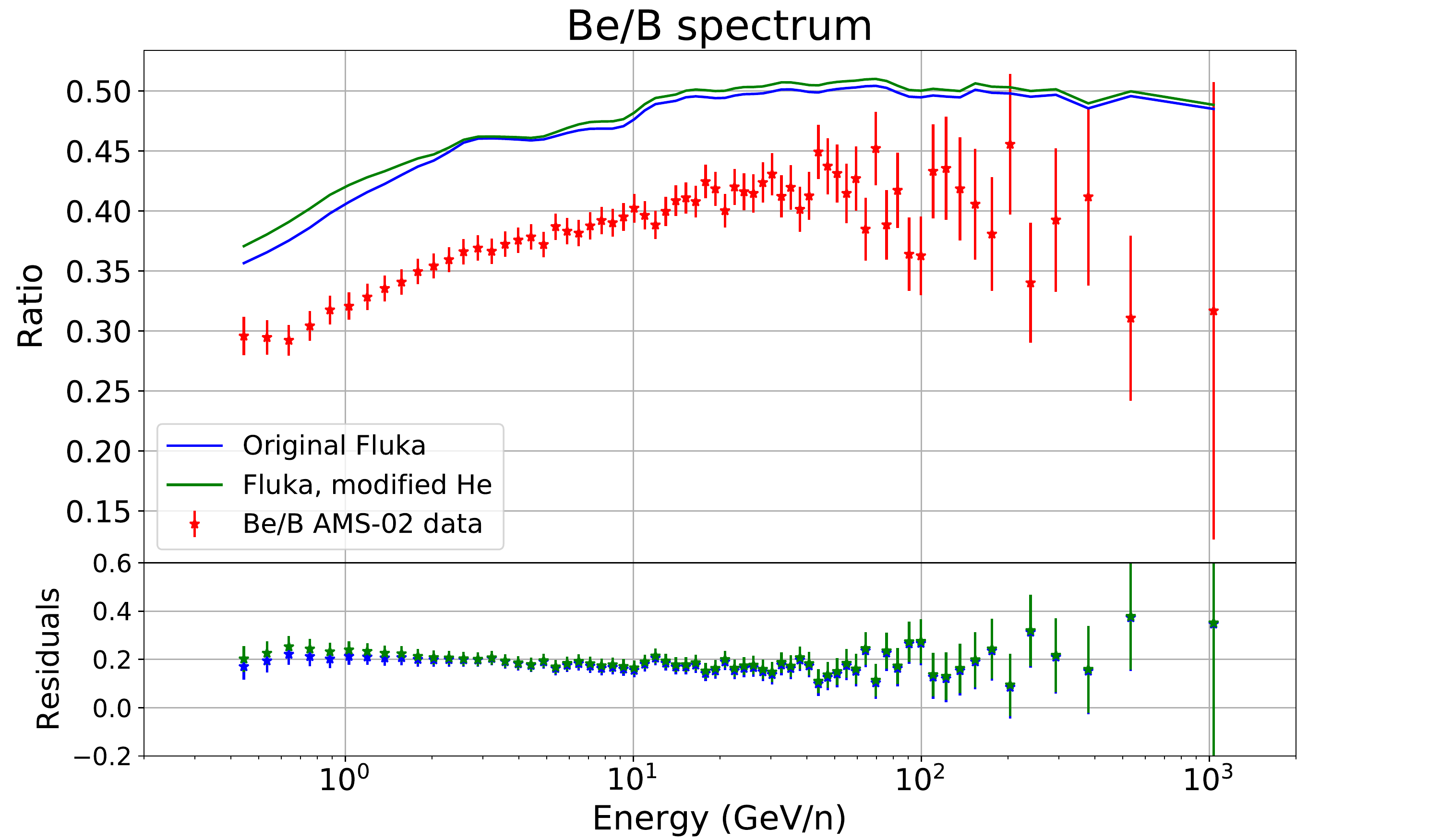} \hspace{-0.25cm}
\includegraphics[width=0.335\textwidth,height=0.19\textheight,clip] {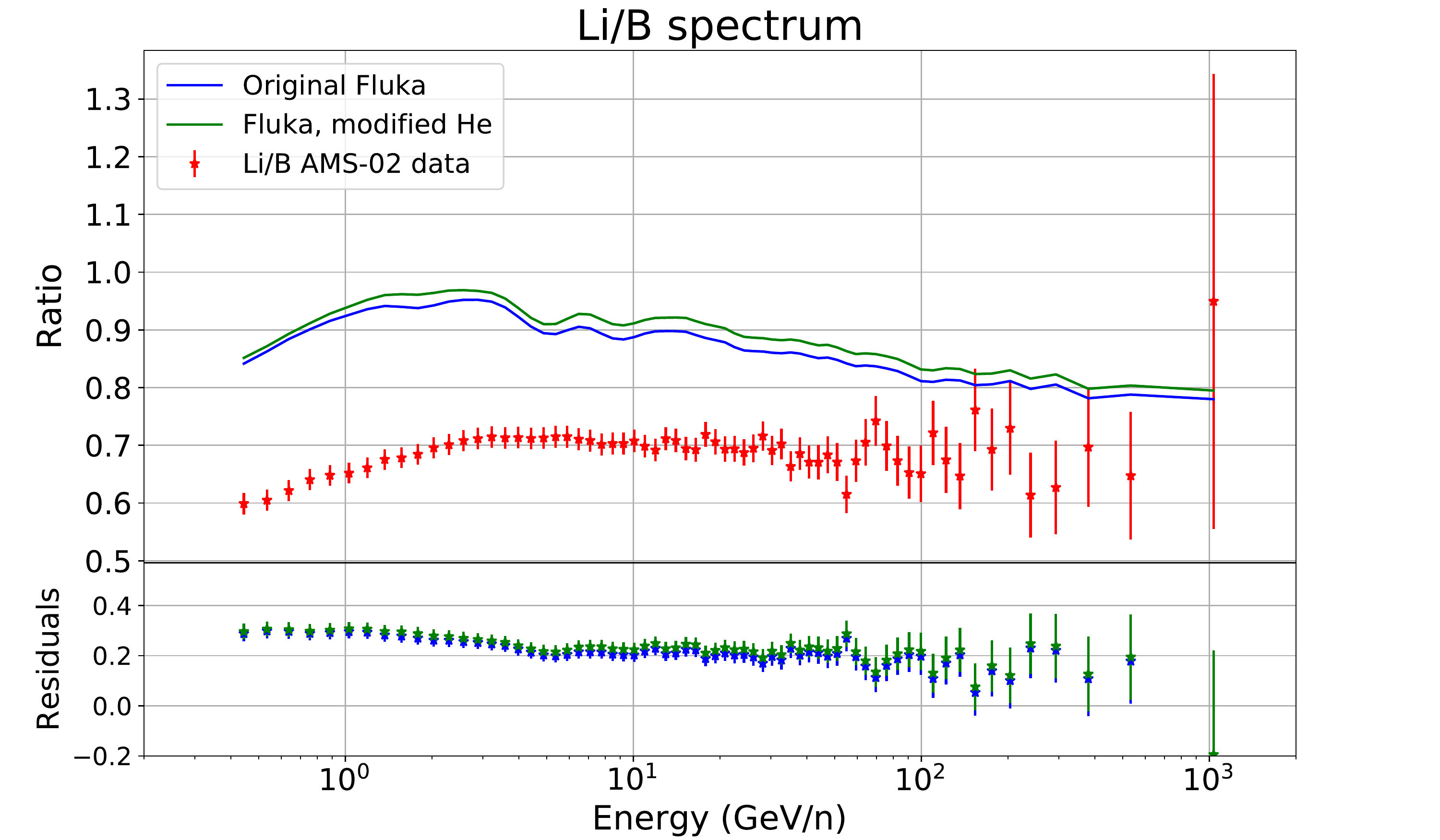} 
\hspace{-0.25cm}
\includegraphics[width=0.335\textwidth,height=0.19\textheight,clip] {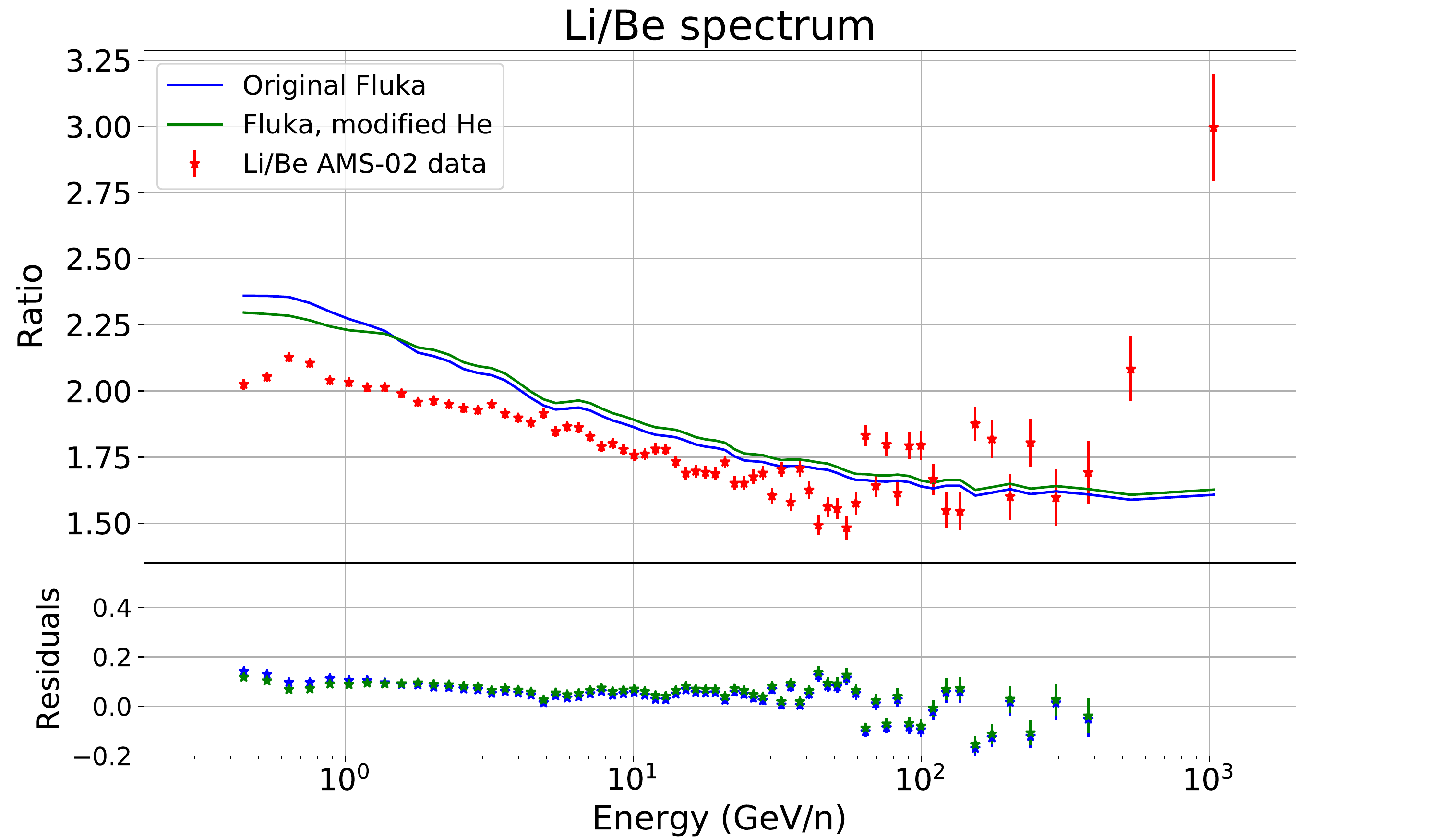}
\end{center}
\caption{\footnotesize Comparison of the predicted Be/B, Li/B and Li/Be spectra with the spallation cross sections for He as target calculated with FLUKA and calculated using the typical parametrisation of the ratio $\sigma_{H}/\sigma_{He}$ \cite{ferrando1988measurement}. Variations between both predictions are always below 5\%.}
\label{fig:SecHe_F}
\end{figure*}

Finally, we studied the effects of varying the inclusive cross sections of spallation with He as target and the total inelastic cross sections. We compared the predicted secondary-over-secondary ratios derived with the FLUKA model to those obtained when using the typical procedure for the calculation of the He cross sections in Figure~\ref{fig:SecHe_F} and to those obtained varying the inelastic cross section by $10\%$ (much above the discrepancies seen between the FLUKA and CROSEC computations for B) in Figure~\ref{fig:Sec_IneF}.

The effect of changes in the inelastic cross sections is not important for the primary CR particles, as C or O, since these changes are compensated with the source term (as discussed in \cite{Genoliniranking}). However, the inelastic (associated to the destruction term of eq. \ref{eq:caprate}) cross sections are also relevant for the predictions of secondary CR fluxes. To illustrate how variations on these cross sections affect the predictions of the fluxes of secondary CRs, a $10\%$ increase (which means more fragmentation, and therefore, slightly smaller flux of the secondary CR particles) and decrease (less fragmentation and larger fluxes) has been applied to the FLUKA inelastic cross sections. The results are shown in Figure~\ref{fig:Sec_IneF}, where the fluxes with the subscript ``up'' are those resulting from the simulations with the $10\%$ decrease in the inelastic cross sections and those with the subscript ``down'' are those with $10\%$ increase in the fragmentation, while those without subscript are the fluxes obtained with the original FLUKA model. The residuals in these plots are calculated with respect to the original ratios (FLUKA 2020 in the legend), since we want to compare the possible deviations in the fluxes from the original predictions.

We show, together with the AMS-02 data and the original secondary-over-secondary flux ratios, different combinations of these fluxes, corresponding to the ratio of both species with increased cross sections (down/down), or decreased cross sections (up/up), the ratio of the flux of the element in the numerator with decreased cross sections over the original flux of the element in the denominator (up/original), the ratio of the original flux of the element in the numerator over the flux of the element in the denominator with decreased cross sections (original/up) and the ratio of the flux of the element in the numerator with increased cross sections over the flux of the element in the denominator with decreased cross sections (down/up). 

For the two first cases (up/up and down/down) the cross sections of both CR elements have same deviations, which correspond to the cases where the FLUKA inelastic cross sections are biased from the "real" ones by the same amount for both elements. The third and fourth cases (up/original and original/up) correspond to the cases in which the cross sections of one of the elements is deviated, while the other one remains as predicted. Finally, the down/up case corresponds to the case in which the FLUKA inelastic cross sections of both elements are biased, but in the opposite directions (i.e. the inelastic cross sections of one of them are overestimated while for the other one are underestimated), which represent the highest possible change in the predicted ratios with respect to the original prediction.
As we see, the three ratios show essentially the same behaviour: the up/up and down/down cases do not show any difference from the original simulation in any of the three ratios. In turn, the up/original and original/up are nearly symmetrical with respect to the original FLUKA prediction (almost symmetrical residuals around 0), showing maximum deviations of around $5\%$ at $\sim 1.5 \units{GeV/n}$, going to $\sim 2.5\%$ at $30 \units{GeV/n}$. Finally, the down/up case shows maximum discrepancies with respect to the original predictions of $10\%$ at $\sim 1.5 \units{GeV/n}$, going to $\sim 5\%$ at $30 \units{GeV/n}$. We remark that, although we used a variation of $10\%$ to illustrate the effect of changes in the inelastic cross sections, the usual uncertainties reported are about $3-5\%$ \cite{Genoliniranking}.

Finally, in Figure~\ref{fig:SecHe_F} we show a comparison of the predicted Be/B, Li/B and Li/Be spectra with the spallation cross sections for He as target calculated with FLUKA and calculated using the typical parametrisation of the ratio $\sigma_{H}/\sigma_{He}$ \cite{ferrando1988measurement}. The simulation is performed with the same diffusion coefficient that fits the B/C data, in table~\ref{tab:Fdiff_params}. In general, they imply very small changes, of around $2-3\%$, due to the small amount of He in the ISM gas, in comparison with the amount of H. 

\section{Conclusions}
\label{sec:Fconc}

In this chapter, the FLUKA calculations for the inelastic and inclusive cross sections have been presented and widely discussed. We found that the inelastic cross sections for He, $^1H$, C and B are in good agreement with data and are quite similar to the CROSEC predictions at the level of a few percent. Nevertheless, elements with atomic number $Z\geq10$ (from Ne) start to be significantly different in both computations. In any case, these discrepancies are not important for studies of the diffusion and production of light secondary CRs.

Then, the inclusive cross sections involving the production of Be, Li and B isotopes have been checked with current data, showing a good agreement in general, except for the case of some channels of production of $^9$Be. %It is argued that in some of these channels, as well as happens with the channels of $^{10}$Be the large error bars and the discrepancies between experimental data are too big usually to .

Moreover, FLUKA allows the computation of the spallation cross sections of reactions with He, while in the current parametrisations they are usually evaluated by extrapolating results from the proton reaction channels. These cross sections have been compared, and are found to be consistent with the parametrisations too. 

Also the effect of ghost nuclei has been revised. We stress here the importance of these results, achieved for first time in a nuclear code as FLUKA. 

In the second part of this chapter, the FLUKA inelastic and inclusive cross sections have been implemented in the DRAGON propagation code to demonstrate that every observable can be reproduced with high accuracy.
In general, after checking all these different features, we do not see significant differences with respect to other parametrisations used, and it seems that the predictions using the FLUKA cross sections can be perfectly matched to the most recent CR experimental data.

Furthermore, the secondary-over-secondary ratios have been carefully studied taking into account possible deviations of the inelastic and He spallation cross sections, finding small variations on these ratios, which could be about 2\% at high energies. Large values of the halo size seem to be favoured when using these cross sections, although, due to the large uncertainties on its determination, this is something not very relevant at the current level of precision.

In conclusion, the cross sections obtained with the FLUKA code seem to be consistent with the current parametrisations and with experimental data, and are able to reproduce every CR observable with a low level of uncertainty.

In the next chapter, the optimal diffusion parameters are searched and discussed for the DRAGON2, GALPROP and FLUKA cross sections in different configurations of the propagation.
\chapter{Markov Chain Monte Carlo analyses of the diffusion parameters: discussion of propagation models and cross sections}
\label{sec:4}

As we discussed in chapter~\ref{sec:1}, the basics of the confinement of CRs in the Galaxy can be easily explained from their resonant interaction with alfvenic waves generated in the plasma by sudden instabilities, caused by phenomena like supernovae, mergers of massive bodies or even magnetic reconnection events. The turbulence generated in the interstellar plasma is accompanied by strong magnetic fields, with large impact on the dynamics of the turbulent cascade, following the magneto-hydrodynamic (MHD) laws \cite{MHDLazarian}. The complexity of the turbulence and its interactions with charged particles make us take a diffusion approximation, using a simple parametrisation for the diffusion coefficient that, even if it does not catch the full phenomenology in place, has demonstrated to be suitable enough to explain experimental CR data and diffuse emissions. At this point, the importance of the secondary-to-primary ratios is evinced, since they are the main bridge between the diffusion coefficient and the CR fluxes ($\frac{J_{sec}}{J_{prim}}(E) \propto \sigma(E)/D(E)$).

At the moment, we have discussed the uncertainties related to the spallation and inelastic cross sections using always the parametrisation of the diffusion coefficient given in equation \ref{eq:indepdiff}, which is an extension of the typical power-law with the $\beta ^{\eta}$ term added to better follow the low-energy behaviour of the secondary-over-primary ratios. In addition, reacceleration (regulated by the Alfvén velocity, $V_A$) is included, and convection is not taken into account, while expected to be possible. Moreover, the hardening observed in the CR spectra above $200 \units{GeV/n}$ is explained by the local source hypothesis instead of the currently more accepted break in the power-law index of the diffusion coefficient. This chapter aims at motivating these choices, as well as comparing the different parametrisations nowadays discussed in front of the modern theory on CR interactions with turbulent MHD plasma waves in the Milky Way.

On the other hand, in the determination of the propagation parameters from the secondary-over-primary ratios, the widely discussed systematic uncertainties involving production cross sections are usually compensated by the parameters of the diffusion coefficient and $V_A$, since these are only constrained by the ratios themselves. In this chapter we will develop a procedure that allows the determination of these propagation parameters, taking into account the possible defects on the spallation cross sections for the light secondary CRs Li, Be and B.

Accurate predictions on the CR propagation are crucial to better understand complex phenomena in the Galaxy, like its evolution, its chemical composition, the formation of structures, the interstellar magnetic fields and even the elusive dark matter component \cite{Bergstrom:2012fi, Bertone:2010at}.

In section \ref{sec:CRWAVES} we will review the theory on CR transport in view of modern advances in the understanding of MHD turbulence, in the quasi-linear theory framework. The most consistent approximations that lead to power-law parametrisations of the diffusion coefficient are reviewed and discussed in order to understand their differences and how they accommodate the phenomenology behind. In this section, we will motivate our choice of the diffusion coefficient used along the thesis. %Also some caveats of these approximations are commented in this chapter.
Then, in section \ref{sec:Det} we will carry out the determination of the diffusion parameters for different diffusion models, taking into account the uncertainties related to spallation cross sections of Be, B and Li production by means of the use of a Markov-Chain Monte Carlo (MCMC) algorithm, analysing the ratios of each of these secondaries to carbon and oxygen (B/C, B/O, Be/C, Be/O, Li/C and Li/O). Finally, in section \ref{sec:CXSanal}, we will try to refine the determination of the diffusion parameters for each of the cross sections parametrisations, to get solid conclusions from the secondary Be, Li and B.

This work will be soon submitted to a peer-review journal~\cite{Luque:MCMC}.

\section{Advanced theory on CR scattering and parametrisations of the diffusion}
\label{sec:CRWAVES}

The theory on the CR transport conventionally relies on the gyro-resonant interaction of CRs with Alfvén waves, which are circularly polarised and transverse waves propagating along the direction of the pre-existing magnetic field lines. The dynamical nature of the ISM and the active evolution of structures give rise to spatial variations of the properties of the interstellar plasma. Instabilities are continuously triggered by supernova explosions \cite{NF_Turbulent, Ferriere_SN_turbulence}, merger events, magnetic reconnections like solar flares and other events \cite{Subram, Cluster_turb} generating turbulence in the medium that cascades from large structures (sizes of the order of kiloparsecs) down to microscopical scales. The diversity of driving mechanisms inject turbulence at different scales, following the MHD laws which couple the quasi-neutral plasma medium with the magnetic field which is embedded in. In addition, the partial ionization of plasma changes the properties of MHD turbulence, increasing the scale at which it is damped \cite{3D_turbul_rev}. 

In the last years, better understanding of the MHD turbulence generated in the Galaxy has been achieved, demonstrating the need to update the simple (although effective) power-law parametrisation approach to get a better description of the global phenomenology of CR transport. In fact, considering compressible MHD turbulence, three modes arise instead of only one, generated in superposition \cite{Alfven_turbul, Alfven_Win}. They are the incompressible Alfvén mode and two magnetosonic (compressible) modes, namely fast and slow modes. These modes can be easily obtained as a solution of the dispersion relation of perturbations in compressible warm plasmas (see, for example, \url{http://farside.ph.utexas.edu/teaching/plasma/Plasmahtml/node65.html} for a complete derivation):

\begin{equation}
(\omega^2 - k_{\parallel}^2V_A^2) \left[\omega^4 - k^2(c_s^2 + V_A^2)\omega^2 + k^2k_{\parallel}^2c_s^2V_A^2)\right] = 0
\label{eq:Disp_rel}
\end{equation}
where $\omega$ is the angular frequency of the wave, $k$ and $k_{\parallel}$ are the total wavenumber and the wavenumber of the wave in the direction of the magnetic field lines respectively, $V_A = \frac{B_0}{\sqrt{\mu_0 \rho_0}}$ ($B_0$ is the unperturbed magnetic field, $\mu_0$ the magnetic permeability and $\rho_0$ is the unperturbed plasma density) is the Alfvén velocity and $c_s = \sqrt{\frac{\Gamma P_0}{\rho_0}}$ ($\Gamma$ is the adiabatic constant of the plasma as gas and $P_0$ is the unperturbed pressure) is the sound velocity in the plasma. In Figure \ref{fig:Modes} we represent the basic properties of these waves, together with their dispersion relation:

\begin{figure}[!bh]
	\centering
	\includegraphics[width=0.6\textwidth, height=0.18\textheight]{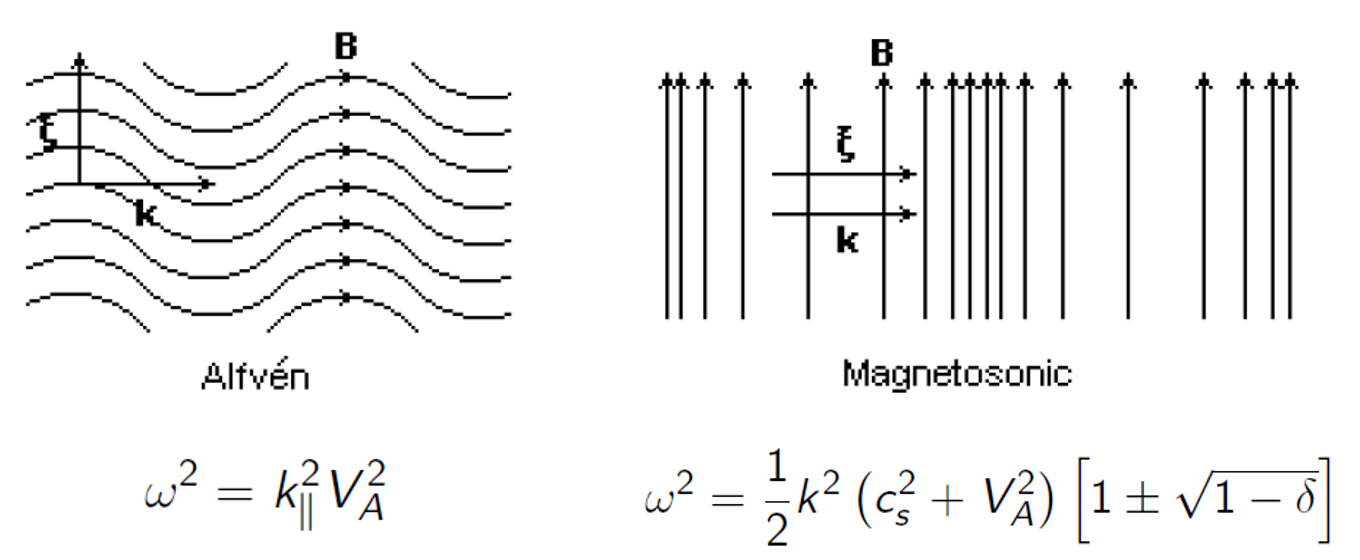}
	\hspace{1mm}
	\caption{\footnotesize Representation of the Alfvén and magnetosonic waves together with their dispersion relation. Here $\xi$ represents the fluctuation of any of the properties of the gas (pressure, magnetic field, density, etc). The Alfvén wave is a transverse wave ($\xi \bot k$) propagating along the magnetic field direction, while the magnetosonic wave is longitudinal ($\xi \parallel k$) and propagates perpendicular to the original magnetic field direction. Here, $\delta$ is a value between 0 and 1, defined as $\delta \equiv \left(2k_{\parallel}c_sV_A/k(c_s^2 + V_A^2)\right)^2$. }
	\label{fig:Modes}
\end{figure}

Plasma pressure and magnetic pressure perturbations may work together or in opposition, which originates the magnetosonic mode to be decoupled into slow (with these two effects out of phase) and fast waves (in phase). The Alfvén waves are the result of the tension of the magnetic lines, in analogy to the tension of a string in a stringed instrument. Furthermore, the phase velocity of magnetosonic waves depends on the angle of propagation with respect to the magnetic field and the plasma $\beta$ value, causing that, while the fast modes are propagated quasi-isotropically \cite{makwana2019properties}, the slow modes cannot propagate orthogonal to the magnetic field lines \cite{MHD_AlfTurbulence}.

Alfvén waves are energetically dominant in the$\units{GeV}$ region, which justifies the power-law approximation of the diffusion coefficient. In fact, in the solar wind, they contain about 90$\%$ of the total energy of plasma and fields (see  \cite{Alfven_Win}). This is due to the severe kinetic (viscous) damping suffered by the magnetosonic modes \cite{ModesPolariz}. Nevertheless, for certain scales or depending on the state of ionization of the plasma, where this damping is not dominant, magnetosonic modes can start to be more important \cite{Lynn2004, MSvsAlf}. In fact, recent calculations show that the fast modes get dominant at high energies ($\mathcal{O}(TeV)$) over the Alfvén modes, as demonstrated in a recent work from the thesis' author \cite{Fornieri2021}. This implies important deviations from the traditional power law that should be taken into account.

In a partially ionized medium, damping mainly arises from collisional friction (viscous damping), mainly between ions and neutrals \cite{Li_2008, Lazarian_2004, Lithwick_2001}. The damping scale where the MHD turbulence is dissipated can be determined by comparing the turbulent energy cascade rate with the damping rate (see  \cite{3D_turbul_rev}).
In addition, wave dissipation (and growth) can occur due to the resonant interaction with CR particles and can result in sizeable changes in the spectra of CRs, as demonstrated in ref.~\cite{Ptuskin_2006} for a Kraichnan-type cascade. This dissipation is important for CR at energies below a few GeV/n, and may result in a considerable increase of the particle diffusion coefficient. On top of this, total dissipation starts to be effective when the frequencies of the wave approach the plasma frequency, which causes that very low energy particles (with energies $\sim$ thermal energy of the plasma particles) do not get scattered, so that $ lim_{E\rightarrow0} D(E) = \infty $, which means that their motion becomes ballistic (particles will follow the magnetic lines without scattering and the diffusion approximation will not be valid any more), as already shown in ref.~\cite{Reichherzer:2019dmb}. 

The most recent CR data are indicating that the usual approximations employed need to incorporate all this phenomenology. In particular, the discrepancies found in the predictions on the amplitude of spatial anisotropies, those found in the low-energy part of CR spectra and the hardening of the CR spectra at high energies suggest that we should upgrade our diffusion modes to account for these physical processes. Fresh ideas have arisen in the last years to incorporate a spatial variability of the diffusion coefficient. Here we would like to highlight two different approaches: the first one consists of adding a spatially dependent term as a function of the radial distance, $f(r)$, or of the height, $f(z)$, in the diffusion coefficient; the other one takes into account different properties of the plasma in the halo and in the disk, which are described with two different diffusion coefficients (two-zone diffusion). 

For the former approach, the radial dependence was studied in ref. \cite{evoli2008cosmic} as an alternative explanation of the CR gradient problem observed in gamma rays. In turn, a height-dependent term was studied in ref. \cite{di2013cosmic}, to take into account the spatial variability of the turbulent component of the magnetic field ($D^{-1} \propto B_{ran} \propto exp[z/z_t]$ where $z_t$ is the scale height). Nevertheless, although these terms are important when studying the Galaxy in gamma or radio waves, no new features are expected to be observed in the local flux of CRs at Earth when including this factors. 

The second approach has been explored, for example, in ref. \cite{Tomassetti_delta} and \cite{Tomassetti_MC}. It assumes the Galaxy to be formed by the magnetised halo and a warm extended disk, using a different power-law parametrisation for the two zones. With this set-up, and, more precisely, by taking two very different spectral indices of the diffusion coefficient in the halo and in the disk, the hardening observed in the CR spectrum was reproduced. This theory explains the hardening as the transition from CR local spectra dominated by the diffusion in the halo at low energies to CR local spectra dominated by the diffusion in the disk at very high energies. Although such combination of spectral indices seems to be difficult to be motivated, these works showed that the spectral hardening may be a consequence of spatial variations of the diffusion coefficient. 

In fact, the hardening of the CR spectrum can be explained in several other ways: by source effects, as, for example, from the local source hypothesis \cite{Local_Sources} - that explains the hardening as a consequence of detecting the injected CR particles from nearby sources - and other interesting hypotheses (different accelerations mechanisms, \cite{Biermann_2010}, or different populations of CR sources, \cite{Source_mechanisms} and \cite{Yuan2011}), or by a diffusion phenomenon, whose better explanation seems to be related to the turbulence generated by streaming instability of CRs interacting with waves \cite{Aloisio:2013tda}. 

The impact of the different hypotheses has been studied for various CR species \cite{Lavalle2011, DonatoSerpico2011, Vladimirov:2011rn} favoring the diffusion hypothesis, mainly because of the change in the slope observed in secondary CR spectra \cite{Aguilar:2018njt}. This is why currently most of the studies use a change in the spectral index of the diffusion coefficient at high energies.  Nevertheless, a combination of both effects can not be discarded, and the fact that the break position inferred by different CR species is slightly different (as in  \cite{Genolini:2019ewc} and  \cite{genolini2017indications}) can support this hypothesis. Here we will perform analyses for the diffusion parameters for both hypotheses (see ref. \cite{Niu:2020qnh}).

On the other hand, some works have also implemented variations of the diffusion coefficient to better reproduce the CR spectra at low energies. One of the first modifications was the addition of the term $\beta^\eta$, with $\eta$ as a free parameter instead of being fixed to 1 or 0, as traditionally done. More recently, the appearance of a change of the spectral index at low energies (low energy break in the diffusion coefficient) has also been studied \cite{Weinrich:2020cmw, Weinrich:2020ftb, Genolini:2019ewc}. Nevertheless, it is not clear whether the diffusion coefficient should follow a power-law behaviour in the sub-GeV regime. 

\begin{figure}[!htpb]
	\centering
	\includegraphics[width=0.58\textwidth, height=0.24\textheight]{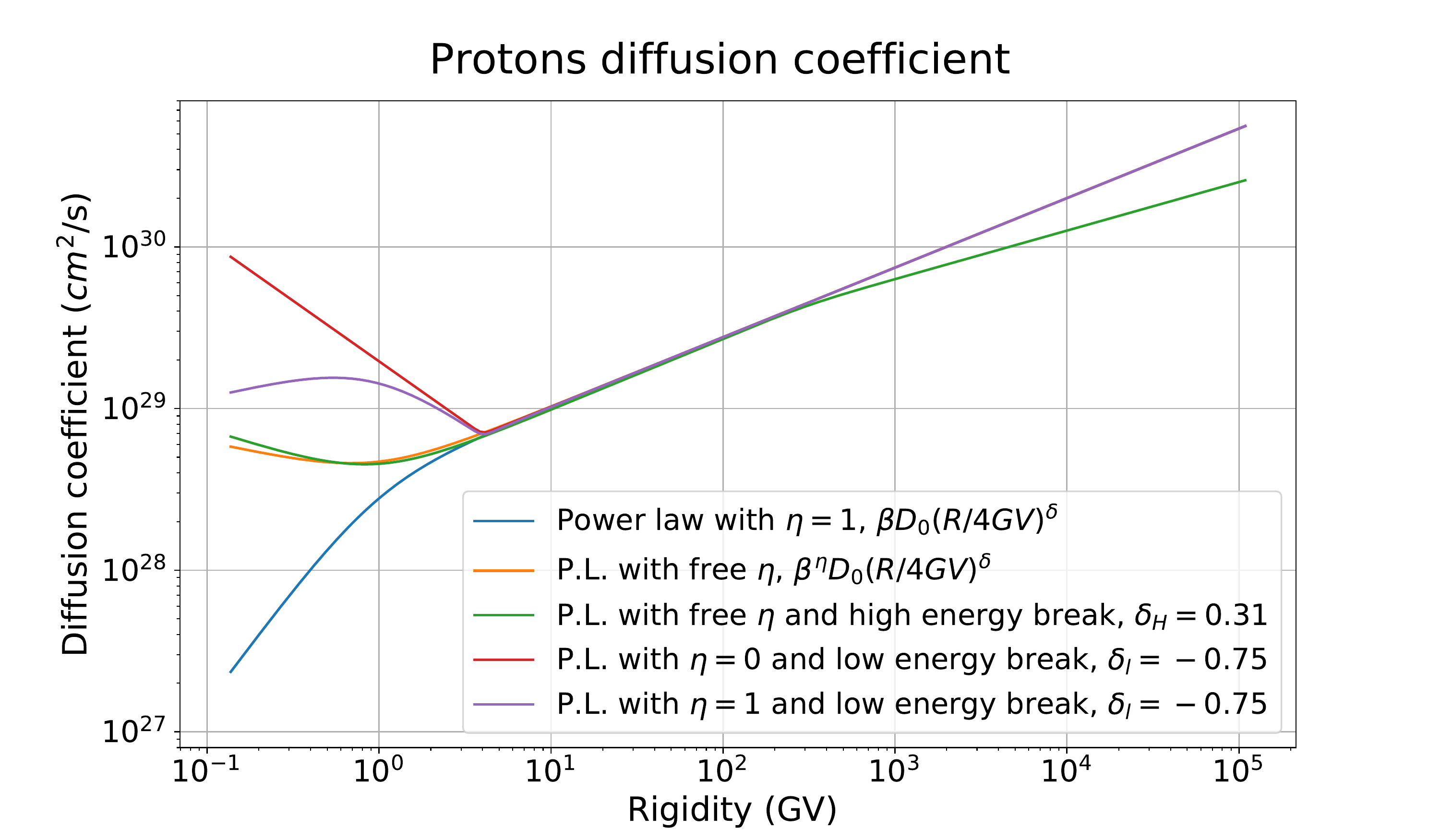}
	\hspace{1mm}
	\caption{\footnotesize Different diffusion coefficient parametrisations derived from the simplest power law in rigidity and normalized at $4 \units{GV}$, $D_0(R/4\units{GV})^{\delta}$. The traditionally used parametrisation includes the $\beta$ factor of the particle (blue line) and variations of this dependence are those with $\eta \neq 1$ (yellow line, with $\eta = -0.66$ and green line with $\eta = -0.78$). Finally, most recent parametrisations include a change of the spectral index at high energy (green line; with break position set at $312 \units{GV}$, as found in  \cite{genolini2017indications}) and at low energies (violet line; with break position set at $4 \units{GV}$, as found in  \cite{Weinrich:2020cmw}). The diffusion parameters used are those found with the DRAGON2 cross sections in the B/C fit under the source hypothesis given in table~\ref{fig:boxplot_Source}, except for the red line, which uses the B/C fit under the diffusion hypothesis, given in table~\ref{fig:boxplot_Diff}.}
	\label{fig:Diff_coef}
\end{figure}

Both approaches provide a very similar shape for the diffusion coefficient at low energy (see Figure~\ref{fig:Diff_coef}) and they explain the data with similar quality as demonstrated in table 1 of \cite{Weinrich:2020cmw}. In our simulations, we preferred to include the $\beta^\eta$ term instead of the low energy break, since it requires less free parameters in the fit (and thus less degeneracy) and it produces a softer transition between the two regimes. 

Figure \ref{fig:Diff_coef} displays the different diffusion coefficients discussed above, from a simple power-law function to the variations of this parametrisation that allow to reproduce recent CR data, in order to visualize the main differences among them. 

As shown in the figure, the simplest case, with a power-law with $\eta = 1$, quickly decays at low energies, which means that the confinement vastly increases for low energy particles. This is unexpected, as commented below, because of possible dissipation of waves or non-resonant interactions \cite{Reichherzer:2019dmb} at those energies. Then, by using negative values for the $\eta$ parameter (yellow and green lines, $\eta = -0.66$ and $\eta = -0.78$ respectively) the trend of the diffusion coefficient tends to change, with a smooth increase for lower energies (similar to the mirror regime proposed in  \cite{Reichherzer:2019dmb}). This is the same behaviour shown for the case of a low-energy break with $\eta = 0$, but this transition is much more abrupt. Finally, in the case with a low-energy break and $\eta = 1$, as in ref. \cite{Weinrich:2020cmw}, \cite{Weinrich:2020ftb} or \cite{Genolini:2019ewc}, the diffusion coefficient smoothly increases above the break position ($4 \units{GV}$), but decreases at lower energies since $\beta \rightarrow 0$, which means that very low-energy particles are suffering very strong scattering with waves.

On the other hand, the other propagation parameters that can affect the shape of the simulated secondary-to-primary ratios are the Alfvén velocity of waves, the convection velocity caused by galactic (vertically outwards) winds and the convective velocity gradient in the z direction. The most recent analyses of secondary-to-primary ratios have demonstrated that, when including reacceleration, the predictions show no need to include convection to explain AMS-02 data \cite{Weinrich:2020cmw, AMSemcee}. In turn, the authors of ref. \cite{Boschini:2019gow} found a best fit value for the convective velocity $> 0$ when analyzing very low-energy data to fit the local interstellar spectra (LIS), although it is subject to many uncertainties, mainly related to the solar modulation. Considering this, we can conclude that the information about the convection can be partially masked into the solar modulation and the $\eta$ value, the reacceleration, and other observables are needed to better constrain convection. Hence, convective velocity is not included in our scheme, allowing us to save computational time in the simulations. 

%In sum, with this discussion we have motivated the propagation parameters being used along the thesis, with a diffusion coefficient including the $\beta^{\eta}$ term with negative $\eta$ to match the expected trend of the CR scattering with MHD waves, and without inclusion of convection. 
In the next section we proceed to perform an extensive MCMC analysis of the ratios of Li, Be and B to C and O, taking into account the uncertainties in their production cross sections and considering both, the source and diffusion hypotheses to explain the hardening of CR fluxes at high energy.

\section{Determination of the propagation parameters}
\label{sec:Det}

As already mentioned, the most direct way to determine the diffusion parameters is by means of the ratio of the fluxes of secondary CRs over primary CRs. It is well known that the ratio of the flux of a pure secondary CR, formed by the spallation reaction of a primary CR, over the flux of that primary CR element (strictly a pure primary, with no secondary component) is $\frac{J_{sec}(E)}{J_{pri}(E)} \sim \sigma(E) \tau_{diff}(E) = \frac{\sigma(E)}{m_p}X(E)$, where $X(E) \equiv m_p n_{gas} c \tau_{diff} \propto 1/D(E)$ is the grammage traversed by the particle as defined in section~\ref{sec:1}. 

Nevertheless, it should be stressed that not all the particles of a given CR species can be considered as pure primaries, in the sense that an appreciable amount of their flux can be originated not only from the CR sources. For instance, around $20\%$ of the carbon flux at low energies comes from spallation reactions, thus meaning that carbon cannot be considered a purely primary CR nuclei. Here we will call ``primaries'' all the particles which are injected at the sources, keeping in mind that a fraction of these particles can be produced in spallation reactions. Therefore, the correct expression for the secondary-over-primary flux ratio will be given, at high energies, by the following equation:
\begin{equation}
\frac{J_{sec}(E)}{J_{pri}(E)} \propto \frac{\tau_{diff}(E) Q_{sec}(E)}{J_{pri}(E)} \sim \frac{\tau_{diff}(E) \sum^{\alpha \rightarrow sec}Q_{\alpha}(E)\sigma_{\alpha \rightarrow sec}(E)}{Q_{pri}(E) + \sum^{\alpha \rightarrow pri}Q_{\alpha}(E)\sigma_{\alpha \rightarrow pri}(E)} 
\label{eq:sec/prim}
\end{equation}
where, in the denominator, $Q_{pri}(E)$ is the source term (injection) and the second term is the contribution from any other CR species involved in the reaction network which can produce the primaries through spallation processes. The second term in the denominator is always subdominant, and it could only be neglected for oxygen.

Even though the analytical expression gets more complicated when considering the actual situation (showing that source terms can also be important), these ratios have demonstrated to be hardly sensitive to adjustments in the source terms of other nuclei (i.e. nuclei not involved in the ratio), at the level of a few percent, behaving essentially as in the classical approximation $\frac{J_{sec}(E)}{J_{pri}(E)} \sim \sigma(E) \tau_{diff}(E)$.

In this section we analyse the ratios of boron, beryllium and lithium over carbon and oxygen to constrain the diffusion coefficient within two different hypothesis to reproduce the hardening in CR spectra (namely, the source and diffusion hypotheses), considering the effects of cross sections uncertainties already described (chapter~\ref{sec:XSecs}).

\begin{figure}[!htpb]
	\centering
	\hspace{1. cm} \includegraphics[width=0.82\textwidth, height=0.173\textheight]{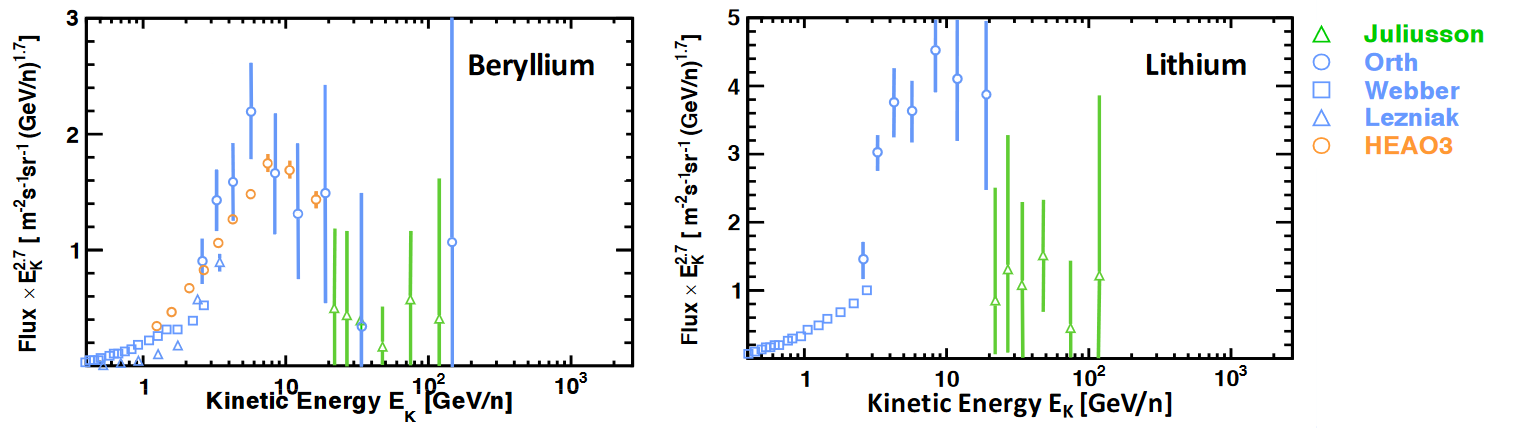}
	\caption{\footnotesize Li and Be measured fluxes in the pre-AMS-02 era. Plots taken from \cite{Jia:2018aze}.}
	\label{fig:LiBe_preAMS}
\end{figure}

There are very few works involving the ratios of secondary Li and Be nuclei. This is mainly due to the fact that experimental fluxes of Li and Be were poorly measured before AMS-02 (see Figure~\ref{fig:LiBe_preAMS}). In addition, the spallation cross sections uncertainties are larger for the Li and Be production. Moreover, the uncertainties on the measured Mg, Si and Ne fluxes were also large until very recently \cite{AMS_Ne}, thus yielding even larger uncertainties on the calculated fluxes of Li and Be.

Some researchers have claimed the need to use other secondary-to-primary ratios to cross check the validity of the cross section set \cite{maurin2010systematic}, concluding that this cross check will help reducing systematic uncertainties. 

The first work which analyzed the ratios of Li, Be and B to a primary (C in this case) was \cite{de2006observations}, in which the models were compared to CRIS data in the very low energy region of the spectrum. Other works have used a combination of the ratios of B over C (B/C), B over O (B/O) and C over O (C/O) - see \cite{Evoli:2019wwu} - or the Sc + Ti + V over Fe (sub-Fe/Fe), which are subject to larger uncertainties, with the B/C. Very recently, the authors of ref.~\cite{Weinrich:2020cmw} have done an extensive analysis on the ratios of Li, Be and B to C and O (including also $^3$He and N) taking into account cross sections uncertainties. Nevertheless, they did not use the AMS-02 data for Ne, Mg and Si and did not take into account all the observables that can be considered for this study (namely, secondary-over-secondary ratios).

In the next section we explain the algorithms used and the analyses performed to extract the propagation parameters from the LiBeB/CO ratios.

\subsection{Analysis Procedure and fitting}
\label{sec:proc}

Basically, the strategy we follow consists of an iterative procedure, which starts taking a set of propagation parameters - $D_0$, $V_A$, $\eta$ and $\delta$ - as first guess, and mainly performs a fit of the ratios of Li, Be and B to C and O (since they are the main contributors to the formation of those secondary nuclei via spallation) to the AMS-02 data in order to determine these parameters and their associated uncertainties. Previous to the fit, the injection spectra (source terms) are adjusted by fitting the AMS-02 data for carbon, nitrogen, oxygen, neon, magnesium and silicon. If the guessed diffusion parameters are very different (out of 1 $\sigma$ uncertainty) from those predicted by the fit, the injection spectra are adjusted again with the new predicted set of diffusion parameters (as primary spectra slightly depend on them too) and the fit is performed again to obtain a more refined prediction on the diffusion parameters. Convergence is reached when the output diffusion parameters after the last iteration are consistent (within 1 $\sigma$ uncertainty) with those used in the previous iteration to adjust the injection spectra.

Two diffusion parametrisations have been tested, that will be referred to as source hypothesis and diffusion hypothesis. These parametrisations differ in the physical interpretation of the high energy part of the CR spectra, as explained above, and are given by the following equations: 
\begin{equation}
D = D_0 \beta^{\eta}\left(\frac{R}{R_0} \right)^{\delta} \text{ \hspace{1.8 cm} \textbf{Source hypothesis}}
\label{eq:sourcehyp}
\end{equation}
\begin{equation}
 D = D_0 \beta^{\eta}\frac{\left(R/R_0 \right)^{\delta}}{\left[1 + \left(R/R_b\right)^{\Delta \delta / s}\right]^s}
 \text{  \hspace{0.7 cm} \textbf{Diffusion hypothesis}}
\label{eq:breakhyp}
\end{equation}
The expression used to parametrise the injection spectra of primary nuclei is the doubly broken power law of equation~\ref{eq:powerlaw} when we treat the diffusion parameters in the source hypothesis, and a simple broken power law when we treat the diffusion hypothesis.

As in the previous chapters, the reference rigidity is $R_0 = 4\units{GV}$. In equation~\ref{eq:breakhyp} there are three new parameters, that correspond to the rigidity break ($R_b$), to the change in spectral index ($\Delta\delta = \delta - \delta_h $, with $\delta_h$ as the spectral index for rigidities above the break) and a smoothing parameter $s$, used to allow a soft transition around the hardening position. In our analyses, we fix these parameters to the values found in ref. \cite{genolini2017indications}, since they are determined from the AMS-02 fluxes of protons and helium (with less experimental uncertainties than heavier nuclei at high energies). Their values are: $\Delta\delta = 0.14 \pm 0.03$, $R_b = (312 \pm 31) \units{GV}$ and $s = 0.040 \pm 0.0015$. Different break parameters slightly change the rest of the diffusion parameters obtained in the fit, but the fact that they are fixed from the primary fluxes allows us to avoid degeneracies with the main diffusion parameters, leading to a more rigid determination.

We follow same philosophy (i.e. extract information about the propagation from primary CRs to get less uncertainties in the determination of the diffusion parameters from secondary-over-primary ratios) for the determination of the Fisk potential, setting it to the value used throughout the thesis and argued in chapter~\ref{sec:XSecs}, corresponding to $\phi = 0.61 \units{GV}$ (obtained from the NEWK neutron monitor experiment in combination with Voyager-1 data). The force-field approximation for the solar modulation is adequate to the level of precision or models have at low energy. 

In order to perform the fit of the ratios, a Markov chain Monte Carlo (MCMC) procedure relying on Bayesian inference is developed in order to get the probability distribution functions for a set of diffusion parameters ($D_0$, $V_A$, $\eta$ and $\delta$) to describe the data (in this case the AMS-02 data) and their confidence intervals. The technical details are presented in the next section.

\subsubsection{MCMC and minimization algorithm}
\label{sec:bayes}

In this work, Bayesian inference is used to get the posterior probability distribution functions (PDFs), $\mathscr{P}$, for a set of diffusion parameters to explain the AMS-02 data. To evaluate it, the prior PDF, $\Pi$, and the likelihood, $\mathscr{L}$, must be defined. These three terms are related by:
\begin{equation}
\mathscr{P}(\vec{\theta}|\vec{D}) \propto  \mathscr{L}(\vec{D}|\vec{\theta}) \Pi(\vec{\theta})
\label{eq:bayesian} 
\end{equation}
where $\vec{\theta} = \{\theta_1, \theta_2, ... , \theta_m\}$ is the set of parameters, $\vec{D}$ is the data set used (CR flux ratios measured by AMS-02) and $\mathscr{P}(\vec{\theta}|\vec{D})$ is the posterior PDF for the parameters. 

The posterior probability is calculated from equation~\ref{eq:bayesian} by a MCMC algorithm which uses, instead of the classical \textit{Metropolis-Hastings} algorithm, a modified version of the \textit{Goodman \& Weare} algorithm \cite{Goodman_Weare}. This method is directly implemented in the \textit{emcee} module of Python (see ref.  \cite{emcee} for full technical information).

To optimize the algorithm and make the convergence faster, it is convenient to have a good initial guess of the diffusion parameters rather than a random one. To find this initial guess and make it as much precise as possible, another simpler and much faster minimization was used. This was done using the python package \textit{gp\_minimize} from the \textit{Scikit-Optimize} module, which can be used to minimize the $\chi^2$ value of a fit by a Gaussian optimization process. It is able to provide similar results to the MCMC procedure frequently, but in much less time although with some other caveats that make it unsuitable for obtaining the final conclusions (mainly local minima stacking).

In this procedure, as it is computationally impossible to have a simulation for each of all possible combinations of propagation parameters, a ``grid'' of simulations is built, so that simulations for (around) a thousand of regularly spaced combinations of values are carried out, and the other combinations are interpolated by using the python tool \textit{RegularGridInterpolator} from the \textit{Scipy} module. In other words, we are creating kind of a four-dimensional ``matrix'' (a dimension for each diffusion parameter) and, to each of the points of the matrix we associate a vector containing the simulated flux ratios at different energies. The diffusion equation was integrated in 171 points with equal spacing in a logarithmic scale, from $10 \units{MeV/n}$ to $\sim100 \units{TeV/n}$.

The errors coming from this interpolation were estimated comparing different simulated fluxes of different nuclei with the interpolated fluxes, obtaining errors always much smaller than 1\%. Errors in the interpolation of fluxes in different energies are completely negligible. % are much smaller than 1\%. 

%The range of diffusion parameters, for the spatially-dependent and spatially-independent models, is shown in the table~\ref{tab:range}.
%\begin{table}[h]
%\centering
%\resizebox*{0.27\columnwidth}{0.097\textheight}{
%\begin{tabular}{|lc|}
%  \multicolumn{2}{c}{\hspace{0.3cm}\large\textbf{ \ \ \ \ \ }} \\ \hline & \textbf{Parameters range}\\ 
%  \hline
%{$E_{\text{b}}$} ($GeV$)  & [0.01, 10000]\\
%{$D_0$} ($10^{28} cm^{2} s^{-1}$) & [3.2 , 6.2] \\
%{$v_A$} ($km/s$) & [0 , 40] \\
%{$\eta$}    & [-3. , 1.] \\       
%{$\delta$}    & [0.33, 0.6] \\     
%\hline
%\end{tabular}}
%\caption{\footnotesize Parameter space considered in the analyses of both diffusion models. }
%\label{tab:range}
%\end{table}

The prior PDF is defined as a uniform distribution for all the parameters:
\begin{equation}
\Pi(\theta) = \left\{
\begin{array}{ll}
\prod_{i} \cfrac{1}{\theta_{i, max} - \theta_{i, min}}  &  
\text{for $\theta_{i, min} < \theta_i < \theta_{i, max}$} \\
 0    & \text{elsewhere} \\ 
\end{array}
\right.
\label{eq:prior} 
\end{equation}

On the other hand, the likelihood $\mathscr{L}$ is set to be a Gaussian function (as already used in many studies, e.g.  \cite{Trotta:2010mx} or \cite{Niu:2018waj}):
\begin{equation}
\mathscr{L}(\vec{D}|\vec{\theta}) = \prod_{i}\frac{1}{\sqrt{2\pi\sigma_i^2}} \exp \left[- \frac{(\Phi_i(\vec{\theta}) - \Phi_{i, data})^2}{2\sigma^2_i} \right]
\label{eq:likelihood} 
\end{equation}
where $\Phi_i(\vec{\theta})$ is the flux ratio computed in the simulation, $\Phi_{i, data}$ is the corresponding flux ratio measured by AMS-02 and $\sigma_i$ is its associated error for the $i$-th energy bin. 

In this approach we do not take into account possible correlations among data, as the public data from the AMS-02 Collaboration do not include the full covariance matrix. The treatment of data with correlated errors is described in appendix~\ref{sec:covariance}.

%In this analysis, the normalization term $(\sqrt{2\pi \sigma_i^2})^{-1})$ is not included in the likelihood, since we use different kind of data sets, such that the neperian logarithm of our likelihood is equivalent to the $\chi ^2$ value.

Then, in order to take into account cross sections uncertainties, we make use of a nuisance parameter for each of the secondary CRs analysed (i.e. B, Be and Li), which enables a renormalization (or scaling) of the parametrisation used for their production cross sections. The associated nuisance term is usually defined as (eq. 4 of ref. \cite{Weinrich:2020cmw}):
\begin{equation}
\mathscr{N}_X = \sum_i\frac{\left(y_{i,X} - \hat{y}_{i,X}\right)^2}{\sigma_{i,X}^2}
\label{eq:nuisance} 
\end{equation}
where $\hat{y}_{i,X}$ is the experimental cross section value at the $i$-th point of energy and $\sigma_{i,X}$ its associated experimental uncertainty referred to the secondary CR species $X$, that can be B, Be or Li. The term $y_{i,X}$ is defined as $y_{i,X} \equiv \Delta_X \times \hat{y}_{i,X}$, where $\Delta_X$ is the nuisance parameter to be adjusted, referred to the species X, and independent of energy. %Independent of or dependent on https://forum.wordreference.com/threads/independent-from-of.1268255/

As the nuisance parameters are independent of energy, eq. \ref{eq:nuisance} can be rewritten as:  
\begin{equation}
\mathscr{N}_X = (\Delta_X - 1)^2 
\sum_i \left( \frac{\hat{y}_{i,X}}{\sigma_{i,X}}\right)^{2} = 
(\Delta_X - 1)^2 N \left\langle \hat{y}/\sigma \right\rangle_X^{2}
\label{eq:nuisance2}
\end{equation}
Here $N$ stands for the number of experimental data points and $\langle \sigma/\hat{y}\rangle_X$ is the average relative error for a given channel of production of $X$. Given the lack of cross sections experimental data, we use the reaction channels with $^{12}$C and $^{16}$O as projectiles. Then, the final expression for the nuisance term associated to a secondary CR species $X$ is:
\begin{equation}
    \mathscr{N}_X = (\Delta_X - 1)^2 \sum_{X'}(N_{C \rightarrow X'} \langle \sigma/\hat{y} \rangle_{C \rightarrow X'}^{-2} + N_{O \rightarrow X'} \langle \sigma/\hat{y} \rangle_{O \rightarrow X'}^{-2})
    \label{eq:myNuisance}
\end{equation}
where $X'$ stands for all possible isotopes of the CR species $X$ (for example, for B it means the sum for the isotopes $^{10}$B and $^{11}$B). The nuisance terms are therefore originated from Gaussian distributions with mean 1 and sigma given by $\langle \sigma/\hat{y} \rangle$, and can be interpreted as penalty terms that prevent the scaling factors to be far from 1.

The fact of applying the same amount of renormalization (scaling) to every channel seems reasonable, since if we do it unevenly (i.e. different scaling factors for different channels) we are re-weighting the importance of the channels (in addition, in the algorithms used to extrapolate parametrisations from a channel to another, these scalings are also extrapolated, supporting this idea). On the other hand, leaving unchanged the energy dependence of cross sections implies no further degeneracies with the diffusion coefficient. A different approach to take into account cross sections uncertainties is reported in ref. \cite{derome2019fitting}, where they notice that the degrees of freedom used to model the cross sections uncertainties can bias the determination of the transport parameters.

With this strategy we are including the normalization of the cross sections in the analysis, thus enabling the cross sections to be scaled up or down, with each secondary being more penalized as the scaling associated to it deviates from 1. The effect of scaling all the production cross sections causes the fluxes to be scaled by the same factor. Therefore the fluxes of the CR species $X$ in the likelihood will be scaled of a factor $\Delta_X$. We argue here that, although a change in the spallation cross sections with a given projectile implies a change of its inelastic cross sections, these cross sections are mostly dominated by the proton production, resulting in changes below 1\%, which are very small in comparison to the current uncertainties. In addition, this change in the inelastic cross sections would only affect primary CRs, whose flux is fixed to that of AMS-02 data in the analysis.

Nevertheless, when studying each ratio independently, the cross sections uncertainties are mostly absorbed by the normalization of the diffusion coefficient, as also reported in ref. \cite{Weinrich:2020cmw}, which means that $\Delta_X$ will remain equal to 1 and no conclusions on these uncertainties can be obtained.

Combining the secondary-to-primary ratios of different species allows accessing to this information. In order to include the full correlation between cross sections and CR fluxes, the secondary-over-secondary ratios are included in this combined analysis. As we have seen, they are a perfect tool to study cross sections parametrisations with very small uncertainty at high energies. Therefore including these ratios in the analysis can make more robust the determination of the diffusion coefficient, providing information about the correlations between the secondary species. In this way, to avoid biasing the results (since we do not include the halo size), we use the likelihood function defined using the secondary-over-secondary ratios above $30 \units{GeV}$.

In conclusion, the logarithm of the likelihood function used to carry out the combined analysis of Li, B and Be is given by:
\begin{equation}
     \ln \mathscr{L}^{Total} = \sum^{Li,Be,B/(C,O,Li,Be,B)}_R \ln(\mathscr{L}(R)) + \sum^{B, Be, Li}_X \mathscr{N}_X
    \label{eq:mylikelihood}
\end{equation}
where $R$ indicates the flux ratios (six secondary-over-primary ratios and three secondary-over-secondary ratios) and $X$ indicates the cross sections channels included (those coming from $^{12}$C and $^{16}$O projectiles only). %The term $\mathscr{N(X)}$ in the r.h.s. represents the contribution from the nuisance parameters.

We remark here that other ratios, as those involving nitrogen, or the C/O ratio, are not included in the analysis given the degeneracy with the source terms, which can significantly bias the determination of the propagation parameters.

\subsection{Results}
\label{sec:results}

The algorithm explained above has been applied to find the optimal diffusion parameters that best fit the spectra of each of the ratios of B, Be and Li to C and O (B/C, B/O, Be/C, Be/O, Li/C, Li/O) independently, for both diffusion models (eqs.~\ref{eq:sourcehyp} and~\ref{eq:breakhyp}). In addition, the combined analysis, in which the secondary-over-primary and secondary-over-secondary ratios (only at $E > 30 \units{GeV}$) as well as cross sections uncertainties are included, has also been performed. It must be highlighted that for each of the different cross sections, the value of the halo size is different, and set to the value found in section~\ref{sec:size}.

\begin{figure}[!t]
	\centering
	\includegraphics[width=\textwidth, height=0.6\textheight]{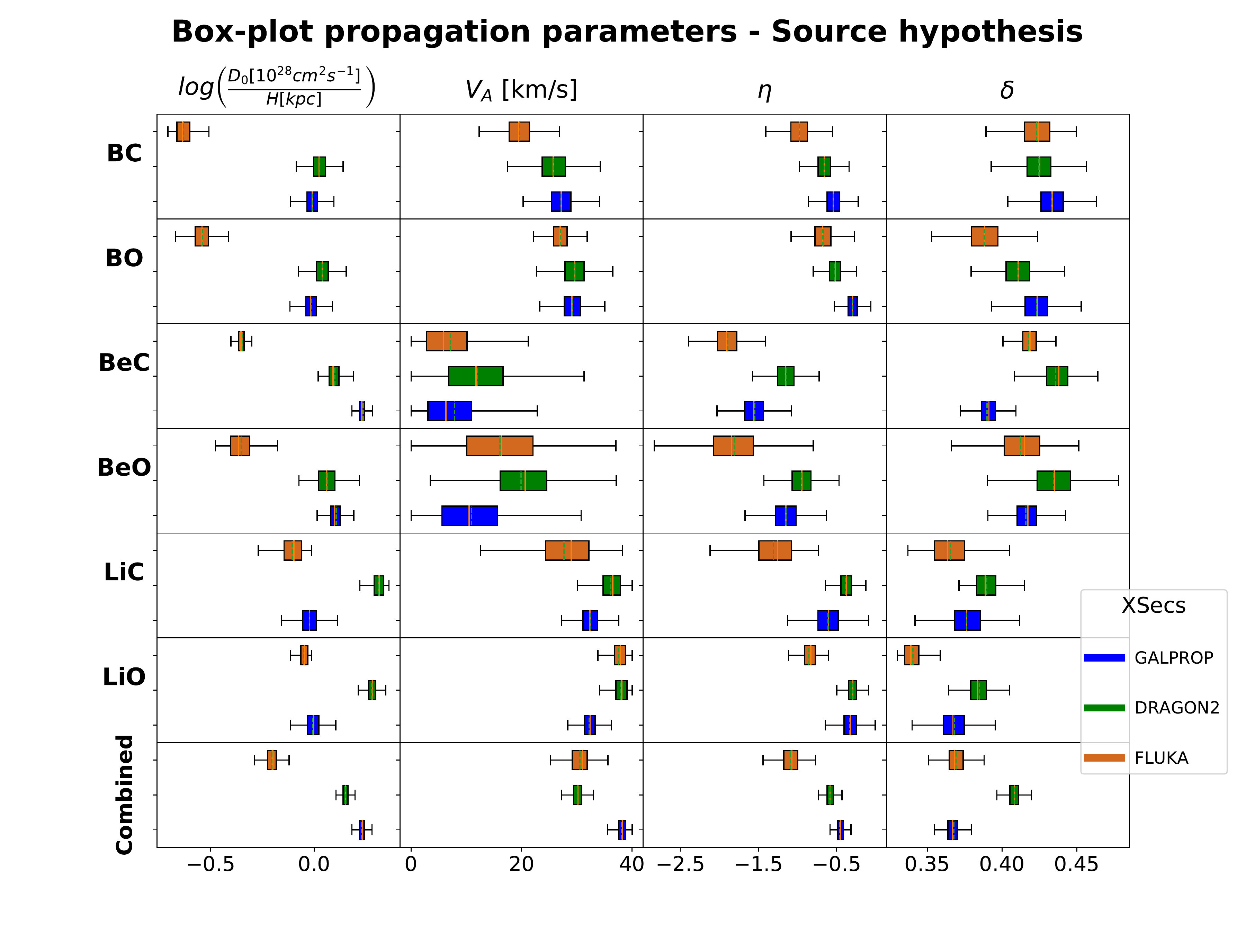}
	\caption{\footnotesize Box-plots representing the probability distribution resulting from the MCMC algorithm used to find the best propagation parameters to describe AMS-02 results with the diffusion coefficient of eq.~\ref{eq:sourcehyp}, and with the FLUKA, GALPROP and DRAGON2 cross sections.}
	\label{fig:boxplot_Source}
\end{figure}

\begin{figure}[!t]
	\centering
	\includegraphics[width=0.99\textwidth, height=0.6\textheight]{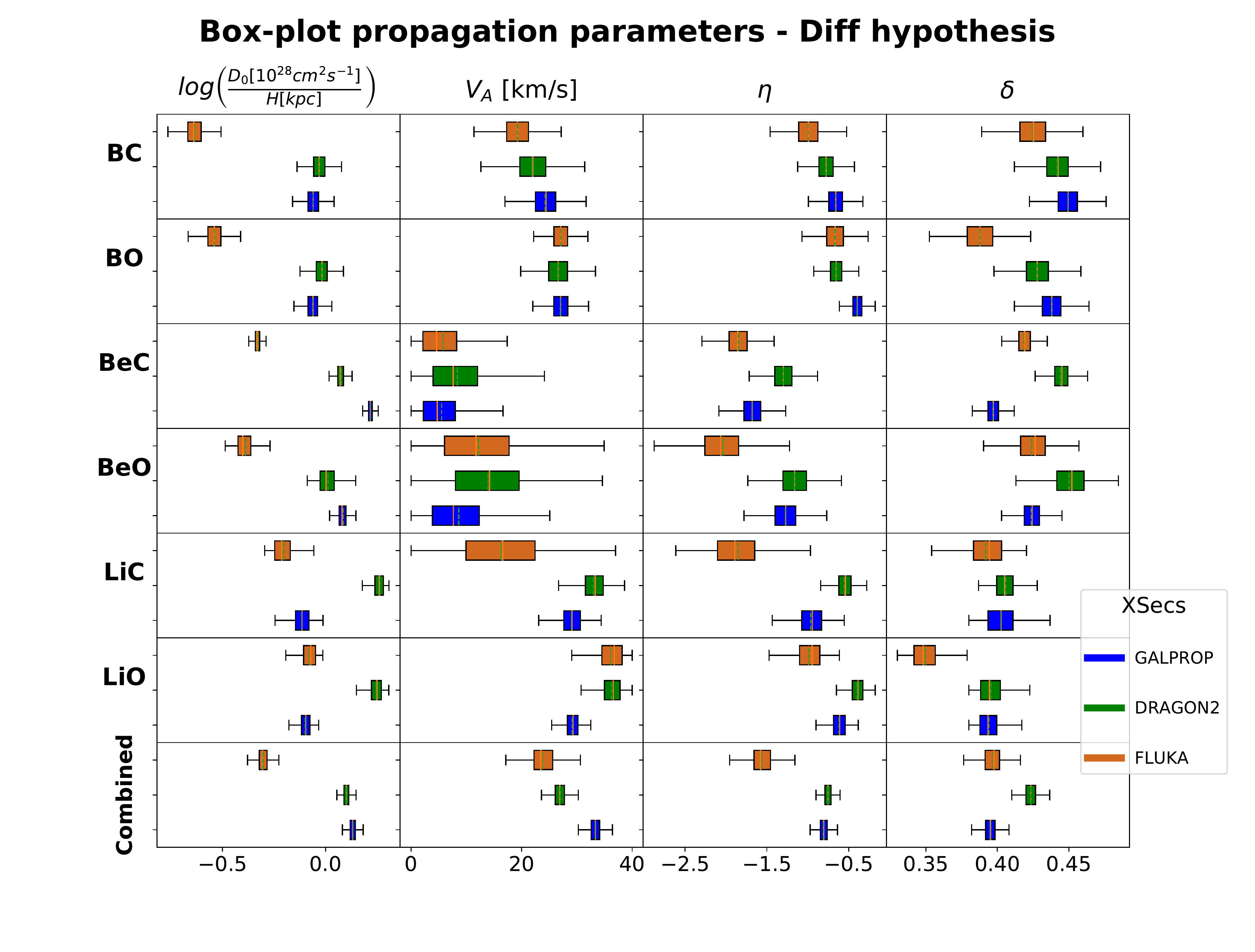}
	\caption{\footnotesize Same as in figure~\ref{fig:boxplot_Source} but for the diffusion hypothesis of the hardening in the CR spectra.}
	\label{fig:boxplot_Diff}
\end{figure}

The posterior PDF of each parameter is usually very close to a Gaussian distribution function, with the exception of the $V_A$ parameters in the Be/C and Be/O ratios and, in a few cases, of the $\eta$ parameter in the FLUKA and GALPROP cross sections, whose PDFs look more similar to an inverse log-normal distribution. These results are summarized in Figure~\ref{fig:boxplot_Source} and~\ref{fig:boxplot_Diff}.% and are summarised in tables, together with the PDFs of each analysis, in appendix~\ref{sec:appendixC}.

A box-plot representation is used since it can better represent the statistical information behind. Each box in Figures \ref{fig:boxplot_Source} and \ref{fig:boxplot_Diff} displays the median (orange solid line in the middle of the boxes), the mean (dashed cyan line; which, in most of the cases, is coincident with the median), the interquartile range IQR (range between Q1 and Q3 quartiles, i.e. from the 25th to the 75th percentiles, ranging $0.6745 \sigma$ around the median in the case of Gaussian PDFs), represented by the boxes and the "minimum" and "maximum" values (Q1 - 1.5$\cdot$IQR and Q3 + 1.5$\cdot$IQR, which contains the 99.3\% of the probability in the case of Gaussian PDFs). %The boxes represent the IQR, ranging $0.6745 \sigma$ around the median and the arms show the parameters range covering almost 3$\sigma$ ($\pm 2.698$) around the median. %https://towardsdatascience.com/understanding-boxplots-5e2df7bcbd51

An explicit summary of the median values and of the ranges of propagation parameters within the 68 and 95\% of their PDFs is given in form of tables in the appendix~\ref{sec:appendixC} for each of the cross sections used and diffusion coefficient parametrisations. In addition, the corner plots with the PDFs for the combined analyses are also shown in the appendix, for both the source and diffusion hypotheses.

Very similar results (always inside 1-2 $\sigma$) are found for the propagation parameters analysed in both hypotheses for the CR hardening. In general, the diffusion hypothesis favors smaller values of $D_0$, $V_A$ and $\eta$ and larger values of $\delta$. This is something expected, since the slope (controlled by the $\delta$ parameter) in the source hypothesis tends to be smaller, to compensate the hardening observed in CR spectra at the highest energies. Therefore, if $\delta$ is larger, $D_0$ must be smaller to balance this change (see the contour plots in appendix~\ref{sec:appendixC} to see the correlations between parameters).

%\newpage

\begin{figure}[!htpb]
	\centering
	\includegraphics[width=0.47\textwidth, height=0.23\textheight]{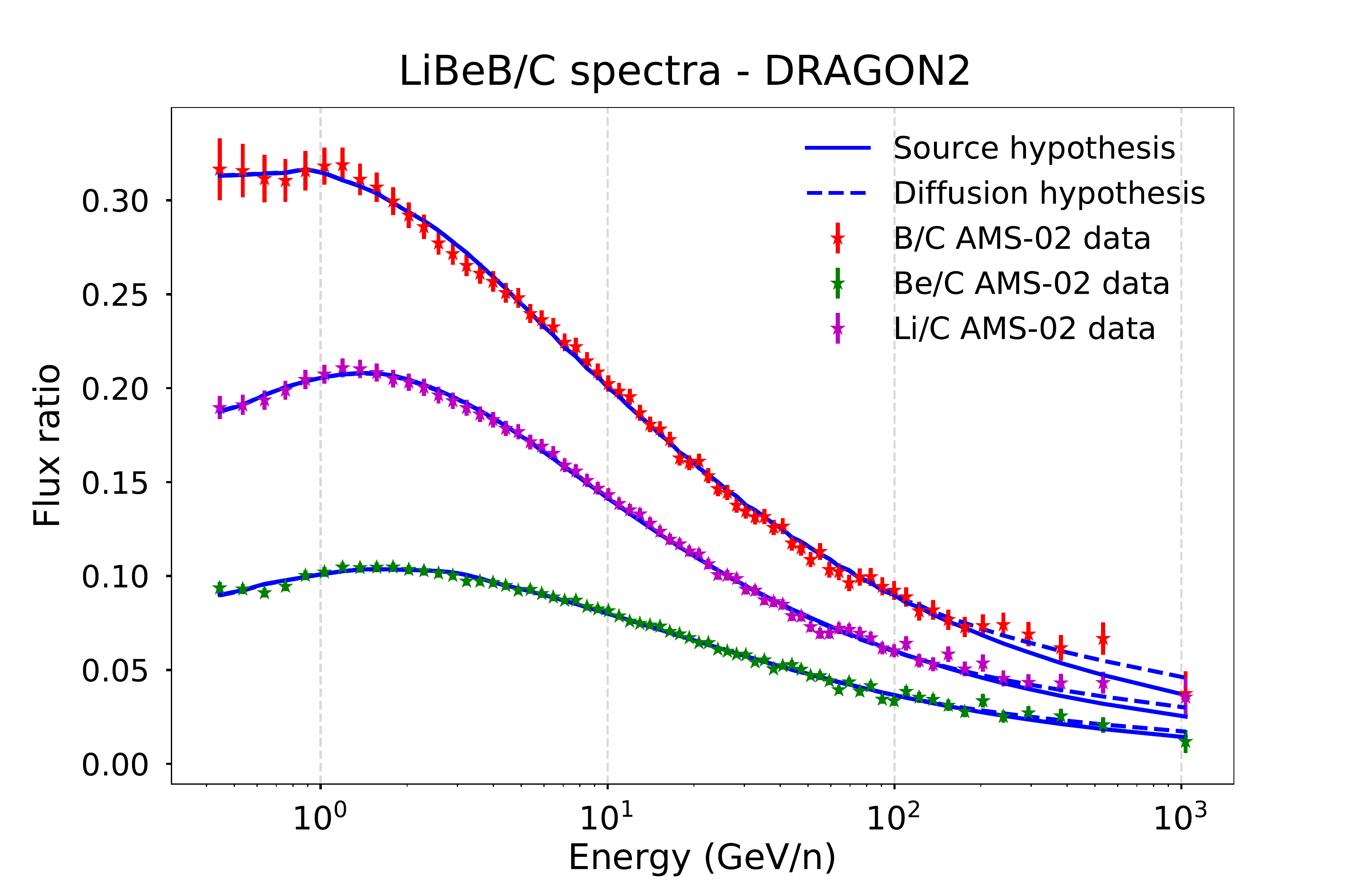}
	\includegraphics[width=0.47\textwidth, height=0.23\textheight]{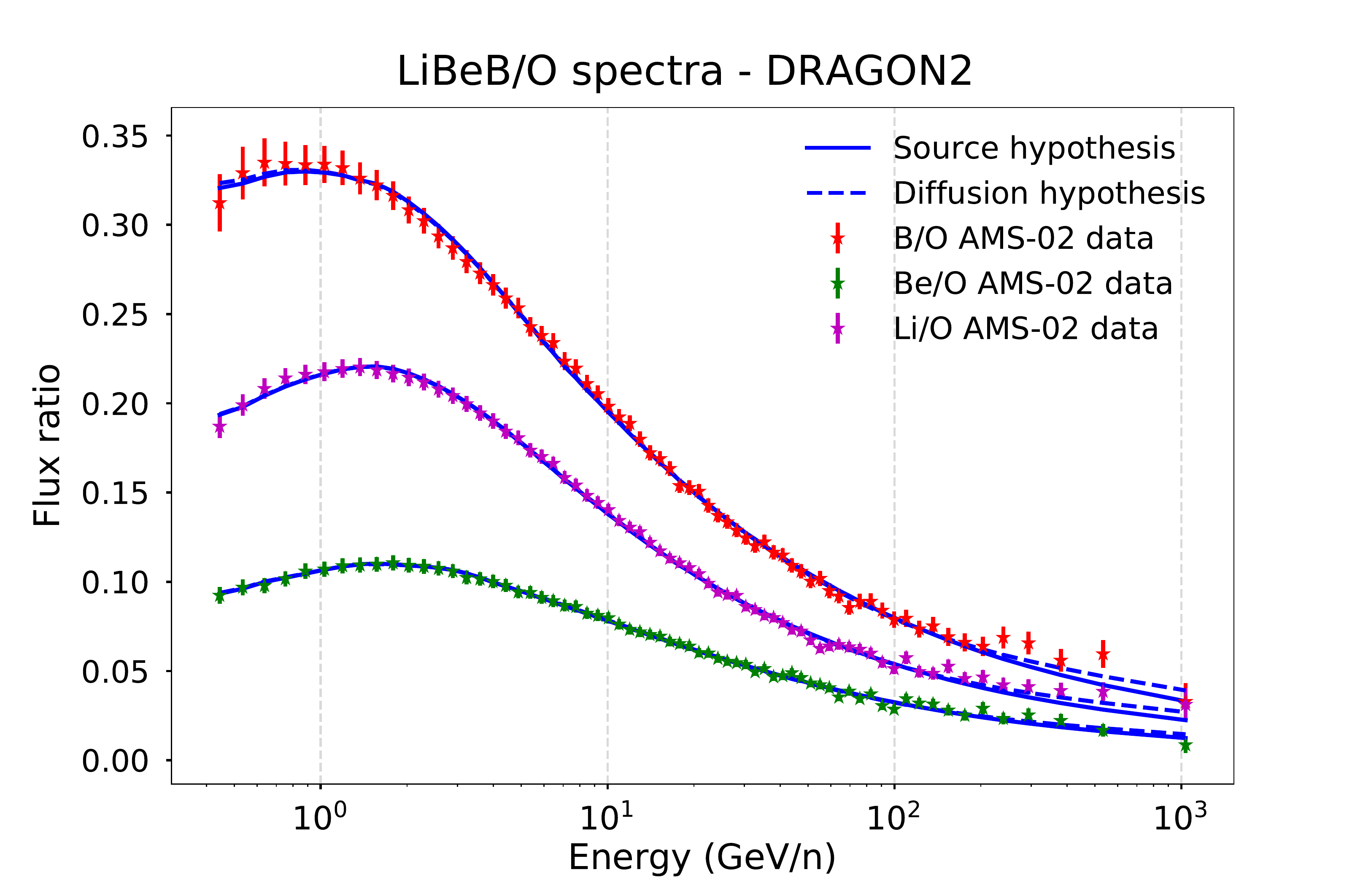}
	\caption{\footnotesize Simulated secondary-over-primary ratios with the propagation parameters determined independently in the MCMC analysis for the DRAGON2 cross sections parametrisations and for both hypothesis of the CR fluxes hardening. As we see, the parametrisation of the diffusion coefficients allows perfect reproduction of the shapes of the ratios. Fits with same quality are achieved for the FLUKA and GALPROP cross sections too.}
	\label{fig:Single_ratios}
\end{figure}

\hspace{-0.3cm}Some important remarks must be made about these results:

\textbf{Comparison among the trend of different ratios}
\begin{itemize}

    \item In general, the relative uncertainties in the determination of propagation parameters are larger in the $\eta$ and $V_A$ parameters, due to their degeneracy and the need of more data points at the low energy region.

    \item The propagation parameters obtained for each pair of ratios involving the same secondary (e.g. B/C and B/O) are always compatible with each other. Small discrepancies can arise, mainly due to the production of secondary carbon, which is important at low energies, meaning that the cross sections of carbon production can be also important for the determination of the propagation parameters at these energies. 
    
    \item As expected, each of the ratios (also in the combined analysis and for every cross sections set used) favors a negative value of $\eta$, theoretically motivated by dissipation of MHD waves or appearance of non-resonant interactions at low energies. The Be ratios show clearly smaller values of this parameter for every  cross sections set used, but this seems to be due to the halo size used in the simulations. The degeneracy between these two parameters makes one compensate the defects of the other, thus requiring much lower $\eta$ values.
    
    \item Interestingly, the value of $\delta$ is usually around 0.42-0.45 for the B and Be ratios, while a lower value is favored for the lithium ratios (between 0.36 and 0.40). This fact is even clearer for the FLUKA cross sections of Li production. 
    
    \item The values of $V_A$ obtained are smaller than 30 km/s for the B ratios ($\sim 26\units{km/s}$), while the Li ratios larger values are found (around $35 \units{km/s}$). On the other hand, the $V_A$ values for Be ratios are much smaller, usually compatible with $V_A=0$. This is again due to the degeneracy with the halo size employed in the simulations. Therefore, excluding the Be ratios, an average value of $V_A \sim 30 \units{km/s}$ is found, which seems to be in agreement with the expected values (see refs. \cite{Spanier_2005, Spangler:2010nu, Lerche_2001}). %This must be due to the degeneracy at low energies between parameters, even the convection velocity. %Hence, this $V_A$ value must be taken as an effective value instead of a real physical parameter.
    
    \item Finally, As we see from Figure~\ref{fig:Single_ratios}, the secondary-over-primary flux ratios are perfectly reproduced when making each fit individually. This means that the degrees of freedom available in the parametrisations of the diffusion coefficient are enough to suitably match observations. On the other hand, it should be mentioned that the uncertainties appreciably change when taking into account correlated errors from the AMS-02 experiment, as found in the appendix~\ref{sec:covariance}, where we revise the analysis of the B/C ratio taking into account these correlated errors.
    
\end{itemize}

\begin{figure}[!htpb]
	\centering
	\textbf{\large{Cross sections scaling factors}}
		
	\includegraphics[width=0.62\textwidth, height=0.385\textheight]{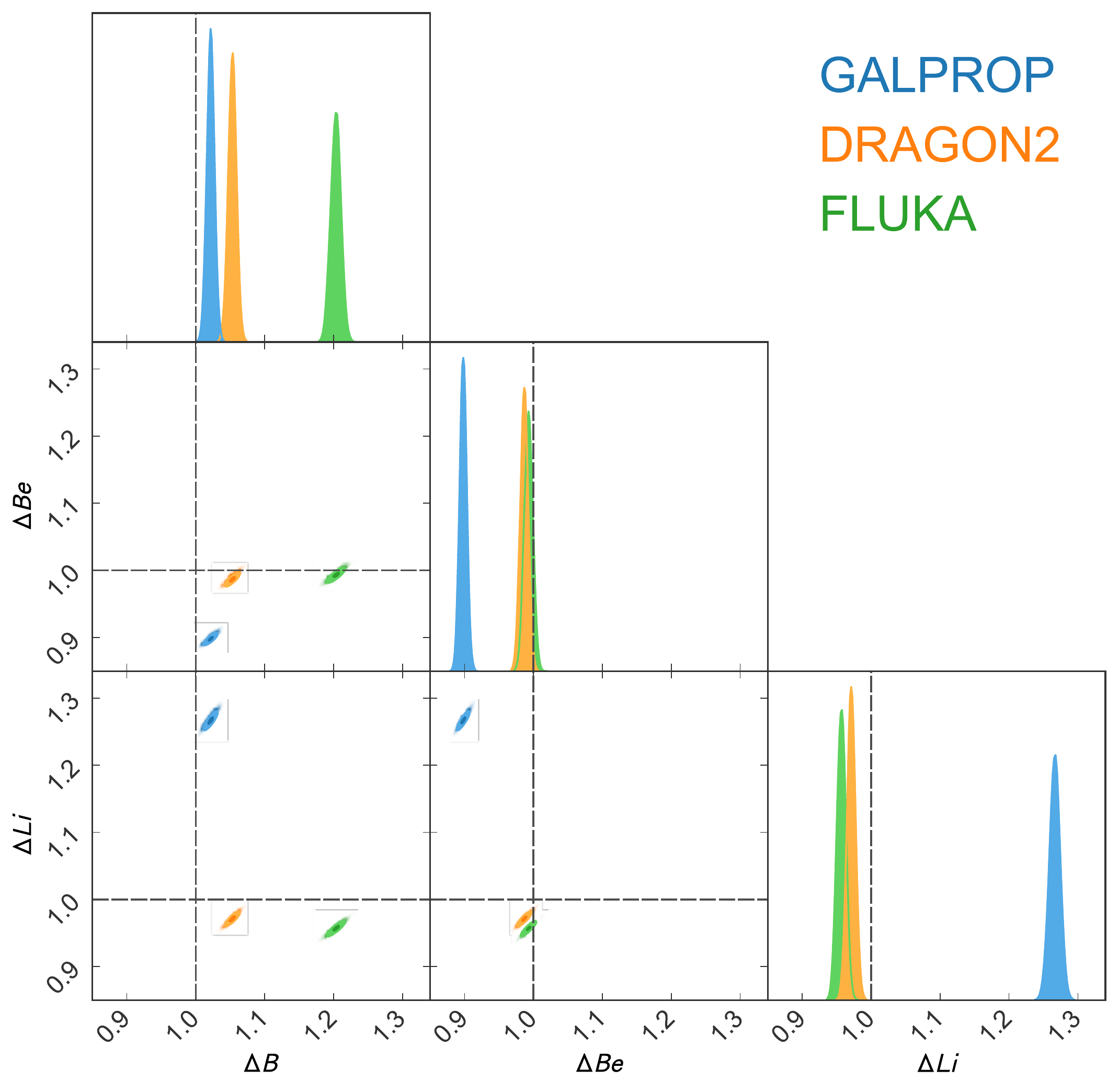}
	\caption{\footnotesize Corner plot of the scaling (nuisance) parameters determined by the MCMC combined analysis. A dashed line indicating scaling = 1 is added as reference for each of the panels.}
	\label{fig:Nuisance_combined}
\end{figure}

\textbf{Comparison among different cross sections sets}
\begin{itemize}
    \item One of the main differences between the results obtained with the two approaches for calculating the cross sections (i.e., from the FLUKA cross sections to the parametrisations) concerns the ratio $D_0/H$. While the parametrisations yield values close to unity, the FLUKA cross sections yields a value significantly lower than 1. This is not an effect only due to the different cross sections sets, but mainly to the (very) different halo size value found to fit the $^{10}$Be ratios with the FLUKA cross sections. This was observed when adjusting the diffusion coefficient for different halo size values with same cross sections, in section~\ref{sec:size}, and must not be confused with an exclusive effect of the different cross sections, but to a deviation from the assumption that $D_0/H$ should remain nearly constant.
    
    \item In general, the different cross sections give very similar predictions. One of the main differences are the $\eta$ and $V_A$ values obtained for the Be ratios with the DRAGON2 cross sections, which are larger (more consistent with the Li and B ratios) than those obtained with other cross sections. However, this might be also related to the degeneracy with the halo size value used and suggests that the determination of the halo size performed for the DRAGON2 cross sections seems to be more accurate than for the other cross sections.
    
    \item Another important point to stress is that, for all the cross sections sets, the value of $\delta$ determined by the Li ratios is significantly smaller (often more than $2\sigma$) than the one determined by the other ratios. In particular, the FLUKA cross sections predict a much smaller value, specially the Li/O ratio. This points to the fact that, in both, parametrisations and Monte Carlo approaches, the overall slope of the cross sections for Li production at high energies tends to be softer than it should be. Nevertheless, the best fit $\delta$ value for the Li ratios with the DRAGON2 cross sections is closer to the value inferred from the Be and B ratios, which might indicate that updating cross sections measurements can help to solve this issue.
    
    \item Finally, it is worth noticing that the $D_0$ values obtained are slightly different in the independent analyses of different secondary CRs, which is a consequence of the cross sections used. This is prevented by adding an overall scaling factor ($\Delta$) on the cross sections in the combined analysis. The relative differences of $D_0$ obtained and the secondary-over-secondary ratios for each of the cross sections sets reveal the relative degree of scaling on the production cross sections needed for each of the secondary CRs. These scaling factors, determined as nuisance parameters, are shown in Figure~\ref{fig:Nuisance_combined} and discussed in detail down below and in section~\ref{sec:CXSanal}.
\end{itemize}

\textbf{Comparing combined vs independent ratio analyses}
\begin{itemize}
    \item The most remarkable fact seems to be that in the combined analysis there is a clear trend to smaller $\delta$ values, usually below 0.40 which suggests that the most constraining ratios are those involving Li. Only for the combined ratios with the DRAGON2 cross sections the median value of $\delta$ is higher than 0.40, but also in this case it is below the median for B and Be ratios. This reflects the defects on the shapes of the cross sections and may cause an important bias in our determination of the propagation parameters (and also in the determination of the nuisance parameters). Probably, it could be fixed by adding more degrees of freedom in the analysis to reshape the cross sections (mainly Li production, since they present the largest uncertainties). %More interestingly, it may be linked to a primary source of Li, such that its spectrum is more difficult to be reproduced assuming pure secondary origin, although, as we saw in chapter~\ref{sec:XSecs}, the uncertainties in the Li production cross sections are high, thus making plausible to explain the discrepancy by means of a change in the Li cross sections.

    \item The results of the combined analysis are shown in Figs.~\ref{fig:SecPrim_DRAGON2},~\ref{fig:SecPrim_GALPROP} and~\ref{fig:SecPrim_FLUKA} for the DRAGON2, GALPROP and FLUKA cross sections, respectively. As we see, the results of the combined analysis with the DRAGON2 cross sections show an almost perfect fit of all the ratios, having the largest discrepancies for the Be ratios, which could be due the the halo size used. In turn, with the GALPROP cross sections the predictions are noticeably discrepant for Be and B ratios at low energies while only the Li ratios are well reproduced within experimental errors. This is due to the defects in the shape describing the cross sections. In addition, the fact that the parameters slightly favors those found in the independent analyses of the Li ratios is also observed with the DRAGON2 cross sections (although in a smaller extent) and it could be related to the lack of experimental data on cross sections of Li production, leading to a poorer parametrisation of the Li production cross sections. On the other hand, The results with the FLUKA cross sections are quite remarkable since one would expect larger discrepancies associated to the cross sections shape.

    \begin{figure}[!ht]
	\centering
	\includegraphics[width=0.47\textwidth, height=0.23\textheight]{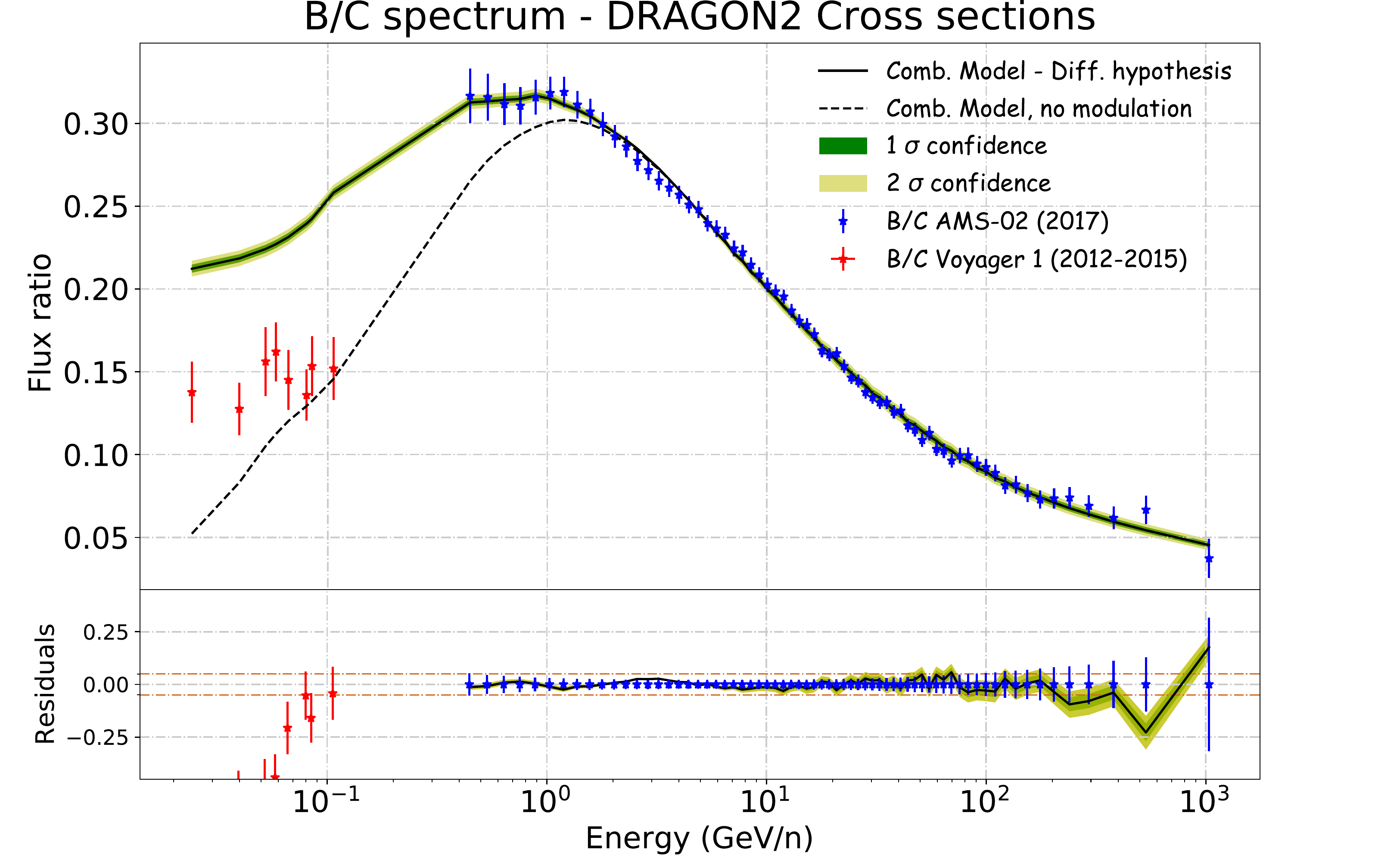}
	\includegraphics[width=0.47\textwidth, height=0.23\textheight]{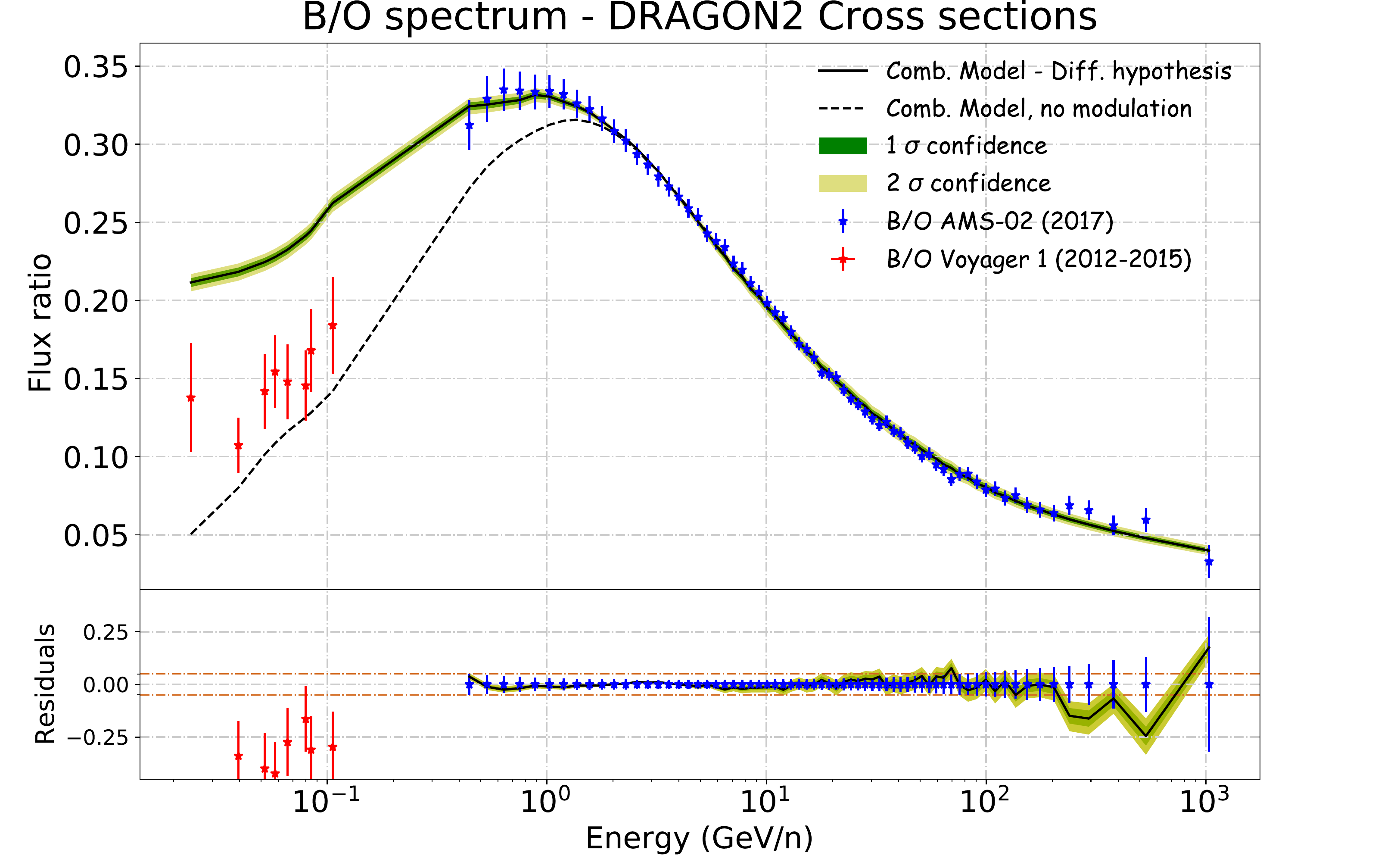}
	
	\vspace{0.4cm}
	
	\includegraphics[width=0.47\textwidth, height=0.23\textheight]{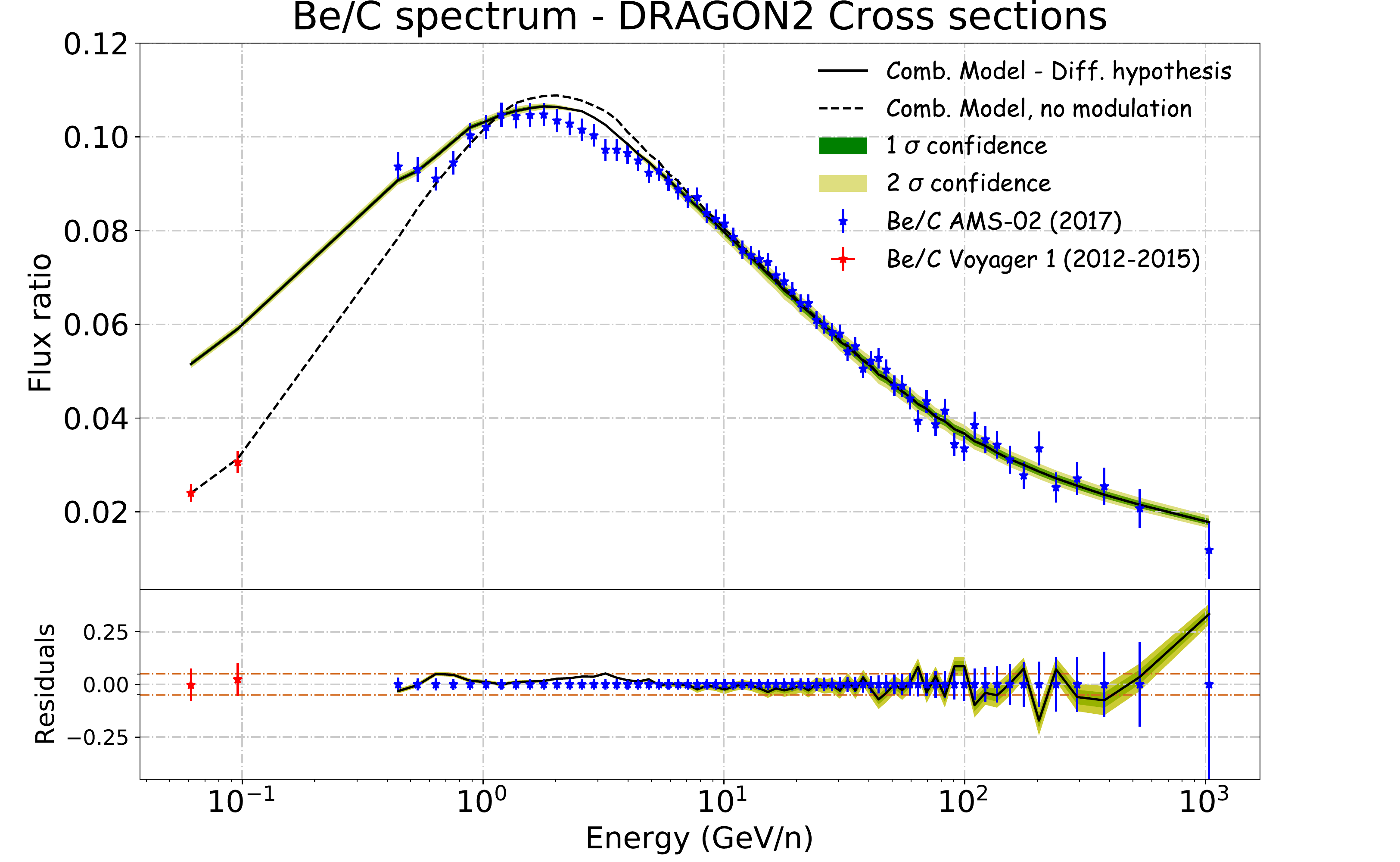}
	\includegraphics[width=0.47\textwidth, height=0.23\textheight]{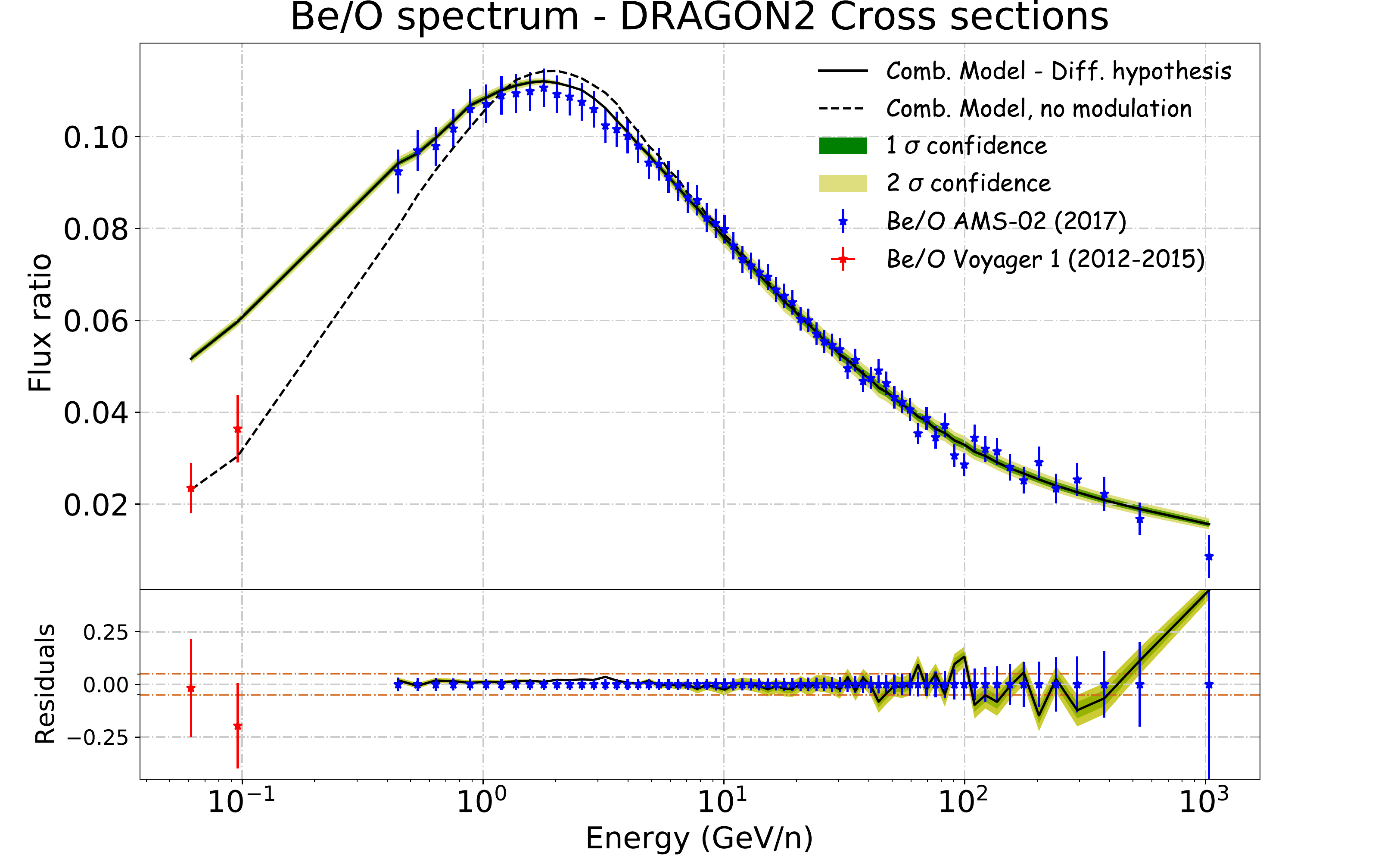}
	
	\vspace{0.4cm}
	
	\includegraphics[width=0.47\textwidth, height=0.23\textheight]{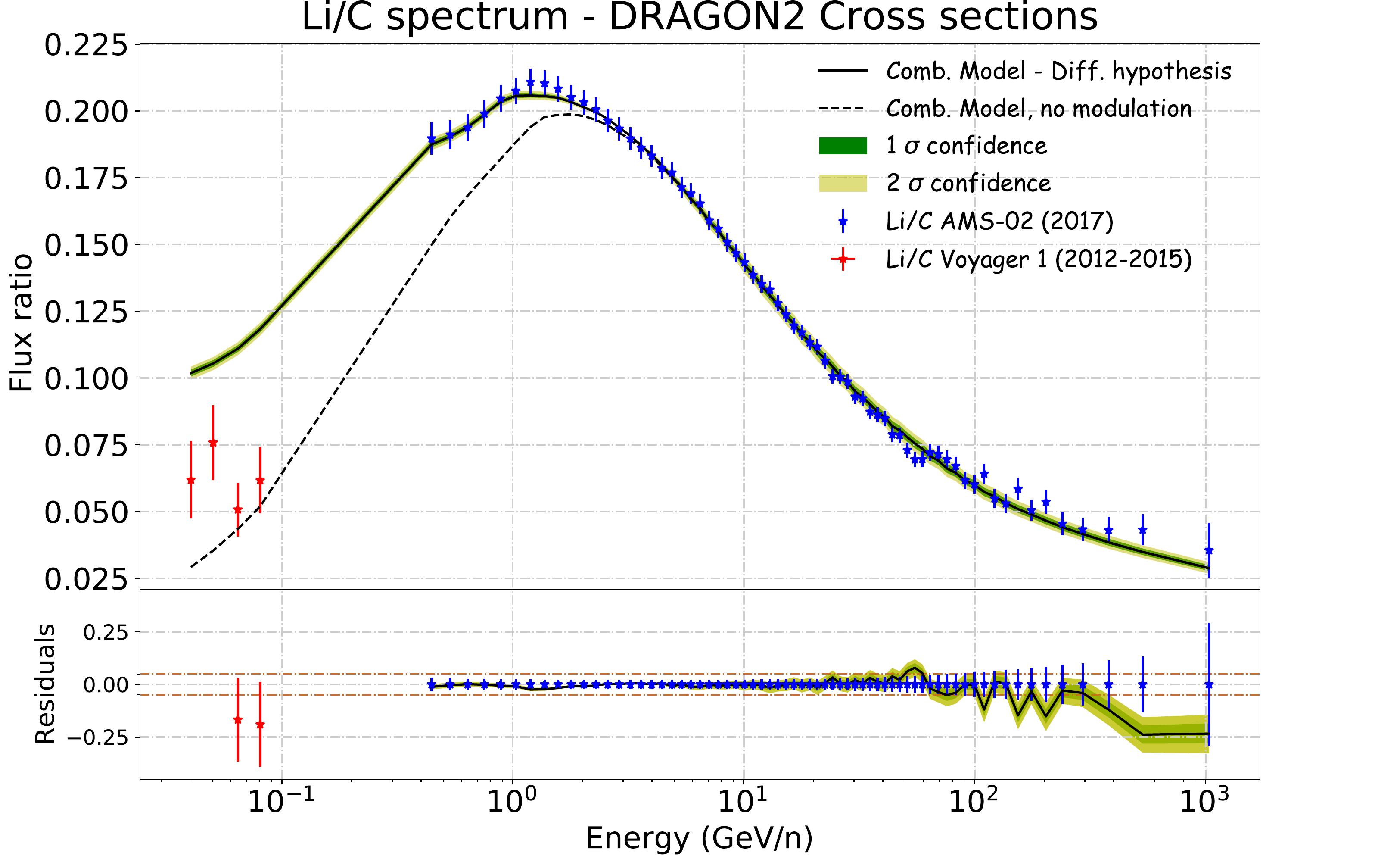}
	\includegraphics[width=0.47\textwidth, height=0.23\textheight]{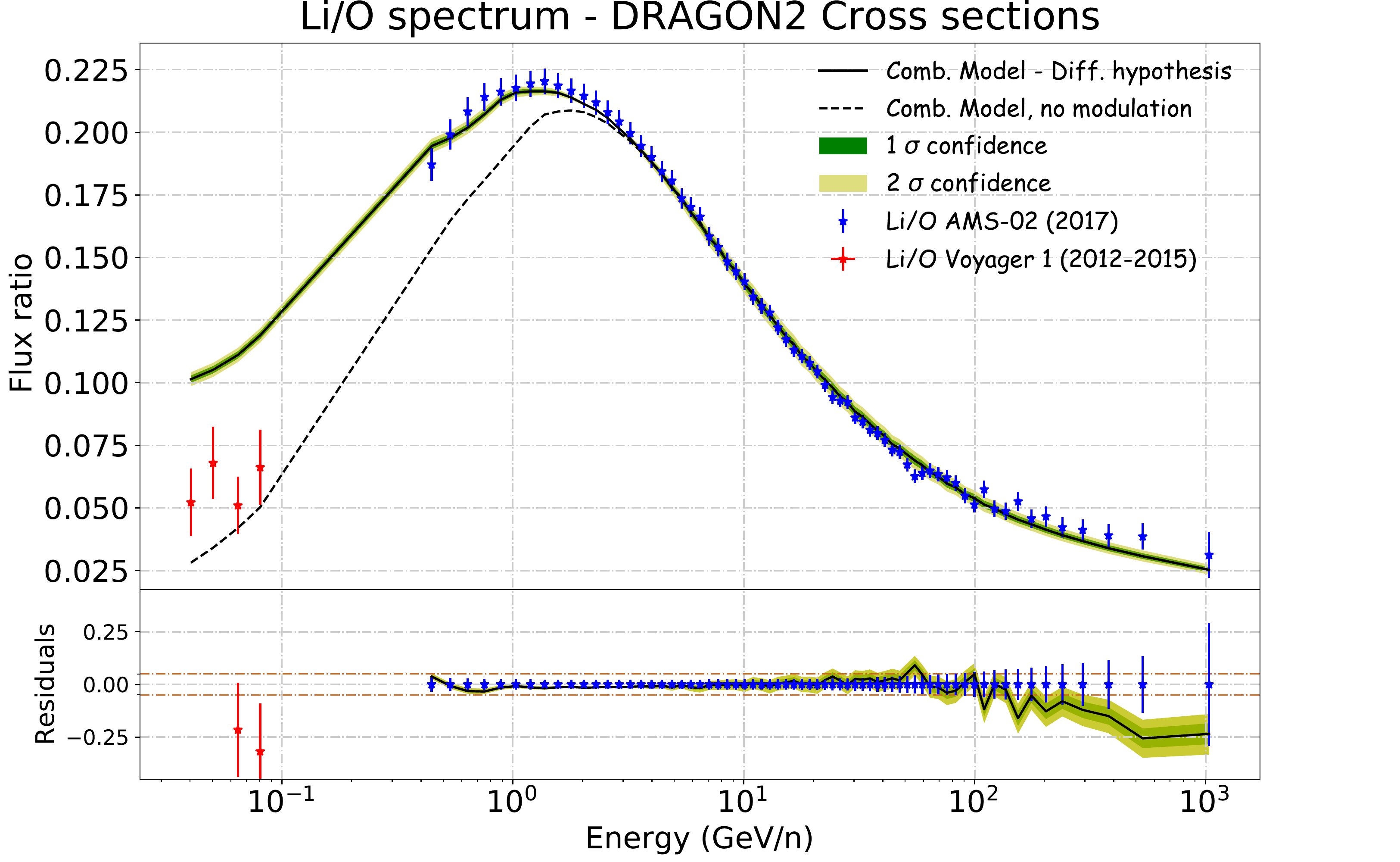}
	\caption{\footnotesize Secondary-over-primary spectra (B/C, B/O, Be/C, Be/O, Li/C, Li/O) predicted with the parameters determined by the MCMC combined analysis for the DRAGON2 cross sections. The grid lines at the level of 5\% residuals are highlighted in the lower panel in a different color for clarity. In addition, the Voyager-1 data are also included for completeness.}
	\label{fig:SecPrim_DRAGON2}
\end{figure}

\begin{figure}[!ht]
	\centering
	\includegraphics[width=0.47\textwidth, height=0.23\textheight]{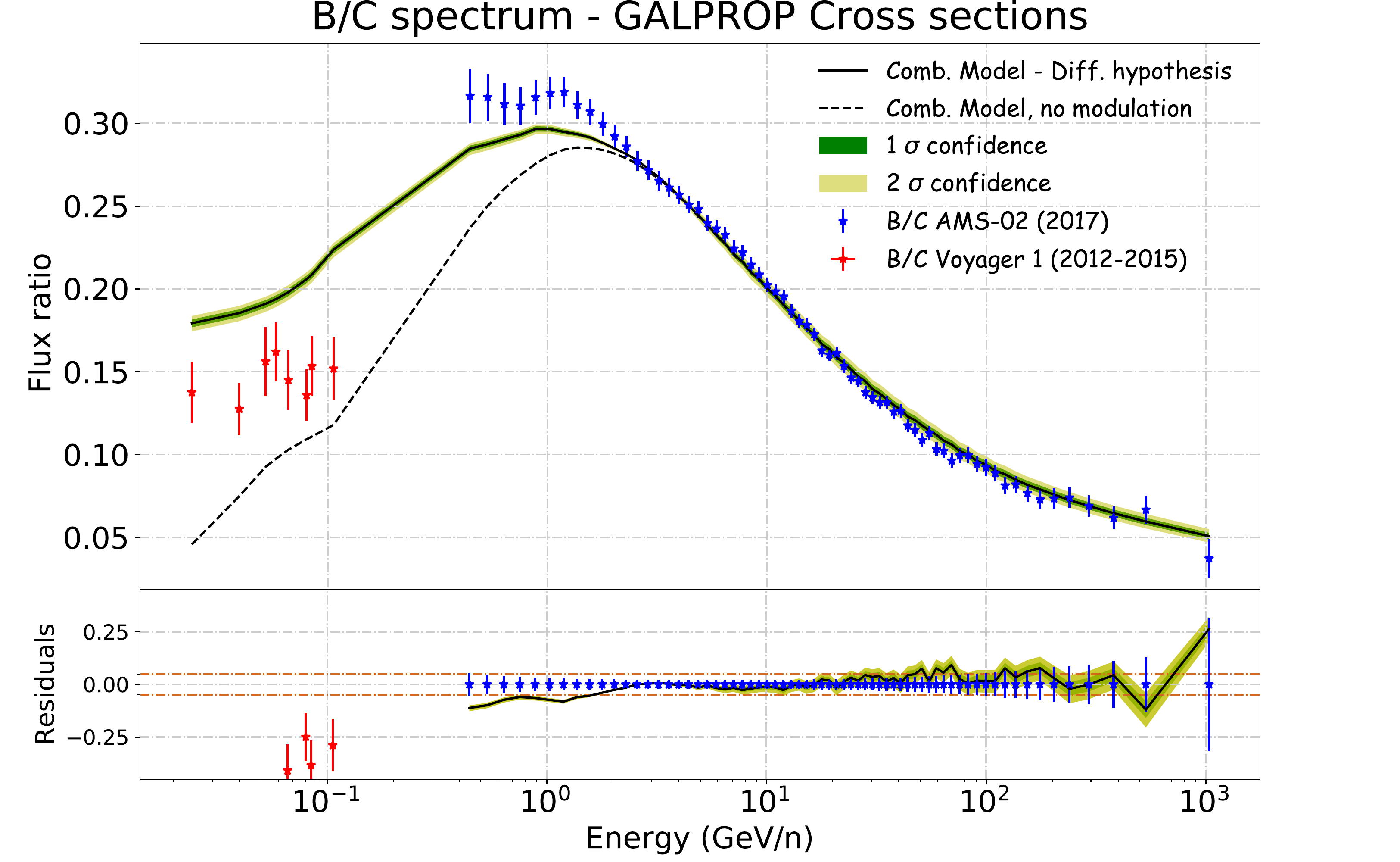}
	\includegraphics[width=0.47\textwidth, height=0.23\textheight]{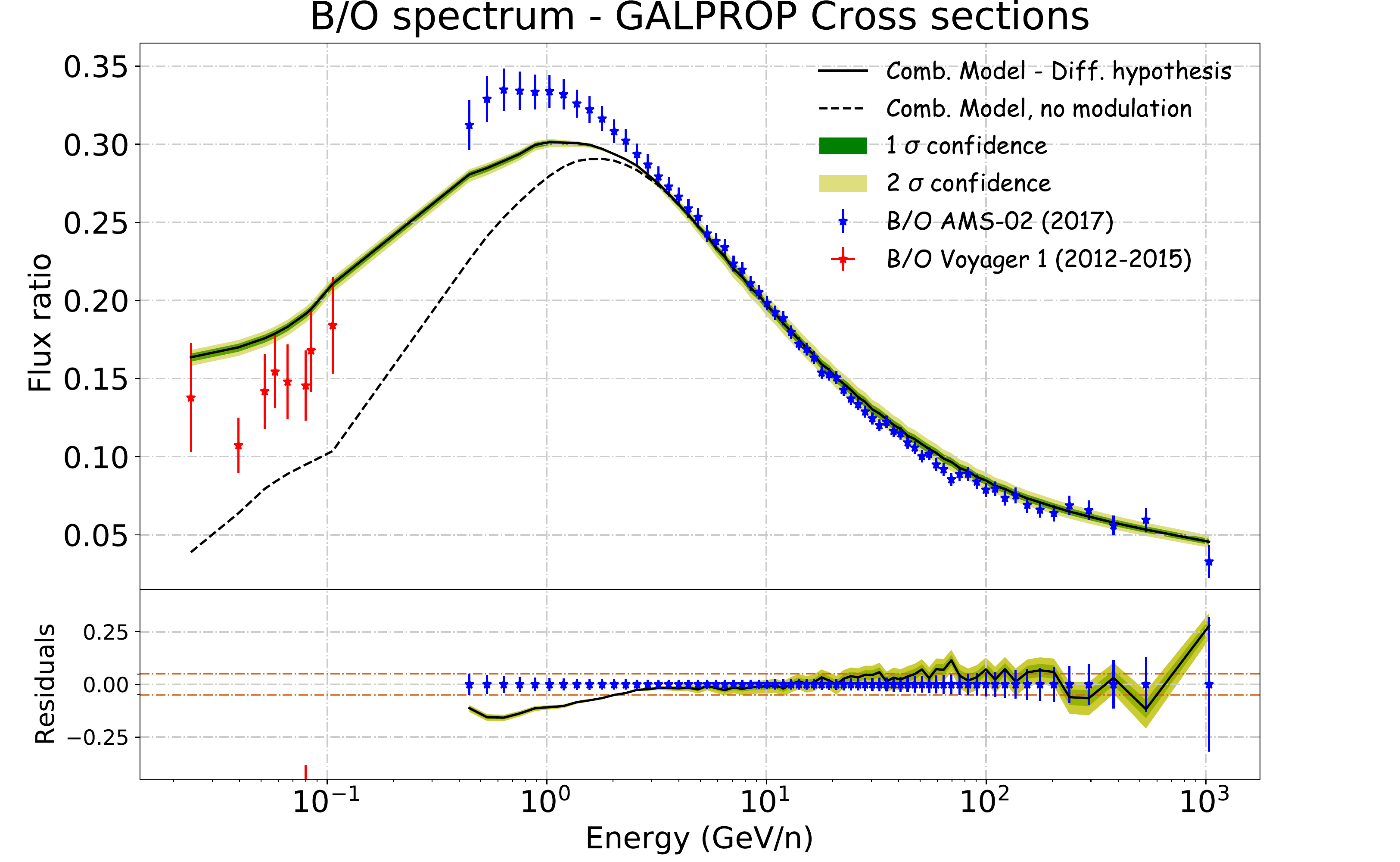}
	
	\vspace{0.4cm}
	
	\includegraphics[width=0.47\textwidth, height=0.23\textheight]{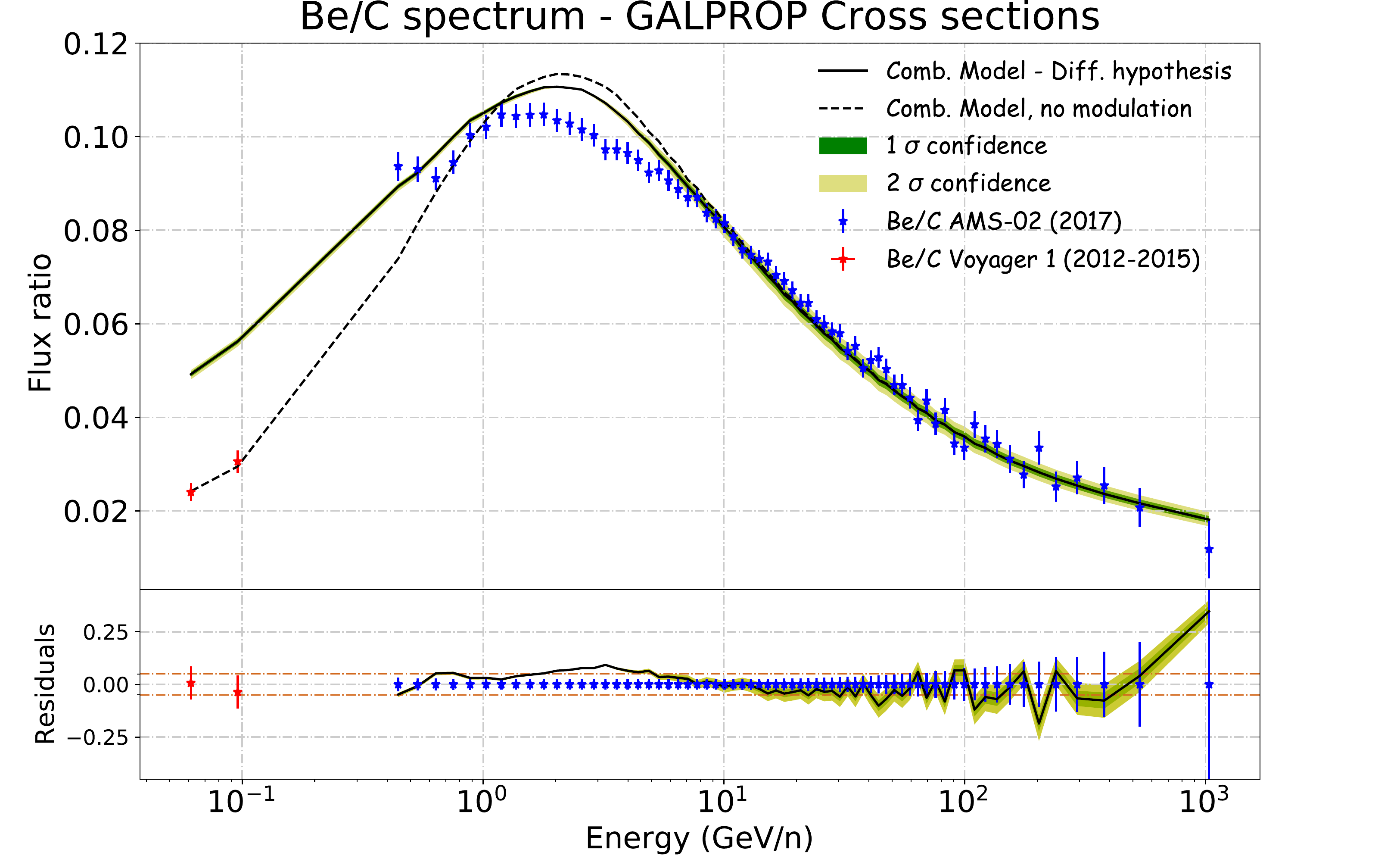}
	\includegraphics[width=0.47\textwidth, height=0.23\textheight]{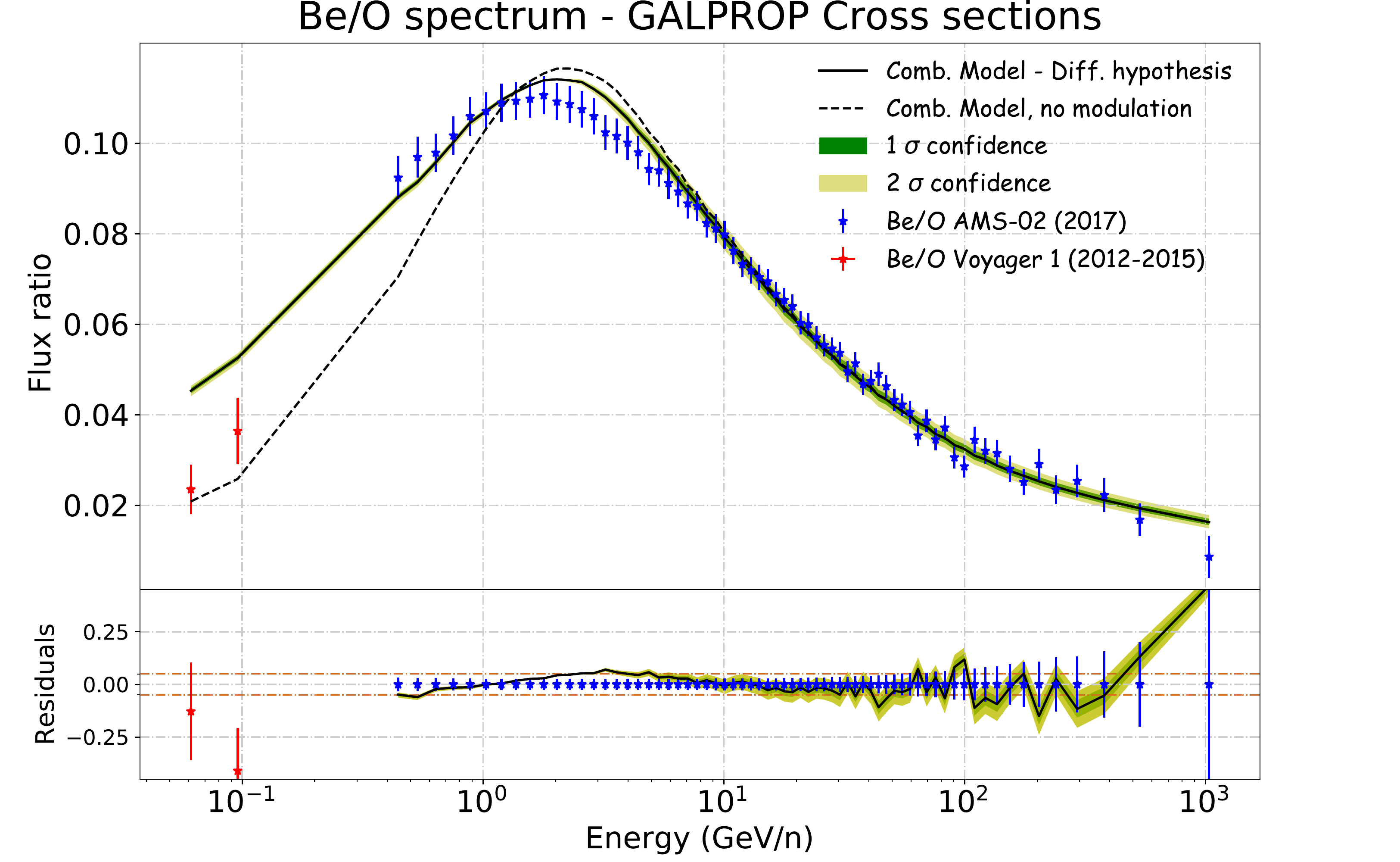}
	
	\vspace{0.4cm}
	
	\includegraphics[width=0.47\textwidth, height=0.23\textheight]{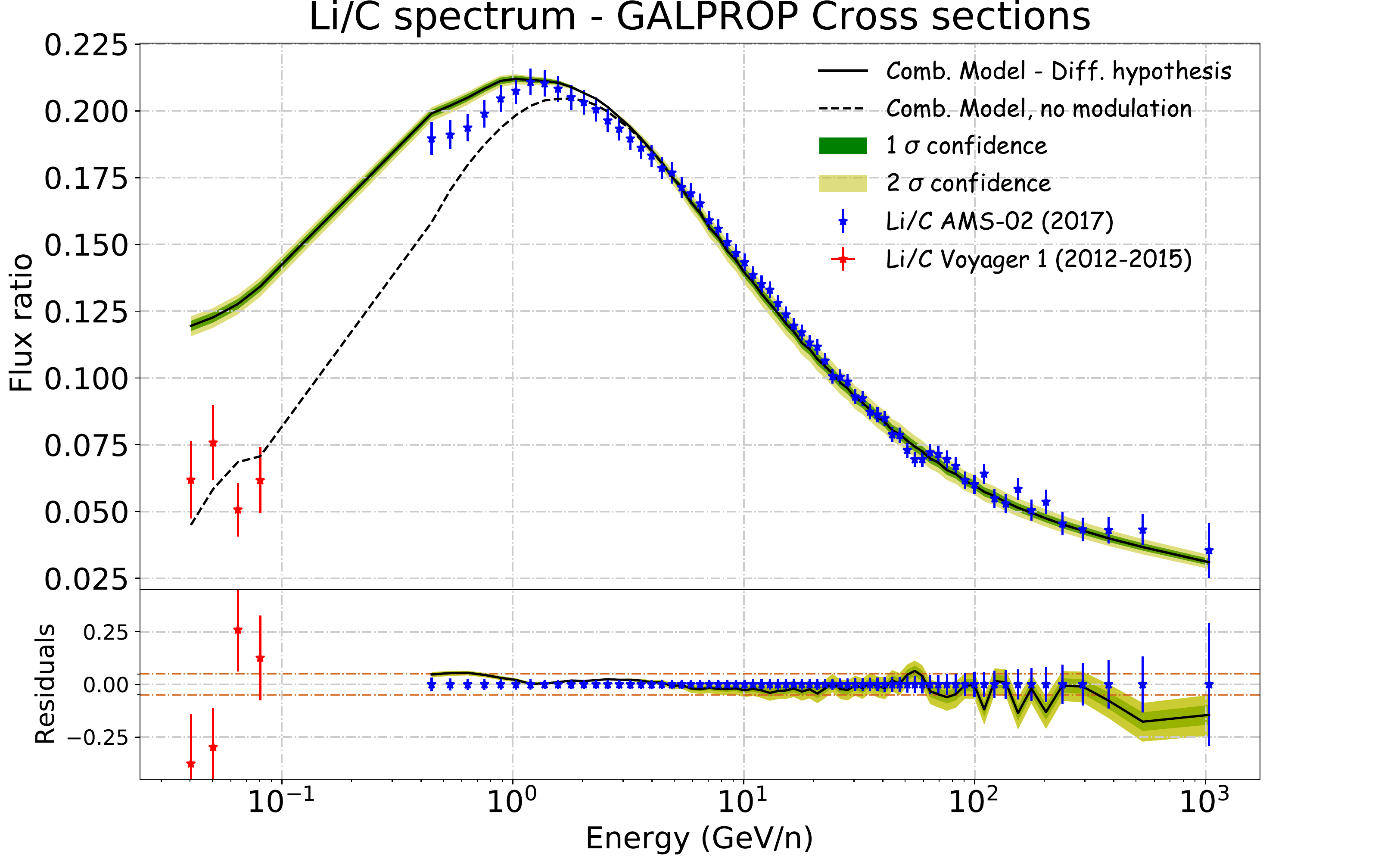}
	\includegraphics[width=0.47\textwidth, height=0.23\textheight]{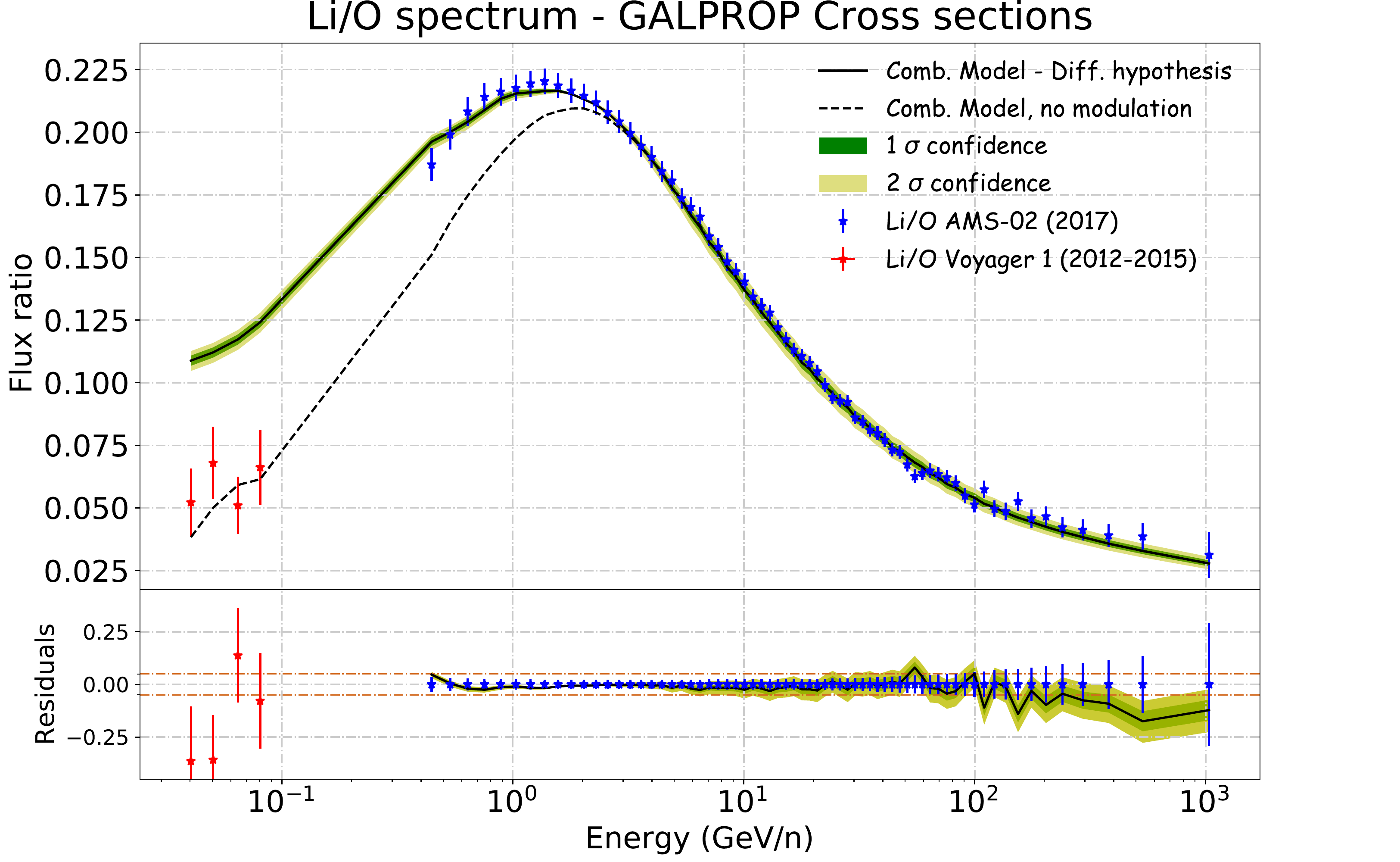}
	\caption{\footnotesize Same as in Fig. \ref{fig:SecPrim_DRAGON2} but with the GALPROP cross sections.}
	\label{fig:SecPrim_GALPROP}
\end{figure}

\begin{figure}[!ht]
	\centering
	\includegraphics[width=0.47\textwidth, height=0.23\textheight]{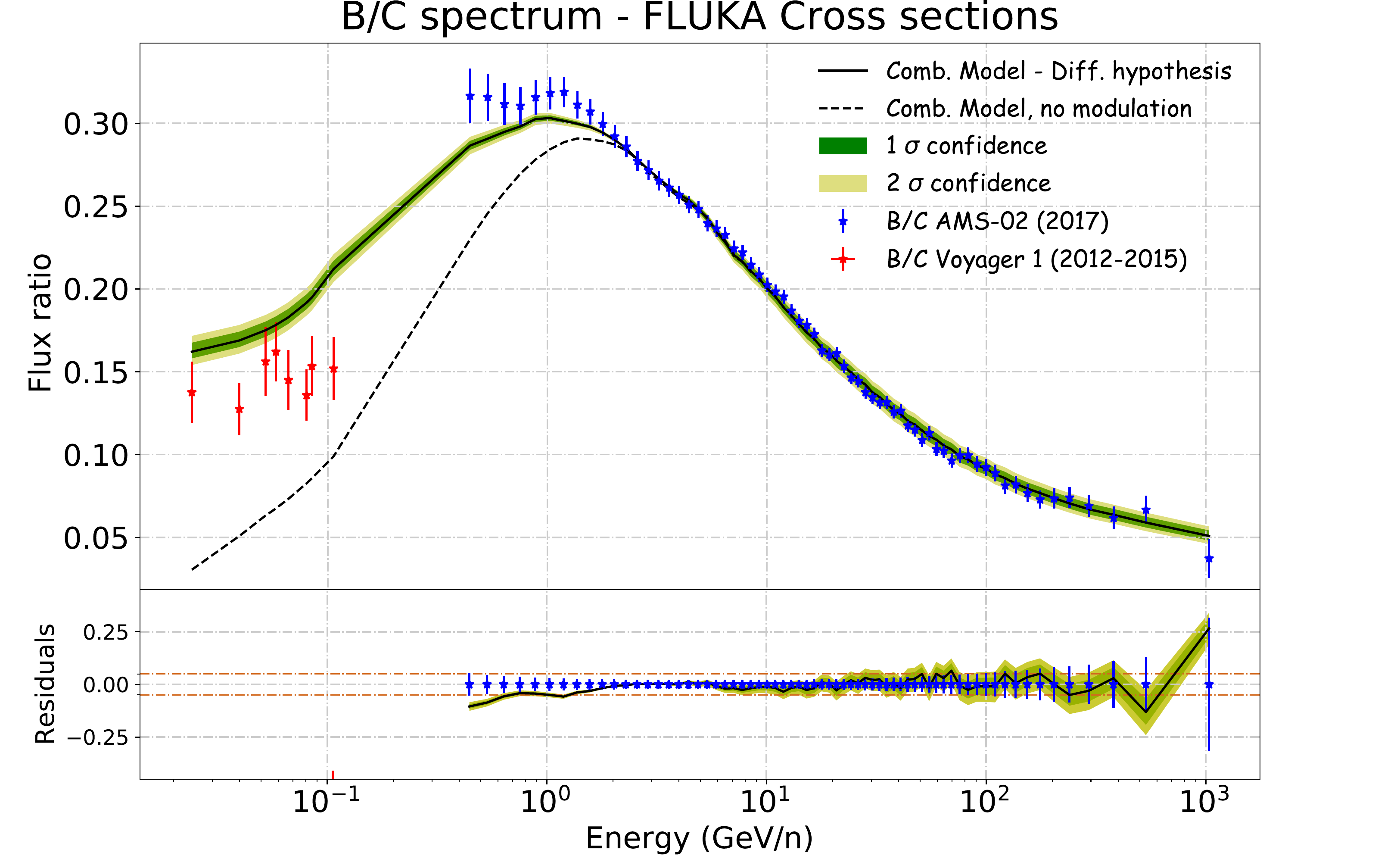}
	\includegraphics[width=0.47\textwidth, height=0.23\textheight]{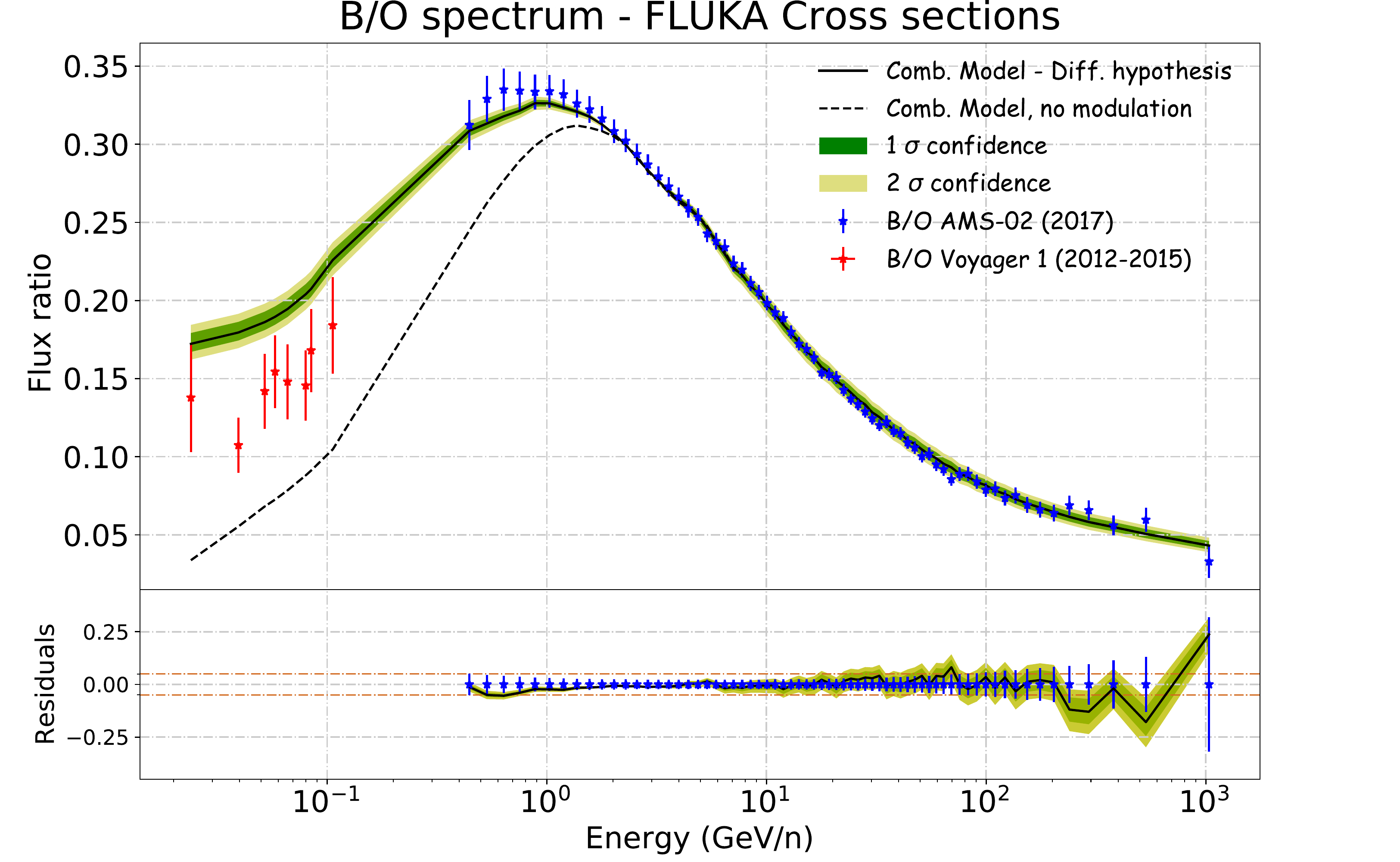}
	
	\vspace{0.4cm}
	
	\includegraphics[width=0.47\textwidth, height=0.23\textheight]{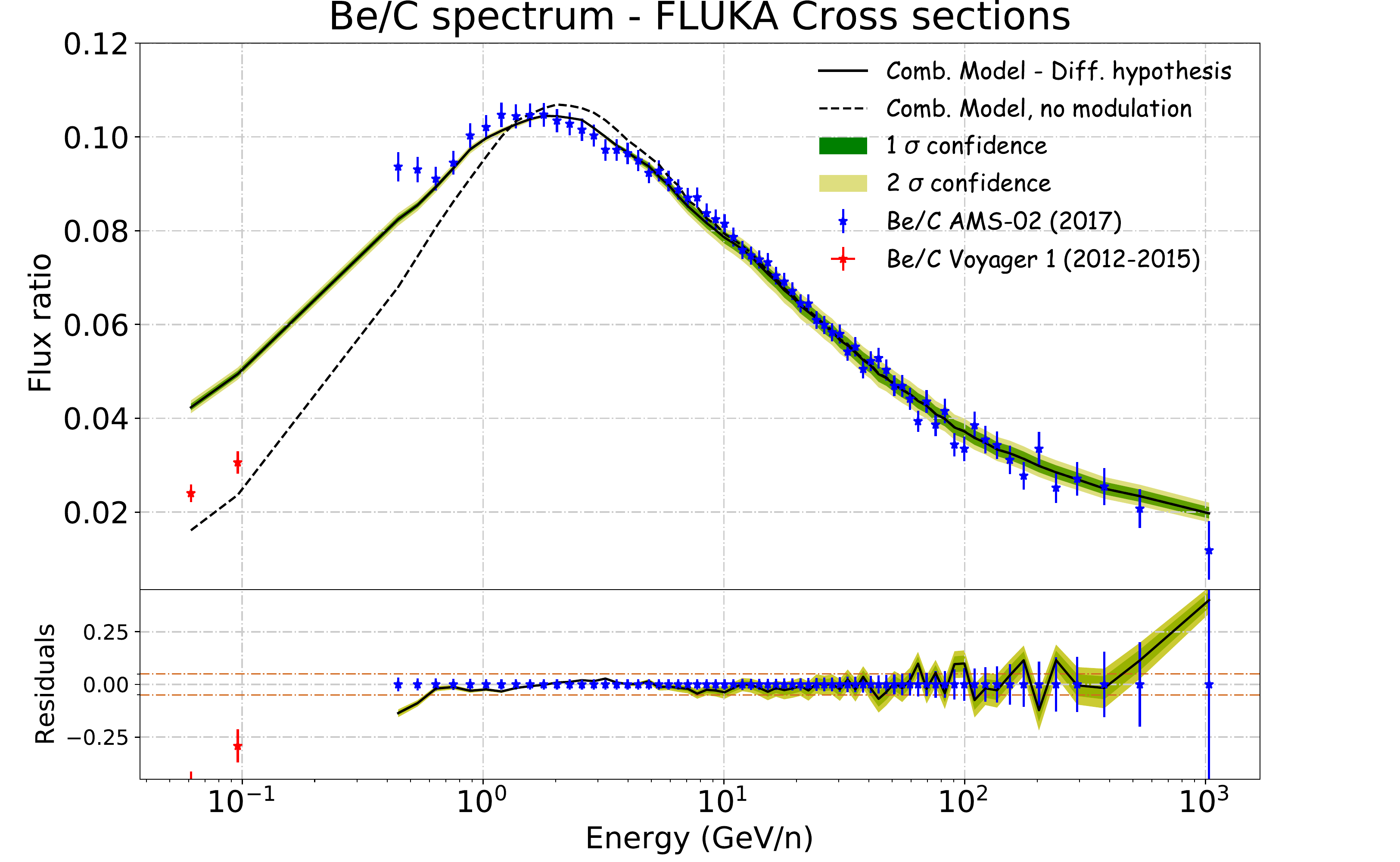}
	\includegraphics[width=0.47\textwidth, height=0.23\textheight]{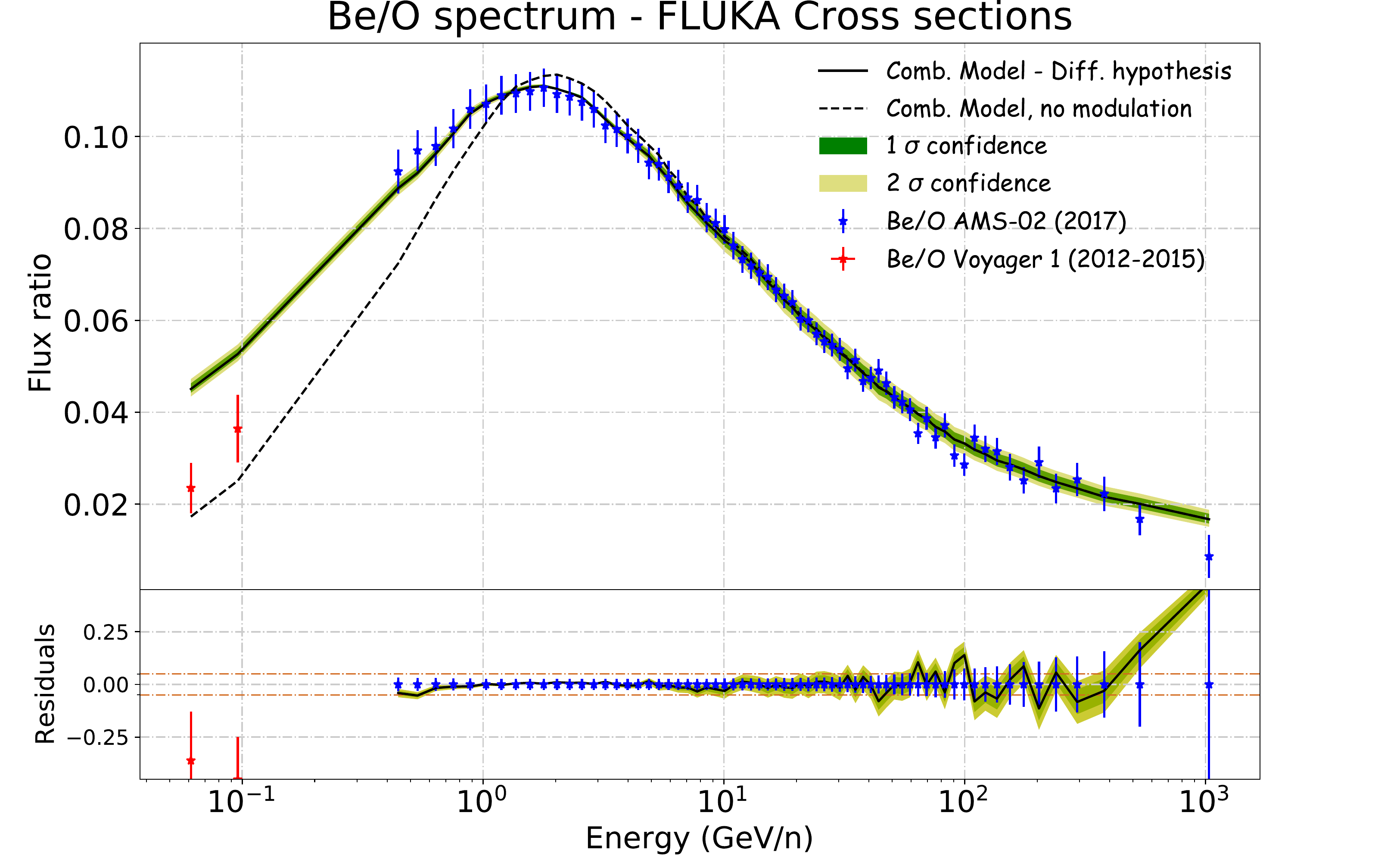}
	
	\vspace{0.4cm}
	
	\includegraphics[width=0.47\textwidth, height=0.23\textheight]{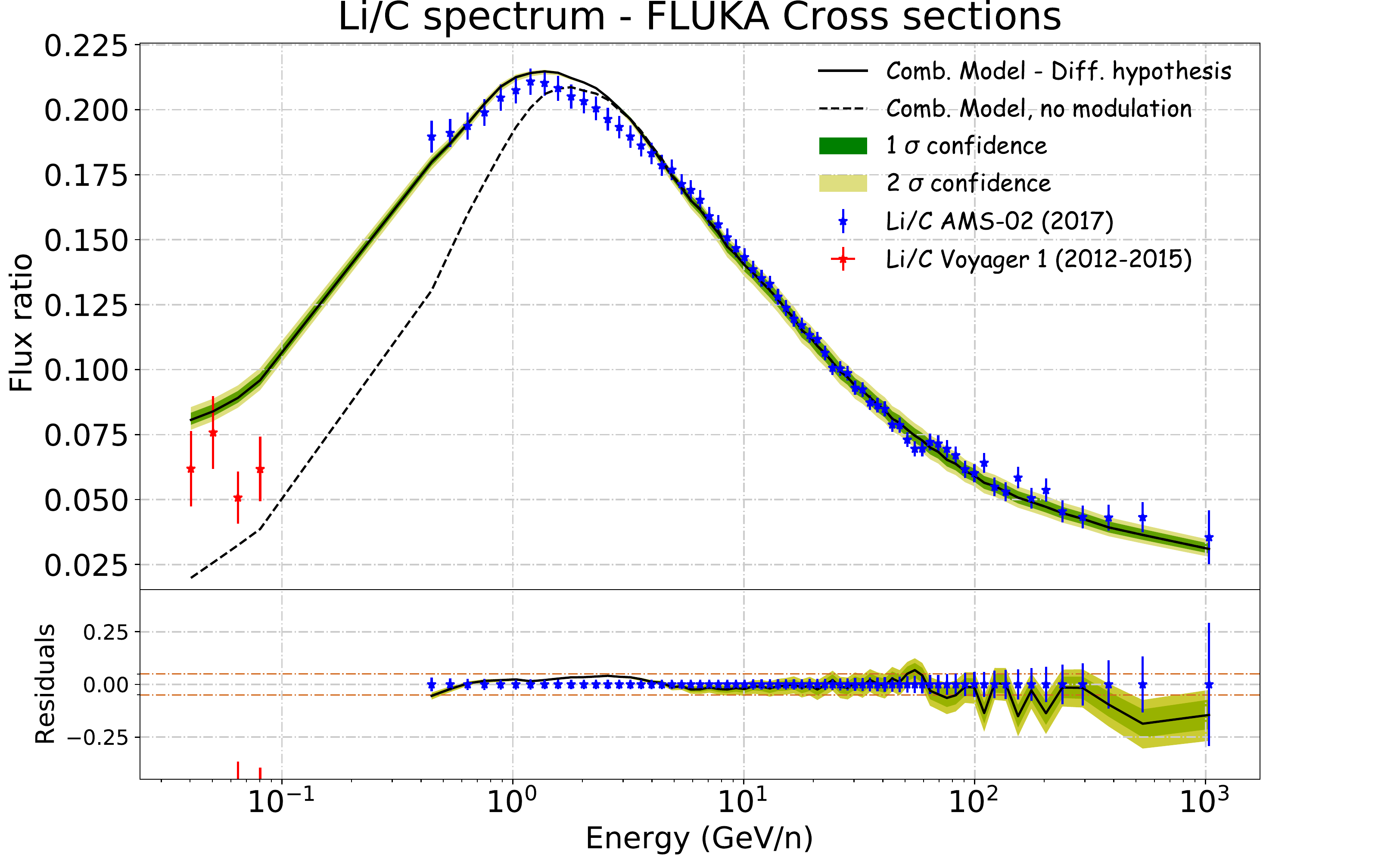}
	\includegraphics[width=0.47\textwidth, height=0.23\textheight]{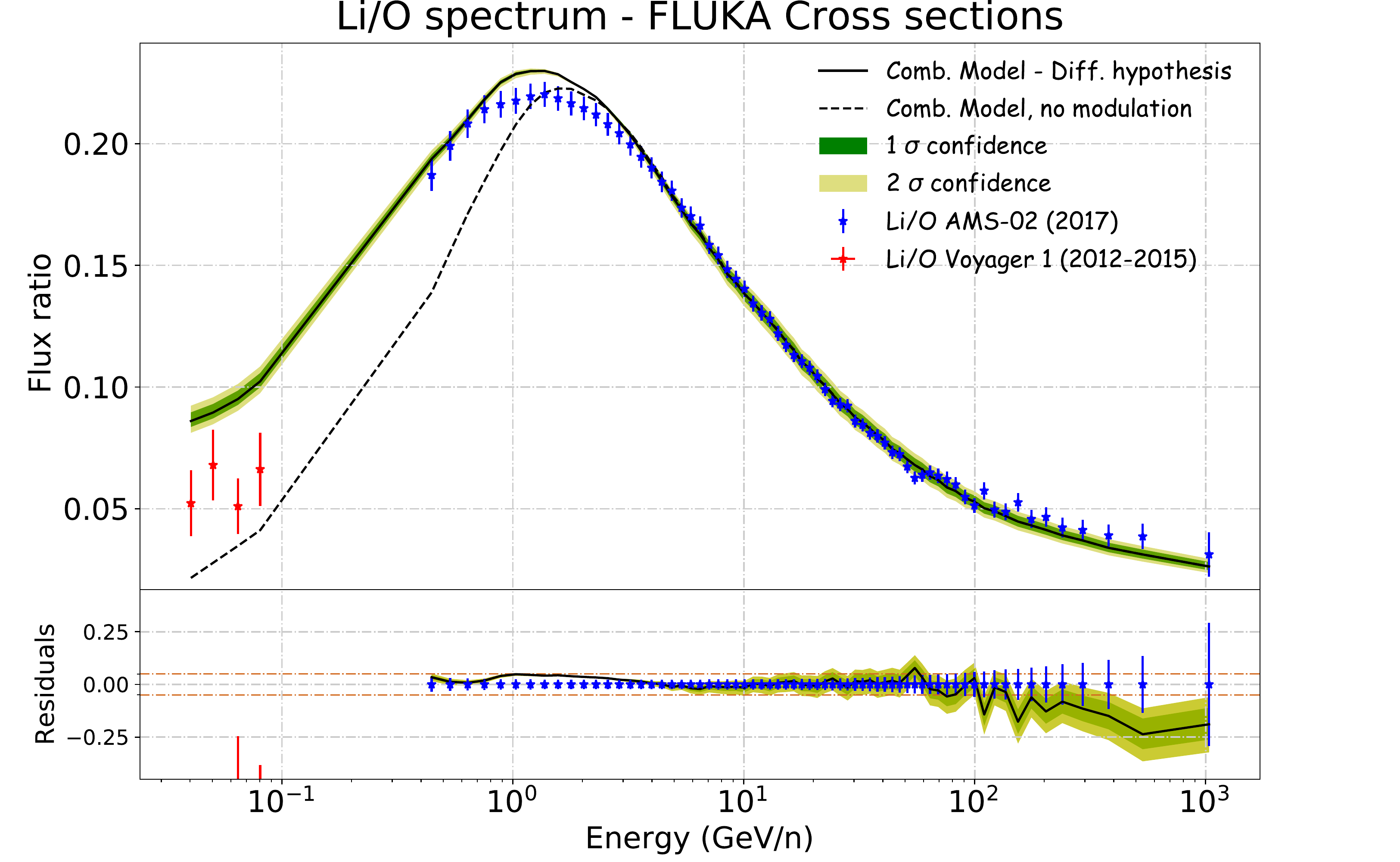}
	\caption{\footnotesize \footnotesize Same as in Figs. \ref{fig:SecPrim_DRAGON2} and \ref{fig:SecPrim_GALPROP} but with the FLUKA cross sections.}
	\label{fig:SecPrim_FLUKA}
\end{figure}

   \item From Figure~\ref{fig:Nuisance_combined} we see for the GALPROP cross sections that, while the scaling factors for Li and Be are the largest ($\sim$26 and 10\% respectively), the scaling for B is the smallest (1-2\% up) and it is even compatible with no scaling, according to the results obtained in chapter~\ref{sec:XSecs}. However, the DRAGON2 cross sections parametrisations determine a scaling for B, Be and Li of 5\%, 1\% and 3\% ($\pm$1\%), respectively. Finally, as discussed in chapter~\ref{sec:3}, the scaling determined with the FLUKA cross sections for the B production is the largest, of about 20\%, while the scaling factors for Li and Be remain well below 10\%.
   
   \item To make a rough estimation of the total uncertainty in the determination of these scaling factors we must consider that a 2\% uncertainty, due to inelastic (annihilation) cross sections, is reasonable for the flux of secondary CRs at high energies (as discussed in section~\ref{sec:Fluka_effect}). This makes the total uncertainties in the determination of the scaling factors to be at least of $\pm (4 - 6)\%$, taking into account the statistical uncertainties associated to the diffusion parameters ($\sim (3 - 4)\%$ as discussed in chapter~\ref{sec:XSecs}). Nevertheless, due to uncertainty associated to the shape of the cross sections, the uncertainty on these parameters from the combined analysis may increase of an additional $(3 - 4)\%$. Therefore, we estimate that the uncertainty in the determination of the scaling factors can be as high as $\pm (7 - 9)\%$, which is around half of the experimental uncertainty associated to the cross sections measurements. 

\end{itemize}

\begin{figure}[!hb]
	\centering
	\includegraphics[width=0.51\textwidth, height=0.233\textheight]{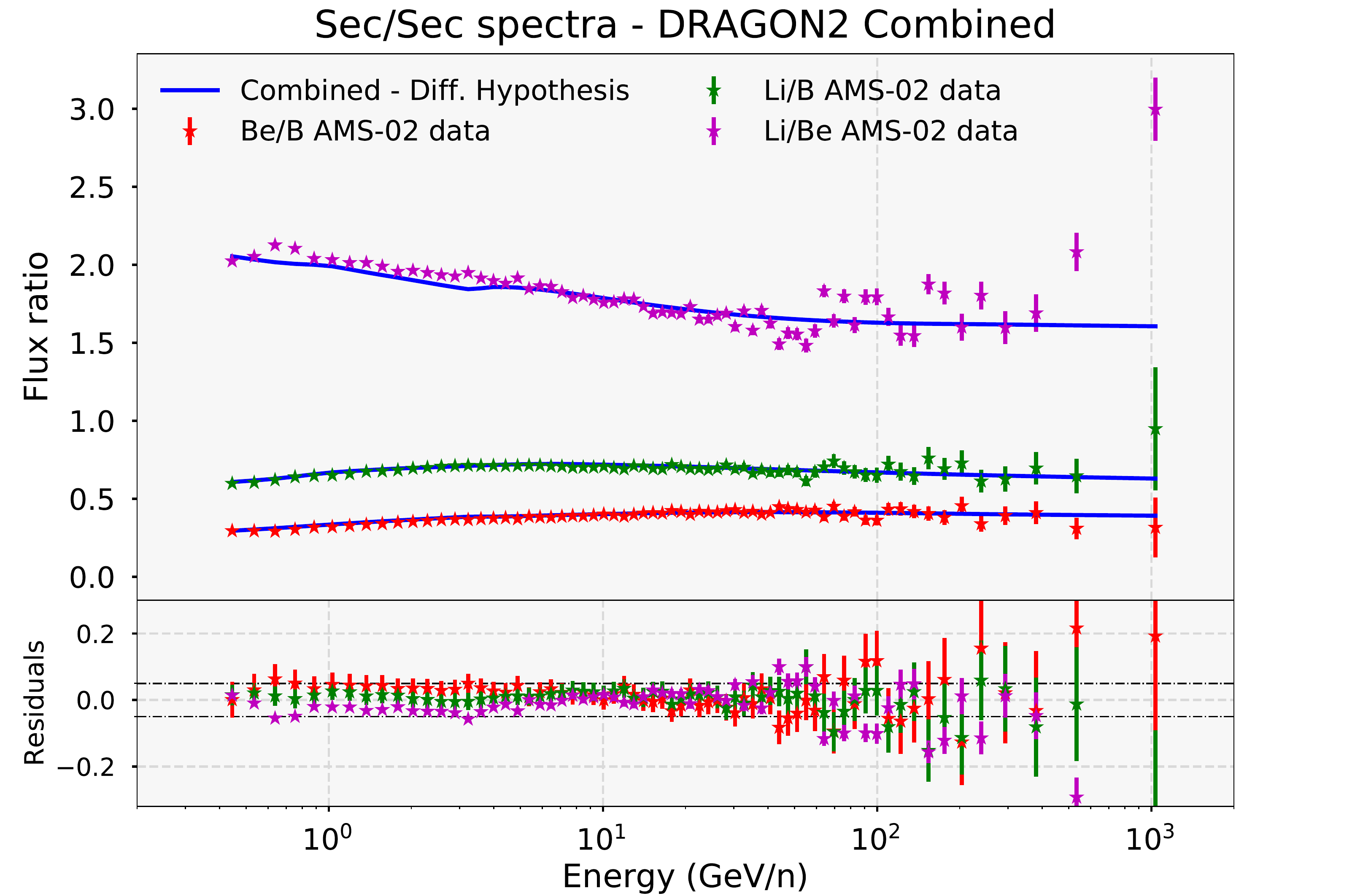} \hspace{-0.6cm}
	\includegraphics[width=0.51\textwidth, height=0.233\textheight]{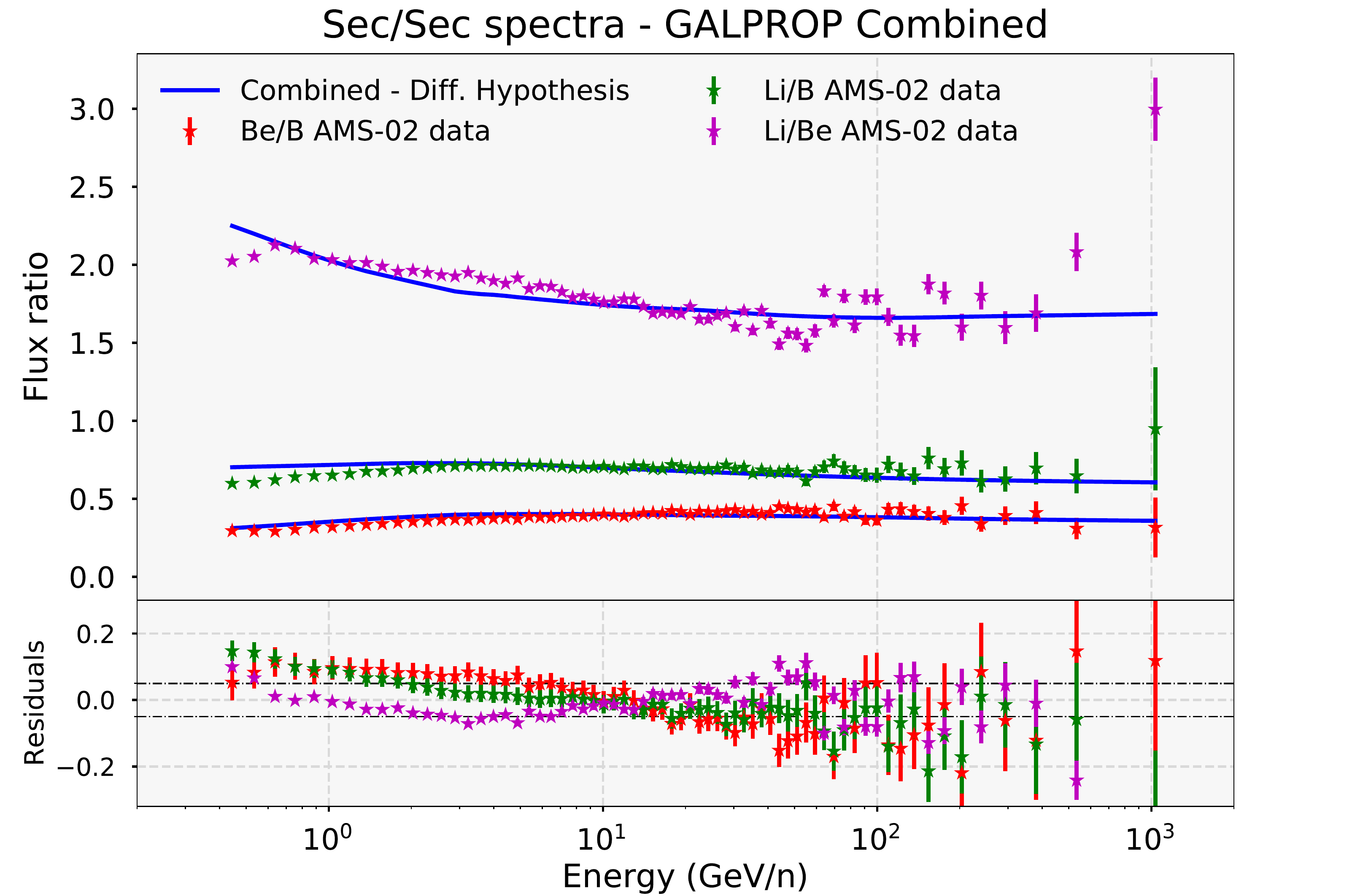}
	\includegraphics[width=0.51\textwidth, height=0.233\textheight]{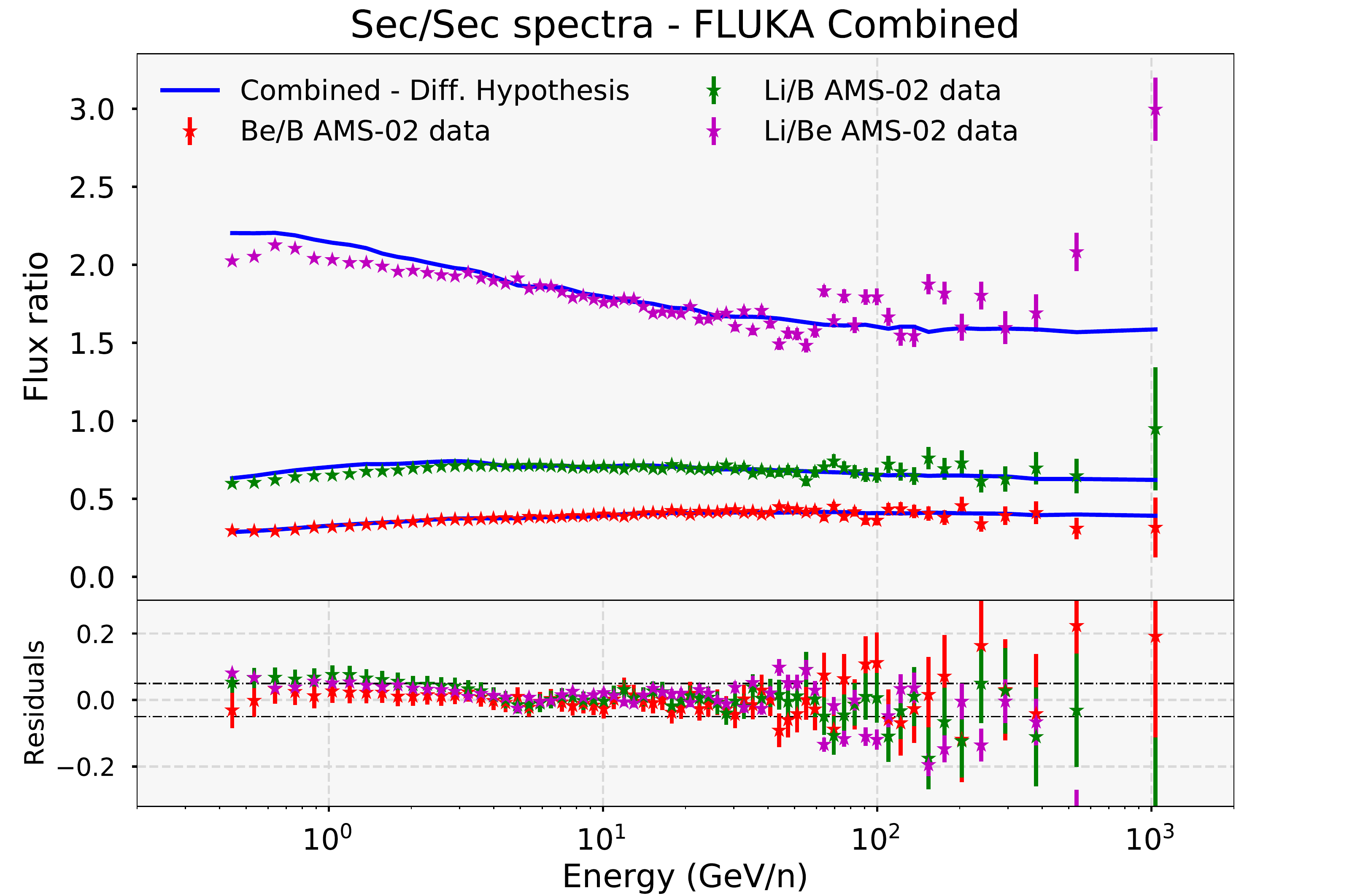}
	\caption{\footnotesize Secondary-over-secondary flux ratios of Li, Be and B computed using the propagation parameters and scaling factors found in the combined analysis for each of the cross sections sets compared to the AMS-02 experimental data. The grid lines at the level of 5\% residuals are highlighted in the lower panel in a different color in order to guide the eye.}
	\label{fig:SecSec_combined}
\end{figure}

In Figure~\ref{fig:SecSec_combined} we show a comparison of the predicted secondary-to-secondary spectra of Li, Be and B with the AMS-02 data using the propagation parameters and scaling factors obtained in the combined analysis for each of the cross sections sets.

As we see, the secondary-over-secondary ratios predicted with the DRAGON2 cross sections are consistent with data within the level of 5\% (highlighted residual grid lines). The Li/Be and Be/B ratios are the ones with slightly larger discrepancies with respect to experimental data, which indicates that the halo size could need a readjustment. In the case of the GALPROP parametrisations, the discrepancies are larger (often above 5\% at low energy) and, more relevantly, the shapes of these ratios seem also to be rather inconsistent, specially the Be/B ratio. Finally, the FLUKA cross sections show a good agreement with data at high energies and in the full energy range approximately inside 5\% residuals, although the low energy predicted shapes are slightly discrepant. 

\subsection{Main conclusions on the standard analyses}

In general, we see that the propagation parameters obtained are very similar for the cross sections studied, favoring a $\delta \sim 0.39$-$0.45$, specially hard for Li. In addition, we have demonstrated that the secondary-over-primary ratios of B, Be and Li to C and O can be reproduced with low discrepancies including a renormalization of the production cross sections. We underline that the DRAGON2 cross sections show an almost perfect fit of all these ratios with respect to the AMS-02 and Voyager-1 data at the same time with very low corrections on the normalization of the B, Be and Li production cross sections.

On the other hand, the uncertainties on the shape of the cross sections seem to be significant in the secondary-over-secondary flux ratios in the GALPROP and, in smaller extent, the FLUKA cross sections. Particularly, we have shown that $\delta$ value found in the combined analysis tend to be similar to those obtained in the independent analyses of the Li/C and Li/O ratios, even for the DRAGON2 cross sections. This can be fixed including more degrees of freedom in the cross sections spectrum of Li production, expected since the spallation channels of Li production are the most uncertain.  However, this means that the estimation of the propagation parameters from this combined analysis may be biased, specially because of the cross sections of Li production. This bias is also present in the scaling factors obtained in the combined analyses. 

In order to investigate this bias in the determination of the propagation parameters we explore a different procedure that could allow us to have a more reliable set of propagation parameters in the next section.

\section{``Corrected'' cross sections analysis}
\label{sec:CXSanal}

To prevent the propagation parameters from being biased, especially for the GALPROP and DRAGON2 parametrisations, we proceed in a different way to perform the analysis. This procedure will allow us to use the propagation parameters extracted from the B ratios, since the amount of data for the cross sections of B production allows a better description of the parametrisations in these channels. First, we fit the AMS-02 secondary-over-secondary ratios only adjusting the scaling factors of the B, Be and Li production cross sections, using the diffusion parameters obtained in the B/O analysis of each cross sections set (see figures~\ref{fig:boxplot_Source} and~\ref{fig:boxplot_Diff}). Then, we apply the same independent MCMC analysis to each of the ratios and compare the propagation parameters to observe the differences obtained for each secondary.

The idea behind the first step is that all possible scalings that fit the AMS-02 secondary-over-secondary ratios and lay inside cross sections data uncertainties are allowed candidates, but the preferred one is that combination of scaling factors which involves the minimum change from the original parametrisation. In order to carry out this analysis we use again a MCMC procedure, in which the free parameters are the scaling factors, $\Delta_B$, $\Delta_{Be}$ and $\Delta_{Li}$, and the data to be compared with are just the secondary-over-secondary ratios above $30 \units{GeV}$, with a penalty factor proportional to the nuisance term of eq.~\ref{eq:myNuisance}.

Since this strategy relies on adjusting the secondary-over-secondary ratios from the diffusion parameters derived from the B ratios, we focus on the cross sections parametrisations (i.e. GALPROP and DRAGON2) in this part of the study, given that we are confident that the predicted B spectrum must be more accurate than those of Li and Be, as explained in previous chapters.

This will avoid biasing the scaling factors obtained by the cross sections parametrisations. In table \ref{tab:Scaling_comp}, a comparison among the scaling factors obtained with both analysis approaches is shown.
\begin{table}[htb!]
\centering
\resizebox*{0.46\columnwidth}{0.09\textheight}{
\begin{tabular}{|c|c|c|c|c|c|c|c|}

\cline{2-7} \multicolumn{1}{c|}{ } & \multicolumn{2}{|c|}{\textbf{\large{$\Delta_{B}$}} }& \multicolumn{2}{|c|}{\textbf{\large{$\Delta_{Be}$}} } & \multicolumn{2}{|c|}{\textbf{\large{$\Delta_{Li}$}} } \\ 
\cline{2-7}
\multicolumn{1}{c|} { }& \textbf{Fit} & \textbf{Comb} & \textbf{Fit} & \textbf{Comb} & \textbf{Fit} & \textbf{Comb} \\ \hline

\textbf{DRAGON2} & 1.04 & 1.05  &  0.975  & 0.985  & 0.945  &  0.97 \\ \hline 

\textbf{GALPROP} & 0.94 & 1.01  &  0.865  & 0.895 & 1.18  &  1.26  \\ \hline 

\textbf{FLUKA} & 1.15 & 1.20  &  0.94  & 0.99  & 0.90  & 0.955 \\ \hline 
\end{tabular}
}
\caption{\footnotesize Comparison between the scaling factors inferred in the combined analyses (Comb) and in the fit of the secondary-over-secondary ratios of AMS-02 (Fit) for the cross sections sets used. These scaling factors are the same for both diffusion coefficients used. The statistical errors on this determination are $\sim 1\%$.}
\label{tab:Scaling_comp}
\end{table}

As we see, while the scaling factors from both procedures are quite similar for the DRAGON2 cross sections, they are considerably different for the FLUKA and GALPROP cross sections. This indicates that the bias introduced because of the shape of the cross sections can be relevant. In fact, these differences are higher than the $\pm$4\% uncertainties related to variations in the diffusion coefficient that we assume, although they are lower than $\pm$8\%. 

We stress the fact that scaling factors for the DRAGON2 cross sections are <$5.5\%$ and that, as expected, the B scaling is the closest to 1 in the DRAGON2 and GALPROP cross sections parametrisations, while FLUKA cross sections of B production require a much larger scaling factor.

\begin{figure}[!htpb]
	\centering
	\includegraphics[width=0.495\textwidth, height=0.23\textheight]{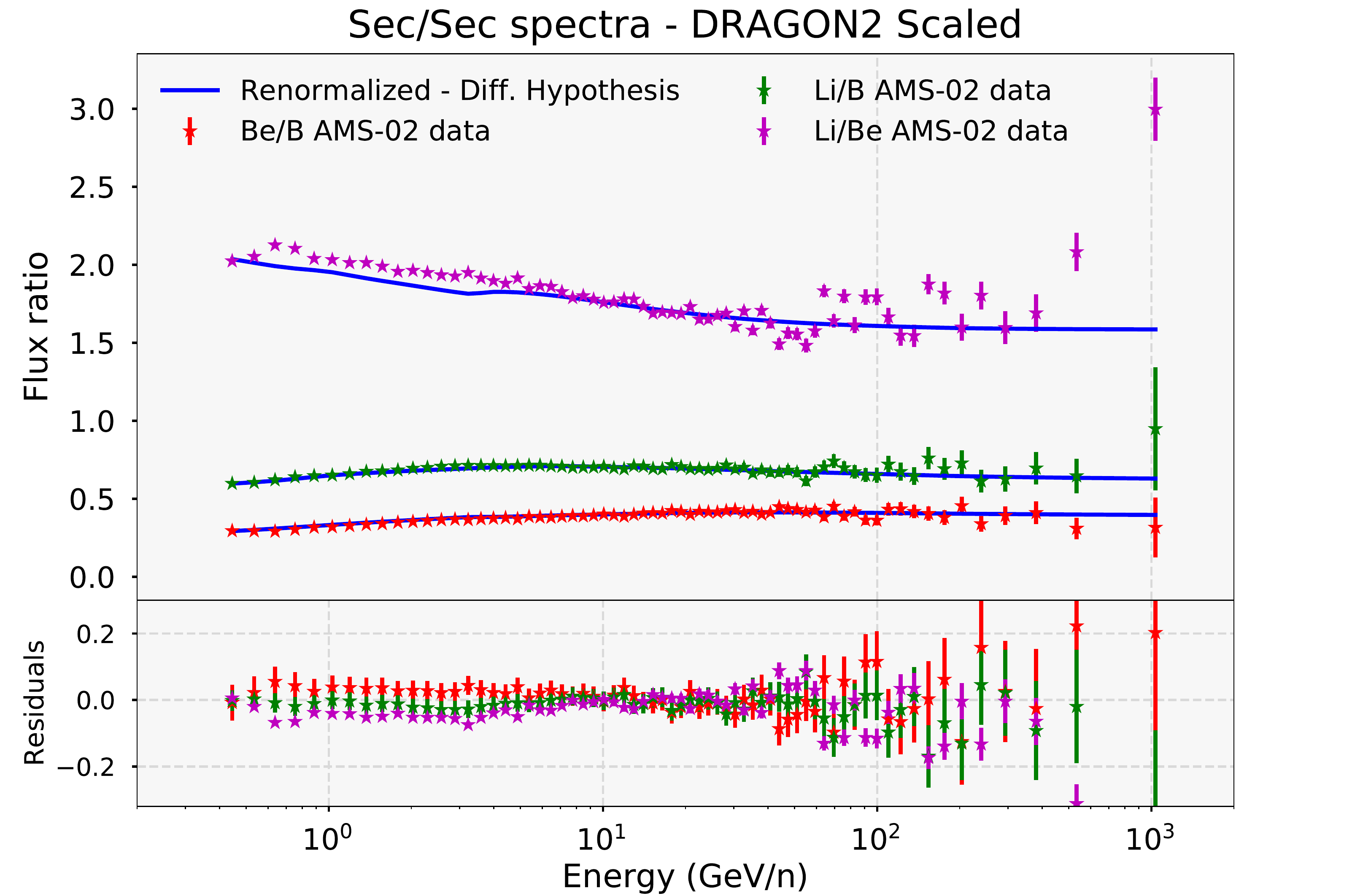}
	\includegraphics[width=0.495\textwidth, height=0.23\textheight]{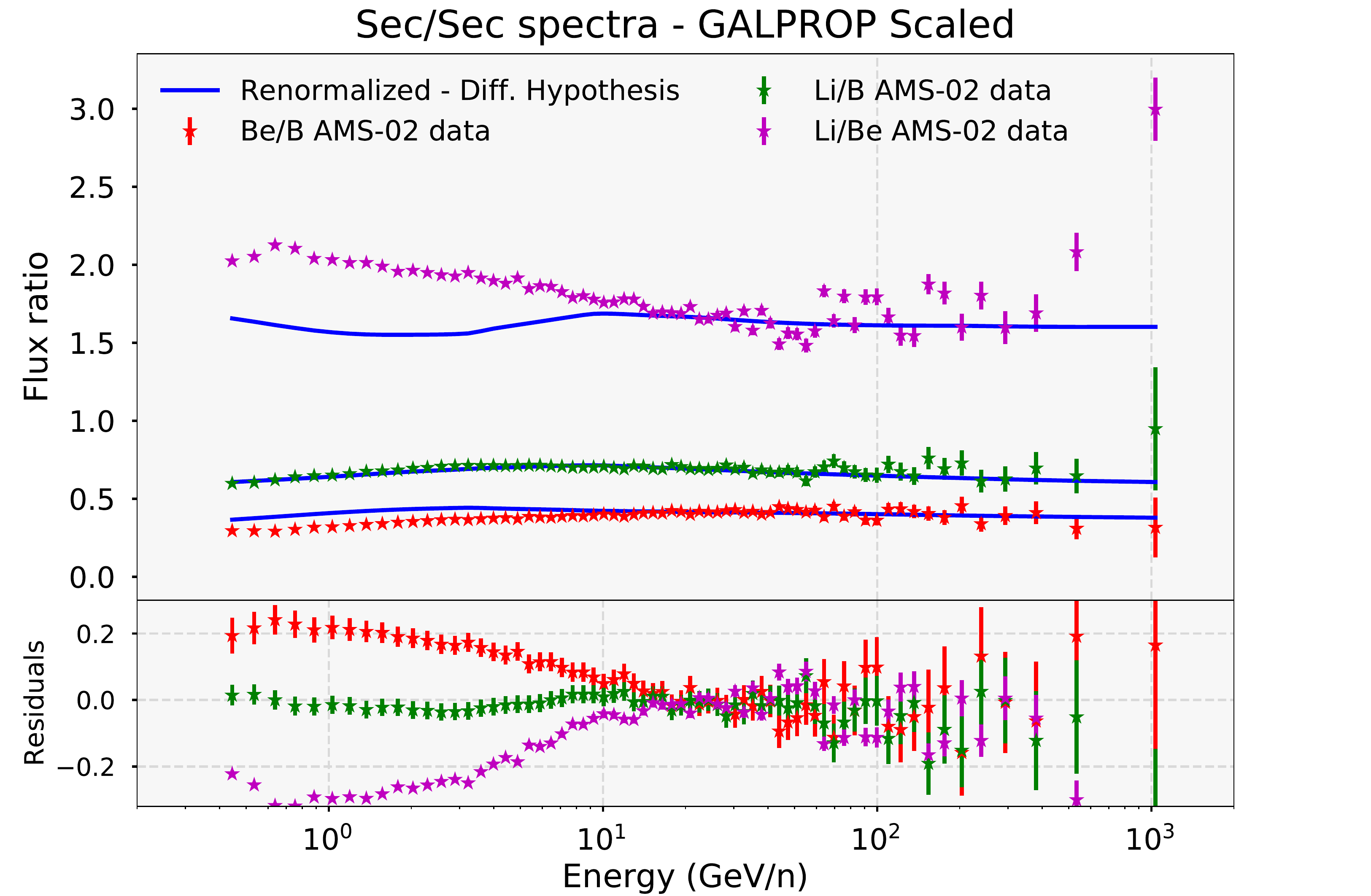}
	\caption{\footnotesize Secondary-over-secondary ratios predicted by the scaled DRAGON2 and GALPROP cross sections parametrisations compared to AMS-02 data. They are fitted to reproduce the high energy part of the spectra, while the low energy part of the Li/Be and Be/B ratios mainly depends on the halo size value used.}
	\label{fig:SecSec_DR}
\end{figure}

These scaled secondary-over-secondary flux ratios are expected to show larger discrepancies at low energies in the Be/B and Li/Be spectra, mainly due to the effect of the halo size, that may be further adjusted, while the Li/B spectrum must show good agreement with data. They are shown in Figure~\ref{fig:SecSec_DR}.

With these scaling factors, the independent analysis of each of the ratios is performed for the DRAGON2 and GALPROP parametrisations and the results are shown in Fig.~\ref{fig:boxplot_Renorm}. Again, the results are summarised in tables in the appendix~\ref{sec:appendixC}, also for the FLUKA cross sections.

It is worth underlying here that analysing the ratios after scaling the cross sections does not lead to just a change in the predicted normalization of the diffusion coefficient ($D_0$), but also to slight changes in the slope ($\delta$) and in the low-energy parameters (although never by more than 2$\sigma$).

\begin{figure}[!htb]
	\centering
	\includegraphics[width=\textwidth, height=0.59\textheight]{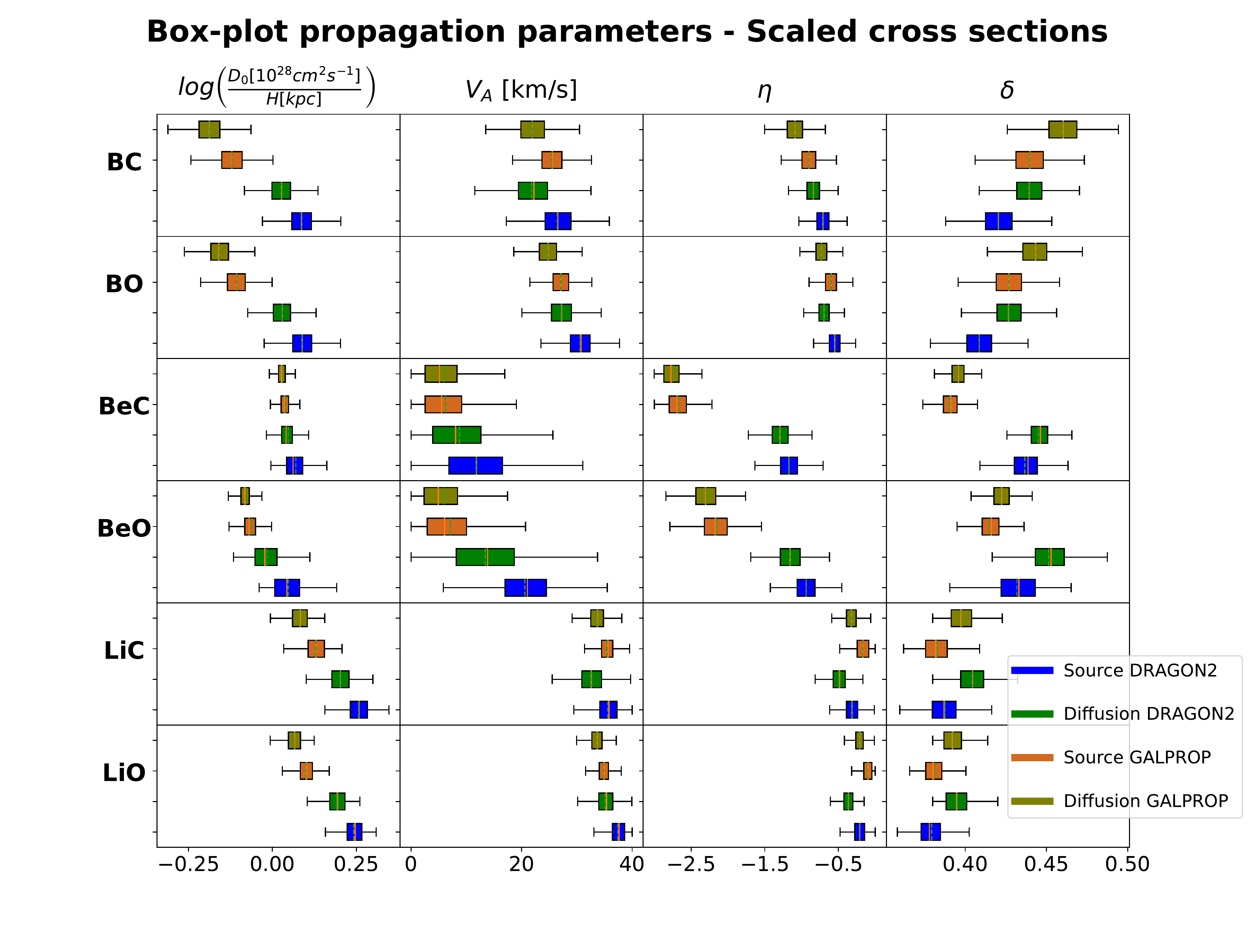}
	\caption{\footnotesize Box-plots representation of the diffusion parameters obtained in the independent analyses of the Li,Be,B/C,O AMS-02 ratios for the scaled GALPROP and DRAGON2 cross sections for both hypothesis on the diffusion coefficient.}
	\label{fig:boxplot_Renorm}
\end{figure}
As we can see, the GALPROP cross sections predict different propagation parameters for each secondary, meaning that the analysis must be performed including more degrees of freedom for the shape of their cross sections energy spectrum. This is probably related to the fact that the Be and Li scaling factors deviate more than $10\%$, given the cross sections at low energy are always better constrained than at high energy. As expected, the diffusion parameters inferred from the ratios of B are the most reliable. On the other hand, we see that for the DRAGON2 parametrisations, both the $D_0$ and $\delta$ values are near to be identical for the B and Be ratios, while they are significantly different for the Li ratios. The differences in the predicted $\eta$ and $V_A$ values are mainly due to the halo size used and could be prevented by adjusting it properly. From these results, we can see that a $\delta$ value of 0.43-0.45 is favored when excluding the Li predictions, and this value goes down to an average of $\sim$0.41 when including it.

\section{Final conclusions}
\label{MCMC_conc}

In this chapter, we have studied the two parametrisations of the diffusion coefficient that best fit with the theoretical expectations and are suitable to reproduce the experimental data for the two most accurate cross sections parametrisations (namely DRAGON2 and GALPROP cross sections) and for the FLUKA cross sections developed in chapter~\ref{sec:3}.

We have presented the MCMC analyses of the secondary-over-primary and secondary-over-secondary flux ratios, for the Li, Be and B, in order to determine the propagation parameters that best reproduce AMS-02 data for the three cross sections sets and both diffusion coefficients (with and without high energy break). This combination of different secondary CRs provides a way to get rid of cross sections uncertainties, which are mostly dominated by their normalization in the cross sections parametrisations.

It has been observed that, when including the normalization of the production cross sections in the analysis, the secondary-over-primary ratios can be all well described inside $\sim 5\%$ for the DRAGON2 parametrisations. On the other hand, the GALPROP parametrisations show important discrepancies that are due to the shape of the cross sections energy spectra. This analysis has also shown that the FLUKA cross sections can successfully be used for reproducing the CR observables within very small discrepancies, which is something totally new for a Monte Carlo based code in the CR context. Nevertheless, the propagation parameters determined in this combined analysis seem to be biased by the shape of the cross sections, specially favoring the Li ratios. In order to investigate and prevent this bias, which might be particularly important for the cross sections parametrisations because of the uncertainty on the cross sections of Li production, an alternative procedure has been used to obtain a more reliable determination of the diffusion parameters.

After doing the second analysis on the GALPROP and DRAGON2 cross sections, we have determined that the shape of the energy spectrum for the cross sections of Li production needs to be slightly adjusted in order to get the same prediction for the diffusion parameters in all the ratios. For the DRAGON2 spallation cross sections, the inferred diffusion parameters are almost coincident for the Be and B elements with $\delta \sim 0.43-0.45$. In turn, the GALPROP cross sections infer more disperse values of $\delta \sim 0.37-0.46$. It seems that a change on the shape of the GALPROP cross sections of Be and Li production is needed, which must be related to the fact that both, Li and Be cross sections energy spectra need scaling factors deviating of more than $10\%$ to reproduce the secondary-over-secondary ratios. When excluding the predictions from the Li ratios, both the DRAGON2 and GALPROP cross sections favor a $\delta$ value around $0.44$. Isotopic cosmic-ray data will be valuable in order to adjust the different spallation cross section parametrisations in the various isotopic channels.

A more detailed analysis is under preparation to also include the halo size as a free parameter, in order to adjust the full propagation parameters from the Be and B species.
\chapter{Secondary emissions from cosmic ray interactions: gamma rays and antiprotons}
\label{sec:5}

As we have described, secondary particles produced in CR collisions are used to keep track of the interactions they undergo along their way throughout the Galaxy. In addition to the direct information about their transport, which can be extracted from secondary nuclei like Li, Be or B, other secondaries - namely leptons, antinuclei, gamma rays and neutrinos - are formed and may provide essential information about acceleration mechanisms, magnetic field configurations, environment of sources, etc. Studying these messengers at the same time can provide  correlated information to constrain our models with high precision and to possibly find out unexpected phenomena.

An important fraction of the particles produced from spallation reactions of CR nuclei with the ISM gas are short-living mesons (mainly pions), which are the responsible of the further formation of leptons and gamma rays from their decays. Secondary electrons and positrons are mainly produced via the decay of charged pions (accompanied by neutrinos) and, since roughly the same amount of $e^{-}$ and $e^+$ is expected to be produced from such reactions, the observed excess of $e^{-}$ over $e^+$ in CRs indicates that most of electrons have a primary origin (injected at the sources). As a matter of fact, the lepton component ($e^+$ + $e^{-}$) accounts for around 1\% of the CRs and only a very small fraction of it has a secondary origin. 

While the vast majority of CR electrons are accelerated in SNRs, CR positrons seem to be emitted from pulsar wind nebulae (PWN), expected to be symmetric $e^+$ and $e^{-}$ pairs emitters \cite{Amato:2013fua, Serpico_efraction}. PWN are structures found inside the shells of SNRs, which emit a broad-band spectrum of non-thermal radiation powered by magnetized neutron stars, typically detected in the radio and gamma-ray band as pulsars. This non-thermal emission (mainly detected in radio and X-ray frequencies) is well characterized \cite{Jankowski:2017yje} and seems to require a broken power law in the lepton spectrum with a break at $200-400 \units{GeV}$ (see \cite{Bykov:2017xpo}, \cite{PWN_Bucc}, \cite{Blasi:2010de}). 

On the other hand, gamma rays may be produced by the decays of neutral mesons, mainly pions, and by the interactions of CR leptons with magnetic and radiation fields in the ISM, carrying important information about the environmental conditions, such as the gas density and radiation field. Gamma rays are frequently produced near CR sources, which appear as gamma-ray point-like sources. Since photons can travel several$\units{kpc}$ through the universe almost undisturbed, they can be directly associated with their sources. However, gamma rays are also produced during the propagation of CRs, and appear as a diffuse radiation. The study of the galactic gamma-ray diffuse emission is important to constrain models and opens a new window to unveil the mechanisms behind CR interactions and acceleration \cite{Ackermann_2012, Acero_2016}.

Gamma-ray measurements have contributed to several studies in association with gravitational waves \cite{GWGR} or with cosmic neutrinos from flaring blazars \cite{IceCube:2018dnn}. In addition, new physics can be tested with gamma rays and can manifest itself as an excess of gamma rays, for example in dark matter (DM) searches \cite{GRDM, Dario_Serpico_2009, BRINGMANN2012194}.

Furthermore, a useful tool for identifying the nature of the evasive dark matter is the study of antinuclei produced in inelastic hadronic interactions of CRs - mainly protons and helium - with the interstellar gas. CR antinuclei, i.e. antiprotons, antideuterons, and antihelium, can be a valuable probe to search for annihilating and decaying DM signatures due to their low expected background. For instance, low-energy anti-deuterons would provide a “smoking gun” signature of DM annihilation or decay, almost completely free of astrophysical background \cite{vonDoetinchem:2020vbj, Baer_2005}. This enables to probe a variety of DM models that currently evade or complement collider experiments and other direct or indirect searches. Important efforts are currently made in order to get precise detection of antinuclei, mainly based on balloon-borne experiments (see, e.g. \cite{Dimiccoli:2020coh} and \cite{Bird:2019zzc}), although only antiproton measurements have satisfactorily been achieved by experiments like BESS \cite{Bess_AP, YAMAMOTO2013227}, CAPRICE \cite{Bergstr_m_2000}, Pamela \cite{Adriani:2012paa} and, recently, AMS-02 \cite{AMS_Ap}.

In this chapter we will test the diffusion models obtained in the previous section in order to predict the fluxes of antiprotons (in section~\ref{sec:Antip}) and the gamma-ray local emissivity (in section~\ref{sec:Emiss}). We will also analyse the uncertainties around these predictions to make them more robust.

\section{Study of cosmic-ray antiprotons}
\label{sec:Antip}

Antiproton experiments have been running from the 1970s \cite{Bogomolov:1979hu, Ap_old} with increasing precision and have become an important tool to deliver leading constraints on DM models, as well as astrophysical production and propagation scenarios \cite{Bergstrom_1999, Simon_1998, CHEN2015495, Kappl_2015, Boudaud_2015, Cerde_o_2014, Fornengo_2013}. Indeed, antiproton data can also be used, together with CR nuclei, to model and constrain the propagation parameters (see, e.g., \cite{Maurin:2002ua}) governing CR transport.

Although the bulk of the antiproton flux was seen to be generally consistent with the predictions assuming their secondary origin (equation~\ref{eq:sec_AP_Sourceterm}), the predictions' huge systematic uncertainties (mainly related to the antiproton production cross sections) still allow the addition of primary components (as in eq.~\ref{eq:DM_AP_Sourceterm}), either from astrophysical origin \cite{Ap_from_SNR} or from other exotic scenarios, like DM decay or annihilation \cite{Salati_Donato_Fornengo_2010}. 

The total source term for secondary production of antiprotons from spallation reactions of CRs with the interstellar gas has the following form:
\begin{equation}
Q_{CR+ISM\longrightarrow\bar{p}}(E_{\bar{p}}) = \sum_{CR}4 \pi~ n_{ISM} \int^{\infty}_0 dE \phi_{CR}(E)\frac{d\sigma_{CR+ISM\longrightarrow\bar{p}}}{dE_{\bar{p}}}(E, E_{\bar{p}})
\label{eq:sec_AP_Sourceterm}
\end{equation}
where the key ingredients are the CR fluxes and the $\bar{p}$ production cross sections. Antiprotons can interact with the interstellar gas as they propagate through the Galaxy, resulting either in annihilation or loss of a fraction of their energy. Non-annihilation inelastic interactions of antiprotons with interstellar protons yield an extra source of lower energy antiprotons, which we refer to as a "tertiary" source of antiprotons \cite{Evoli:2017vim}. 

On the other hand, the source term of antiprotons from DM annihilation (or decay) mainly depends on the DM density profile in the Galaxy $\rho (\vec{x})$, on the velocity-averaged DM annihilation cross section (or on the decay rate) into pairs of Standard Model particles ($q\bar{q}$ pairs, $W^+W^-$, etc.), and on the DM particle mass, as shown in the following equation:
\begin{equation}
Q^{Ann}_{DM\longrightarrow\bar{p}}(\vec{x}, E) = \frac{1}{2}\left(\frac{\rho (\vec{x})}{m_{DM}}\right)^2 \sum_f \left< \sigma v \right>_f \frac{dN_{\bar{p}}^f}{dE}
\label{eq:DM_AP_Sourceterm}
\end{equation}

%%%%%%%%%%%  EQ. DESCRIBING DM DECAY RATE  %%%%%%%%%%%%
%\begin{equation}
%Q^{Dec}_{DM\longrightarrow\bar{p}}(\vec{x}, T) = left(\frac{\rho (\vec{x})}{m_{DM}}\right) \sum_f \Gamma_f \frac{dN_{\bar{p}}^f}{dT}
%\label{eq:DM_AP_Sourceterm}
%\end{equation}

%Primordial black holes also can annihilate into particles as their temperature goes down (less mass). See MacGibbon, 1990 Emission of particles from primordial black holes
%Idea introducted in https://www.nature.com/articles/353807a0.pdf

Among the most used DM profile densities, derived from N-body simulations and cosmological studies \cite{Ascasibar:2003mm, Lu:2005tu}, the most common one is the Navarro-Frenk-White (NFW) density profile \cite{Navarro:1995iw}, $\rho _{NFW}(\vec{r}) = \rho_h r_h/r (1 + r/r_h)^{-2}$ with a characteristic halo radius $r_h\sim20 \units{kpc}$ , and a characteristic halo density $\rho_h$, used to normalize it from observations \cite{Salucci:2010qr}.

Currently the debate is focused on whether a primary antiproton component like that of eq.~\ref{eq:DM_AP_Sourceterm} is really needed, or a purely secondary production of antiprotons is adequate to describe the experimental data. The AMS-02 antiproton flux is measured in an energy range from around $500 \units{MeV}$ to $\sim350 \units{GeV}$, reaching a precision of around 5\% at $50 \units{GeV}$. However, the main uncertainties are related to the antiproton predictions from models, essentially in the cross sections of antiproton production and in the selection of the propagation parameters.

The total antiproton cross sections do not include only the prompt production from reactions like $p + p \longrightarrow \bar{p} + p + p + p$, which is one of the main channels, but also antineutron decay ($p + p \longrightarrow [\bar{n} \longrightarrow \bar{p}] + X$) and hyperon decay (mainly from the $\bar{\Lambda}$ and $\bar{\Sigma}$ baryons, $p + p \longrightarrow [\bar{\Lambda}, \bar{\Sigma} \longrightarrow \bar{p}] + X$) as well. Therefore, the total antiproton production cross sections can be calculated as (eq. 17 of \cite{reinert2018precision}):
\begin{equation}
\left(E \frac{d^3\sigma}{dp^3} \right)_{CR\longrightarrow\bar{p}} = \left(E \frac{d^3\sigma}{dp^3} \right)_{CR\longrightarrow\bar{p}}^{prompt} \times (2 + \Delta_{IS} + 2\Delta_{Hyp})
\label{eq:TotalAP_XS}
\end{equation}
Where the factor (2 + $\Delta_{IS}$) takes into account the antineutron decay contribution (antineutron production is expected to be roughly the same as the prompt antiproton production, but isospin asymmetry causes different production of the two species, which is accounted for by means of the $\Delta_{IS}$ term). Finally, the hyperon term, $2\Delta_{Hyp}$ is included to account for the formation of $\bar{\Lambda}$ and $\bar{\Sigma}$ antibaryons, whose production is expected to be equal.

\begin{figure}[!tpb]
	\centering
	\includegraphics[width=0.99\textwidth, height=0.3\textheight]{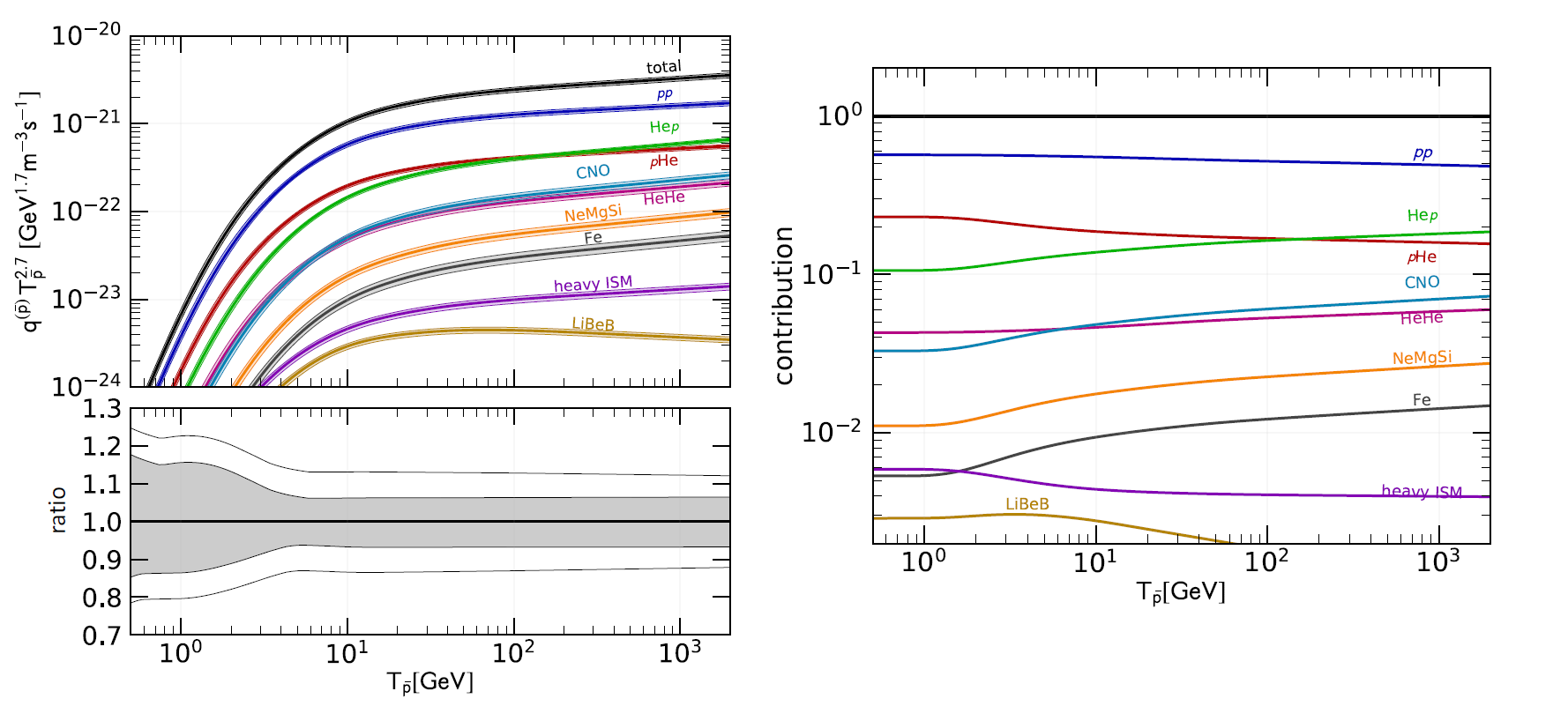}
	\caption{\footnotesize Left: Source terms of the main reaction channels with the $1\sigma$ uncertainties associated to the production cross sections. In the lower panel the uncertainties involved in the prompt antiproton production (shaded area) and the total cross section uncertainties (delimited by the outer lines) are displayed. Right: Relative contribution of the main reaction channels to the total antiproton source spectrum. Both pictures are taken from ref. \cite{Korsmeier:2018gcy}-}
	\label{fig:Ap_contrib}
\end{figure}

In Figure~\ref{fig:Ap_contrib}, the relative contribution of the main reaction channels for antiproton secondary production is shown, together with the source terms of each reaction channel and the typical uncertainties associated to the cross sections parametrisation, taken from \cite{Korsmeier:2018gcy}. In the lower panel of the left plot, the shaded region indicates the uncertainties on the prompt cross sections evaluated from their fit to data, while the full region enclosed by the outer solid lines marks the total cross sections uncertainties. 

As we see, the uncertainty associated to the formation of CR antiprotons is around 14\%, at energies above $7 \units{GeV}$, while it is larger than 20\% for lower energies. Nuclei heavier than protons and helium only contribute at a few percent level, as already stated by older estimations \cite{Moskalenko:1998id, Donato:2001ms}. The dominant channel is the p-p reaction, involving the 50-60\% of the total spectrum, followed by the p-He and He-p reaction channels, involving the $\sim$10-20\% of the total spectrum each. Finally, the He-He and CNO+ISM reactions contribute with < 8\% each. Full information on the derivation of these results and the data measurements can be found in refs. \cite{Korsmeier:2018gcy} and \cite{reinert2018precision}.

\begin{figure}[!tpb]
	\centering
	\includegraphics[width=0.47\textwidth, height=0.245\textheight]{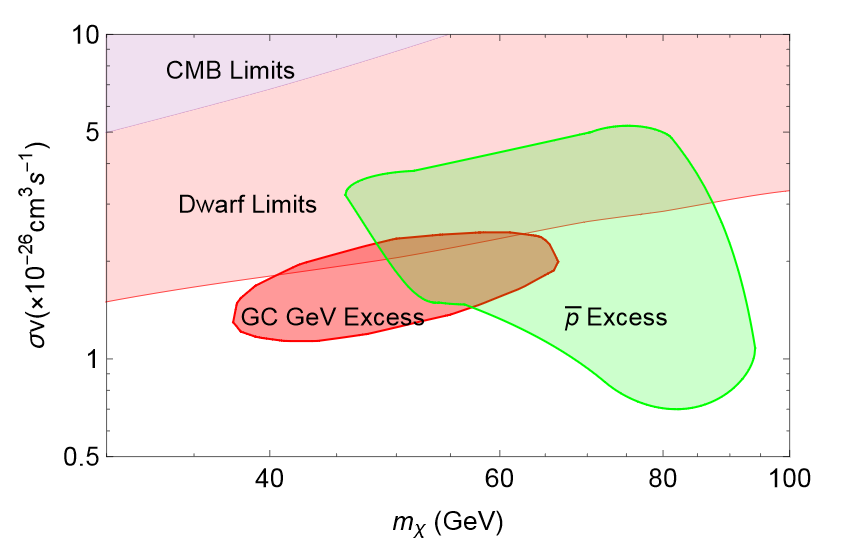}
	\caption{\footnotesize DM favored parameters from different constraints (CMB, dwarf galaxies, Galactic Center excess and antiprotons) in the decay channel to a pair of $b\bar{b}$ quarks, taken from \cite{Cholis:2019ejx}.}
	\label{fig:DM_constraints}
\end{figure}

Recently, various authors used the latest cross sections measurements from collider experiments to build more refined predictions. Most of these studies have revealed a possible excess of the predictions with respect to the AMS-02 antiproton spectrum between $10-20 \units{GeV}$ (for instance, see refs. \cite{AMS_Ap_signal}, \cite{reinert2018precision}, \cite{Heisig:2020nse}, \cite{Cuoco:2016eej} and \cite{Cuoco:2019kuu}), consistent with a $30$–$90 \units{GeV}$ DM particle with a velocity-averaged cross section $\left< \sigma v\right> \sim 2\times 10^{-26}$ $cm^3s^{-1}$, which seems to be not in contradiction with the constraints from the Galactic Center excess of$\units{GeV}$ gamma rays and from dwarf galaxies, as shown in Fig.~\ref{fig:DM_constraints}, taken from \cite{Cholis:2019ejx}.

However, the significance of this excess depends on the interpretation of the theoretical and experimental systematic uncertainties (interestingly, new analyses taking into account the correlation of AMS-02 systematic errors have shown that the significance of the signal is considerably reduced, as in ref. \cite{Heisig:2020nse}), and it seems that the full systematic uncertainties are compatible with the antiproton AMS-02 spectrum \cite{boudaud2020ams}, with no need to add any extra component.

\subsection{Testing antiproton predictions}
\label{sec:ApPred}

In this sections we will examine the consistency of the antiproton predictions derived from the diffusion parameters found in the MCMC analyses performed for the different nucleon spallation cross sections with the AMS-02 data. In this way we will be able to inspect the antiproton spectra for different diffusion parameters and the systematic uncertainties related to the use of different nucleon spallation cross sections, affecting both the shape of the spectrum and its normalization.

In order to perform this test, the predicted antiproton spectrum will be calculated for the propagation parameters obtained for the diffusion hypothesis (eq.~\ref{eq:breakhyp}) in the combined analysis and the from the B/O analyses without scaling and after scaling the nucleon spallation cross sections with the procedure followed in section~\ref{sec:CXSanal}. The choice of testing the prediction from the B/O ratio is due to the fact that the boron production cross sections are subject to less uncertainties, and we can avoid uncertainties related to the secondary C production. The treatment of tertiary antiprotons follows the one explained in ref. \cite{Evoli:2017vim}. %Furthermore, it must be noticed that when using the FLUKA predicted propagation parameters, the inelastic cross sections of H and He elements are the ones derived with the FLUKA nuclear code, unlike the inelastic cross sections used for the GALPROP and GRAGON2 options.

\begin{figure}[!bph]
	\centering
	\includegraphics[width=0.99\textwidth, height=0.32\textheight]{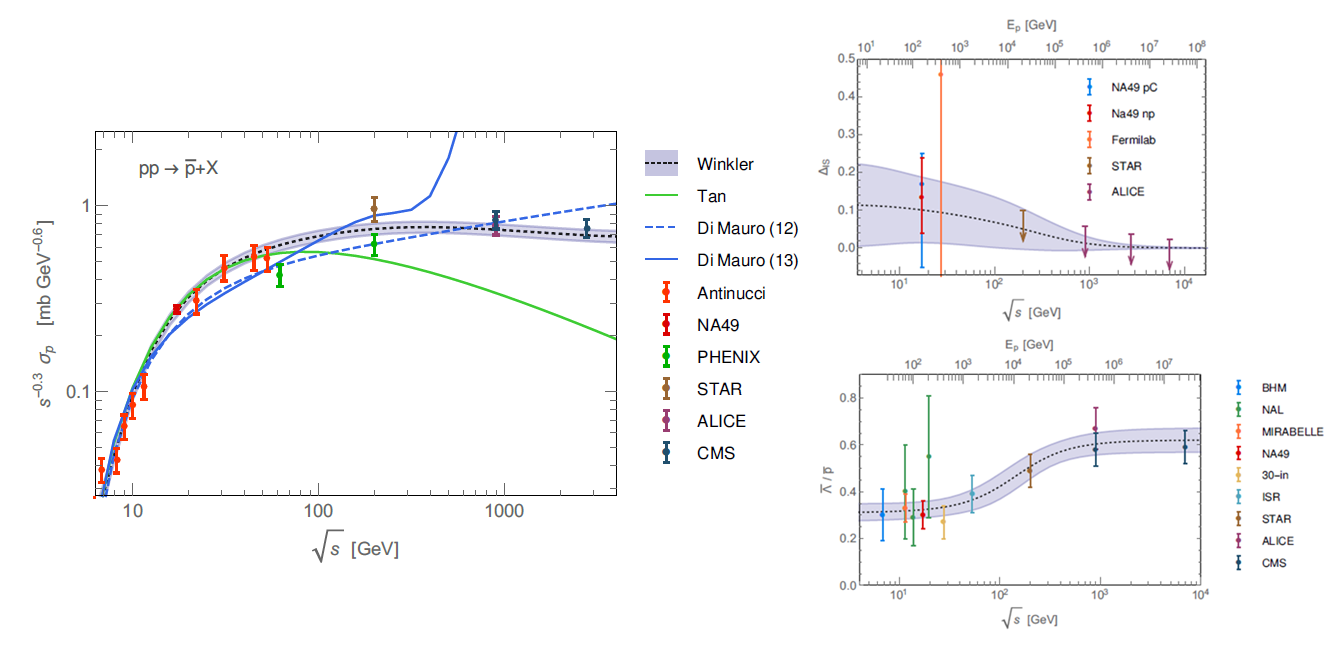}
	\caption{\footnotesize Prompt production (left panel), isospin asymmetry (top-right panel) and hyperon  (lower-right panel) contributions to the $p+p\longrightarrow\bar{p}$ total cross sections spectrum. The fits to data are shown with the $1\sigma$ uncertainties. In the left panel other antiproton production cross sections are shown for a comparison: the Tan\&Ng parameterization \cite{Tan_Ng} and the Di Mauro parametrisations \cite{DiMauro_Antip} with two different extrapolations at higher energies (see \cite{reinert2018precision} for full details).}
	\label{fig:Winkler_XS}
\end{figure}

We will use the cross sections for antiproton production derived in ref. \cite{reinert2018precision} (hereafter Winkler cross sections), which were added to the DRAGON code in its last version. The different contributions for the total production cross sections (eq.~\ref{eq:TotalAP_XS}) are shown in Figure~\ref{fig:Winkler_XS} together with the uncertainties associated with the fits to experimental data. In order to save computation time, only the proton and helium reactions are computed for the antiproton production, which will produce a systematic underestimation of the predicted flux of around $3-7\%$, in agreement with the right panel of Figure~\ref{fig:Ap_contrib}. Thus, the predicted spectra are scaled by a 5\% factor, since this is the contribution of the heavier CR species to the antiproton production in the region between $10-100 \units{GeV}$. Therefore, when we take the total uncertainties, a 2\% uncertainty associated to this scaling will be included.

Finally, charge-sign modulation is applied to antiprotons, following ref. \cite{Cholis:2015gna}. In this case, the solar modulation is modified to have the form:
\begin{equation}
\phi^{\pm} (t, R) = \phi_0(t) + \phi_1^{\pm}(t) F(R/R_0)
\label{eq:Charge-sign_Modul}
\end{equation}
choosing $F(R/R_0) \equiv \frac{R_0}{R}$ and $R_0 = 1 \units{GV}$. $\phi_0$ is the Fisk potential, set to be $0.61 \units{GV}$ as along all the thesis. Following \cite{reinert2018precision}, we choose $\phi^+_{1,AMS-02} = 0 $ and $\phi^-_{1} = 0.55 \units{GV}$.

\begin{figure}[!htb]
	\centering
	\includegraphics[width=0.55\textwidth, height=0.26\textheight]{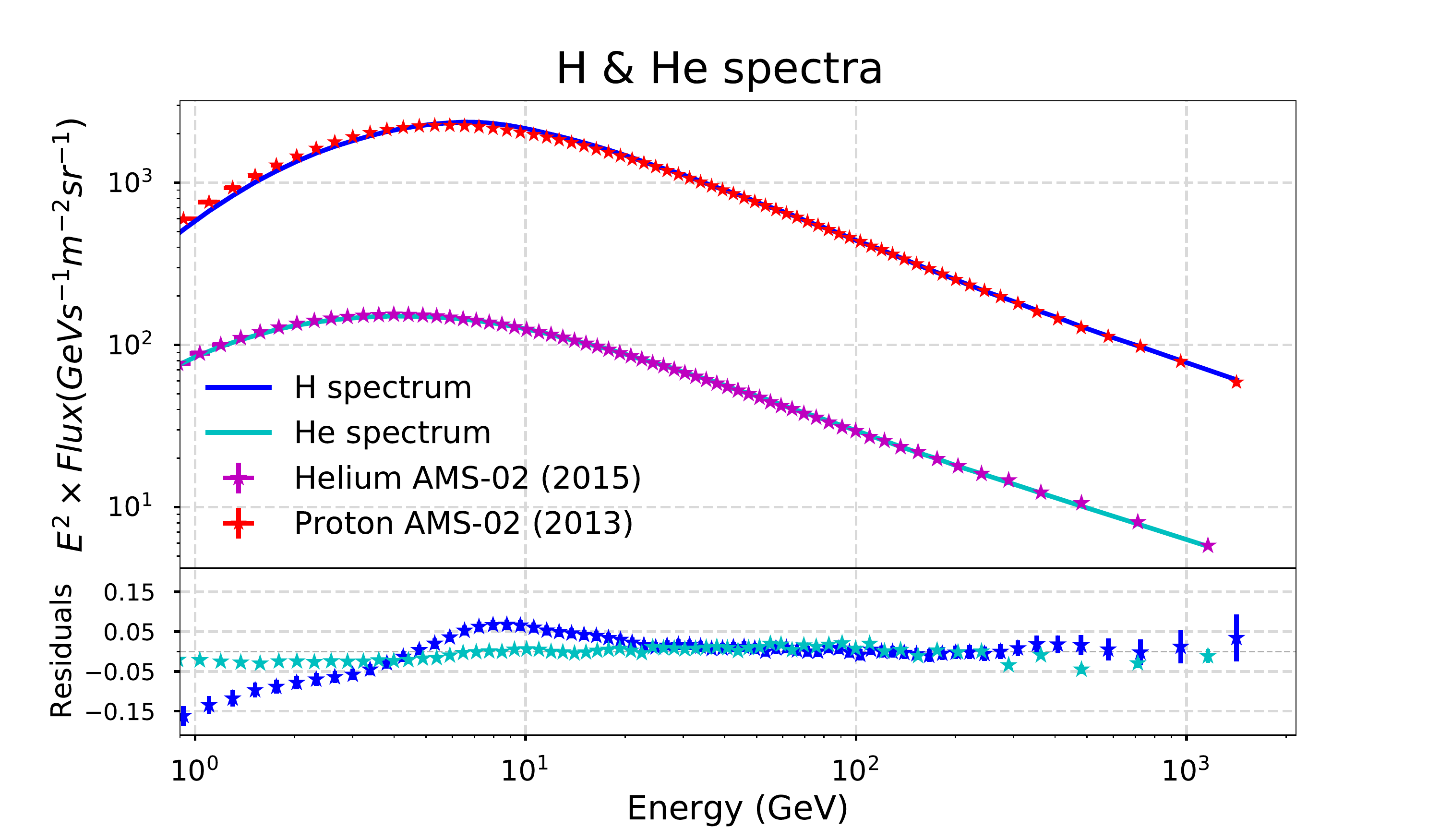}
	\caption{\footnotesize Derived proton and helium spectra computed to match AMS-02 data as nearly as possible. Residuals  are also shown.}
	\label{fig:Prim_Ap}
\end{figure}

The proton and helium fluxes are computed to match the AMS-02 spectra and are similar for all the diffusion parameters. The predicted spectra using the propagation parameters obtained in the B/O analysis with the original DRAGON2 cross sections are shown in Figure~\ref{fig:Prim_Ap}, where they are compared with the AMS-02 data. As we see, there is a maximum discrepancy of around 5\% in the proton spectrum, peaked at around $8 \units{GeV}$. This is due to the simple force field approximation used.

In order to fix the antiproton flux from the low energy discrepancies in the fit of the H spectrum, antiprotons are corrected by multiplying their flux by the factor $f \equiv \Delta_H C_H + \Delta_{He} C_{He}$, where $\Delta$ indicates the residual from the calculated H or He spectra to the AMS-02 data, and $C_H$ and $C_{He}$ are the fractional contributions from H and He to the total antiproton flux ($C_H + C_{He} = 1$). The He contribution to the antiproton flux is computed by making the simulation with no amount of He, and calculating the change in the antiproton production. In this way, after applying this correction, the antiproton flux is the one expected as if there were no discrepancies in the predicted H and He spectra with respect to AMS-02 data. In the calculation of the $\bar{p}/p$ spectrum no correction is applied, since the discrepancies related to the primary CRs are mitigated by the ratio.

\begin{figure}[!htpb]
	\centering
	\includegraphics[width=0.49\textwidth, height=0.24\textheight]{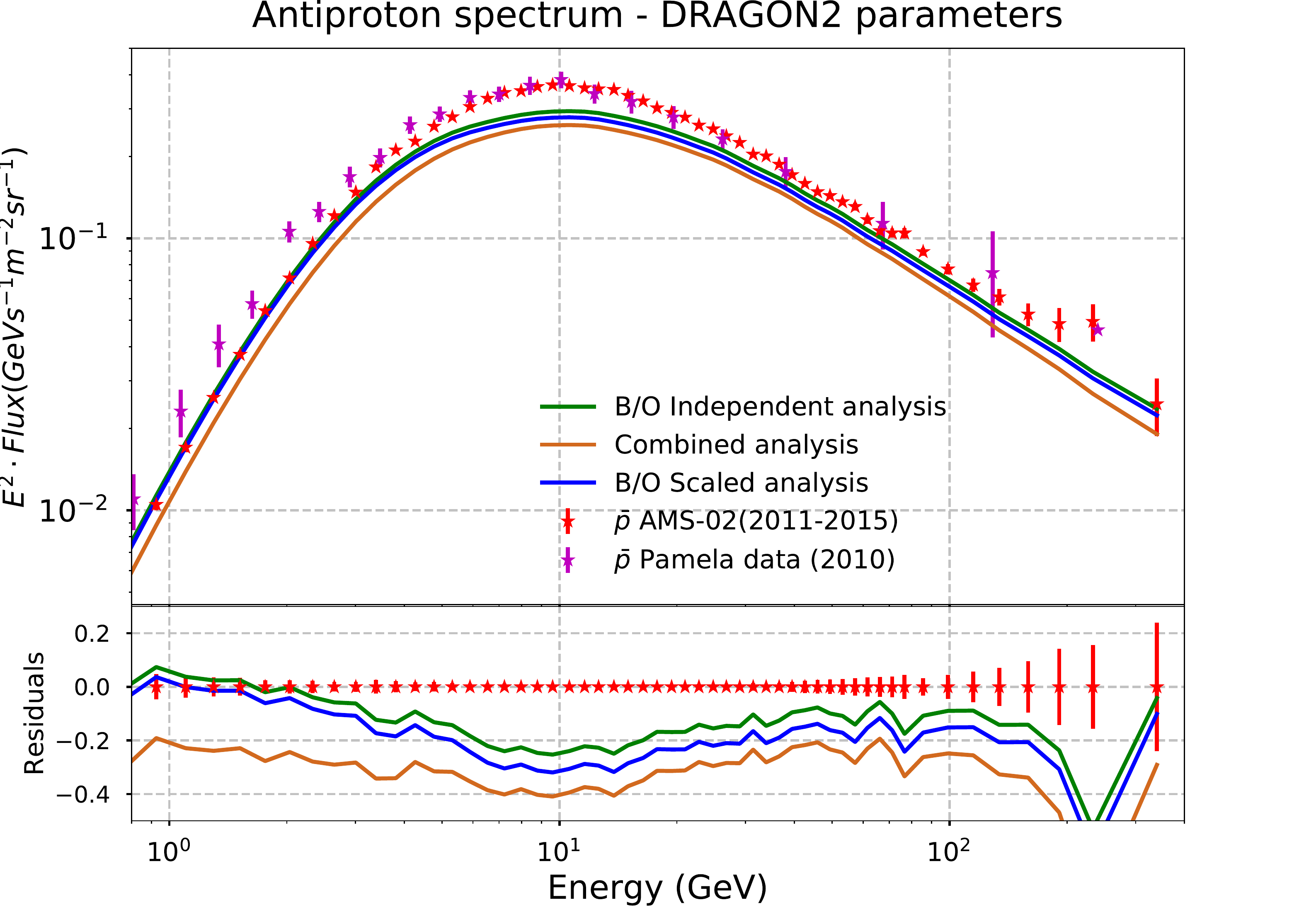}
	\includegraphics[width=0.49\textwidth, height=0.24\textheight]{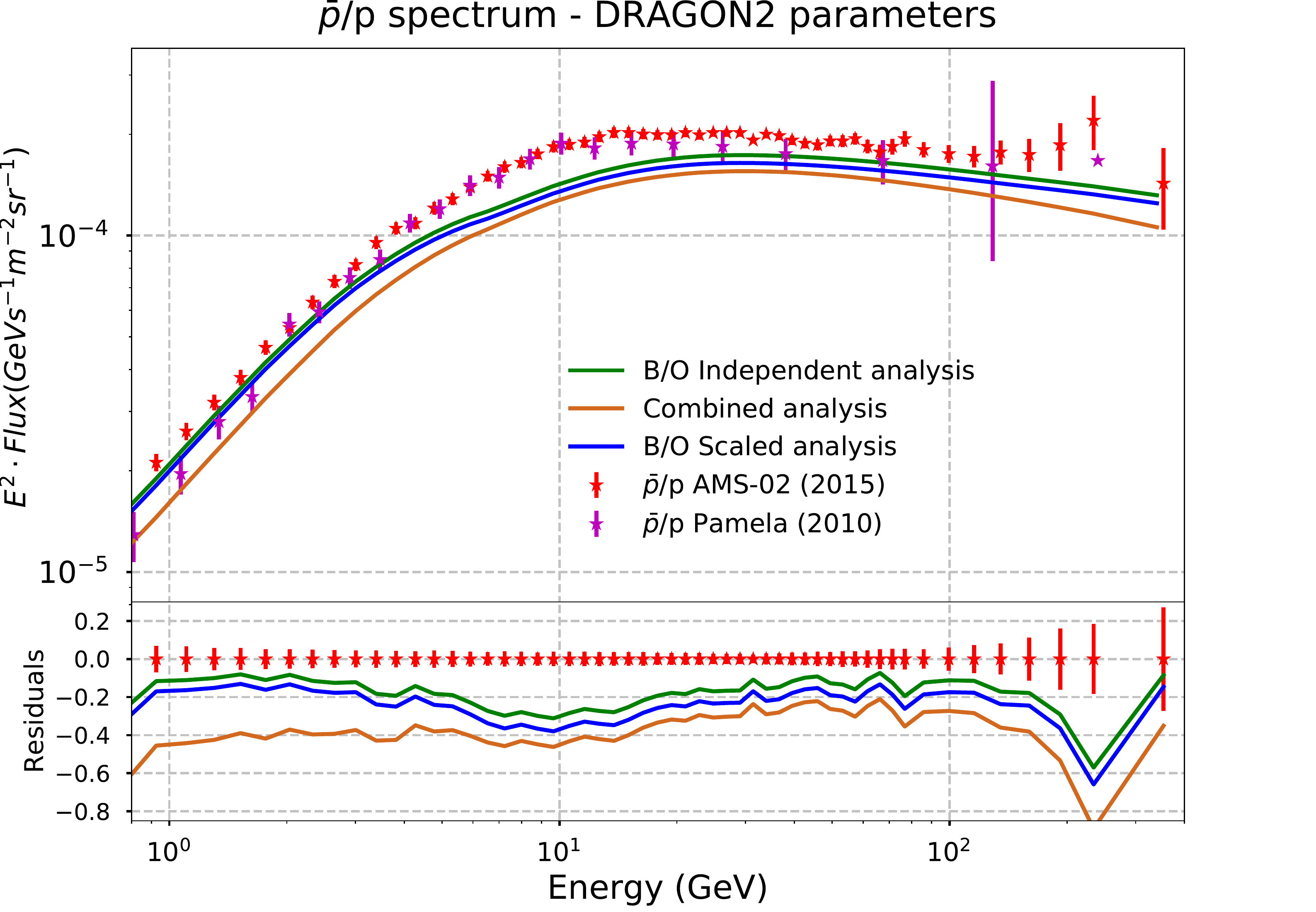}
	
	\vspace{0.6 cm}
	
	\includegraphics[width=0.49\textwidth, height=0.24\textheight]{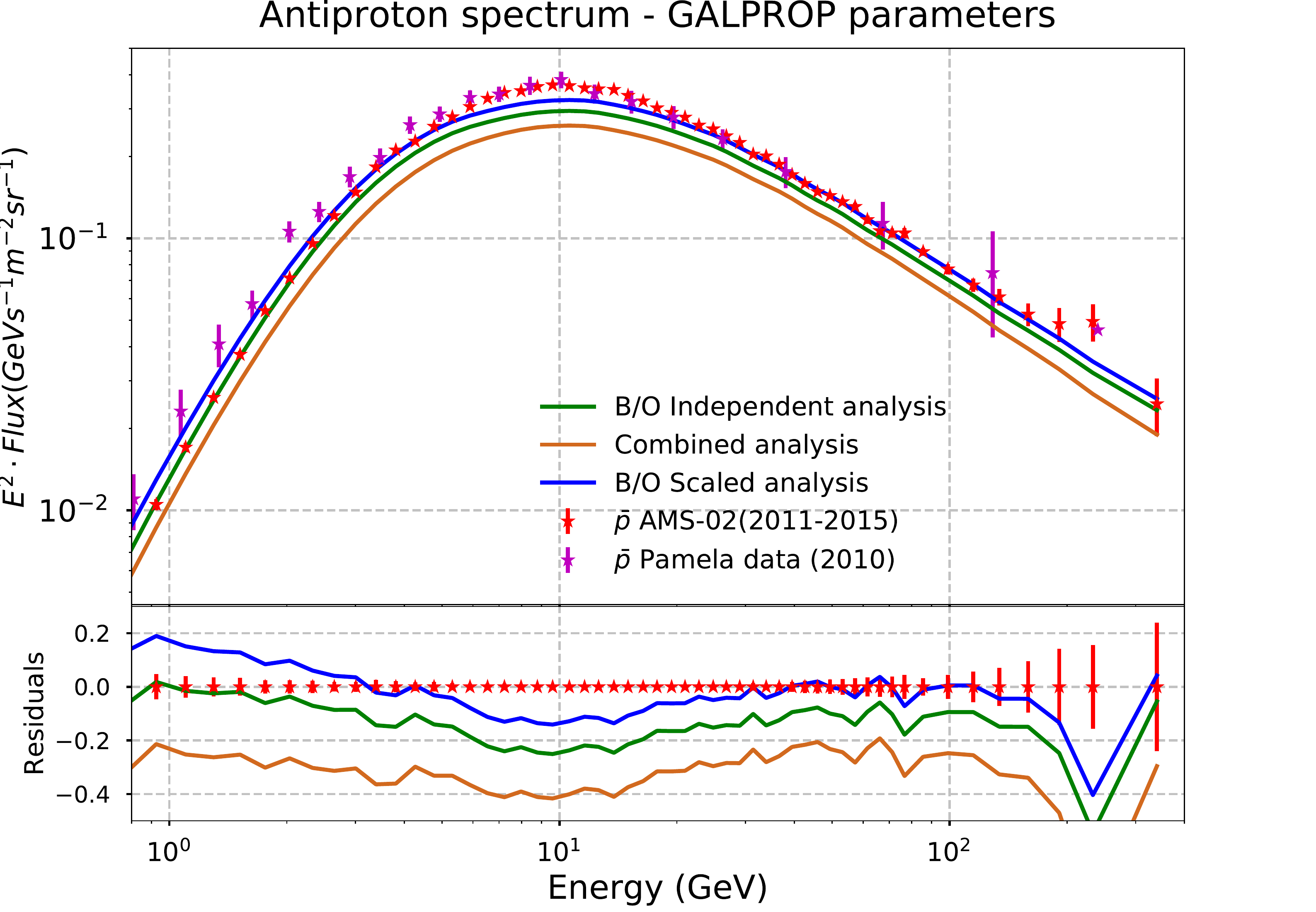}
	\includegraphics[width=0.49\textwidth, height=0.24\textheight]{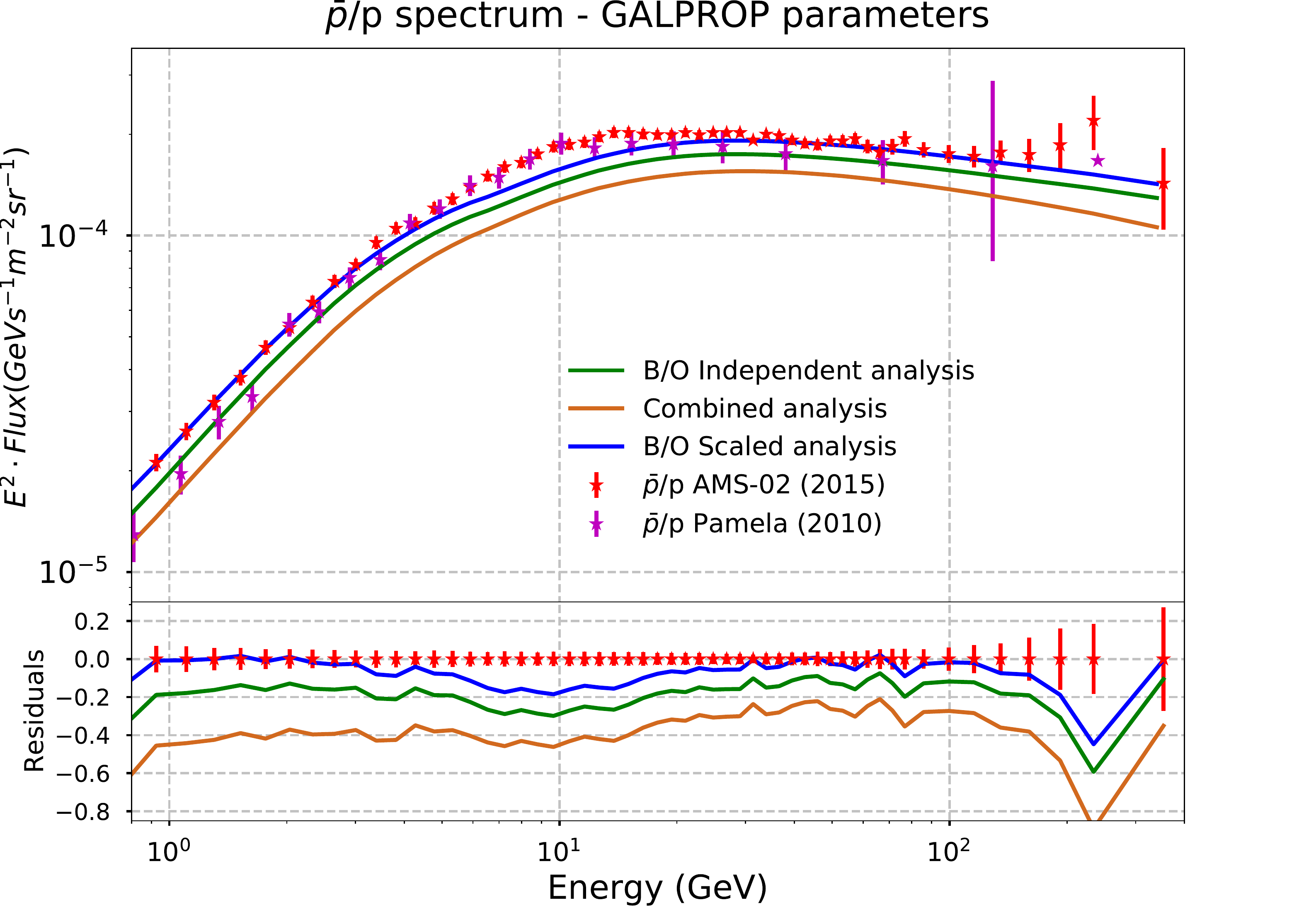}
	
	\vspace{0.6 cm}
	
	\includegraphics[width=0.49\textwidth, height=0.24\textheight]{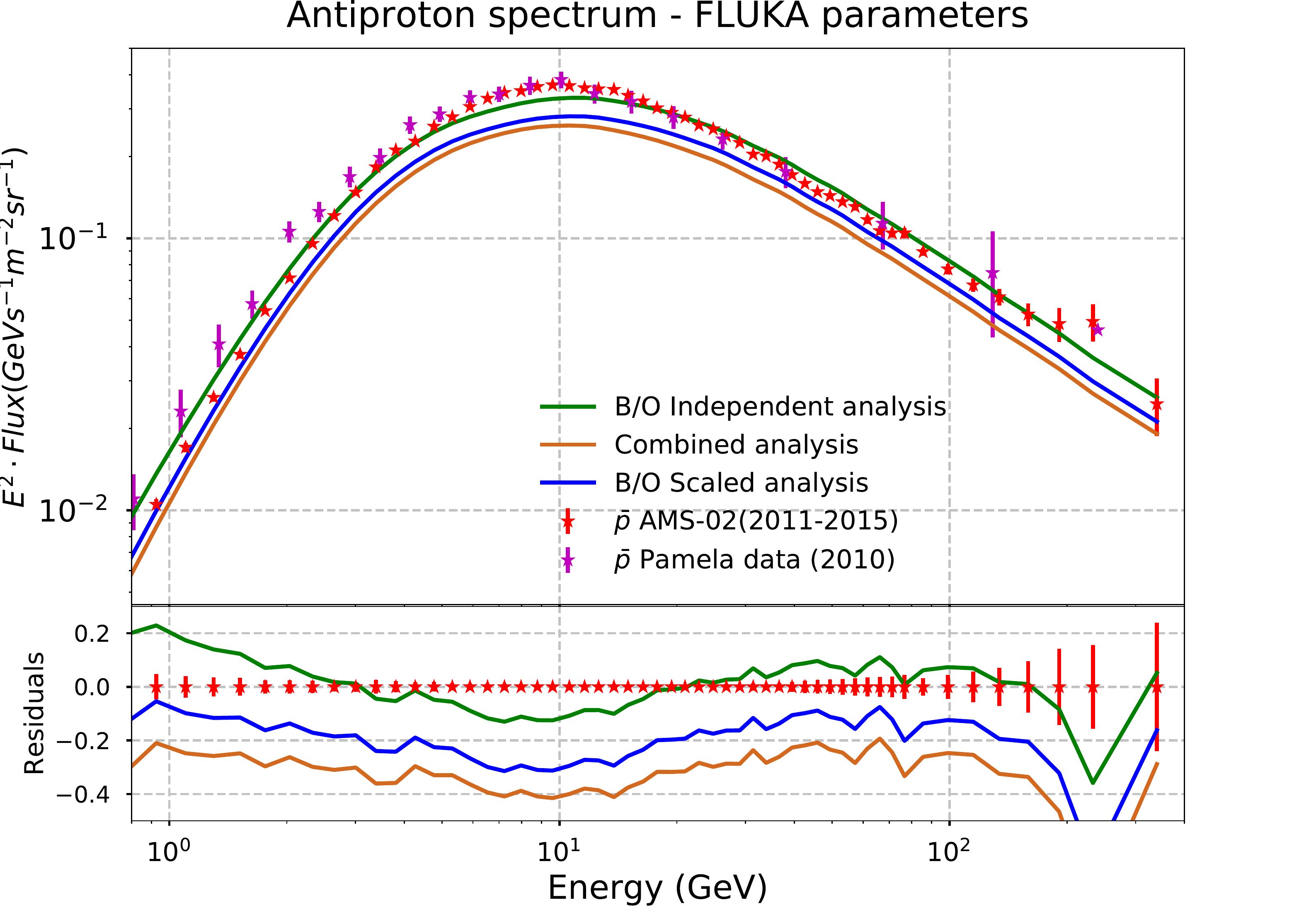}
	\includegraphics[width=0.49\textwidth, height=0.24\textheight]{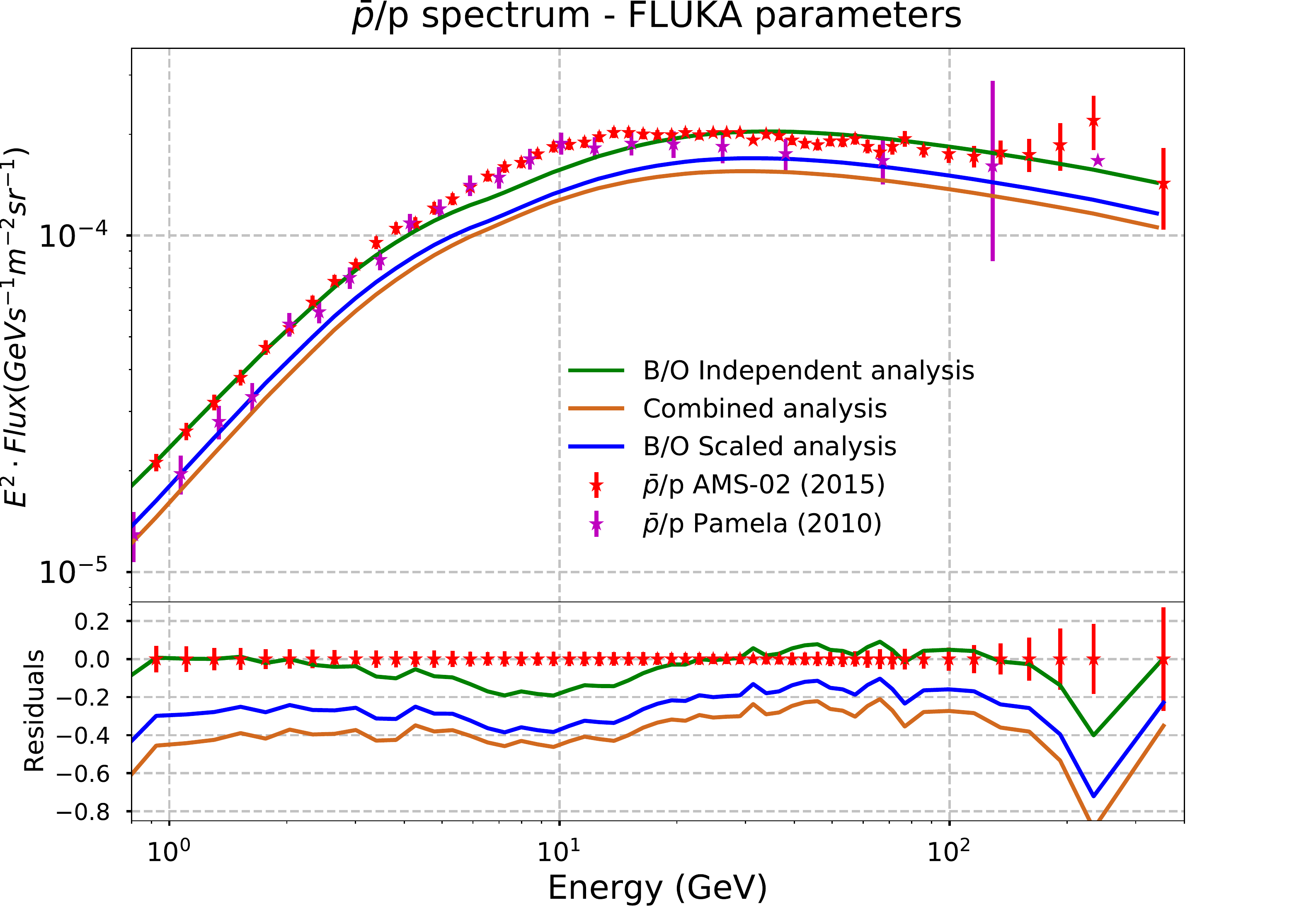}
	\caption{\footnotesize Antiproton (left panels) and $\bar{p}/p$ (right panels) spectra computed with the propagation parameters determined in the B/O, scaled B/O and combined MCMC analyses for the DRAGON2 (upper row), GALPROP (middle row) and FLUKA (lower row) spallation cross sections. These predicted spectra are compared to AMS-02 data and their residuals are shown in the lower panels.}
	\label{fig:Ap_results}
\end{figure}

The predicted antiproton and $\bar{p}/p$ spectra are compared to data for the the different diffusion parameters found with the DRAGON2, GALPROP and FLUKA nucleon spallation cross sections and they are shown in Figure~\ref{fig:Ap_results}. These comparisons allow us to visualize the effect of different diffusion parameters in the predicted antiproton flux and the importance of adding a scaling factor in the nucleon spallation cross sections of secondary CRs. We remind here that there are other sources of uncertainties, like those associated with the choice of diffusion parameterization and the solar modulation, that can be important for the antiproton fluxes at energies below $\sim 7 \units{GeV}$ and above $\sim200 \units{GeV}$. Within this range of energies these systematic uncertainties are expected to be negligible.

As we see from these plots, all the predicted spectra have very similar shapes, with a common feature easily observed from the residual plots as a dip, at energies between $7$ to $11 \units{GeV}$, with a peak of excess of data over predicted flux at $10 \units{GeV}$. This is the signal that has recently been associated with a signature of possible DM annihilation. It is very significant the fact the this deficit prevails with roughly the same shape for all the different propagation parameters. In fact, in the energy region from $7$ to $100 \units{GeV}$ the predicted fluxes approximately only differ by a normalization factor. 

On the other hand, it can be seen that the fluxes predicted using the combined analysis are those which exhibit the larger discrepancies with respect to data. This fact suggests that, while for the B/O analyses we must only take into account the uncertainties related with the cross sections of boron production, when using the combined analysis we must also take into account the uncertainties in the Li and Be production, which can be as large as 20 \% as we have seen. %In fact, this can be taken as another clue to discard the propagation parameters found in the combined analysis.

Other important conclusions can be reached when comparing the predicted flux obtained from the propagation parameters found in the B/O analyses (which differ in the normalization due to the scaling factor used for the cross sections of B production): a 1\% change of the scaling factor in the production cross sections leads to a set of propagation parameters which do not translate into a 1\% change in the antiproton flux. For example, in the case of the GALPROP cross sections, the scaling factor in the B production cross sections used in the scaled analyses was 0.94 (i.e. 6\% percentage of scaling), while the difference of antiproton fluxes derived from the scaled and original analyses is about 10\%. In the case of the FLUKA cross sections, while the scaling factor was 1.15 (15\% percentage of scaling), the difference in the antiproton flux between the scaled and original models is about 23\%. In turn, for the DRAGON2 cross sections, the scaling factor was 1.04 (4\% percentage of scaling) and the difference between the scaled and original model is about 5\%. 

This means that the uncertainties in the antiproton flux related to the nucleon spallation cross sections are more important than it is usually expected, but they mainly seem to act as a normalization factor.

\begin{figure}[!bp]
	\centering
	\includegraphics[width=0.68\textwidth, height=0.28\textheight]{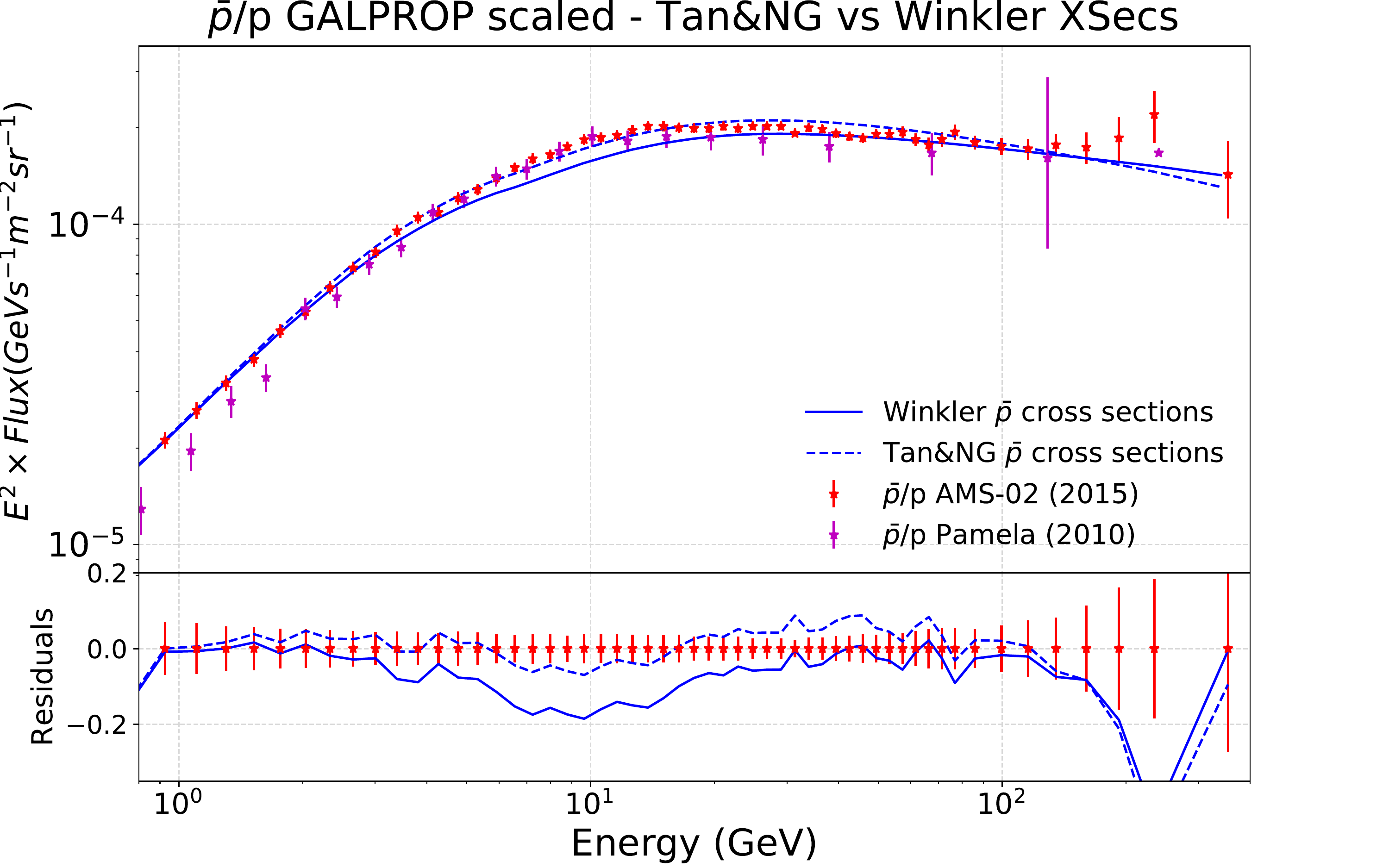}
	\caption{\footnotesize Comparison between the predicted $\bar{p}/p$ using the Winkler and Tan and Ng antiproton production cross sections for the diffusion parameters found in the scaled B/O analysis with the GALPROP spallation cross sections. Residuals with respect to AMS-02 data are shown in the lower panel.}
	\label{fig:Tan}
\end{figure}

Nevertheless, it must be highlighted the fact that these predictions share another common point: the cross sections of antiproton production. While we have seen that different diffusion parameters that explain the secondary-to-primary CR flux ratios roughly do not change the shape of the predicted antiproton spectrum (they practically act as a normalization factor) the cross sections of antiproton production may alter the shape of the predicted spectra, modifying the significance of any possible excess. As an example, the antiproton flux predicted by the propagation parameters derived from the scaled B/O analysis of the GALPROP cross sections is computed with the Tan and Ng cross sections of antiproton production \cite{Tan_Ng} to show the change in the predicted shape in Figure~\ref{fig:Tan}.

As we can see from this comparison, the shape of the predicted antiproton flux has significantly changed, making the discrepancies at around $10 \units{GeV}$ disappear. In fact, the predicted flux using these cross sections seems to be highly compatible with AMS-02 data. It implies that, although the Winkler antiproton production cross sections are the most updated ones, the uncertainties associated with experimental data are large and can considerably affect the significance of any feature.

\subsection{Associated uncertainties}
\label{sec:Ap_unc}

In this section we proceed to re-evaluate the uncertainties associated with the antiproton production, namely statistical uncertainties related to the determination of the diffusion parameters, to the solar modulation, those associated to the cross sections of antiproton production, the contribution of heavy elements to the antiproton flux and, for the first time, the uncertainty related to the spallation cross sections of secondary CR nuclei production. 

\begin{figure}[!tpb]
	\centering
	\includegraphics[width=0.49\textwidth, height=0.25\textheight]{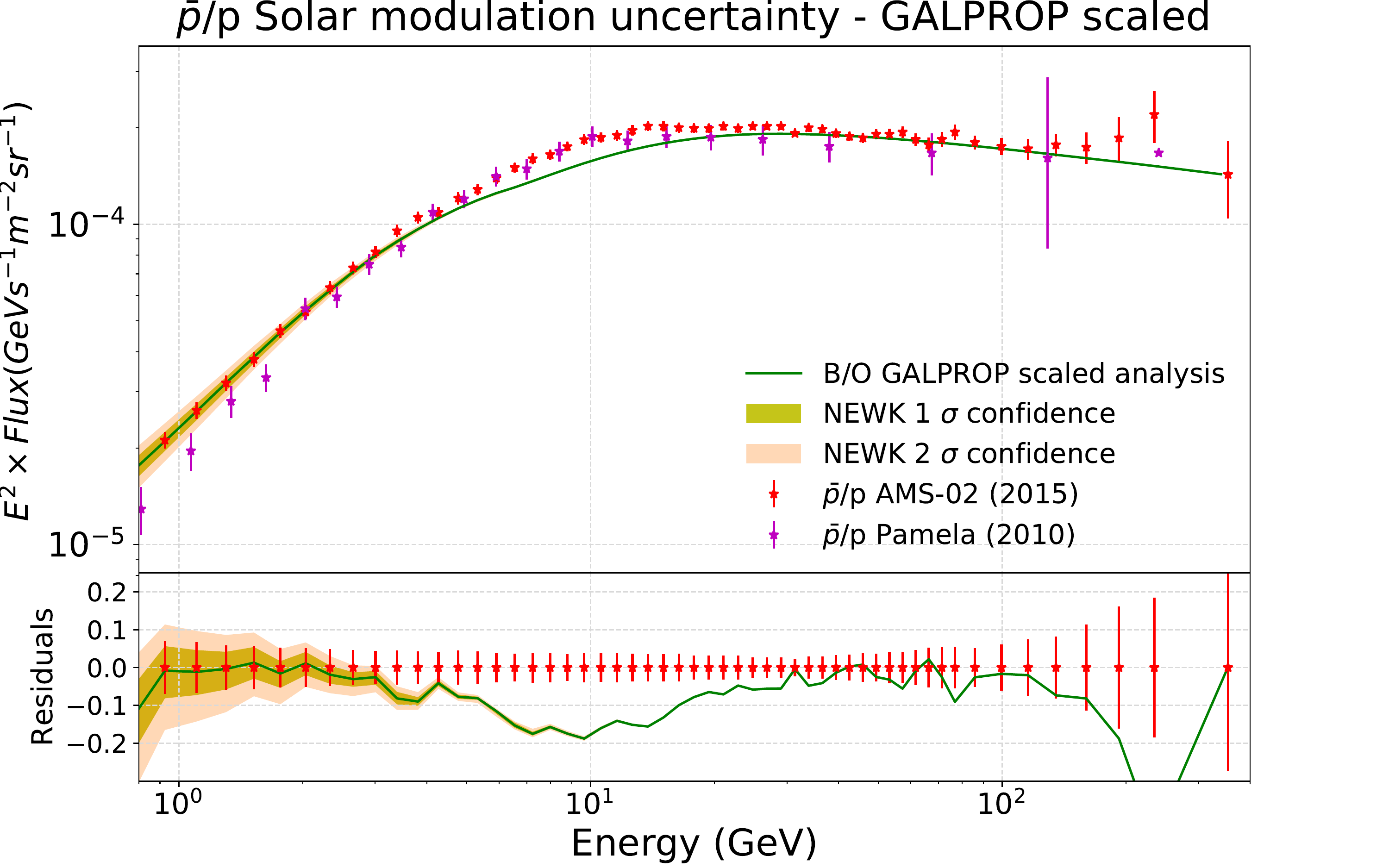}
	\includegraphics[width=0.49\textwidth, height=0.25\textheight]{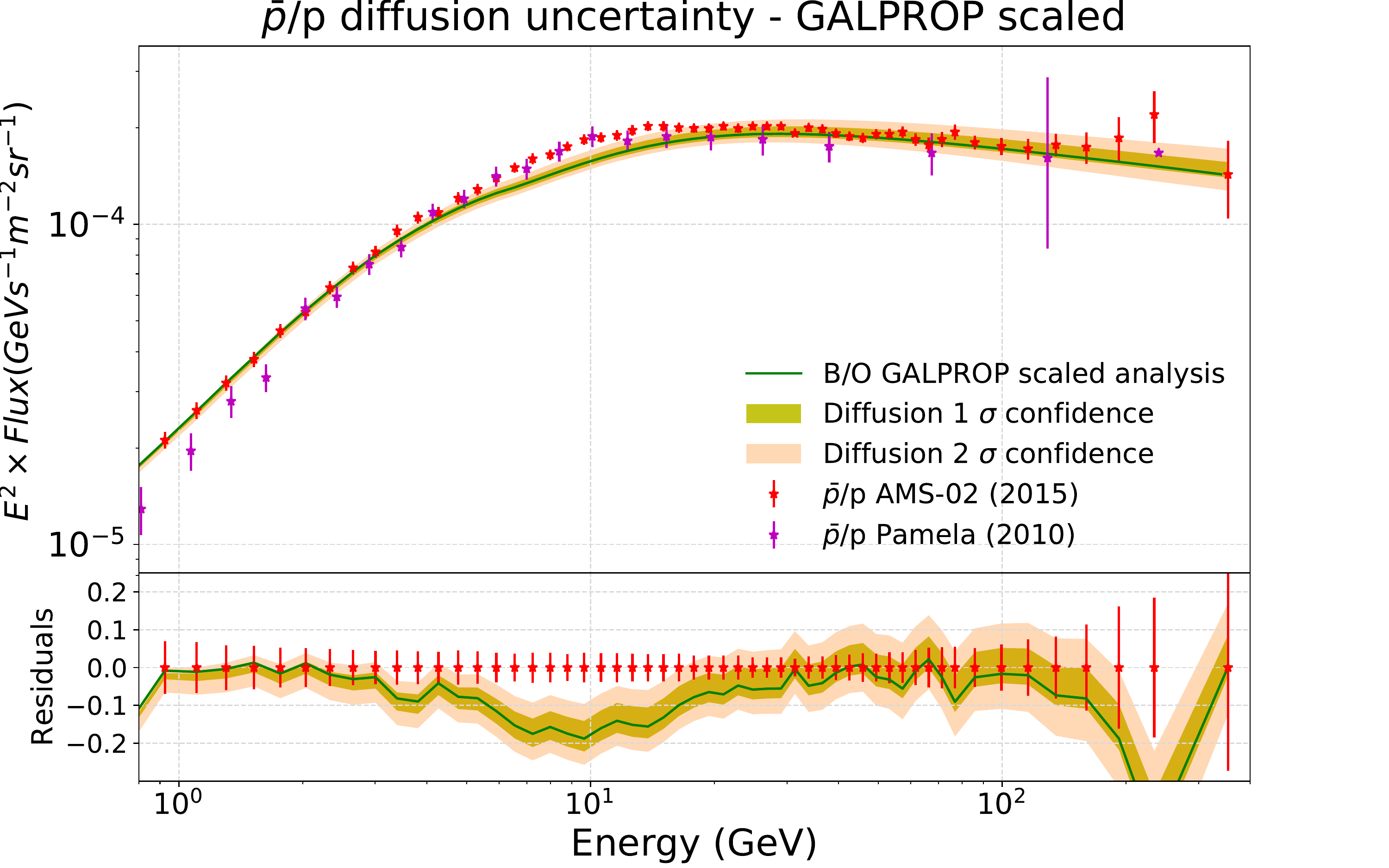}
	
	\vspace{0.6 cm}
	
	\includegraphics[width=0.49\textwidth, height=0.25\textheight]{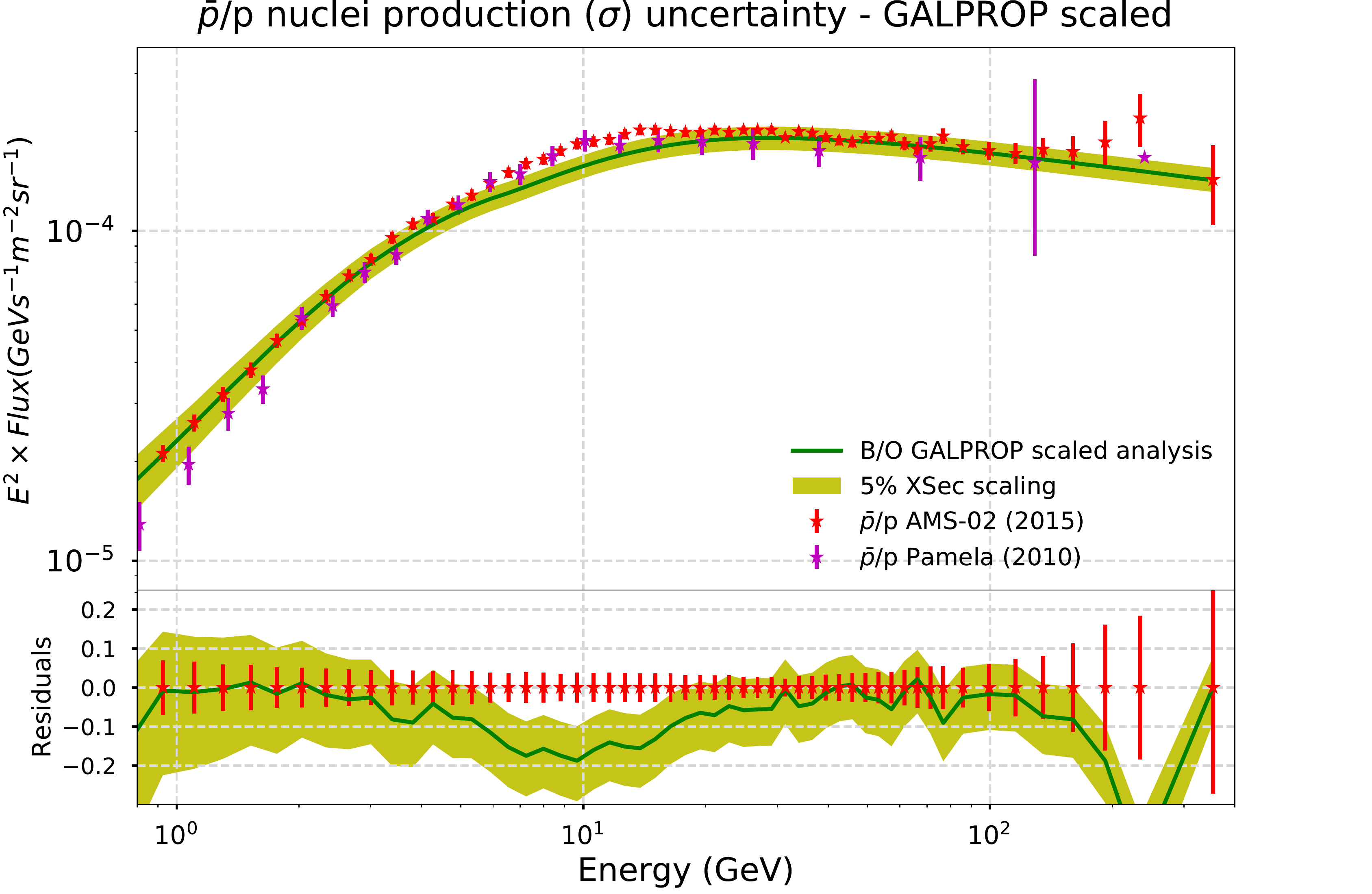}
	\includegraphics[width=0.49\textwidth, height=0.25\textheight]{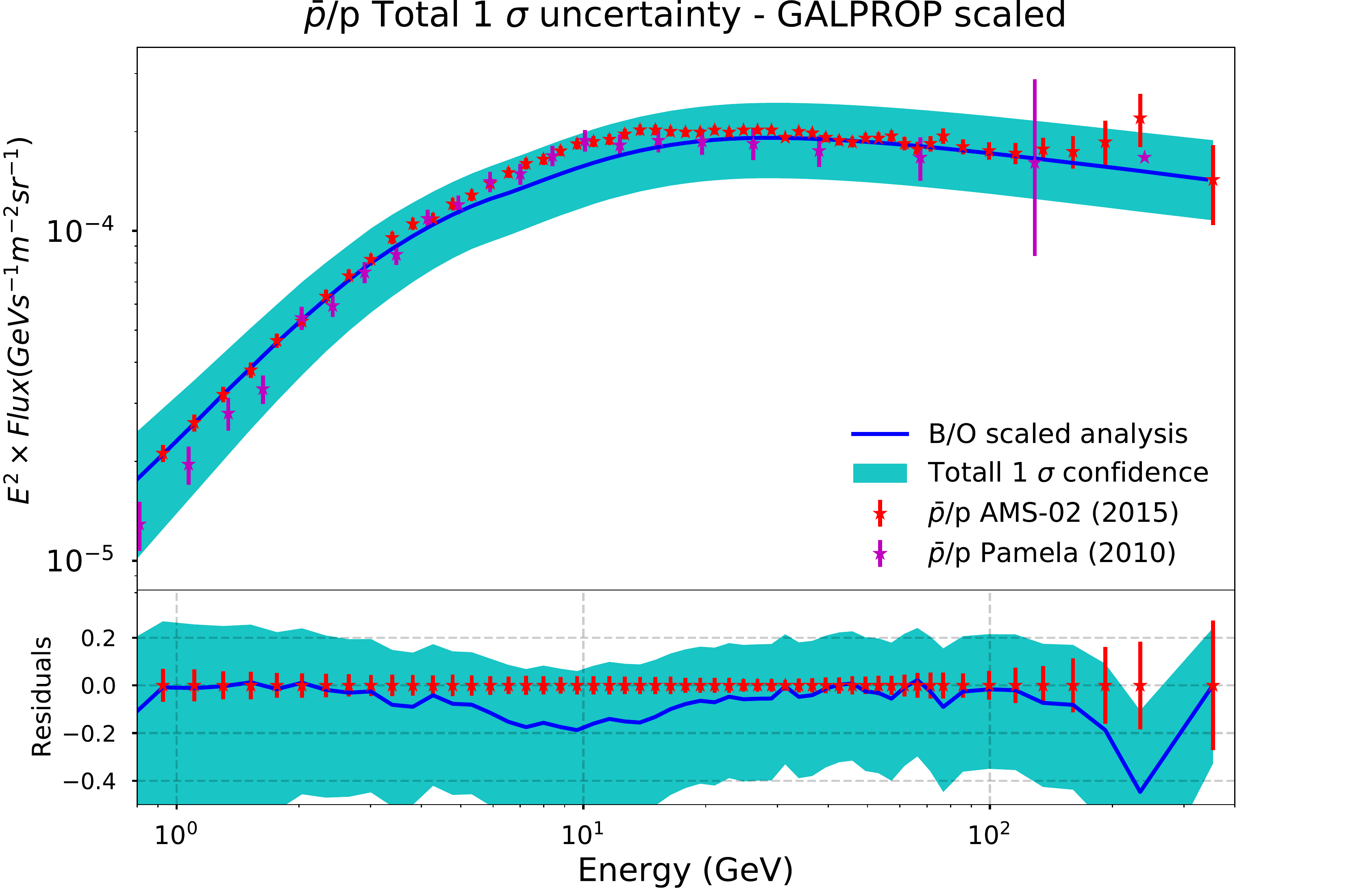}
	\caption{\footnotesize Uncertainties associated to the $\bar{p}/p$ flux ratio evaluated with the diffusion parameters inferred from the B/O scaled analysis with the GALPROP cross sections. The 1$\sigma$ and 2$\sigma$ uncertainties associated with the solar modulation (upper-left panel), with the propagation parameters (upper-right panel), and with a 5\% scaling of the GALPROP cross sections of B production (lower-left panel) are also shown in the plots. The combined (total) 1$\sigma$ uncertainties are also shown in the lower-right panel.}
	\label{fig:Uncertainties}
\end{figure}

These uncertainties are shown in Figure \ref{fig:Uncertainties} superimposed to the $\bar{p}/p$ flux ratio derived from the diffusion parameters inferred from the B/O scaled analysis with the GALPROP cross sections. In particular, Figure~\ref{fig:Uncertainties} shows the uncertainties associated with the Fisk potential (upper-left panel), those associated to diffusion (upper-right panel), those associated to cross sections of secondary nuclei production (lower-left panel) and finally the total uncertainties (lower-right) evaluating by summing all the individual contributions. We stress again that, when calculating the $\bar{p}/p$ flux ratio, the uncertainties associated to the fit of the spectra of primary CRs can be safely neglected. 

The 1$\sigma$ and 2$\sigma$ diffusion uncertainties are obtained for the determination of propagation parameters in the MCMC analysis. They are very similar for all the different analyses (see~\ref{tab:PDF_single_Diff}) except for the combined analysis, which exhibits lower uncertainties because of the larger number of constraints. The uncertainties on the Fisk potential are those reported by the NEWK experiment, in the period of time associated to the AMS-02 operations, and are of $\sim 0.055\units{GV}$ (1$\sigma$) and $\sim 0.12\units{GV}$ (2$\sigma$). Other important uncertainties, related to the solar modulation, are those of the $\phi_1^{\pm}$ parameter, since it depends on the charge-sign of the CR, but they are very difficult to evaluate and are not included in this study. Then, the uncertainties associated to the cross sections of nuclei production used to determine the propagation parameters are computed by scaling of 5\% the original GALPROP cross sections of boron production and determining again the diffusion parameters. The 5\% scaling is taken since, as explained in chapters~\ref{sec:XSecs} and~\ref{sec:3}, this is compatible with the uncertainties expected in the scaled analysis.%to fit the LiBeB secondary-over-secondary flux ratios of AMS-02. 
However, we must remember that the experimental error bars in the main channels of B production were always larger than 10\% (see figures~\ref{fig:LUmodelsC} and~\ref{fig:LUmodelsO}), and therefore this factor is indeed very conservative. In this case, this 5\% scale translates into a $\sim 9\%$ difference in the derived antiproton fluxes, almost constant at energies between $5-100 \units{GeV}$.

In addition, as we scaled the derived antiproton fluxes by 5\% to account for the contribution of nuclei heavier than helium in the antiproton production, an extra 2\% uncercertainty is added (since the heavy CR nuclei contribution is estimated to be from 3 to 7\% according to Figure~\ref{fig:Ap_contrib}, right).

Finally, a very conservative 12\% uncertainty is added to account for the uncertainties in the cross sections of antiproton production (they are estimated to be between 12-20 \% in ref. \cite{Korsmeier:2018gcy}).

\begin{figure}[!hb]
	\centering
	\includegraphics[width=0.75\textwidth, height=0.33\textheight]{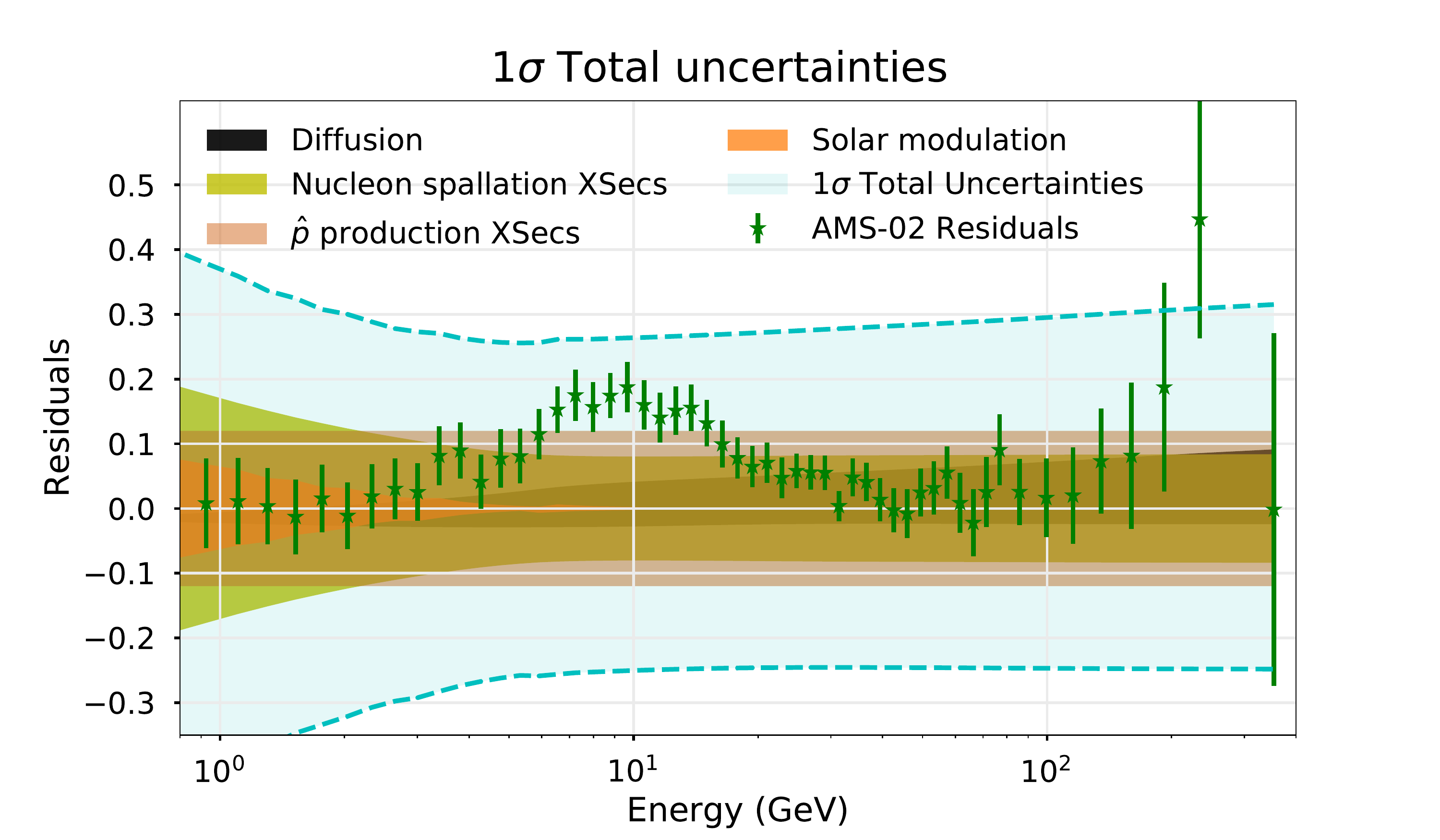}
	\caption{\footnotesize Combined total 1$\sigma$ uncertainties associated to the antiproton production along with each of the sources of uncertainty. Residuals in this plot are defined as (data-predicted)/predicted. The excess of data with respect to the predicted flux around $10 \units{GeV}$, that was associated with a signature of DM annihilation (or decay) into antiprotons, is evident, but it is within the total 1$\sigma$ uncertainty band.}
	\label{fig:Unc_comp}
\end{figure}

From the lower-right panel of Figure~\ref{fig:Uncertainties} we can see that the residuals of the model with respect the AMS-02 data lie inside the $1\sigma$ uncertainty band. Each of the sources of uncertainty is displayed with more detail in Figure~\ref{fig:Unc_comp}. We can see that the total 1$\sigma$ uncertainties are of $\sim 27\%$ at $10 \units{GeV}$ and around 30\% at $100 \units{GeV}$, and satisfactorily contain the excess of data over predicted flux, which indicates that this discrepancy could be explained with more refined data (mainly related to the cross sections). 

A related study has been already presented in \cite{ICPPA_Pedro}, focused on the $10 \units{GeV}$ excess in the antiproton spectrum to show that a consistent fit (within AMS-02 experimental errors) can be achieved for the secondary-over-primary ratios of B, Be and Li and the $\hat{p}/p$ spectra simultaneously, assuming that the only source of antiprotons is CR collisions.
%\newpage

\subsection{Conclusions}
\label{sec:Ap_conc}

In this section, we have derived the antiproton flux predicted by different sets of propagation parameters that reproduce secondary-over-primary ratios, with combined information from the Li, Be and B CRs,  finding that their different predictions roughly just tend to change the normalization of the predicted antiproton spectrum, with no relevant variations on the shape. The same feature (excess of data over predicted flux) at around $10 \units{GeV}$, which appears in every propagation model and may be explained by adding a primary source of antiprotons, was related to the annihilation (or decay) of a DM particle of around $30-90 \units{GeV}$ into a $b\bar{b}$ pair, which is also able to explain the Galactic Center gamma-ray excess.

Nevertheless, considering the different sources of uncertainty associated to the antiproton production allowed us to conclude that, while the diffusion parameters mostly affect the normalization of the predicted antiproton fluxes, the antiproton production cross sections are able to alter appreciably the antiproton spectrum shape within the existing experimental uncertainties ($\sim (12 - 20)\%$). 

Concretely, we have propagated the uncertainties from the spallation cross sections of B production for the first time and added up to the rest of uncertainty sources to see that the total 1$\sigma$ uncertainties at $10 \units{GeV}$ can be as large as 30\% and, hence, most of the derived antiproton spectra could be compatible with the AMS-02 data. This determination of the total uncertainties is actually very conservative and can be larger by at least a 10\% or even 20\% when including the AMS-02 correlated errors in the analysis (see App.~\ref{sec:covariance}), thus, altering considerably the significance of the excess, as was concluded in ref.~\cite{Heisig:2020nse}. This means that precision tests to study the possible contribution of DM decay (or annihilation) into antiprotons are difficult to be reliably performed, given these large uncertainties.

\section{Study of diffuse galactic gamma-ray emissions}
\label{sec:Emiss}

The Milky Way looks bright in the gamma-ray sky thanks to a superposition of many different processes. This gamma-ray emission can be firstly divided into the contribution from point-like sources (which helps in recognizing the sources of CR injection) and that from the diffuse emissions, directly related to the CR interactions with the ISM gas and the interstellar radiation field (ISRF). Above $50 \units{MeV}$ the diffuse intensity averaged over the sky is 5 times larger than the intensity produced by resolved sources. % https://fermi.gsfc.nasa.gov/ssc/data/access/lat/Model_details/FSSC_model_diffus_reprocessed_v12.pdf

In order to understand the diffuse emission, a complete picture of the structure and contents of the Galaxy (ISRF, interstellar molecular and ionized gas distributions, magnetic fields) must be combined with the knowledge on the propagation and interactions of CRs.

\begin{figure}[!thpb]
	\centering
	\includegraphics[width=0.62\textwidth, height=0.394\textheight]{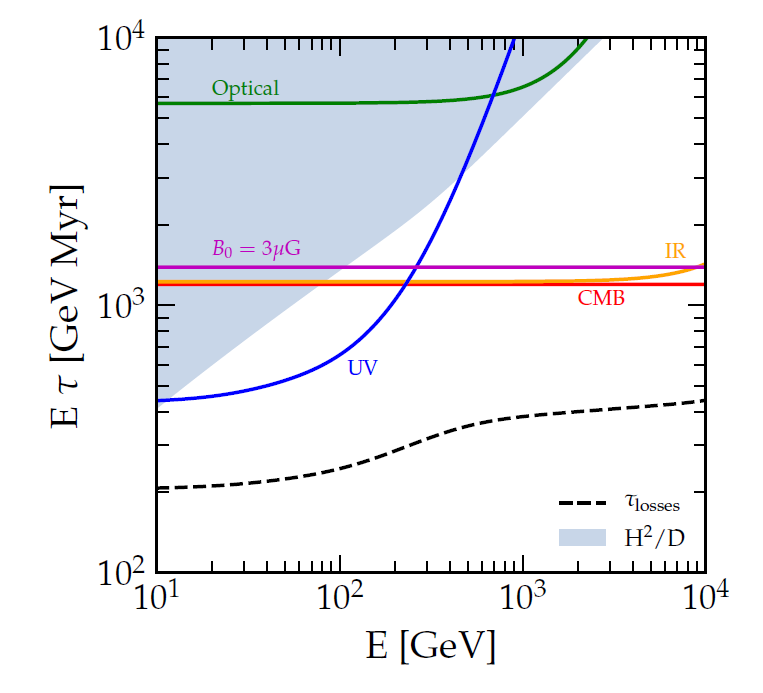}
	\caption{\footnotesize Time scales of energy loss (multiplied by energy) of CR electrons during their propagation in the Milky Way. The dashed line represents the total timescale, while the solid lines refer to contributions of the magnetic field (in magenta) or ISRF components. The shadow region highlights the escape timescale from the Galaxy due to diffusion. Taken from ref. \cite{Evoli:2020ash}.}
	\label{fig:ISFR_Elosses}
\end{figure}

The physical processes related to gamma-ray diffuse emission in the Galaxy are:
\begin{itemize}
    \item \textbf{Pion decay:} The decay of neutral pions formed via hadronic interactions of CRs with ISM gas is the dominant mechanism of gamma-ray emission at high energies. Although all scalar neutral mesons can decay in the same way, pions are produced more abundantly, mainly from reactions involving CR protons.
    %, since their flux is much higher than that of heavier CR particles.
    In this way, the intensity of the gamma-ray emission depends on the pion production rate, which is dependent on the flux of CR protons. This mechanism produces a characteristic signature, known as the ``pion bump'', which served to find the first evidence that CR protons are accelerated in SNRs \cite{ackermann2013detection} by the Fermi-LAT collaboration. 
    
    \item \textbf{Bremsstrahlung:} also known as braking emission, in the astrophysical context bremsstrahlung gamma rays are  mainly emitted in the interactions of electrons (whose emission is 1000 times more intense than from protons) with Coulomb fields in the ambient plasma. This happens in general when CR electrons travel through the ISM plasma, generating this diffuse emission along their path. It is the dominant gamma-ray emission at sub-GeV energies.
    
    \item \textbf{Inverse-Compton scattering:} The effect by which relativistic electrons transfer part of their energy to photons via collisions is known as Inverse-Compton effect. %In order for this process to be favorable, these electrons must be relativistic. 
    The intensity of the radiation is proportional to the energy density of the radiation field. In the Galaxy this effect permanently occurs when high energy electrons collide with photons from the Cosmic Microwave Background (CMB), starlight, ultraviolet (UV) or infrared (IR) photons from the ISRF (see~\ref{fig:ISFR_Elosses}).
\end{itemize}

In some cases, like in Active Galactic Nuclei (AGNs), synchrotron emission can contribute significantly to the gamma-ray emission (concretely through a mechanism known as Synchrotron Self Compton emission), but in general it does not play an important role at gamma-ray frequencies unlike in radio and X-ray frequencies. 

These processes also play a major role in the propagation of CR leptons, since these emissions are important sources of energy losses. This causes a transition from a diffusion-dominated to loss-dominated regime in the spectrum of electrons. As a consequence of these losses, CR electrons are able to travel shorter distances than CR nuclei, therefore probing a more local region of the Galaxy before being detected. They are the main responsible of IC and bremsstrahlung emissions. Also synchrotron emission is a very important source of energy loss.  

The recent measurements performed by AMS-02 \cite{Aguilar:2019ksn} up to $\sim \units{TeV}$ energies have shown the presence of a smooth hardening in the electron spectrum at around $42.1\pm5.4 \units{GeV}$. This feature does not seem to be related with the rising positron fraction \cite{Evoli:2020ash} and has been assumed to arise because of the overlap of two different components (sources) of CR electrons, explained as closer sources becoming dominant (see \cite{Fornieri:2019ddi}). Nevertheless, very recently it has been also explained by carefully taking into account the transition to the Klein-Nishina regime on the UV light background in ref. \cite{Evoli:2020ash}.

\begin{figure}[!ptb]
	\centering
	\includegraphics[width=0.65\textwidth, height=0.25\textheight]{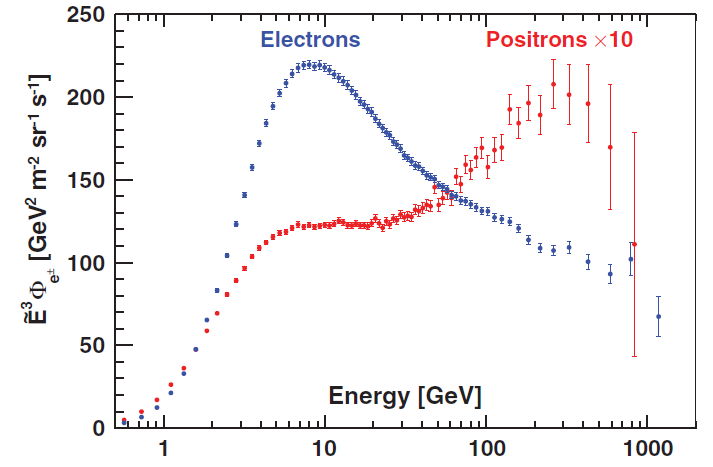}
	\caption{\footnotesize AMS-02 electron and positron spectra. Taken from ref. \cite{Aguilar:2019ksn}.}
	\label{fig:Leptons_AMS}
\end{figure}

In addition, the DAMPE \cite{Ambrosi:2017wek}), HESS \cite{Aharonian:2009ah} and CALET \cite{CALET_electrons} experiments measured the total CR electron spectrum (i.e. the e$^-$+e$^+$ spectrum, hereafter CRE spectrum) at energies around $20 \units{TeV}$, reporting a sharp softening at $\sim1\units{TeV}$. This is expected to be the cut-off of energy acceleration of CRs in the sources, but it is not totally clear yet. A similar cut-off has been reported by the AMS-02 collaboration at about $300 \units{GeV}$. These features can be seen in Figure~\ref{fig:Leptons_AMS}.

The rise of the positron spectrum above $10\units{GeV}$ was first detected by Pamela \cite{Adriani:2010ib, Pam_posit} and Fermi \cite{FermiLAT:2011ab} and then confirmed by AMS-02 \cite{AMS_posit, AMS_positb}. This is a feature not expected if positrons were purely secondary particles, and suggests that the high energy part of their spectrum is remarkably dominated by production of positrons in sources, most likely PWN, as explained above.

Some CR detectors are not able to distinguish electrons from positrons, since a magnetic field is needed to determine the charge sign. This is, for example, the case of Cherenkov telescopes, such as H.E.S.S. \cite{Hinton:2004eu} and MAGIC \cite{Magic}, since the air showers created by these leptons are impossible to be distinguished. These experiments report the total CRE spectrum. 

Nowadays, instruments studying the gamma-ray sky are able to also detect CR electrons, since they make use of the detection of the electromagnetic showers with calorimeters (in the case of satellite experiments) or with the Cherenkov light emitted (ground-based telescopes) - which could also be considered a calorimetric detection, with the atmosphere as calorimeter. 

At the present time, there are few experiments dedicated to the gamma-ray detection in the energy range from $\sim 200 \units{MeV}$ to $1 \units{TeV}$. The best example is the Fermi-LAT experiment, able to measure for first time the diffuse emission at energies above $10\units{GeV}$. It has an unprecedented sensitivity compared to its predecessor, EGRET \cite{Thompson:2015nda} on board the Compton Gamma Ray Observatory \cite{Schonfelder:2002mw, Hunter:1995vz}, and a very large field of view, close to one fifth of the sky. Its angular resolution is around 0.8º at $1\units{GeV}$ and better than 0.2º above $10\units{GeV}$.

The Galactic diffuse emission is specially intense at$\units{GeV}$ energies \cite{Thompson:2011zzb} and it was observed from the earliest gamma-ray experiments \cite{Dingus:1995us}. Nevertheless, also the background (for example due to extra-galactic light) is intense at these energies. On the other hand, the diffuse emission in the$\units{TeV}$ sky is almost exclusively produced from CR interactions, and its intensity is significantly larger than the background at $10-100 \units{TeV}$. At these energies, the extra-galactic gamma-ray flux is strongly suppressed by pair production of the extra-galactic background light, bremsstrahlung emission is negligible and the inverse Compton flux from CR electrons is suppressed by the softening of the electron spectrum in the TeV range and by the onset of Klein-Nishina suppression of the Compton-scattering cross-section \cite{Beischer:2020rts}.

\subsection{The predicted diffuse gamma-ray sky}
\label{sec:gamma_sky}

Together with CRs, the key inputs needed for the prediction of the gamma-ray emission in the sky are the ISRF and the 3D distributions of gas in the galaxy. Although both are approximately known, they generally represent the main source of uncertainties in gamma-ray studies.

\begin{figure}[!hptb]
	\centering
	\includegraphics[width=0.92\textwidth, height=0.32\textheight]{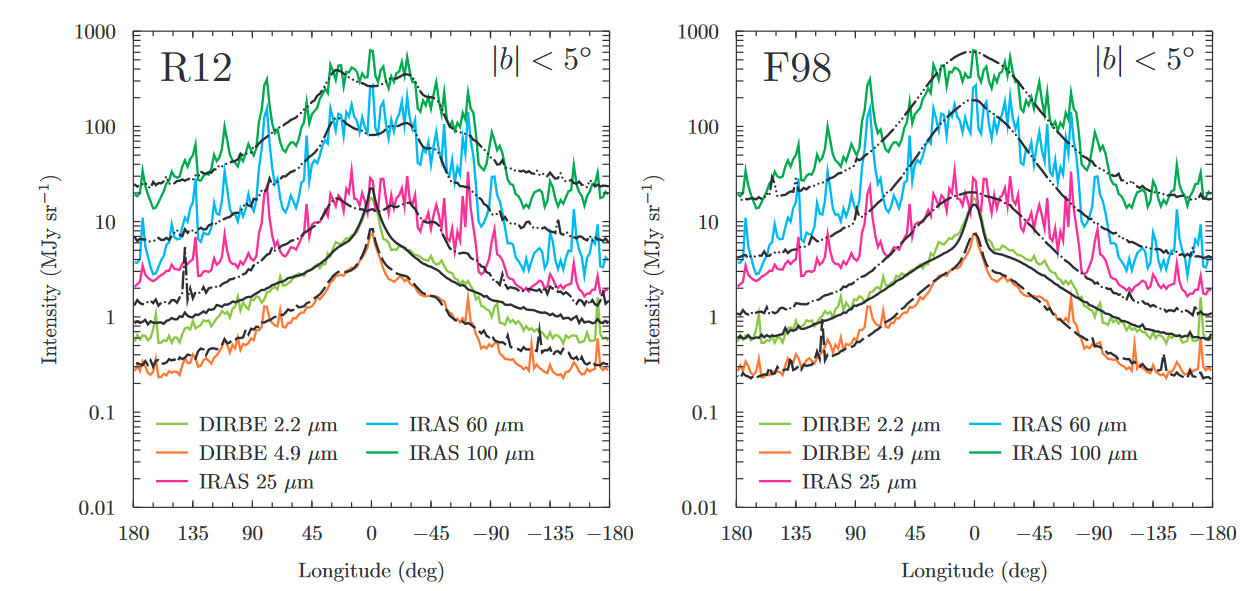}
	\caption{\footnotesize Models of Robitaille \cite{Robitaille:2012kg} and Freudenreich \cite{Freudenreich:1997bx} compared to IR data from COBE/DIRBE \cite{1997ApJ...480..173S}, IRAS \cite{2005ApJS..157..302M}, as studied in ref. \cite{Porter:2017vaa}. }
	\label{fig:ISRF_dist}
\end{figure}

The ISRF is modeled based on surveys of stellar populations combined with measurements of the dust and its emissivity. The first calculation combining the stellar emission with dust scattering, absorption and re-emission of radiation was performed using the FRANKIE code in ref. \cite{Porter:2005qx}. Nevertheless, until very recently, 3D spatial models including asymmetric elements (such as spiral arms or a central bulge or bar) were not used \cite{Porter:2017jsz}. The predicted emissivity distribution of ISRF from two common models are shown in~\ref{fig:ISRF_dist} for the central region of the Galaxy (galactic latitude, |b| < 5º).

The almost fully isotropic CMB does not present such problems and exhibits a perfect black-body spectrum. An interesting review of the structure of the ISRF is provided in ref. \cite{Porter:2017vaa}. 

\begin{figure}[!ptb]
	\centering
	\includegraphics[width=0.92\textwidth, height=0.23\textheight]{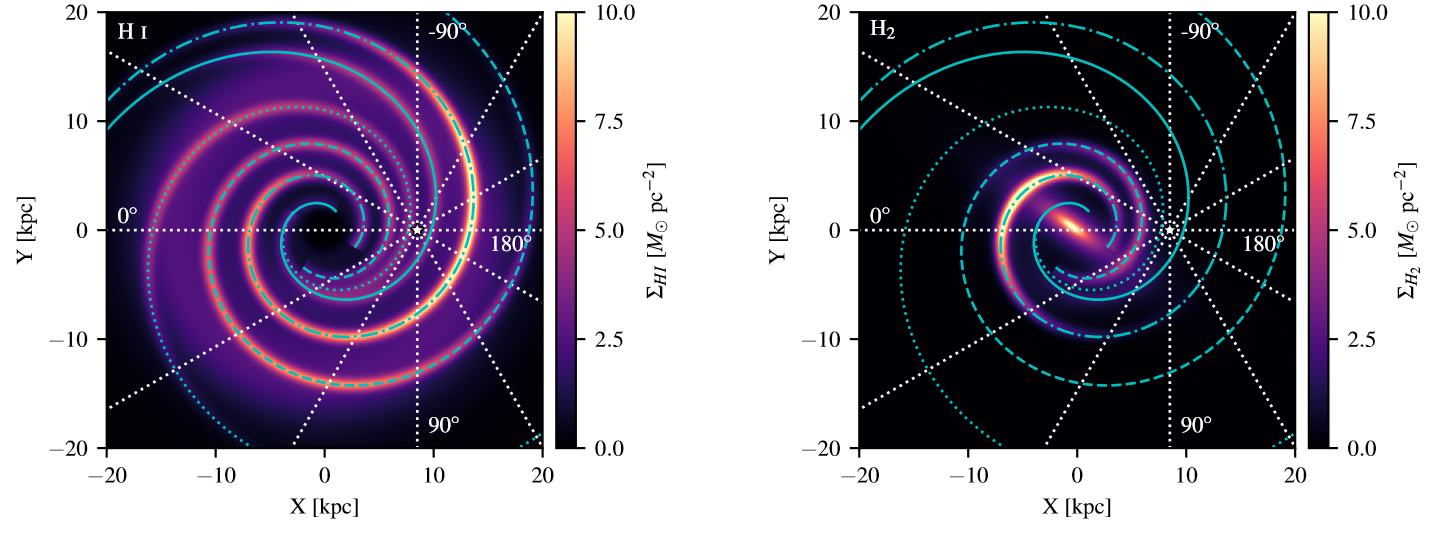}
	\caption{\footnotesize 3D distribution of HI (left) and $H_2$ (right) gas in the Milky Way derived in ref. \cite{Johannesson:2018bit}. The white dashed lines illustrate constant galactic longitude, l. Cyan curves on the density distribution depict the locations of the four spiral arms in the model. }
	\label{fig:Gas_distrib}
\end{figure}

On the other hand, models for the 3D distribution of gas in the Milky Way suffer from even larger unknowns. Hydrogen gas can be in the form of molecular gas (denoted as $H_2$), atomic gas ($HI$) and ionized gas ($HII$), each of them related to different parts of the Galactic structure: $H_2$ gas dominates the central part of the galaxy forming the Central Molecular Zone (CMZ), $HI$ is dominant in the Galactic disc and $HII$ closely follows the distribution of atomic gas, being very disperse (conforming a 10\% of the amount of $HI$ gas).
The study of $HI$ gas requires analysis of the emission line of H at $21 \units{cm}$, while $H_2$ is not directly observed, being traced from $^{12}$CO surveys and, thus, being the less reliable. Figure~\ref{fig:Gas_distrib} displays the distribution of atomic and molecular gas from the North Galactic Pole view.

Our purpose in this section is to derive gamma-ray sky maps and determine how their energy distribution changes for the propagation models studied. Concretely, we will use the propagation parameters derived for the original and scaled FLUKA, DRAGON2 and GALPROP cross sections sets. In this way, we can illustrate the difference between the maps not only for different cross sections sets used to infer the propagation parameters, but also the effect of scaling these cross sections. 

The first step consist of reproducing the electron and positron spectra, and therefore we need to start by choosing a configuration of the magnetic field and of the radiation fields. The DRAGON code allows the parametrisation of three components of the magnetic field (see \cite{pshirkov2011deriving}): one for the disc, one for the halo and a turbulent one. We set the magnetic field parameters to $B_{disc}$ = $2\mu G$, $B_{halo}$ = $4\mu G$ and $B_{turb}$ = $7.5\mu G$ as in ref. \cite{Fornieri:2019ddi}.  The energy density of the radiation fields, important for the energy losses related to IC scattering, were taken from \cite{porter2008inverse}. 

%Since the total CRE spectrum is normalized to data (at 100 GeV), changes on these magnetic field values do not affect significantly the spectrum, causing just a deformation of it around the normalization point.

The electron, positron and total CRE spectra allow us to estimate the bremsstrahlung and IC emission. %, which is the dominant emission in the diffuse gamma-ray emission below 100 MeV, as well as the IC emission.%, dominant \textcolor{red}{above 10 GeV (up to 10 TeV)}.
For this computation, the positron spectrum is explained as the sum of the positrons formed as secondaries after nuclear interactions plus the positrons injected at sources. These sources are accounted for by means of an extra-component, that emits evenly electrons and positrons, and can be included in the computation with DRAGON. We choose it to reproduce the high energy positron data of AMS-02 (concretely, at $300\units{GeV}$), following a single power-law distribution in energy. The cross sections for lepton production from p-ISM and He-ISM reactions are taken from ref. \cite{Kamae:2006bf} (hereafter Kamae cross sections). Reactions of heavier CR nuclei are not included, as their contribution to electron and positron production is supposed to be negligible, although it can be important at the level of a few percent.

\begin{figure}[!ptb]
	\centering
	\includegraphics[width=0.82\textwidth, height=0.34\textheight]{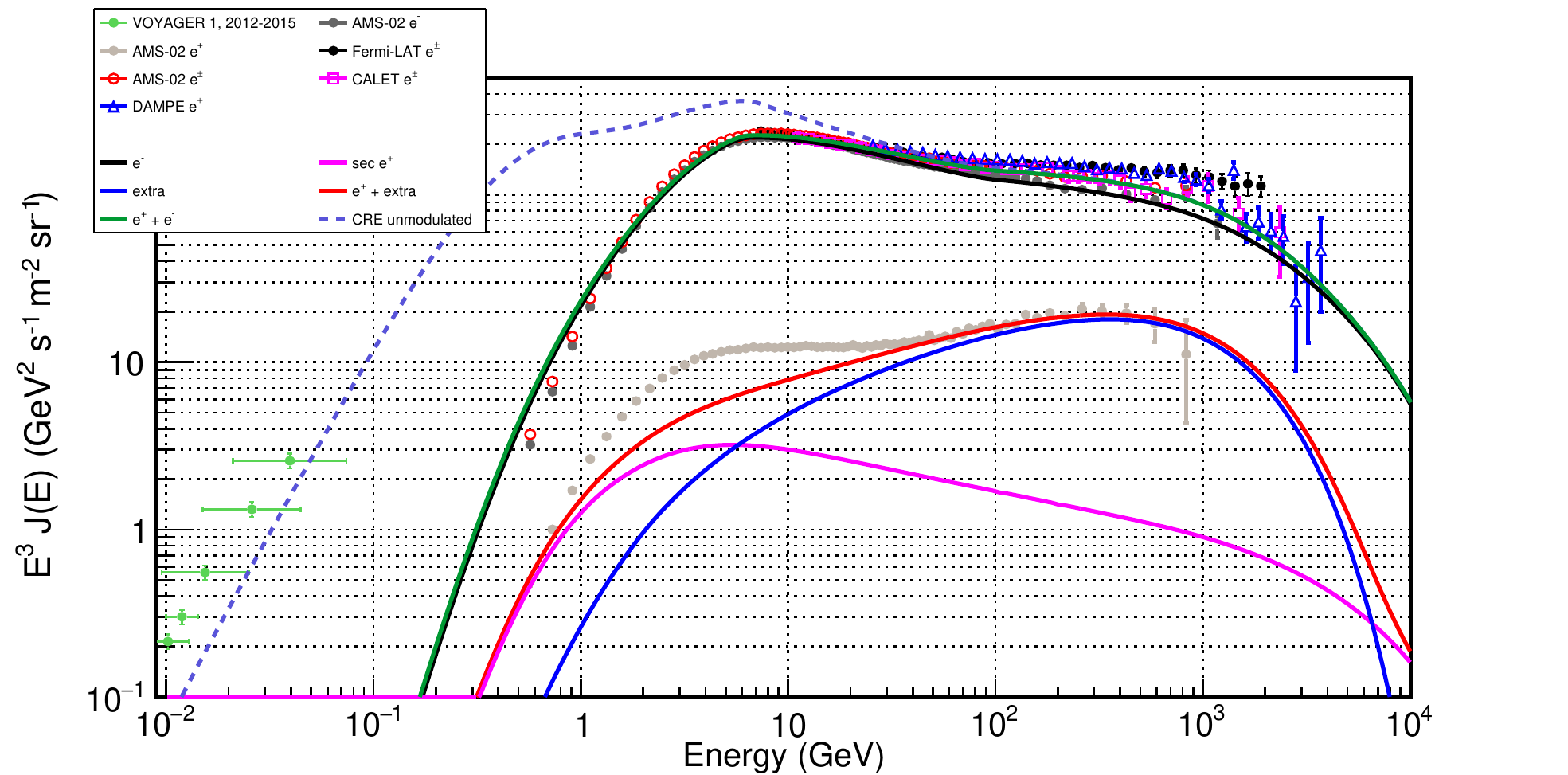}
	\caption{\footnotesize Derived electron, positron and total CRE spectra compared to AMS-02, Fermi-LAT, DAMPE and CALET data. The blue line represents an extra source, injecting evenly electrons and positrons. Also the component of secondary positrons is displayed as a magenta line.}
	\label{fig:Electrons_posit}
\end{figure}

The computed electron, positron and total CRE spectra are compared to data in Figure~\ref{fig:Electrons_posit}, for the propagation parameters inferred from the B/O analysis with the original DRAGON2 cross sections in the diffusion approximation. Differences in the propagation parameters do not induce significant differences in the spectra, but only small variations in the very low energy region of the positron spectrum.
The difficulties in reproducing the very-low energy part of the spectra are partially due to the use of the force-field approximation to account for the solar modulation. In fact, the positron fraction is time-dependent at low energies, because solar modulation affects electrons and positrons differently \cite{modul_posfraction}. Moreover, the lepton production cross sections are very important at those energies, where the contribution from the secondary production is considerable. However, a more detailed study about the diverse sources of uncertainties will be performed in a further work.

%GammaSky, a dedicated code used  to simulate diffuse γ-ray maps. This package features, among other options, the gas maps included in the public version of GALPROP (Moskalenko et al. 2002; Ackermann et al. 2012; GALPROP-web 2015). We adopt the emissivities given in Kamae et al. (2006), accounting for the energy dependence of the pp inelastic cross-section (significant above the TeV). We disregard γ-ray opacity due to the interstellar radiation field, since it is negligible up to a few tens of TeV (Ahlers & Murase 2014).

\begin{figure}[!t]
	\centering
	\includegraphics[width=0.495\textwidth, height=0.22\textheight]{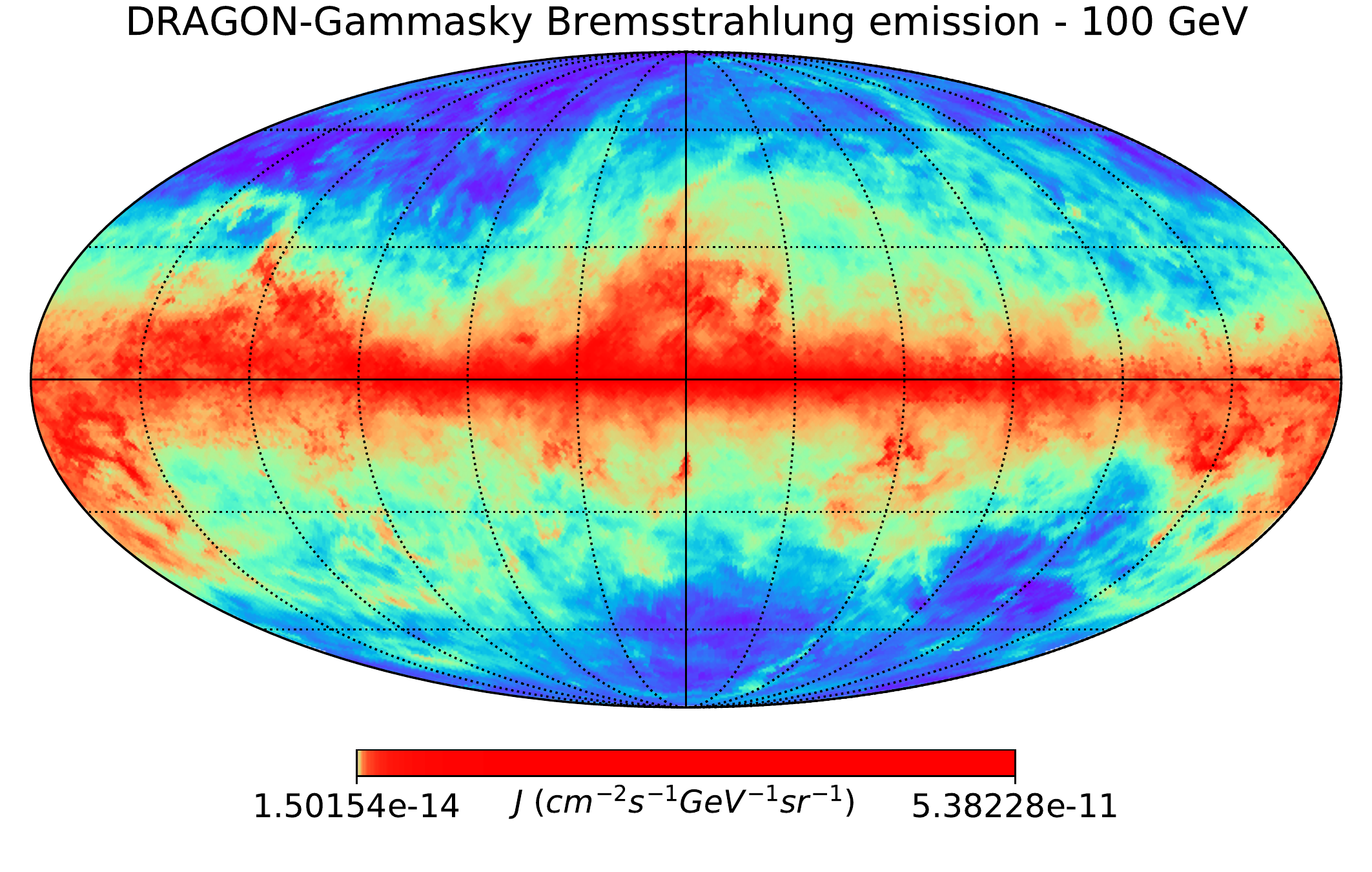}
	\includegraphics[width=0.495\textwidth, height=0.22\textheight]{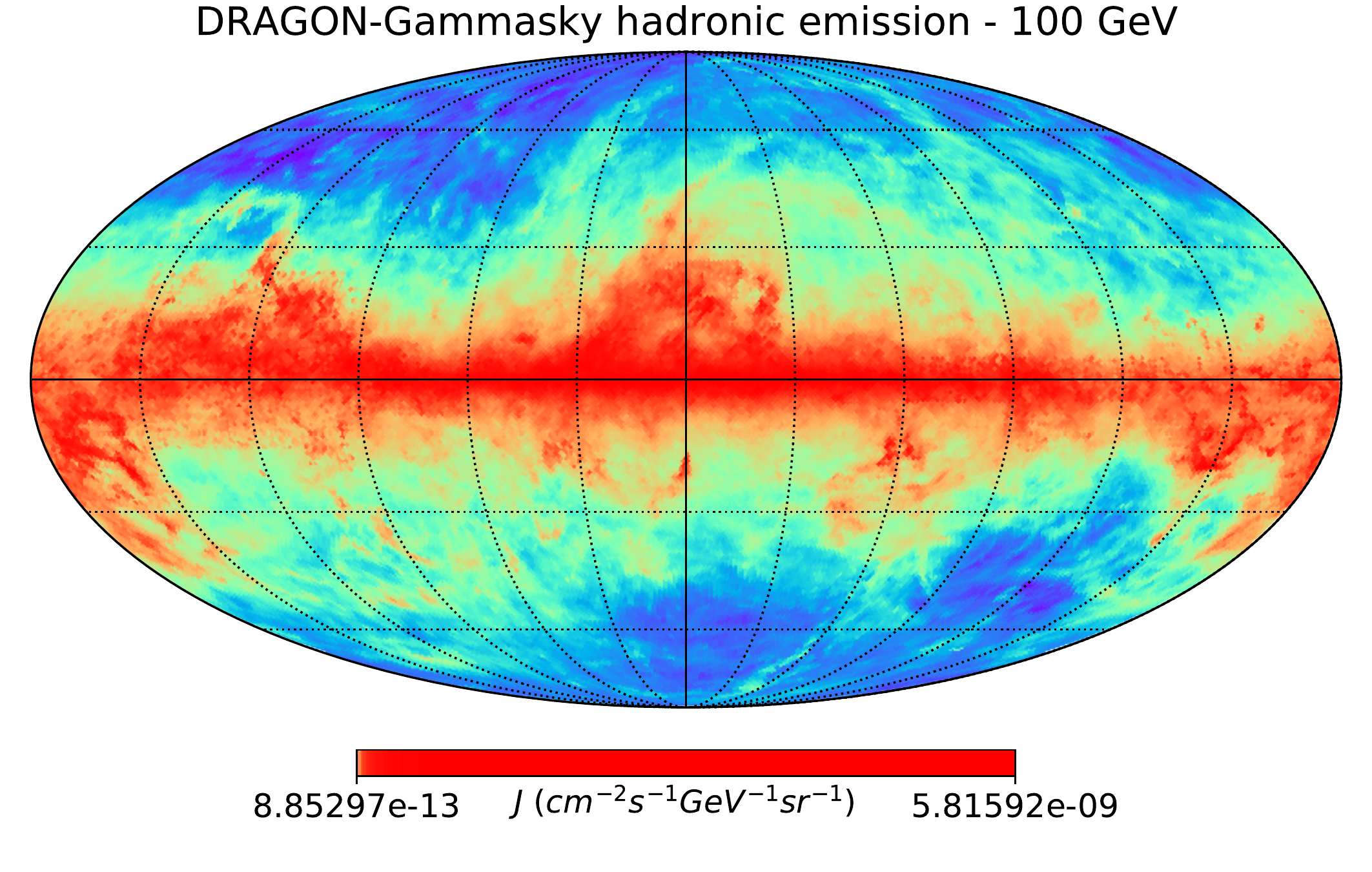}
	
	\vspace{0.35cm}
	
	\includegraphics[width=0.495\textwidth, height=0.22\textheight]{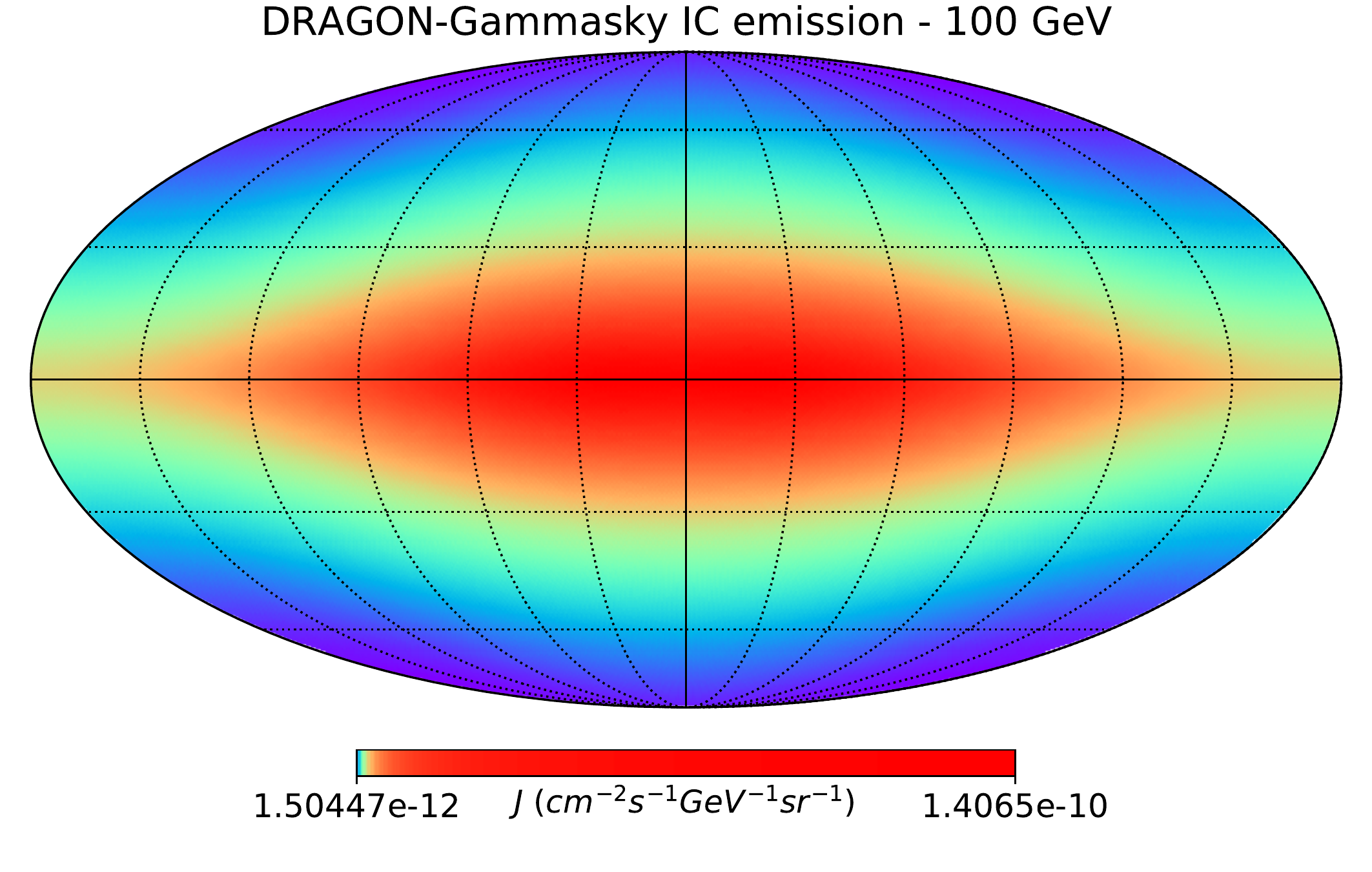}
	\includegraphics[width=0.495\textwidth, height=0.22\textheight]{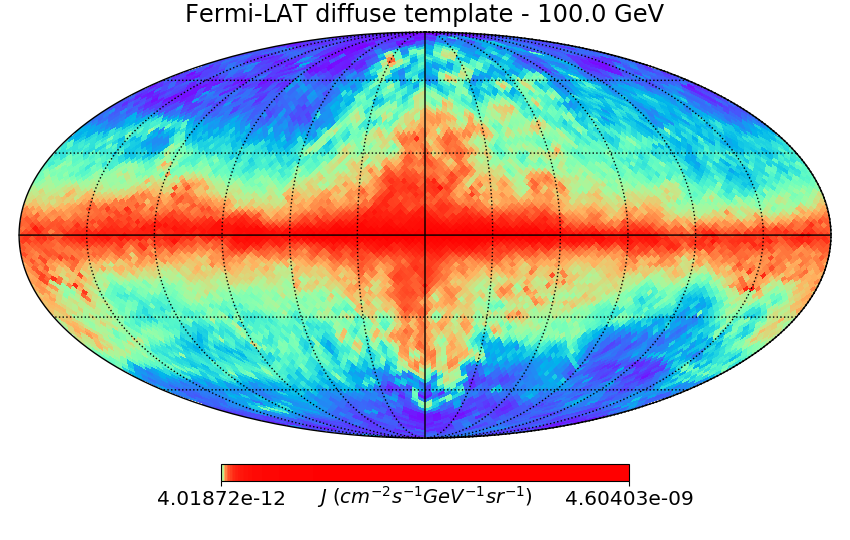}
	\caption{\footnotesize Gamma-ray intensity maps of the Galaxy, in Mollweide projection, for the propagation parameters inferred from the B/O analysis with the original DRAGON2 cross sections. The upper-right map displays the gamma-ray emission from $\pi_0$ decays, the upper-left map the bremsstrahlung emission, and the lower-left map the IC emission. The lower-right map is the Fermi-LAT template for the diffuse emission, derived from the 8 first years of operation (PASS 8 data; $P8\_R3$, taken from {\tt gll\_iem\_v07.fits} \url{https://fermi.gsfc.nasa.gov/ssc/data/access/lat/BackgroundModels.html}).}
	\label{fig:sky_maps}
\end{figure}

Once the derived CRE spectrum matches the observational data, gamma-ray intensity maps of the Galaxy can be created. In order to compute the gamma-ray diffuse emission we used the GammaSky code, a dedicated code able to calculate sky maps of various radiative processes in our Galaxy, used for a variety of different researches as in ref. \cite{Cirelli:2014lwa, Gaggero:2015xza, Mazziot}. Our computation is performed using the gas maps included in the public version of GALPROP \cite{Moskalenko:2001ya, Ackermann:2012pya}. Each of the components and the total gamma-ray intensity predicted at $100\units{GeV}$ for the propagation parameters inferred from the B/O analysis with the original DRAGON2 cross sections under the diffusion approximation are shown in Figure~\ref{fig:sky_maps}, together with the Fermi-LAT most updated template of the gamma-ray diffuse emission (lower-right plot).

Following the calculations performed in ref.~\cite{Mazziot} for the IC, bremsstrahlung and hadronic cross sections of gamma-ray production are calculated with FLUKA, %(v. 15.10.2016)
assuming an ISM composition with relative abundance that is H : He : C : N : O : Ne : Mg : Si = $1:0.096:4.65\times 10^{-4}:8.3\times 10^{-5}:8.3\times 10^{-4}:1.3\times 10^{-4}:3.9\times 10^{-5}:3.69\times 10^{-5}$.

As we can see, the hadronic emission and bremsstrahlung intensity maps mainly keep track of the gas composition, with the IC component principally probing the ISRF, which is more intense at the Galactic Center. Only the maps computed for the propagation parameters inferred from the B/O analysis with the original DRAGON2 cross sections are shown, since these distributions look very similar for any propagation model. The most recent Fermi-LAT template for the total diffuse emission {\tt gll\_iem\_v07.fits} (\url{https://fermi.gsfc.nasa.gov/ssc/data/access/lat/BackgroundModels.html}) is also displayed for comparison. In this template, the contributions from large scale structures, such as the gamma-ray Loop-I \cite{Casandjian:2009wq} and the Fermi Bubbles \cite{Fermi-LAT:2014sfa} are also added, which explains the differences observed around the Galactic Center.

\begin{figure}[!hpbt]
	\centering
	\includegraphics[width=0.49\textwidth, height=0.245\textheight]{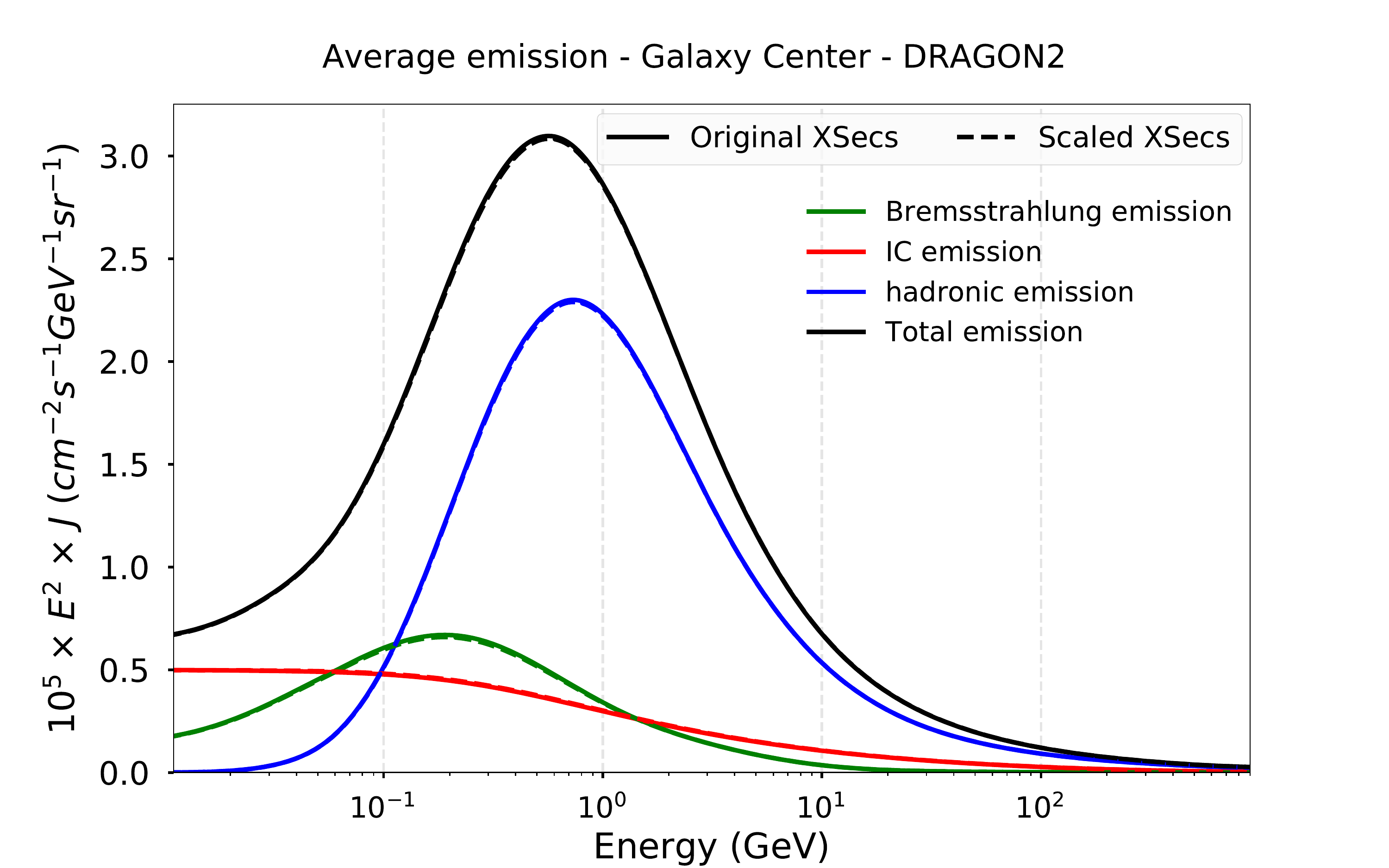}
	\includegraphics[width=0.49\textwidth, height=0.245\textheight]{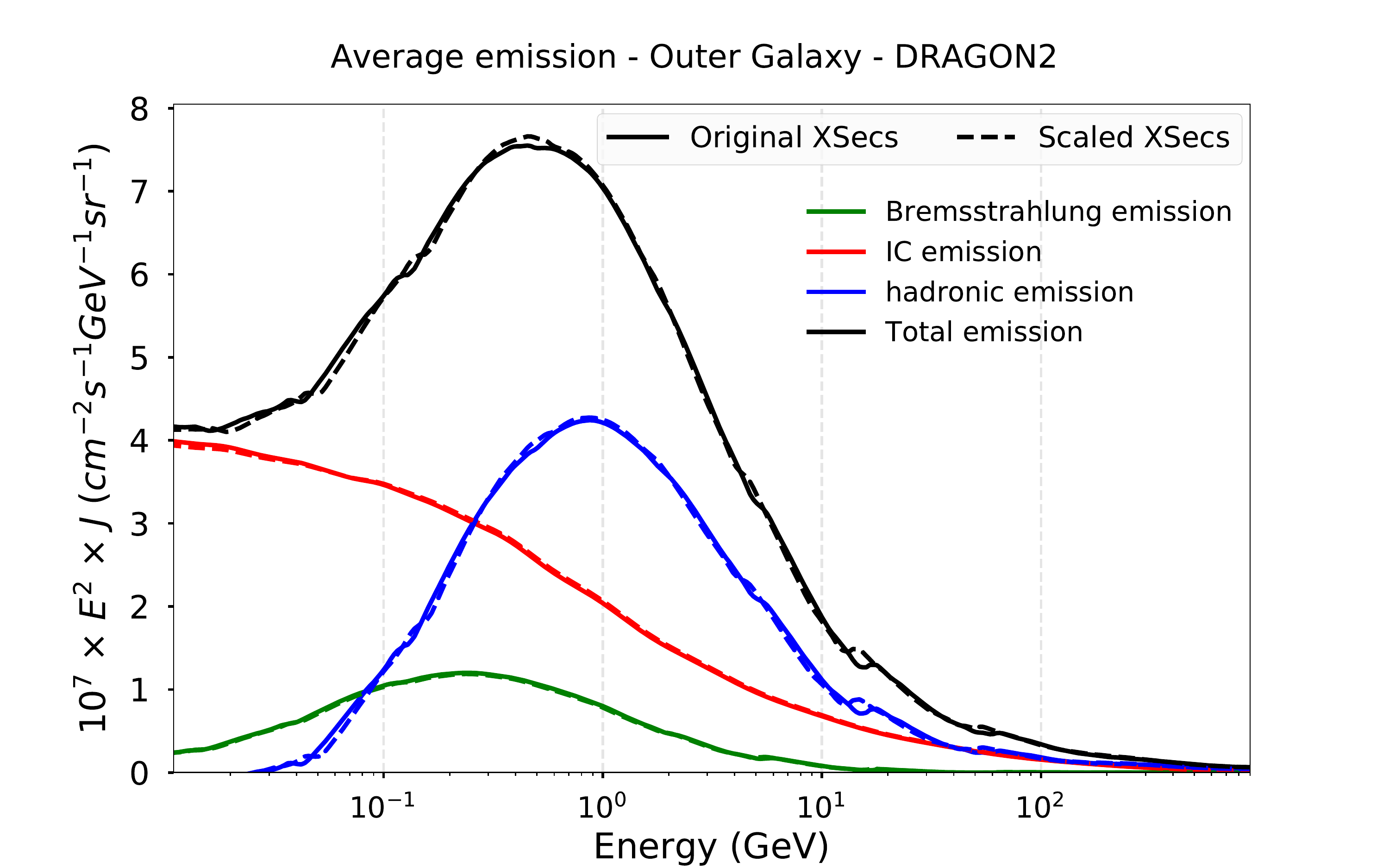}
	
	\includegraphics[width=0.49\textwidth, height=0.245\textheight]{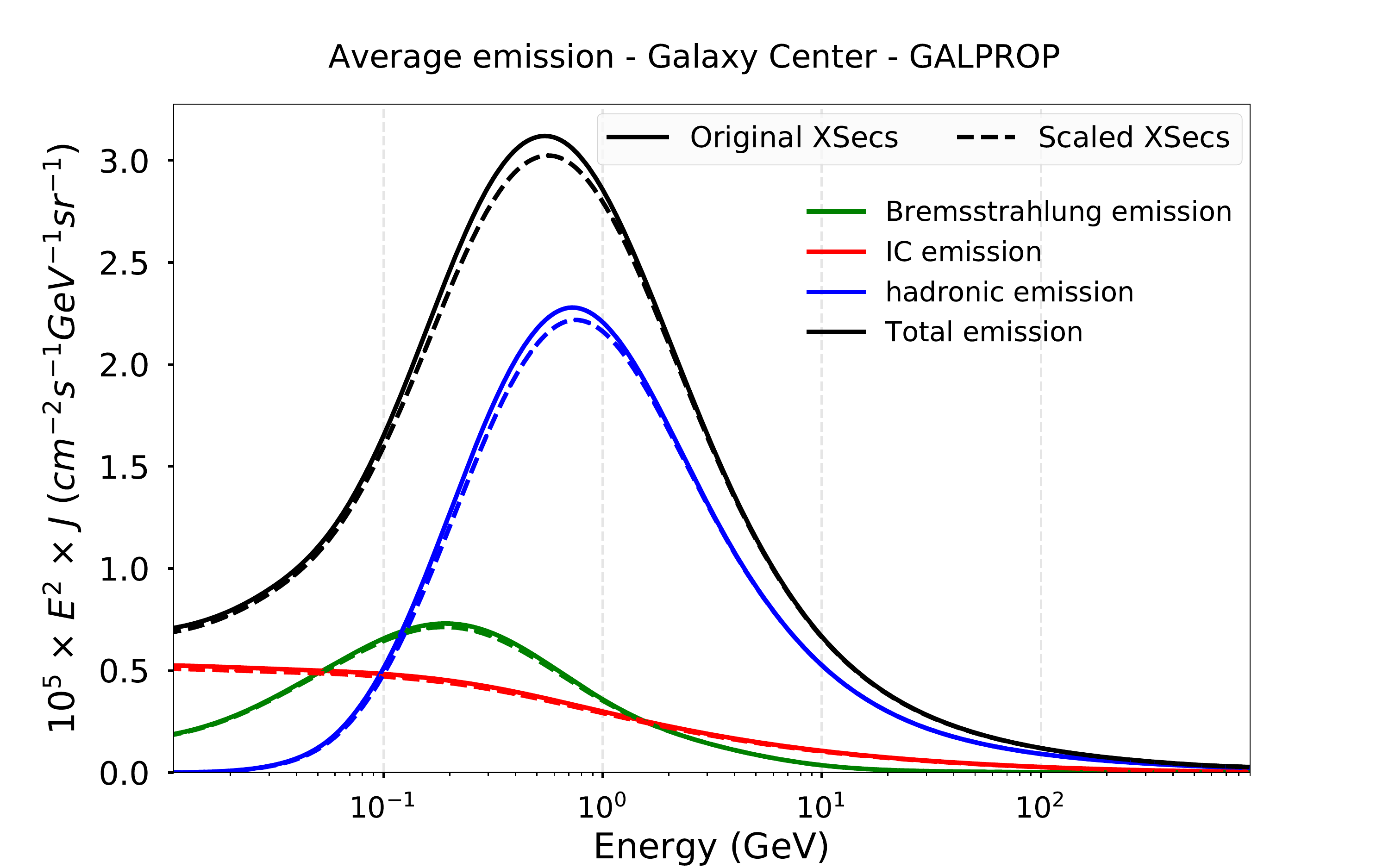}
	\includegraphics[width=0.49\textwidth, height=0.245\textheight]{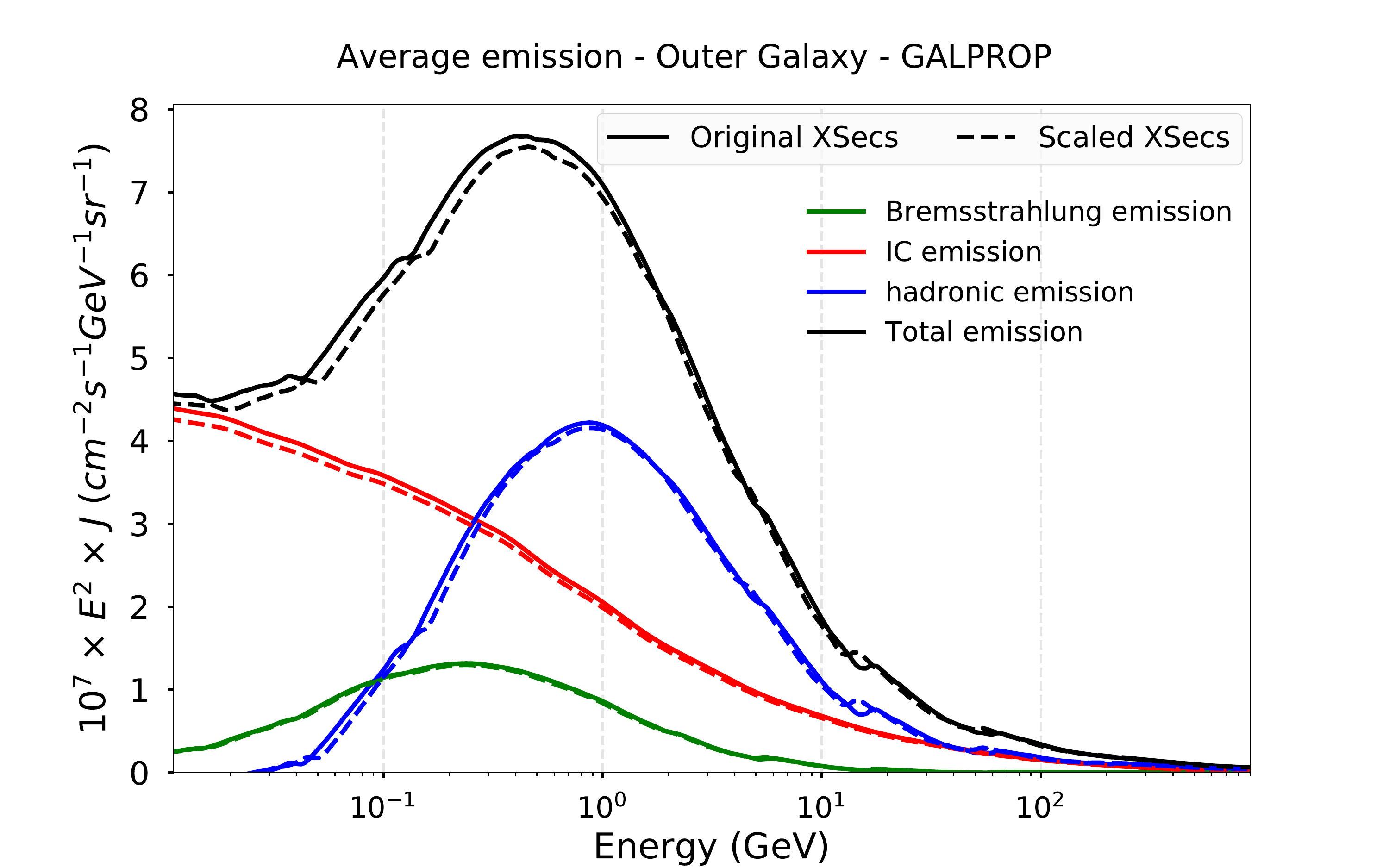}
	
	\includegraphics[width=0.49\textwidth, height=0.245\textheight]{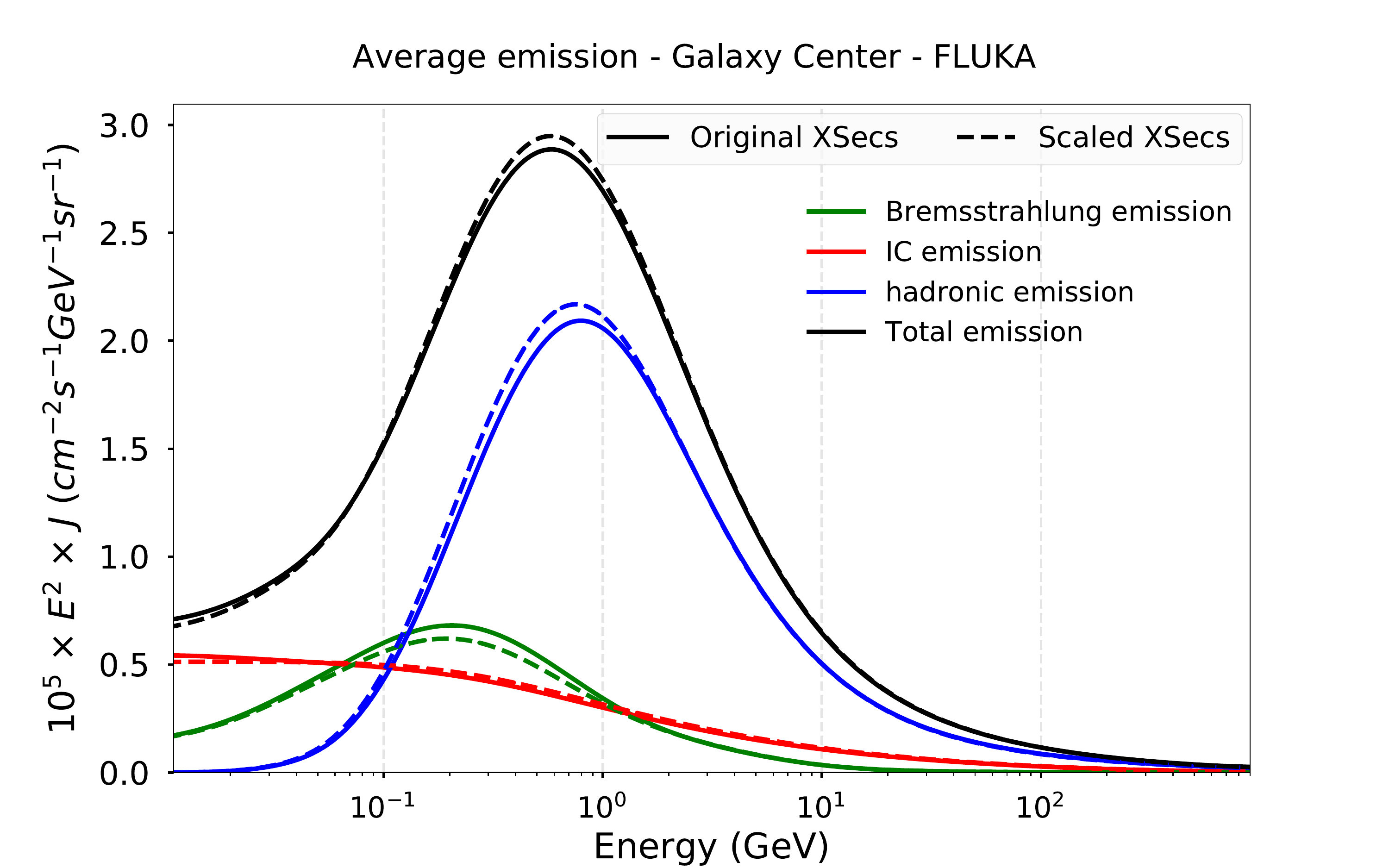}
	\includegraphics[width=0.49\textwidth, height=0.245\textheight]{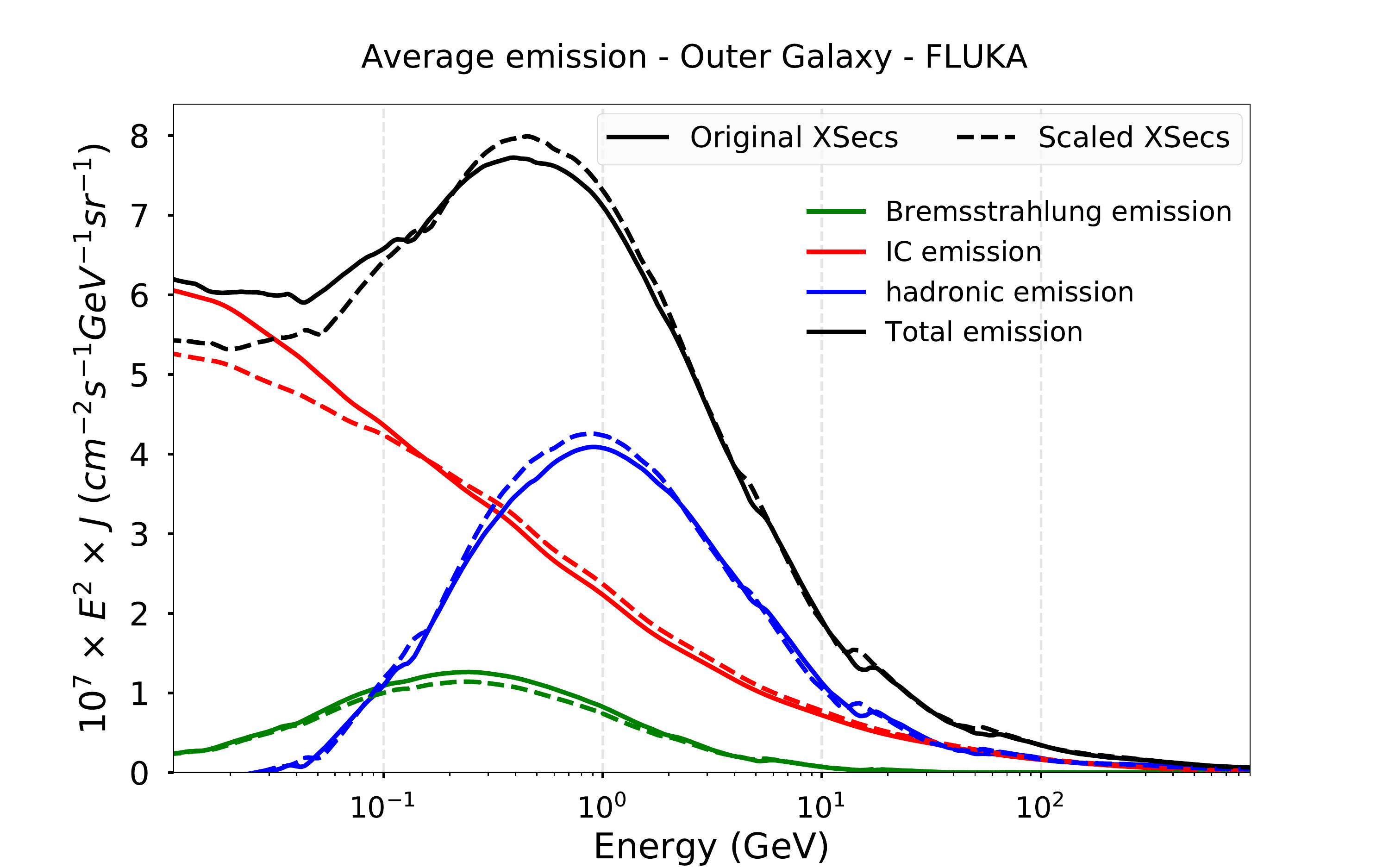}
	\caption{\footnotesize Average diffuse gamma-ray intensity spectrum for the propagation parameters derived from the B/O analysis for the original and scaled FLUKA, GALPROP and DRAGON2 cross sections in a region of 35º around the Galactic Center (left) and the emission outside this 35º region (called Outer Galaxy, at the right). Notice that for the Galaxy Center average diffuse emission the intensity is multiplied by $10^5$, while for the Outer Galaxy by $10^7$. }
	\label{fig:gamma_spectrum}
\end{figure}

In order to compare the different propagation models we have calculated the average gamma-ray intensity spectrum in a region of 35º around the Galactic Center and the emission outside this region (hereafter Outer Galaxy). These results can be seen in Figure~\ref{fig:gamma_spectrum}.

%\newpage
In these plots we can observe the pion bump signature around $0.7 \units{GeV}$ and the dominance of the hadronic emission (mainly from $\pi_0$ decay) over the other components at high energies. This Figure also illustrates the fact that different diffusion models lead to very small differences in every component of the diffuse gamma-ray emission predictions which are found in the low energy part of the spectra. Interestingly, there are no significant differences in the emission components for the different predictions, except for the IC emission in the Outer Galaxy from the diffusion model derived with the FLUKA cross sections. These differences are more relevant at lower energies, so that this can be a meaningful signature to be searched for in future gamma-ray experiments.

In addition to the gas distributions and radiation fields energy densities, the cross sections of production of pions, the bremsstrahlung and the IC processes are important here. In order to study these emissions without the repercussion of the galactic gas we will study the local gamma-ray emissivity in the next section.

In a further work, the gamma-ray extra-galactic diffuse background (which accounts for $\sim14\%$ of the total diffuse emission) and other structures like the Fermi bubbles or the Loop-I will be added to produce a complete map of the diffuse emission.

%\newpage 
\subsection{Local gamma-ray emissivity}
\label{sec:emissivity}

In order to avoid the uncertainties related to the gas and radiation fields, we turn our attention to the gamma-ray emissivity produced in the local ISM (LISM) by protons and Helium nuclei. In this context, emissivity is defined as the diffuse gamma-ray flux, coming from CR interactions with the interstellar gas per unit of gas atom. This quantity is inferred from the Fermi-LAT gamma-ray data \cite{Casandjian:2015hja} and its determination is useful to help constraining different propagation models.

The main contribution to the local emissivity at low energy is originated from Coulomb scattering of CRs with gas and electron bremsstrahlung; a further contribution, dominant above $\sim 100 \units{MeV}$, comes from the gamma-ray emissions of unstable particles formed via nuclear reactions (hadronic emissions).

Formally, this emissivity can be calculated by means of the following equation \cite{Mazziot}:
\begin{equation}
\frac{Q_S (E_S)}{n_{gas}} = 4\pi \int J(E) \frac{d\sigma(E_S | E)}{dE_S} dE
    \label{eq:emissivity}
\end{equation}
where $n_{gas}$ is the number of target gas atoms per unit volume and $\sigma (E_S | E)$ the cross section of production of any secondary particle, S, as for example neutrinos, gamma rays, positrons, etc. The ratio $Q_S ( E_S )/ n_{gas}$ is expressed in units of $\units{GeV} s^{-1}$. 

\begin{figure}[!htb]
	\centering
	\includegraphics[width=0.495\textwidth, height=0.205\textheight]{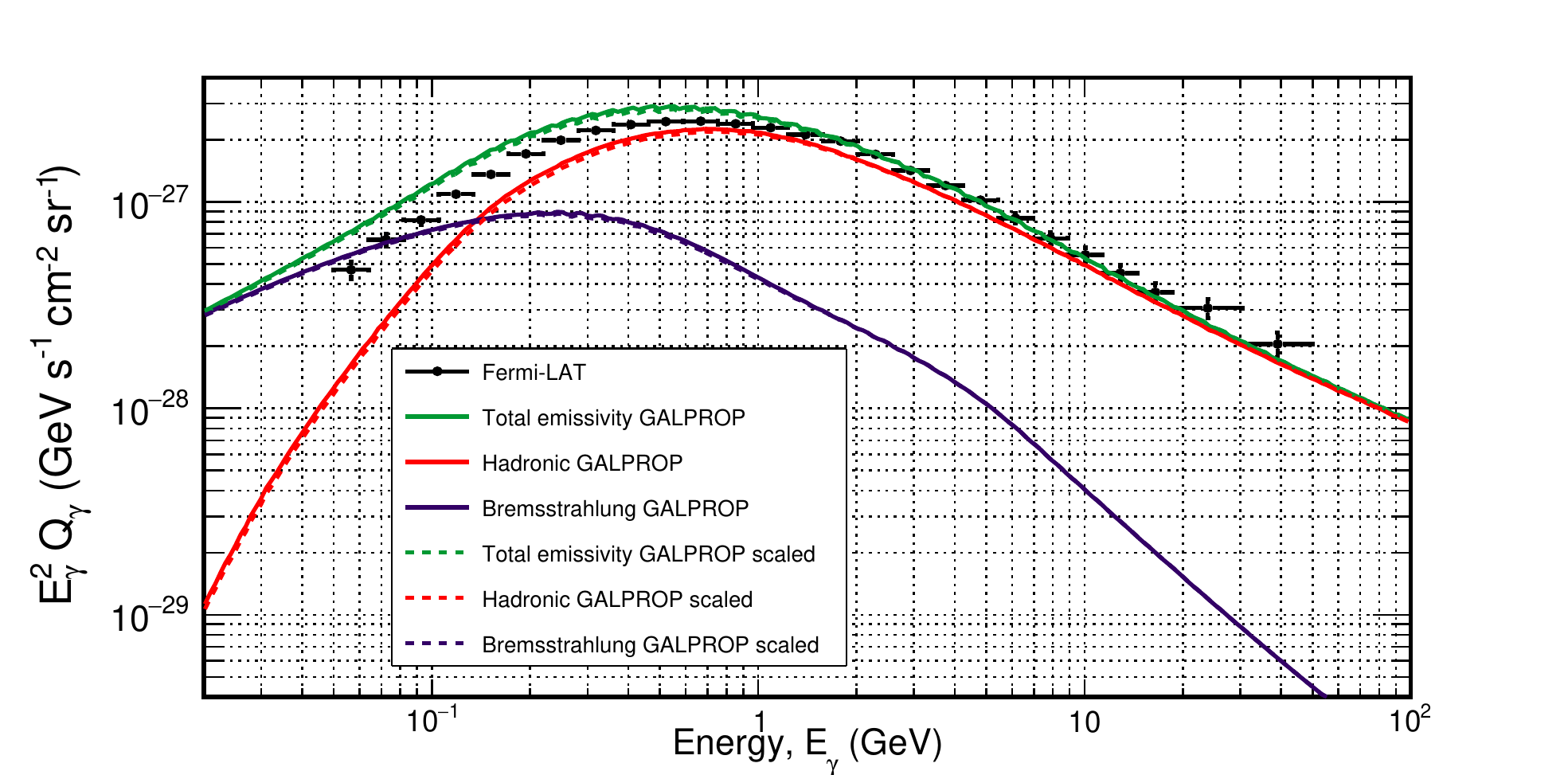}
	\includegraphics[width=0.495\textwidth, height=0.205\textheight]{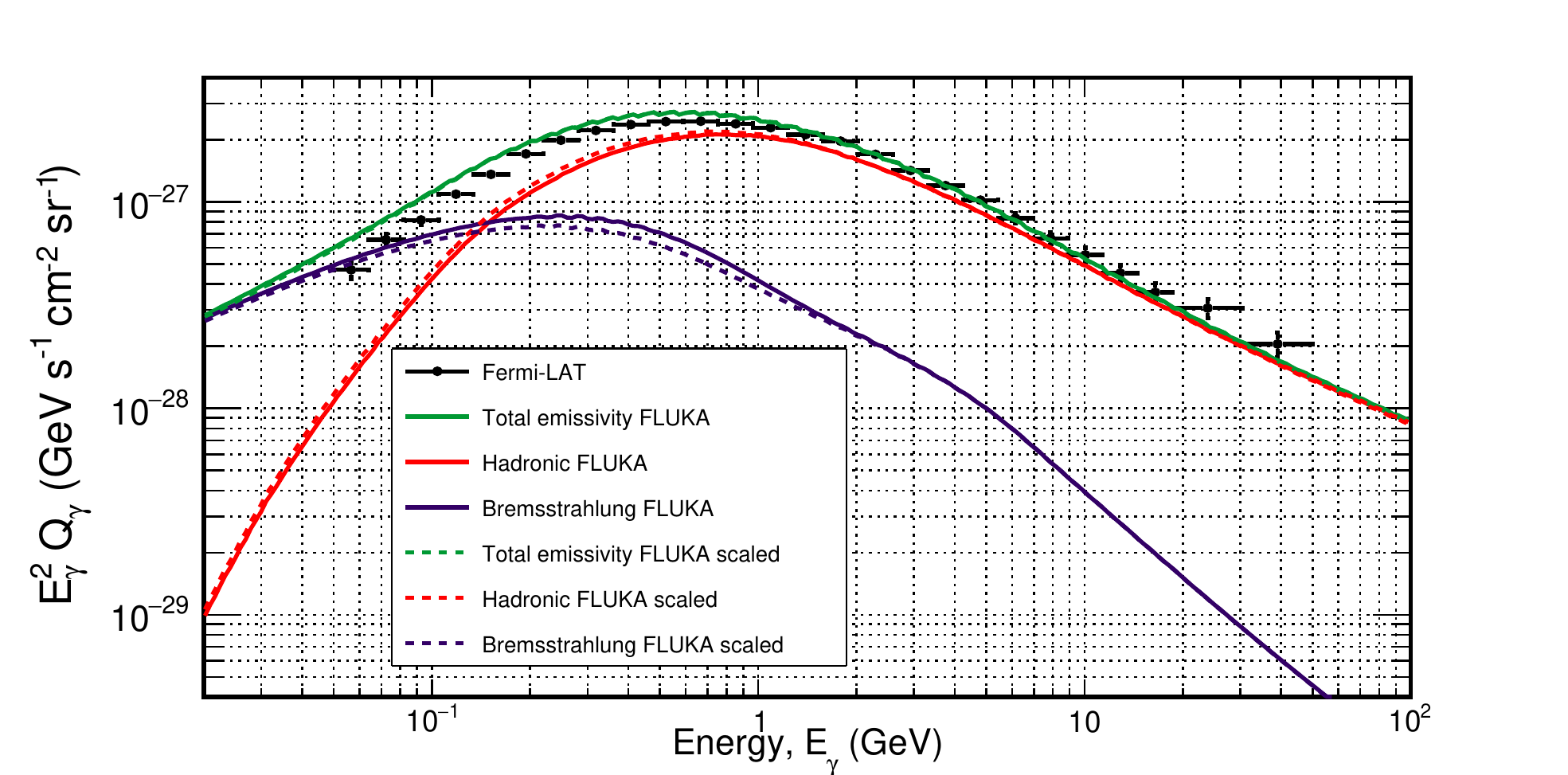}
	
	\includegraphics[width=0.52\textwidth, height=0.215\textheight]{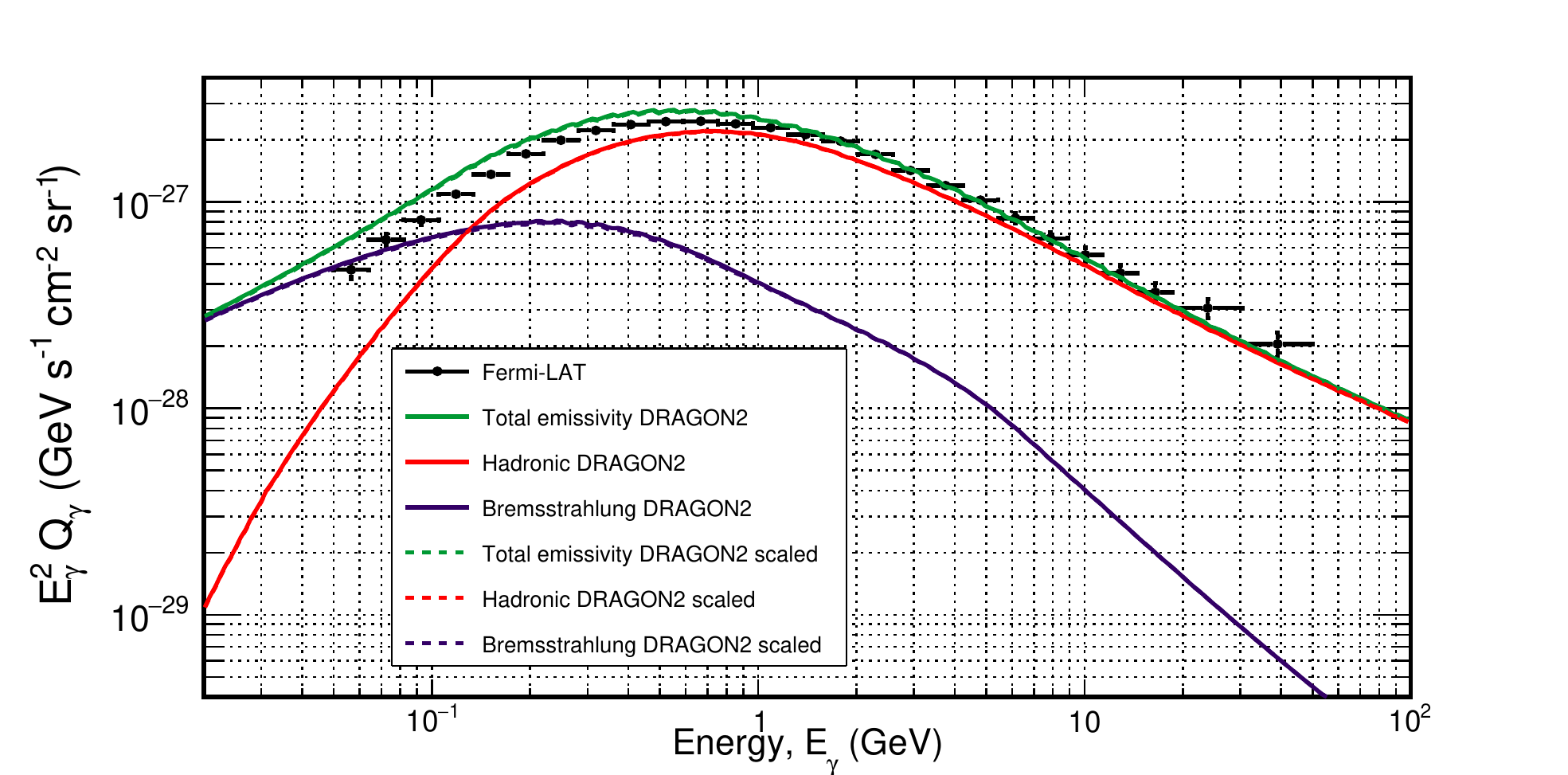}
	\caption{\footnotesize Local gamma-ray emissivity spectrum for the propagation parameters derived from the B/O analysis for the original and scaled FLUKA, GALPROP and DRAGON2 cross sections. Bremsstrahlung and hadronic emissions are also displayed to visualize the region of dominance of each component of the total emissivity. The experimental data is taken from ref. \cite{Casandjian:2015hja}}
	\label{fig:Emiss_spectrum}
\end{figure}

The results of these computations for the propagation parameters, derived from the B/O analysis for the original and scaled FLUKA, GALPROP and DRAGON2 cross sections, are shown for the diffusion coefficient of equation~\ref{eq:breakhyp} (diffusion hypothesis) in Figure~\ref{fig:Emiss_spectrum}. They are compared to the emissivity observed by the Fermi-LAT, inferred from the recent PASS8 data set\footnote{\footnotesize These are public data taken from the  \href{https://fermi.gsfc.nasa.gov/ssc/data/access/lat/BackgroundModels.html}{Fermi background models website.}}.

As we can see, all the models are very close to match the Fermi data, and they are completely consistent with data in the energy region above $1 \units{GeV}$, the energy at which the hadronic emission starts to be dominant. Below this energy, the models slightly overestimate the experimental data, which is probably due to a wrong adjustment of the CRE spectra at sub-GeV energy, difficult to be precisely tuned because of the solar modulation and of the few CR data points in this region. We see a small difference (almost negligible) between the scaled and original models, although there are slight, but significant, differences between the predictions from different cross sections sets, with the FLUKA models yielding the closest prediction. Probably this means that the low energy shape of the spallation cross sections, where nuclear resonances dominate, becomes important for this determination.

Finally, thanks to the use of the FLUKA nuclear code to calculate the gamma-ray production cross sections, we are able to study the gamma-ray emission lines coming from the hadronic interactions of CR protons and helium with ISM nuclei (even being found as traces) such as $p + p \longrightarrow \gamma + X$ or $p + ^{14}$N$ \longrightarrow \gamma + X$. These lines in the MeV region could be detected by the future space experiments like e-Astrogam \cite{DEANGELIS20181} and have been explored in very few works \cite{gamma_lines}.

\begin{figure}[!ptb]
	\centering
	\includegraphics[width=0.83\textwidth, height=0.32\textheight]{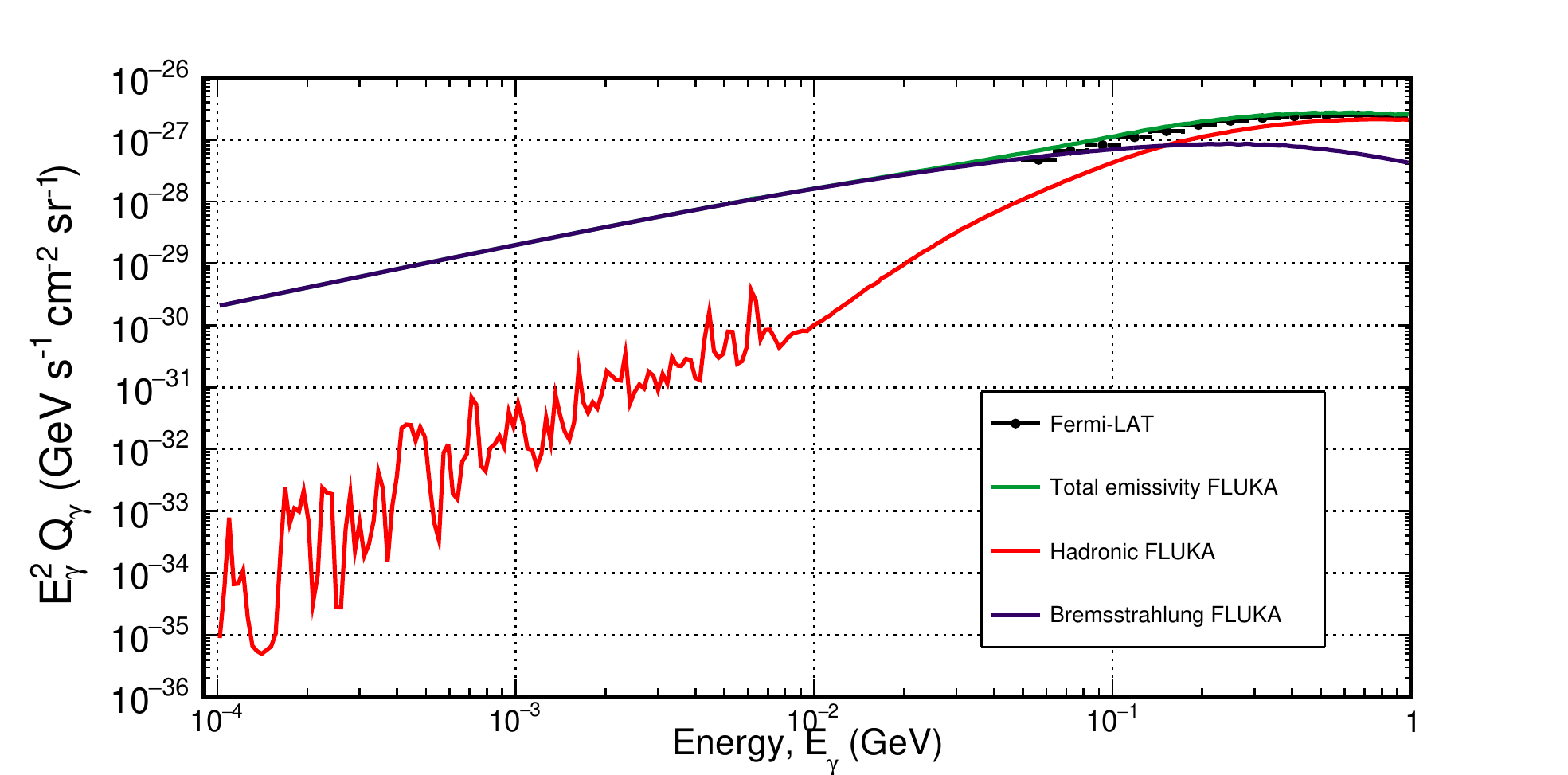}
	\caption{\footnotesize Local gamma-ray emissivity spectrum derived from the diffusion parameters obtained in the B/O analysis with the original FLUKA cross sections in the energy region between $0.1 \units{MeV}$ and $1 \units{GeV}$, showing the gamma-ray emission lines corresponding to the hadronic interactions between CR protons and helium with the ISM gas with the detailed composition previously reported.}
	\label{fig:gamma_lines}
\end{figure}

In Figure~\ref{fig:gamma_lines} we show the calculated local gamma-ray emissivity spectrum derived from the diffusion parameters obtained in the B/O analysis with the FLUKA cross sections in the energy region between $0.1 \units{MeV}$ and $1\units{GeV}$.

For a future work, a fine tuning of the low energy part of the emissivity and of the CR local spectra of protons, helium and electrons will be performed. In addition, the cross sections of gamma-ray production will be also updated with the last version of FLUKA, although no meaningful changes are expected since the changes incorporated in this last version have no direct repercussion on these calculations (see \url{http://www.fluka.org/fluka.php?id=release_notes}). In addition, thanks to the GammaSky code, this kind of analysis can be easily performed for making predictions of the neutrino emission and its distribution in the Galaxy, which is left to a further work.

\subsection{Conclusions}
\label{sec:emis_conc}
In this section, we have deeply studied the diffuse gamma-ray emissions coming from CR interactions with the ISM and interstellar radiation fields for various diffusion models derived from the different spallation cross sections sets tested along the thesis. 

We have seen that the distribution of the gamma-ray intensity in the Galaxy is mostly dominated by the gas distribution, although at low energies the radiation fields play a relevant role. We examined separately the average gamma-ray emission in the Galactic Center region and in the outer Galaxy region for the different diffusion models, finding no differences in the distributions of each gamma-ray emission component, although revealing that small differences arise just in the sub-GeV energy region. This means that, the high energy gamma-ray emissions are quite insensitive to changes in the diffusion coefficient, while at low energy we can see meaningful differences. In this average emission, the most important inputs are the gamma-ray production cross sections. Moreover, the spatial distribution of the diffusion coefficient is important in order to evaluate precise gamma-ray sky maps.

In order to study the local gamma-ray emission without the repercussion of the interstellar gas, we have also investigated the local gamma-ray emissivity, which can be compared with experimental data from the Fermi-LAT. We found that, taking into account the detailed composition of the gas in the LISM, the predicted emissivity is in good agreement with data for the diffusion models derived from the GALPROP, DRAGON2 and FLUKA cross sections sets. In particular, the diffusion models derived for the FLUKA cross sections match the low energy data with higher precision. This is probably related to the difference in the spallation cross sections in the region around a few hundreds of$\units{MeV}$, which is dominated by nuclear resonances that are substantially different for the cross sections parametrisations with respect to the nuclear Monte Carlo models used in FLUKA. In addition, a significant difference is observed at low energy for the IC emission predicted by the diffusion models derived for the FLUKA cross sections for the outer region of the sky, due to the same reason. This means that we could use gamma-ray information to better test our spallation cross sections at low energies.

Finally, the gamma-ray lines expected in the very low-energy part of the local emissivity spectrum have been shown to emphasize the importance of using a nuclear code for the calculation of cross sections, instead of relying on parametrisations limited to the data energy range.

\chapter{Summary and Conclusions}
\label{sec:6}

CRs provide the possibility to shed light on new physics in the fields of particle physics, astrophysics or cosmology. They are invaluable probes of the properties of the interstellar medium, and carry information about galactic magnetic fields, gas structures and stellar formation rates. In addition, they form the pivot point for multi-messenger studies and can provide crucial information about the elusive DM component of the universe. 

After a comprehensive introduction on the main physical and technical aspects needed to understand their phenomenology, in chapter~\ref{sec:XSecs} we have studied the predicted fluxes of the secondary CRs B, Be and Li with the most widespread cross sections parametrisations (Webber, GALPROP and DRAGON2). This comparison enables us to see %, using the diffusion coefficient that allows the fit of the boron-over-carbon flux ratio (since B is subject to less uncertainties from these cross sections) to AMS-02 data, 
that Be and Li fluxes do not match the experimental data, being the derived Be flux usually overestimated in every case. In turn, while the Webber and DRAGON2 parametrisations lead to an overestimation of the Li flux, the GALPROP spallation cross sections parametrisations underestimate it. This is widely explained by the larger uncertainty associated to the channels of Li production. %, with much less data than the others. 
At this point, the use of secondary-over-secondary fluxe ratios is proposed and investigated as a tool to directly study these cross sections, since they are roughly unaffected by the propagation parameters at high energies and offer the possibility to combine different secondary CRs in order to get rid of the systematic uncertainties associated to the spallation cross sections, mainly due to the normalization factor of the parametrisations.

In order to understand the extent of these uncertainties we have propagated the mean errors associated to experimental cross sections data into the secondary-over-secondary flux ratios of these elements. In this way we can evaluate whether the flux predictions are consistent with  experimental data from AMS-02 within the propagated cross sections uncertainties, given that recent studies claimed that the underestimation of Li is due to a primary component of Li (i.e. injected at the source) not accounted for in the analysis. Here, we demonstrate that these uncertainties account largely for this excess without the need of adding a primary component of Li. We stress that isotopic CR composition measurements would allow improving independently the cross sections of formation of different isotopes and will provide crucial insights on the origin of secondary CRs.

%we have traced two bracketing models associated to the 1 $\sigma$ mean errors of the experimental cross sections data for the main production channels of production of these secondary nuclei (namely $^{12}C$ and $^{16}O$). While channels of Be production are the most abundant in terms of data, their associated uncertainties are, in average, the lowest. In turn, in the case of Li, additionally to the scarcity of existent data, these mean errors are larger than 30\% for the channels with $^{16}O$ as projectile. Then, these uncertainties are propagated to the secondary-over-secondary ratios of these elements to evaluate if the flux predictions are compatible to experimental data from AMS-02 within the propagated cross sections uncertainties, given that recent studies claiming that the underproduction of Li flux is due to a primary component of Li (i.e. injected at the source) not accounted for in the analysis. Here we demonstrate that these uncertainties account largely for this overproduction without necessity of adding a primary component of Li.

%In fact, a fit on these ratios is achieved appropriately scaling the spallation cross sections, stating this strategy can be a very powerful tool to reduce the uncertainty coming from these cross sections in the determination of the diffusion parameters. 
Finally, the Galactic halo size value, averaged for the different cross sections used, has been determined from the Be isotopes ratios, obtaining a reasonable value of $6.8 \pm 1$ kpc. Having more and more precise data in the $1-10 \units{GeV/n}$ region would mean an important improvement on our models of propagation. At the moment, the best we can do is to combine information from different CR ratios (mainly $^{10}$Be isotope ratios and the Be/B and Be/primary ratios) and other radio, X-ray and gamma-ray measurements to obtain better constraints on the halo size value. This will improve our predictions on antiproton and lepton fluxes.

In chapter~\ref{sec:3} we have developed and fully tested the cross sections, both inelastic and inclusive, computed with the FLUKA Monte Carlo nuclear code and its compatibility with CR data. First, inelastic cross sections have been compared to experimental data and the CROSEC parametrisations, finding good agreement with data and the parametrisations up to Ne ($Z=10$), when the differences start to be important (although, with no important implications on CR computations). The effect of inelastic cross sections changes, propagated to the fluxes of secondary CRs, have been also assessed, finding expected uncertainties from these cross sections of around 2\% at high energies. 

Furthermore, the direct spallation cross sections have been compared to data, finding a good agreement in general, except for the $^{9}$Be isotope, which shows significant differences with data in various interaction channels. These direct cross sections are then compared to the cumulative ones in order to spot those production channels with an important contribution of ghost nuclei, finding that the most important contributions come from $^{10}$C and $^{11}$C and, probably, $^{9}$Li. On the other hand, the FLUKA cumulative cross sections were compared to GALPROP, Webber and DRAGON2 parametrisations, finding a general overall agreement, except in the case of $^{9}$Be production, and remarking the fact that the $^{10}$Be presents large differences between the different parametrisations. Also a small, but important, systematic underestimation of the B isotopes production cross sections was pointed out.

The inclusive and inelastic cross sections were implemented in the DRAGON code to show that, using these cross sections, the spectra of every element can be reproduced as well as the secondary-over-primary flux ratios (after adequately tuning the diffusion coefficient). Both, a comparison of the predicted secondary-over-secondary flux ratios to data and a comparison of the predicted fluxes for a fixed diffusion coefficient between the DRAGON2 parametrisations and the FLUKA cross sections lead to the conclusion that the cross sections of B production require larger scaling than those of Be and Li production, probably larger than 10\%. Finally, the discrepancies on the production of Be isotopes lead to a very large prediction of the halo size value, of $16 \units{kpc}$.

Chapter~\ref{sec:4} has been dedicated to a detailed study of the diffusion parametrisations and the diffusion parameters needed to reproduce data, by means of a Monte Carlo algorithm performed individually on the B/C, B/O, Be/C, Be/O, Li/C and Li/O flux ratios and applied to their combined analysis including scaling factors for their production cross sections. These analyses have been performed for the FLUKA, DRAGON2 and GALPROP cross sections sets using the diffusion coefficient parametrisation from the 'Source' and 'Diffusion' hypotheses. In this chapter, we have demonstrated that this combined analysis is affected by the uncertainties associated to the overall shape of the production cross sections (mainly the Li production cross sections), which influence our determination of the propagation parameters leading to a bias more pronounced in the GALPROP cross sections. In turn, the DRAGON2 cross sections, although still affected by the uncertainties in the Li production cross sections, reproduces the flux ratios within the experimental uncertainties of the AMS-02 data.

To prevent from this bias, a two-step analysis, consisting of scaling first the cross sections to reproduce the high energy part of the B, Be and Li secondary-over-secondary flux ratios, and then applying the MCMC analysis to each secondary-over-primary flux ratio has been carried out. From this analysis, roughly the same diffusion parameters have been found for the ratios of B and Be species with the DRAGON2 cross sections, in addition to scaling factors smaller than 5\%, while the Li ratios require slightly different diffusion parameters. On the other hand, for the GALPROP cross sections, the predicted diffusion parameters are significantly different for B, Be and Li, indicating that more degrees of freedom are needed to adjust the shape of their cross sections. These analyses lead to a $\delta$ value around 0.44, in agreement with magneto-hydrodynamic simulations. Provided that we have reached experimental precision at the level of $<5$\% in CR measurements, novel studies like those presented here are crucial to identify the characteristics of CR propagation and avoid being limited by the cross sections measurements precision.

The diffusion parameters inferred for the three cross sections sets studied and from the three kind of analyses performed in chapter~\ref{sec:4} have been applied to test their predictions for reproducing the gamma-ray and antiproton fluxes.

Generally, the antiproton predicted fluxes tends to an underestimation with respect to the AMS-02 data for the tested diffusion models. In addition, a prominent feature is found for all the models around $10 \units{GeV}$, which is coincident with a possible signature of a DM particle recently reported. Nevertheless, when studying these models with another parametrisation of $\bar{p}$ production cross sections, this feature disappears, which may indicate that it arises from these cross sections and the large uncertainty related to them. Here, we have also demonstrated that antiprotons could help constraining the propagation parameters, as they show that the combined analysis models produce very vague predictions. Finally, the total uncertainties are derived, including for first time the uncertainties associated with the cross sections of secondary CR production. This source of uncertainty can account for even more than 10\% of the $\bar{p}$ flux when considering the total error bars in the cross sections data of the main channels. Finally, we estimate a conservative 1$\sigma$ uncertainty at 
$10\units{GeV}$, that can be of 27\%,  without including neither the correlated errors of the AMS-02 experiment nor the total uncertainties related to the production cross sections.

Finally, gamma-ray sky maps have been produced and the Galactic gamma-ray intensity and local gamma-ray emissivity have been studied with the different diffusion models previously derived. These computations show no important differences between the predictions from the different diffusion models, except for the low energy part of the emissivity, which is significantly lower for the models derived from the FLUKA cross sections, and the averaged Inverse Compton intensity in the outer part of the Galaxy, which is larger for the model derived from the FLUKA cross sections. In conclusion, we see that the average gamma-ray emission and emissivity are rather insensitive to the diffusion coefficient used at high energies, while it can significantly affect them at very low energy. This is meaningful, since possible signals of dark matter can be revealed from gamma rays and all sources of possible uncertainty must be studied.

%\newpage
\section{Plans for future research}
\begin{itemize}
	\item A straightforward extension of the MCMC analyses consists of incorporating the Galactic halo size value. In fact, particularly interesting is the analysis including also this parameter together with the scaling factor of the spallation cross sections combining the ratios of B and Be, and eventually the $\bar{p}$/p ratio too. At this point, we could also investigate the possibility of including any extra degree of freedom for the $\bar{p}$ production cross sections in form of nuisance parameters. This is something not already done and can help to improve the precision of our predictions.
	
	\item Integrating a more dedicated code for the simulation of the solar modulation effect would be very interesting, specially for the predictions on the gamma-ray production at low energies. Adding charge-sign dependent modulation to the computation of the electron spectra will be a first step before implementing a fully dedicated code, which is very time consuming.
	
	\item An undergoing work involving the theoretical calculation of the spatial diffusion coefficient taking into account magnetosonic modes in the interstellar plasma in addition to alfvenic modes will offer a unique opportunity to improve our diffusion coefficient parametrisations and extend them to energies not yet covered by data. As a preliminary result, we observe that fast magnetosonic modes are dominant at high energies and that the diffusion coefficient calculated including these modes does not follow a perfect power-law in energy.
\end{itemize}

%=====================================================================
% APPENDIX
%  Appendices, if any, must precede the cited literatures.
%  Appendices shall be numbered in Roman Capitals (e.g. Appendix IV)

\begin{appendices}
	\chapter{Cross sections parametrisations in the main channels}
\label{sec:appendixA}

In this appendix we are showing a comparison between the different cross section parametrisations used in this study and the experimental data in the most important reaction channels for the light secondary CRs Li, B, Be. Experimental data are taken from various experiments and authors: some can be found in EXFOR (Experimental Nuclear Reaction Data)\footnote{\url{https://www-nds.iaea.org/exfor/exfor.dhtm}} others in the GALPROP database of cross sections ($isotope\_cs.dat$) and the rest come from various publications and experiments (Bodemann1993, Davids1970, Fontes1977, Korejwo1999, Korejwo2002, Moyle1979, Olson1983, Radin1979, Read1984, Roche1976, W90, W98a and Zeitlin2011). This data is available upon request. More information can be found in section 5 of \cite{Evoli:2017vim} and mainly in the appendix of \cite{Evoli:2019wwu}.
\newpage
\vspace{0.5cm}
\textbf{\small Secondary nuclei from $^{12}$C}
%\vspace{1cm}
\begin{figure*}[!hbt]
\begin{center}
\includegraphics[width=0.42\textwidth,height=0.19\textheight,clip] {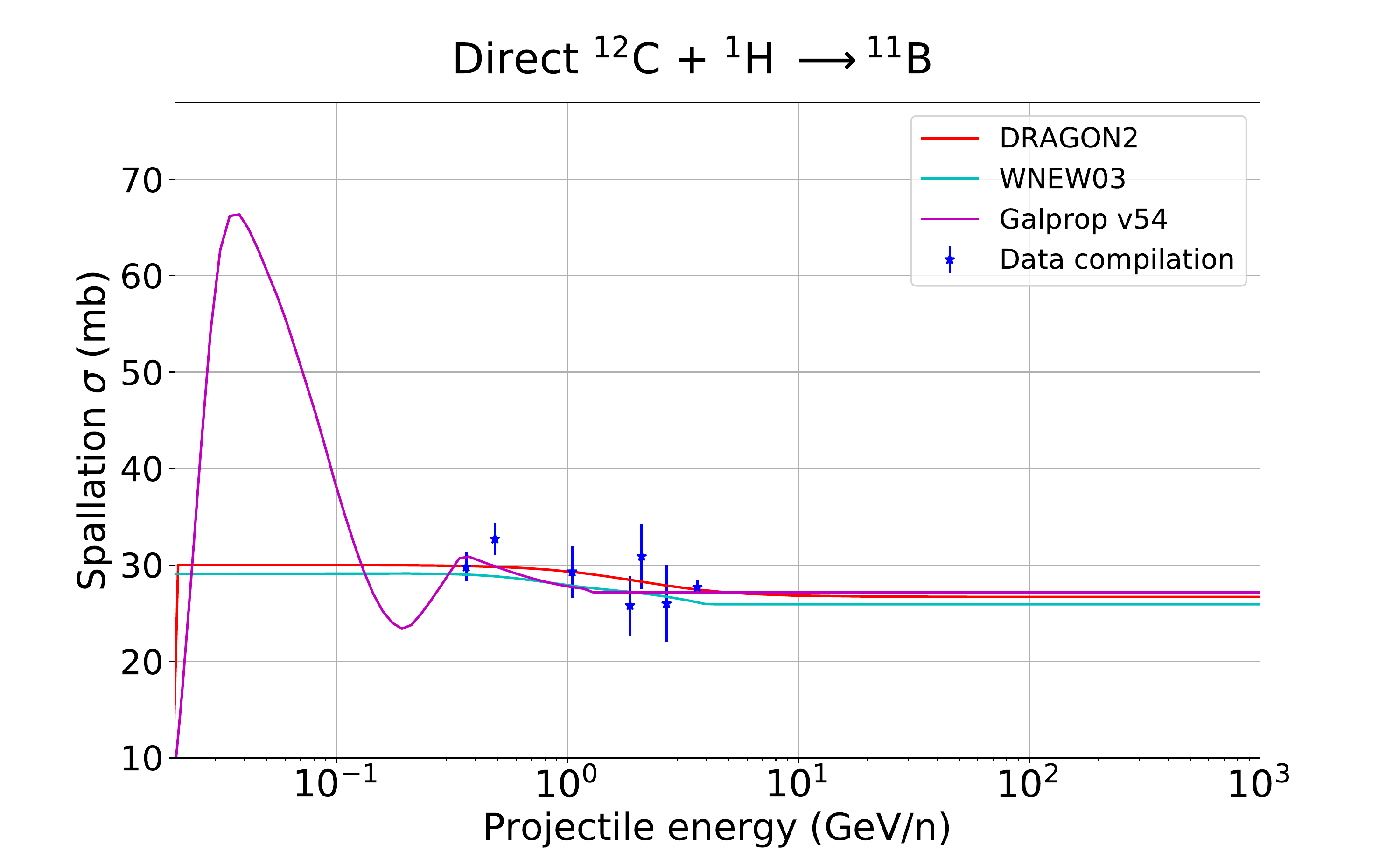} \hspace{0.5cm}
\includegraphics[width=0.42\textwidth,height=0.19\textheight,clip] {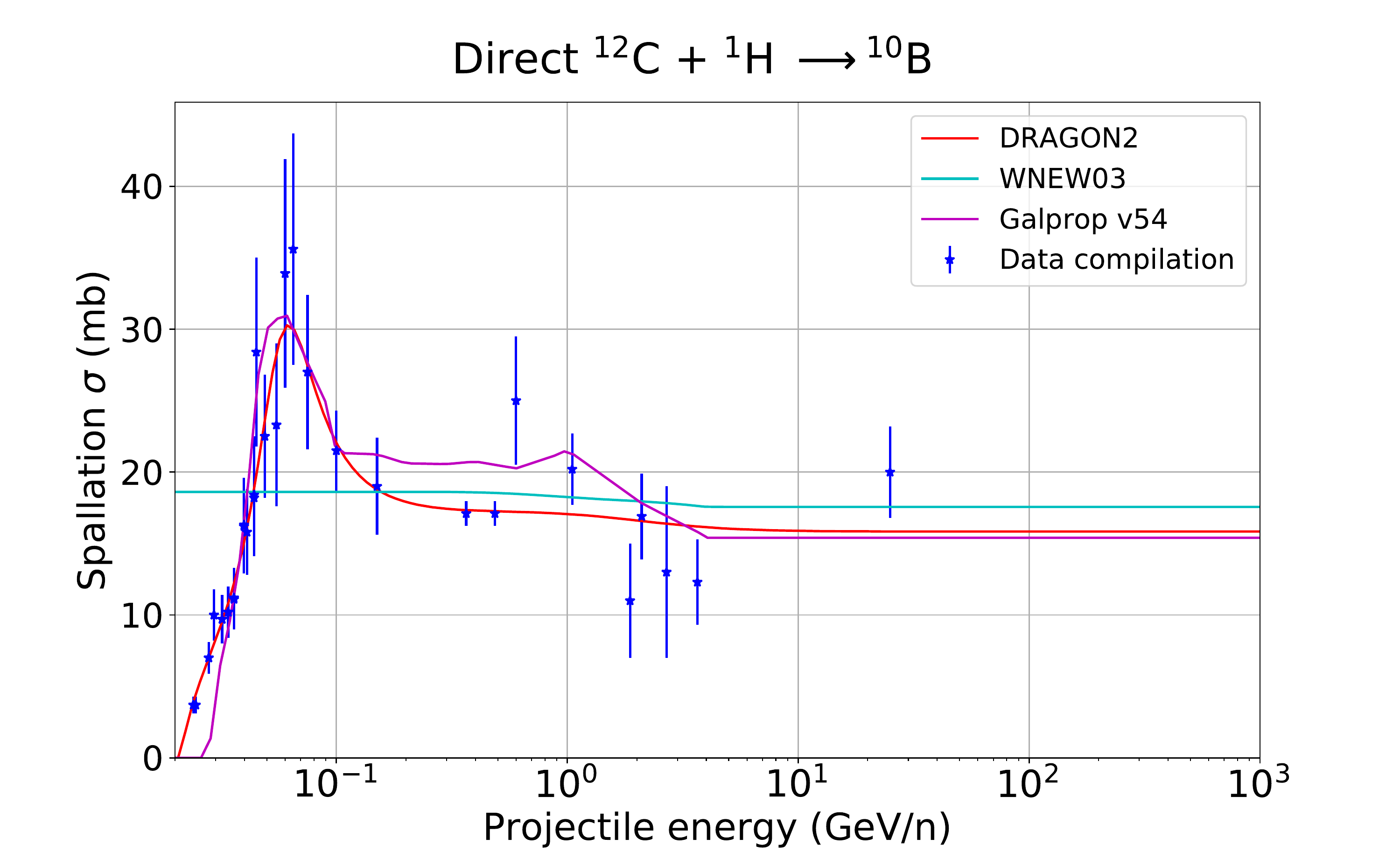} 
\vspace{0.5cm}

\includegraphics[width=0.325\textwidth,height=0.172\textheight,clip] {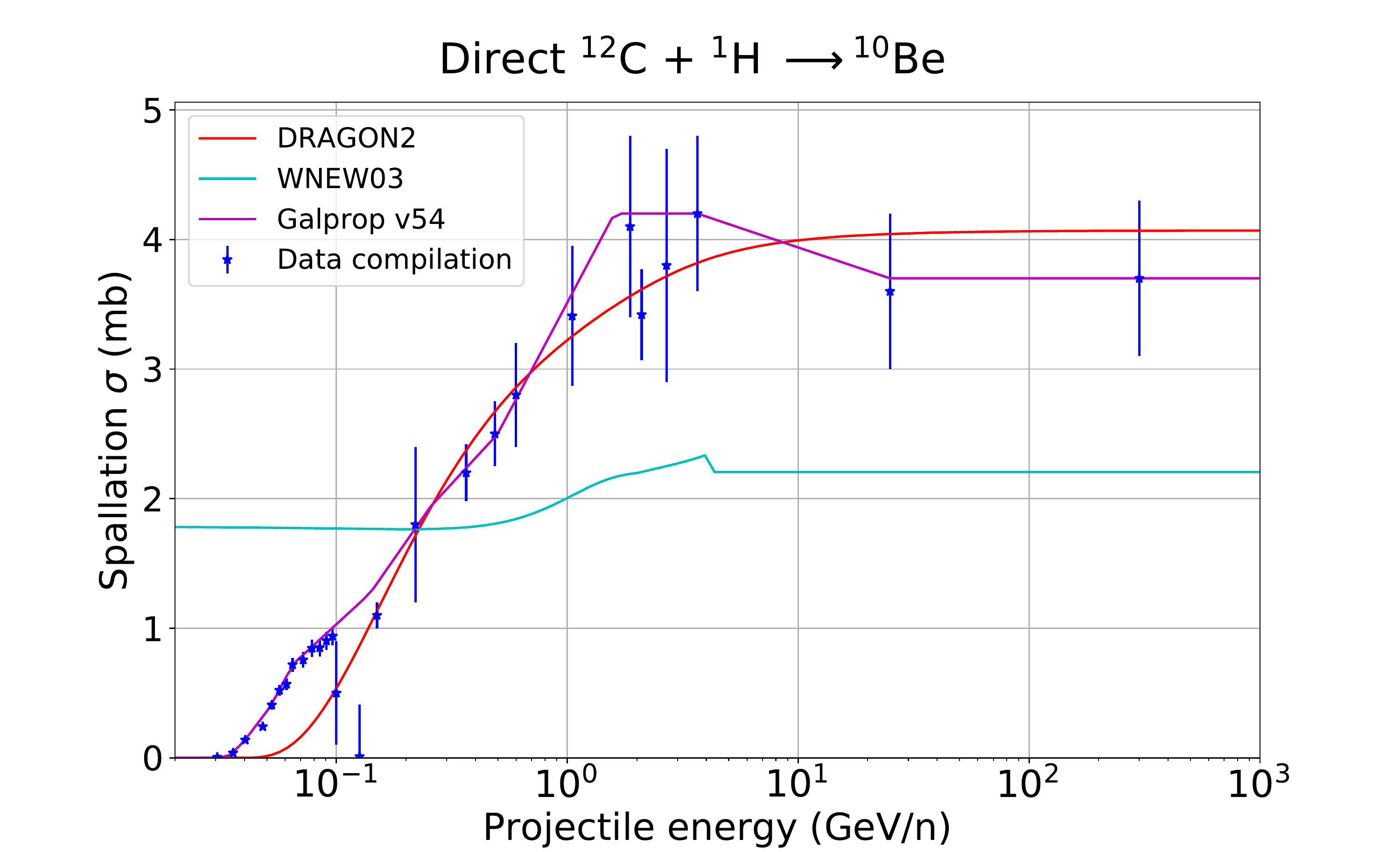} %\hspace{0.5cm}
\includegraphics[width=0.325\textwidth,height=0.172\textheight,clip] {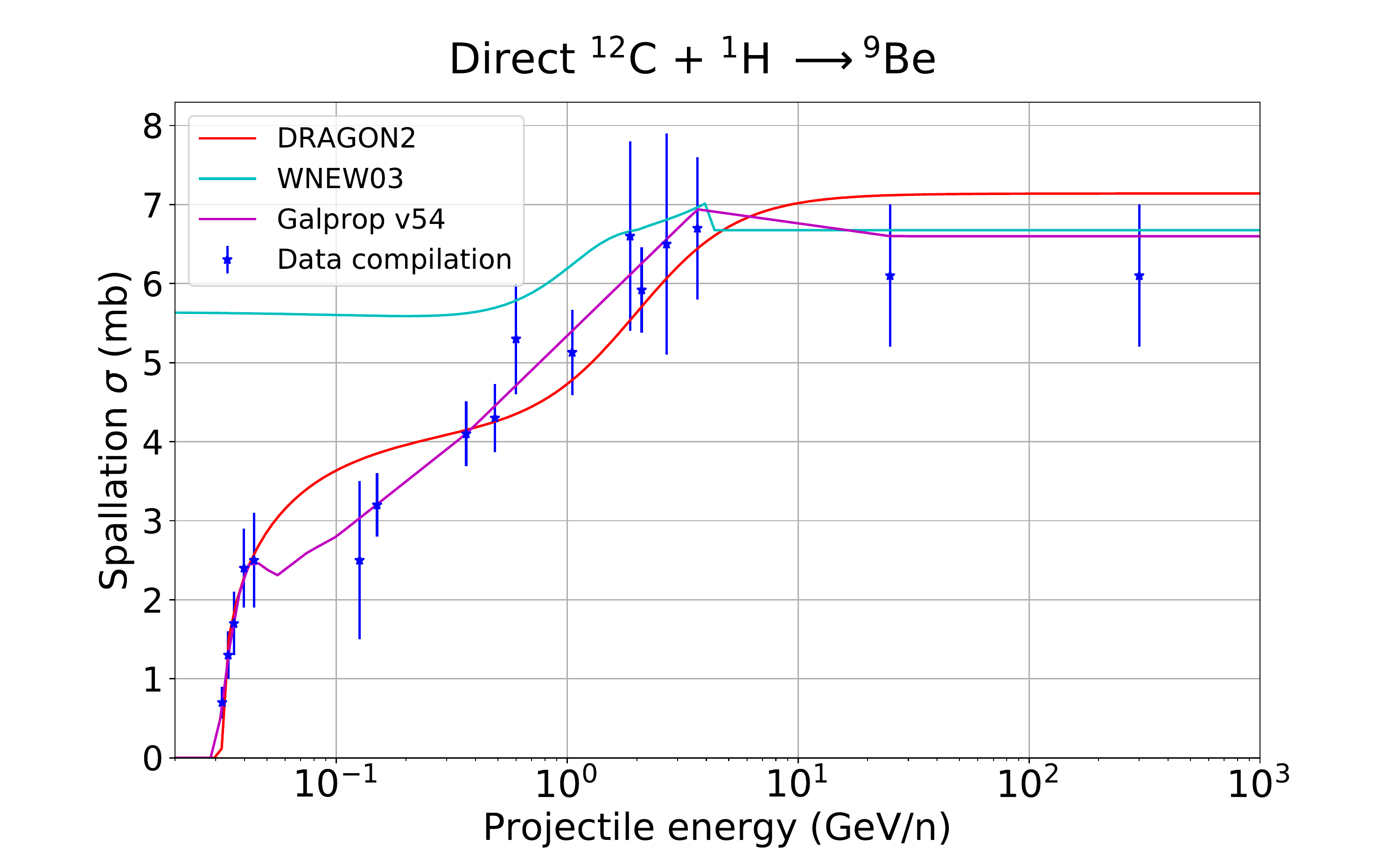}
\includegraphics[width=0.325\textwidth,height=0.172\textheight,clip] {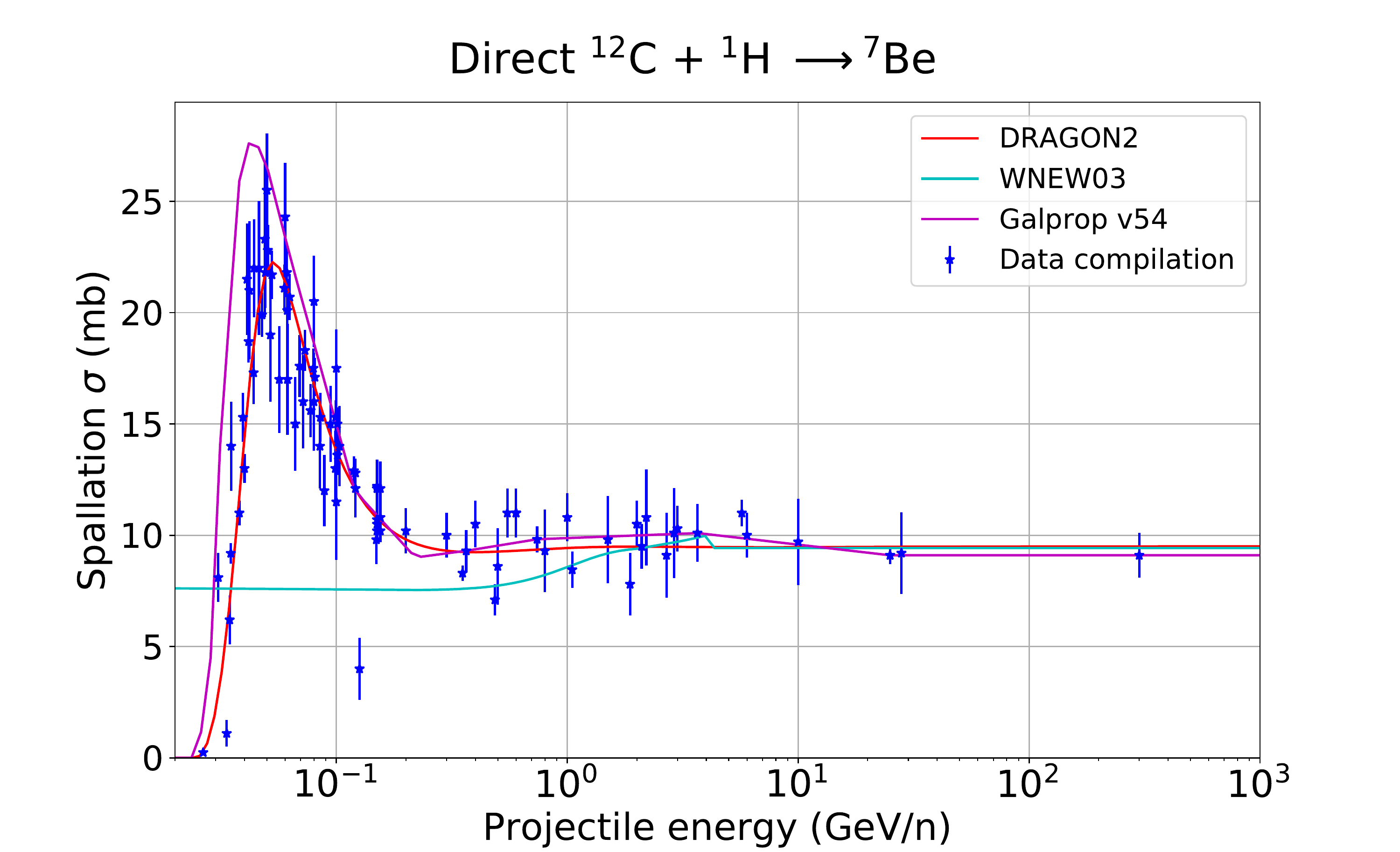}
\vspace{0.5cm}

\includegraphics[width=0.42\textwidth,height=0.19\textheight,clip] {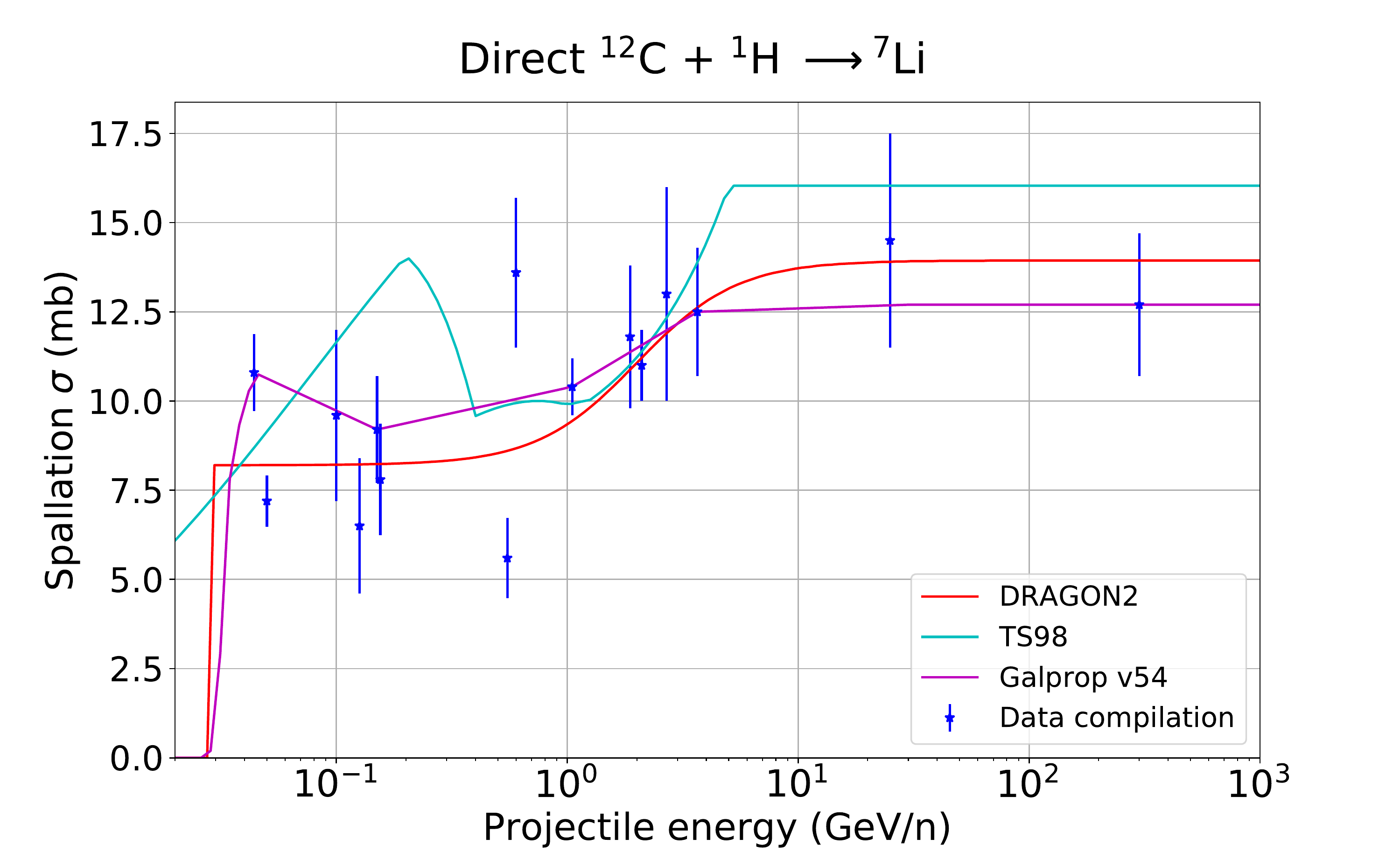} \hspace{0.5cm}
\includegraphics[width=0.42\textwidth,height=0.19\textheight,clip] {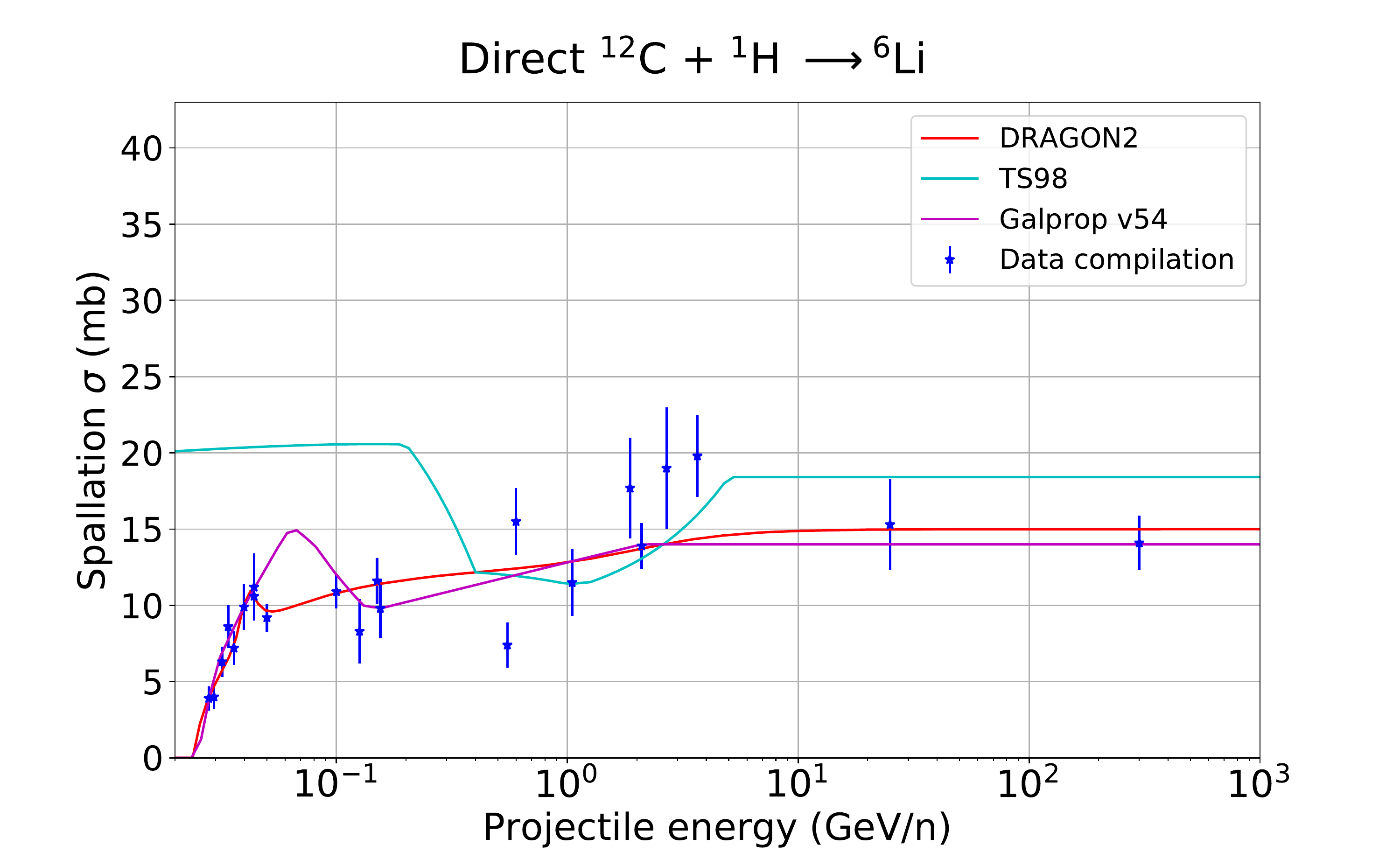}    
\end{center}
\caption{\footnotesize Comparison on the public cross sections to the available experimental data for various channels coming from $^{12}$C.}
\label{fig:XSC}
\end{figure*} 

\vspace{1cm}
\newpage
\textbf{\small Secondary nuclei from $^{16}$O}
\vspace{0.6cm}

\begin{figure*}[!hbt]
\label{fig:XSO}
\begin{center}
\includegraphics[width=0.42\textwidth,height=0.19\textheight,clip] {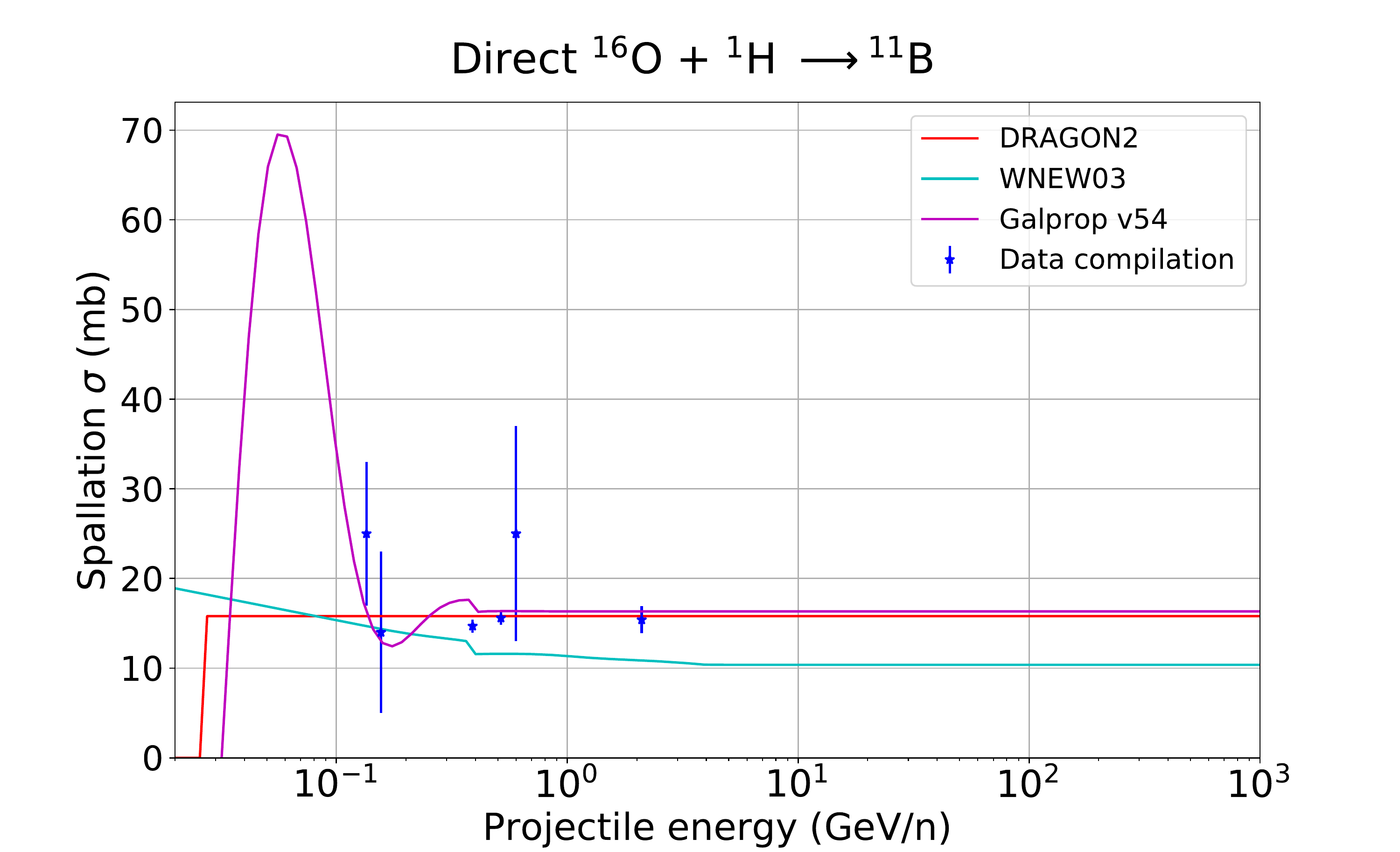} \hspace{0.5cm}
\includegraphics[width=0.42\textwidth,height=0.19\textheight,clip] {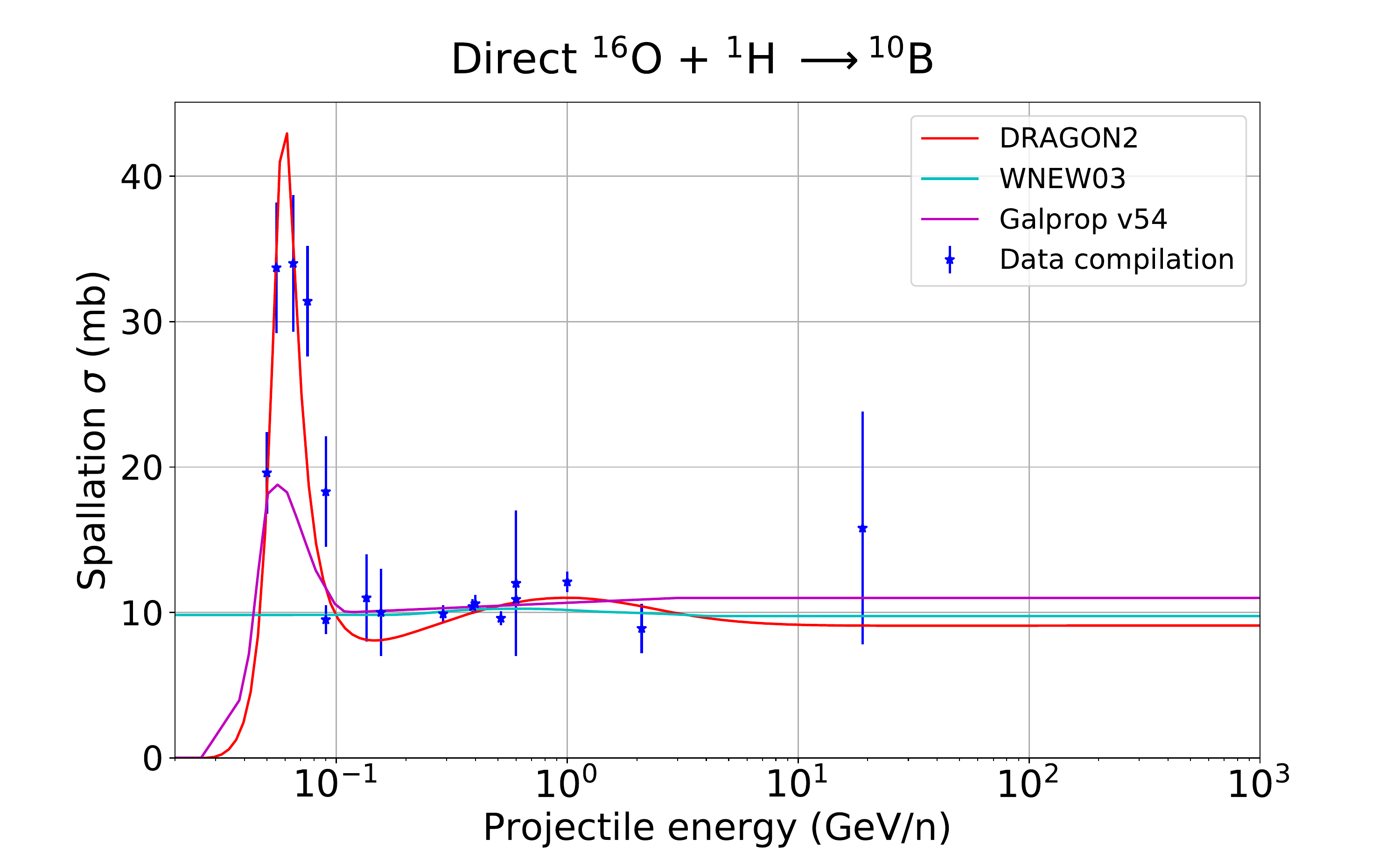} 
\vspace{0.5cm}

\includegraphics[width=0.325\textwidth,height=0.172\textheight,clip] {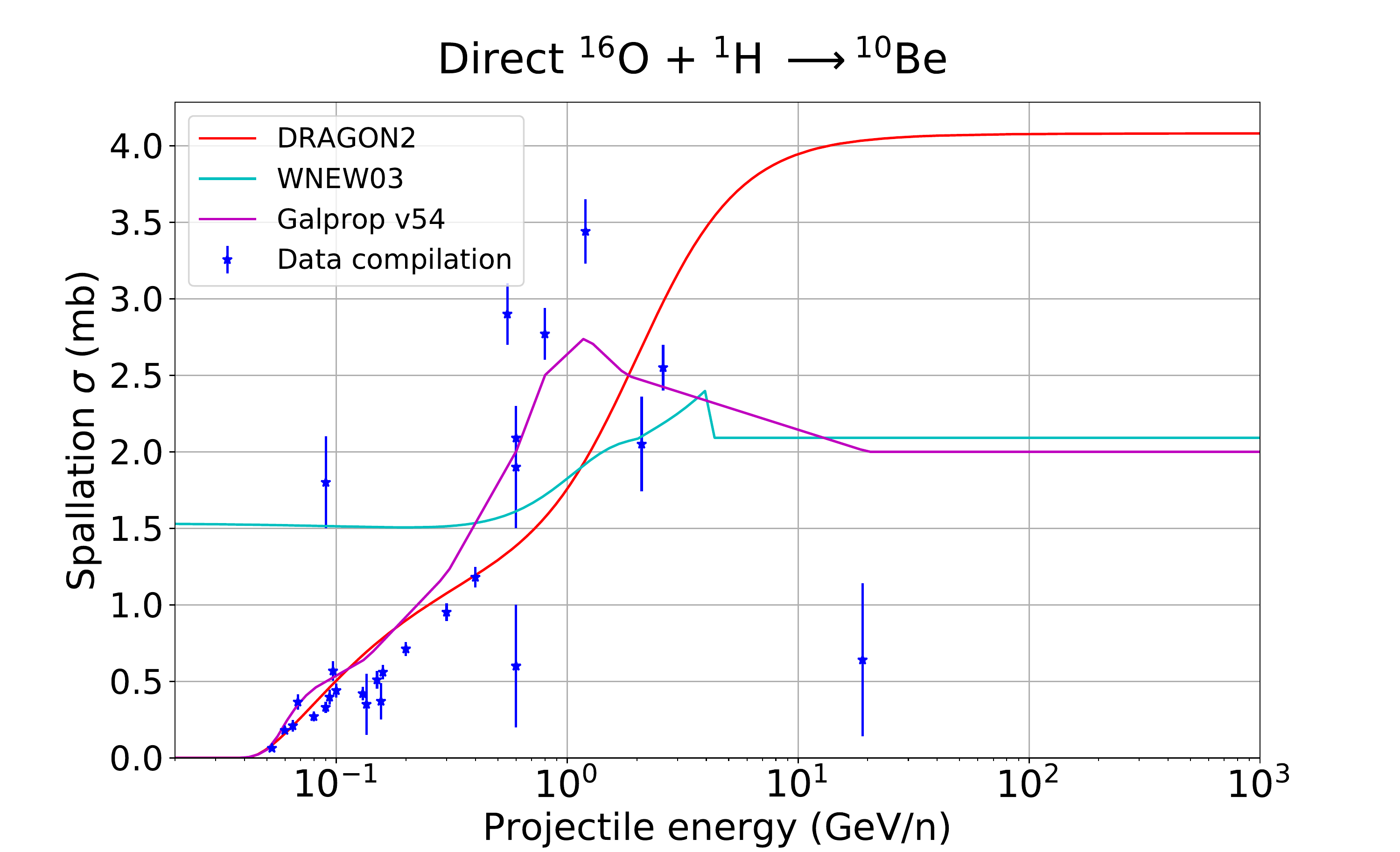} %\hspace{0.5cm}
\includegraphics[width=0.325\textwidth,height=0.172\textheight,clip] {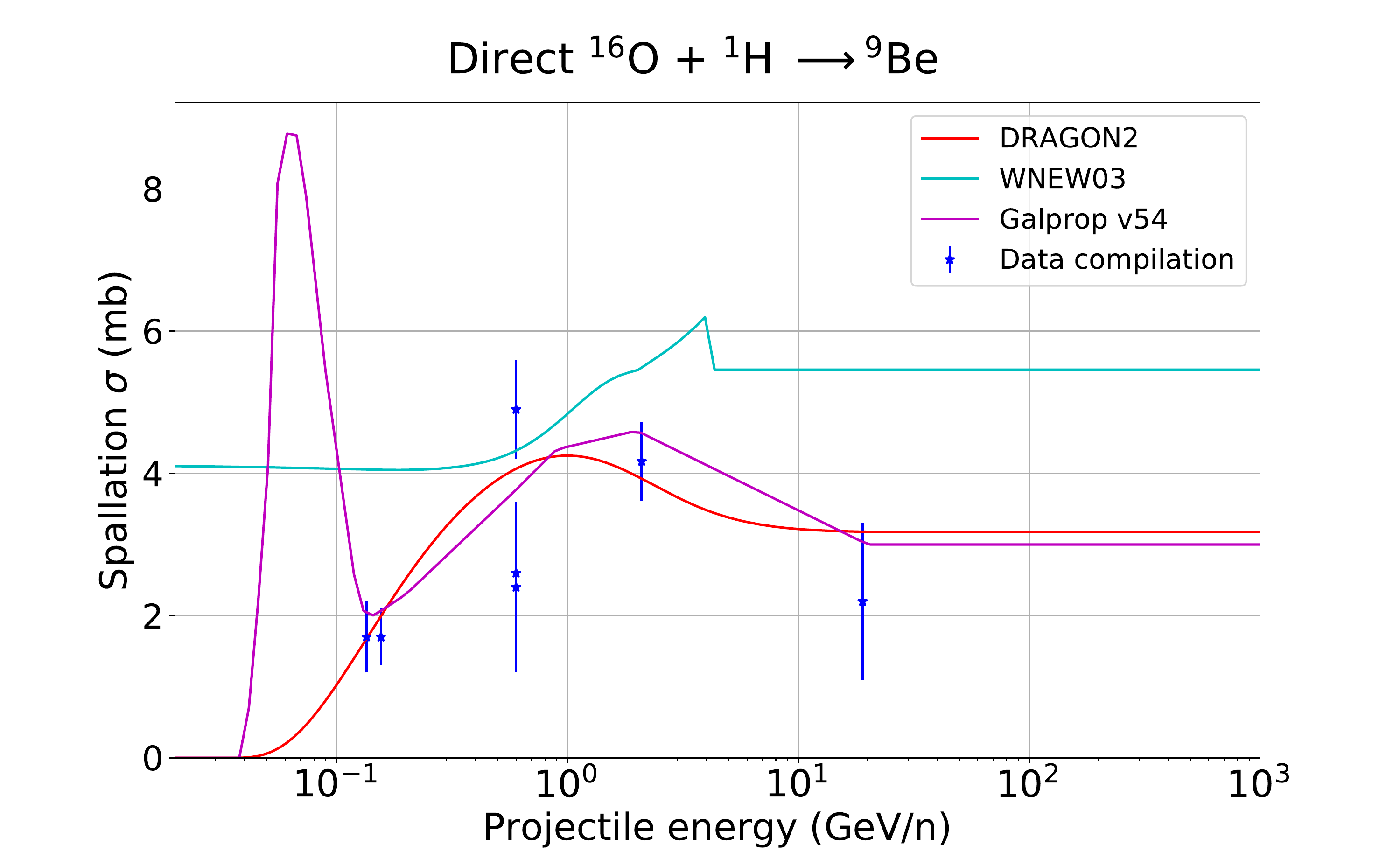}
\includegraphics[width=0.325\textwidth,height=0.172\textheight,clip] {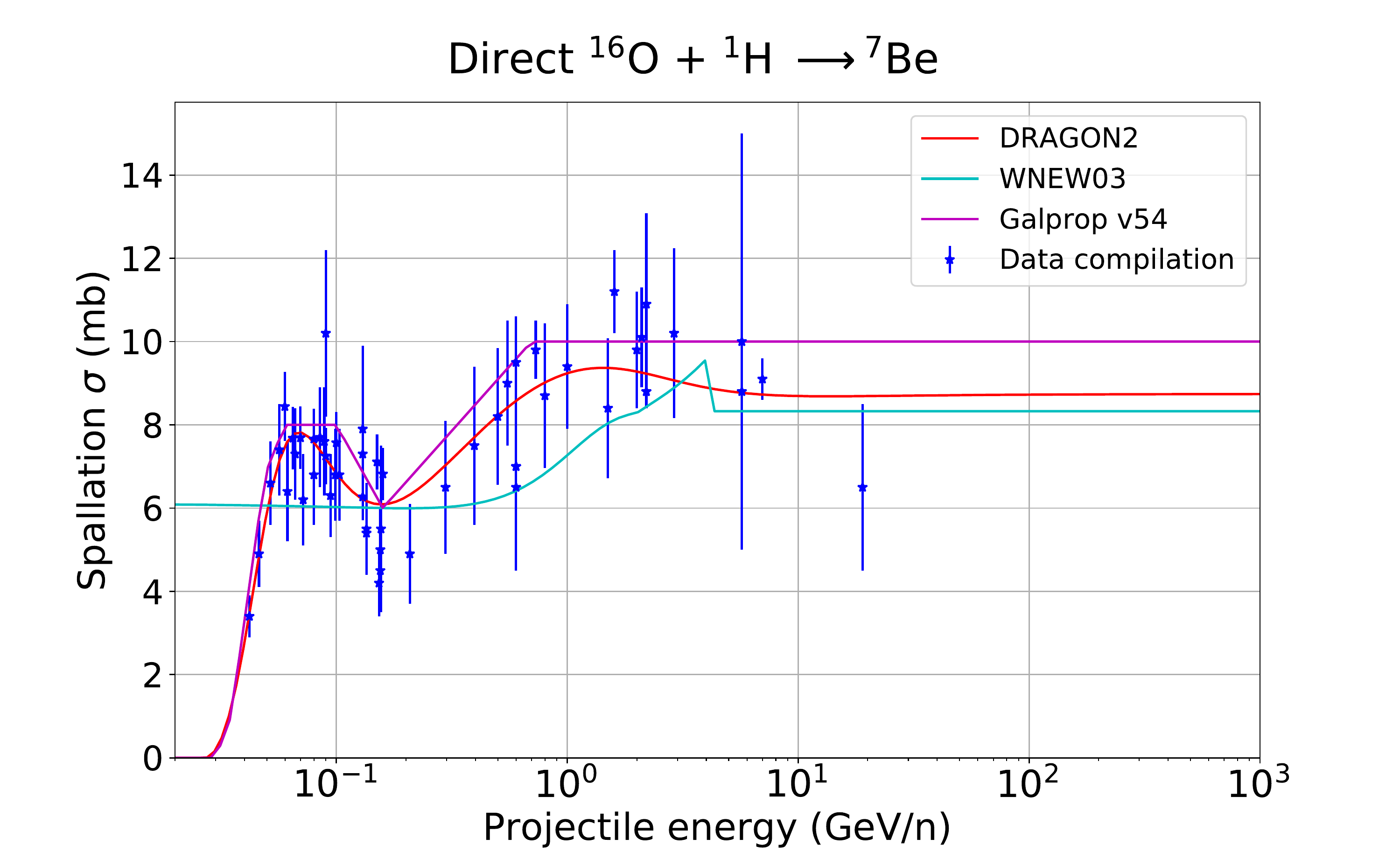}

\vspace{0.5cm}
\includegraphics[width=0.42\textwidth,height=0.19\textheight,clip] {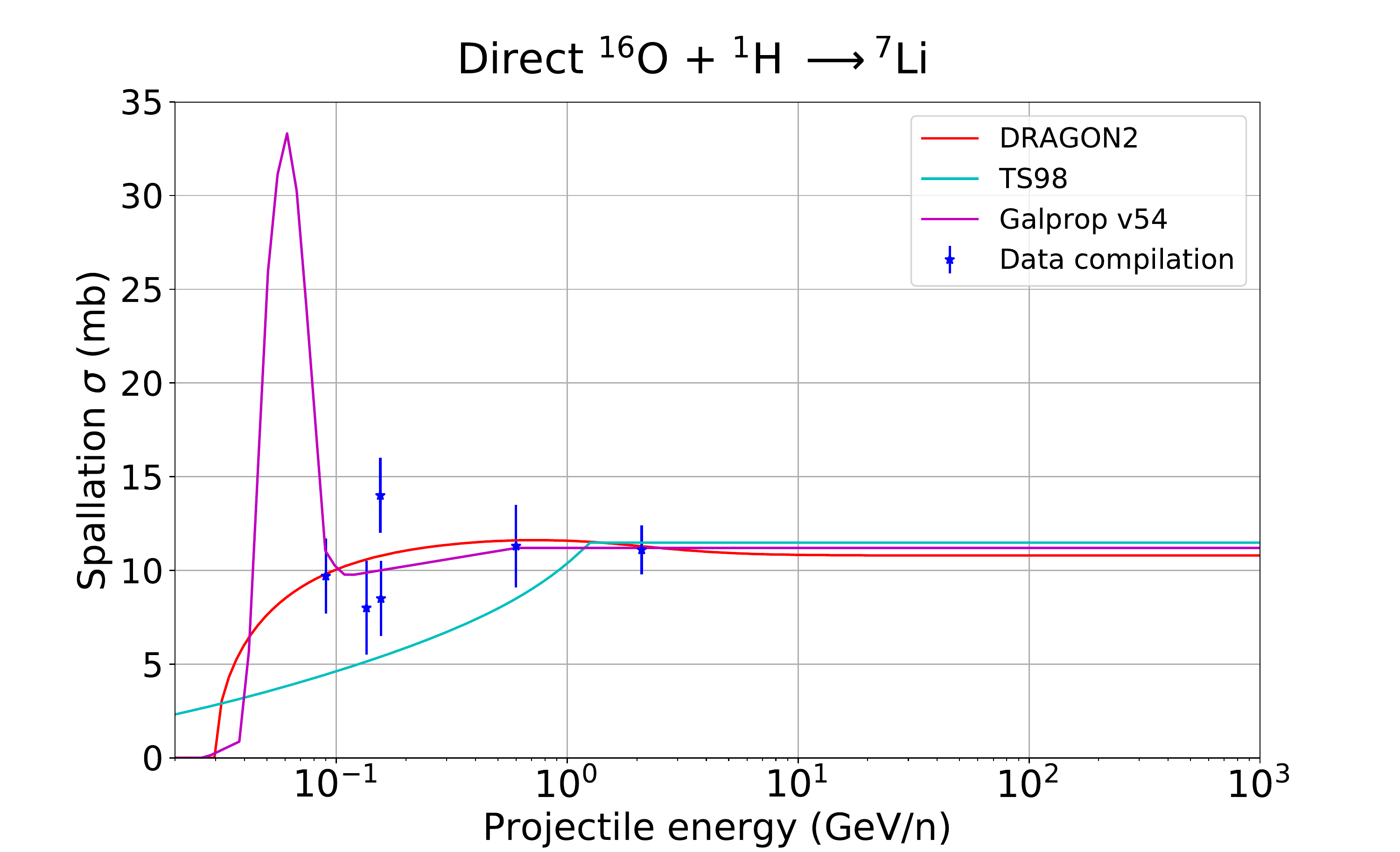} \hspace{0.5cm}
\includegraphics[width=0.42\textwidth,height=0.19\textheight,clip] {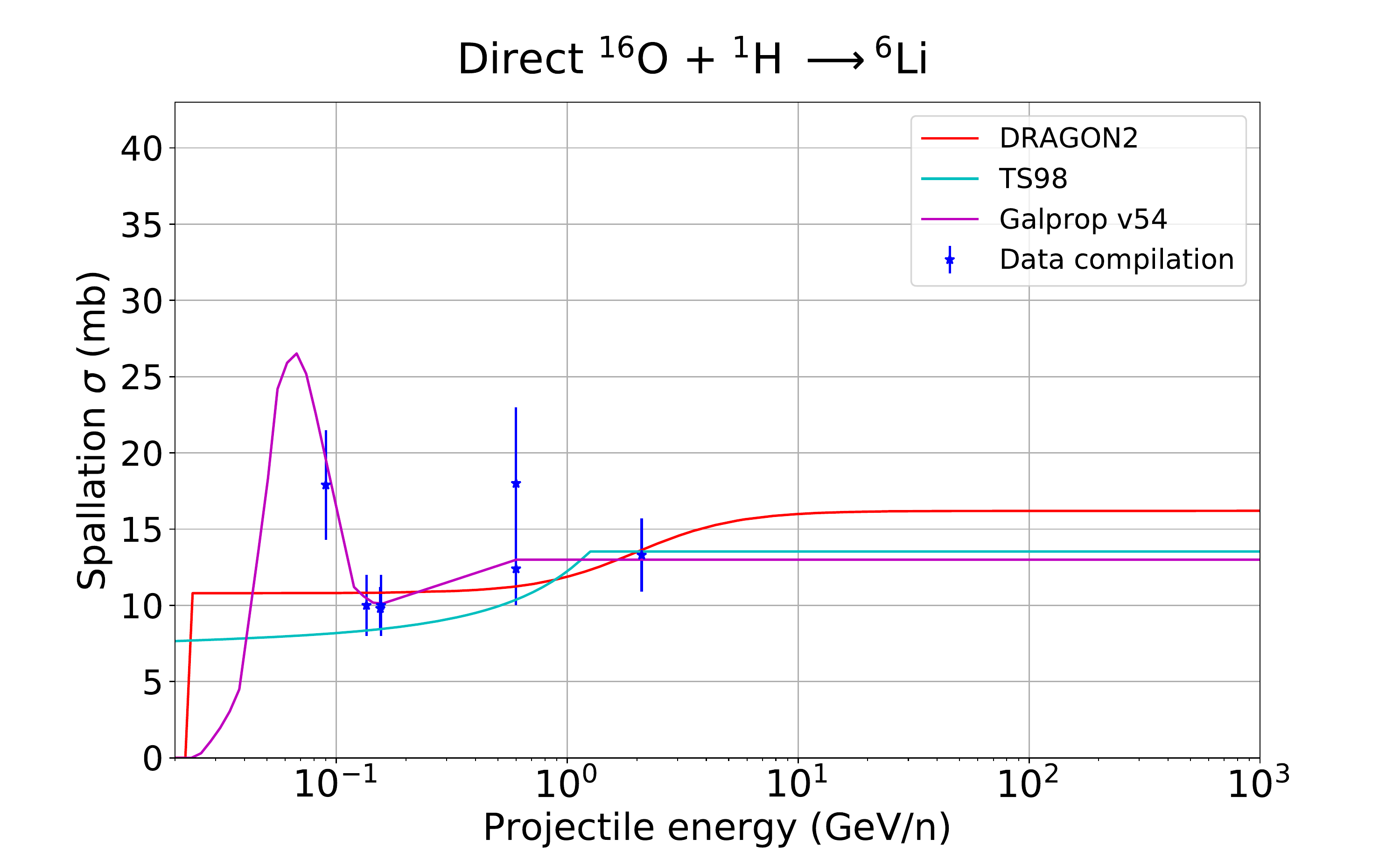}
\end{center}
\caption{\footnotesize Comparison on the public cross sections to the available experimental data for various channels coming from $^{16}$O.}
\end{figure*} 

\newpage

	\chapter{Gas density distribution effect on the $^{10}$Be spectrum}
\label{sec:appendixB}

This appendix aims at pointing out the importance of the density distribution of the interstellar gas in unstable cosmic rays, namely the $^{10}$Be isotope. Concretely, we are showing that there are differences in the flux of this isotope at Earth when using the common 2D model of the Galaxy and using a more detailed model taking into account the spiral arms and shape of the Milky Way, which we call 3D model. This is something that has not been noticed in the past because of the difficulty and time needed for performing detailed 3D simulations, which are possible with DRAGON. These simulations need to specify the number of arms in the disc and their main properties.

\begin{figure*}[!hbt]
\begin{center}
\includegraphics[width=0.85\textwidth,height=0.32\textheight,clip] {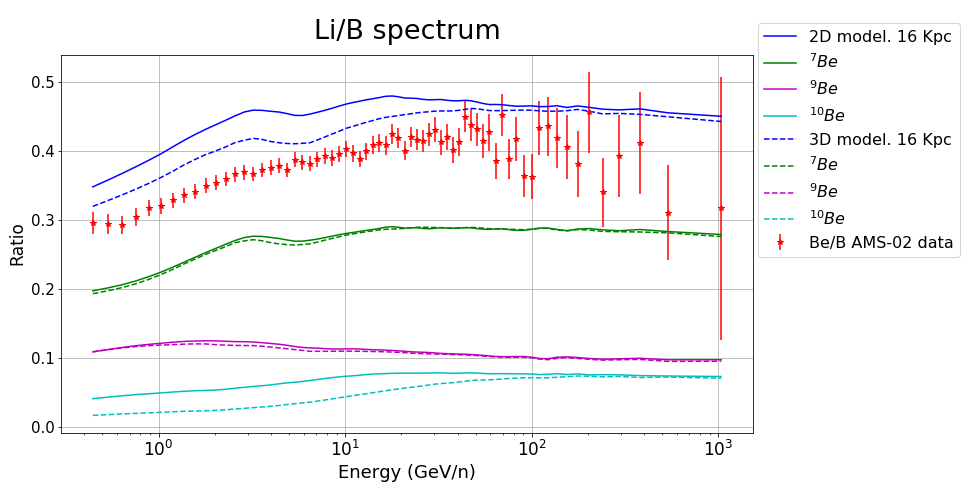}     

\caption{\footnotesize Comparison of the simulated flux of isotopes of Be reaching Earth using a 2D and 3D model for the shape of the Galaxy. To make this comparison, the B/B ratio is shown together with the ratios of the isotopes of Be to B (i.e. $^{10}$Be/B, $^{9}$Be/B and $^{7}$Be/B) for the two different configurations of the gas. While we observe no appreciable differences between the stable isotopes, $^{10}$Be/B shows a significant discrepancy below 30 GeV, which is degenerated with the effect of the halo size in the ratios.}
\label{fig:Density effect}
\end{center}
\end{figure*}

In fig. \ref{fig:Density effect} we show the Be/B ratio and the ratios of the isotopes of Be to B (i.e. $^{10}$Be/B, $^{9}$Be/B and $^{7}$Be/B) for two different configurations of the gas. One, using a simple uniform distribution, a 2D model, and another one using a simple 3D model. There is a very important difference in the flux of this element below $30 \units{GeV}$, that leads to a discrepancy of around 30\%. This discrepancy can be due to the fact that the $^{10}$Be decay length at low energy is of the order of a few hundreds of parsecs and the amount of the isotope reaching Earth is restricted to be formed very close to it. Also the gas profile in the z direction can be very important.

This affects to both, the halo size determination and the shape of the ratios secondary-over-secondary and, in general, it is expected to be important in the propagation of any relatively short-lived particle, and can have a small impact also in the propagation of leptons. 

Nevertheless, a dedicated study on how different distributions can affect the flux of unstable particles can be very complex and time-consuming. Here, we just considered worth pointing this effect out.

	\chapter{Other applications to astrophysical problems with FLUKA}
\label{sec:appendixFluka}

In this appendix we review an interesting astrophysical study carried out with the FLUKA code. In this work, we estimate the fluxes at Earth of secondary particles (gamma rays, neutrinos, leptons and neutrons) produced from the interactions of CRs with the solar surface.

CRs entering the Solar system can reach the planets and the Sun, producing emission of secondary particles due to their interactions with the surfaces or the atmospheres of the celestial bodies. The Moon \cite{Gamma_Moon_1, Gamma_Moon_2}, the Earth \cite{Gamma_Earth} and the Sun \cite{Gamma_Sun} are bright and well known sources of gamma rays. Lunar and terrestrial gamma rays are originated from the hadronic interactions of CR nuclei with the lunar surface and with the upper layers of the Earth's atmosphere, respectively, while the solar gamma-ray emission consists of two components: the disk emission, which is due to CRs interacting with the solar surface \cite{Gamma_Sun_2, Gamma_Sun-Stars} and is localized around the solar disk; the diffuse emission, due to the inverse Compton scatterings of CR leptons with the solar optical photons, that extends up to tens of degrees from the Sun \cite{Gamma_Sun_Disk-Halo, Gamma_Sun-Stars}.

The effects of Interplanetary and heliospheric magnetic fields is a key point of this work. In particular, the total path length of charged CRs increases with the intensity of the magnetic field, and this corresponds to an increase of the interaction probability and, consequently, to an increase of the cascades of secondary particles. In addition, the profile of the solar atmosphere needs to be accounted in detail, since the cascades usually develop from a low-density medium toward a denser medium. The modelization of the Interplanetary magnetic field is achieved using the Parker model \cite{Parker_model}. Additionally, we adopted the potential field source surface (PFSS) model \cite{PFSS} and a modification of this model, which enhances the magnetic field profile near the Sun, following the BIFROST model \cite{BIFROST}, to describe the solar magnetic field. Finally, to evaluate the CR intensities at the Sun we have used a customized version of the HelioProp code \cite{Helioprop}. 

\begin{figure*}[!bp]
\begin{center}
\includegraphics[width=0.7\textwidth,height=0.3\textheight,clip] {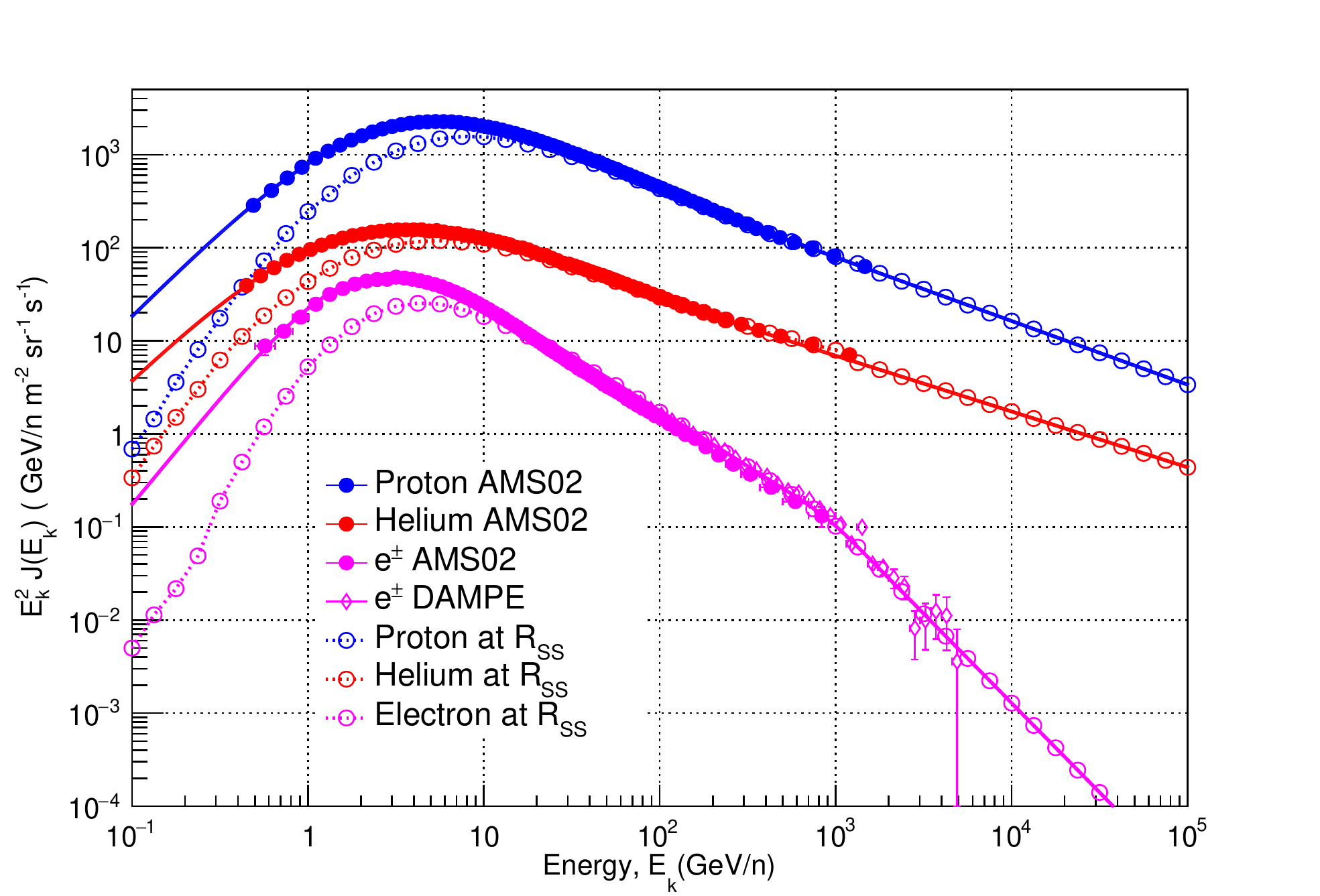} 
\end{center}
\caption{\footnotesize Results for the proton, helium and electron intensities as function of the kinetic energy (kinetic energy per nucleon in the case of helium) compared to experimental data. The full circles correspond to the experimental data measured by AMS-02 at Earth during the Carrington Rotation 2111; the open circles with dashed lines indicate the intensities near the Sun, evaluated by scaling the intensities at Earth with the survival probabilities
obtained from the Helioprop code.}
\label{fig:Prop_Fluk}
\end{figure*}

With this configuration of the local magnetic field, the full propagation of CR protons, helium and electrons in the heliosphere was simulated and adjusted to the existent data, as shown in Fig. \ref{fig:Prop_Fluk}. This has important implications for the spectra of CRs at low energy and lie the foundations for more detailed treatments of the solar modulation.

\begin{figure*}[!htb]
\begin{center}
\includegraphics[width=0.7\textwidth,height=0.3\textheight,clip] {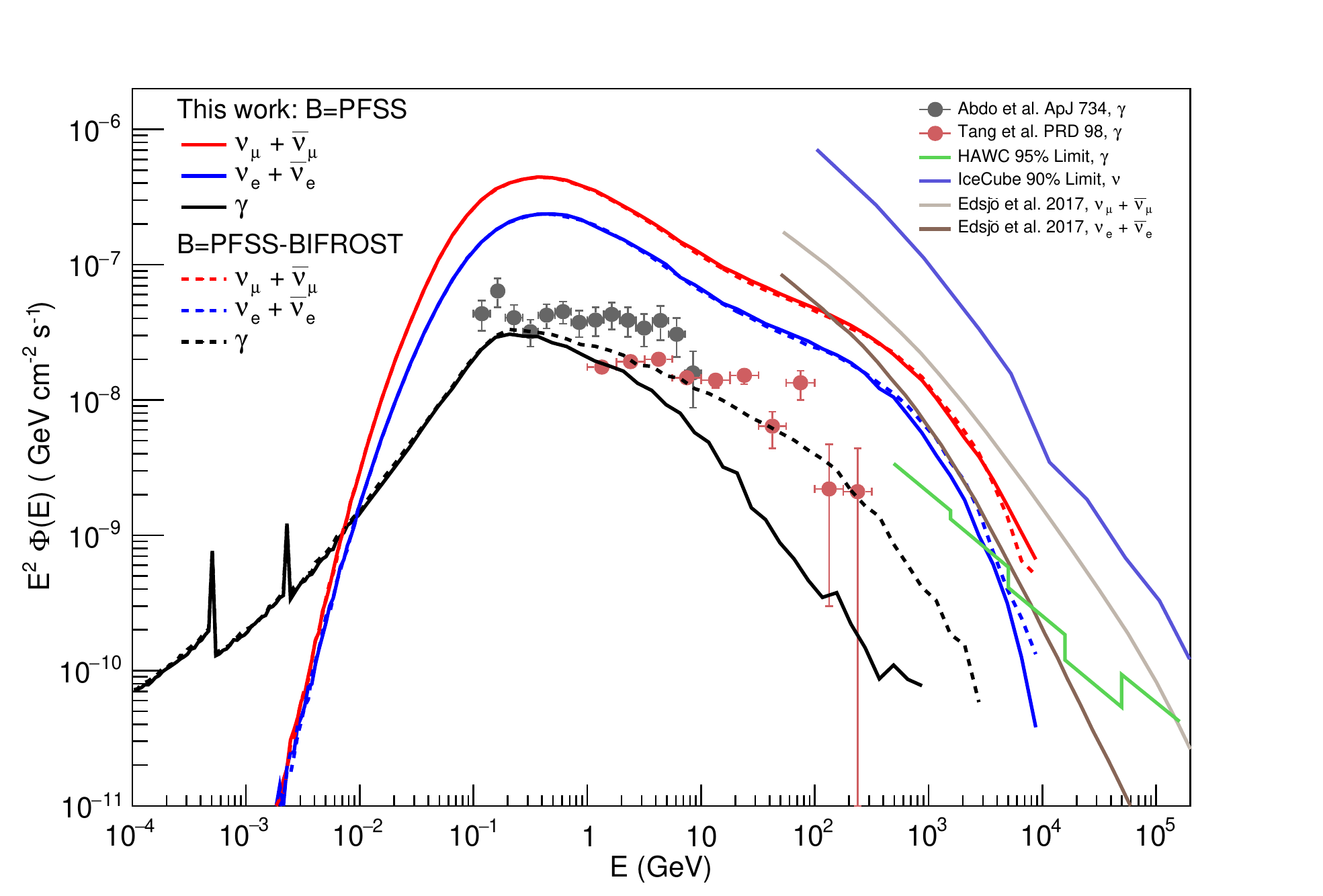} 
\end{center}
\caption{\footnotesize Gamma-ray and neutrino 
fluxes at the Earth calculated for the Carrington rotation 2111 with two models for the Sun magnetic field. The black lines refer to the gamma-ray flux, the red lines to muonic neutrinos (i.e. neutrinos and antineutrinos) and the blue lines to the electronic neutrinos. The prediction of neutrinos by Ref. \cite{Edsj__2017} is also shown. In addition, we include results from gamma-ray observations from Fermi-LAT \cite{Gamma_Sun}, gamma-ray HAWC's 95\% limit \cite{HAWK_Sun} and neutrinos IceCube's 90\% limit \cite{Aartsen:2019avh}.}
\label{fig:Sun_Fluk}
\end{figure*}

In Fig. \ref{fig:Sun_Fluk} we show the predictions of our simulation about the fluxes at Earth of gamma rays and neutrinos in the Carrington Rotation 2111 for both, the original (B$=$PFSS) and the enhanced (B$=$PFSS-BIFROST) PFSS model. This figure illustrates well one the main conclusions of the paper: while the gamma-ray production strongly depends on the surrounding magnetic field, the repercussion of the field in the neutrino emission seems negligible.

This study shows an interesting way to evaluate different models of the solar magnetic field and its structure from a multi-messenger approach, in addition to a detailed evaluation of the solar modulation effect on the fluxes of CRs.

%The magnetic field should also affect the inverse Compton gamma-ray emission close to the Sun, since the electrons should move along curved trajectories, whose lengths determine the interaction probability with the intense optical photon field. In this way the inverse Compton emission should also be peaked close to the solar surface, and might be not well separated by the disk emission due to the interaction of cosmic rays with the solar atmosphere. 

	\chapter{Including AMS-02 error correlations into the analysis}
\label{sec:covariance}

Given the level of precision required to perform these analyses, some authors (from \cite{derome2019fitting}) have recently argued the importance of including the systematic errors of AMS-02 data in the analysis in order to prevent from energy correlations. Nevertheless, the AMS-02 collaboration does not provide this correlation matrix. To every one of the sources of systematic errors a correlation matrix is associated, such that we need to define covariance matrices for any possible error. These errors can be related to the effective acceptance of AMS-02, geomagnetic errors, scale errors, etc. A complete explanation of these errors is given in \cite{Heisig:2020nse}.

This correlation matrix can be thus estimated following the expression:
\begin{equation}
(C_{rel}^{\alpha})_{ij} = \sigma_i^{\alpha} \sigma_j^{\alpha} exp\left(-0.5\frac{(log(R_i/R_j))^2}{(l^{\alpha}_{\rho})^2}\right)
\label{eq:Cov_matrix}
\end{equation}
where the main element intended to be estimated is the correlation length associated to the error $\alpha$, $l^{\alpha}_{\rho}$ and $(C_{rel}^{\alpha})_{ij}$ is the ij-th element built from the relative errors $\sigma_i$ and $\sigma_j$ at rigidity bins $R_i$ and $R_j$.

In order to show the differences in the errors associated to the propagation parameters obtained when including these correlation matrices we analyse the B/C ratio using the correlation matrices derived in \cite{Heisig:2020nse}.

In the next table the results of the B/C analysis for the diffusion hypothesis including the systematic covariance are compared to the ones assuming no correlations:
\begin{table}[htb!]
\centering
\resizebox*{0.59\columnwidth}{0.08\textheight}{
\begin{tabular}{|c|c|c|c|c|}
\hline &\textbf{$D_0$ $\left(\frac{10^{28} cm^{2}}{s \times kpc} \right)$} & \textbf{$V_A$} & \textbf{$\eta$} & \textbf{$\delta$} \\ 
\hline
\textbf{Covariance} & 6.69 $\pm$ 0.36 & 23.30 $\pm$ 4.36   &  -0.69 $\pm$ 0.31 &  0.44 $\pm$ 0.02 \\ \hline 

\textbf{No Covariance} & 6.55  $\pm$ 0.27 & 22.02 $\pm$ 3.54 &  -0.78. $\pm$ 0.13  &  0.44 $\pm$ 0.01  \\ \hline 
\end{tabular}
}
\caption{Comparison of the propagation parameters obtained in the analysis of the B/C ratio including and without including the covariance matrices of AMS-02 data. These analyses are carried out under the diffusion hypothesis, eq. \ref{eq:breakhyp}}
\label{tab:CovvsNoCov}
\end{table}

\begin{figure*}[!bh]
\begin{center}
\includegraphics[width=0.495\textwidth,height=0.22\textheight,clip] {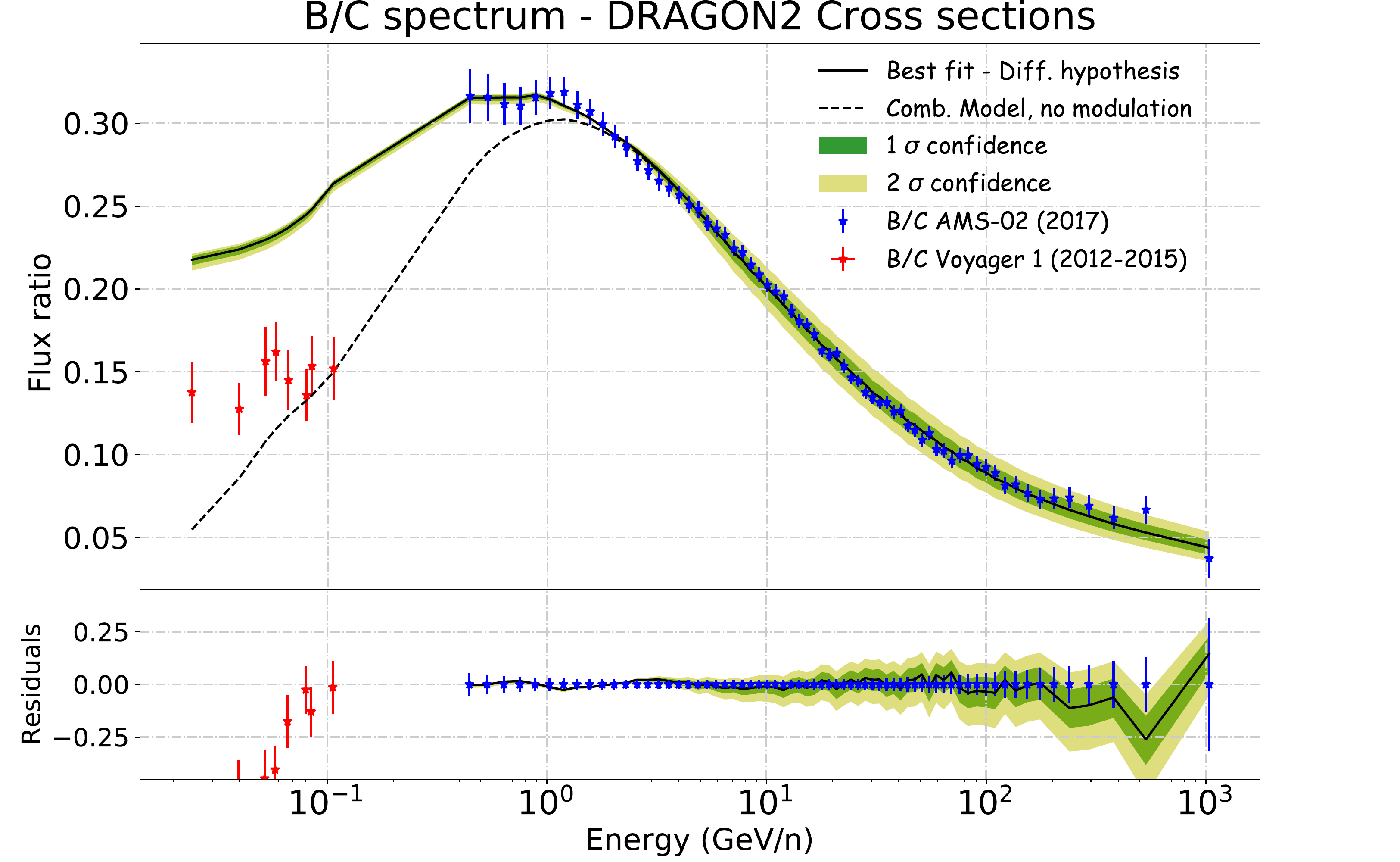} 
\hspace{-0.05cm}
\includegraphics[width=0.495\textwidth,height=0.22\textheight,clip] {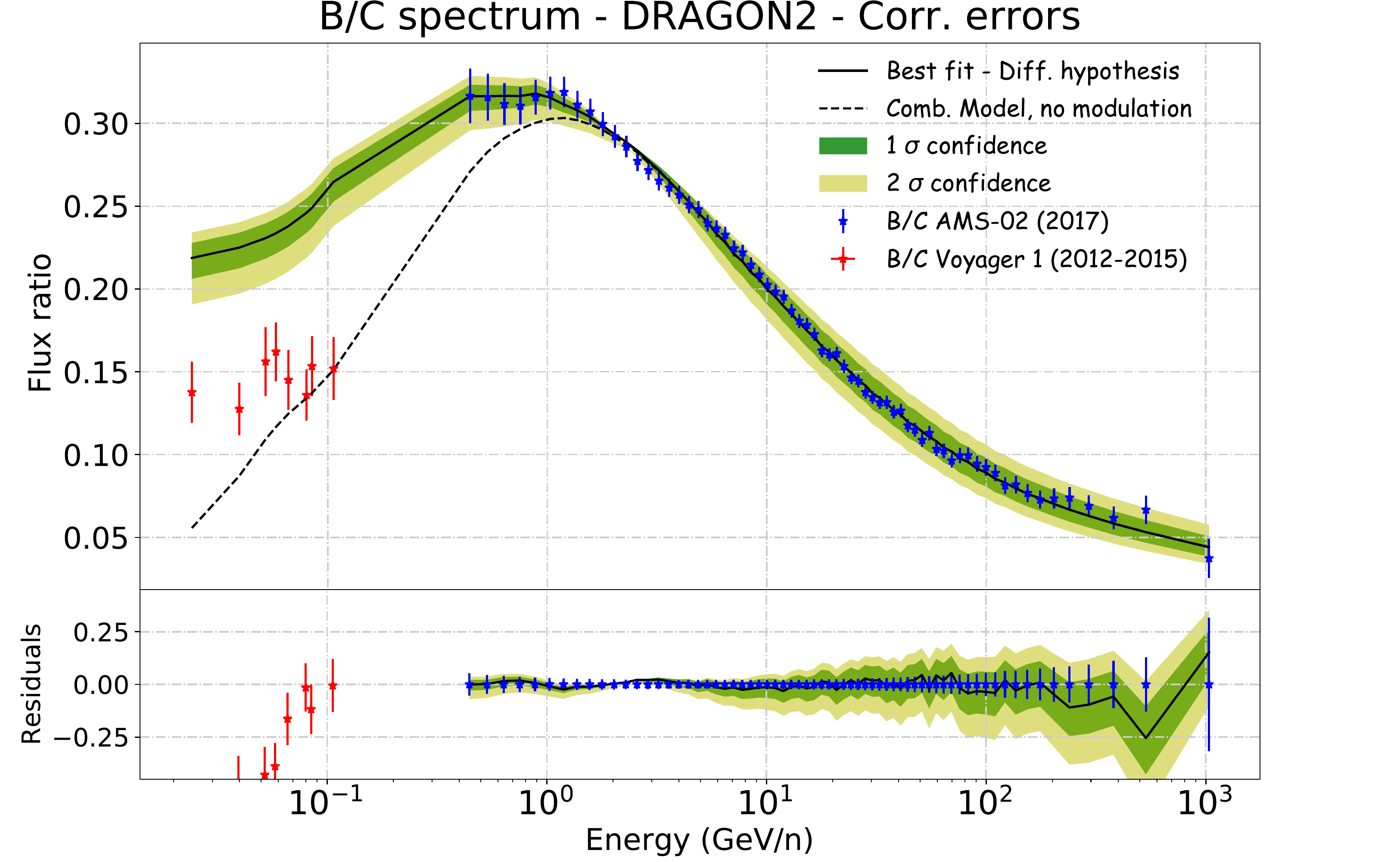}
\end{center}
\caption{\footnotesize $1\sigma$ and $2\sigma$ uncertainty bands in the predictions of the B/C spectrum for the analysis with and without including correlations in the AMS-02 data with the DRAGON2 cross sections.}
\label{fig:CorrvsUncorr}
\end{figure*}

As expected, the propagation parameters are roughly equal in both analyses, presenting just negligible (statistically speaking) deviations. However, the statistical uncertainties associated to their determination are considerably larger than the derived without taking into account correlations. The uncertainty bands for the predictions of the B/C spectrum are shown in Fig. \ref{fig:CorrvsUncorr} for the analysis with and without including error correlations in the AMS-02 data and with the DRAGON2 cross sections. %Furthermore, the inclusion of the covariance matrix can be important when studying antiprotons, as was demonstrated in \cite{Heisig:2020nse}. 
\vspace{0.6cm}

	\chapter{Summary of the MCMC results}
\label{sec:appendixC}

In this appendix, the corner plots obtained from the MCMC algorithm of the combined analysis for each cross sections set and both diffusion parametrisations used ('Source' and 'Diffusion' hypotheses) are shown to visualize the probability distribution functions (PDFs) of each propagation parameter and the correlations between them. These corner plots contain the diffusion parameters ($D_0$, $\delta$, $\eta$ and $V_A$) as well as the scaling factors included as nuisance parameters.

Furthermore, we are summarizing the results for the PDFs of each of the propagation parameters found in form of tables. First, we show the results for the independent analyses of each secondary-to-primary ratio (B/C, B/O, Be/C, Be/O, Li/C, Li/O), then, the parameters found in the combined analysis of these ratios and, finally, the analysis of each of the ratios after scaling the original cross sections according to the secondary-over-secondary ratios. The tables contain the median value (coinciding with the maximum posterior probability value) $\pm$ the $1\sigma$ uncertainty related to its determination, assuming their PDFS to be Gaussian, and the actual range of values contained in the 95\% probability of the distribution.

%\newpage
\vspace{1cm}
\begin{center}
    \textbf{\large{Propagation parameters PDF - Source hypothesis}}
\end{center} 
\begin{figure}[!htpb]
	\centering
	\includegraphics[width=0.88\textwidth, height=0.54\textheight]{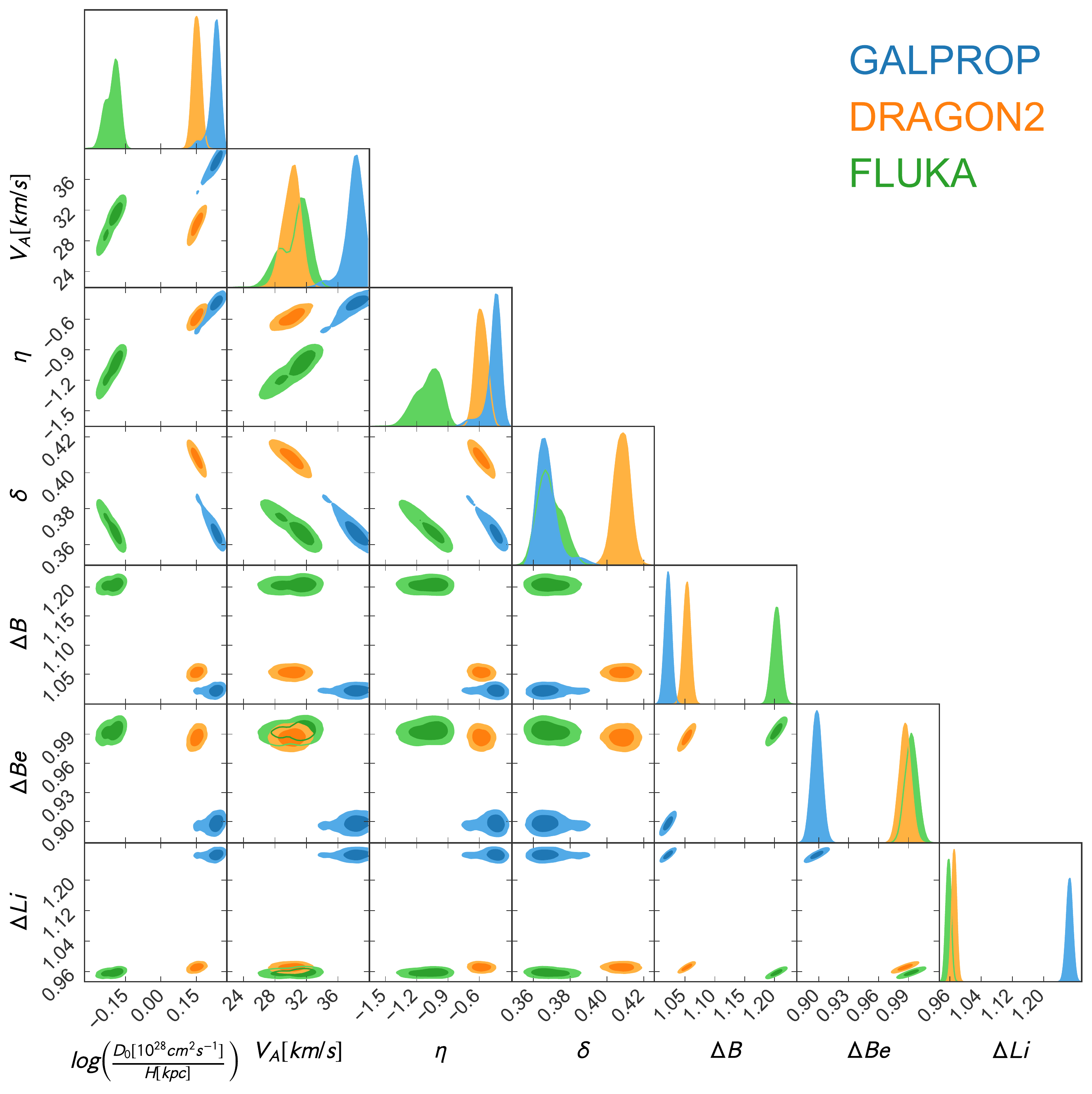}
	\caption{\footnotesize Corner plot representing the PDF of every parameter included in the combined MCMC analysis with the diffusion coefficient parametrisation of eq. \ref{eq:sourcehyp} for the GALPROP, DRAGON2 and FLUKA cross sections sets. The inner panels represent the correlations between parameters with the 1 and 2 $\sigma$ surface probabilities highlighted.}
	\label{fig:Corner_Source}
\end{figure}

\newpage 

\vspace{3cm}

\begin{center}
    \textbf{\large{Propagation parameters PDF - Diffusion hypothesis}}
\end{center} 

\begin{figure}[!htpb]
	\centering
	\includegraphics[width=0.95\textwidth, height=0.55\textheight]{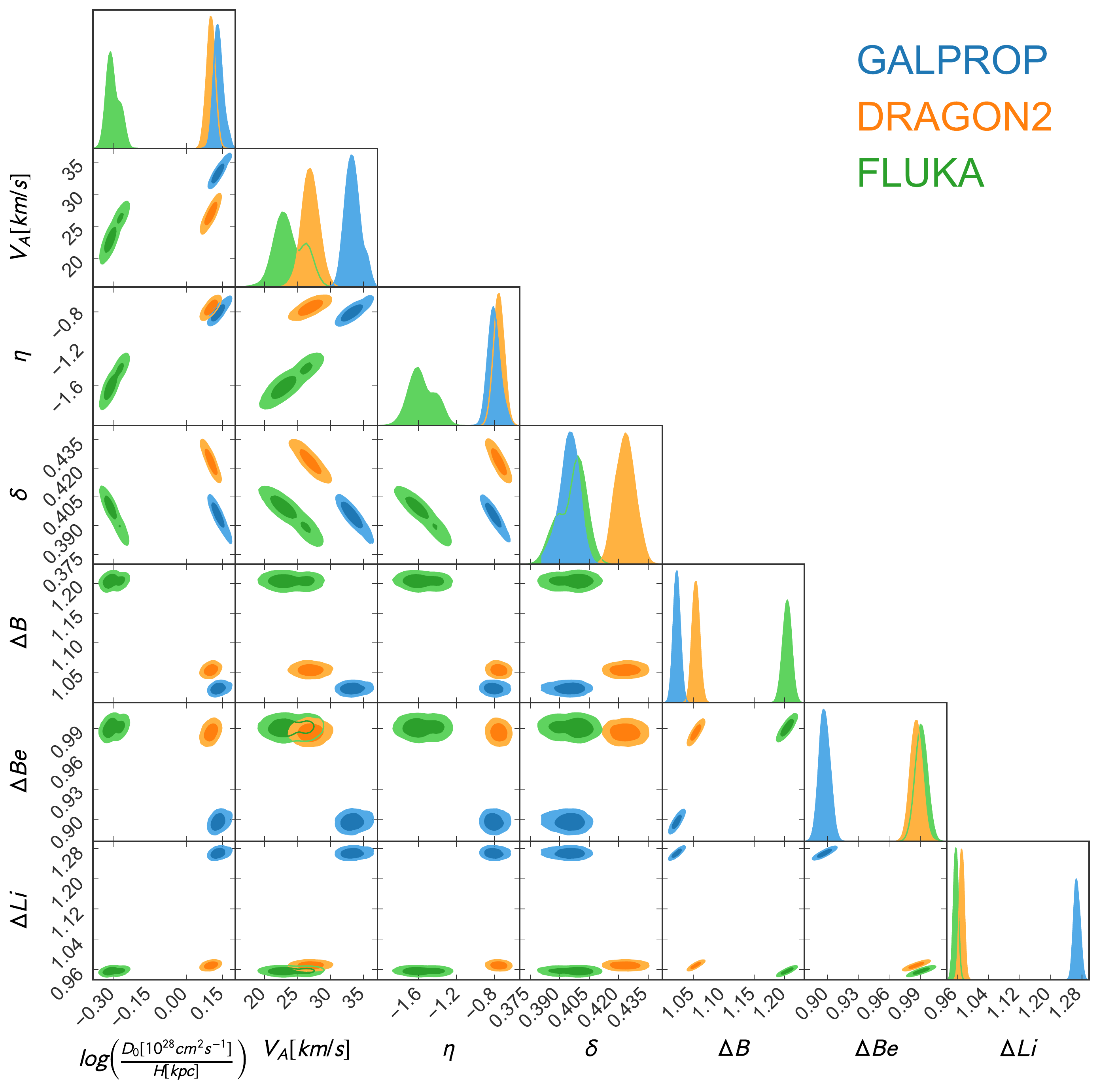}
	\caption{\footnotesize Corner plot as in figure \ref{fig:Corner_Source} but for the diffusion coefficient parametrisation of eq. \ref{eq:breakhyp}.}
	\label{fig:Corner_Diff}
\end{figure}

\begin{table}[htb!]
\centering
%\small
\textbf{\Large{Independent analyses - Source hypothesis}}

\vspace{0.4cm}

\resizebox*{0.81\columnwidth}{0.17\textheight}{
%\rowcolors{1}{}{yellow!90!white!50!} 
\begin{tabular}{|c|c|c|c|c|c|c|c|}
\hline
\multicolumn{7}{|c|}{\textbf{\large{DRAGON2 parameters - Source hypothesis}}} \\
\hline &\textbf{B/C} & \textbf{B/O} & \textbf{Be/C} & \textbf{Be/O} & \textbf{Li/C} & \textbf{Li/O} \\ 
\hline
%\cellcolor{white}
\multirow{2}{6em}{\centering$D_0/H$ $\left(\frac{10^{28} cm^{2}}{\units{s} \cdot \units{kpc}} \right)$} & 1.02 $\pm$ 0.05 & 1.04$\pm$ 0.05   &  1.09 $\pm$ 0.05 &  1.06 $\pm$ 0.07 & 1.37  $\pm$ 0.05 &  1.33 $\pm$ 0.04  \\  &  [0.94., 1.12]  &   [0.95, 1.14] &   [1.03, 1.19]  &  [0.96, 1.19] &   [1.27, 1.45] &   [1.25, 1.39] \\ \hline 

\multirow{2}{6em}{\centering\large{$v_A$ ($\units{km/s}$)}}  & 25.72  $\pm$ 3.22 & 20.65 $\pm$ 2.65 &  11.74 $\pm$ 7.24  &  29.65 $\pm$ 7.19 &  36.49 $\pm$ 2.68 &  38.12 $\pm$ 1.66 \\  &  [19.72, 32.16.]  &   [24.85, 34.95] &   [0., 26.26]  &  [15.27, 40.75] &   [31.07, 40.31] &   [34.9, 40.66] \\ \hline 

\multirow{2}{6em}{\centering\large{$\eta$}}    & -0.66 $\pm$ 0.12 & -0.51 $\pm$ 0.11 &  -1.15 $\pm$ 0.16 &  -0.94 $\pm$ 0.19 & -0.37 $\pm$ 0.11 &  -0.29 $\pm$ 0.08  \\ &  [-0.80, -0.42]  &   [-0.72, -0.33] &   [-1.47, -0.83]  &  [-1.32, -0.6] &   [-0.59, -0.15] &   [-0.45, -0.15] \\ \hline 

\multirow{2}{6em}{\centering\large{$\delta$}}  & 0.43 $\pm$ 0.01 & 0.41 $\pm$ 0.01 &  0.44 $\pm$ 0.01  &  0.43 $\pm$ 0.02 & 0.39 $\pm$ 0.01 &  0.38 $\pm$ 0.01  \\  &  [0.41, 0.45]  &   [0.39, 0.43] &   [0.41, 0.46]  &  [0.39, 0.47] &   [0.37, 0.41] &   [0.36, 0.40] \\ \hline 
\end{tabular}
}

\vspace{0.8 cm}
\resizebox*{0.81\columnwidth}{0.17\textheight}{
%\rowcolors{1}{}{yellow!90!white!50!} 
\begin{tabular}{|c|c|c|c|c|c|c|c|}
\hline
\multicolumn{7}{|c|}{\textbf{\large{GALPROP parameters - Source hypothesis}}} \\
\hline &\textbf{B/C} & \textbf{B/O} & \textbf{Be/C} & \textbf{Be/O} & \textbf{Li/C} & \textbf{Li/O} \\ 
\hline
%\cellcolor{white}
\multirow{2}{6em}{\centering$D_0/H$ $\left(\frac{10^{28} cm^{2}}{\units{s} \cdot \units{kpc}} \right)$} & 0.99 $\pm$ 0.04 & 0.98 $\pm$ 0.04   &  1.26 $\pm$ 0.03 &  1.10 $\pm$ 0.05 & 0.98 $\pm$ 0.05 &  1.00 $\pm$ 0.05  \\  &  [0.92, 1.07]  &   [0.92, 1.07] &   [1.22, 1.32]  &  [1.05, 1.15] &   [0.88, 1.08] &   [0.91, 1.07] \\ \hline 

\multirow{2}{6em}{\centering\large{$v_A$ ($\units{km/s}$)}}  & 27.17  $\pm$ 2.61 & 29.11 $\pm$ 2.14 &  6.31 $\pm$ 7.20  &  10.50 $\pm$ 7.48 &  32.37 $\pm$ 2.04 &  32.40 $\pm$ 1.66 \\  &  [21.95, 32.25]  &   [25.07, 33.39] &   [0., 15.05]  &  [0., 25.46] &   [28.65, 36.45] &   [29.28, 35.04] \\ \hline 

\multirow{2}{6em}{\centering\large{$\eta$}}    & -0.54 $\pm$ 0.12 & -0.29 $\pm$ 0.09 &  -1.56 $\pm$ 0.19 &  -1.15 $\pm$ 0.19 & -0.6 $\pm$ 0.20 &  -0.32 $\pm$ 0.14  \\ &  [-0.78, -0.30]  &   [-0.45, -0.11] &   [-1.94, -1.18]  &  [-1.53, -0.78] &   [-1.0, -0.44] &   [-0.60, -0.10] \\ \hline 
\multirow{2}{6em}{\centering\large{$\delta$}}  & 0.43 $\pm$ 0.01 & 0.42 $\pm$ 0.01 &  0.39 $\pm$ 0.01  &  0.42 $\pm$ 0.01 & 0.38 $\pm$ 0.01 &  0.37 $\pm$ 0.01  \\  &  [0.41, 0.45]  &   [0.40, 0.45] &   [0.37, 0.41]  &  [0.40, 0.44] &   [0.36, 0.40] &   [0.36, 0.39] \\ \hline 
\end{tabular}
}

\vspace{0.8 cm}
\resizebox*{0.81\columnwidth}{0.17\textheight}{
%\rowcolors{1}{}{yellow!90!white!50!} 
\begin{tabular}{|c|c|c|c|c|c|c|c|}
\hline
\multicolumn{7}{|c|}{\textbf{\large{FLUKA parameters - Source hypothesis}}} \\
\hline &\textbf{B/C} & \textbf{B/O} & \textbf{Be/C} & \textbf{Be/O} & \textbf{Li/C} & \textbf{Li/O} \\ 
\hline
%\cellcolor{white}
\multirow{2}{6em}{\centering$D_0/H$ $\left(\frac{10^{28} cm^{2}}{\units{s} \cdot \units{kpc}} \right)$} & 0.53 $\pm$ 0.03 & 0.58$\pm$ 0.03   &  0.70$\pm$ 0.02 &  0.69 $\pm$ 0.06 & 0.91  $\pm$ 0.06 &  0.96 $\pm$ 0.03  \\  &  [0.47, 0.57]  &   [0.52, 0.64] &   [0.67, 0.74]  &  [0.62, 0.8] &   [0.78, 1.0] &   [0.9, 1.0] \\ \hline 

\multirow{2}{6em}{\centering\large{$v_A$ ($\units{km/s}$)}}  & 19.46  $\pm$ 2.80 & 26.98 $\pm$ 1.94 &  5.82 $\pm$ 6.52  &  16.36 $\pm$ 9.28 &  28.98 $\pm$ 7.32 &  37.9 $\pm$ 1.78 \\  &  [14.46, 25.06]  &   [23.34, 30.86] &   [0., 18.86]  &  [0., 34.72] &   [14.34, 37.46] &   [34.34, 40.32] \\ \hline 

\multirow{2}{6em}{\centering\large{$\eta$}}    & -0.97 $\pm$ 0.16 & -0.68 $\pm$ 0.15 &  -1.90 $\pm$ 0.20 &  -1.75 $\pm$ 0.21 & -1.26 $\pm$ 0.39 &  -0.83 $\pm$ 0.13  \\ &  [-1.29, -0.67]  &   [-0.98, -0.38] &   [-2.14, -1.50]  &  [-2.3, -1.3] &   [-2.04, -0.76] &   [-1.09, -0.67] \\ \hline 

\multirow{2}{6em}{\centering\large{$\delta$}}  & 0.42 $\pm$ 0.01 & 0.39 $\pm$ 0.01 &  0.42 $\pm$ 0.01  &  0.41 $\pm$ 0.02 & 0.36 $\pm$ 0.02 &  0.34 $\pm$ 0.01  \\  &  [0.40, 0.44]  &   [0.37, 0.42] &   [0.40, 0.44]  &  [0.38, 0.43] &   [0.34, 0.40] &   [0.32, 0.37] \\ \hline 
\end{tabular}
}
\caption{Results obtained in the independent analysis of the flux ratios B/C, Be/C, Li/C, B/O, Be/O and Li/O for the diffusion coefficient parametrisation of eq. \ref{eq:sourcehyp} for each of the cross sections sets analysed. Maximum posterior probability value with its standard deviation (assuming the PDF to be a Gaussian) and the range of parameters within $95\%$ of confidence level are shown for each parameter and ratio. }
\label{tab:PDF_single_Source}
\end{table}

\begin{table}[htb!]
\centering
%\small
\textbf{\Large{Independent analyses - Diffusion hypothesis}}

\vspace{0.4cm}

\resizebox*{0.81\columnwidth}{0.18\textheight}{
%\rowcolors{1}{}{yellow!90!white!50!} 
\begin{tabular}{|c|c|c|c|c|c|c|c|}
\hline
\multicolumn{7}{|c|}{\textbf{\large{DRAGON2 parameters - Diffusion hypothesis}}} \\
\hline &\textbf{B/C} & \textbf{B/O} & \textbf{Be/C} & \textbf{Be/O} & \textbf{Li/C} & \textbf{Li/O} \\ 
\hline
%\cellcolor{white}
\multirow{2}{6em}{\centering$D_0/H$ $\left(\frac{10^{28} cm^{2}}{\units{s} \cdot \units{kpc}} \right)$} & 0.97 $\pm$ 0.04 & 0.98$\pm$ 0.04   &  1.07 $\pm$ 0.03 &  1.00 $\pm$ 0.05 & 1.30  $\pm$ 0.04 &  1.28 $\pm$ 0.05  \\  &  [0.90., 1.05]  &   [0.91, 1.06] &   [1.03, 1.14]  &  [0.92, 1.11] &   [1.21, 1.37] &   [1.18, 1.36] \\ \hline 

\multirow{2}{6em}{\centering\large{$v_A$ ($\units{km/s}$)}}  & 22.02  $\pm$ 3.54 & 26.53 $\pm$ 2.54 &  7.60 $\pm$ 6.70  &  14.19 $\pm$ 8.76 &  33.31 $\pm$ 2.72 &  36.60 $\pm$ 2.47 \\  &  [14.94, 28.86.]  &   [21.45, 31.43] &   [0., 20.00]  &  [0, 29.27] &   [27.87, 37.49] &   [31.64, 40.00] \\ \hline 

\multirow{2}{6em}{\centering\large{$\eta$}}    & -0.78 $\pm$ 0.13 & -0.65 $\pm$ 0.10 &  -1.30 $\pm$ 0.16 &  -1.16 $\pm$ 0.21 & -0.54 $\pm$ 0.12 &  -0.38 $\pm$ 0.11  \\ &  [-1.04, -0.54]  &   [-0.75, -0.55] &   [-1.60, -0.98]  &  [-1.58, -0.74] &   [-0.78, -0.36] &   [-0.60, -0.22] \\ \hline 

\multirow{2}{6em}{\centering\large{$\delta$}}  & 0.44 $\pm$ 0.01 & 0.43 $\pm$ 0.01 &  0.45 $\pm$ 0.01  &  0.45 $\pm$ 0.02 & 0.41 $\pm$ 0.01 &  0.39 $\pm$ 0.01  \\  &  [0.42, 0.46]  &   [0.41, 0.45] &   [0.43, 0.47]  &  [0.41, 0.47] &   [0.39, 0.43] &   [0.37, 0.41] \\ \hline 
\end{tabular}
}

\vspace{0.8 cm}
\resizebox*{0.81\columnwidth}{0.18\textheight}{
%\rowcolors{1}{}{yellow!90!white!50!} 
\begin{tabular}{|c|c|c|c|c|c|c|c|}
\hline
\multicolumn{7}{|c|}{\textbf{\large{GALPROP parameters - Diffusion hypothesis}}} \\
\hline &\textbf{B/C} & \textbf{B/O} & \textbf{Be/C} & \textbf{Be/O} & \textbf{Li/C} & \textbf{Li/O} \\ 
\hline
%\cellcolor{white}
\multirow{2}{6em}{\centering$D_0/H$ $\left(\frac{10^{28} cm^{2}}{\units{s} \cdot \units{kpc}} \right)$} & 0.94 $\pm$ 0.04 & 0.94 $\pm$ 0.03   &  1.24 $\pm$ 0.02 &  1.08 $\pm$ 0.03 & 0.89 $\pm$ 0.04 &  0.91 $\pm$ 0.03  \\  &  [0.87, 1.04]  &   [0.88, 1.00] &   [1.21, 1.28]  &  [1.04, 1.15] &   [0.81, 0.97] &   [0.85, 0.96] \\ \hline 

\multirow{2}{6em}{\centering\large{$v_A$ ($\units{km/s}$)}}  & 24.44  $\pm$ 2.97 & 27.08 $\pm$ 1.91 &  4.74 $\pm$ 4.99  &  7.64 $\pm$ 7.23 &  29.11 $\pm$ 2.20 &  29.32 $\pm$ 1.52 \\  &  [18.90, 29.54]  &   [23.26, 30.86] &   [0., 14.72]  &  [0., 22.10] &   [24.71, 33.21] &   [26.28, 31.76] \\ \hline 

\multirow{2}{6em}{\centering\large{$\eta$}}    & -0.66 $\pm$ 0.12 & -0.39 $\pm$ 0.09 &  -1.68 $\pm$ 0.16 &  -1.27 $\pm$ 0.19 & -0.94 $\pm$ 0.20 &  -0.61 $\pm$ 0.11  \\ &  [-0.90, -0.18]  &   [-0.57, -0.23] &   [-1.98, -1.36]  &  [-1.65, -0.91] &   [-1.34, -0.62] &   [-0.83, -0.41] \\ \hline 
\multirow{2}{6em}{\centering\large{$\delta$}}  & 0.45 $\pm$ 0.01 & 0.44 $\pm$ 0.01 &  0.40 $\pm$ 0.01  &  0.42 $\pm$ 0.01 & 0.40 $\pm$ 0.01 &  0.39 $\pm$ 0.01  \\  &  [0.43, 0.47]  &   [0.42, 0.46] &   [0.38, 0.43]  &  [0.40, 0.44] &   [0.37, 0.42] &   [0.37, 0.41] \\ \hline 
\end{tabular}
}

\vspace{0.8 cm}
\resizebox*{0.81\columnwidth}{0.18\textheight}{
%\rowcolors{1}{}{yellow!90!white!50!} 
\begin{tabular}{|c|c|c|c|c|c|c|c|}
\hline
\multicolumn{7}{|c|}{\textbf{\large{FLUKA parameters - Diffusion hypothesis}}} \\
\hline &\textbf{B/C} & \textbf{B/O} & \textbf{Be/C} & \textbf{Be/O} & \textbf{Li/C} & \textbf{Li/O} \\ 
\hline
%\cellcolor{white}
\multirow{2}{6em}{\centering$D_0/H$ $\left(\frac{10^{28} cm^{2}}{\units{s} \cdot \units{kpc}} \right)$} & 0.52 $\pm$ 0.03 & 0.58 $\pm$ 0.03   &  0.72 $\pm$ 0.01 &  0.67 $\pm$ 0.04 & 0.80  $\pm$ 0.06 &  0.93 $\pm$ 0.05  \\  &  [0.46, 0.56]  &   [0.52, 0.54] &   [0.7, 0.75]  &  [0.63, 0.71] &   [0.73, 0.92] &   [0.84, 0.99] \\ \hline 

\multirow{2}{6em}{\centering\large{$v_A$ ($\units{km/s}$)}}  & 19.25  $\pm$ 2.94 & 27.03 $\pm$ 1.90 &  4.62 $\pm$ 5.66  &  11.75 $\pm$ 8.58 &  16.57 $\pm$ 9.59 &  36.78 $\pm$ 3.37 \\  &  [13.54, 25.19]  &   [23.48, 30.86] &   [0., 14.94]  &  [0., 28.43] &   [0., 33.79] &   [32.04, 40.64] \\ \hline 

\multirow{2}{6em}{\centering\large{$\eta$}}    & -0.99 $\pm$ 0.17 & -0.67 $\pm$ 0.15 &  -1.9 $\pm$ 0.19 &  -2.06 $\pm$ 0.33 & -1.89 $\pm$ 0.38 &  -0.95 $\pm$ 0.23  \\ &  [-1.23, -0.65]  &   [-0.97, -0.37] &   [-2.17, -1.51]  &  [-2.6, -1.4] &   [-2.43, -1.23] &   [-1.41, -0.74] \\ \hline 

\multirow{2}{6em}{\centering\large{$\delta$}}  & 0.43 $\pm$ 0.01 & 0.39 $\pm$ 0.01 &  0.42 $\pm$ 0.01  &  0.43 $\pm$ 0.02 & 0.39 $\pm$ 0.02 &  0.35 $\pm$ 0.01  \\  &  [0.39, 0.43]  &   [0.37, 0.42] &   [0.40, 0.44]  &  [0.39, 0.46] &  [0.36, 0.41] &   [0.33, 0.39] \\ \hline 
\end{tabular}
}
\caption{Same as in \ref{tab:PDF_single_Source} but for the diffusion coefficient parametrisation of eq. \ref{eq:breakhyp}.}
\label{tab:PDF_single_Diff}
\end{table}

\begin{table}[htb!]
\centering
\textbf{\Large{Combined analysis}}

\vspace{0.4cm}

%\small
\resizebox*{0.81\columnwidth}{0.17\textheight}{
%\rowcolors{1}{}{yellow!90!white!50!} 
\begin{tabular}{|c|c|c|c|c|c|c|c|}
\hline
 \multicolumn{1}{|c|} { } & \multicolumn{3}{c|}{\textbf{\large{Diffusion hypothesis}}} & \multicolumn{3}{c|}{\textbf{\large{Source hypothesis}}}  \\
\cline{2-7} &\textbf{DRAGON2} & \textbf{GALPROP} & \textbf{FLUKA} & \textbf{DRAGON2} & \textbf{GALPROP} & \textbf{FLUKA} \\ 
\hline
%\cellcolor{white}
\multirow{2}{6em}{\centering$D_0/H$ $\left(\frac{10^{28} cm^{2}}{\units{s} \cdot \units{kpc}} \right)$} & 1.11 $\pm$ 0.02 & 1.14 $\pm$ 0.03   &  0.74 $\pm$ 0.02 &  1.16 $\pm$ 0.02 & 1.26  $\pm$ 0.03 &  0.82 $\pm$ 0.02  \\  &  [1.06, 1.30]  &   [1.08, 1.30] &   [0.70, 0.79]  &  [1.11, 1.15] &   [1.21, 1.32] &   [0.79, 0.91] \\ \hline 

\multirow{2}{6em}{\centering\large{$v_A$ ($\units{km/s}$)}}  & 26.92  $\pm$ 1.28 & 33.32 $\pm$ 1.23 &  23.47 $\pm$ 2.91  &  30.20 $\pm$ 1.28 &  38.21 $\pm$ 1.11 &  31.03 $\pm$ 2.56 \\  &  [24.52, 29.48]  &   [31.03, 35.78] &   [19.99, 29.29]  &  [27.80, 32.17] &   [35.99, 40.01] &   [25.91, 34.16] \\ \hline 

\multirow{2}{6em}{\centering\large{$\eta$}}    & -0.75 $\pm$ 0.06 & -0.80 $\pm$ 0.07 &  -1.58 $\pm$ 0.17 &  -0.58 $\pm$ 0.06 & -0.44 $\pm$ 0.7 &  -1.07 $\pm$ 0.15  \\ &  [-0.87, -0.65]  &   [-0.92, -0.66] &   [-1.80, -1.24]  &  [-0.68, -0.46] &   [-0.58, -0.36] &   [-1.37, -0.87] \\ \hline 

\multirow{2}{6em}{\centering\large{$\delta$}}  & 0.42 $\pm$ 0.01 & 0.40 $\pm$ 0.01 &  0.40 $\pm$ 0.01  &  0.41 $\pm$ 0.01 & 0.37 $\pm$ 0.01 &  0.37 $\pm$ 0.01  \\  &  [0.39, 0.44]  &   [0.38, 0.42] &   [0.38, 0.42]  &  [0.38, 0.43] &   [0.35, 0.39] &   [0.35, 0.39] \\ \hline 
\end{tabular}
}
\caption{Results obtained for the combined species analysis for each cross sections sets and diffusion coefficient parametrisation shown as in the former tables.}
\label{tab:PDF_Combined}
\end{table}

\begin{table}
\centering

\textbf{\Large{Independent scaled analyses - Source hypothesis}}

\vspace{0.4cm}

\resizebox*{0.81\columnwidth}{0.17\textheight}{
%\rowcolors{1}{}{yellow!90!white!50!} 
\begin{tabular}{|c|c|c|c|c|c|c|c|}
\hline
\multicolumn{7}{|c|}{\textbf{\large{DRAGON2 parameters - Source hypothesis}}} \\
\hline &\textbf{B/C} & \textbf{B/O} & \textbf{Be/C} & \textbf{Be/O} & \textbf{Li/C} & \textbf{Li/O} \\ 
\hline
%\cellcolor{white}
\multirow{2}{6em}{\centering$D_0/H$ $\left(\frac{10^{28} cm^{2}}{\units{s} \cdot \units{kpc}} \right)$} & 1.09 $\pm$ 0.05 & 1.09$\pm$ 0.05   &  1.06 $\pm$ 0.02 &  1.04 $\pm$ 0.06 & 1.30  $\pm$ 0.05 &  1.28 $\pm$ 0.05  \\  &  [1.0, 1.19]  &   [1.01, 1.19] &   [1.01, 1.16]  &  [0.94, 1.16] &   [1.20, 1.40] &   [1.18, 1.35] \\ \hline 

\multirow{2}{6em}{\centering\large{$v_A$ ($\units{km/s}$)}}  & 26.58  $\pm$ 3.54 & 30.72 $\pm$ 2.78 &  11.80 $\pm$ 7.05  &  21.06 $\pm$ 6.06 &  35.81 $\pm$ 2.55 &  37.63 $\pm$ 1.82 \\  &  [18.5, 33.16]  &   [25.16, 35.58] &   [0., 25.48]  &  [8.96, 30.15] &   [30.71, 40.01] &   [32.99, 40.50] \\ \hline 

\multirow{2}{6em}{\centering\large{$\eta$}}    & -0.71 $\pm$ 0.12 & -0.55 $\pm$ 0.11 &  -1.17 $\pm$ 0.18 &  -0.94 $\pm$ 0.18 & -0.32 $\pm$ 0.11 &  -0.21 $\pm$ 0.09  \\ &  [-0.95, -0.57]  &   [-0.77, -0.35] &   [-1.30, -0.58]  &  [-2.58, -1.67] &   [-0.54, -0.10] &   [-0.41, -0.03] \\ \hline 

\multirow{2}{6em}{\centering\large{$\delta$}}  & 0.42 $\pm$ 0.01 & 0.41 $\pm$ 0.01 &  0.44 $\pm$ 0.01  &  0.43 $\pm$ 0.02 & 0.39 $\pm$ 0.01 &  0.38 $\pm$ 0.01  \\  &  [0.40, 0.44]  &   [0.39, 0.43] &   [0.41, 0.45]  &  [0.39, 0.45] &   [0.37, 0.41] &   [0.36, 0.41] \\ \hline 
\end{tabular}
}

\vspace{0.8 cm}
\resizebox*{0.81\columnwidth}{0.17\textheight}{
%\rowcolors{1}{}{yellow!90!white!50!} 
\begin{tabular}{|c|c|c|c|c|c|c|c|}
\hline
\multicolumn{7}{|c|}{\textbf{\large{GALPROP parameters - Source hypothesis}}} \\
\hline &\textbf{B/C} & \textbf{B/O} & \textbf{Be/C} & \textbf{Be/O} & \textbf{Li/C} & \textbf{Li/O} \\ 
\hline
%\cellcolor{white}
\multirow{2}{6em}{\centering$D_0/H$ $\left(\frac{10^{28} cm^{2}}{\units{s} \cdot \units{kpc}} \right)$} & 0.89 $\pm$ 0.04 & 0.93$\pm$ 0.04   &  0.90 $\pm$ 0.02 &  0.93 $\pm$ 0.03 & 1.14  $\pm$ 0.04 &  1.11 $\pm$ 0.03  \\  &  [0.81, 0.98]  &   [0.86, 0.94] &   [1.01, 1.08]  &  [0.87, 1.01] &   [1.06, 1.22] &   [1.05, 1.17] \\ \hline 

\multirow{2}{6em}{\centering\large{$v_A$ ($\units{km/s}$)}}  & 25.56  $\pm$ 2.80 & 27.09 $\pm$ 2.16 &  5.50 $\pm$ 5.33  &  6.06 $\pm$ 6.28 &  35.61 $\pm$ 1.71 &  34.94 $\pm$ 1.30 \\  &  [19.96, 30.70.]  &   [22.93, 31.41] &   [0., 16.16]  &  [0, 18.72] &   [32.19, 37.17] &   [32.34, 37.0] \\ \hline 

\multirow{2}{6em}{\centering\large{$\eta$}}    & -0.90 $\pm$ 0.14 & -0.60 $\pm$ 0.11 &  -2.69 $\pm$ 0.20 &  -2.17 $\pm$ 0.25 & -0.17 $\pm$ 0.11 &  -0.10 $\pm$ 0.08  \\ &  [-1.17, -0.62]  &   [-0.82, -0.38] &   [-2.99, -2.29]  &  [-2.58, -1.67] &   [-0.39, +0.05] &   [-0.26, +0.04] \\ \hline 

\multirow{2}{6em}{\centering\large{$\delta$}}  & 0.44 $\pm$ 0.01 & 0.43 $\pm$ 0.01 &  0.39 $\pm$ 0.01  &  0.42 $\pm$ 0.02 & 0.38 $\pm$ 0.01 &  0.38 $\pm$ 0.01  \\  &  [0.41, 0.46]  &   [0.41, 0.45] &   [0.37, 0.41]  &  [0.40, 0.44] &   [0.36, 0.41] &   [0.36, 0.41] \\ \hline 
\end{tabular}
}

\vspace{0.8 cm}
\resizebox*{0.81\columnwidth}{0.17\textheight}{
%\rowcolors{1}{}{yellow!90!white!50!} 
\begin{tabular}{|c|c|c|c|c|c|c|c|}
\hline
\multicolumn{7}{|c|}{\textbf{\large{FLUKA parameters - Source hypothesis}}} \\
\hline &\textbf{B/C} & \textbf{B/O} & \textbf{Be/C} & \textbf{Be/O} & \textbf{Li/C} & \textbf{Li/O} \\ 
\hline
%\cellcolor{white}
\multirow{2}{6em}{\centering$D_0/H$ $\left(\frac{10^{28} cm^{2}}{\units{s} \cdot \units{kpc}} \right)$} & 0.63 $\pm$ 0.03 & 0.73 $\pm$ 0.04   &  0.65 $\pm$ 0.02 &  0.65 $\pm$ 0.05 & 0.84  $\pm$ 0.05 &  0.83 $\pm$ 0.03  \\  &  [0.58, 0.69]  &   [0.65, 0.80] &   [0.61, 0.68]  &  [0.57, 0.75] &   [0.76, 0.93] &   [0.78, 0.86] \\ \hline 

\multirow{2}{6em}{\centering\large{$v_A$ ($\units{km/s}$)}}  & 18.98  $\pm$ 4.06 & 31.10 $\pm$ 2.87 &  5.52 $\pm$ 5.94  &  18.59 $\pm$ 7.65 &  32.52 $\pm$ 3.53 &  36.41 $\pm$ 1.88 \\  &  [10.86, 26.60]  &   [25.44, 35.86] &   [0., 17.40]  &  [3.29, 31.71] &   [25.46, 39.24] &   [32.65, 38.51] \\ \hline 

\multirow{2}{6em}{\centering\large{$\eta$}}    & -1.35 $\pm$ 0.18 & -0.64 $\pm$ 0.16 &  -2.13 $\pm$ 0.20 &  -1.58 $\pm$ 0.36 & -1.07 $\pm$ 0.25 &  -0.59 $\pm$ 0.15  \\ &  [-1.71, -0.99]  &   [-0.96, -0.34] &   [-2.45, -1.93]  &  [-2.14, -0.86] &   [1.57, -0.61] &   [-0.9, -0.43] \\ \hline 

\multirow{2}{6em}{\centering\large{$\delta$}}  & 0.42 $\pm$ 0.01 & 0.37 $\pm$ 0.01 &  0.42 $\pm$ 0.01  &  0.42 $\pm$ 0.02 & 0.35 $\pm$ 0.01 &  0.34 $\pm$ 0.01  \\  &  [0.39, 0.43]  &   [0.35, 0.40] &   [0.40, 0.44]  &  [0.38, 0.45] &   [0.33, 0.38] &   [0.32, 0.36] \\ \hline 
\end{tabular}
}
\caption{Results obtained in the independent analysis of the flux ratios B/C, Be/C, Li/C, B/O, Be/O and Li/O for the diffusion coefficient parametrisation of eq. \ref{eq:sourcehyp} with the scaled cross sections sets analysed. }
\label{tab:PDF_single_scaled_Source}
\end{table}

\begin{table}[htb!]
\centering
%\small
\textbf{\Large{Independent scaled analyses - Diffusion hypothesis}}

\vspace{0.4cm}

\resizebox*{0.81\columnwidth}{0.18\textheight}{
%\rowcolors{1}{}{yellow!90!white!50!} 
\begin{tabular}{|c|c|c|c|c|c|c|c|}
\hline
\multicolumn{7}{|c|}{\textbf{\large{DRAGON2 parameters - Diffusion hypothesis}}} \\
\hline &\textbf{B/C} & \textbf{B/O} & \textbf{Be/C} & \textbf{Be/O} & \textbf{Li/C} & \textbf{Li/O} \\ 
\hline
%\cellcolor{white}
\multirow{2}{6em}{\centering$D_0/H$ $\left(\frac{10^{28} cm^{2}}{\units{s} \cdot \units{kpc}} \right)$} & 1.03 $\pm$ 0.04 & 1.03$\pm$ 0.04   &  1.04 $\pm$ 0.03 &  0.98 $\pm$ 0.05 & 1.22  $\pm$ 0.04 &  1.22 $\pm$ 0.04  \\  &  [0.95, 1.11]  &   [0.95, 1.11]  &   [1.0, 1.11]  &  [0.92, 1.09] &   [1.14, 1.30] &   [1.14, 1.30] \\ \hline 

\multirow{2}{6em}{\centering\large{$v_A$ ($\units{km/s}$)}}  & 22.20  $\pm$ 4.29 & 27.25 $\pm$ 2.76 &  8.60 $\pm$ 6.77  &  13.84 $\pm$ 8.22 &  32.61 $\pm$ 2.66 &  35.36 $\pm$ 2.12 \\  &  [13.62, 27.32]  &   [21.73, 32.47] &   [0., 22.14]  &  [0, 29.27] &   [27.29, 37.83] &   [31.08, 38.56] \\ \hline 

\multirow{2}{6em}{\centering\large{$\eta$}}    & -0.84 $\pm$ 0.13 & -0.69 $\pm$ 0.10 &  -1.29 $\pm$ 0.16 &  -1.16 $\pm$ 0.20 & -0.48 $\pm$ 0.13 &  -0.37 $\pm$ 0.09  \\ &  [-1.10, -0.6]  &   [-0.89, -0.49] &   [-1.61, -0.97]  &  [-1.36, -0.78] &   [-0.74, -0.26] &   [-0.75, -0.19] \\ \hline 

\multirow{2}{6em}{\centering\large{$\delta$}}  & 0.44 $\pm$ 0.01 & 0.43 $\pm$ 0.01 &  0.45 $\pm$ 0.01  &  0.45 $\pm$ 0.02 & 0.40 $\pm$ 0.01 &  0.39 $\pm$ 0.01  \\  &  [0.42, 0.46]  &   [0.41, 0.45] &   [0.43, 0.47]  &  [0.42, 0.47] &   [0.39, 0.42] &   [0.36, 0.41] \\ \hline 
\end{tabular}
}

\vspace{0.8 cm}
\resizebox*{0.81\columnwidth}{0.18\textheight}{
%\rowcolors{1}{}{yellow!90!white!50!} 
\begin{tabular}{|c|c|c|c|c|c|c|c|}
\hline
\multicolumn{7}{|c|}{\textbf{\large{GALPROP parameters - Diffusion hypothesis}}} \\
\hline &\textbf{B/C} & \textbf{B/O} & \textbf{Be/C} & \textbf{Be/O} & \textbf{Li/C} & \textbf{Li/O} \\ 
\hline
%\cellcolor{white}
\multirow{2}{6em}{\centering$D_0/H$ $\left(\frac{10^{28} cm^{2}}{\units{s} \cdot \units{kpc}} \right)$} & 0.83 $\pm$ 0.04 & 0.85 $\pm$ 0.04   &  1.02 $\pm$ 0.03 &  0.92 $\pm$ 0.03 & 1.08 $\pm$ 0.03 &  1.07 $\pm$ 0.03  \\  &  [0.78, 0.92]  &   [0.79, 0.92] &   [0.98, 1.07]  &  [0.88, 0.96] &   [1.02, 1.14] &   [1.01, 1.12] \\ \hline 

\multirow{2}{6em}{\centering\large{$v_A$ ($\units{km/s}$)}}  & 21.95  $\pm$ 3.14 & 24.82 $\pm$ 2.31 &  5.14 $\pm$ 4.83  &  4.91 $\pm$ 5.46 &  33.77 $\pm$ 1.85 &  33.67 $\pm$ 1.39 \\  &  [15.75, 28.23]  &   [19.20, 29.04] &   [0., 14.80]  &  [0., 15.83] &   [30.07, 36.73] &   [30.89, 36.15] \\ \hline 

\multirow{2}{6em}{\centering\large{$\eta$}}    & -1.09 $\pm$ 0.16 & -0.73 $\pm$ 0.11 &  -2.70 $\pm$ 0.17 &  -2.31 $\pm$ 0.21 & -0.32 $\pm$ 0.10 &  -0.21 $\pm$ 0.07  \\ &  [-1.44, -0.79]  &   [-0.57, -0.23] &   [-2.96, -2.36]  &  [-2.69, -1.90] &   [-0.52, -0.12] &   [-0.35, -0.07] \\ \hline 

\multirow{2}{6em}{\centering\large{$\delta$}}  & 0.46 $\pm$ 0.01 & 0.44 $\pm$ 0.01 &  0.40 $\pm$ 0.01  &  0.42 $\pm$ 0.01 & 0.40 $\pm$ 0.01 &  0.39 $\pm$ 0.01  \\  &  [0.44, 0.48]  &   [0.42, 0.47] &   [0.38, 0.41]  &  [0.40, 0.43] &   [0.38, 0.42] &   [0.37, 0.40] \\ \hline 
\end{tabular}
}

\vspace{0.8 cm}
\resizebox*{0.81\columnwidth}{0.18\textheight}{
%\rowcolors{1}{}{yellow!90!white!50!} 
\begin{tabular}{|c|c|c|c|c|c|c|c|}
\hline
\multicolumn{7}{|c|}{\textbf{\large{FLUKA parameters - Diffusion hypothesis}}} \\
\hline &\textbf{B/C} & \textbf{B/O} & \textbf{Be/C} & \textbf{Be/O} & \textbf{Li/C} & \textbf{Li/O} \\ 
\hline
%\cellcolor{white}
\multirow{2}{6em}{\centering$D_0/H$ $\left(\frac{10^{28} cm^{2}}{\units{s} \cdot \units{kpc}} \right)$} & 0.59 $\pm$ 0.03 & 0.68 $\pm$ 0.03   &  0.64 $\pm$ 0.01 &  0.61 $\pm$ 0.04 & 0.74  $\pm$ 0.07 &  0.83 $\pm$ 0.04  \\  &  [0.55, 0.66]  &   [0.62, 0.74] &   [0.63, 0.65]  &  [0.57, 0.69] &   [0.64, 0.87] &   [0.75, 0.81] \\ \hline 

\multirow{2}{6em}{\centering\large{$v_A$ ($\units{km/s}$)}}  & 12.72  $\pm$ 5.88 & 27.65 $\pm$ 2.65 &  4.81 $\pm$ 5.13  &  11.94 $\pm$ 7.52 &  25.22 $\pm$ 7.08 &  35.14 $\pm$ 2.90 \\  &  [0.96, 23.16]  &   [22.35, 32.75] &   [0., 15.07]  &  [0., 26.98] &   [11.06, 37.50] &   [29.34, 38.86] \\ \hline 

\multirow{2}{6em}{\centering\large{$\eta$}}    & -1.62 $\pm$ 0.19 & -0.87 $\pm$ 0.15 &  -2.21 $\pm$ 0.19 &  -1.92 $\pm$ 0.33 & -1.67 $\pm$ 0.44 &  -0.75 $\pm$ 0.20  \\ &  [-2.0, -1.24]  &   [-1.27, -0.59] &   [-2.53, -1.83]  &  [-2.46, -1.26] &   [-2.55, -0.81] &   [-1.15, -0.43] \\ \hline 

\multirow{2}{6em}{\centering\large{$\delta$}}  & 0.44 $\pm$ 0.01 & 0.40 $\pm$ 0.01 &  0.43 $\pm$ 0.01  &  0.43 $\pm$ 0.02 & 0.38 $\pm$ 0.02 &  0.35 $\pm$ 0.01  \\  &  [0.41, 0.45]  &   [0.37, 0.42] &   [0.41, 0.44]  &  [0.38, 0.45] &   [0.34, 0.42] &   [0.33, 0.47] \\ \hline 
\end{tabular}
}
\caption{Same as in \ref{tab:PDF_single_scaled_Source} but for the diffusion coefficient parametrisation of eq. \ref{eq:breakhyp}.}
\label{tab:PDF_scaled_Diff}
\end{table}

\end{appendices}

%=====================================================================
% BIBLIOGRAPHY
\newpage
\setlength{\parskip}{5mm}
\titlespacing{\chapter}{0cm}{0mm}{0mm}
\titleformat{\chapter}[display]
  {\normalfont\huge\bfseries}
  {\chaptertitlename\ \thechapter}{20pt}{\Huge}

\bibliographystyle{unsrt}  

\listoftables
\listoffigures

% Add the bib file
%\addcontentsline{toc}{chapter*}{\bibname} %This must be commented for the pdf copy!

\bibliography{References/mybibfile}

%=====================================================================
% PUBLICATIONS
%  publications if any may be listed after the literature cited.
%\addcontentsline{toc}{chapter}{List of Publications}
%\include{chap_07_List_of_Publications}

%=====================================================================
% ACKNOWLEDGMENTS
%   This is the last item in the thesis. It should be signed by
%   author, with date.

\newpage
\setlength{\parskip}{3mm}
\titlespacing{\chapter}{0cm}{0mm}{0mm}
\titleformat{\chapter}[display]
  {\normalfont\huge\bfseries \centering}
  {\chaptertitlename\ \thechapter}{20pt}{\Huge}

\chapter*{Acknowledgments}
\label{ch:Acknowledgments}
\addcontentsline{toc}{chapter}{\nameref{ch:Acknowledgments}}
\vspace{10mm}
This work means to me something else than the culmination of my career as student or the beginning of my walk as researcher. This is the end of what, undoubtedly, has been my best life stage, where I have had the opportunity of learning, enjoying and discovering the wonders of this world and which will steer the path to be followed in my future.

Along this phase I have met lots of new people, people who have taught me to be different and evolve, to overcome my prejudices and to remove barriers, from whom I feel very happy to have together with me or simply to passingly have got to know, since, if I have something clear is that those moments, where friendship and joy make me full of happiness, will not last forever and have made me become what I am today.

Certainly, I can not avoid implicitly mentioning my family, since they have played the most important role in my evolution and have represented the pillars where to lean on. I know I will never have time enough to thank them and give back everything they made for me, but they will never exit from this privileged place in my soul. Thanks mum for continuously repeat me what to do, thanks Ale for the inspiration I need, thanks Ángel to be such roller-model and thanks to all my uncles/aunts, cousins, grandparents for all the moments shared.
Here I would like to dedicate you a short line, grandpa: I have you so present everywhere... I love you so much.

But the most special person, who has really been my friend, relative and everything I have needed, is my beloved Yoli. What to say about the nights spent on Skype, the surprises in the most unexpected moments, the adventures lived together... simply your presence makes me feel so full and happy.

Of course, I can not overlook that other family which was together with me along my way: My friends. Guys, thank you so much for those moments allowing me to forget all possible problems and please, never stop being so wonderful.

\newpage
Finally, I wish to record a deep sense of gratitude to my supervisors and colleagues in Bari, for their valuable guidance and constant support at all stages of my Ph.D study and related research. They have absolutely taught me what is making science.

So, for all those moments, the most sincere thanks to all those who stayed on my side.

\end{document}